\magnification=\magstep1
\hsize=16truecm
\vsize=24truecm
\parindent=0pt
\nopagenumbers
\mathsurround=2pt
\topskip=20pt
\lineskiplimit=1pt
\lineskip=2pt


\catcode`\@=11
\font\tenmsx=msam10
\font\sevenmsx=msam7
\font\fivemsx=msam5
\font\tenmsy=msbm10
\font\sevenmsy=msbm7
\font\fivemsy=msbm5
\newfam\msxfam
\newfam\msyfam
\textfont\msxfam=\tenmsx  \scriptfont\msxfam=\sevenmsx
  \scriptscriptfont\msxfam=\fivemsx
\textfont\msyfam=\tenmsy  \scriptfont\msyfam=\sevenmsy
  \scriptscriptfont\msyfam=\fivemsy

\def\hexnumber@#1{\ifnum#1<10 \number#1\else
 \ifnum#1=10 A\else\ifnum#1=11 B\else\ifnum#1=12 C\else
 \ifnum#1=13 D\else\ifnum#1=14 E\else\ifnum#1=15 F\fi\fi\fi\fi\fi\fi\fi}

\def\msx@{\hexnumber@\msxfam}
\def\msy@{\hexnumber@\msyfam}

\mathchardef\varnothing="0\msy@3F

\def\Bbb{\ifmmode\let\next\Bbb@\else
  \def\next{\errmessage{Use \string\Bbb\space only in math mode}}\fi\next}
\def\Bbb@#1{{\Bbb@@{#1}}}
\def\Bbb@@#1{\fam\msyfam#1}

\catcode`\@=\active



\font\fourteenbf=cmbx12 scaled \magstep1
\font\twelvebf=cmbx12

\font\smalltt=cmtt8
\font\sansser=cmss10
\font\smallsansser=cmss8

\font\tentt=cmtt10
\font\eighttt=cmtt8

\newfam\ttfam
\textfont\ttfam=\tentt  \scriptfont\ttfam=\eighttt
  \scriptscriptfont\ttfam=\eighttt
  
\def\ftt#1{\leavevmode\hbox{\tt #1}}
\def\fstt#1{\leavevmode\hbox{\smalltt #1}}
\def\frm#1{\enskip \leavevmode\hbox{\rm #1} \enskip}
\def\fsss#1{\leavevmode\hbox{\smallsansser #1}}
\def\fss#1{\leavevmode\hbox{\sansser #1}}



\def\index#1#2{\write\IndexFile{\string\indexitem{#1} {#2} {\folio}}}

\def\indexdef#1#2#3{\index{#2}{#3}{\it #1\/}}
\def\indexddef#1#2#3#4#5{\index{#2}{#3}\index{#4}{#5}{\it #1\/}}

\def\hfline#1#2{
       \count255=#1
       \pageno=\count255
       \headline={\tenrm \ifnum\pageno=\count255\hss
                         \else #2 \hss \folio \fi}
       \footline={\tenrm \ifnum\pageno=\count255\hss \folio 
                         \else \hss\fi}
               }

\def\tsym#1{{\mathsurround=0pt $\hskip 1pt #1$}}

\def\mdisp#1{\par \nobreak \medskip \nobreak
             \vtop{\line{\hskip\parindent \hss $#1$ \hss}}
             \nobreak \medskip \nobreak}

\def\ldisp#1{\par \nobreak \medskip \nobreak
             \line{\hskip\parindent \hskip 5pt $#1$ \hss}
             \nobreak \medskip \nobreak}
             
\def\rdisp#1{\par \nobreak \medskip \nobreak
             \line{\hskip\parindent \hss $#1$ \hskip 5pt}
             \nobreak \medskip \nobreak}
                    
\def\frame#1#2{\vtop{\hrule
	              \hbox{\vrule\vbox{{\leftskip=#1
	                                 \rightskip=#1 #2}}\vrule}%
                      \hrule}}

\def\sectionhead #1 #2 {{\twelvebf #1 \enspace #2}}
\def\definition#1{{\it #1\/}}

\def\proclaim  #1 #2\endpro{\bigbreak
   \noindent{\bf#1 \enspace}{\sl#2}\par\bigbreak}
\def\proof{{\it Proof \enspace}}
   
\def\eop{\ \vbox{\hrule
                       \hbox{\vrule
                             \hskip 6pt
                             \vrule height 6pt width 0pt
                             \vrule}%
                       \hrule}%
                }



\def\Bool{{\Bbb B}}
\def\Nat{{\Bbb N}}
\def\Int{{\Bbb Z}}

\def\lcurl{\{\,}
\def\rcurl{\,\}}

\def\oneto #1 #2 {#1 = 1,\, \ldots ,\, #2}

\def\nullto #1 #2 {#1 = 0,\, \ldots ,\, #2}

\def\vector #1 #2 {#1_1, \ldots , #1_{#2}}

\def\lvector #1 #2 {#1_1,\, \ldots , \, #1_{#2}}

\def\list #1 #2 {#1_1\ \cdots\ #1_{#2}}

\def\inflist #1 {#1_1\ #1_2\ \cdots}



\def\sem#1{\lsem #1 \rsem}
\def\lsem{\, \vcenter{\vbox{\hrule
                \hbox{\vrule
                   \hskip 1pt
                      \vrule
                      \hskip 2pt
                      \vrule height 9pt width 0pt}
                \hrule}}%
                }

\def\rsem{\vcenter{\vbox{\hrule
                \hbox{\vrule height 9pt width 0pt
                   \hskip 2pt
                      \vrule
                      \hskip 1pt
                      \vrule} 
                \hrule}} \,
                }

\def\Oneptset{{\Bbb I}}
\def\onept{\varepsilon}

\def\SynInt{{\underline {Int}}}

\def\lcb{\char`\{\thinspace}
\def\rcb{\thinspace\char`\}}

\def\sqle{\sqsubseteq}

\def\lub{\bigsqcup}
\def\directed#1{\fss{d}(#1)}
\def\Bot#1{\fss{bot}\left(#1\right)}
\def\dom{\mathop{\rm dom}}
\def\bdom{\mathop{\bf dom}}

\def\Im{\mathop{\rm Im}}
\def\Hom{\mathop{\rm Hom}}

\def\ass #1 #2 {{#2}_{\scriptscriptstyle \diamond}^{#1}} 
\def\assb #1 #2 {(#2_{\scriptscriptstyle \diamond})^{#1}} 
\def\totale #1 #2 {\langle #1 \nobreak \to \nobreak #2 \rangle}
\def\total #1 #2 {#2^{#1}}
\def\mono #1 #2 {( #1 \nobreak \to \nobreak #2 )}
\def\cont #1 #2 {[ #1 \nobreak \to \nobreak #2 ]}

\def\kterm#1{{\scriptstyle K}\hskip -1pt#1}
\def\aterm#1{{\scriptstyle A}#1}
\def\bterm#1{{\scriptstyle H}\hskip -0.5pt#1}

\def\Bf{B_{\fsss{f}}}
\def\fo{{\odot}}


\def\blob{\hskip 1pt \vbox{\hrule
                       \hbox{\vrule
                             \hskip 6pt
                             \vrule height 4pt width 0pt
                             \vrule}%
                       \hrule}%
           \hskip 1pt }


\font\bigbf=cmbx12 scaled \magstep2
\font\vbigbf=cmbx12 scaled \magstep3
\vbox{
\bigskip\bigskip\bigskip\bigskip\bigskip\bigskip\bigskip\bigskip
{\vbigbf Computing with Equations}
\bigskip\bigskip\bigskip\bigskip\bigskip\bigskip\bigskip\bigskip
{\bigbf Chris Preston}
\bigskip
{\twelvebf Fakult\"at f\"ur Mathematik}
\medskip
{\twelvebf Universit\"at Bielefeld}}

\vfill\eject

\hfline{-1}{CONTENTS}

\def\sline{\hbox to 339.37pt}
\def\lline{\hbox to 380pt}

{\fourteenbf Contents}
\bigskip\bigskip
{\parindent=20pt
\medskip
\lline{{\bf Preface} \hss }
\bigskip\medskip
\lline{{\bf 1.\enskip  Introduction} \hss 1}
\medskip
{\leftskip=20pt
\item{1.1} \sline {\enskip Computing with equations: Some examples \hss 1}
\smallskip
\item{1.2} \sline{\enskip A procedure for interpreting equations \hss 11}
\smallskip
\item{1.3} \sline {\enskip Notes \hss 17}
\bigskip
}

\bigskip
\lline{{\bf 2.\enskip  Some universal algebra} \hss 18}
\medskip
{\leftskip=20pt
\item{2.1} \sline{\enskip Typed sets \hss 20}
\smallskip
\item{2.2} \sline{\enskip Algebras and homomorphisms \hss 25}
\smallskip
\item{2.3} \sline{\enskip Initial and free algebras \hss 33}
\smallskip
\item{2.4} \sline{\enskip Term algebras \hss 40} 
\smallskip
\item{2.5} \sline{\enskip Extensions of signatures and algebras \hss 46}
\smallskip
\item{2.6} \sline{\enskip Tree algebras \hss 49} 
\smallskip
\item{2.7} \sline{\enskip Notes \hss 54} 
\bigskip
}

\bigskip
\lline{{\bf 3.\enskip  Data objects} \hss 55}
\medskip
{\leftskip=20pt
\item{3.1} \sline{\enskip The basic data objects and the ground term algebra \hss 55}
\smallskip
\item{3.2} \sline{\enskip Bottomed algebras and bottomed extensions \hss 57}
\smallskip
\item{3.3} \sline{\enskip Regular bottomed algebras and the trace \hss 64}
\smallskip
\item{3.4} \sline{\enskip Cored bottomed algebras \hss 70}
\smallskip
\item{3.5} \sline{\enskip Monotone bottomed algebras \hss 75}
\smallskip
\item{3.6} \sline{\enskip Notes \hss 80}
\bigskip
}

\bigskip
\lline{{\bf 4.\enskip  Completion of bottomed extensions} \hss 81}
\medskip
{\leftskip=20pt
\item{4.1} \sline{\enskip Associated orderings for regular extensions \hss 83}
\smallskip
\item{4.2} \sline{\enskip Complete partially ordered sets \hss 89}
\smallskip
\item{4.3} \sline{\enskip Initial completions \hss 96}
\smallskip
\item{4.4} \sline{\enskip Initial completion of monotone regular extensions  \hss 104}
\smallskip
\item{4.5} \sline{\enskip Notes  \hss 108}
\bigskip
}

}
\bigskip

\vfill\eject

{\parindent=20pt
\bigskip
\lline{{\bf 5.\enskip  Functions} \hss 109}
\medskip
{\leftskip=20pt
\item{5.1} \sline{\enskip Concrete cartesian closed categories \hss 111}
\smallskip
\item{5.2} \sline{\enskip Functional types and functional algebras \hss 123}
\smallskip
\item{5.3} \sline{\enskip Primitive functions \hss 132}
\smallskip
\item{5.4} \sline{\enskip The basic set-up \hss 139}
\smallskip
\item{5.5} \sline{\enskip Notes \hss 144}
\bigskip
}
\bigskip
\lline{{\bf 6.\enskip  Functionally free term algebras} \hss 145}
\medskip
{\leftskip=20pt
\item{6.1} \sline{\enskip Functional and functionally free algebras \hss 147}
\smallskip
\item{6.2} \sline{\enskip The basic term algebras \hss 156}
\smallskip
\item{6.3} \sline{\enskip Support systems and the trace \hss 164}
\smallskip
\item{6.4} \sline{\enskip Notes \hss 168}
\bigskip
}
\bigskip
\lline{{\bf 7.\enskip  Equations} \hss 169}
\medskip
{\leftskip=20pt
\item{7.1} \sline{\enskip Equations and their solutions \hss 171}
\smallskip
\item{7.2} \sline{\enskip Computing values\hss 179}
\smallskip
\item{7.3} \sline{\enskip An algorithm for computing values\hss 184}
\smallskip
\item{7.4} \sline{\enskip Replacement rules\hss 200}
\smallskip
\item{7.5} \sline{\enskip Notes\hss 211}
\bigskip
}

\bigskip
\lline{{\bf 8.\enskip  Completeness of the algorithm} \hss 212}
\medskip
{\leftskip=20pt
\item{8.1} \sline{\enskip Solutions as fixed-points \hss 212}
\smallskip
\item{8.2} \sline{\enskip The Kleene sequence \hss 215}
\smallskip
\item{8.3} \sline{\enskip Logical relations and functionally free algebras\hss 221}
\smallskip
\item{8.4} \sline{\enskip An application of the method of logical relations \hss 224}
\smallskip
\item{8.5} \sline{\enskip Notes \hss 228}
\bigskip
}

\bigskip
\lline{{\bf Bibliography} \hss 229}
\bigskip
\lline{{\bf Index} \hss 232}
\bigskip
}

\vfill\eject

\hfline{-3}{PREFACE}

{\fourteenbf Preface}
\bigskip\bigskip
About ten years ago the Mathematics Department at which I am employed was 
forced to get involved in teaching a first year course in Computer Science.
One half of the course was based on the book
{\it Introduction to Functional Programming}\footnote{\tsym{^1}}
{Prentice Hall, Hemel Hempstead (1988)} by Richard Bird and Philip Wadler
and used David Turner's {\it Miranda}\footnote{\tsym{^2}}
{{\it Miranda} is a trademark of Research Software Limited.}
as programming language. 
I taught this course several times, the first time at least with
an appalling lack of knowledge of what Computer Science is all about.
The present study has grown out of what I have learnt from this undertaking.
Its intention is to give a mathematical account of how I believe
students could be taught to think about functional programming languages
and to explain how such languages work. 
\medskip
For those who understand something of the terminology used to describe 
the semantics of programming languages, this intention can be stated more 
precisely as follows: I present a certain denotational semantics for 
a rudimentary functional programming language and show that this is 
equivalent to the operational semantics defined in terms of an explicit 
algorithm. 
\medskip
This account is certainly not meant as a text for the kind of course 
from which it arose, and in fact it has nothing to say about the art of
functional programming. It endeavours rather to provide a mathematical basis for
such a course, and is aimed at those who would regard such a basis as being important, 
i.e, at an audience of mathematicians.
\medskip
The kind of functional programming languages to be dealt with are those
like {\it Miranda\/}, {\it Haskell\/} and {\it ML\/}, in which objects, 
usually functions, are defined by equations. The meaning of such 
equations will be given directly without resorting to the lambda calculus
(and so the study is not really suitable for dealing with a language such as {\it Lisp\/}).
In fact, this avoidance of the lambda calculus should not pass without comment.
It seems that almost all the accounts offering a mathematical explanation of
functional programming are written either by computer scientists 
or logicians. They are therefore written in a language foreign 
to the very large majority of `ordinary' mathematicians. 
(Just as most arts graduates are proud of never having understood
any mathematics, so it seems that most mathematicians are just as proud of 
the fact that they do not really know what logic is all about.) 
I hope that the present study is at least written in a language with which this 
group of logically challenged people is familiar.
\medskip
The mathematical framework which will be employed is to some extent
determined by the languages the study seeks to explain.
For example, data types in {\it Miranda\/}, {\it Haskell\/} and {\it ML\/} 
are defined by specifying a signature, and the equations in these languages 
are built up out of terms taken from corresponding term algebras.
This means that some elementary concepts from universal algebra
can hardly be avoided. Moreover, one is then more-or-less
forced to use the fact that term algebras are initial as a basis for giving
equations a meaning. This can be seen as an instance of 
the `initial algebra semantics' 
philosophy propagated by the ADJ group\footnote{\tsym{^3}}
{consisting of J.A.\ Goguen, J.W.\ Thatcher, E.G.\ Wagner 
and J.B.\ Wright, for example in {\it Initial algebra semantics 
and continuous algebras}, Journal ACM, 24, 68--95, 1977}.
\medskip
There are two topics which are not treated here, and whose
omission should at least be commented upon. The first concerns
local definitions. These have not been considered in order to keep the 
presentation as simple as possible, but in principle their inclusion
would not cause any difficulty and requires no additional mathematical
concepts. The second topic is that of polymorphism.
It could be claimed that this has also been omitted to simplify the 
presentation. However, its inclusion would require a new, and much
less elementary, mathematical framework. 
\medskip
As already indicated, the main result presented here is a statement 
about the equivalence of operational and denotational semantics. 
The proof of the hard part of this result (the completeness of the
operational semantics) has been strongly influenced by the proofs 
of similar statements in Glynn Winskel's book 
{\it The Formal Semantics of Programming Languages}\footnote{\tsym{^4}}
{The MIT Press, Cambridge, Massachusetts (1993)}. I highly recommend
Winskel's book.
\medskip
I have tried to keep the account self-contained, and have thus included 
all the standard results -- together with their proofs --
which are needed from universal algebra, category theory
(which is only used in a very rudimentary and superficial way)
and domain theory. However, although
the formal mathematical prerequisites are minimal and regardless of what my
original intention was, it is probably realistic
to suppose that the presentation has ended up at the graduate level.
It is not assumed that the reader
knows anything about functional programming, but some experience
of this topic would, of course, not be amiss. 
\medskip
The usual thanks to secretaries and the suchlike for typing the manuscript are
in the present case redundant: I typed it all myself. 
This means that all errors occurring here, regardless of what kind, are 
entirely due to me.

\bigskip\bigskip
Chris Preston

Bielefeld

May 2001

\vfill\eject

\newwrite\IndexFile
\immediate\openout\IndexFile=indall_r

\hfline{1}{1.1 \ COMPUTING WITH EQUATIONS: SOME EXAMPLES}

{\fourteenbf Chapter 1\quad Introduction}
\bigskip\bigskip
\sectionhead {1.1} {Computing with equations: Some examples}
\bigskip\medskip
Our intention is to explain how and why functional programming 
languages work and in particular what it means for functions to be 
defined by equations. In the introduction we present and analyse some
simple examples with the hope of giving the reader a rough idea of what this
is all about. This chapter can be omitted or skimmed through by anyone having 
a solid knowledge of functional programming.
The reader should be warned that it contains some statements which are vague
and  others which are only a first approximation to the truth. Moreover, the examples
have been chosen to illustrate points which will be important later, and not
as gems from the Art of Programming.

\bigskip
The examples involve lists, and the following notation will be employed
for such objects:
If $X$ is a non-empty set then the set $\bigcup_{n \ge 0} X^n$ will be
denoted by $X^*$, this being considered as the set of all finite lists of elements
from $X$.
If $m \ge 1$ then the element $(\vector x m )$ of $X^m \subset X^*$ will always be
denoted simply by $\list x m $. The single element in $X^0$ is denoted by
$\onept$ and represents the empty list. If $m = 0$ then an expression such as
$\list x m $ is to be interpreted as $\onept$. Lists can be constructed using
either of the mappings $\triangleleft : X \times X^* \to X^*$ and 
$\triangleright : X^* \times X \to X^*$ defined by
\medskip
\mdisp{
x \triangleleft s\ =\ \cases{
x\ \list x m  & if $\,s = \list x m \,$ with $\,m \ge 1\,$,\cr
\noalign{\smallskip}
x             & if $\,s = \onept\,$.}
}
\mdisp{
s \triangleright x\ =\ \cases{
\list x m \ x & if $\,s = \list x m \,$ with $\,m \ge 1\,$,\cr
\noalign{\smallskip}
x             & if $\,s = \onept\,$.}
}
\medskip
Each element of $X^* \setminus \{\onept\}$ has a unique
representation of the form $x \triangleleft s$ with $x \in X$ and $s \in X^*$,
as well as a unique representation
of the form ${\breve s} \triangleright {\breve x}$ with ${\breve x} \in X$ and 
${\breve s} \in X^*$.
\medskip
Now let $\Nat$ denote the set $\{0,1,2,\ldots\,\}$ of natural numbers (and
note that we regard $0$ as being a natural number),
and consider the function $\fss{sqs} : \Nat \to \Nat^*$ defined by
\medskip
\mdisp{\fss{sqs}\,(n)
\ =\ \cases{ \onept & if $\,n = 0\,$,\cr
\noalign{\smallskip}
         1\ 4\ \cdots\ n^2 & if $\,n \ge 1\,$.}}
\medskip
Clearly this function satisfies the relationship (or equation)
\medskip
\mdisp{\fss{sqs}\,(n)
\ =\ \cases{ \onept & if $\,n = 0\,$,\cr
\noalign{\smallskip}
         \fss{sqs}\,(n-1) \,\triangleright\,  n^2 & if $\,n \ge 1\,$.}}
\medskip
Moreover, using induction on $n$, it is easy to see that
$\fss{sqs}$ is the unique function having this property.
This gives a simple characterisation of $\fss{sqs}$ as the unique solution
of an equation involving $\onept$ and
$\triangleright$. In order to make things a little less simple, 
the problem will now be posed of finding a similar
equation for $\fss{sqs}$, but using $\triangleleft$ instead of $\triangleright$.
This new characterisation will be arrived at by first
looking at some auxiliary functions: Let
$\fss{shunt} : \Nat^* \times \Nat^* \to \Nat^*$ be the function defined by
\mdisp{
\fss{shunt}\,(\list m p , \list n q )
\ =\ m_p\ \cdots\ m_1\ \list n q \ ; }

in particular $\,\fss{shunt}\,(a,\onept) \,=\, \fss{rev}\,(a)\,$ for all $a \in \Nat^*$, 
with $\fss{rev} : \Nat^* \to \Nat^*$ defined by
\mdisp{
\fss{rev}\,(\list m p )
\ =\ m_p\ \cdots\ m_1\ .}
Furthermore, let $\fss{sqrs} : \Nat \to \Nat^*$ be the function defined by
\mdisp{
\fss{sqrs}\,(n)\ =\ n^2\ (n-1)^2\ \cdots\ 1}
(with of course $\fss{sqrs}\,(0) = \onept$).
Then $\fss{sqs}$ can clearly be obtained as the composition of the functions $\fss{sqrs}$
and $\fss{rev}$. The reason for this indirect and somewhat roundabout
approach is that
the functions $\fss{shunt}$ and $\fss{sqrs}$ have simple characterisations in
terms of $\onept$ and $\triangleleft$.
On the one hand, the following relationships are valid:
\smallskip
\mdisp{\fss{sqrs}\,(n)\ =\ \cases{ \onept & if $\,n = 0\,$, \cr
      n^2 \,\triangleleft\, \fss{sqrs}\,(n-1) & if $\,n \in \Nat \setminus \{0\}\,$, \cr} }
\smallskip
\smallskip
\mdisp{\fss{shunt}\,(a,b)\ =\ \cases{ b & if $\,a = \onept\,$, \cr
      \fss{shunt}\,(c, m \triangleleft b)
                   & if $\,a = m \triangleleft c\,$. \cr} }
\smallskip
On the other hand, it is easily seen that
$\fss{shunt}$ and $\fss{sqrs}$ are the unique functions having
these properties. (In the case of the function
$\fss{shunt}$ this follows by induction on the length of the list $a$.) 
Putting things together, and adding an auxiliary function
$\fss{sq} : \Nat \to \Nat$ which just squares its argument,
now gives us the four equations:
\medskip
\mdisp{\eqalign{\fss{sqrs}\,(n)\ &=\ \cases{ \onept & if $\,n = 0\,$, \cr
      \fss{sq}\,(n) \,\triangleleft\, \fss{sqrs}\,(n-1) & 
                  if $\,n \in \Nat \setminus \{0\}\,$, \cr} \cr
\noalign{\smallskip}
        \fss{sq}\,(n)\ &=\ n \times n\ ,\cr
\noalign{\smallskip}
    \fss{shunt}\,(a,b)\ &=\ \cases{ b & if $\,a = \onept\,$, \cr
      \fss{shunt}\,(c, m \triangleleft b)
                   & if $\,a = m \triangleleft c\,$, \cr} \cr
\noalign{\smallskip}
     \fss{sqs}\,(n)\ &=\ \fss{shunt}\,(\fss{sqrs}\,(n),\onept)\ .
}}
\medskip
The functions $\fss{sqrs} : \Nat \to \Nat^*$, $\fss{sq} : \Nat \to \Nat$,
$\fss{shunt} : \Nat^*\times \Nat^* \to \Nat^*$
and $\fss{sqs} : \Nat \to \Nat^*$
can be considered to be defined by these equations
in as much as they are their unique solution, i.e., they are 
the unique set of functions satisfying these equations.
(Saying that the functions satisfy the equations of course means
that they satisfy them for all $n \in \Nat$ and all $a, b \in \Nat^*$.) 
\medskip
But there is another, more practical, sense in which $\fss{sqrs}$, 
$\fss{sq}$, $\fss{shunt}$ and $\fss{sqs}$ are defined by the equations:
The equations can be seen as replacement (or substitution) rules for actually 
computing the values of the functions. If an expression or a sub-expression
`matches' the left-hand side of an equation then this expression
or sub-expression can be replaced with the corresponding right-hand side of
the equation. Such substitutions are performed until the names of the functions
have all disappeared.
For example, $\fss{sqs}\,(3)$ can be computed via the steps
\medskip
\ldisp{\quad\fss{sqs}\,(3) 
\ =\ \fss{shunt}\,(\fss{sqrs}\,(3),\onept)  
\ =\ \fss{shunt}\,(\fss{sq}(3) \,\triangleleft\, \fss{sqrs}\,(2),\,\onept)
}
\vskip-\medskipamount
\ldisp{\qquad\quad
\ =\  \fss{shunt}\,(\fss{sqrs}\,(2),\,\fss{sq}(3) \triangleleft \onept)
\ =\  \fss{shunt}\,(\fss{sq}(2) \,\triangleleft\, \fss{sqrs}\,(1),\,\fss{sq}(3) 
                                                             \triangleleft \onept)}
\vskip-\medskipamount
\ldisp{\qquad\qquad
\ =\  \fss{shunt}\,(\fss{sqrs}\,(1),\,\fss{sq}(2) \triangleleft (\fss{sq}(3) 
                                                           \triangleleft \onept))}
\vskip-\medskipamount
\ldisp{\qquad\qquad\qquad
\ =\ \fss{shunt}\,(\fss{sq}(1) \,\triangleleft\, \fss{sqrs}\,(0),
                     \,\fss{sq}(2) \triangleleft (\fss{sq}(3) \triangleleft \onept))}
\vskip-\medskipamount
\ldisp{\qquad\qquad\qquad\qquad
\ =\  \fss{shunt}\,(\fss{sqrs}\,(0),
                   \,\fss{sq}(1) \triangleleft (\fss{sq}(2) \triangleleft 
                           (\fss{sq}(3) \triangleleft \onept)))}
\vskip-\medskipamount
\ldisp{\qquad\qquad\qquad\qquad\qquad
\ =\  \fss{shunt}\,(\onept,
                   \,\fss{sq}(1) \triangleleft (\fss{sq}(2) \triangleleft 
                    (\fss{sq}(3) \triangleleft \onept)))
}
\vskip-\medskipamount
\rdisp{
\ =\ \fss{sq}(1) \triangleleft (\fss{sq}(2) \triangleleft 
                              (\fss{sq}(3) \triangleleft \onept))
\ =\ (1\times 1) \triangleleft ((2 \times 2) \triangleleft 
                              ((3 \times 3) \triangleleft \onept))
\ =\ 1\ 4\ 9
\quad}
\medskip
In the first step, the expression
$\fss{sqs}\,(3)$ matches the left-hand side of the equation for $\fss{sqs}$
with $n = 3$, and so it can be replaced with
$\fss{shunt}\,(\fss{sqrs}\,(3),\onept)$, which is the right-hand side of this
equation with $n = 3$.
Similarly, in the next step
the sub-expression $\fss{sqrs}\,(3)$ of the expression
$\fss{shunt}\,(\fss{sqrs}\,(3),\onept)$ matches
the left-hand side of the equation of $\fss{sqrs}$ with
$n = 3$, and so it can be replaced with
$\fss{sq}(3) \,\triangleleft\, \fss{sqrs}\,(2)$,
which is the corresponding right-hand side. This results in
the new expression
$\fss{shunt}\,(\fss{sq}(3) \,\triangleleft\, \fss{sqrs}\,(2),\,\onept)$.
\medskip

It is easy to see that the value of $\fss{sqs}\,(n)$ 
can be computed in the above manner for any $n \in \Nat$.

\medskip
In most modern functional programming languages the equations for the 
functions $\fss{sqrs}$, $\fss{sq}$, $\fss{shunt}$ and $\fss{sqs}$ can be 
written more-or-less exactly as they appear above.  For example, in the 
programming language {\it Haskell\/} they can be written as

\bigskip
{\leftskip=75pt
$\ftt{sqrs n \ = \ case n == 0 of}$
\vskip 2.0pt
$\ftt{ \ \ \ \ \ \ \ \ \ \ \ \ \ \  True -> []}$
\vskip 2.0pt
$\ftt{ \ \ \ \ \ \ \ \ \ \ \ \ \ \  False -> sq n :\ sqrs (n-1)}$ 
\vskip 4.0pt
$\ftt{sq n \ = \ n*n}$
\vskip 4.0pt
$\ftt{shunt a b \ = \ case a of}$
\vskip 2.0pt
$\ftt{ \ \ \ \ \ \ \ \ \ \ \ \ \ \  [] -> b}$
\vskip 2.0pt
$\ftt{ \ \ \ \ \ \ \ \ \ \ \ \ \ \  m :\ c -> shunt c (m :\ b)}$ 
\vskip 4.0pt
$\ftt{sqs n \ = \ shunt (sqrs n, [])}$
\bigskip
}
(although most programmers would probably prefer to write the equations for $\fss{sqrs}$ 
and $\fss{shunt}$ somewhat differently using guards and pattern matching).
\medskip
Evidently 
the construction operator $\triangleleft$ is represented in {\it Haskell\/} by $\ftt{:}$
and the empty list $\onept$ by $\ftt{[]}$. Note  
that the arguments of functions are not enclosed in brackets (except to avoid
ambiguities) and they are not
separated by commas; this is typical of modern functional programming
languages.

\medskip
The evaluation mechanism in {\it Haskell\/} would use these equations as
replacement rules in essentially the same way as above to compute values.
For example, it would reduce the term $\,\ftt{sqs 3}\,$
to the constant term $\,\ftt{1:4:9:[]}\,$ (which would then be presented in
the more legible form $\ftt{[1,4,9]}$) via the same steps which were used 
to deduce that $\,\fss{sqs}\,(3) = 1\ 4\ 9\,$.
\medskip

Based on the above discussion, let us make the following points:
\medskip\smallskip
1.\enskip The {\it Haskell\/} equations for the names
$\ftt{sqrs}$,  $\ftt{sq}$, $\ftt{shunt}$ and  $\ftt{sqs}$
are a particular representation of the
corresponding `mathematical' equations for the functions 
$\fss{sqrs}$,  $\fss{sq}$, $\fss{shunt}$ and  $\fss{sqs}$. (The form of the
{\it Haskell\/} equations is in part, of course, dictated by the
restrictive ASCII-code alphabet in which they are written.)
\medskip
2.\enskip The Haskell equations can be seen as defining the functions
$\fss{sqrs}$,  $\fss{sq}$, $\fss{shunt}$ and  $\fss{sqs}$, in that
these functions are the unique solutions of the corresponding
`mathematical' equations. This is the way that mathematically inclined humans 
usually look at the equations, and it
is referred to as their 
\indexdef{denotational semantics}{denotational semantics}{}{semantics}{denotational}.
\medskip
3.\enskip The Haskell equations can also be seen as 
replacement rules for computing the values obtained by
applying the functions to specific values of their arguments.
This is the way that a machine (or perhaps a computer scientist)
can be made to obtain information from the equations, and it is referred to
as their 
\indexdef{operational semantics}{operational semantics}{}{semantics}{operational}.
\medskip
4.\enskip When the equations are used as replacement rules then the values that 
they compute are correct: They are the corresponding values of the
functions which are the unique solutions of the `mathematical' equations.
Conversely, any such value can be computed using an appropriate sequence of
replacements. In this sense, the denotational and the operational semantics
are equivalent.
\medskip\smallskip
This is all rather vague. Moreover, the situation considered above is not
at all typical in that the equations have a unique solution. Nevertheless,
the fourth point gives the first inkling of the main theme of our study, which
is to describe a framework for dealing with equations in which the
denotational and the operational semantics are equivalent.
\medskip
To lead into the next topic we want to discuss, let us note
that the {\it Haskell\/} equations
for $\fss{sqrs}$, $\fss{sq}$, $\fss{shunt}$ and $\fss{sqs}$ encode only a part
of the information contained in the original equations.
To be more specific, if only the {\it Haskell\/} equations are given
then it is not possible, without additional information, to reconstruct the
set-up started with above. For example, if $\Int$ rather than $\Nat$ 
is chosen as the basic set of data objects for numbers then
they could just as well be 
describing functions $\fss{sqrs} : \Int \to \Int^*$, 
$\fss{sq} : \Int \to \Int$, 
$\fss{shunt} : \Int^* \times \Int^* \to \Int^*$ 
and $\fss{sqs} : \Int \to \Int^*$ satisfying the equations
\medskip
\mdisp{\eqalign{\fss{sqrs}\,(n)\ &=\ \cases{ \onept & if $\,n = 0\,$, \cr
      \fss{sq}(n) \,\triangleleft\, \fss{sqrs}\,(n-1) & if 
                $\,n \in \Int \setminus \{0\}\,$, \cr} \cr
\noalign{\smallskip}    	
    \fss{sq}\,(n)\ &=\ n \times n\ , \cr
\noalign{\smallskip}
    \fss{shunt}\,(a,b)\ &=\ \cases{ b & if $\,a = \onept\,$, \cr
      \fss{shunt}\,(c, m \triangleleft b)
                   & if $\,a = m \triangleleft c\,$, \cr} \cr
\noalign{\smallskip}
     \fss{sqs}\,(n)\ &=\ \fss{shunt}\,(\fss{sqrs}\,(n),\onept)
}}
\medskip
for all $n \in \Int$ and all $a,\,b \in \Int^*$. If this is the case
then there is a problem, since these new equations no longer have a
solution. More precisely, there is no way of defining $\fss{sqrs}\,(n)$
for $n < 0$ so that the equation for $\fss{sqrs}$ holds:
The list $\fss{sqrs}\,(-m)$ would have to have one more component than the
list $\fss{sqrs}\,(-m-1)$ for each $m \ge 1$, hence the
list $\fss{sqrs}\,(-1)$ would have to have $p$ more components than the
list $\fss{sqrs}\,(-1-p)$ for each $p \ge 1$. 
In particular, $\fss{sqrs}\,(-1)$ would have to have at least $p$ components
for each $p \ge 1$, which is impossible.
\medskip
This situation is not very satisfactory. The requirement that
$\fss{sqrs}\,(n)$ be defined for all $n \in \Int$, and the impossibility of
meeting this requirement when $n < 0$, means that the values $\fss{sqrs}\,(n)$
for $n \in \Nat$ are also lost, although in some sense they are obviously
just as much defined as they were before. Of course, in the particular
example being considered the problem could be solved by simply changing the condition
$n = 0$ in the definition of $\fss{sqrs}$ to $n \le 0$. In general, however, 
the problem cannot be solved in such an {\it ad hoc\/} manner. For instance, even if 
a set of equations has a solution, this can be `spoilt' by adding an new equation
for a function which does not have a solution (for example, a function
$\fss{imp} : \Int \to \Int$ which should satisfy the equation
\mdisp{\fss{imp}\,(n)\ =\ 1 \,+\,\fss{imp}\,(n)}
for all $n \in \Int$). The enlarged set of equations then has no solution, even
though the additional equation is really completely irrelevant.
\medskip
The first step in solving the above problem is to explicitly allow functions to be undefined
for certain values of their arguments.
To make this more precise for the equations we are looking at, let $\bot_\Int$ be
some element not in $\Int$
(to be thought of as an `undefined' element of $\Int$) and put
$\Int^\bot = \Int \cup \{\bot_\Int\}$.
Moreover, put ${\Bbb L} = \Int^*$, let
$\bot_{\Bbb L}$ be some element not in ${\Bbb L}$
(to be thought of as an `undefined' element of ${\Bbb L}$) and let
${\Bbb L}^\bot = {\Bbb L} \cup \{\bot_{\Bbb L}\}$.
Extend the mapping $\triangleleft : \Int \times {\Bbb L} \to {\Bbb L}$
to a mapping from $\Int^\bot \times {\Bbb L}^\bot$  to ${\Bbb L}^\bot$ by
putting $\,n \,\triangleleft\, a \,=\, \bot_{\Bbb L}\,$ for all
$(n,a) \in (\Int^\bot \times {\Bbb L}^\bot) \setminus (\Int \times {\Bbb L})$.
Note that if $a \in {\Bbb L}^\bot \setminus \{\bot_{\Bbb L}\}$ then either
$a = \onept$ or $a$ has a unique representation of the form
$m \triangleleft c$ with $(m,c) \in \Int\times {\Bbb L}$.
\medskip
The {\it Haskell\/} equations will now be regarded as describing functions 
$\fss{sqrs} : \Int^\bot \to {\Bbb L}^\bot$, 
$\fss{sq} : \Int^\bot \to \Int^\bot$,
$\fss{shunt} : {\Bbb L}^\bot \times {\Bbb L}^\bot \to {\Bbb L}^\bot$ and 
$\fss{sqs} : \Int^\bot \to {\Bbb L}^\bot$ satisfying 
\medskip
\mdisp{\eqalign{\fss{sqrs}\,(n)\ &=\ \cases{ \bot_{\Bbb L} & if $\,n = \bot_\Int\,$, \cr
             \onept & if $\,n = 0\,$, \cr
 \fss{sq}(n) \,\triangleleft\, \fss{sqrs}\,(n-1) & 
           if $\,n \in \Int \setminus \{0\}\,$, \cr} \cr
\noalign{\smallskip}
    \fss{sq}\,(n)\ &=\ n \times n\ , \cr
\noalign{\smallskip}
\fss{shunt}\,(a,b)\ &=\ \cases{ \bot_{\Bbb L} & if $\,a = \bot_{\Bbb L}\,$, \cr
       b & if $\,a = \onept\,$, \cr
       \fss{shunt}\,(c, m \triangleleft b)
                   & if $\,a = m \triangleleft c\,$ with 
		   $\,(m,c)  \in \Int \times {\Bbb L}\,$, \cr} \cr
\noalign{\smallskip}    	
     \fss{sqs}\,(n)\ &=\ \fss{shunt}\,(\fss{sqrs}\,(n),\onept)
}}
\medskip
for all $n \in \Int^\bot$, $a,\, b \in {\Bbb L}^\bot$
(where $m \times n$ is defined to be $\bot_\Int$ if $(m,n) \notin \Int \times \Int$).
It is easily checked that these equations have a unique solution: For $n \in \Nat$
the values $\fss{sqrs}\,(n)$ and $\fss{sqs}\,(n)$ are the same as before
and $\,\fss{sqrs}\,(n) \,=\, \fss{sqs}\,(n) \,=\, \bot_{\Bbb L}\,$ for all $n < 0$.
Moreover, $\fss{shunt}$ is also defined in essentially the same
way as before but with $\fss{shunt}(a,b) = \bot_{\Bbb L}$
whenever $(a,b) \notin {\Bbb L}\times {\Bbb L}$.
\medskip
This second interpretation of the {\it Haskell\/} equations can be seen as
introducing a new denotational semantics. However, although the
operational semantics of the equations remains unchanged, the
denotational and operational semantics are still equivalent.
More precisely, if the value of one of the functions (occurring as the unique
solution of the equations) is defined for a particular value of
its argument, then this value can be computed using the equations
as replacement rules. Conversely, if the value is not defined then
no value can be computed. For example, the value of the function $\fss{sqs}$
is defined with its argument equal to $3$, and just as before
this value can be computed to be $\,1\ 4\ 9\,$. On the other hand,
the value of $\fss{sqs}$ is undefined when its argument is $-1$, and 
an attempt to use the equations as replacement rules to compute a `value' will
result in something similar to the following non-terminating process:
\medskip
\ldisp{\quad\fss{sqs}\,(-1) 
\ =\ \fss{shunt}\,(\fss{sqrs}\,(-1),\onept)  
\ =\ \fss{shunt}\,(\fss{sq}(-1) \,\triangleleft\, \fss{sqrs}\,(-2),\,\onept)
}
\vskip-\medskipamount
\ldisp{\qquad\quad
\ =\  \fss{shunt}\,(\fss{sqrs}\,(-2),\,\fss{sq}(-1) \triangleleft \onept)
\ =\  \fss{shunt}\,(\fss{sq}(-2) \,\triangleleft\, 
     \fss{sqrs}\,(-3),\,\fss{sq}(-1) \triangleleft \onept)}
\vskip-\medskipamount
\ldisp{\qquad\qquad
\ =\  \fss{shunt}\,(\fss{sqrs}\,(-3),\,\fss{sq}(-2) \triangleleft 
               (\fss{sq}(-1) \triangleleft \onept))}
\vskip-\medskipamount
\ldisp{\qquad\qquad\quad
\ =\ \fss{shunt}\,(\fss{sq}(-3) \,\triangleleft\, \fss{sqrs}\,(-4),
                     \,\fss{sq}(-2) \triangleleft (\fss{sq}(-1) \triangleleft \onept))}
\vskip-\medskipamount
\ldisp{\qquad\qquad\qquad
\ =\ \fss{shunt}\,(\fss{sqrs}\,(-4),
                     \,\fss{sq}(-3) \triangleleft 
		     (\fss{sq}(-2) \triangleleft (\fss{sq}(-1) \triangleleft \onept)))}
\vskip-\medskipamount
\mdisp{\vdots}
\smallskip
Allowing undefined values,
the same holds true for the equation for $\fss{imp}$. 
This is now an equation for a function $\fss{imp} : \Int^\bot \to \Int^\bot$
which should satisfy
\mdisp{\fss{imp}\,(n)\ =\ 1 \,+\,\fss{imp}\,(n)}
for all $n \in \Int^\bot$. Here $1 + \bot_\Int$ has to be given a
meaning, and the only reasonable choice is to define it to be $\bot_\Int$.
In this case the equation has a unique solution with $\fss{imp}\,(n) = \bot_\Int$ for all
$n \in \Int^\bot$. For each $n \in \Int$ the value $\fss{imp}\,(n)$ is thus
undefined, which agrees with the fact that no value can be computed using the
equation as a replacement rule.
\medskip
By introducing undefined values, our equations again had a unique solution in the 
case where the basic set of data objects for numbers was $\Int$ rather than $\Nat$. 
The next point which has to be discussed deals with the opposite
extreme, and is the problem of having
more than one solution. Let us look at the {\it Haskell} equation
\medskip\smallskip
{\leftskip=75pt
$\ftt{fone n \ = \ case n == 0 of}$
\vskip 2.0pt
$\ftt{ \ \ \ \ \ \ \ \ \ \ \ \ \ \  True -> 1}$
\vskip 2.0pt
$\ftt{ \ \ \ \ \ \ \ \ \ \ \ \ \ \  False -> fone (n-1)}$ 
\medskip
}
\smallskip
considered as  corresponding to the `mathematical' equation
\smallskip
\mdisp{\fss{fone}\,(n)\ =\ \cases{ \bot_\Int & if $\,n = \bot_\Int\,$, \cr
\noalign{\smallskip}
     1 & if $\,n = 0\,$, \cr
\noalign{\smallskip}
      \fss{fone}\,(n-1) & if $\,n \in \Int \setminus \{0\}\,$, \cr} }
\smallskip
for a function $\fss{fone} : \Int^\bot \to \Int^\bot$. 
For each $p \in \Int^\bot$ let
$\fss{fone}_p : \Int^\bot \to \Int^\bot$ be given by
$\fss{fone}_p\,(\bot_\Int) = \bot_\Int$,
$\fss{fone}_p\,(n) = 1$ if $n \in \Nat$ and
$\fss{fone}_p\,(n) = p$ for all $n < 0$.
Then it is easily checked that
$\lcurl \fss{fone}_p \,:\, p \in \Int^\bot \rcurl$ is exactly the set of
solutions of the equation, and in what follows it is convenient to just write
$\fss{fone}_\bot$ instead of $\fss{fone}_{\bot_\Int}$.
\medskip
What should be regarded here as the denotational semantics,
since it does not make sense to speak of
{\it the\/} solution of the equation?
One possibility is to declare $\fss{fone}_\bot$ to be {\it the\/}
solution: It is the least defined solution
in that $\fss{fone}_\bot\,(n)$ is only defined
(i.e., $\fss{fone}_\bot\,(n) \in \Int$) for $n \in \Nat$, whereas
$\fss{fone}_p\,(n)$ is defined for all $n \in \Int$ for each $p \in \Int$.
It can thus be argued that $\fss{fone}_\bot$ is the most natural solution,
since if $n \notin \Nat$ then nothing can be deduced from the equation about the value of
$\fss{fone}\,(n)$; this value should therefore be left undefined.
\medskip
Taking $\fss{fone}_\bot$ to be {\it the\/} solution also fits in with the
operational semantics, since the value $\fss{fone}\,(n)$ can clearly be computed
using the equation as a replacement rule if and only if $n \in \Nat$, i.e.,
if and only if $\fss{fone}_\bot\,(n)$ is defined.
\medskip
An alternative point of view is to regard all the solutions as having equal
weight, but to consider that $\fss{fone}\,(n)$ is defined for the argument 
$n \in \Int^\bot$ if and only if there exists $m \in \Int$ such that 
$\fss{fone}_p\,(n) = m$ for all $p \in \Int^\bot$, i.e., if and only if 
$\fss{fone}\,(n)$ does not depend on which solution is being used.
\medskip
In the example being considered these two points of view coincide. Our study
will tend to take the alternative point of view, because it allows
things to be formulated without having to worry about the existence of a
least defined solution. However, the existence of a least defined solution
will play an important role in Chapter~8.
\medskip
The final topic to be looked at in this section starts with the {\it Haskell\/}
equations

\bigskip
{\leftskip=75pt
$\ftt{hd a \ = \ case a of}$
\vskip 2.0pt
$\ftt{ \ \ \ \ \ \ \ \ \ \ \ \ \ \  Nil -> und}$
\vskip 2.0pt
$\ftt{ \ \ \ \ \ \ \ \ \ \ \ \ \ \  Cons m c -> m}$ 
\vskip 4.0pt
$\ftt{sone \ = \ hd (Cons 1 (Cons und Nil))}$
\vskip 4.0pt
$\ftt{und \ = \ und}$
\medskip
}
\medskip

These can be regarded as describing a function
$\fss{hd} : {\Bbb L}^\bot \to \Int^\bot$ together with elements
$\fss{sone}$ and $\fss{und}$ of $\Int^\bot$ satisfying the equations
\medskip
\mdisp{\eqalign{
\fss{hd}\,(a)\ &=\ \cases{ \bot_\Int & if $\,a = \bot_{\Bbb L}\,$, \cr
       \fss{und} & if $\,a = \onept\,$, \cr
       m & if $\,a = m \triangleleft c\,$ with 
		   $\,(m,c)  \in \Int \times {\Bbb L}\,$, \cr} \cr
\noalign{\smallskip}    	
     \fss{sone} \ &=\ \fss{hd}\,( 1 \triangleleft (\fss{und} \triangleleft
                                  \onept)) \cr
\noalign{\smallskip}    	
     \fss{und} \ &=\ \fss{und}
}}
\smallskip
for all $a \in {\Bbb L}^\bot$. For each $p \in \Int^\bot$ let
$\fss{hd}_p : {\Bbb L}^\bot \to \Int^\bot$ be given by
\smallskip
\mdisp{
\fss{hd}_p\,(a)\ =\ \cases{ \bot_\Int & if $\,a = \bot_{\Bbb L}\,$, \cr
       p & if $\,a = \onept\,$, \cr
       m & if $\,a = m \triangleleft c\,$ with 
		   $\,(m,c)  \in \Int \times {\Bbb L}\,$, \cr} }
\smallskip
let $\fss{und}_p = p$, let $\fss{sone}_p = 1$ for all $p \in \Int$ and
$\fss{sone}_{\bot_\Int} = \bot_\Int$. Then it is easily checked that
$\lcurl (\fss{hd}_p,\fss{sone}_p,\fss{und}_p) \,:\, p \in \Int^\bot \rcurl$ is exactly 
the set of solutions of these equations. In particular, the value of $\fss{sone}$ is 
not defined by these equations, since 
$\fss{sone}_{\bot_\Int} = \bot_\Int$ (or, alternatively,
$\fss{sone}_p$ is not independent of 
$p \in \Int^\bot$).
\medskip
Now it could be claimed that $\fss{sone}$ should have the value $1$, since
the function $\fss{hd}$ extracts the first element from a list, and the first element
of the list $1 \triangleleft (\fss{und} \triangleleft \onept)$ is $1$.
But if $\fss{und}\, = \bot_\Int$ then this is not true, since the list
$1 \triangleleft (\fss{und} \triangleleft \onept)$ is then equal to $\bot_{\Bbb L}$; this
is a consequence of how the mapping 
$\triangleleft : \Int \times {\Bbb L} \to {\Bbb L}$ was extended to a mapping
from $\Int^\bot \times {\Bbb L}^\bot$  to ${\Bbb L}^\bot$.
Thus it must either be accepted that the value of
$\fss{sone}$ is undefined, or the set-up must be changed so that
$1 \triangleleft (\fss{und} \triangleleft \onept)$ is always a list having
$1$ as its first element.
\medskip
This leads to a third interpretation of lists occurring in Haskell equations.
Again let $\Int^\bot = \Int \cup \{\bot_\Int\}$ be as before, but now replace the set 
${\Bbb L}^\bot$ with 

\mdisp{{\Bbb L}^\diamond \  =\ (\Int^\bot)^* \cup  \Bot{(\Int^\bot)^*}}

where $\Bot{(\Int^\bot)^*}$ is a disjoint copy of $(\Int^\bot)^*$, and where
for each $z \in (\Int^\bot)^*$ the corresponding element in
$\Bot{(\Int^\bot)^*}$  denoted by $z^\bot$.

\medskip
Note that an element of ${\Bbb L}^\diamond$ has either the form
$\list x n $ or the form $(\list x n )^\bot$, where $n \ge 0$ and
$x_j \in \Int^\bot$ for each $\oneto j n $.
The element $\list x n $ describes a `real' list with $n$ components 
(although some or all of these components may be `undefined'). The element 
$(\list x n )^\bot$, on the other hand, should be thought of as a `partial' 
list containing at least $n$ components, of which the first $n$ 
components are `known' to be $\lvector x n $.
In particular, this means that the element $\varepsilon^\bot$ gives no information
about a list, and it will thus be thought of as the `undefined' list.

\medskip
Since $\Int \subset \Int^\bot$, the set of real lists is a subset of 
${\Bbb L}^\diamond$, i.e.,
${\Bbb L} = \Int^* \subset {\Bbb L}^\diamond$.
The mapping $\triangleleft : \Int \times {\Bbb L} \to {\Bbb L}$ will be extended to a 
mapping from $\Int^\bot \times {\Bbb L}^\diamond$  to ${\Bbb L}^\diamond$
by letting
\smallskip
\mdisp{x \triangleleft z \ =\ \cases{ 
           x \triangleleft s & if $\,z = s\,$ for some
                   $\,s \in  (\Int^\bot)^*\,$, \cr
   \noalign{\smallskip}
        (x \triangleleft s)^\bot & if $\,z = s^\bot\,$ for some
                 $\,s \in  (\Int^\bot)^*\,$. \cr }}

\smallskip
Note that if $a \in {\Bbb L}^\diamond \setminus \{\varepsilon^\bot\}$ then either
$a = \onept$ or $a$ has a unique representation of the form
$x \triangleleft z$ with $x \in \Int^\bot$ and $z \in {\Bbb L}^\diamond$.
\medskip

The {\it Haskell\/} equations for $\ftt{hd}$, $\ftt{sone}$ and $\ftt{und}$ can now be 
regarded as describing a function
$\fss{hd} : {\Bbb L}^\diamond \to \Int^\bot$ together with elements
$\fss{sone}$ and $\fss{und}$ of $\Int^\bot$ satisfying 
\medskip
\mdisp{\eqalign{
\fss{hd}\,(a)\ &=\ \cases{ \bot_\Int & if $\,a = \varepsilon^\bot\,$, \cr
       \fss{und} & if $\,a = \onept\,$, \cr
       x & if $\,a = x \triangleleft z\,$ with 
		   $\,(x,z)  \in \Int^\bot \times {\Bbb L}^\diamond\,$, \cr} \cr
\noalign{\smallskip}    	
     \fss{sone} \ &=\ \fss{hd}\,( 1 \triangleleft (\fss{und} \triangleleft
                                  \onept)) \cr
\noalign{\smallskip}    	
     \fss{und} \ &=\ \fss{und}
}}
\medskip
for all $a \in {\Bbb L}^\bot$. As in the previous interpretation the set of
solutions has the form
$\lcurl (\fss{hd}_p,\fss{sone}_p,\fss{und}_p) \,:\, p \in \Int^\bot \rcurl$,
however, the function $\fss{hd}_p : {\Bbb L}^\diamond \to \Int^\bot$ and the elements 
$\fss{sone}_p,\,\fss{und}_p \in \Int^\bot$ are here given by
$\fss{sone}_p = 1$, $\fss{und}_p = p$ and
\medskip
\hrule height 0pt
\vskip-10pt
\mdisp{
\fss{hd}_p\,(a)\ =\ \cases{ \bot_\Int & if $\,a = \varepsilon^\bot\,$, \cr
       p & if $\,a = \onept\,$, \cr
       x & if $\,a = x \triangleleft z\,$ with 
		   $\,(x,z)  \in \Int^\bot \times {\Bbb L}^\diamond\,$, \cr} 
}
\medskip
In this interpretation $\fss{sone}$ does have the value $1$,
since $\fss{sone}_{\bot_\Int} = 1$ (or, alternatively, since
$\fss{sone}_p = 1$ for all $p \in \Int^\bot$),
and $1 \triangleleft (\fss{und} \triangleleft \onept)$ is always a list having
$1$ as its first element, even though $\fss{und}$ is undefined.

\medskip
Note that the second and third interpretations of the above equations give different
values to $\fss{sone}$. This would seem to pose a problem when the equations are used
as replacement rules. In fact, at first sight the replacement

\mdisp{\fss{sone}\ =\ \fss{hd}\,( 1 \triangleleft (\fss{und} \triangleleft
                                  \onept))\ =\ 1}

would appear to show that $\fss{sone}$ has the value $1$. However, although
the replacement of
$\fss{hd}\,( 1 \triangleleft (\fss{und} \triangleleft\onept))$
by $1$ is valid in the third interpretation, it is not in the second.
In the second interpretation the only valid replacement is to use the equation 
for $\fss{sone}$ to replace $\fss{sone}$ in an expression by itself,
which of course leaves the expression unchanged. 
Thus a valid attempt to use the equations as replacement rules in this case
would result in the following non-terminating process:
\smallskip
\ldisp{\quad \fss{sone}
\ =\ \fss{hd}\,( 1 \triangleleft (\fss{und} \triangleleft \onept))
\ =\ \fss{hd}\,( 1 \triangleleft (\fss{und} \triangleleft \onept))
}
\vskip-\medskipamount
\rdisp{
\ =\ \fss{hd}\,( 1 \triangleleft (\fss{und} \triangleleft \onept))
\ =\ \fss{hd}\,( 1 \triangleleft (\fss{und} \triangleleft \onept))
\ =\ \cdots \qquad
}
\smallskip
which can be seen as confirming the fact that the value of $\fss{sone}$ is undefined
in the second interpretation.
Two lessons can be learnt from this. The first is that the validity of the
replacement rules will depend on which interpretation is being used. The second is that
the `correct' choice of valid replacements must be made in order to arrive at the value:
For instance, replacing $\fss{sone}$ by itself is still valid in the third 
interpretation, but this alone will not give the answer $1$.
\medskip
What happens now if the third interpretation is used in the equations
for $\fss{sqrs}$, $\fss{shunt}$, $\fss{sq}$ and $\fss{sqs}$?
They will then be regarded as describing functions 
$\fss{sqrs} : \Int^\bot \to {\Bbb L}^\diamond$, 
$\fss{sq} : \Int^\bot \to \Int^\bot$, 
$\fss{shunt} : {\Bbb L}^\diamond \times {\Bbb L}^\diamond \to {\Bbb L}^\diamond$ and 
$\fss{sqs} : \Int^\bot \to {\Bbb L}^\diamond$ satisfying 
\medskip
\mdisp{\eqalign{\fss{sqrs}\,(n)\ &=\ \cases{ \varepsilon^\bot & if $\,n = \bot_\Int\,$, \cr
             \onept & if $\,n = 0\,$, \cr
      \fss{sq}(n) \,\triangleleft\, \fss{sqrs}\,(n-1) & if 
            $\,n \in \Int \setminus \{0\}\,$, \cr} \cr
\noalign{\smallskip}    	
    \fss{sq}\,(n)\ &=\ n \times n, \cr
\noalign{\smallskip}
\fss{shunt}\,(a,b)\ &=\ \cases{ \varepsilon^\bot & if $\,a = \varepsilon^\bot\,$, \cr
       b & if $\,a = \onept\,$, \cr
       \fss{shunt}\,(c, m \triangleleft b)
                   & if $\,a = m \triangleleft c\,$ with 
		   $\,(m,c)  \in \Int^\bot \times {\Bbb L}^\diamond\,$, \cr} \cr
\noalign{\smallskip}
     \fss{sqs}\,(n)\ &=\ \fss{shunt}\,(\fss{sqrs}\,(n),\onept)
}}
\medskip
for all $n \in \Int^\bot$, $a,\, b \in {\Bbb L}^\diamond$.
Unfortunately, however, these equations do not have a solution.
The problem here is essentially the same as that which occurred when $\Nat$ was
replaced by $\Int$: There is no way of defining $\fss{sqrs}\,(n)$
for $n < 0$ so that the equation for $\fss{sqrs}$ holds; as before
the list $\fss{sqrs}\,(-1)$ would have to have at least $p$ components
for each $p \ge 1$, which is still impossible.
(Note that in the second interpretation this problem does not arise because
$n \triangleleft \bot_{\Bbb L} = \bot_{\Bbb L}$ for all $n \in \Int$, whereas here
a list of the form $x \triangleleft z$ is never equal to the undefined list
$\varepsilon^\bot$.)
Of course, as mentioned before, the problem could be solved in this particular case
simply by changing the condition
$n = 0$ in the definition of $\fss{sqrs}$ to $n \le 0$; but this is again only an
{\it ad hoc\/} solution.
\medskip
A better way to solve the above problem is suggested by the `fact' that
the list $\fss{sqrs}\,(-1)$ should have at least $p$ components
for each $p \ge 1$, which becomes possible if infinite lists are allowed.
This leads to a fourth (and for the present section final) interpretation of lists 
occurring in Haskell equations.
Extend the set ${\Bbb L}^\diamond$ to the set

\mdisp{{\Bbb L}^\infty \  =\ {\Bbb L}^\diamond \cup (\Int^\bot)^{**}} 

where 
$(\Int^\bot)^{**}$ denotes the set of all infinite lists of elements from $\Int^\bot$.
The mapping $\triangleleft : \Int^\bot \times {\Bbb L}^\diamond \to {\Bbb L}^\diamond$ 
is then extended to a mapping from $\Int^\bot \times {\Bbb L}^\infty$  to 
${\Bbb L}^\infty$ in the obvious way: If $x \in \Int^\bot$ and $z \in (\Int^\bot)^{**}$
then $x \triangleleft z$ is the infinite list whose first component is $x$ 
and whose remaining components are the components of $z$, each being shifted one place
to the right.
Note that if $a \in {\Bbb L}^\infty \setminus \{\varepsilon^\bot\}$ then either
$a = \onept$ or $a$ has a unique representation of the form
$x \triangleleft z$ with $x \in \Int^\bot$ and $z \in {\Bbb L}^\infty$.
\medskip
The fourth interpretation of the {\it Haskell\/} equations describes functions 
$\fss{sqrs} : \Int^\bot \to {\Bbb L}^\infty$, 
$\fss{sq} : \Int^\bot \to \Int^\bot$, 
$\fss{shunt} : {\Bbb L}^\infty \times {\Bbb L}^\infty \to {\Bbb L}^\infty$
and 
$\fss{sqs} : \Int^\bot \to {\Bbb L}^\infty$ satisfying the same equations
as above, but with ${\Bbb L}^\infty$ replacing ${\Bbb L}^\diamond$.
It is easily checked that these equations now  have a unique solution, and that
$\fss{sqrs}\,(-n)$ is the infinite list 
\mdisp{-n\ \ -(n+1)\ \ -(n+2)\ \ -(n+3)\ \ \cdots}
for each $n > 1$.
For all practical purposes (for example, 
computing $\fss{sqrs}\,(n)$ for some $n \ge 0$) there is 
no real difference between the second and the fourth interpretation of these equations.
\medskip
Note that if the fourth interpretation is applied to the equations for
$\fss{hd}$, $\fss{sone}$ and $\fss{und}$ then the situation is essentially the same as in
the third interpretation.
The set of solutions still has the form
$\lcurl (\fss{hd}_p,\fss{sone}_p,\fss{und}_p) \,:\, p \in \Int^\bot \rcurl$
with $\fss{sone}_p = 1$, $\fss{und}_p = p$ and
the function $\fss{hd}_p : {\Bbb L}^\infty \to \Int^\bot$ given by

\smallskip
\mdisp{
\fss{hd}_p\,(a)\ =\ \cases{ \bot_\Int & if $\,a = \varepsilon^\bot\,$, \cr
       p & if $\,a = \onept\,$, \cr
       x & if $\,a = x \triangleleft z\,$ with 
		   $\,(x,z)  \in \Int^\bot \times {\Bbb L}^\infty\,$. \cr} 
}
\smallskip
In particular, in this interpretation $\fss{sone}$ again has the value $1$.
\medskip
The second and fourth interpretations introduced above seem to have the advantage
over the third that the equations are more likely to have a solution. This somewhat
vague suspicion will later be confirmed.
Whether the fourth interpretation is to be preferred to the second is, to
some extent, just a matter of taste. However, it is a fact that, whereas the `older'
functional programming languages (such as the original `eager' version
of {\it ML\/}) use the second interpretation, the `newer' languages 
(such as {\it Haskell\/}) tend to use the fourth.

\vfill\eject

\hfline{11}{1.2 \ A PROCEDURE FOR INTERPRETING EQUATIONS}

\sectionhead {1.2} {A procedure for interpreting equations}
\bigskip\medskip
In this section we outline a procedure which, in particular, can be used to 
obtain the second, third and fourth interpretations of the {\it Haskell\/} equations
for $\fss{sqrs}$, $\fss{shunt}$, $\fss{sqr}$ and $\fss{sqs}$
introduced in the previous section. 
In its general form the procedure will then constitute the basis of our study.
\medskip
First, however, it is convenient to rewrite the {\it Haskell\/} equations using
a uniform prefix notation. More precisely,
the infix operator names $\ftt{-}$, $\ftt{*}$ and $\ftt{==}$ will be replaced by
corresponding prefix operator names $\ftt{sub}$, $\ftt{mul}$ and
$\ftt{eq}$, and the built-in constructor name
$\ftt{:}$ for list-building by a prefix constructor name
$\ftt{Cons}$; moreover, $\ftt{Nil}$ will be used instead of $\ftt{[]}$ to
denotes the empty list. This results in the equations
taking on the following guise:

\bigskip
{\leftskip=55pt
$\ftt{sqrs n \ = \ case eq n 0 of}$
\vskip 2.0pt
$\ftt{ \ \ \ \ \ \ \ \ \ \ \ \ \ \  True -> Nil}$
\vskip 2.0pt
$\ftt{ \ \ \ \ \ \ \ \ \ \ \ \ \ \  False -> Cons (sq n) (sqrs (sub n 1))}$ 
\vskip 4.0pt
$\ftt{sq n \ = \ mul n n}$
\vskip 4.0pt
$\ftt{shunt a b \ = \ case a of}$
\vskip 2.0pt
$\ftt{ \ \ \ \ \ \ \ \ \ \ \ \ \ \  Nil -> b}$
\vskip 2.0pt
$\ftt{ \ \ \ \ \ \ \ \ \ \ \ \ \ \  Cons m c -> shunt c (Cons m b)}$ 
\vskip 4.0pt
$\ftt{sqs n \ = \ shunt (sqrs n) Nil}$
\bigskip
}

The above equations are essentially built up out of names, which  
can be classified into the following four groups:
\medskip\smallskip
{\parindent=25pt
\item{---} The names $\ftt{sqrs}$, $\ftt{sq}$, $\ftt{shunt}$ and $\ftt{sqs}$, 
which are those of the functions which the equations are supposed to define.
\smallskip
\item{---} The names $\ftt{n}$, $\ftt{m}$,
$\ftt{a}$, $\ftt{b}$ and $\ftt{c}$, which play the role 
of `local variables'. 
\smallskip
\item{---} The names $\ftt{sub}$, $\ftt{mul}$ and $\ftt{eq}$, which are 
those of `primitive' operators.
\smallskip
\item{---} The names $\ftt{True}$, $\ftt{False}$, $\ftt{Nil}$, $\ftt{Cons}$,
$\ftt{0}$ and $\ftt{1}$ of `data constructors'.
\medskip\smallskip
}
In order to understand how these names are allowed to fit together each name
must be first assigned to a type. There are three basic
data types involved here which will be denoted by $\ftt{int}$ (for {\it integer\/}),
$\ftt{list}$ (for {\it integer list\/}), and
$\ftt{bool}$ (for {\it boolean\/}).
Each name which refers to a data object is assigned the corresponding type:
The names $\ftt{n}$, $\ftt{m}$, $\ftt{0}$ and $\ftt{1}$ are assigned the
type $\ftt{int}$, $\ftt{Nil}$, $\ftt{a}$, $\ftt{b}$ 
and $\ftt{c}$ the type $\ftt{list}$ and $\ftt{True}$ and $\ftt{False}$ 
are assigned the type $\ftt{bool}$. The remaining names are
names of functions, and each of these is assigned a functional type
of the form $\list {\beta} n \to \beta$, where 
$\lvector {\beta} n ,\, \beta $ are basic data types, i.e., elements of the set
$\{\ftt{int},\ftt{list},\ftt{bool}\}$.
For example, $\ftt{shunt}$ is assigned the type 
$\ftt{list}\ \ftt{list}\to \ftt{list}$, since it is the name of a function
having two arguments, both taking values of type $\ftt{list}$,
and which produces values also of 
type $\ftt{list}$. Similarly, the name $\ftt{sqrs}$ is assigned the type
$\ftt{int}\to \ftt{list}$, $\ftt{sq}$ the type
$\ftt{int}\to \ftt{int}$, $\ftt{sqs}$ the type
$\ftt{int}\to \ftt{list}$, $\ftt{sub}$ and $\ftt{mul}$ are assigned the type
$\ftt{int}\ \ftt{int} \to \ftt{int}$, $\ftt{eq}$ the type
$\ftt{int}\ \ftt{int} \to \ftt{bool}$
and, finally, $\ftt{Cons}$ is assigned the type
$\ftt{int}\ \ftt{list} \to \ftt{list}$.
\medskip
With these assignments it is easily checked that each of the individual
components of an equation is a `syntactically correct' term:
The `syntactically correct' terms of type $\ftt{list}$ occurring in the 
equations are the ten terms
\medskip
{\parindent=15pt
$\ftt{sqrs n}$,\quad $\ftt{Nil}$,\quad $\ftt{Cons (sq n) (sqrs (sub n 1))}$,
\quad 
$\ftt{shunt a b}$, \quad $\ftt{a}$,
\par
$\ftt{b}$, \quad$\ftt{Cons m c}$, \quad 
$\ftt{shunt c (Cons m b)}$, \quad 
$\ftt{sqs n}$, \quad $\ftt{shunt (sqrs n) Nil}$,
}
\medskip
those of type $\ftt{bool}$ are $\,\ftt{eq n 0}\,$, $\,\ftt{True}\,$ and 
$\,\ftt{False}\,$, while those of type $\ftt{int}$ are 
$\,\ftt{sq n}\,$ and $\,\ftt{mul n n}\,$.
\medskip
Note that both sides of the equation
for $\ftt{sq}$ are terms of type $\ftt{int}$.
Moreover, the left-hand sides of the remaining equations are all terms of type 
$\ftt{list}$, as is the right-hand side of the equation for $\ftt{sqs}$.
The same is also true of the two possible right-hand sides 
$\,\ftt{Nil}\,$ and $\,\ftt{Cons n (sqrs (sub n 1))}\,$
in the equation for $\ftt{sqrs}$, as well as of the
two possible right-hand sides 
$\,\ftt{b}\,$ and $\,\ftt{shunt c (Cons m b)}\,$ 
in the equation for $\ftt{shunt}$.
\medskip
Associated with the set $B = \{\ftt{int},\ftt{list},\ftt{bool}\}$ of basic 
data types needed in the {\it Haskell\/} 
equations is a corresponding set 
\mdisp{ K\ =\ \{\ftt{Nil},\ftt{Cons},\ftt{True},\ftt{False}\} \,\cup\, \SynInt }
of `data constructor' names, where here
$\SynInt$ is the set containing for each integer $n \in \Int$ its 
standard representation ${\underline n}$ as a string of characters. 
(Of course, the names $\ftt{Nil}$, $\ftt{Cons}$, $\ftt{True}$, $\ftt{False}$, 
$\ftt{0}$ and $\ftt{1}$ 
from $K$ already occurred in the equations.) Moreover, extending what was
done above, each element of $K$ is assigned a type: 
$\ftt{True}$ and $\ftt{False}$ are of type $\ftt{bool}$, 
${\underline n}$ is of type $\ftt{int}$ for each $n \in \Int$,
$\ftt{Nil}$ is of type $\ftt{list}$
and $\ftt{Cons}$ is of type $\ftt{int}\ \ftt{list} \to \ftt{list}$.
\medskip
The pair $\Lambda = (B,K)$, together with the assignment of the appropriate type 
to each element
of $K$, is a particular instance of what is known as a 
\indexdef{signature}{signature}{}.
\medskip
The general procedure as it would be applied
to interpreting the {\it Haskell\/}
equations for $\fss{sqrs}$, $\fss{sq}$, $\fss{shunt}$ and $\fss{sqs}$
will now be described. It involves taking the following steps:
\medskip
1.\enskip The first step is to choose what is called a 
\indexdef{\tsym{\Lambda}-algebra}{algebra}{}.
This is any pair $(X_B,p_K)$ consisting of a family of sets
$X_B = \lcurl X_\beta \,:\, \beta \in B \rcurl$, 
elements $p_{\fstt{True}}$, $p_{\fstt{False}}$ of $X_{\fstt{bool}}$,
for each $n \in \Int$ an element $p_{\underline n}$ of $X_{\fstt{int}}$,
an element $p_{\fstt{Nil}}$ of $X_{\fstt{list}}$,
and a mapping 
$p_{\fstt{Cons}} : X_{\fstt{int}}\times X_{\fstt{list}}\to X_{\fstt{list}}$.
\medskip
In all the interpretations of the {\it Haskell\/} equations
the `natural' choice of the \tsym{\Lambda}-algebra $(X_B,p_K)$ was made.
This is with: 
\medskip\smallskip
{\parindent=25pt
\item{---} $X_{\fstt{int}} = \Int$, 
$X_{\fstt{list}} = {\Bbb L} = \Int^*$ and 
$X_{\fstt{bool}} = \Bool = \{T,F\}$,
\smallskip
\item{---} $p_{\fstt{True}} = T$ and $p_{\fstt{False}} = F$, 
$p_{\fstt{Nil}} = \onept$,
\smallskip
\item{---} the mapping
$p_{\fstt{Cons}} : X_{\fstt{int}}\times X_{\fstt{list}}\to X_{\fstt{list}}$
given by $p_{\fstt{Cons}}(n,a) = n \triangleleft a$,
\smallskip
\item{---} $p_{\underline n} = n$ for each $n \in \Int$. 
\medskip\smallskip
}
In fact, this particular \tsym{\Lambda}-algebra $(X_B,p_K)$ has a certain 
universal property, that of being 
\indexddef{initial}{initial algebra}{}{algebra}{initial}, 
which essentially characterises it in the class of all \tsym{\Lambda}-algebras.
\medbreak
2.\enskip The next step is to fix for each
$\beta \in B$ an element $\bot_\beta$ not in the set $X_\beta$ (to be
thought of as an `undefined' object of type $\beta$) and then to extend
the \tsym{\Lambda}-algebra $(X_B,p_K)$ to a new
\tsym{\Lambda}-algebra $(D_B,f_K)$ with $\bot_\beta \in D_\beta$ 
for each $\beta \in B$.
By `extend' is meant that $X_\beta \subset D_\beta$ for each 
$\beta \in B$, $f_\kappa = p_\kappa$ for each 
$\kappa \in K \setminus \{\ftt{Cons}\}$ and that $p_{\fstt{Cons}}$ is the
restriction of $f_{\fstt{Cons}}$ to $X_{\fstt{int}} \times X_{\fstt{list}}$.
\medskip
In all three of the interpretations we took $D_{\fstt{int}} = \Int^\bot$ with
$\bot_{\fstt{int}} = \bot_{\Int}$, and implicitly also $D_{\fstt{bool}} = \Bool^\bot$ with
$\bot_{\fstt{bool}} = \bot_{\Bool}$. 
In the second interpretation the choice was $D_{\fstt{list}} = {\Bbb L}^\bot$ with
$\bot_{\fstt{list}} = \bot_{\Bbb L}$, in the third
$D_{\fstt{list}} = {\Bbb L}^\diamond$ with $\bot_{\fstt{list}} = \varepsilon^\bot$,
and in the fourth $D_{\fstt{list}} = {\Bbb L}^\infty$, again with
$\bot_{\fstt{list}} = \varepsilon^\bot$. In all three cases
the mapping $f_{\fstt{Cons}} : D_{\fstt{int}}\times D_{\fstt{list}}\to D_{\fstt{list}}$
was given by $\,f_{\fstt{Cons}}(x,z) \,=\, x \triangleleft z\,$ for all
$x \in D_{\fstt{int}}$, $z \in D_{\fstt{list}}$, but with
the definition of $\triangleleft$ depending on the interpretation.
\medskip\smallskip
3.\enskip The third step is to decide which kinds of functions are going to be 
allowed to occur as the solutions of equations.  
This is done by choosing for each functional type 
$\sigma = \list {\beta} n \to \beta$ an appropriate subset $D_\sigma$ 
of the set $\total {D_{\beta_1}\times \cdots \times D_{\beta_n}} {D_\beta} $
of all functions from $D_{\beta_1}\times \cdots \times D_{\beta_n}$ to 
$D_\beta$. The only functions of type $\sigma$ which will be allowed as 
solutions are then those lying in the set $D_\sigma$.
In this way the family of sets $D_B$ is extended to a family 
$D_S = \lcurl D_\sigma \,:\, \sigma \in S \rcurl$, where $S = B \cup F$ and where
$F$ is the set of functional types.
\medskip
There were no restrictions made on the functions in our
interpretations of the {\it Haskell\/} equations.
In other words, the implicit choice was to take
$D_\sigma = \total {D_{\beta_1}\times \cdots \times D_{\beta_n}} {D_\beta} $
for each functional type $\sigma = \list {\beta} n \to \beta$.
However, more typical is the
situation to be considered later in which the functions are restricted 
to being, in a certain sense, either `monotone' or `continuous'.
\medskip\smallskip
4.\enskip The next step is to interpret the names $\ftt{sub}$, $\ftt{mul}$ 
and $\ftt{eq}$ of the `primitive' operators occurring in the equations. This 
means that mappings $\fss{add}$ and $\fss{sub}$ must be chosen from 
$D_{\fstt{int}}\times D_{\fstt{int}}$ to $D_{\fstt{int}}$ as well as a mapping
$\fss{eq} : D_{\fstt{int}}\times D_{\fstt{int}} \to D_{\fstt{bool}}$.
But in practice there is really no choice here: It will always be the case that
$X_{\fstt{int}} = \Int$ and $X_{\fstt{bool}} = \Bool$ and then
$\fss{sub}$, $\fss{mul}$ and $\fss{eq}$ more-or-less have to be chosen to be the 
strict extensions of the 
operators $-$, $\times$ and $=$: The strict extension of a mapping
$p : X_{\fstt{int}} \times X_{\fstt{int}} \to X_\beta$ is the mapping
${\breve p} : D_{\fstt{int}} \times D_{\fstt{int}} \to D_\beta$ defined by
\smallskip
\mdisp{{\breve p}(m,n)\ =\ \cases{
   p(m,n) & if $\,(m,n) \in X_{\fstt{int}} \times X_{\fstt{int}}\,$,\cr
     \bot_\beta & otherwise.}}
\smallskip
(Of course, in general the rest of 
the usual arithmetical and relational operators are allowed to occur, and then these must 
also be interpreted in the same way.)
\bigskip

Suppose appropriate choices for the 
\tsym{\Lambda}-algebras $(X_B,p_K)$ and $(D_B,f_K)$ and for the family 
$D_S$ have been made. 
In particular, we assume that
$X_{\fstt{int}} = \Int$ with $p_{\underline n} = n$ for each $n \in \Int$,
$X_{\fstt{bool}} = \Bool$ with
$p_{\fstt{True}} = T$ and $p_{\fstt{False}} = F$, and that
$D_{\fstt{int}} = \Int \cup \{\bot_{\fstt{int}}\}$ and 
$D_{\fstt{bool}} = \Bool \cup \{\bot_{\fstt{bool}}\}$.
Then the interpretation of the {\it Haskell\/} equations is that they  
describe elements
$\fss{sqrs} \in D_{\fstt{int} \to \fstt{list}}$, 
$\fss{sq} \in D_{\fstt{int} \to \fstt{int}}$, 
$\fss{shunt} \in D_{\fstt{list}\ \fstt{list} \to \fstt{list}}$
and $\fss{sqs} \in D_{\fstt{int} \to\fstt{list}}$ 
(and so in particular they describe functions
$\fss{sqrs} : D_{\fstt{int}} \to D_{\fstt{list}}$,
$\fss{sq} : D_{\fstt{int}} \to D_{\fstt{int}}$,
$\fss{shunt} : D_{\fstt{list}} \times  D_{\fstt{list}}\to D_{\fstt{list}}$
and $\fss{sqs} : D_{\fstt{int}} \to D_{\fstt{list}}$)
satisfying the equations

\medskip
\mdisp{\eqalign{\fss{sqrs}\,(n)\ &=\ \cases{
      \bot_{\fstt{list}} & if $\,\fss{eq}(n,0) = \bot_{\fstt{bool}}\,$, \cr
             f_{\fstt{Nil}} & if $\,\fss{eq}(n,0) = T\,$, \cr
      f_{\fstt{Cons}}(\fss{sq}(n),\fss{sqrs}\,(\fss{sub}(n,1)) & 
           if $\,\fss{eq}(n,0) = F\,$, \cr
      } \cr
\noalign{\medskip}
    \fss{sq}\,(n)\ &=\ \fss{mul}\,(n,n)\ , \cr
\noalign{\medskip}    	
\fss{shunt}\,(a,b)\ &=\ \cases{
		   \bot_{\fstt{list}} & if $\,a = \bot_{\fstt{list}}\,$, \cr
       b & if $\,a = f_{\fstt{Nil}}\,$, \cr
       \fss{shunt}\,(c, f_{\fstt{Cons}}(m,b))
                   & if $\,a \ne \bot_{\fstt{list}}\,$ and \cr
                   & \qquad\qquad  $\,a = f_{\fstt{Cons}}(m,c)\,$, \cr
} \cr
\noalign{\smallskip}
     \fss{sqs}\,(n)\ &=\ \fss{shunt}\,(\fss{sqrs}\,(n),f_{\fstt{Nil}})
}}
\medskip
for all $n \in D_{\fstt{int}}$, $a,\, b \in D_{\fstt{list}}$.
Note that in general certain conditions must be imposed 
on the \tsym{\Lambda}-algebra $(D_B,f_K)$ for the above formulation to make sense:
The equation for $\fss{shunt}$ requires that
$f_{\fstt{Cons}}(m,c) \ne f_{\fstt{Nil}}$ for all
$(m,c) \in D_{\fstt{int}} \times D_{\fstt{list}}$
and that for each $a \in D_{\fstt{list}} \setminus \{\bot_{\fstt{list}},f_{\fstt{Nil}}\}$
there exist unique elements
$m\in D_{\fstt{int}}$, $c \in D_{\fstt{list}}$ such that
$a = f_{\fstt{Cons}}(m,c)$.
This requirement will later lead to the concept of what will be called a 
\indexddef{regular}{regular bottomed extension}{}{bottomed extension}{regular}
extension of the \tsym{\Lambda}-algebra $(X_B,p_K)$.
\medskip
The reader is left to check that the second, third and fourth interpretations of the 
{\it Haskell\/} equations introduced in the previous section
can all be obtained as special cases of the procedure outlined above.
\medskip
In general equations will be defined, as in the particular
example above, by starting with a suitable signature $\Lambda = (B,K)$,
which provides the data types that can occur in the equations 
and also the names and the types of the  corresponding 
`data constructors'. An equation then consists of a left- and a right-hand 
side, each of which has to be a `syntactically correct' term of the same type.
The terms which occur as left-hand sides have to have a particularly
simple form and there are various other conditions which have to be satisfied,
for example that the only `local variables' which can occur on the right-hand
side of an equation are those occurring on the left-hand side.
\medskip
The procedure for interpreting the equations is in principle the same as that 
described above: First choose an initial \tsym{\Lambda}-algebra
$(X_B,p_K)$, the elements in the set $X_\beta$ being the basic data objects
of type $\beta$ for each $\beta \in B$, then choose a suitable regular
extension $(D_B,f_K)$ of $(X_B,p_K)$ to take account of `undefined' and 
possibly also `partially defined' data objects, and finally choose
a family $D_S$ to determine which kinds of functions can occur as solutions. 
The interpretation of a system of equations is then
that they describe functions, each of which is a member of the appropriate set
in the family $D_S$, and which satisfy the equations obtained in the 
same way as in the particular example above. 
(Note that in general it will be possible for a system of equations 
not to have a solution.)
\medskip
The fact that $(X_B,p_K)$ is taken to be an initial \tsym{\Lambda}-algebra
implies it is essentially uniquely determined by the signature
$\Lambda$. The only real choice in the interpretation thus
lies in that of the extension $(D_B,f_K)$ and that of the family $D_S$.
\medskip
Equations and their solutions will be treated in Chapter~7. 
In the preceding chapters the necessary mathematical machinery is developed:
Chapter~2 deals with some standard topics from universal algebra, 
Chapter~3 looks at a framework for 
specifying data objects, 
Chapter~4 with a selection of elementary results from 
domain theory, Chapter~5 with the construction of function hierarchies
(which involves some very elementary category theory)
and Chapter~6 with properties of term algebras.
\medskip
In Chapter~7 an algorithm will be introduced for computing values.
This means that the equations are considered as replacement rules and an
efficient procedure for choosing the appropriate rule is specified.
Chapter~8 then shows that there is a reasonable framework in which the
denotational and operational semantics are equivalent for each system of equations. 
\medskip

The theory will be applied to a primitive language which essentially looks like
the part of {\it Haskell} employed in the examples. In this language
the equations for $\fss{sqrs}$, $\fss{sq}$, $\fss{shunt}$ and $\fss{sqs}$ 
appear in the form:
\bigskip
{\leftskip=35pt
$\ftt{sqrs n }\ =\ \ftt{ case (eq n 0) of \lcb True -> Nil,}$ 
\vskip 2.5pt
$\qquad\qquad\qquad\qquad\qquad\quad
 \ftt{False -> Cons (sq n) (sqrs (sub n 1))\rcb}$
\vskip 5.0pt
$\ftt{sq n }\ =\ \ftt{ mul n n}$
\vskip 5.0pt
$\ftt{shunt a b }\ = \ \ftt{ case a of \lcb Nil -> b, Cons -> shunx b\rcb}$
\vskip 5.0pt
$\ftt{shunx a n b }\ =\ \ftt{ shunt b (Cons n a)}$
\vskip 5.0pt
$\ftt{sqs n }\ =\ \ftt{ shunt (sqrs n) Nil}$
\bigskip
}

The additional equation is for an auxiliary function $\ftt{shunx}$ which is needed 
because, unlike in {\it Haskell\/}, in the framework to be employed here anonymous
functions cannot be defined using abstraction (which is essentially
what 
\mdisp{\ftt{Cons m c -> shunt c (Cons m b)}}

denotes). The names occurring to the left of the arrows $\,\ftt{->}\,$
are just `data constructor' names, and need not be thought of as terms.
Note that the term $\,\ftt{shunx b}\,$ appearing in the equation for $\fss{shunt}$
is an example of a partial application: $\ftt{shunx}$ is the name of a function with
three arguments and here it occurs with only one;
$\,\ftt{shunx b}\,$ is thus a term of type $\ftt{int list} \to \ftt{list}$,
where $\ftt{int list}$ (left of the arrow $\to$) give the types of the 
missing second and third arguments.
Partial applications play an essential role in situations such as this.
\medskip
The final point we want to discuss corrects the impression which has been made so far
about the kinds of functions which will occur in the equations. 
We have stated that each functional type is of the form
$\sigma = \list {\beta} n \to \beta$ with $\list {\beta} n ,\,\beta$ basic data
types, and that the corresponding set of functions $D_\sigma$ is chosen to be
an appropriate subset
of the set of all functions from $D_{\beta_1}\times \cdots \times D_{\beta_n}$ to 
$D_\beta$. This is sufficient for the examples considered above, but in general
it is more than useful to allow so-called higher order types, i.e., functions
whose arguments can also be functions. 
The functional types to be introduced in Chapter~5 will certainly include 
higher-order types. Example~1.2.1 on the following page
gives a typical instance of a function having a higher-order type.

\vfill\eject

\bigskip\medskip
\frame{20pt}{\bigskip 
{\it Example~1.2.1\enspace} 
The equations for $\fss{sqrs}$, $\fss{sq}$, $\fss{shunt}$ and $\fss{sqs}$ 
can be replaced by the following more-or-less equivalent equations:
\bigskip
{\leftskip=35pt
$\ftt{revs p n }\ =\ \ftt{ case (eq n 0) of \lcb True -> Nil,}$ 
\vskip 2.5pt
$\qquad\qquad\qquad\qquad\qquad\quad
 \ftt{False -> Cons (p n) (revs p (sub n 1))\rcb}$
\vskip 5.0pt
$\ftt{sq n }\ =\ \ftt{ mul n n}$
\vskip 5.0pt
$\ftt{shunt a b }\ = \ \ftt{ case a of \lcb Nil -> b, Cons -> shunx b\rcb}$
\vskip 5.0pt
$\ftt{shunx a n b }\ =\ \ftt{ shunt b (Cons n a)}$
\vskip 5.0pt
$\ftt{sqs n }\ =\ \ftt{ shunt (revs sq n) Nil}$
\bigskip
}
The equation for $\ftt{revs}$ describes a function of
type $(\ftt{int} \to \ftt{int})\ \ftt{int} \to \ftt{int}$, i.e., a function
from $D_{\fstt{int}\to \fstt{int}} \times D_{\fstt{int}}$ to $D_{\fstt{int}}$. 
\medskip
In fact these equations describe elements
\medskip
{\leftskip=30pt
$\fss{revs} \in D_{(\fstt{int}\to \fstt{int})\ \fstt{int} \to \fstt{list}}$, 
$\fss{sq} \in D_{\fstt{int} \to \fstt{int}}$, 
\smallskip
$\fss{shunt} \in D_{\fstt{list}\ \fstt{list} \to \fstt{list}}$, \smallskip
$\fss{shunx} \in D_{\fstt{list}\ \fstt{int}\ \fstt{list} \to \fstt{list}}$
and $\fss{sqs} \in D_{\fstt{int} \to\fstt{list}}$, 
\medskip
}
and so in particular they describe functions
\medskip
{\leftskip=30pt
$\fss{revs} : D_{\fstt{int}\to \fstt{int}} \times D_{\fstt{int}} \to D_{\fstt{int}}$,
$\fss{sq} : D_{\fstt{int}} \to D_{\fstt{int}}$,
\smallskip
$\fss{shunt} : D_{\fstt{list}} \times  D_{\fstt{list}}\to D_{\fstt{list}}$,
\smallskip
$\fss{shunx} : D_{\fstt{list}} \times  D_{\fstt{int}}\times  D_{\fstt{list}}
\to D_{\fstt{list}}$
and $\fss{sqs} : D_{\fstt{int}} \to D_{\fstt{list}}$
\medskip
}
satisfying the equations
\medskip
\mdisp{\eqalign{\fss{revs}\,(p,n)\ &=\ \cases{
      \bot_{\fstt{list}} & if $\,\fss{eq}(n,0) = \bot_{\fstt{bool}}\,$, \cr
             f_{\fstt{Nil}} & if $\,\fss{eq}(n,0) = T\,$, \cr
      f_{\fstt{Cons}}(p(n),\fss{revs}\,(p,\fss{sub}(n,1)) & 
           if $\,\fss{eq}(n,0) = F\,$, \cr
      } \cr
\noalign{\medskip}
    \fss{sq}\,(n)\ &=\ \fss{mul}\,(n,n)\ , \cr
\noalign{\medskip}    	
\fss{shunt}\,(a,b)\ &=\ \cases{
		   \bot_{\fstt{list}} & if $\,a = \bot_{\fstt{list}}\,$, \cr
       b & if $\,a = f_{\fstt{Nil}}\,$, \cr
       \fss{shunx}\,(b,m,c)
                   & if $\,a \ne \bot_{\fstt{list}}\,$ and \cr
                   & \qquad\qquad  $\,a = f_{\fstt{Cons}}(m,c)\,$, \cr
} \cr
\noalign{\smallskip}
     \fss{shunx}\,(a,n,b)\ &=\ \fss{shunt}\,(b, f_{\fstt{Cons}}(n,a)),\cr
\noalign{\smallskip}
     \fss{sqs}\,(n)\ &=\ \fss{shunt}\,(\fss{revs}\,(\fss{sq},n),f_{\fstt{Nil}})
}}
\medskip
for all $n \in D_{\fstt{int}}$, $a,\, b \in D_{\fstt{list}}$, 
$p \in D_{\fstt{int}\to\fstt{int}}$.
\medskip
\bigskip\medskip
}

\vfill\eject

\hfline{17}{1.3 \ NOTES}

\sectionhead {1.3} {Notes}
\bigskip\medskip
The reader with no previous experience of functional programming is   
recommended to look at Bird and Wadler (1988) before proceeding any further. 
Other books offering good introductions to functional programming from 
various viewpoints are
Field and Harrison (1988), Glaser, Henkin and Till (1984),
Henderson (1980), Holyer (1993), MacLennan (1990), Paulson (1991), 
Reade (1989) and Wikstr\"om (1987). 
\medskip
The approach described in this chapter has similarities with
that taken in Rosen (1973). The equivalence of the denotational and operational
semantics corresponds to Rosen's {\it Validity Theorem\/} (Theorem~8.4), a result
which was conjectured in Morris (1968). Rosen
points out, moreover, the connections with Kleene's first recursion theorem
(Kleene (1952), \S 66), which can be seen as the fore-runner of the material
to be presented in Chapter~7. 

\vfill\eject

\hfline{18}{}

\bigskip
{\fourteenbf Chapter 2 \quad Some universal algebra}
\bigskip\medskip
In this chapter those basic facts from universal algebra are presented
which will be needed in the subsequent chapters. The emphasis here is on initial algebras 
in their various forms.

\medskip
In order to give a foretaste of the kind of universal algebra which will occur,
it is perhaps instructive to start with a very familiar situation 
and serve it up in a somewhat unusual guise.
(What follows is, however, certainly not new: It appears in essentially this 
form in Dedekind's book {\it Was sind und was sollen die Zahlen?} first 
published in 1888.)
\medskip
Consider the set 
$\Nat = \{0,1,2,\ldots\}$ of natural numbers together with its first element
$0$ and the
successor function $\fss{s} : \Nat \to \Nat$ (defined by
$\fss{s}(n) = n + 1$ for each $n \in \Nat$).
It can then be asked what is special about the triple $(\Nat,0,\fss{s})$.
To make things a  bit more precise, define a triple
$(X,e,p)$ consisting of a non-empty set $X$, an element $e \in X$ and
a mapping $p : X \to X$ to be a 
\indexdef{natural number triple}{natural number triple}{}, 
and then ask
how $(\Nat,0,\fss{s})$ can be characterised in the class of all
such triples.
\medskip
To answer this question the  appropriate structure-preserving mappings must first be 
introduced: If
$(X,e,p)$ and $(X',e',p')$ are natural number triples 
then a mapping $\pi : X \to X'$ is called a
\definition{homomorphism} from $(X,e,p)$ to $(X',e',p')$
if $\pi(e) = e'$ and $p'\,\pi = \pi\,p$ (i.e., $p'(\pi(x)) = \pi(p(x))$ 
for all $x \in X$).
If $\pi$ is such a homomorphism and the mapping $\pi$ is a bijection
then it is easy to see that the inverse mapping
$\pi^{-1} : X' \to X$ is a homomorphism from $(X',e',p')$ to $(X,e,p)$.
In this case $\pi$ is called an \definition{isomorphism}, and
$(X,e,p)$ are $(X',e',p')$ said to be \definition{isomorphic}.
\medskip
The identity mapping is a homomorphism and the composition of two 
homomorphisms is again a homomorphism; being isomorphic thus defines an 
equivalence relation on the class of all natural number triples.
Now if the equivalence class containing  $(\Nat,0,\fss{s})$ can be identified
in some reasonable way then the question asked above can be considered 
to have been answered satisfactorily. Two simple characterisations of this
equivalence class are given below.
\medskip
A natural number triple $(X,e,p)$ is said to be 
\indexddef{initial}{initial natural number triple}{}{natural number triple}{initial}
if for each such triple
$(X',e',p')$ there is a unique homomorphism from
$(X,e,p)$ to $(X',e',p')$. Then $(\Nat,0,\fss{s})$ is initial, since
given $(X',e',p')$, a homomorphism from
$(\Nat,0,\fss{s})$ to $(X',e',p')$ can be defined by induction,
and its uniqueness also follows by induction.
It follows that a triple is isomorphic to
$(\Nat,0,\fss{s})$ if and only if it is initial. In other words,
the equivalence class containing 
$(\Nat,0,\fss{s})$ consists of exactly all the initial
natural number triples. This was the first characterisation.
\medskip
Here is the second characterisation: A natural number triple
$(X,e,p)$ will be called a 
\indexdef{Peano triple}{peano triple}{}
if the following three conditions
are satisfied:
\medskip\smallskip
{\parindent=25pt
\item{(1)} The mapping $p$ is injective.
\medskip
\item{(2)} $p(x) \ne e$ for all $x \in X$.
\medskip
\item{(3)} The only subset $X'$ of $X$ containing $e$ with 
$p(x) \in X'$ for all $x \in X'$ is the set $X$ itself.
\medskip\smallskip
}
Then $(\Nat,0,\fss{s})$ is a Peano triple. 
(This can be seen as one of the possible 
formulations of the Peano axioms. In particular, the statement that
$(\Nat,0,\fss{s})$ satisfies (3) is nothing but the principle of 
mathematical induction.) The reader is left to show that
a triple is isomorphic to
$(\Nat,0,\fss{s})$ if and only if it is itself a Peano triple. In other words,
the equivalence class containing 
$(\Nat,0,\fss{s})$ also consists of exactly all the Peano triples.
\medskip
Of course, a corollary of these two characterisations is that a
natural number triple is initial if and only if it is a Peano triple.
This is a special case of an important result presented in 
Section~2.3 which states that an algebra is initial if and only if
(in the terminology employed there) it is unambiguous and minimal.
For a general algebra
unambiguity corresponds to conditions (1) and (2) in the definition
of a Peano triple, while  minimality corresponds to condition (3).
This characterisation is sometimes expressed by saying that initial algebras
are exactly those for which there is {\it no confusion\/} (unambiguity) and 
{\it no junk\/} (minimality).
\bigskip
Section~2.1, as well as fixing some general notation, introduces the simple notion
of a typed set. 
This takes into account the situation met with
in most modern programming languages in which each name occurring in a 
program is assigned, either explicitly or implicitly, a type.
The kind of object a name can refer to is then determined by its type.
\medskip
In Section~2.2 signatures, algebras and homomorphisms are introduced,  and
invariant families (i.e., subalgebras) are also looked at. Section~2.3 then deals with 
initial 
and free algebras, and in Section~2.4 term algebras are introduced in order to give simple
explicit examples of initial algebras. 
Finally, in Section~2.5 something is said
about extensions of signatures and algebras, 
and Section~2.6 presents a somewhat
idiosyncratic approach to tree algebras. 
\medskip
Except for Section~2.6 (which can be omitted) the choice of material in this chapter
is determined entirely by what will be needed later. The reader interested in a more
balanced account of modern universal algebra should consult the books
referenced at the end of the chapter.

\vfill\eject

\hfline{20}{2.1 \ TYPED SETS}

\sectionhead {2.1} {Typed sets}
\bigskip\medskip
Before beginning in the next section with universal algebra proper 
the simple notion of a typed set will be introduced. 
This takes into account the situation met with
in most modern programming languages in which each name occurring in a 
program is assigned, either explicitly or implicitly, a type.
The kind of object a name can refer to is then determined by its type.
\medskip
The first task, however, is to fix some notation. The empty set will be denoted by
$\varnothing$ and the set $\{\varnothing\}$ by $\Oneptset$; thus $\Oneptset$ is the
`standard' set containing exactly one element. However, to avoid confusion
the single element in $\Oneptset$ will be denoted by $\onept$ (rather than by
$\varnothing$).
\medskip
The words 
\indexdef{function}{function}{}
and 
\indexdef{mapping}{mapping}{}  
are considered to be synonyms. 
Nevertheless, 
in this chapter almost exclusive  use of the word {\it mapping} will be made.
({\it Functions\/} are what the equations are supposed to define.)
If $X$ and $Y$ are sets then $\total X Y $ will be used to 
denote the set of all mappings from $X$ to $Y$. 
In particular, $\total {\varnothing} Y = \Oneptset$ for each set $Y$.
\medskip
Let $f : X \to Y$ be a mapping; then $X$ is called the 
\indexdef{domain}{domain}{} 
of $f$ and will be denoted by $\dom(f)$ and $Y$ is called the 
\indexdef{codomain}{codomain}{}.
For each $A \subset X$ put
\mdisp{f(A)\ =\ \lcurl y \in Y\,:\, y = f(x) \frm{for some} x \in A \rcurl}
and for each $B \subset Y$ put
$f^{-1}(B) = \lcurl x \in X\,:\, f(x) \in B \rcurl$.
The subset $f(X)$ of the codomain $Y$ is called the 
\indexdef{image}{image}{} 
of $f$ and will be denoted by $\Im(f)$. 
\medskip 
Let $S$ be a set and suppose for each $\sigma \in S$ some object
$\alpha_\sigma$ is given (which will usually be a set or a mapping). Then 
by the 
\indexdef{family}{family}{} 
$\lcurl \alpha_\sigma \,:\, \sigma \in S \rcurl$
is  meant the mapping from $S$ to the set consisting of the objects
$\alpha_\sigma$, $\sigma \in S$, which assigns the object $\alpha_\sigma$
to each $\sigma \in S$. Unless something to the contrary is explicitly 
stated this family will be denoted by $\alpha_S$.
If $X_S$ and $Y_S$ are families of sets then $Y_S \subset X_S$ will mean that 
$Y_\sigma \subset X_\sigma$ for each $\sigma \in S$. This will also be indicated by
saying that $X_S$ 
\definition{contains} $Y_S$ or that $Y_S$ 
\indexdef{is contained in}{family}{contained in a family}
$X_S$.
\medskip
The topic announced in the title of this section will now be considered.
A pair $(I,\langle \cdot \rangle)$ consisting of a set 
$I$ and a mapping $\langle \cdot \rangle : I \to S$ is called an 
\indexddef{\tsym{S}-typed set}{typed set}{}{set}{typed} 
with 
\indexdef{typing}{typing}{}
$\langle \cdot \rangle$. The set $S$ should here be 
thought of as a set of `types', $I$ as a set of `names' and 
$\langle \cdot\rangle$ as specifying the type of objects to which the `names' 
can be assigned.
If $\eta \in I$ with $\langle \eta \rangle = \sigma$ then $\eta$ is said to be
\indexdef{of type}{type}{of element in a typed set} 
$\sigma$.
\medskip
The \tsym{S}-typed set $(I,\langle \cdot \rangle)$ will usually just be denoted by
$I$, it being assumed that the typing
$\langle \cdot \rangle$ can be inferred from the context.
In fact, except 
where there is some danger of confusion, $\langle \cdot \rangle$ will always be used to 
denote the typing on a typed set. The empty set $\varnothing$ will be regarded as 
an \tsym{S}-typed set (with the corresponding unique typing). 
\medskip
Note that an ordinary set can (and will) be considered as an 
\tsym{\Oneptset}-typed set, since for any set $I$ there is a unique
typing $\langle\,\cdot\,\rangle : I \to \Oneptset$.
\medskip
If $\alpha_S = \lcurl \alpha_\sigma : \sigma \in S \rcurl$ is a family of objects and
$I$ is an \tsym{S}-typed set then, in order to increase the legibility, it is convenient
to omit the brackets and just use $\alpha_\eta$ to denote the object 
$\alpha_{\langle\eta\rangle}$ for each $\eta \in I$. 

\medskip 
Let $X_S = \lcurl X_\sigma \,:\, \sigma \in S \rcurl$ be a family of sets and
$I$ be an \tsym{S}-typed set. Then $\ass I X $ will denote the set of all 
\indexddef{typed}{typed mapping}{}{mapping}{typed}
mappings from $I$ to 
$\bigcup_{\eta \in I} X_\eta$, i.e., the set of mappings 
$c : I \to \bigcup_{\eta\in I} X_\eta$ such that 
$c(\eta) \in X_\eta$ for each $\eta \in I$. The elements 
of $\ass I X $ are called 
\indexdef{assignments}{assignment}{}.
An assignment $c \in \ass I X $ thus assigns to each `name' $\eta \in I$ an element 
$c(\eta) \in X_\eta$ of the appropriate type. 
Note that $\ass {\varnothing} X = \Oneptset$; moreover,
if $Y_S \subset X_S$ then $\ass I Y $ can clearly be regarded as a subset of 
$\ass I X $.
\medskip
The symbol $\scriptstyle \diamond$ in the expression $\ass I X $ 
is a reminder that the $X$ occurring here only has a meaning
within the context of the family $X_S$. 
It also helps to distinguish $\ass I X $  from the notation $\total I X $ 
used for the set of all mappings from the set $I$ to the set $X$.
(Note that if $J$ and $Y$ are sets, and $J$ is considered as an
\tsym{\Oneptset}-typed set then in fact $\total J Y = \ass J Y $, provided 
$Y$ is identified with the family $Y_\Oneptset$ given by $Y_\onept = Y$.)

\medskip

When dealing with explicit examples, in which case the set $I$ is always finite, an
assignment $c \in \ass I X $ will be specified by  choosing an enumeration 
$\lvector {\eta} m $ of the elements of $I$ and then
writing it as
$\, \{ \eta_1 \to c(\eta_1),\, \eta_2 \to c(\eta_2),\,
               \cdots \,,\,\eta_m \to c(\eta_m) \}\,$.
\medskip

Recall that $\Nat$ is the set $\{0,1,2,\ldots\}$, i.e., $0$ is a natural number. 
For each $n \in \Nat$ let 
$[n] = \{1,2,\ldots,n\}$; in particular $[0] = \varnothing$.
Let $n \in \Nat$ and for each $\oneto j n $ let $X_j$ be a set;
then the 
\indexddef{cartesian product}{cartesian product}{}{product}{cartesian}
$X_1 \times \cdots \times X_n$ is defined to be the set of all mappings 
$\varrho$ from $[n]$ to $\bigcup_{j=1}^n X_j$ such that $\varrho(j) \in X_j$ for
each $j$.
In particular, if $n = 0$ then
$X_1 \times \cdots \times X_n = \Oneptset$.
For each $j$ let $x_j \in X_j$; then
as usual $(\vector x n )$ denotes the the element
$\varrho$ of $X_1 \times \cdots \times X_n$ such that
$\varrho(j) = x_j$ for each $j$; clearly each element of 
$X_1 \times \cdots \times X_n$ has a unique representation of this form.
\medskip
Of course, the cartesian product is just a special case of a set of
assignments: Here the family of sets is
$X_{[n]} = \lcurl X_j \,:\, j \in [n] \rcurl$ and then
$X_1 \times \cdots \times X_n = \ass {[n]} X $, where $[n]$ is 
considered as an \tsym{[n]}-typed set with the identity typing.
\medskip
The simplest case of a cartesian product is when the sets
$\lvector X n $ are all the same:
Let $X$ be a set; then for each $n \in \Nat$ the $n$-fold cartesian product of $X$ 
with itself will be denoted by $X^n$, thus 
$X^n$ is the set of all mappings from $[n]$ to $X$. 
In particular, $X^0 = \Oneptset$; moreover, it is convenient to identify $X^1$ with $X$ in
the obvious way. Note that each element of $X^n$ can be regarded as a finite 
\tsym{X}-typed set:
More precisely, $(\vector x n ) \in X^n$ is the typed set
consisting of the set $[n]$ with the typing which assigns to
each $j$ the type $x_j$.  

\medskip
For each set $X$ the set $\bigcup_{n \ge 0} X^n$ will be denoted by 
$X^*$, which should be thought of as the set of all finite lists of elements 
from $X$. Note that since $X^1$ is being identified with $X$, it follows  that
$X$ is a subset of $X^*$. (In other words, each element $x$ of 
$X$ is identified with the list whose single component is equal to $x$.)
If $m \ge 1$ then the element 
$(\vector x m )$ of $X^m \subset X^*$ will always be denoted simply by $\list x m $. 
Following the remark in the previous paragraph 
each list of elements from $X$ is a finite \tsym{X}-typed set:
The list $\list x n $ is the \tsym{X}-typed set consisting of the
set $[n]$ with the typing which assigns to
each $j$ the type $x_j$.  
\medskip

Let $X_S$ be a family of sets and $\gamma = \list {\sigma} n \in S^*$; 
then, considering $\gamma$ as an \tsym{S}-typed set, 
$\ass {\gamma} X = Y_1 \times \cdots \times Y_n$ with $Y_j = X_{\sigma_j}$
for each $\oneto j n $, which leads to the usual notation 
$X_{\sigma_1} \times \cdots \times X_{\sigma_n}$ for the set $\ass {\gamma} X $.
\medskip
In order to avoid any foundational difficulties it is convenient to fix
once and for all some large set $\Bbb{M}$ of names, and then only to consider
\tsym{S}-typed sets whose underlying sets are subsets of $\Bbb{M}$.
This means that there is the set (and not just the class) of all \tsym{S}-typed sets.
It is assumed that $\Nat \subset \Bbb{M}$ and (in order to give explicit examples)
that $\Bbb{A}^* \setminus \{\onept\}\subset \Bbb{M}$ for some suitable alphabet
$\Bbb{A}$ containing the usual {\it ASCII\/}-symbols. 
\medskip
The set of all finite \tsym{S}-typed sets will be denoted by ${\cal F}_S$. 
Since $\Nat \subset \Bbb{M}$ it follows in particular 
that $S^n \subset {\cal F}_S$ for each $n \in \Nat$ and also $S^* \subset {\cal F}_S$. 
(Beware though that the list $\onept$ with no components is then the
\tsym{S}-typed set $\varnothing$.)
\medskip

Now let $X_S$ and $Y_S$ be families of sets and 
$\varphi_S : X_S \to Y_S$ be a family of mappings, i.e., 
$\varphi_\sigma : X_\sigma \to Y_\sigma$ for each $\sigma \in S$. 
Then for each \tsym{S}-typed set
$I$ there is a mapping $\ass I {\varphi} : \ass I X \to \ass I Y $ 
defined for each $c \in \ass I X $, $\eta \in I$ by

\mdisp{\ass I {\varphi} (c)\,(\eta)\ =\ \varphi_\eta (c(\eta))}

(recalling that $\varphi_\eta = \varphi_{\langle\eta\rangle}$).
Note that if $\gamma = \list {\sigma} n \in S^*$ then 
$\ass {\gamma} {\varphi} $ is just the mapping from 
$\ass {\gamma} X  = X_{\sigma_1} \times \cdots \times X_{\sigma_n}$ to 
$\ass {\gamma} Y = Y_{\sigma_1} \times \cdots \times Y_{\sigma_n}$ defined by

\mdisp{\ass {\gamma} {\varphi} (\vector x n )\ =\ (\varphi_{\sigma_1}(x_1),
                                    \ldots , \varphi_{\sigma_n}(x_n))}

for each $(\vector x n ) \in \ass {\gamma} X $.

\proclaim{Lemma~2.1.1} (1)\enskip Let $X_S$ be a family of
sets and $\fss{id}_S : X_S \to X_S$ be the family of identity mappings
(i.e., with $\fss{id}_\sigma : X_\sigma \to X_\sigma$ the identity mapping
for each $\sigma \in S$). Then $\ass I {\fss{id}} : \ass I X  \to \ass I X $ 
is also the identity mapping for each \tsym{S}-typed set $I$.
\medskip
(2)\enskip
Let $X_S$, $Y_S$ and $Z_S$ be families of
sets and let $\varphi_S : X_S \to Y_S$ and $\psi_S : Y_S \to Z_S$ be
families of mappings. Then for each \tsym{S}-typed set $I$
\mdisp{\ass I {(\psi\varphi)} \ =\ \ass I {\psi} \ass I {\varphi} }
where 
$(\psi\varphi)_S : X_S \to Z_S$  is the family given by
$(\psi\varphi)_\sigma = \psi_\sigma\,\varphi_\sigma$
for each $\sigma \in S$.
\medskip
(3)\enskip Let $X_S$ and $Y_S$ be families of
sets and $\varphi_S : X_S \to Y_S$ be a family of mappings.
If the
mapping $\varphi_\sigma : X_\sigma \to Y_\sigma$ is injective 
(resp.\ surjective) for 
each $\sigma \in S$ then for each \tsym{S}-typed set $I$
the mapping
$\ass I {\varphi} : \ass I X \to \ass I Y $ is also injective 
(resp.\ surjective). In particular, if
$\varphi_\sigma : X_\sigma \to Y_\sigma$ is a bijection for 
each $\sigma \in S$ then the mapping
$\ass I {\varphi} : \ass I X \to \ass I Y $ is a bijection, and in this 
case 
\mdisp{(\ass I {\varphi} )^{-1}\ =\ \assb I {\varphi^{-1}} \  .}
\endpro

\proof (1)\enskip This is clear.  
\medskip
(2)\enskip Let $c \in \ass I X $ and $\eta \in I$; then
\smallskip
\ldisp{\quad
\ass I {(\psi\varphi)} (c)\,(\eta)
\ =\ (\psi_\eta \,\varphi_\eta )\,(c(\eta))
\ =\ \psi_\eta (\varphi_\eta (c(\eta)))}
\vskip-\medskipamount
\rdisp{
\ =\ \psi_\eta (\ass I {\varphi} (c))\,(\eta)
\ =\ \ass I {\psi} (\ass I {\varphi} (c))\,(\eta)
\ =\ (\ass I {\psi} \ass I {\varphi} )(c)\,(\eta)\quad}
\smallskip
and thus
$\ass I {(\psi\varphi)} \, =\, \ass I {\psi} \ass I {\varphi} $.
\medskip
(3)\enskip If $\varphi_\sigma$ is injective for each $\sigma$ 
then there exists a family of mappings $\psi_S : Y_S \to X_S$ such that
$\psi_\sigma\,\varphi_\sigma$ is the identity mapping on $X_\sigma$
for each $\sigma \in S$. Then by (1) 
$\ass I {(\psi\varphi)} $ is the identity mapping on $\ass I X $ and by
(2) $\ass I {\psi} \,\ass I {\varphi} = \ass I {(\psi\varphi)} $. 
Hence $\ass I {\varphi} $ is injective.
The other case is almost identical: 
If $\varphi_\sigma$ is surjective for each $\sigma$ 
then there exists a family of mappings $\psi_S : Y_S \to X_S$ such that
$\varphi_\sigma\,\psi_\sigma$ is the identity mapping on $Y_\sigma$
for each $\sigma \in S$, and as in the first part it then follows that
$\ass I {\varphi} \,\ass I {\psi} $ is the identity mapping on $\ass I Y $.
Finally, if $\varphi_\sigma$ is a bijection for each $\sigma \in S$
then $\assb I {\varphi^{-1}} \ass I {\varphi} $ is the identity mapping on 
$\ass I X $ and 
$\ass I {\varphi} \assb I {\varphi^{-1}} $ is the identity mapping on 
$\ass I Y $, and
thus $(\ass I {\varphi} )^{-1}\, =\, \assb I {\varphi^{-1}} $. \eop
\bigbreak
Some of the mappings which will occur involve the braces 
$\sem{\,\frm{and}\, }$, usually
ornamented with subscripts and often also superscripts. In this case
the value of the mapping is denoted by placing the argument between the braces.
Consider, for instance, a family of mappings $\sem{\cdot}_S$,
with $\sem{\cdot}_\sigma$ a mapping from
$X_\sigma$ to  $Y_\sigma$ for each $\sigma \in S$; 
then the element of $Y_\sigma$ which is obtained
by applying $\sem{\cdot}_\sigma$ to $x \in X_\sigma$ is denoted here by
$\sem{x}_\sigma$. Now if $I$ is an \tsym{S}-typed set then the corresponding mapping from
$\ass I X $ to $\ass I Y $ will be denoted as above by $\ass I {\sem{\cdot}} $, and its
argument will again be placed between the braces. Thus if $c \in \ass I X $ then
$\ass I {\sem{c}} $ is the element of $\ass I Y $ given by
$\,\ass I {\sem{c}} (\eta) \,=\, \sem{c(\eta)}_\eta\,$ for each $\eta \in I$.

\bigskip
Let $X_S$ be a family of sets and let $\sqle_S$ be a corresponding family of partial 
orders, i.e., $\sqle_\sigma$ is a partial order on $X_\sigma$ for each $\sigma \in S$.
Then for each \tsym{S}-typed set $I$ there is a natural partial order $\ass I {\sqle} $ 
on the set $\ass I X $ defined by stipulating that $c \ass I {\sqle} c' $ if and only if
$c(\eta) \sqle_\eta  c'(\eta)$ for each $\eta \in I$.
\medskip
Let $n \ge 2$ and for each $\oneto j n $ let $\sqle_j$ be a partial order on the
set $Y_j$. Then the above construction produces the usual
\indexddef{product partial order}{product partial order}{}{partial order}{product}
$\sqle$ on $Y_1 \times \cdots \times Y_n$ in which 
$(\vector y n ) \sqle (\vector {y'} n )$ if and only if $y_j \sqle_j y'_j$ for 
each $j$.
\bigskip
The final topic in this section is a simple observation which will be applied throughout
the study. Suppose that a number of objects from a given family of sets $X_S$ are to 
be `defined'. Then this can be organised by first choosing an 
appropriate \tsym{S}-typed $I$ consisting of the `names' of these objects, and 
then `constructing' an assignment $c \in \ass I X $ which
gives a meaning to each `name' (and thus `defines' the objects $c(\eta)$, 
$\eta \in I$). In particular, this device will be employed to formulate the equations 
which appear in Chapter~7 so that their solutions
are assignments rather than sets of functions. A  reformulation in this spirit
of the equations considered in Chapter~1 is given in Example~2.1.1 below.
\vfill\eject
\bigskip\medskip
\frame{20pt}{\bigskip 
{\it Example~2.1.1\enspace} Let $(D_B,f_K)$ and the family
$D_S$ be as in Section~1.2 (for the case of the equations for
$\fss{revs}$, $\fss{sq}$, $\fss{shunt}$, $\fss{shuntx}$). Let
\mdisp{I\  =\  \{\,\ftt{revs},\,\ftt{sq},\,\ftt{shunt},\,
                        \ftt{shunx},\, \ftt{sqs}\,\}}
be the \tsym{S}-typed set with 
\medskip
{\leftskip=30pt
$\ftt{revs}$ of type $(\ftt{int}\to\ftt{int})\ \ftt{int} \to \ftt{list}$, \smallskip
$\ftt{sq}$ of type $\ftt{int} \to \ftt{int}$,
$\ftt{shunt}$ of type $\ftt{list}\ \ftt{list} \to \ftt{list}$, \smallskip
$\ftt{shunx}$ of type $\ftt{list}\ \ftt{int}\ \ftt{list} \to \ftt{list}$
and $\ftt{sqs}$ of type $\ftt{int} \to \ftt{list}$.
\medskip
}
\smallskip
If $c \in \ass I D $ and $\eta \in I$ then, since it increases the legibility,
$c(\eta)$ will usually be written as $c_\eta$; this means that
\medskip
{\leftskip=30pt
$c_{\fstt{revs}} \in D_{(\fstt{int}\to\fstt{int})\ \fstt{int}\to\fstt{list}} $, 
$c_{\fstt{sq}} \in D_{\fstt{int}\to\fstt{int}} $,\smallskip
$c_{\fstt{shunt}} \in D_{\fstt{list}\ \fstt{list}\to\fstt{list}} $,\smallskip
$c_{\fstt{shunx}} \in D_{\fstt{list}\ \fstt{int}\ \fstt{list}\to\fstt{list}}$
and $c_{\fstt{sqs}} \in D_{\fstt{int}\to\fstt{list}} $.
\medskip
}
The equations for $\fss{revs}$, $\fss{sq}$, $\fss{shunt}$, $\fss{shunx}$ 
and $\fss{sqs}$ in Example~1.2.1 can now be considered as the following 
equations for an assignment $c \in \ass I D $:
\medskip\smallskip
\mdisp{
\eqalign{
c_{\fstt{revs}}\,(p,n)\ &=\ \cases{
      \bot_{\fstt{list}} & if $\,\fss{eq}(n,0) = \bot_{\fstt{bool}}\,$, \cr
             f_{\fstt{Nil}} & if $\,\fss{eq}(n,0) = T\,$, \cr
      f_{\fstt{Cons}}(p(n), c_{\fstt{rev}}\,(p,\fss{sub}(n,1))) 
                     & if $\,\fss{eq}(n,0) = F\,$, \cr
      } \cr
\noalign{\smallskip}    	
    c_{\fstt{sq}}\,(n)\ &=\ \fss{mul}(n,n)\ , \cr
\noalign{\smallskip}
c_{\fstt{shunt}}\,(a,b)\ &=\ \cases{
      \bot_{\fstt{list}} & if $\,a = \bot_{\fstt{list}}\,$, \cr
       b & if $\,a = f_{\fstt{Nil}}\,$, \cr
       c_{\fstt{shunx}}\,(d,m,b)
                   & if $\,a \ne \bot_{\fstt{list}}\,$ and 
		   $\,a = f_{\fstt{Cons}}(m,d)\,$, \cr
} \cr
\noalign{\smallskip}    	
    c_{\fstt{shunx}}\,(a,n,b)\ &=
      \ c_{\fstt{shunt}}\,(b,f_{\fstt{Cons}}(n,a))\ , \cr
\noalign{\smallskip}
    c_{\fstt{sqs}}\,(n)\ &=\ c_{\fstt{shunt}}\,(c_{\fstt{revs}}\,(c_{\fstt{sq}},n),
             f_{\fstt{Nil}})
}}
\medskip\smallskip
for all $a,\,b \in D_{\fstt{list}}$, $n \in D_{\fstt{int}}$, 
$p \in D_{\fstt{int}\to\fstt{int}}$.  
\bigskip\medskip
}

\vfill\eject

\hfline{25}{2.2 \ ALGEBRAS AND HOMOMORPHISMS}

\sectionhead {2.2} {Algebras and homomorphisms}
\bigskip\medskip
The structure which plays a fundamental role in all of what follows is that 
of an algebra associated with a signature. A 
\indexdef{signature}{signature}{} 
is a quadruple $\Sigma = (S,N,\Delta,\delta)$, where $S$ and 
$N$ are non-empty sets and $\Delta : N \to {\cal F}_S$ and
$\delta : N \to S$ are mappings (recalling that ${\cal F}_S$ is the set of
all finite \tsym{S}-typed sets).
The set $S$ should here be thought of as a set of 
\indexdef{types}{type}{}; 
$N$ can be regarded as a set of 
\indexdef{operator names}{operator name}{}.
\medskip
For each $\sigma \in S$ put
$N_\sigma = \lcurl \nu \in N\,:\, \delta(\nu) = \sigma \rcurl$.
Moreover, if $\nu \in N$ with
$\delta(\nu) = \sigma$ and $\Delta(\nu) = L$ then $\nu$ is said to be
\indexdef{of type}{type}{of operator name} 
$L \to \sigma$. 
This terminology will be used to help avoid mentioning $\Delta$ and $\delta$ explicitly.
\medskip
If $\Sigma = (S,N,\Delta,\delta)$ is a signature then a 
\indexdef{\tsym{\Sigma}-algebra}{algebra}{} 
is any pair $(X_S,p_N)$ consisting of a 
family of sets $X_S = \lcurl X_\sigma \,:\, \sigma \in S \rcurl$ and a 
family of mappings 
$p_N = \lcurl p_\nu \,:\, \nu \in N \rcurl$ such that if $\nu \in N$ is 
of type $L \to \sigma$ then $p_\nu$ is a mapping from 
$\ass L X $ to $X_\sigma$. For each $\sigma \in S$
the set $X_\sigma$ should be thought of as a set of elements of type 
$\sigma$ and for each $\nu \in N$ the mapping 
$p_\nu$ can be thought of as the operator  
corresponding to the operator name $\nu$.
\medskip
A signature $\Sigma = (S,N,\Delta,\delta)$ is said to be
\indexddef{enumerated}{enumerated signature}{}{signature}{enumerated}
if $\Delta(\nu) \in S^*$ for each $\nu \in N$
(recalling that $S^* \subset {\cal F}_S$).
If $\Sigma$ is an enumerated signature then $(X_S,p_N)$ being a
\tsym{\Sigma}-algebra means that if $\nu \in N$ is of type
$\list {\sigma} n \to \sigma$ then
$p_\nu$ is a mapping from 
$X_{\sigma_1} \times \cdots \times X_{\sigma_n}$ to $X_\sigma$.
\medskip
A simple enumerated signature 
$\Lambda = (B,K,\Theta,\vartheta)$ and a `natural'
\tsym{\Lambda}-algebra $(X_B,p_K)$ are given 
in Example~2.2.1 on the following page. The usual way of representing such a
signature is then illustrated in Example~2.2.2.
\medskip
\midinsert
\bigskip\bigskip
\frame{20pt}{\bigskip 
{\it Example 2.2.1\enspace} $\SynInt$ always denotes the subset of
$\,\{\ftt{0},\ftt{1},\ftt{2},\ftt{3},\ftt{4},\ftt{5},\ftt{6},
                                  \ftt{7},\ftt{8},\ftt{9},\ftt{-}\}^*$
containing for each integer $n$ its standard representation
${\underline n}$ as a string of characters. The mapping 
$n \mapsto {\underline n}$ thus maps $\Int$ bijectively onto $\SynInt$.
\medskip
Define an enumerated signature 
$\Lambda = (B,K,\Theta,\vartheta)$ by letting
\medskip\smallskip
{\leftskip=32pt
$B\ =\ \{\ftt{bool},\ftt{nat},\ftt{int},\ftt{pair},\ftt{list}\}$,
\smallskip               
$K\ =\ \{\ftt{True},\ftt{False},\ftt{Zero},\ftt{Succ},\ftt{Pair},
                      \ftt{Nil},\ftt{Cons}\} \cup \SynInt$,
\medskip\smallskip
}
and with $\Theta : K \to B^*$ 
and $\vartheta : K \to B$ given by
\medskip\smallskip
{\leftskip=32pt
$\Theta(\ftt{True})\, =\, \Theta(\ftt{False})\, =\, \onept$, 
\enskip $\vartheta(\ftt{True})\, =\,  \vartheta(\ftt{False})\, =\, 
                                       \ftt{bool}$, 
\smallskip
$\Theta(\ftt{Zero})\, =\, \onept$, \enskip
$\Theta(\ftt{Succ})\, =\, \ftt{nat}$, \enskip
$\vartheta(\ftt{Zero})\, =\,  \vartheta(\ftt{Succ})
\, =\, \ftt{nat}$,
\smallskip
$\Theta(\ftt{Pair})\ =\ \ftt{int}\ \ftt{int}$, \enskip
$\vartheta(\ftt{Pair})\, =\, \ftt{pair}$, 
\smallskip
$\Theta(\ftt{Nil})\, =\, \onept$, \enskip
$\Theta(\ftt{Cons})\ =\ \ftt{int}\ \ftt{list}$, \enskip
$\vartheta(\ftt{Nil})\,=\, \vartheta(\ftt{Cons})
\, =\, \ftt{list}$,
\smallskip
$\Theta({\underline n})\, =\, \onept\,$ 
and $\,\vartheta({\underline n})\, =\, \ftt{int}\,$ 
for each $n \in \Int$.
\medskip\smallskip
}
Thus $\ftt{True}$ and $\ftt{False}$ are of type $\onept \to \ftt{bool}$, 
$\ftt{Zero}$ is of type $\onept \to \ftt{nat}$,
$\ftt{Succ}$ of type $\ftt{nat} \to \ftt{nat}$,
$\ftt{Pair}$ of type $\ftt{int int} \to \ftt{pair}$,
$\ftt{Nil}$ of type $\onept \to \ftt{list}$,
$\ftt{Cons}$ of type $\ftt{int list} \to \ftt{list}$ and
${\underline n}$ is of type $\onept \to \ftt{int}$ for each $n \in \Int$.
\medskip 
Now define a \tsym{\Lambda}-algebra $(X_B, p_K)$ 
with a family of sets $X_B$ and a family of mappings $p_K$
by letting
\medskip\smallskip
{\leftskip=32pt
$X_{\fstt{bool}} \,=\, \Bool\, =\, \{T,F\}$, \enskip 
$X_{\fstt{nat}} \,=\, \Nat$, \enskip
$X_{\fstt{int}} \,=\, \Int$, 
\smallskip
$X_{\fstt{pair}} \,=\, \Int^2$, \enskip
$X_{\fstt{list}} \,=\, \Int^*$, \enskip
\medskip
$p_{\fstt{True}} : \Oneptset \to X_{\fstt{bool}}$\ with\ %
                      $\,p_{\fstt{True}}(\onept) \,=\, T\,$, 
\smallskip
$p_{\fstt{False}} : \Oneptset \to X_{\fstt{bool}}$\ with\ %
                      $\,p_{\fstt{False}}(\onept) \,=\, F\,$, 
\smallskip
$p_{\fstt{Zero}} : \Oneptset \to X_{\fstt{nat}}$\ with\ %
                       $\,p_{\fstt{Zero}}(\onept) \,=\, 0\,$, 
\smallskip
$p_{\fstt{Succ}} : X_{\fstt{nat}} \to X_{\fstt{nat}}$\ with\ %
                       $\,p_{\fstt{Succ}}(n) \,=\, n + 1\,$, 
\smallskip
$p_{{\underline n}} : \Oneptset \to X_{\fstt{int}}$\ with\ %
                         $\,p_{{\underline n}}(\onept) \,=\, n\,$
                         \ for each $n \in \Int$,
\smallskip
$p_{\fstt{Pair}} : X_{\fstt{int}} \times X_{\fstt{int}} \to X_{\fstt{pair}}$\ with\ %
                     $\,p_{\fstt{Pair}}(m,n) \,=\, (m,n)\,$, 
\smallskip
$p_{\fstt{Nil}} : \Oneptset \to X_{\fstt{list}}$\ with\ %
                         $\,p_{\fstt{Nil}}(\onept) \,=\, \onept\,$, 
\smallskip
$p_{\fstt{Cons}} : X_{\fstt{int}} \times X_{\fstt{list}} \to X_{\fstt{list}}$\ with %
                    $\,p_{\fstt{Cons}}(m,s) \,=\, m\, \triangleleft\, s\,$,
\medskip\smallskip
}
where
$m \, \triangleleft\, s$ is here, as in Chapter~1, the element of $\Int^*$ obtained 
by adding $m$ to the beginning of the list $s$, i.e.,
\smallskip
\mdisp{m \, \triangleleft\, s
 \ =\ \cases{
   m\ \list m n   & if $\,s = \list m n \,$ with $\,n \ge 1\,$, \cr
     m            & if $\,s = \onept\,$. \cr}}
\bigskip\medskip
}
\medskip
\endinsert

A general signature can always be replaced by an `equivalent' enumerated signature.
(This really just amounts to fixing an enumeration of the elements in the set
$\Delta(\nu)$ for each $\nu \in N$.) In fact, what is here called an enumerated 
signature corresponds to the usual notion of a signature. The reason
for working with the more general definition introduced above is that in the long run
it turns out to be more natural.
\medskip

\midinsert
\bigskip\bigskip
\frame{20pt}{\bigskip 
{\it Example 2.2.2\enspace} An enumerated signature 
$\Sigma = (S,N,\Delta,\delta)$ with $S$ and $N$
finite can (and in most functional programming languages will) 
be represented in a form similar to the following,
where $\lvector {\sigma} n $ is some enumeration of the elements in the set $S$,
$\nu_{k1},\,\ldots,\,\nu_{km_k}$
an enumeration of the elements of 
$N_{\sigma_k}$ for each $k$ and where $\gamma_{kj} = \Delta(\nu_{kj})$:
\smallskip
\mdisp{
\eqalign{ \sigma_1 
  \ &\ftt{::=}\ \nu_{11}\ \gamma_{11}\ |\ \cdots 
                  \ \ftt{|}\ \nu_{1m_1} \gamma_{1m_1}\cr
   \sigma_2\ &\ftt{::=}
    \ \nu_{21}\ \gamma_{21}\ \ftt{|}\ \cdots\ \ftt{|}
     \ \nu_{2m_2}\ \gamma_{2m_2} \cr
          &\qquad\qquad\vdots \cr
 \sigma_n\ &\ftt{::=}\ \nu_{n1}\ \gamma_{n1}
    \ \ftt{|}\ \cdots\ \ftt{|}\ \nu_{nm_n}\ \gamma_{nm_n}}}
\medskip
The enumerated signature 
$\Lambda = (B,K,\Theta,\vartheta)$ introduced in Example~2.2.1 can thus
be represented in the form
\medskip\smallskip
{\leftskip=110pt
$\ftt{bool ::= True | False}$\smallskip
$\ftt{nat ::= Zero | Succ nat}$\smallskip
$\ftt{int ::= } \cdots\, \ftt{ -2 | -1 | 0 | 1 | 2 }\, \cdots$\smallskip
$\ftt{pair ::= Pair int int}$\smallskip
$\ftt{list ::= Nil | Cons int list}$\bigskip               
}
Of course, there is a problem here with the type
$\ftt{int}$, since $K_{\fstt{int}}$ is infinite, but in all
real programming languages this type is `built-in' and so it does not need to
be included in the part of the signature specified by the programmer.
\bigskip\medskip
}
\bigskip\bigskip
\endinsert

For the remainder of the section let $\Sigma$ denote a fixed signature with
$\Sigma = (S,N,\Delta,\delta)$. 
Note that the sets $N_\sigma$, $\sigma \in S$, form a partition of the
set $N$.
(For instance, in the signature $\Lambda$ in Example~2.2.1 
$\,K_{\fstt{bool}} = \{\ftt{True},\ftt{False}\}$,
$K_{\fstt{nat}} = \{\ftt{Zero},\ftt{Succ}\}$, 
$K_{\fstt{int}} = \SynInt$, 
$K_{\fstt{pair}} = \{\ftt{Pair}\}$ and
$K_{\fstt{list}} = \{\ftt{Nil},\ftt{Cons}\}$.)
\medskip
The next task, of course, is to explain what are
the structure-preserving mappings between algebras. Let 
$(X_S,p_N)$ and $(Y_S,q_N)$ be \tsym{\Sigma}-algebras and let 
$\pi_S : X_S \to Y_S$ be a family of mappings, i.e., 
$\pi_\sigma : X_\sigma \to Y_\sigma$ for each $\sigma \in S$. Then the family 
$\pi_S$ is called a 
\indexdef{\tsym{\Sigma}-homomorphism}{homomorphism}{}
from $(X_S,p_N)$ to $(Y_S,q_N)$ if 

\mdisp{q_\nu \, \ass L {\pi} \ =\ \pi_\sigma \, p_\nu} 

whenever $\nu \in N$ is of type $L \to \sigma$. 
This fact will also be expressed by saying that
$\pi_S : (X_S,p_N) \to (Y_S,q_N)$ is a \tsym{\Sigma}-homomorphism.
\medskip
If $\Sigma$ is enumerated then
$\pi_S : (X_S,p_N) \to (Y_S,q_N)$ being a \tsym{\Sigma}-homomorphism
means that 
if $\nu \in N$ is of type $\list {\sigma} n \to \sigma$ then

\mdisp{q_\nu(\pi_{\sigma_1}(x_1),\ldots,
             \pi_{\sigma_n}(x_n))\ =\ \pi_\sigma(p_\nu(\vector x n ))}

must hold for all $(\vector x n ) \in X_{\sigma_1} \times \cdots \times X_{\sigma_n}$,
this condition being interpreted as $q_\nu(\onept) = \pi_\sigma(p_\nu(\onept))$ 
when $\nu$ is of type $\onept \to \sigma$.

\medskip

If $(X_S,p_N)$ is a \tsym{\Sigma}-algebra then by Lemma~2.1.1~(1) the family 
of identity mappings $\fss{id}_S : X_S \to X_S$ defines a 
\tsym{\Sigma}-homomorphism from $(X_S,p_N)$ to itself. Furthermore, 
the composition of two \tsym{\Sigma}-homomorphisms is again a 
\tsym{\Sigma}-homomorphism:

\proclaim{Proposition~2.2.1} Suppose $(X_S,p_N)$, $(Y_S,q_N)$ and $(Z_S,r_N)$
are \tsym{\Sigma}-algebras and let
$\pi_S : (X_S,p_N) \to (Y_S,q_N)$ and $\varrho_S : (Y_S,q_N) \to (Z_S,r_N)$
be \tsym{\Sigma}-homomorphisms. Then 
$(\varrho\pi)_S
          = \lcurl \varrho_\sigma \, \pi_\sigma \,:\, \sigma \in S \rcurl$
is a \tsym{\Sigma}-homomorphism from $(X_S,p_N)$ to $(Z_S,r_N)$.
\endpro

\proof Let $\nu \in N$ be of type $L \to \sigma$; then
by Lemma~2.1.1~(2)
\mdisp{
r_\nu\,\ass L {(\varrho\pi)}
\ =\ r_\nu\,\ass L {\varrho} \,\ass L {\pi}
\ =\ \varrho_\sigma\,q_\nu\,\ass L {\pi}
\ =\ \varrho_\sigma\,\pi_\sigma\,p_\nu
\ =\ (\varrho\pi)_\sigma\,p_\nu}
and hence
$(\varrho\pi)_S$
is a \tsym{\Sigma}-homomorphism from $(X_S,p_N)$ to $(Z_S,r_N)$. \eop
\bigbreak

\proclaim{Proposition~2.2.2} Let $\pi_S : (X_S,p_N) \to (Y_S,q_N)$ be a 
\tsym{\Sigma}-homomorphism. Suppose for each $\sigma \in S$ that 
$\pi_\sigma : X_\sigma \to Y_\sigma$ is a bijection and let 
$\pi^{-1}_\sigma : Y_\sigma \to X_\sigma$ be the inverse mapping. Then 
$\pi^{-1}_S : (Y_S,q_N) \to (X_S,p_N)$ is also a \tsym{\Sigma}-homomorphism.
\endpro

\proof Let $\nu \in N$ be of type $L \to \sigma$. Then
$q_\nu\,\ass L {\pi} = \pi_\sigma\,p_\nu$, and therefore by Lemma~2.1.1~(3) 
it follows that 
$p_\nu\,\assb L {\pi^{-1}} \,=\, p_\nu\,(\ass L {\pi} )^{-1} 
\,=\, \pi_\sigma^{-1}\,q_\nu$,
which implies that
$\pi^{-1}_S$ is a \tsym{\Sigma}-homomorphism. \eop
\bigbreak
A \tsym{\Sigma}-homomorphism $\pi_S : (X_S,p_N) \to (Y_S,q_N)$ is a said to be a 
\indexdef{\tsym{\Sigma}-isomorphism}{isomorphism}{}
if the mapping 
$\pi_\sigma : X_\sigma \to Y_\sigma$ is a bijection for each $\sigma \in S$.
(Proposition~2.2.2 implies that this definition is `correct'.)
If it is clear which signature is involved then the name of the signature can be omitted,
i.e., homomorphism (resp. isomorphism) will be written instead of 
\tsym{\Sigma}-homomorphism (resp. \tsym{\Sigma}-isomorphism) when $\Sigma$ can be
deduced from the context.  
\medskip
The \tsym{\Sigma}-algebras $(X_S,p_N)$ and $(Y_S,q_N)$ are said to be 
\indexddef{isomorphic}{isomorphic algebras}{}{algebras}{isomorphic} 
if there exists an isomorphism from $(X_S,p_N)$ to $(Y_S,q_N)$; it is easily checked 
that the property of being isomorphic defines an equivalence relation on the 
class of all \tsym{\Sigma}-algebras.
\medskip\smallskip
In what follows consider the \tsym{\Sigma}-algebra $(X_S,p_N)$ as being fixed.
Let ${\grave X}_S \subset X_S$
(i.e.,  ${\grave X}_\sigma \subset X_\sigma$ for each $\sigma \in S$). 
Then the family of sets ${\grave X}_S$ is said to be 
\indexdef{invariant in $(X_S,p_N)$}{invariant family}{}, 
or just 
\indexdef{invariant}{family}{invariant},
if $p_\nu(\ass L {\grave X} ) \subset {\grave X}_\sigma$ 
whenever $\nu \in N$ is of type $L \to \sigma$. 
In particular, the family $X_S$ is itself trivially invariant.
The family ${\grave X}_S$ 
being invariant means exactly that the following two conditions 
have to be satisfied:
\medskip\smallskip
{\parindent=25pt
\item{(1)} $p_\nu(\onept) \in {\grave X}_\sigma$ for each $\nu \in N$ of type 
$\varnothing \to \sigma$.
\medskip
\item{(2)} If $\nu \in N$ is of type $L  \to \sigma$ with 
$L \ne \varnothing$ and $c \in \ass L X $ is such that
$c(\eta) \in {\grave X}_\eta $ for each $\eta \in L$ then
$p_\nu(c) \in {\grave X}_\sigma$.
\bigskip
}
If $\Sigma$ is an enumerated signature then these two conditions become the following:
\medskip\smallskip
{\parindent=25pt
\item{(1)} $p_\nu(\onept) \in {\grave X}_\sigma$ for each $\nu \in N$ of type 
$\onept \to \sigma$.
\medskip
\item{(2)} If $\nu \in N$ is of type $\list {\sigma} n  \to \sigma$ with 
$n \ge 1$ and $x_j \in {\grave X}_{\sigma_j}$ for each $j$ then
$p_\nu(\vector x n ) \in {\grave X}_\sigma$.
\bigskip
}
A related notion is that of a  subalgebra:
A \tsym{\Sigma}-algebra $(Y_S,q_N)$ is said to be a 
\indexdef{subalgebra}{subalgebra}{} 
of $(X_S,p_N)$ if $Y_S \subset X_S$ and $q_\nu$ is the restriction of $p_\nu$ to 
$\dom(q_\nu)$ for each $\nu \in N$. In this case
the family $Y_S$ is clearly invariant.
Conversely, let ${\grave X}_S$ be any invariant family and for each 
$\nu \in N$ of type $L \to \sigma$
let ${\grave p}_\nu$ denote the restriction of $p_\nu$ to 
$\ass L {\grave X} $, so ${\grave p}_\nu$ is a mapping from 
$\ass L {\grave X} $ to ${\grave X}_\sigma$. Then
$({\grave X}_S,{\grave p}_N)$ is a subalgebra of $(X_S,p_N)$.
This means that there is a one-to-one correspondence between 
invariant families and subalgebras of $(X_S,p_N)$. (The former tend to dominate, however,
in this study.) If ${\grave X}_S$ is an invariant family 
then the corresponding subalgebra 
$({\grave X}_S,{\grave p}_N)$ is called the 
\indexddef{subalgebra associated with ${\grave X}_S$}
{associated subalgebra}{}{subalgebra}{associated with invariant family}.

\proclaim{Lemma~2.2.1} Let $\pi_S$ be a homomorphism from $(X_S,p_N)$
to a \tsym{\Sigma}-algebra $(Y_S,q_N)$.
\medskip
(1)\enskip If the family ${\grave X}_S$ is invariant in $(X_S,p_N)$ and 
${\grave Y}_\sigma = \pi_\sigma({\grave X}_\sigma)$ for each $\sigma \in S$ 
then ${\grave Y}_S$ is invariant in $(Y_S,q_N)$.
\medskip
(2)\enskip If the family ${\grave Y}_S$ is invariant in $(Y_S,q_N)$ and 
${\grave X}_\sigma = \pi^{-1}_\sigma({\grave Y}_\sigma)$ for each 
$\sigma \in S$ then ${\grave X}_S$ is invariant in $(X_S,p_N)$. 
\endpro

\proof (1)\enskip It must be shown that if $\nu \in N$ is of type $L \to \sigma$
then $q_\nu(c) \in {\grave Y}_\sigma$ for each $c \in \ass L {\grave Y} $. 
But if $c \in \ass L {\grave Y} $ then there
exists $c' \in \ass L {\grave X} $ with $\ass L {\pi} (c') = c$ (since 
Lemma~2.1.1~(3) implies that 
$\ass L {\pi} (\ass L {\grave X} ) = \ass L {\grave Y} $) and hence 
\mdisp{q_\nu(c) \ =\  q_\nu(\ass L {\pi} (c')) 
\ =\  \pi_\sigma(p_\nu(c')) \,\in\,
\pi_\sigma({\grave X}_\sigma) \ =\ {\grave Y}_\sigma\  .}
\medskip
(2)\enskip This time it must be shown that if $\nu \in N$ is of type 
$L \to \sigma$ and $c \in \ass L {\grave X} $ then
$p_\nu(c) \in {\grave X}_\sigma$, i.e., that
$\pi_\sigma(p_\nu(c)) \in {\grave Y}_\sigma$.  
But if $c \in \ass L {\grave X} $ then it is easy to check that
$\ass L {\pi} (c) \in \ass L {\grave Y} $ and therefore
$\,\pi_\sigma(p_\nu(c)) 
 \,=\, q_\nu(\ass L {\pi} (c)) \,\in\, {\grave Y}_\sigma\,$. 
\eop
\bigbreak
Lemma~2.2.1 says that both the image and the pre-image of a subalgebra 
under a homomorphism are again subalgebras.

\proclaim{Lemma~2.2.2} Let $A_S \subset X_S$. 
Then the family ${\breve X}_S$ defined by 
\mdisp{ {\breve X}_\sigma
\  =\ A_\sigma \,\cup \bigcup\limits_{\nu \in N_\sigma} \Im(p_\nu)}
for each $\sigma \in S$ is invariant in $(X_S,p_N)$. \endpro

\proof If $\nu \in N$ is of type $L \to \sigma$ then 
$p_\nu(c) \in p_\nu(\ass L X ) = \Im(p_\nu) \subset {\breve X}_\sigma$ for all 
$c \in \ass L X $, and so in
particular for all $c \in \ass L {\breve X} $. \eop

\proclaim{Lemma~2.2.3} Let $A_S \subset X_S$. 
Then there exists a minimal invariant  family containing $A_S$
(i.e., an invariant family ${\hat X}_S$ with $A_S \subset {\hat X}_S$ such that if 
$Y_S$ is any invariant family containing $A_S$ then ${\hat X}_S \subset Y_S$). 
\endpro

\proof As already noted, the family $X_S$ is itself invariant, and it contains of course
$A_S$. Moreover, it is easy to see that an arbitrary intersection of 
invariant families is again invariant. 
(More precisely, if $X_S^{t}$ is invariant for each $t \in T$ and 
${\grave X}_\sigma = \bigcap_{t \in T} X_\sigma^{t}$ for each 
$\sigma \in S$ then ${\grave X}_S$ is also invariant.) 
The intersection of all the invariant families containing $A_S$ is thus the required 
minimal family. In fact, this minimal family 
${\hat X}_S$ can be given somewhat more explicitly:
For each $n \in \Nat$ define a family ${\hat X}^{n}_S \subset X_S$
by putting ${\hat X}^{0}_S = A_S$ and for each $n \in \Nat$, 
$\sigma \in S$ letting
\mdisp{
{\hat X}^{n+1}_\sigma\ =\ {\hat X}^{n}_\sigma 
   \ \cup \bigcup\limits_{\nu \in N_\sigma} \Im(p^{n}_\nu)\ ,}
where if $\nu$ is of type $L \to \sigma$ then
$p^{n}_\nu$ is the restriction of $p_\nu$ to $\assb L {{\hat X}^{n}} $. Then it is
straightforward to check that 
$\,{\hat X}_\sigma\, =\, \bigcup_{n \in \Nat} {\hat X}^{n}_\sigma\,$
for each $\sigma \in S$. 
This shows that each element of ${\hat X}_\sigma$ can be `constructed' in a finite 
number of steps out of elements from the family $A_S$ and elements which have
already been `constructed'. \eop
\bigbreak
The main interest here is in the minimal invariant family (i.e., the family
given by Lemma~2.2.3 with $A_\sigma = \varnothing$ for each $\sigma \in S$).
The subalgebra associated with this family will be referred to as the
\indexddef{minimal subalgebra of $(Y_S,q_N)$}
{minimal subalgebra}{}{subalgebra}{minimal}.
\medskip
It is often the case that a \tsym{\Sigma}-algebra $(X_S,p_N)$ is given 
and then $Z_S$ is defined to be the minimal invariant family. This will 
then be accompanied by the statement that $Z_S$ is defined by the following 
three rules:
\medskip\smallskip
{\parindent=25pt
\item{(1)} If $\nu \in N$ is of type $\varnothing \to \sigma$ then
$p_\nu(\onept)$ is an element of $Z_\sigma$.
\medskip
\item{(2)} If $\nu \in N$ is of type $L  \to \sigma$ with 
$L \ne \varnothing$ and $c \in \ass L X $ is such that
$c(\eta) \in Z_\eta $ for each $\eta \in L$ then
$p_\nu(c)$ is an element of $Z_\sigma$.
\medskip
\item{(3)} The only elements in $Z_\sigma$ are those which can be obtained
using (1) and (2). 
\medskip\smallskip
}  
Rules (1) and (2) say that $Z_S$ is invariant. Rule (3)
should be regarded as a somewhat imprecise reformulation of
the final statement in the proof of Lemma~2.2.3.
Of course, such a statement is really redundant, but it usually helps to
clarify how the mappings in the family $p_N$ operate in the particular case 
under consideration. (It is often convenient to
divide up rules (1) and (2) into various sub-cases, ending
up with not three but four or more rules.) 
\medbreak
If $\Sigma$ is enumerated then the above rules take on the
following form:
\medskip\smallskip
{\parindent=25pt
\item{(1)} If $\nu \in N$ is of type $\onept \to \sigma$ then
$p_\nu(\onept)$ is an element of $Z_\sigma$.
\medskip
\item{(2)} If $\nu \in N$ is of type $\list {\sigma} n  \to \sigma$ for
some $n \ge 1$ and $x_j \in Z_{\sigma_j}$ for $\oneto j n $ then
$p_\nu(\vector x n )$ is an element of $Z_\sigma$.
\medskip
\item{(3)} The only elements in $Z_\sigma$ are those which can be obtained
using (1) and (2). 
\bigskip
}  
The \tsym{\Sigma}-algebra $(X_S,p_N)$ is now said to be 
\indexddef{minimal}{minimal algebra}{}{algebra}{minimal} 
if $X_S$ is the only invariant family. Such algebras play an important role, and so some
of their elementary properties will now be presented.
Note that the minimal subalgebra of $(X_S,p_N)$ is always a minimal \tsym{\Sigma}-algebra. 

\proclaim{Proposition~2.2.3} Let $(X_S,p_N)$ and $(Y_S,q_N)$ 
be \tsym{\Sigma}-algebras. 
\medskip
(1)\enskip If $(X_S,p_N)$ is minimal then there exists at most one homomorphism
from $(X_S,p_N)$ to $(Y_S,q_N)$.
\medskip
(2)\enskip If $(Y_S,q_N)$ is minimal then any homomorphism 
$\pi_S : (X_S,p_N) \to (Y_S,q_N)$ is surjective (i.e., the  mapping 
$\pi_\sigma : X_\sigma \to Y_\sigma$ is surjective for each $\sigma \in S$).
\endpro

\proof (1)\enskip Let $\pi_S$ and $\varrho_S$ be homomorphisms from $(X_S,p_N)$
to $(Y_S,q_N)$ and for each $\sigma \in S$ let 
${\grave X}_\sigma 
= \lcurl x \in X_\sigma \,:\, \pi_\sigma(x) = \varrho_\sigma(x) \rcurl$. 
Consider $\nu \in N$ of type $L \to \sigma$ and let
$c \in \ass L {\grave X} $; then 
$c(\eta) \in {\grave X}_\eta $ for each $\eta \in L$ and hence
\mdisp{\ass L {\pi} (c)\,(\eta)\ =\ \pi_\eta (c(\eta))
\ =\ \varrho_\eta (c(\eta)) \ =\ \ass L {\varrho} (c)\,(\eta)}
which implies that $\ass L {\pi} (c) = \ass L {\varrho} (c)$. Therefore
\mdisp{
\pi_\sigma(p_\nu(c))
\ =\ q_\nu(\ass L {\pi} (c))\ =\ q_\nu(\ass L {\varrho} (c))
\ =\ \varrho_\sigma(p_\nu(c))\ ,}
i.e., $p_\nu(c) \in {\grave X}_\sigma$. This shows that
the family ${\grave X}_S$ is invariant and thus 
${\grave X}_S = X_S$, since $(X_S,p_N)$ is minimal.
In other words, $\pi_S = \varrho_S$. 
\medskip
(2)\enskip This follows immediately from Lemma~2.2.1~(1). \eop

\proclaim{Proposition~2.2.4} If $(X_S,p_N)$ is minimal then 
$\bigcup\limits_{\nu \in N_\sigma} \Im(p_\nu) = X_\sigma$ for each 
$\sigma \in S$.
\endpro
\proof This follows immediately from Lemma~2.2.2. \eop
\bigbreak
The converse of Proposition~2.2.4 does not hold in general. However,
the condition occurring there can be combined with
a second condition to give a useful sufficient criterion for minimality:

\proclaim{Proposition~2.2.5} Suppose there exists a family of mappings
$\#_S$ with $\#_\sigma : X_\sigma \to \Nat$ for each $\sigma$
such that if $\nu \in N$ is of type $L \to \sigma$ 
then 
\mdisp{ \#_\eta (c(\eta))\ < \ \#_\sigma(p_\nu(c))}
holds for all $c \in \ass L X $, $\eta \in L$.
For each $\sigma \in S$ let $A_\sigma$ be a subset of $X_\sigma$ such that
$\,A_\sigma \,\cup\, \bigcup_{\nu \in N_\sigma} \Im(p_\nu) \,=\, X_\sigma\,$.
Then $X_S$ is the only invariant family containing $A_S$. 
In particular, if $\,\bigcup\limits_{\nu \in N_\sigma} \Im(p_\nu) \,=\, X_\sigma\,$ 
for each $\sigma \in S$ then $(X_S,p_N)$ is minimal.  \endpro

\proof Let ${\hat X}_S$ be the minimal invariant family containing $A_S$ and suppose
${\hat X}_S \ne X_S$. There thus exists $\sigma \in S$ and 
$x \in X_\sigma \setminus {\hat X}_\sigma$ such that 
$\#_\sigma(x) \le \#_\tau(x')$ whenever
$x' \in X_\tau \setminus {\hat X}_\tau$ for some $\tau \in S$. 
Then $x \in \Im(p_\nu)$ for some $\nu \in N_\sigma$, since 
$A_\sigma \subset {\hat X}_\sigma$. Let $\nu$ be of type $L \to \sigma$, 
so there exists $c \in \ass L X $ with $x = p_\nu(c)$.
But it then follows that $\#_\eta (c(\eta)) < \#_\sigma(x)$ and hence that 
$c(\eta) \in {\hat X}_\eta $ for each $\eta \in L$
(by the minimality of $\#_\sigma(x)$).
However, this implies $x \in {\hat X}_\sigma$, since the family ${\hat X}_S$ is 
invariant, which is a contradiction. \eop
\bigbreak

If $\Sigma$ is enumerated then the condition involving the
family $\#_S$ in Proposition~2.2.5 is that 
whenever $\nu \in N$ is of type $\list {\sigma} n \to \sigma$ then
\mdisp{ \#_{\sigma_j}(x_j) 
          \ < \ \#_\sigma(p_\nu(\vector x n ))}
must hold for each $\oneto j n $ 
for each $(\vector x n ) \in X_{\sigma_1} \times \cdots \times X_{\sigma_n}$.
It is easy to check that the 
\tsym{\Lambda}-algebra $(X_B,p_K)$ defined in
Example~2.2.1 is minimal. 
\bigskip

A subset $S' \subset S$ is said to be
\definition{closed in $\Sigma$} or simply \definition{closed}
if $\sigma \in S'$ whenever there exists $\nu \in N$ of type 
$L \to \sigma$ for some $L$ and $\langle \eta \rangle \in S'$
for each $\eta \in L$. 
(This means in particular that $\sigma \in S'$ if there 
exists an element $\nu \in N$ of type $\varnothing \to \sigma$.) 
The signature $\Sigma$ will be called 
\indexddef{pervasive}{pervasive signature}{}{signature}{pervasive} 
if the only closed subset of $S$ is $S$ itself. 
\medskip
It is easily checked that the signature $\Lambda$ in 
Example~2.2.1 is pervasive. (Note that
if $\Sigma$ is enumerated then $S'$ being 
closed means exactly that $\sigma \in S'$ must hold whenever 
there exists $\nu \in N$ of type 
$\list {\sigma} n \to \sigma$ and $\sigma_j \in S'$ for each $\oneto j n $.)

\proclaim{Proposition~2.2.6} If $\Sigma$ is pervasive then 
$X_\sigma \ne \varnothing$ for all $\sigma \in S$ 
for any \tsym{\Sigma}-algebra $(X_S,p_N)$. Conversely, if $(X_S,p_N)$
minimal and $X_\sigma \ne \varnothing$ for all $\sigma \in S$ then $\Sigma$ 
is pervasive. \endpro

\proof The set $S_o = \lcurl \sigma \in S : X_\sigma \ne \varnothing \rcurl$ is 
clearly closed. Thus if $\Sigma$ is pervasive then 
$S_o = S$, i.e., $X_\sigma \ne \varnothing$ for each $\sigma \in S$. 
Conversely, if $S'$ is closed and the family $X'_S$ is defined 
by $X'_\sigma = X_\sigma$ for $\sigma \in S'$ and $X'_\sigma = \varnothing$ 
for $\sigma \in S \setminus S'$ then $X'_S$ is invariant. Therefore if 
$(X_S,p_N)$ is minimal and 
$X_\sigma \ne \varnothing$ for each $\sigma \in S$ then $S' = S$. \eop
\bigbreak

Note that the set 
$S' = \lcurl \sigma \in S \,:\, N_\sigma \ne \varnothing \rcurl$ is always 
closed, and hence a necessary condition for pervasiveness is that
$N_\sigma \ne \varnothing$ for each $\sigma \in S$.
\bigskip
To end the section the special case of a 
\indexddef{single-sorted}{single-sorted signature}{}{signature}{single-sorted}
signature will be looked at, i.e., a signature of the form 
$(\Oneptset, N,\Delta,\delta)$.
In this case there is no choice for $\delta$ (since there is only one mapping 
possible from $N$ to $\Oneptset$) and so a single-sorted
signature can be regarded as being a pair $(N,\Delta)$ consisting of a set $N$ and 
a mapping $\Delta : N \to {\cal F}_\Oneptset$ (recalling that ${\cal F}_\Oneptset$ 
is just considered to be the set  
of all finite subsets of the basic set of names $\Bbb{M}$). 
\medskip
Let $\Sigma = (N,\Delta)$ be a single-sorted
signature; then (identifying a family $Z_\Oneptset$ with the single
object $Z_\onept$ it contains) a \tsym{\Sigma}-algebra is here a pair
$(X,p_N)$ consisting of a set $X$ and a family of mappings $p_N$
with $p_\nu : \total {\Delta(\nu)} X \to X$ for each $\nu \in N$.
\medskip
Consider the very special case of an enumerated single-sorted signature
$\Sigma = (N,\Delta)$. Then, since $\Oneptset^*$ can clearly be 
identified with the set of natural numbers $\Nat$, 
$\Delta$ can here be regarded as a mapping from $N$ to $\Nat$. If 
$(X,p_N)$ is a  \tsym{\Sigma}-algebra then $p_\nu$ is a mapping
from the cartesian product $X^{\Delta(\nu)}$ to $X$, so 
$\Delta(\nu)$ is just the number of arguments taken by the operator $p_\nu$.
\medskip
If $\Sigma = (S,N,\Delta,\delta)$ is an arbitrary signature
then a single-sorted signature $\Sigma^o = (N,\Delta^o)$ 
can be defined by letting $\Delta^o(\nu)$ be the underlying set 
involved in the \tsym{S}-typed set $\Delta(\nu)$.
This means that $\Sigma^o$ is obtained from $\Sigma$ by no longer distinguishing
between the various types.
Now let $(X,p_N)$ be a \tsym{\Sigma^o}-algebra
and for each $\sigma \in S$ put $X_\sigma = X$.
Then, since any mapping from $L$ to $\bigcup_{\eta \in L} X_\eta = X$
is automatically typed, it follows that 
$\ass {\Delta(\nu)} X = \total {\Delta^o(\nu)} X $
for each $\nu \in N$, which means that $(X_S,p_N)$ is a
\tsym{\Sigma}-algebra.
This almost trivial method of obtaining \tsym{\Sigma}-algebras
turns out to be surprisingly useful.

\vfill\eject

\hfline{33}{2.3 \ INITIAL AND FREE ALGEBRAS}

\bigskip
\sectionhead {2.3} {Initial and free algebras}
\bigskip\medskip
For the whole of the section let $\Sigma = (S,N,\Delta,\delta)$ be a signature.
A \tsym{\Sigma}-algebra $(X_S,p_N)$ is said to be 
\indexddef{initial}{initial algebra}{}{algebra}{initial}
if for each \tsym{\Sigma}-algebra $(Y_S,q_N)$ there exists a unique 
homomorphism from $(X_S,p_N)$ to $(Y_S,q_N)$. (The terminology `initial' 
is the standard one used in category theory to describe such a situation.) 
\medskip
In Proposition~2.3.2 initial algebras are characterised as those that are
minimal and possess a further property, here called unambiguity.
In Proposition~2.3.3 it is shown there is a unique isomorphism class of initial 
\tsym{\Sigma}-algebras. This essentially amounts to showing that an initial 
\tsym{\Sigma}-algebra exists, which follows from Proposition~2.3.2 and the 
existence of a minimal unambiguous \tsym{\Sigma}-algebra. 
\medskip
A \tsym{\Sigma}-algebra $(X_S,p_N)$ is said to be 
\indexddef{unambiguous}{unambiguous algebra}{}{algebra}{unambiguous}
if the following hold:
\medskip\smallskip
{\parindent=25pt
\item{(1)} The mapping $p_\nu$ is injective for each $\nu \in N$.
\medskip
\item{(2)} For each $\sigma \in S$ the sets
$\Im(p_\nu)$, $\nu \in N_\sigma$, are disjoint subsets of $X_\sigma$.
\medskip\smallskip
}
In particular, the \tsym{\Lambda}-algebra $(X_B,p_K)$ in 
Example~2.2.1 is clearly unambiguous. Related to unambiguity is the 
following property: A \tsym{\Sigma}-algebra $(X_S,p_N)$ is said to be 
\indexddef{regular}{regular algebra}{}{algebra}{regular} 
if for each $x \in X_\sigma$ there exists a unique 
$\nu \in N_\sigma$ and a unique element $c \in \dom(p_\nu)$ such that 
$p_\nu(c) = x$. Thus $(X_S,p_N)$ is regular if and only if 
the mapping $p_\nu$ is injective for each $\nu \in N$
and for each $\sigma \in S$ the sets
$\Im(p_{\nu})$, $\nu \in N_\sigma$, form a partition of $X_\sigma$.

\proclaim{Lemma~2.3.1} A minimal \tsym{\Sigma}-algebra is regular
if and only if it is unambiguous. \endpro

\proof This follows immediately from Proposition~2.2.4. \eop
\bigbreak
Before going any further consider again the natural number triples 
introduced at the beginning of the chapter. These are really the algebras 
corresponding to the enumerated signature
consisting of a single type
$\ftt{nat}$ and two operator names $\ftt{Zero}$ and $\ftt{Succ}$ with
$\ftt{Zero}$ of type $\onept \to \ftt{nat}$ and $\ftt{Succ}$ of 
type $\ftt{nat} \to \ftt{nat}$; however, it was more convenient
to represent an algebra 
$(\{X_{\fstt{nat}}\},\{p_{\fstt{Zero}},p_{\fstt{Succ}}\})$
using the natural number triple 
$(X_{\fstt{nat}},p_{\fstt{Zero}}(\onept),p_{\fstt{Succ}})$.
The equivalence of initial and Peano triples is easily seen to be
just a special case of Proposition~2.3.2.
\medskip
In order to get started an unambiguous
\tsym{\Sigma}-algebra is needed, and for this the following trivial observation
is useful: Let $\Sigma^o = (N,\Delta^o)$ be the single-sorted signature
defined at the end of the previous section (so
$\Delta^o(\nu)$ is just the underlying set involved in the
\tsym{S}-typed set $\Delta(\nu)$). Let $(X,p_N)$ be a 
\tsym{\Sigma^o}-algebra and $(X_S,p_N)$ be the
\tsym{\Sigma}-algebra with $X_\sigma = X$ for each $\sigma \in S$.

\proclaim{Lemma~2.3.2} If the \tsym{\Sigma^o}-algebra $(X,p_N)$ is
unambiguous then so is the  \tsym{\Sigma}-algebra $(X_S,p_N)$. \endpro

\proof This is clear. \eop
\bigbreak

\proclaim{Lemma~2.3.3} There exists an unambiguous \tsym{\Sigma}-algebra. \endpro

\proof By Lemma~2.3.2 it is enough to show that
if $\Sigma' = (N,\Delta)$ is a single-sorted
signature then there exists a unambiguous \tsym{\Sigma'}-algebra. 
The construction given below 
is just one of many possibilities of defining such an algebra.
\medskip
Let $M = M_o \cup N$, where $M_o$ is the set of all pairs of the form $(\nu,\eta)$ with
$\nu \in N$ and $\eta \in \Delta(\nu)$, and let $X$ be the set of all non-empty
finite subsets of $M^*$. Now if $\nu \in N$ with $\Delta(\nu) = \varnothing$ then define
$p_\nu : \Oneptset \to X$ by letting $p_\nu(\onept) = \{\nu\}$ (where here $\nu$ is the 
list consisting of the single component $\nu$), and if  
$\nu \in N$ with $\Delta(\nu) = L \ne \varnothing$ then define a mapping
$p_\nu : \total L X \to X$ by letting
\mdisp{
p_\nu(c) \ =\ \{\nu\}\, \cup \,\bigcup\limits_{\eta \in L} 
  \lcurl (\nu,\eta) \triangleleft s \,:\, s \in c(\eta) \rcurl }
for each $c \in X^L$, where $\triangleleft : M \times M^* \to M^*$ is the
(infix) operation of adding an element to the beginning of a list.
This gives  a \tsym{\Sigma'}-algebra $(X,p_N)$.
But it is easily checked  that $\Im(p_{\nu_1})$ and $\Im(p_{\nu_2})$ are 
disjoint subsets of $X$ whenever 
$\nu_1 \ne \nu_2$, and also that $p_\nu$ is injective 
for each $\nu \in N$. Hence $(X,p_N)$ is unambiguous. \eop

\proclaim{Proposition~2.3.1} There exists a minimal unambiguous 
\tsym{\Sigma}-algebra. \endpro

\proof By Lemma~2.3.3 there exists an unambiguous \tsym{\Sigma}-algebra
$(X_S,p_N)$. But then any subalgebra of 
$(X_S,p_N)$ is also unambiguous.
In particular, the minimal subalgebra is minimal and unambiguous. \eop

\proclaim{Proposition~2.3.2} The following statements are equivalent for a
\tsym{\Sigma}-algebra $(X_S,p_N)$:
\medskip\smallskip
{\parindent=25pt
\item{(1)} $(X_S,p_N)$ is initial.
\medskip
\item{(2)} $(X_S,p_N)$ is minimal and unambiguous.
\medskip
\item{(3)} $(X_S,p_N)$ is minimal and regular.
\medskip\smallskip
} \endpro

\proof The equivalence of (2) and (3) is just Lemma~2.3.1.
To show their equivalence to (1) a couple of simple facts will be needed:

\proclaim{Lemma~2.3.4} Let $(X_S,p_N)$ be a minimal regular
\tsym{\Sigma}-algebra. Then there exists a unique family 
$\#_S$ with $\#_\sigma : X_\sigma \to \Nat$ for each 
$\sigma \in S$ such that $\,\#_\sigma(p_\nu(\onept)) = 0\,$ if
$\nu \in N$ is of type $\varnothing \to \sigma$ and such that 
\mdisp{\#_\sigma(p_\nu(c))
  \ =\ 1 \,+\, \max \lcurl \#_\eta (c(\eta)) \,:\, \eta \in L \rcurl}
for all $c \in \ass L X $
whenever $\nu \in N$ is of type $L \to \sigma$ 
with $L \ne \varnothing$. \endpro

\proof For each $m \in \Nat$ a family of mappings
$\#^m_S$ will be defined with $\#^m_\sigma : X_\sigma \to \Nat$ for each 
$\sigma \in S$, and then $\#_S$ will be obtained as the limit of the
sequence $\{\#^m_S\}_{m \ge 0}$.
The definition is by induction on $m$:
First define $\#^0_\sigma = 0$ for each $\sigma \in S$.
Next suppose that the family $\#^m_S$ has already been defined for some
$m \in \Nat$. Then since $(X_S,p_N)$ is regular there is a unique family of
mappings $\#^{m+1}_S$ such that
$\,\#^{m+1}_\sigma(p_\nu(\onept)) = 0\,$ if
$\nu \in N$ is of type $\varnothing \to \sigma$ and such that
\mdisp{\#^{m+1}_\sigma(p_\nu(c))
  \ =\ 1 \,+\, \max \lcurl \#^m_\eta (c(\eta)) \,:\, \eta \in L \rcurl}
for all $c \in \ass L X $ whenever $\nu \in N$ is of type $L \to \sigma$ 
with $L \ne \varnothing$. 
\medskip
Then $\#^m_S \le \#^{m+1}_S$ for each $m \in \Nat$ (i.e., 
$\#^m_\sigma(x) \le \#^{m+1}_\sigma(x)$ for all
$x \in X_\sigma$, $\sigma \in S$): This follows by induction on $m$,
since $\#^0_S \le \#^1_S$ holds by definition and if 
$\#^m_S \le \#^{m+1}_S$ for some $m \in \Nat$ and
$\nu \in N$ is of type $L \to \sigma$ 
with $L \ne \varnothing$ then
\smallskip
\ldisp{\quad\#^{m+1}_\sigma(p_\nu(c))
  \ =\ 1 \,+\, \max \lcurl \#^m_\eta (c(\eta)) \,:\, \eta \in L \rcurl}
\vskip-\medskipamount
\rdisp{
  \ \le\ 1 \,+\, \max \lcurl \#^{m+1}_\eta (c(\eta)) \,:\, \eta \in L \rcurl
\ =\ \#^{m+2}_\sigma(p_\nu(c))
\quad}
for all $c \in \ass L X $, which implies that $\#^{m+1}_S \le \#^{m+2}_S$.
\medskip
The sequence $\{\#^m_\sigma(x)\}_{m \ge 0}$ is bounded for each
$x \in X_\sigma$, $\sigma \in S$: Let ${\grave X}_\sigma$ denote the set
of those elements $x \in X_\sigma$ for which this
is the case. Then it is easily 
checked that the family ${\grave X}_S$ is invariant, and
hence ${\grave X}_S = X_S$, since $(X_S,p_N)$ is minimal. 
\medskip
Let $x \in X_\sigma$; then by the above
$\{\#^m_\sigma(x)\}_{m \ge 0}$ is a bounded 
increasing sequence from $\Nat$, and so there exists an element
$\#_\sigma(x) \in \Nat$ such that $\#^m_\sigma(x) = \#_\sigma(x)$
for all but finitely many $m$. This defines a mapping
$\#_\sigma : X_\sigma \to \Nat$ for each $\sigma \in S$, and it
immediately follows that the family $\#_S$ has the required property.
\medskip
It remains to show the uniqueness, so
suppose $\#'_S$ is another family of mappings with this property. For 
each $\sigma \in S$ let 
$X'_\sigma = \lcurl x \in X_\sigma \,:\, \#'_\sigma(x) = \#_\sigma(x) \rcurl$;
then the family $X'_S$ is clearly invariant and hence 
$X'_S = X_S$, since $(X_S,p_N)$ is minimal. \eop
\bigbreak

\proclaim{Lemma~2.3.5} Let $(X_S,p_N)$ be a \tsym{\Sigma}-algebra isomorphic
to a minimal unambiguous \tsym{\Sigma}-algebra. Then $(X_S,p_N)$ is also
minimal and unambiguous. \endpro

\proof Let $\pi_S : (X_S,p_N) \to (Y_S,q_N)$ be an isomorphism, with
$(Y_S,q_N)$ a minimal unambiguous \tsym{\Sigma}-algebra.
If $X'_S$ is an invariant family in $(X_S,p_N)$ then Lemma~2.2.1~(1) implies 
that 
$\pi_\sigma(X'_\sigma) = Y_\sigma$ for each $\sigma \in S$, since
$Y_S$ is the only invariant family in $(Y_S,q_N)$. Thus 
$\,X'_\sigma \, =\, \pi^{-1}_\sigma(\pi_\sigma(X'_\sigma)) 
        \, =\, \pi^{-1}_\sigma(Y_\sigma) \, =\, X_\sigma\,$
for each $\sigma \in S$, 
i.e., $X'_S = X_S$, and this shows that $(X_S,p_N)$ 
is minimal. Now if $\nu \in N_\sigma$ and $x \in \Im(p_\nu)$ then
by the definition of a homomorphism $\pi_\sigma(x) \in \Im(q_\nu)$.
It immediately follows that if $\nu_1,\, \nu_2 \in N_\sigma$ with 
$\nu_1 \ne \nu_2$ then 
$\Im(p_{\nu_1}) \cap \Im(p_{\nu_2}) = \varnothing$.
Finally, if $\nu \in N$ is of type $L \to \sigma$ then 
$q_\nu \, \ass L {\pi} \,=\, \pi_\sigma \, p_\nu$;
moreover, $q_\nu$ is injective, $\pi_\sigma$ is a bijection
and by Lemma~2.1.1~(3) $\ass L {\pi} $ is also a bijection. 
Thus $p_\nu$ is injective. \eop

\proclaim{Lemma~2.3.6} Any two initial \tsym{\Sigma}-algebras are isomorphic.
\endpro

\proof Let $(X_S,p_N)$ and $(Y_S,q_N)$ be initial \tsym{\Sigma}-algebras.
There then exists a unique homomorphism $\pi_S$ from $(X_S,p_N)$ to 
$(Y_S,q_N)$ and a unique homomorphism $\pi'_S$ from $(Y_S,q_N)$ to 
$(X_S,p_N)$. However, by Proposition~2.2.1 
$\lcurl \pi'_\sigma \, \pi_\sigma \,:\, \sigma \in S \rcurl$ is then a 
homomorphism from $(X_S,p_N)$ to itself, and by assumption there is only one 
such homomorphism. Hence $\pi'_\sigma (\pi_\sigma(x)) = x$ for each
$x \in X_\sigma$ (since the identity mappings also define a homomorphism from
$(X_S,p_N)$ to itself). In the same way $\pi_\sigma (\pi'_\sigma(y)) = y$ for 
each $y \in Y_\sigma$, and therefore $\pi_S$ is an isomorphism. 
(This proof is quite general: It is just the proof that in any 
category possessing initial objects these objects are, up to isomorphism, unique.) \eop 
\bigbreak
The proof of Proposition~2.3.2 can now be commenced. Suppose first that $(X_S,p_N)$
is minimal and regular and let $(Y_S,q_N)$ be any \tsym{\Sigma}-algebra. 
It is enough to construct a homomorphism $\pi_S$ from $(X_S,p_N)$ to $(Y_S,q_N)$, 
since Proposition~2.2.3~(1) then implies that this
homomorphism is unique.
\medskip
Let $\#_S$ be the family of mappings given by Lemma~2.3.4
(with $\#_\sigma : X_\sigma \to \Nat$ for each 
$\sigma \in S$) and for each $\sigma \in S$, $m \in \Nat$ let
$X^m_\sigma = \lcurl x \in X_\sigma \,:\, \#_\sigma(x) = m \rcurl$. Define
$\pi_S$ by defining $\pi_\sigma$ on $X^m_\sigma$ for each $\sigma \in S$ 
using induction on $m$. Let $x \in X^0_\sigma$; then, since $(X_S,p_N)$
is regular, there exists a unique 
$\nu \in N_\sigma$ and a unique element $c \in \dom(p_\nu)$ such that 
$p_\nu(c) = x$, and here $\nu$ must be of type $\varnothing \to \sigma$
and $x = p_\nu(\onept)$. Thus put $\pi_\sigma(x) = q_\nu(\onept)$,
which defines $\pi_\sigma$ on $X^0_\sigma$ for each $\sigma \in S$.
Now let $m > 0$ and suppose $\pi_\tau$ has already been defined on
$X^k_\tau$ for all $k < m$ and all $\tau \in S$.
Let $x \in X^m_\sigma$; again using the regularity of  $(X_S,p_N)$ there 
exists a unique $\nu \in N_\sigma$ and a unique $c \in \dom(p_\nu)$
such that $x = p_\nu(c)$. Let $\nu$ be of type $L \to \sigma$, so
here $L \ne \varnothing$ and
$c \in \ass L X $. Then by the defining property of $\#_S$ it follows that
$\#_\eta (c(\eta)) < m$ for each $\eta \in L$,
which means $\pi_\eta (c(\eta))$ is already defined
for each $\eta \in L$ and hence that $\ass L {\pi} (c)$ is already defined
(i.e., $\ass L {\pi} (c)$ is the element $c'$ of $\ass L Y $ given by
$c'(\eta) =  \pi_\eta (c(\eta))$ for each $\eta \in L$).
Thus here put $\pi_\sigma(x) = q_\nu(\ass L {\pi} (c))$.
In this way $\pi_\sigma$ is defined on $X^m_\sigma$ for each $\sigma \in S$
and each $m \in \Nat$, and the family $\pi_S$ is a homomorphism
more-or-less by construction. 
\medskip
This shows that any minimal regular (and thus any
minimal unambiguous) \tsym{\Sigma}-algebra is initial.
The converse is a special case 
of a more general situation: Suppose a property $P$ of 
\tsym{\Sigma}-algebras is given satisfying the following three conditions:
\bigbreak
{\parindent=25pt
\item{(1)} There exists a \tsym{\Sigma}-algebra having property $P$.
\medskip
\item{(2)} Every \tsym{\Sigma}-algebra having property $P$ is initial.
\medskip
\item{(3)} A \tsym{\Sigma}-algebra isomorphic to a
\tsym{\Sigma}-algebra having property $P$ also has property $P$.
\medskip\smallskip
}
Then every initial \tsym{\Sigma}-algebra has property $P$, and so
having property $P$ is equivalent to being initial.
(Let $(X_S,p_N)$ be initial; by (1) there exists a
\tsym{\Sigma}-algebra $(Y_S,q_N)$ having property $P$, and by (2)
and Lemma~2.3.6 $(X_S,p_N)$ and $(Y_S,q_N)$ are isomorphic; thus by (3)
$(X_S,p_N)$ has property $P$.)
Now by Proposition~2.3.1 there exists a minimal 
unambiguous \tsym{\Sigma}-algebra, and the first part of the proof 
shows that each minimal 
unambiguous \tsym{\Sigma}-algebra is initial.
Finally, by Lemma~2.3.5 a \tsym{\Sigma}-algebra isomorphic to a 
minimal unambiguous \tsym{\Sigma}-algebra is minimal and unambiguous,
and therefore each initial 
\tsym{\Sigma}-algebra is minimal and unambiguous. \eop
\bigbreak

Proposition~2.3.2 implies that the \tsym{\Lambda}-algebra 
$(X_B,p_K)$ in Example~2.2.1 is initial.

\proclaim{Proposition~2.3.3} There is exactly one isomorphism class of initial
\tsym{\Sigma}-algebras. (This means, more precisely, that an initial 
\tsym{\Sigma}-algebra $(X_S,p_N)$ exists and that a \tsym{\Sigma}-algebra 
$(Y_S,q_N)$ is initial if and only if it is isomorphic to $(X_S,p_N)$.) \endpro

\proof Propositions 2.3.1 and 2.3.2 imply that an initial 
\tsym{\Sigma}-algebra exists, and by Lemma~2.3.6 any two initial 
\tsym{\Sigma}-algebras are isomorphic.
Finally, it is also clear that a \tsym{\Sigma}-algebra isomorphic to an initial
\tsym{\Sigma}-algebra is itself initial. \eop
\bigbreak
A type $\sigma \in S$ will be said to be 
\indexddef{primitive}{primitive type}{}{type}{primitive}
if $\nu$ is of type $\varnothing \to \sigma$ for each 
$\nu \in N_\sigma$. (In the
signature $\Lambda$ in Example~2.2.1
the primitive types are therefore $\ftt{bool}$ 
and $\ftt{int}$.)
Note that if $(X_S,p_N)$ is an initial \tsym{\Sigma}-algebra and 
$\sigma \in S$ is 
a primitive type then the mapping $\nu \mapsto p_\nu(\onept)$ maps $N_\sigma$
bijectively onto $X_\sigma$. (For the \tsym{\Lambda}-algebra 
$(X_B,p_K)$ in Example~2.2.1 this gives the obvious bijections
between $K_{\fstt{bool}} = \{\ftt{True},\ftt{False}\}$ and
$X_{\fstt{bool}} = \Bool = \{T,F\}$ and between
$K_{\fstt{int}} = \SynInt$ and $X_{\fstt{int}} = \Int$.)
\bigskip

\proclaim{Proposition~2.3.4} The minimal subalgebra of an unambiguous 
\tsym{\Sigma}-algebra is initial. \endpro

\proof This follows from Proposition~2.3.2, since, as already
noted, any subalgebra of an unambiguous algebra is unambiguous. \eop
\bigskip
The concept of being initial can be  generalised and this leads to what is called 
a free algebra. 
For the rest of the section let $I$ be an \tsym{S}-typed set and define a family
$I_S$ by putting
$\,I_\sigma  \,=\, \lcurl \eta \in I : \langle \eta \rangle = \sigma \rcurl\,$
for each $\sigma \in S$,
\medskip
A \tsym{\Sigma}-algebra $(X_S,p_N)$ is said to 
\indexdef{contain}{algebra}{containing a typed set}
$I$ if $I_S \subset X_S$ (which is of course equivalent to having
$\eta \in X_\eta$ for each $\eta \in I$).
A \tsym{\Sigma}-algebra $(X_S,p_N)$ is then called  
\indexddef{\tsym{I}-free}{free algebra}{}{algebra}{free}
if it contains $I$ and 
for each \tsym{\Sigma}-algebra $(Y_S,q_N)$ and each $c \in \ass I Y $ there 
exists a unique \tsym{\Sigma}-homomorphism 
$\pi^c_S : (X_S,p_N) \to (Y_S,q_N)$ such that
$\pi^c_\eta (\eta) = c(\eta)$ for all $\eta \in I$.
In Proposition~2.3.6 it will be shown that an \tsym{I}-free \tsym{\Sigma}-algebra
exists and that it is, in an appropriate sense, unique.
\medskip
Note that a \tsym{\Sigma}-algebra $(X_S,p_N)$ is initial if and only if
it is \tsym{\varnothing}-free. Moreover, it will turn out that 
a \tsym{\Sigma}-algebra $(X_S,p_N)$ containing $I$ is \tsym{I}-free if 
and only if a related algebra $(X_S,p_{N_I})$ is initial with respect 
to an extended signature $\Sigma_I$. 
\medskip
For each $\eta \in I$ let 
$\diamond_\eta$ be some element not in $N$ and such that 
$\diamond_{\eta_1} \ne \diamond_{\eta_2}$ whenever $\eta_1 \ne \eta_2$.
Put $N_I = N \cup \lcurl \diamond_\eta \,:\, \eta \in I \rcurl$ and define 
a signature $\Sigma_I = (S,N_I,\Delta_I,\delta_I)$ with 
mappings $\Delta_I : N_I \to {\cal F}_S$ and
$\delta_I : N_I \to S$ given by
\smallskip
\mdisp{\Delta_I(\nu) \ =\ \cases{
      \Delta(\nu) & if $\,\nu \in N\,$, \cr
       \varnothing & if $\,\nu = \diamond_\eta\,$ for some 
                              $\,\eta \in I\,$, \cr }}
\vskip-\smallskipamount
\mdisp{\delta_I(\nu) \ =\ \cases{
      \delta(\nu) & if $\,\nu \in N\,$, \cr
        \langle \eta \rangle & if $\,\nu = \diamond_\eta\,$ for some 
                              $\,\eta \in I\,$. \cr }}
\smallskip
Thus in $\Sigma_I$ each $\nu \in N$ has the same type as it had in
$\Sigma$ and $\diamond_\eta$ is of type 
$\varnothing \to \langle \eta \rangle$ for each $\eta \in I$.
\medskip
Let $(X_S,p_N)$ be a \tsym{\Sigma}-algebra containing $I$ and for each
$\eta \in I$ define a mapping 
$p_{\diamond_\eta} : \Oneptset \to X_\eta$ 
by letting $p_{\diamond_\eta}(\onept) = \eta$. This results in a
\tsym{\Sigma_I}-algebra $(X_S,p_{N_I})$ which will be called the 
\indexddef{\tsym{\Sigma_I}-algebra associated with $(X_S,p_N)$}
{associated algebra}{}{algebra}{associated}.

\proclaim{Lemma~2.3.7} Let $(Y_S,q_{N_I})$ be any \tsym{\Sigma_I}-algebra and 
$\pi_S : X_S \to Y_S$ be a family of mappings. Then 
$\pi_S : (X_S,p_{N_I}) \to (Y_S,q_{N_I})$ is a \tsym{\Sigma_I}-homomorphism 
if and only if $\pi_S$ is a \tsym{\Sigma}-homomorphism from
$(X_S,p_N)$ to $(Y_S,q_N)$ with
$\pi_\eta(\eta) = q_{\diamond_\eta}(\onept)$ for each $\eta \in I$. 
\endpro

\proof This follows directly from the definition of $(X_S,p_{N_I})$. \eop

\proclaim{Proposition~2.3.5} $(X_S,p_N)$ is an \tsym{I}-free 
\tsym{\Sigma}-algebra if and only if 
$(X_S,p_{N_I})$ is an initial \tsym{\Sigma_I}-algebra. \endpro

\proof Suppose first that $(X_S,p_N)$ is \tsym{I}-free and let
$(Y_S,q_{N_I})$ be any \tsym{\Sigma_I}-algebra. Then, putting
$c(\eta) = q_{\diamond_\eta}(\onept)$ for each $\eta \in I$ defines an assignment
$c \in \ass I Y $, and thus there exists a unique \tsym{\Sigma}-homomorphism 
$\pi_S : (X_S,p_N) \to (Y_S,q_N)$ with
$\pi_\eta(\eta) = q_{\diamond_\eta}(\onept)$ for all $\eta \in I$.
Therefore by Lemma~2.3.7 $\pi_S$ is 
a \tsym{\Sigma_I}-homomorphism from $(X_S,p_{N_I})$ to $(Y_S,q_{N_I})$.
Consider any
\tsym{\Sigma_I}-homomorphism 
${\hat \pi}_S : (X_S,p_{N_I}) \to (Y_S,q_{N_I})$.
Then, again using Lemma~2.3.7, ${\hat \pi}_S$ is a
\tsym{\Sigma}-homomorphism from $(X_S,p_N)$ to $(Y_S,q_N)$ with
${\hat \pi}_\eta(\eta) = q_{\diamond_\eta}(\onept)$
for all $\eta \in I$, and hence ${\hat \pi}_S = \pi_S$.
This shows that $(X_S,p_{N_I})$ is an initial \tsym{\Sigma_I}-algebra.
\medskip
Suppose conversely $(X_S,p_{N_I})$ is an initial 
\tsym{\Sigma_I}-algebra;
let $(Y_S,q_N)$ be a \tsym{\Sigma}-algebra and $c \in \ass I Y $.
For each $\eta \in I$ define $q_{\diamond_\eta} : \Oneptset \to Y_\eta$ 
by $q_{\diamond_\eta} (\onept) = c(\eta)$; then $(Y_S,q_{N_I})$ is a
\tsym{\Sigma_I}-algebra and hence there is a unique
\tsym{\Sigma_I}-homomorphism 
$\pi^c_S : (X_S,p_{N_I}) \to (Y_S,q_{N_I})$.
But by Lemma~2.3.7 $\pi^c_S$ is then a
\tsym{\Sigma}-homomorphism from $(X_S,p_N)$ to $(Y_S,q_N)$ with
$\pi^c_\eta(\eta) = q_{\diamond_\eta}(\onept)= c(\eta)$ 
for all $\eta \in I$. Finally, consider any \tsym{\Sigma}-homomorphism 
${\hat \pi}^c_S$ from  $(X_S,p_N)$ to $(Y_S,q_N)$ 
with ${\hat \pi}^c_\eta(\eta) = c(\eta)$ for all
$\eta \in I$. Then Lemma~2.3.7 implies that
${\hat \pi}^c_S$ is also a
\tsym{\Sigma_I}-homomorphism from $(X_S,p_{N_I})$ to $(Y_S,q_{N_I})$
and thus ${\hat \pi}^c_S = \pi^c_S$.
This shows that $(X_S,p_N)$ is \tsym{I}-free. \eop
\bigbreak

\proclaim{Proposition~2.3.6} There exists an 
\tsym{I}-free \tsym{\Sigma}-algebra $(X_S,p_N)$. Moreover, if 
$(X'_S,p'_N)$ is any \tsym{I}-free \tsym{\Sigma}-algebra then there exists a unique
\tsym{\Sigma}-isomorphism $\pi_S$ from $(X_S,p_N)$ to $(X'_S,p'_N)$ 
such that $\pi_\eta(\eta) = \eta$ for each $\eta \in I$. \endpro

\proof By Proposition~2.3.3 there exists an initial \tsym{\Sigma_I}-algebra
$(X_S,p_{N_I})$, and clearly there then exists such an algebra in which
the sets in the family $X_S$ are disjoint. This implies in particular 
that $p_{\diamond_\xi}(\onept) \ne p_{\diamond_\eta}(\onept)$ whenever $\xi$ and 
$\eta$ are different elements of $I$. The set 
$\lcurl p_{\diamond_\eta}(\onept) : \eta \in I \rcurl$ can thus be identified 
with the set $I$ and so $I_S \subset X_S$.
In this way $(X_S,p_N)$ can be regarded as a \tsym{\Sigma}-algebra
containing $I$. Moreover, 
$(X_S,p_{N_I})$ is then the \tsym{\Sigma_I}-algebra associated with
$(X_S,p_N)$, and hence Proposition~2.3.5 implies that $(X_S,p_N)$ is 
\tsym{I}-free. There therefore exists an \tsym{I}-free \tsym{\Sigma}-algebra, and
the final statement follows directly from the definition of being
\tsym{I}-free. \eop 
\bigbreak
Let $(X_S,p_N)$ be a \tsym{\Sigma}-algebra containing $I$.
Then $(X_S,p_N)$ is said to be 
\indexddef{\tsym{I}-minimal}{minimal algebra}{}{algebra}{minimal}
if $X_S$ is the only
invariant family in $(X_S,p_N)$ containing $I_S$.
Moreover, $(X_S,p_N)$ is said to be 
\indexddef{\tsym{I}-unambiguous}{unambiguous algebra}{}{algebra}{unambiguous} 
if the mapping $p_\nu$ is injective for each $\nu \in N$ and for
each $\sigma \in S$ the sets $\Im(p_\nu)$, $\nu \in N_\sigma$, are disjoint subsets of 
$X_\sigma \setminus I_\sigma$. Finally, $(X_S,p_N)$ is called 
\indexddef{\tsym{I}-regular}{regular algebra}{}{algebra}{regular} 
if the mapping $p_\nu$ is injective for each $\nu \in N$ and for
each $\sigma \in S$ the sets
$\Im(p_\nu)$, $\nu \in N_\sigma$, form a partition of 
$X_\sigma \setminus I_\sigma$.

\proclaim{Proposition~2.3.7} The following statements are equivalent:
\medskip\smallskip
{\parindent=25pt
\item{(1)} $(X_S,p_N)$ is \tsym{I}-free.
\medskip
\item{(2)} $(X_S,p_N)$ is \tsym{I}-minimal and \tsym{I}-unambiguous.
\medskip
\item{(3)} $(X_S,p_N)$ is \tsym{I}-minimal and \tsym{I}-regular.
\medskip
} \endpro

\proof This follows from Propositions 2.3.2 and 2.3.5. \eop
\bigbreak
It is useful to express the equivalence between statements (1) and (3)
in Proposition~2.3.7 somewhat more explicitly:

\proclaim{Proposition~2.3.8} A \tsym{\Sigma}-algebra $(X_S,p_N)$
containing $I$ is \tsym{I}-free if and only if the following three conditions hold:
\medskip\smallskip
{\parindent=25pt
\item{(1)} $p_\nu$ is injective for each $\nu \in N$. 
\medskip
\item{(2)} The sets $\Im(p_\nu)$, $\nu \in N_\sigma$, form a partition of 
$X_\sigma \setminus I_\sigma$ for each $\sigma \in S$. 
\medskip
\item{(3)} $X_S$ is the only invariant family containing $I_S$. 
\bigskip
} \endpro

\proof Conditions (1) and (2) are those occurring in the definition
of being \tsym{I}-regular, and
condition (3) is that occurring in the definition of being \tsym{I}-minimal.
Thus by Proposition~2.3.7 $(X_S,p_N)$ 
is \tsym{I}-free if and only if (1), (2) and (3) hold. \eop

\vfill\eject

\hfline{40}{2.4 \ TERM ALGEBRAS}

\bigskip
\sectionhead {2.4} {Term algebras}
\bigskip\medskip
This section introduces what are called term algebras. 
These provide the simplest explicit examples of initial algebras,
and they can be used as the basic components of a `real' programming 
language. Until further notice 
assume that $\Sigma = (S,N,\Delta,\delta)$ is an enumerated signature.
(The general case is dealt with at the end of the section.)

\medskip

Let $Z$ be a set; then the concatenation of two lists $\ell,\, \ell' \in Z^*$ 
will be denoted by $\ell\ \ell'$. Thus $\,\ell\ \onept \,=\, \onept\ \ell \,=\, \ell\,$ 
and if $\ell = \list z m $, $\ell' = \list {z'} n $ with $m,\, n \ge 1$ then

\mdisp{\ell \ \ell'\  =\ \list z m \ \list z' n \ .} 

Concatenation is clearly associative, and hence if $p \ge 1$ and
$\lvector {\ell} p \in Z^*$ then {\it the\/} concatenation of
$\lvector {\ell} p $ can be denoted simply by $\list {\ell} p $. This 
notation is
clearly compatible with the notation being employed for the elements of
$Z^*$ (in the sense that $\ell = \list z n \in Z^*$ can be considered as the 
concatenation of the $n$ one element lists $\lvector z n $).
If $z \in Z$ and $\ell \in Z^*$ then in particular there is  the list
$z\ \ell$ obtained by adding $z$ to the beginning of the list $\ell$
(a list which is also being denoted by $z \triangleleft\, \ell$).
\medskip
Let $A$ be a set and $\Gamma : N \to A$ be a mapping. 
A \tsym{\Sigma}-algebra $(Y_S,q_N)$ can then be obtained as follows: For 
each $\sigma \in S$ put $Y_\sigma = A^*$; if $\nu \in N$ is of type 
$\list {\sigma} n \to \sigma$ (with $n \ge 0$) then let
$q_\nu : Y_{\sigma_1} \times \cdots \times Y_{\sigma_n} \to Y_\sigma$ be
the mapping defined by

\mdisp{q_\nu(\vector {\alpha} n )\ =\ \Gamma(\nu)\ \list {\alpha} n }
\smallskip
for each
$(\vector {\alpha} n ) \in Y_{\sigma_1} \times \cdots \times Y_{\sigma_n}$,
i.e., $q_\nu(\vector {\alpha} n )$ is the list obtained by
concatenating the element
$\Gamma(\nu)$ and the lists $ \lvector {\alpha} n $.
In particular, if $\nu \in N$ is of type $\onept \to \sigma$ then
$q_\nu(\onept)$ is just the list consisting of the single component
$\Gamma(\nu)$. 
(This is really just an instance of the construction described at the
end of Section~2.2.)
\medskip
Now let $(E_S,\blob_N)$ be the minimal subalgebra of $(Y_S,q_N)$, i.e., 
$E_S$ is the minimal invariant family and $\blob_\nu$ is the 
corresponding restriction of 
$q_\nu$ for each $\nu \in N$. This minimal 
\tsym{\Sigma}-algebra $(E_S,\blob_N)$ is called the 
\indexddef{term \tsym{\Sigma}-algebra specified by $\Gamma$}
{term algebra}{}{algebra}{term},  
and $\Gamma$ is then referred to as a 
\indexdef{term algebra specifier}{term algebra specifier}{}.
Thus $E_\sigma \subset A^*$ for each $\sigma \in S$ and the family 
$E_S$ can be thought of as being defined by the following rules:
\medskip\smallskip
{\parindent=25pt
\item{(1)} If $\nu \in N$ is of type $\onept \to \sigma$ then 
the list consisting of the single component $\Gamma(\nu)$ is an element of 
$E_\sigma$.
\medskip
\item{(2)} If $\nu \in N$ is of type $\list {\sigma} n \to \sigma$ with 
$n \ge 1$ and $e_j \in E_{\sigma_j}$ for each $\oneto j n $ then
$\,\Gamma(\nu)\ \list e n \,$ is an element of $E_\sigma$.
\medskip
\item{(3)} The only elements in $E_\sigma$ are those which can be obtained
using (1) and (2). 
\bigskip
}
By Proposition~2.2.4 it is clear that each element of $E_\sigma$ contains 
at least one component, i.e., $\onept \notin E_\sigma$ for each 
$\sigma \in S$. 
\medskip
The simplest example of this construction is with
$A = N$ and with $\Gamma : N \to N$ the identity mapping:
The term \tsym{\Sigma}-algebra specified by this mapping is called the
\indexddef{standard term \tsym{\Sigma}-algebra}
{standard term algebra}{}{term algebra}{standard}.
\medskip
In the special case when $(E_S,\blob_N)$ is the standard term 
\tsym{\Sigma}-algebra then $E_\sigma \subset N^*$ for each 
$\sigma \in S$ and the family $E_S$ is defined by the following rules:
\medskip\smallskip
{\parindent=25pt
\item{(1)} If $\nu \in N$ is of type $\onept \to \sigma$ then the list
consisting of the single component $\nu$ is an 
element of $E_\sigma$.
\medskip
\item{(2)} If $\nu \in N$ is of type $\list {\sigma} n \to \sigma$ with 
$n \ge 1$ and $e_j \in E_{\sigma_j}$ for each $\oneto j n $ then
$\ \nu\ e_1\ \cdots \ e_n\ $ is an element of $E_\sigma$.
\medskip
\item{(3)} The only elements in $E_\sigma$ are those which can be obtained
using (1) and (2). 
\bigskip
} 

\bigskip
\frame{20pt}{\bigskip 
{\it Example~2.4.1\enspace} Consider the standard term 
\tsym{\Lambda}-algebra
$(E_B, \blob_K)$ arising from the enumerated signature 
$\Lambda$ in Example~2.2.1. Then:
\medskip
{\leftskip=26pt
$E_{\fstt{bool}}$ consists of the two elements $\,\ftt{True}\,$ and 
$\,\ftt{False}\,$ of $K^*$.
\medskip
$E_{\fstt{nat}}$ consists of exactly the following elements of $K^*$:
\mdisp{\ftt{Zero}\,,\ \ftt{Succ Zero}\,,\ \ftt{Succ Succ Zero}\,, 
                               \ \ftt{Succ Succ Succ Zero}\,,\ \ldots\ .}
\smallskip
$E_{\fstt{int}}$ consists of all elements of $K^*$ having the form
$\,{\underline n}\,$ with $n \in \Int$.
\medskip
$E_{\fstt{pair}}$ consists of all elements of $K^*$ of the form 
$\ \ftt{Pair}\ {\underline m}\ {\underline n}\ $ with $m, \, n \in \Int$. 
(Each element of $E_{\fstt{pair}}$ thus has exactly three components.)
\medskip
$E_{\fstt{list}}$ consists of the element $\,\ftt{Nil}\,$ plus all 
elements of $K^*$ having the form $\ \ftt{Cons}\ {\underline n}\ e\ $
with $n \in \Int$ and $e \in E_{\fstt{list}}$. For instance, 
\mdisp{\ftt{Cons 42 Cons -128 Cons 0 Cons -21 Nil}} 
is an element of $E_{\fstt{list}}$.
\bigskip
}}
\bigskip\bigskip

It turns out that the standard term \tsym{\Sigma}-algebra is initial. 
This is a special case of Proposition~2.4.1 below.
\medskip
Let $\Gamma : N \to A$ be a term algebra specifier and 
$(E_S,\blob_N)$ be the term \tsym{\Sigma}-algebra specified by $\Gamma$.
For each $\nu \in N$ let 
$\chi_\nu : \dom(\blob_\nu) \times A^* \to A^*$ be defined by 
\mdisp{\chi_\nu(v,\alpha) \ =\ \blob_\nu(v)\ \alpha\ .} 

\proclaim{Lemma~2.4.1} Suppose 
$\Im(\chi_{\nu_1})$ and $\Im(\chi_{\nu_2})$ are disjoint subsets of 
$A^*$ whenever 
$\sigma \in S$ and $\nu_1,\, \nu_2 \in N_\sigma$ with $\nu_1 \ne \nu_2$.
Then $(E_S,\blob_N)$ is initial. 
\endpro

\proof By Proposition~2.3.2 it is enough to show that $(E_S,\blob_N)$ is 
unambiguous. But

\mdisp{\Im(\blob_\nu)\ =\ \chi_\nu(\dom(\blob_\nu) \times \{\onept\})
\ \subset\ \Im(\chi_\nu)}

for each $\nu \in N$; thus if
$\nu_1,\, \nu_2 \in N_\sigma$ with $\nu_1 \ne \nu_2$ then 
$\dom(\blob_{\nu_1})$ and 
$\dom(\blob_{\nu_2})$ are clearly disjoint. It therefore remains to 
show that $\blob_\nu$ is injective for each $\nu \in N$, and for this 
the following lemma is required:

\proclaim{Lemma~2.4.2} Suppose the assumption in the statement
of Lemma~2.4.1 is satisfied and let
$\chi_\sigma : E_\sigma \times A^* \to A^*$ be 
defined by $\chi_\sigma(e,\alpha)\ =\ e\ \alpha$. 
Then for each $\sigma \in S$ the mapping $\chi_\sigma$ is injective.
\endpro

\proof For each $\sigma \in S$ define a subset $G_\sigma$ of $A^*$ by

\mdisp{G_\sigma \ =\ \lcurl e \in E_\sigma \,:\, e\ \alpha \in E_\sigma 
     \frm{for some} \alpha \in A^* \frm{with} \alpha \ne \onept \rcurl}  

and put $G = \bigcup_{\sigma \in S} G_\sigma$. Then, since $\chi_\sigma$ is 
injective if and only if $G_\sigma = \varnothing$, it follows that $G \ne \varnothing$
if and only if $\chi_\sigma$ is not injective for some $\sigma \in S$. 
\medskip
Suppose that $G \ne \varnothing$ and let
$m = \min \lcurl |\alpha| \,:\, \alpha \in G \rcurl$, where $|\alpha|$ denotes
the number of components in the list $\alpha \in A^*$. There thus exists 
$\sigma \in S$, $e \in E_\sigma$ and $\alpha \in A^* \setminus \{\onept\}$
such that $|e| = m$ and $e' = e\ \alpha \in E_\sigma$. Now by 
Proposition~2.2.4 there exist $\nu,\,\nu' \in N_\sigma$ and 
$v \in \dom(\blob_\nu)$,
$v' \in \dom(\blob_{\nu'})$ such that $e = \blob_\nu(v)$ and 
$e' = \blob_{\nu'}(v')$,
and this implies that
$\chi_\nu(v,\alpha) \,=\, e\ \alpha \,=\, e' 
                           \,=\, e'\ \onept \,=\, \chi_{\nu'}(v',\onept)$, 
which by assumption is only possible if $\nu' = \nu$.
Let $\nu$ be of type $\list {\sigma} n \to \sigma$ and put
$\Gamma(\nu) = \alpha'$.
Thus if $v = (\vector e n )$ and $v' = (\vector {e'} n )$ then
$e\ =\ \alpha'\ \list e n $
and $e\ \alpha\ =\ \alpha'\ \list{e'} n $. 
There must therefore exist $1 \le j \le n$ with $e_j \ne e'_j$ (and so in 
particular $n > 0$). Let $i$ be the least index with $e_i \ne e'_i$ and let 
${\hat e}$ be the shorter of the lists $e_i$ and $e'_i$. But then 
${\hat e} \in G_{\sigma_i}$ and $|{\hat e}| < |e|$, which by the minimality
of $|e|$ is not possible.
\medskip
This shows that $G = \varnothing$ and hence that $\chi_\sigma$ is injective 
for each $\sigma \in S$. \eop 
\bigbreak
The proof of Lemma~2.4.1 can now be completed.
Let $\nu \in N$ be of type $\list {\sigma} n \to \sigma$ 
and let $v ,\, v' \in \dom(\blob_\nu)$ with 
$\blob_\nu(v) = \blob_\nu(v')$. This means
that if $v = (\vector e n )$ and $v' = (\vector {e'} n )$ then 
$\,\Gamma(\nu)\ \list e n \, =\, \Gamma(\nu)\ \list {e'} n \,$.
Consider $1 \le j \le n$ and assume $e_i = e'_i$ for each $i < j$. Then 
\smallskip
\ldisp{\quad \chi_{\sigma_j}(e_j,\, e_{j+1}\ \cdots\ e_n\,)
  \ =\ e_j\ e_{j+1}\ \cdots\ e_n}
\vskip-\medskipamount
\rdisp{\ =\ e'_j\ e'_{j+1}\ \cdots\ e'_n
 \ =\ \chi_{\sigma_j}(e'_j,\,e'_{j+1}\ \cdots\ e'_n \,)\quad}
\smallskip
and so by Lemma~2.4.2 $e_j = e'_j$. Therefore by $n$ applications of 
Lemma~2.4.2 it follows that $e_j = e'_j$ for each $\oneto j n $; i.e., $v = v'$.
This shows that $\blob_\nu$ is injective. \eop
\bigbreak
\proclaim{Proposition~2.4.1} Suppose that the restriction of $\Gamma$
to $N_\sigma$ is injective for each $\sigma \in S$. Then the term
\tsym{\Sigma}-algebra $(E_S,\blob_N)$ specified by $\Gamma$ is initial. 
\endpro

\proof This follows immediately from the fact that $\Gamma(\nu)$ is the 
first component of each element of $\Im(\chi_\nu)$, and hence the 
hypothesis of Lemma~2.4.1 is satisfied. \eop
\bigbreak

\proclaim{Proposition~2.4.2} The standard term \tsym{\Sigma}-algebra is initial. 
\endpro

\proof This follows from Proposition~2.4.1, since
the identity mapping used to specify the standard term \tsym{\Sigma}-algebra 
is trivially injective. \eop
\bigbreak

It will now no longer be assumed that $\Sigma$ is  enumerated. 
In order to apply the above results to $\Sigma$ 
the problem of replacing the general signature 
$\Sigma = (S,N,\Delta,\delta)$ 
with an `equivalent' enumerated signature $\Sigma' = (S,N,\Delta',\delta)$ must be
considered. In order to carry this out fix for each $\nu \in N$ an enumeration
of the elements in the set $\Delta(\nu)$: More precisely, choose
a bijective mapping $i_\nu$ from $[n_\nu]$ to the underlying set 
$\Delta(\nu)$, where $n_\nu$ is the cardinality of $\Delta(\nu)$.
There is then a mapping $\Delta' : N \to S^*$ given by
\mdisp{\Delta'(\nu) 
\ =\ \langle i_\nu(1) \rangle\ \cdots
 \ \langle i_\nu(n_\nu) \rangle}
for each $\nu \in N$. This defines an 
enumerated signature $\Sigma' = (S,N,\Delta',\delta)$, which will be called the
\indexdef{signature obtained from $\Sigma$ and the family 
of enumerations $i_N$}{signature}{obtained from family of enumerations}.

\proclaim{Lemma~2.4.3} Let $X_S$ be a family of sets and let
$\nu \in N$ with $\Delta'(\nu) = \list {\sigma} {n_\nu} $. Then the mapping
$i_\nu^* : \ass {\Delta(\nu)} X 
 \to X_{\sigma_1} \times \cdots \times X_{\sigma_{n_\nu}}$
given by 
\mdisp{i_\nu^*(c)
\ =\ (c(i_\nu(1)),\ldots,c(i_\nu(n_\nu)))}
for each $c \in \ass {\Delta(\nu)} X $ is a bijection. \endpro

\proof This is clear. \eop
\bigbreak

\proclaim{Proposition~2.4.3} Let $(X_S,p'_N)$ be a \tsym{\Sigma'}-algebra
and for each $\nu \in N$ put $p_\nu = p'_\nu\,i^*_\nu$. Then
$(X_S,p_N)$ is a \tsym{\Sigma}-algebra.
Conversely, if
$(X_S,p_N)$ is any \tsym{\Sigma}-algebra then there exists a unique
\tsym{\Sigma'}-algebra $(X_S,p'_N)$ such that 
$p_\nu = p'_\nu\,i^*_\nu$ for each $\nu \in N$. \endpro

\proof This follows immediately from Lemma~2.4.3. \eop
\bigbreak
If $(X_S,p'_N)$ is a \tsym{\Sigma'}-algebra
and $p_\nu = p'_\nu\,i^*_\nu$ for each $\nu \in N$ then
$(X_S,p_N)$ is called the
\indexddef{\tsym{\Sigma}-algebra associated with $(X_S,p'_N)$ and $i_N$}
{associated algebra}{}{algebra}{associated}.

Proposition~2.4.3 says that there is a one-to-one correspondence
between \tsym{\Sigma}-algebras and \tsym{\Sigma'}-algebras.
Moreover, Proposition~2.4.4 implies
that a \tsym{\Sigma'}-algebra has a property (such as being minimal or
initial) if and only if the corresponding  \tsym{\Sigma}-algebra
also has this property.
In this sense the signatures $\Sigma'$ and $\Sigma$ can be regarded as being
equivalent.
\bigskip

\proclaim{Proposition~2.4.4}  Let $(X_S,p'_N)$ be a \tsym{\Sigma'}-algebra
and let $(X_S,p_N)$ be the \tsym{\Sigma}-algebra 
associated with $(X_S,p'_N)$ and $i_N$. Then:
\medskip
(1)\enskip $(X_S,p_N)$ is minimal if and only if
$(X_S,p'_N)$ is a minimal \tsym{\Sigma'}-algebra. 
\medskip
(2)\enskip $(X_S,p_N)$ is initial if and only if
$(X_S,p'_N)$ is an initial \tsym{\Sigma'}-algebra. \endpro

\proof (1)\enskip This follows from the easily verified fact that
a family ${\grave X}_S$ is invariant in 
$(X_S,p'_N)$ if and only if it is invariant in $(X_S,p_N)$. 
\medskip
(2)\enskip This follows from (1), Lemma~2.3.3 and Proposition~2.3.2. \eop
\bigbreak
Now let $\Gamma : N \to A$ be a mapping, which will still be referred
to as a 
\indexdef{term algebra specifier}{term algebra specifier}{}.
Then $\Gamma$ is also a term algebra specifier for the signature
$\Sigma'$, so let $(E_S,\blob'_N)$ be the term \tsym{\Sigma'}-algebra
specified by $\Gamma$.
For each $\nu \in N$ put $\blob_\nu = \blob'_\nu\,i^*_\nu$
(with $i^*_\nu$ given by Lemma~2.4.3). 
This means that $E_\sigma \subset A^*$ for each $\sigma \in S$ and if 
$\nu \in N$ is of type $L \to \sigma$ then 
$\blob_\nu : \ass L E \to E_\sigma$ is the mapping given by

\mdisp{ \blob_\nu(c)
\ =\ \Gamma(\nu)\ \ c(i_\nu(1))\,\ \cdots\ \,c(i_\nu(m))}

for each $c \in \ass L E $, where $m = |L|$ (the cardinality of $L$).
Moreover, the family $E_S$ can be regarded as being defined by the following 
rules:
\medskip\smallskip
{\parindent=25pt
\item{(1)} If $\nu \in N$ is of type $\varnothing \to \sigma$ then the list
consisting of the single component $\Gamma(\nu)$ is an 
element of $E_\sigma$.
\medskip
\item{(2)} If $\nu \in N$ is of type $L \to \sigma$
with $L \ne \varnothing$
and $e_j \in E_{\sigma_j}$ for $\oneto j m $, where 
$\sigma_j = \langle i_\nu(j) \rangle$ and $m = |L|$,
then $\ \Gamma(\nu)\ e_1\ \cdots \ e_m\ $ is an element of $E_\sigma$.
\medskip
\item{(3)} The only elements in $E_\sigma$ are those which can be obtained
using (1) and (2). 
\bigskip
} 
The algebra $(E_S,\blob_N)$ is called the 
\indexddef{term \tsym{\Sigma}-algebra specified 
by $\Gamma$ and the family of enumerations $i_N$}
{term algebra}{}{algebra}{term},
or simply the 
\indexddef{standard term \tsym{\Sigma}-algebra defined by the family $i_N$}
{standard term algebra}{}{term algebra}{standard}
in the special case when $\Gamma : N \to N$ is the identity mapping.

\proclaim{Lemma~2.4.4} The \tsym{\Sigma}-algebra $(E_S,\blob_N)$ is initial 
if and only if $(E_S,\blob'_N)$ is an initial 
\tsym{\Sigma'}-algebra. \endpro

\proof This follows from Proposition~2.4.4~(2), since
by definition $(E_S,\blob_N)$ is the \tsym{\Sigma}-algebra associated with
$(E_S,\blob'_N)$ and $i_N$. \eop
\bigbreak

\proclaim{Proposition~2.4.5} If the restriction of $\Gamma$ to $N_\sigma$ is
injective for each $\sigma \in S$ then $(E_S,\blob_N)$ is an initial 
\tsym{\Sigma}-algebra. In particular, the standard term \tsym{\Sigma}-algebra
defined by any family of enumerations is initial. \endpro

\proof This follows immediately from Proposition~2.4.1 and Lemma~2.4.4. \eop
\bigbreak

Suppose that the term \tsym{\Sigma}-algebra $(E_S,\blob_N)$ specified by 
$\Gamma : N \to A$ and the family of enumeration $i_N$ 
is initial. Then for each \tsym{\Sigma}-algebra $(X_S,p_N)$ 
there exists a unique homomorphism from $(E_S,\blob_N)$ to $(X_S,p_N)$.
This homomorphism $\sem{\cdot}_S$ is 
the unique family of mappings satisfying the following conditions:
\medskip\smallskip
{\parindent=25pt
\item{(1)} If $\nu \in N$ is of type $\varnothing \to \sigma$ then 
$\,\sem{\,\Gamma(\nu)\,}_\sigma \,=\, p_\nu(\onept)\,$.
\medskip
\item{(2)} If $\nu \in N$ is of type $L \to \sigma$
with $L \ne \varnothing$
and $e_j \in E_{\sigma_j}$ for $\oneto j m $, where 
$\sigma_j = \langle i_\nu(j) \rangle$ and $m = |L|$, then 
\mdisp{\sem{\,\Gamma(\nu)\ \list e m \,}_\sigma
          \  =\  p_\nu(\ass L {\sem{c}} )\ ,}
\item{} where $c \in \ass L E $ is the assignment with
$c(i_\nu(j)) = e_j$ for each $\oneto j m $.
\bigbreak
}  
Of course, if the signature $\Sigma$ is enumerated and the above
constructions are applied with the family of identity enumerations $i_N$ then 
the end-result is that nothing happens, since in this case $\Sigma' = \Sigma$.

\bigskip\bigskip
\frame{20pt}{\bigskip 
{\it Example~2.4.2\enspace} There exists a unique homomorphism 
$\sem{\cdot}_B$ from the standard term \tsym{\Lambda}-algebra 
$(E_B,\blob_K)$ to the 
\tsym{\Lambda}-algebra $(X_B,p_K)$ which was introduced in Example~2.2.1.
Thus $\sem{\cdot}_B$ is the unique family of mappings such that: 
\bigskip
{\leftskip=32pt
$\sem{\ftt{True}}_{\fstt{bool}} = T$, \quad 
                                   $\sem{\ftt{False}}_{\fstt{bool}} = F$, 
\medskip
$\sem{\ftt{Zero}}_{\fstt{nat}} = 0$, \quad 
$\sem{\ftt{Succ}\ e}_{\fstt{nat}} = \sem{e}_{\fstt{nat}} + 1\ $ 
    for each $e \in E_{\fstt{nat}}$, 
\medskip
$\sem{{\underline n}}_{\fstt{int}} = n\ $ for each $n \in \Int$,
\medskip
$\sem{\ftt{Pair}\ {\underline m}\ {\underline n}}_{\fstt{pair}} 
                       = (m,n)\ $
    for each $(m,n) \in \Int \times \Int$,
\medskip
$\sem{\ftt{Nil}}_{\fstt{list}} = \onept$,
\medskip
$\sem{\ftt{Cons}\ {\underline n}\ e}_{\fstt{list}} 
                  \ =\ n\, \triangleleft\,\sem{e}_{\fstt{list}}\ $
for all $n \in \Int$, $e \in E_{\fstt{list}}$.
\bigskip
}
Note that $\sem{\cdot}_B$ is in fact an isomorphism, since 
$(X_B,p_K)$ is also initial.
\bigskip
}
\bigskip\bigskip

Term algebras also provide, via Proposition~2.3.6, simple examples of free 
\tsym{\Sigma}-algebras:
Let $I$ be an \tsym{S}-typed set which is assumed to be disjoint from the 
alphabet $A$ and let $\Sigma_I = (S,N_I,\Delta_I,\delta_I)$ be the signature
introduced in Section~2.3. Extend $\Gamma$ to a mapping
$\Gamma_I : N_I \to A \cup I$  by letting $\Gamma_I(\diamond_\eta) = \eta$ for each
$\eta \in I$, and let $i_{N_I}$  be the only extension possible of
$i_N$ to a family of enumerations (i.e., with $i_{\diamond_\eta}$ the unique
mapping from $\varnothing$ to $\varnothing$ for each $\eta \in I$).
Let $(E^I_S,\blob^I_{N_I})$ be the term \tsym{\Sigma_I}-algebra specified by
$\Gamma_I$ and $i_{N_I}$;
the algebra $(E^I_S,\blob^I_N)$ obtained by omitting the mappings
$\,\lcurl \blob^I_{\diamond_\eta} \,:\, \eta \in I \rcurl\,$
will then be called the 
\indexddef{term \tsym{\Sigma}-algebra specified by $\Gamma$, $i_N$ and $I$}
{term algebra}{}{algebra}{term}.
In particular,
$(E^I_S,\blob^I_N)$ contains $I$, since $I \subset (A \cup I)^*$.

\proclaim{Proposition~2.4.6} If the restriction of $\Gamma$ to $N_\sigma$ is
injective for each $\sigma \in S$ then $(E^I_S,\blob^I_N)$ is an \tsym{I}-free 
\tsym{\Sigma}-algebra. \endpro

\proof This follows immediately from Propositions 2.3.6 and 2.4.5, since the
restriction of $\Gamma_I$ to $(N_I)_\sigma$ is
injective for each $\sigma \in S$. \eop
\bigbreak
\vfill\eject

\hfline{46}{2.5 \ EXTENSIONS OF SIGNATURES AND ALGEBRAS}

\sectionhead {2.5} {Extensions of signatures and algebras}
\bigskip\medskip
Let $\Sigma = (S,N,\Delta,\delta)$ be a signature.
A signature $\Sigma' = (S',N',\Delta',\delta')$ is said to be an 
\indexddef{extension}{extension}{of a signature}{signature}{extension of} 
of $\Sigma$ if $S \subset S'$, $N \subset N'$ and
$\Delta$ (resp.\ $\delta$) is the restriction of 
$\Delta'$ (resp.\ $\delta'$) to $N$. Thus if $\nu \in N$ is of
type $L \to \sigma$ in the signature $\Sigma$ then it is still of type 
$L \to \sigma$ in the extension $\Sigma'$. In particular, 
$\Sigma$ is trivially an extension of itself.
\medskip
In what follows let $\Sigma = (S,N,\Delta,\delta)$ and 
$\Sigma' = (S',N',\Delta',\delta')$ be signatures with
$\Sigma'$ an extension of $\Sigma$. 
\medskip
A \tsym{\Sigma'}-algebra $(Y_{S'},q_{N'})$ is now called an 
\indexddef{extension}{extension}{of an algebra}{algebra}{extension of}
of a \tsym{\Sigma}-algebra $(X_S,p_N)$ if 
$X_\sigma \subset Y_\sigma$ for each $\sigma \in S$ and 
$p_\nu$ is the restriction of $q_\nu$ to $\dom(p_\nu)$ for each 
$\nu \in N$. (The case $\Sigma' = \Sigma$ is certainly not 
excluded here, and in fact it will occur frequently in what follows.)
\medskip
Note that if $(Y_{S'},q_{N'})$ is a \tsym{\Sigma'}-algebra then
$(Y_S,q_N)$, obtained by omitting the sets 
$\,\lcurl Y_\sigma \,:\, \sigma \in S' \setminus S \rcurl\,$ from the family $Y_{S'}$
and the mappings
$\,\lcurl q_\nu \,:\, \nu \in N' \setminus N \rcurl\,$ from the family $q_{N'}$,
is a \tsym{\Sigma}-algebra.
Moreover, if $\pi_{S'}$ is a \tsym{\Sigma'}-homomorphism from 
$(Y_{S'},q_{N'})$ to $(Z_{S'},r_{N'})$ 
then the family $\pi_S$, obtained by omitting the mappings
$\,\lcurl \pi_\sigma \,:\, \sigma \in S' \setminus S \rcurl\,$ from the family $\pi_{S'}$,
is a \tsym{\Sigma}-homomorphism from $(Y_S,q_N)$ to $(Z_S,r_N)$. 
\medskip
The following is a variation on Proposition~2.3.3:

\proclaim{Proposition~2.5.1} Let $(X_S,p_N)$ be an initial
\tsym{\Sigma}-algebra. Then there exists an initial \tsym{\Sigma'}-algebra
which is an extension of $(X_S,p_N)$. \endpro

\proof By Proposition~2.3.3 there exists an initial \tsym{\Sigma'}-algebra
$(Y_{S'},q_{N'})$ and $(Y_S,q_N)$ (obtained by omitting the sets 
$\lcurl Y_\sigma \,:\, \sigma \in S' \setminus S \rcurl$  and the mappings
$\lcurl q_\sigma \,:\, \sigma \in N' \setminus N \rcurl$) is then a 
\tsym{\Sigma}-algebra; let $({\hat Y}_S,{\hat q}_N)$ be the minimal 
subalgebra of $(Y_S,q_N)$.
Now $(Y_S,q_N)$ is unambiguous, since
by Proposition~2.3.2 $(Y_{S'},q_{N'})$ is unambiguous.
Hence $({\hat Y}_S,{\hat q}_N)$ is an initial
\tsym{\Sigma}-algebra, and by construction
$(Y_{S'},q_{N'})$ is an extension of $({\hat Y}_S,{\hat q}_N)$.
The unique isomorphism from $(X_S,p_N)$ to 
$({\hat Y}_S,{\hat q}_N)$ can therefore be used to construct an initial 
\tsym{\Sigma'}-algebra which is an extension of $(X_S,p_N)$. \eop
\bigbreak
\proclaim{Proposition~2.5.2} Let $(X'_{S'},p'_{N'})$ be an initial
\tsym{\Sigma'}-algebra which is an extension of an initial 
\tsym{\Sigma}-algebra $(X_S,p_N)$. Moreover, let
$(Y'_{S'},q'_{N'})$ be any
\tsym{\Sigma'}-algebra which is an extension of some 
\tsym{\Sigma}-algebra $(Y_S,q_N)$. Then the unique
\tsym{\Sigma'}-homomorphism $\pi'_{S'}$ from
$(X'_{S'},p'_{N'})$ to $(Y'_{S'},q'_{N'})$ is an extension of the
unique \tsym{\Sigma}-homomorphism $\pi_S$ from $(X_S,p_N)$ to $(Y_S,q_N)$,
i.e., $\pi'_\sigma(x) = \pi_\sigma(x)$ for all $x \in X_\sigma$, 
$\sigma \in S$. \endpro

\proof For each $\sigma \in S$ let $\pi'_\sigma : X_\sigma \to Y'_\sigma$
be the restriction of $\pi'_\sigma$ to $X_\sigma$, and consider 
$\pi_\sigma$ as a mapping from $X_\sigma$ to $Y'_\sigma$.
Then $\pi'_S$ and $\pi_S$ are both \tsym{\Sigma}-homomorphisms from
the initial \tsym{\Sigma}-algebra $(X_S,p_N)$ to the \tsym{\Sigma}-algebra
$(Y'_S,q'_N)$, and hence $\pi'_S = \pi_S$,
i.e., $\pi'_\sigma(x) = \pi_\sigma(x)$ for all $x \in X_\sigma$, 
$\sigma \in S$. \eop
\bigbreak
Now let $I$ be an \tsym{S}-typed set and again let 
$\,I_\sigma  \,=\, \lcurl \eta \in I : \langle \eta \rangle = \sigma \rcurl\,$
for each $\sigma \in S$. A \tsym{\Sigma}-algebra $(Z_S,r_N)$ is said to be 
\indexdef{disjoint}{algebra}{disjoint from a typed set}
from $I$ if $\eta \notin Z_\eta$ for each $\eta \in I$, which is of course equivalent
to having $I_\sigma \cap  Z_\sigma = \varnothing$ for each $\sigma \in S$.

\proclaim{Lemma~2.5.1} Let $(X_S,p_N)$ be an \tsym{I}-free 
\tsym{\Sigma}-algebra which is an extension of an initial 
\tsym{\Sigma}-algebra $(Z_S,r_N)$. Then $(Z_S,r_N)$ is disjoint from $I$. 
\endpro

\proof By Proposition~2.3.7 $(X_S,p_N)$ is \tsym{I}-ambiguous and hence
$\,\Im(p_\nu) \subset X_\sigma \setminus I_\sigma\,$ for all
$\nu \in N_\sigma$, $\sigma \in S$.
But $\Im(r_\nu) \subset \Im(p_\nu)$ for each $\nu \in N$, and so by 
Proposition~2.3.2
$\,Z_\sigma \subset X_\sigma \setminus I_\sigma\,$ for all $\sigma \in S$,
i.e., $I_\sigma \cap  Z_\sigma = \varnothing$ for each $\sigma \in S$. 
\eop 
\bigbreak
In the following two results let $I$ be an \tsym{S}-typed set and
$(Z_S,r_N)$ be an initial \tsym{\Sigma}-algebra disjoint from $I$.
The following is a variation on Proposition~2.5.2:

\proclaim{Proposition~2.5.3} There exists an \tsym{I}-free 
\tsym{\Sigma}-algebra $(X_S,p_N)$ which is an extension of $(Z_S,r_N)$. 
Moreover, if $(X'_S,p'_N)$ is any such
\tsym{I}-free \tsym{\Sigma}-algebra 
and $\pi_S$ is the unique \tsym{\Sigma}-isomorphism from $(X_S,p_N)$ to 
$(X'_S,p'_N)$ such that $\pi_\eta(\eta) = \eta$ for each 
$\eta \in I$ then $\pi_\sigma(z) = z$ for all $z \in Z_\sigma$, $\sigma \in S$.
\endpro

\proof By Proposition~2.3.6 there exists an \tsym{I}-free \tsym{\Sigma}-algebra
$({\bar X}_S,{\bar  p}_N)$. Let $({\bar Z}_S,{\bar r}_N)$ be the minimal 
subalgebra of $({\bar X}_S,{\bar  p}_N)$; then  
by (1) and (2) in Proposition~2.3.8 
$({\bar Z}_S,{\bar r}_N)$ is also
unambiguous and therefore Proposition~2.3.2 implies that
$({\bar Z}_S,{\bar r}_N)$ is 
an initial \tsym{\Sigma}-algebra. Moreover, by Lemma~2.5.1 
$({\bar Z}_S,{\bar r}_N)$ is disjoint from $I$ and by construction
$({\bar X}_S,{\bar p}_N)$ is an extension of $({\bar Z}_S,{\bar r}_N)$. 
Now unique isomorphism from $(Z_S,r_N)$ to 
$({\bar Z}_S,{\bar r}_N)$ can be used to construct an \tsym{I}-free 
\tsym{\Sigma}-algebra $(X_S,p_N)$ which is an extension of $(Z_S,r_N)$
(and note that here the assumption is needed that
$(Z_S,r_N)$ is disjoint from $I$).
Finally, consider any \tsym{\Sigma}-algebra $(X'_S,p'_N)$   
which is an extension of $(Z_S,r_N)$ 
and let $\pi_S$ be any \tsym{\Sigma}-homomorphism from $(X_S,p_N)$ to 
$(X'_S,p'_N)$. Then $\pi_\sigma(z) = z$ for all $z \in Z_\sigma$, 
$\sigma \in S$, since the family $Z'_S$ defined by
$Z'_\sigma = \lcurl z \in Z_\sigma \,:\, \pi_\sigma(z) = z \rcurl$
for each $\sigma \in S$ is clearly invariant in $(Z_S,r_N)$.
\eop
\bigbreak
Now consider a \tsym{\Sigma}-algebra
$(Y_S,q_N)$ and an assignment $c \in \ass I Y $. Then there is both the unique 
\tsym{\Sigma}-homomorphism $\pi^c_S$ from $(X_S,p_N)$ to $(Y_S,q_N)$ 
with $\pi^c_\eta (\eta) = c(\eta)$ for each $\eta \in I$ 
and also the unique \tsym{\Sigma}-homomorphism $\pi_S$ from
$(Z_S,r_N)$ to $(Y_S,q_N)$.

\proclaim{Proposition~2.5.4} $\pi^c_\sigma(z) =  \pi_\sigma(z)$
for all $z \in Z_\sigma$, $\sigma \in S$. \endpro

\proof For each $\sigma \in S$ let $\pi'_\sigma$ denote the restriction of 
$\pi^c_\sigma$ to $Z_\sigma$. Then
$\pi'_S$ is a \tsym{\Sigma}-homomorphism from
$(Z_S,r_N)$ to $(Y_S,q_N)$ and thus $\pi'_S = \pi_S$. \eop
\bigbreak
The final topic of this section is the construction of term algebras.
Suppose in what follows that $\Gamma' : N' \to A'$ is a term algebra specifier 
which is an 
\indexddef{extension}
{extension}{of a term algebra specifier}{term algebra specifier}{extension of}
of a term algebra specifier 
$\Gamma : N \to A$ in the sense that
$A \subset A'$ and $\Gamma(\nu) = \Gamma'(\nu)$ for each $\nu \in N$.

\proclaim{Lemma~2.5.2} Assume the signatures $\Sigma$ and $\Sigma'$ are both 
enumerated. Then the term \tsym{\Sigma'}-algebra 
specified by $\Gamma'$ is an extension of the term \tsym{\Sigma}-algebra 
$(E_S,\blob_N)$ specified by
$\Gamma$. \endpro

\proof This follows from Lemma~2.5.3 below, because the \tsym{\Sigma'}-algebra 
corresponding to the \tsym{\Sigma}-algebra $(Y_S,q_N)$ in the definition of
$(E_S,\blob_N)$ is an extension of $(Y_S,q_N)$. \eop
\bigbreak

\proclaim{Lemma~2.5.3} Let $(Y_{S'},q_{N'})$ be a \tsym{\Sigma'}-algebra 
which is an extension of a \tsym{\Sigma}-algebra
$(X_S,p_N)$, and let ${\hat X}_S$ be the minimal invariant family in
$(X_S,p_N)$ and ${\hat Y}_{S'}$ the minimal invariant family in
$(Y_{S'},q_{N'})$. Then ${\hat X}_S \subset {\hat Y}_S$. \endpro

\proof Let ${\grave Y}_{S'}$ be any invariant family in $(Y_{S'},q_{N'})$ 
and put $Z_\sigma = {\grave Y}_\sigma \cap X_\sigma$ for each $\sigma \in S$.
Then the family $Z_S$ is invariant in $(X_S,p_N)$, since 
\mdisp{p_\nu(\ass L Z ) \ \subset\  q_\nu(\ass L {\grave Y} ) \cap X_\sigma
  \ \subset\  {\grave Y}_\sigma \cap X_\sigma \ =\  Z_\sigma}
for each $\nu \in N$ of type $L \to \sigma$.
Thus ${\hat X}_S \subset Z_S$ and hence ${\hat X}_S \subset {\grave Y}_S$.
In particular, ${\hat X}_S \subset {\hat Y}_S$. \eop

\bigbreak

Lemma~2.5.2 can now be extended to the case where it not assumed 
that $\Sigma$ and $\Sigma'$ are enumerated. Let
$i'_{N'}$ be an enumeration  for $\Sigma'$ which is an 
\indexddef{extension}{extension}{of an enumeration}{enumeration}{extension of} 
of an enumeration $i_{N}$ for $\Sigma$ in that $i'_\nu = i_\nu$ for each
$\nu \in N$. 

\proclaim{Proposition~2.5.5} The term \tsym{\Sigma'}-algebra 
specified by $\Gamma'$ and $i'_{N'}$ is an extension of the
term \tsym{\Sigma}-algebra specified by
$\Gamma$ and $i_N$. \endpro

\proof This follows from Lemma~2.5.2 since the enumerated signature obtained from
$\Sigma'$ and the family of enumerations $i'_{N'}$ is an extension of that obtained from
$\Sigma$ and the family $i_N$. \eop
\bigbreak
In particular Proposition~2.5.5 implies that the
standard term \tsym{\Sigma'}-algebra defined by the family of enumerations $i'_{N'}$
is an extension of the standard term \tsym{\Sigma}-algebra defined by the family $i_N$.

\vfill\eject

\hfline{49}{2.6 \ TREE ALGEBRAS}

\sectionhead {2.6} {Tree algebras}
\bigskip\medskip
It has been seen that
an initial \tsym{\Sigma}-algebra can be obtained explicitly as an
appropriate term algebra. In the present section 
a further explicit method of constructing initial \tsym{\Sigma}-algebras is given,
this time involving labelled trees. The material presented here will not be
needed in the following chapters and so this section can be omitted.
\medskip
For the whole of the section let $\Sigma = (S,N,\Delta,\delta)$ be a signature. 
Let $\,J = \bigcup_{\nu \in N} \Delta(\nu)\,$ (meaning the union 
of the underlying sets, ignoring the
associated typings), and note that if the signature $\Sigma$ is 
enumerated then $J$ is either 
$[n]$ for some $n \in \Nat$ or the set of positive integers
$\Nat_+$. 
In the construction given below an important role is played by certain subsets of
the set of lists $J^*$ and the section thus starts by looking at some simple facts 
about the subsets of $X^*$ for an arbitrary set $X$.
\medskip
As in Chapter~1 let $\triangleleft : X \times X^* \to X^*$
and $\triangleright : X^* \times X \to X^*$ be defined by
\smallskip
\mdisp{x\,\triangleleft\, \list x m 
\ =\ x\ \list x m }
\vskip-\medskipamount
\mdisp{\list x m \,\triangleright\, x 
\ =\ \list x m \ x }
\smallskip
for all $x,\,\lvector x m \in X$. The facts which are
needed about these operations are exactly those listed in the following lemma:

\proclaim{Lemma~2.6.1} (1)\enskip 
$\,x \triangleleft \onept \,=\, \onept \triangleright x\,$ for all $x \in X$.
\medskip
(2)\enskip $\,x \triangleleft (s \triangleright x')
                \,=\,  (x \triangleleft s ) \triangleright x'\,$
for all $x,\,x' \in X$, $s \in X^*$. 
\medskip
(3)\enskip The mappings
$\triangleleft$ and $\triangleright$ are both injective.
\medskip
(4)\enskip $\,\Im(\triangleleft) \,=\, \Im(\triangleright) 
\,=\, X^* \setminus \{\onept\}\,$. 
\medskip
(5)\enskip Let $Z$ be a subset of $X^*$
with $\onept \in Z$ and such that either
$x \,\triangleleft\, s \in Z$ for all $x \in X$, $s \in Z$, or
$s \,\triangleright\, x \in Z$ for all $s \in Z$, $x \in X$. 
Then $Z = X^*$. 
\medskip
(6)\enskip There exists a unique mapping
$|\cdot| : X^* \to \Nat$ with $|\onept| = 0$ such that
$|x \triangleleft s| = 1 + |s|$ for all $x \in X$, $s \in X^*$.
\endpro

\proof Straightforward. Of course in (6) $|s|$ is just the number of 
components in the list $s$. \eop

\proclaim{Lemma~2.6.2} Each finite subset $A$ of $X^*$ containing 
$\onept$ has a unique representation
\mdisp{A \ =\ \{\onept\} \cup \bigcup\limits_{x \in B} x \triangleleft A_x}
with $B$ a finite subset of $X$ and with $A_x$ a non-empty finite subset 
of $X^*$ for each $x \in B$.
Moreover, the sets $x \triangleleft A_x$, $x \in B$, and 
$\{\onept\}$ are all disjoint.
\endpro

\proof If $s \in X^* \setminus \{\onept\}$ then by Lemma~2.6.1~(3) and (4) 
there exists a unique $x \in X$ and a unique element $s' \in X^*$ such that
$s = x \triangleleft s'$. The unique representation is therefore with
$B = \lcurl x \in X \,:\, (x \triangleleft X^*) \cap A \ne \varnothing \rcurl$
and with $A_x = \lcurl s \in X^* \,:\, x \triangleleft s \in A \rcurl$ for each
$x \in B$. \eop
\bigbreak
A subset $A$ of $X^*$ is said to be
\indexddef{\tsym{\triangleright}-preinvariant}
{preinvariant set}{}{set}{preinvariant}
if $s \in A$ whenever $s \triangleright x \in A$ for some $x \in X$.

\proclaim{Lemma~2.6.3} Each non-empty \tsym{\triangleright}-preinvariant 
subset of $X^*$ contains $\onept$.
\endpro
\proof Let $A$ be a \tsym{\triangleright}-preinvariant subset of 
$X^*$ and put $B = X^* \setminus A$; thus 
$s \triangleright x \in B$ for all $s \in B$, $x \in X$.
But if $\onept \in B$ then Lemma~2.6.1~(5) implies that $B = X^*$,
and hence $\onept \in A$ whenever $A \ne \varnothing$. \eop
\bigbreak
The set of all non-empty finite
\tsym{\triangleright}-preinvariant subsets of $X^*$ will be denoted by ${\cal S}$.
By Lemma~2.6.3 each subset in ${\cal S}$ contains the element $\onept$.

\proclaim{Proposition~2.6.1}
(1)\enskip $\{\onept\} \in {\cal S}$
and  ${\cal S}$ is closed under finite unions.
\medskip
(2)\enskip Let $B$ be a finite subset of $X$ and for each $x \in B$ let 
$A_x \in {\cal S}$. Then
\mdisp{\{\onept\} \,\cup\, \bigcup\limits_{x \in B} x \triangleleft A_x} 
is an element of ${\cal S}$.
\medskip
(3)\enskip Let $A \in {\cal S}$ and let 
$\{\onept\} \cup \bigcup_{x \in B} x \triangleleft A_x$ be the unique 
representation of $A$ given in Lemma~2.6.2. Then
$A_x \in {\cal S}$ for each $x \in B$. 
\endpro

\proof (1)\enskip This is clear. 
\medskip
(2)\enskip By (1) it is enough to consider the case when $B = \{x\}$, so let
$A \in {\cal S}$, $x \in X$ and put 
$A' = \{\onept\} \cup (x \triangleleft A)$. Suppose that 
$s \triangleright x' \in A'$ for some 
$s \in X^*$, $x' \in X$. Then, since by Lemma~2.6.1~(4) 
$s \triangleright x' \ne \onept$, it follows that 
$s \triangleright x' = x \triangleleft s'$ for some $s' \in A$.
The first possibility here is that $s' = \onept$; but then by Lemma~2.6.1~(1)
\mdisp{s \triangleright x'\ =\ x \triangleleft \onept 
\ =\ \onept \triangleright x\ ,}
which by Lemma~2.6.1~(3) implies that $x' = x$ and $s = \onept$, and in
particular $s  \in A'$. Assume next that $s' \ne \onept$; then 
by Lemma~2.6.1~(4) $s' = {\breve s} \triangleright {\breve x}$ for some
${\breve s} \in X^*$, ${\breve x} \in X$, and here
by Lemma~2.6.1~(2) 
\mdisp{s \triangleright x' 
\  =\ x \triangleleft ({\breve s} \triangleright {\breve x}) 
\  =\ (x \triangleleft {\breve s}) \triangleright {\breve x}\ ,} 
which by Lemma~2.6.1~(3) implies that
$x' = {\breve x}$ and $s = x \triangleleft {\breve s}$. 
But ${\breve s} \in A$ (because $A$ is \tsym{\triangleright}-preinvariant and 
$s' = {\breve s} \triangleright {\breve x} \in A$), and hence $s \in A'$.
This shows that $A' \in {\cal S}$.
\medskip
(3)\enskip Let $x \in B$ and suppose $s \triangleright x' \in A_x$ for some
$s \in X^*$, $x' \in X$. Then $x \triangleleft s \in A$, since
by Lemma~2.6.1~(2) 
$(x \triangleleft s) \triangleright x' = 
             x \triangleleft (s \triangleright x') \in A$
and $A$ is \tsym{\triangleright}-preinvariant, which by Lemma~2.6.1
(3) and (4) is only possible if $s \in A_x$. Hence $A_x \in {\cal S}$. \eop
\bigbreak

With these preparations made, the construction of an initial 
\tsym{\Sigma}-algebra can now begin.
As at the beginning of the section put
$\,J = \bigcup_{\nu \in N} \Delta(\nu)\,$.
In what follows the set $X$ in the above discussion will be taken to be $J$.
In particular, this means that ${\cal S}$ is now the set of all non-empty 
finite \tsym{\triangleright}-preinvariant subsets of $J^*$. 

\medskip
Denote by ${\cal L}$ the set of all those partial mappings from $J^*$ to 
$N$ whose domains are elements of ${\cal S}$.
For each $\sigma \in S$ put ${\cal L}_\sigma = {\cal L}$
and if $\nu \in N$ is of type $L \to \sigma$ then
define a mapping 
$\amalg_\nu : \ass L {{\cal L}} \to {\cal L}_\sigma$ by letting
$\,\amalg_\nu(c) \,=\, \lambda\,$, where 
\mdisp{\dom(\lambda)
\ =\  \{\onept\} \cup \bigcup\limits_{\eta \in L} 
           \eta \,\triangleleft\, \dom(c(\eta))}
with $\lambda(\onept) = \nu$
and $\lambda(\eta \triangleleft s) = c(\eta)\,(s)$ for all 
$s \in \dom(c(\eta))$, $\eta \in L$ 
(and note that by Lemma~2.6.1 (3) and (4) and Proposition~2.6.1~(2)
this definition makes sense).
In particular, if $\nu$ is of type $\varnothing \to \sigma$ then
$\amalg_\nu : \Oneptset \to {\cal L}_\sigma$ is the mapping with
$\amalg_\nu(\onept) = \lambda$, where $\dom(\lambda) = \{\onept\}$
and $\lambda(\onept) = \nu$.
Thus $({\cal L}_S,\amalg_N)$ is a \tsym{\Sigma}-algebra. 

\proclaim{Proposition~2.6.2} The \tsym{\Sigma}-algebra 
$({\cal L}_S,\amalg_N)$ is unambiguous. \endpro

\proof If $\nu \in N$ and $\lambda \in \Im(\amalg_\nu)$
then $\onept \in \dom(\lambda)$ and $\lambda(\onept) = \nu$, and from this
it immediately follows that
$\Im(\amalg_{\nu_1}) \cap \Im(\amalg_{\nu_2}) = \varnothing$
whenever $\nu_1 \ne \nu_2$. It remains to show that
$\amalg_\nu$ is injective for each $\nu \in N$, so assume
$\nu$ of type $L \to \sigma$ and consider $c,\,c' \in \ass L {{\cal L}} $
with $\amalg_\nu(c) = \amalg_\nu(c')$.
Then
\mdisp{  
\{\onept\} \cup \bigcup\limits_{\eta \in L} 
       \eta \,\triangleleft\, \dom(c(\eta))
\ =\  \dom(\lambda)\ =\ \{\onept\} \cup 
     \bigcup\limits_{\eta \in L} 
       \eta \,\triangleleft\, \dom(c'(\eta))\ ,}
where $\lambda = \amalg_\nu(c)$,
and hence by Lemma~2.6.2 $\dom(c(\eta)) = \dom(c'(\eta))$
for each $\eta \in L$. Let $\eta \in L$ and $s \in \dom(c(\eta))$; then
$c(\eta)\,(s) = \lambda(\eta \triangleleft s) = c'(\eta)\,(s)$, and therefore
$c(\eta) = c'(\eta)$. Thus $c = c'$, and
this shows that $\amalg_\nu$ is injective. \eop
\bigbreak
Now denote the minimal subalgebra of $({\cal L}_S,\amalg_N)$ 
by $({\cal T}_S,\nabla_N)$; thus 
${\cal T}_\sigma \subset{\cal L}_\sigma = {\cal L}$ for 
each $\sigma \in S$, ${\cal T}_S$ is the 
minimal invariant family and for each $\nu \in N$ 
of type $L \to \sigma$ the mapping 
$\nabla_\nu$ is the restriction of $\amalg_\nu$ to 
$\ass L {{\cal T}} $.
The family ${\cal T}_S$ can be thought of as being defined by the following 
rules:
\medskip\smallskip
{\parindent=25pt
\item{(1)} If $\nu \in N$ is of type $\varnothing \to \sigma$ then
the mapping $\lambda$ with $\dom(\lambda) = \{\onept\}$ and with
$\lambda(\onept) = \nu$ is an element of ${\cal T}_\sigma$.
\medskip
\item{(2)} If $\nu \in N$ is of type $L \to \sigma$ with 
$L \ne \varnothing$ and $c \in \ass L {{\cal T}} $ is such that
$c(\eta) \in {\cal T}_\eta $ for each $\eta \in L$ then
the mapping $\lambda$ with
\mdisp{\dom(\lambda) 
\ = \ \{\onept\} \,\cup \,\bigcup\limits_{\eta \in L}
                \eta\,\triangleleft\,\dom(c(\eta))\ ,}
\item{}
$\lambda(\onept) = \nu$
and $\lambda(\eta \triangleleft s) = c(\eta)\,(s)$ for all 
$s \in \dom(c(\eta))$ and all $\eta \in L$ is an element of ${\cal T}_\sigma$.
\medskip
\item{(3)} The only elements in ${\cal T}_\sigma$ are those which can be obtained
using (1) and (2). 
\bigbreak
}

\proclaim{Proposition~2.6.3} $({\cal T}_S,\nabla_N)$ is an initial 
\tsym{\Sigma}-algebra. \endpro

\proof This follows immediately from Propositions 2.3.4 and 2.6.2. \eop
\bigbreak

Because of the explicit description of ${\cal T}_S$ given in 
Proposition~2.6.3 below
$({\cal T}_S,\nabla_N)$ is called the 
\indexddef{tree \tsym{\Sigma}-algebra}{tree algebra}{}{algebra}{tree}.
An element $\lambda \in {\cal L}$ is said to be a 
\indexddef{labelled tree}{labelled tree}{}{tree}{labelled}
if whenever $s \in \dom(\lambda)$ with $\lambda(s) \in N$
of type $L \to \sigma$ then  
$s \triangleright \eta \in \dom(\lambda)$ if and only if 
$\eta \in L$ and, moreover, 
$\lambda(\eta \triangleright s) \in N_\eta $ for each $\eta \in L$.
A labelled tree $\lambda$ is then said to be 
\indexdef{of type $\sigma$}{type}{of tree} 
if $\lambda(\onept) \in N_\sigma$, and the set of all labelled trees of type 
$\sigma$ will be denoted by ${\grave {\cal T}}_\sigma$. 

\proclaim{Proposition~2.6.3} ${\grave {\cal T}}_S = {\cal T}_S$. \endpro

\proof The crucial step in the proof is the following fact:

\proclaim{Lemma~2.6.4} For $\nu \in N$ of type $L \to \sigma$ 
let ${\grave \amalg}_\nu$ denote the restriction of
$\amalg_\nu$ to $\ass L {\grave {\cal T}} $ (as a mapping from 
$\ass L {\grave {\cal T}} $ to ${\cal T}_\sigma$). Then
$\,\bigcup\limits_{\nu \in N_\sigma} \Im({\grave \amalg}_\nu) 
                                           \,=\, {\grave {\cal T}}_\sigma\,$
for each $\sigma \in S$. 
\endpro

\proof It will be shown first that if $\nu \in N_\sigma$ is of type $L \to \sigma$
and $c \in \ass L {\grave {\cal T}} $
then $\amalg_\nu(c) \in {\grave {\cal T}}_\sigma$, and for this 
it is enough to show that $\lambda = \amalg_\nu(c)$ is a labelled tree,
since $\lambda(\onept) = \nu \in N_\sigma$. Hence
consider an element 
$s \,\in\, \dom(\lambda) 
 \,=\, \{\onept\} \cup \bigcup\limits_{\eta \in L} 
            \eta \,\triangleleft\, \dom(c(\eta))$. 
\smallskip
The first possibility is that $s = \onept$; in this case
$\lambda(s) = \nu$ is of type $L \to \sigma$ and 
$s \triangleright \eta  = \eta \triangleleft \onept \in \dom(\lambda)$ 
if and only if $\eta \in L$, and if $\eta \in L$ then
(since $c(\eta) \in {\grave {\cal T}}_\eta$)

\mdisp{\lambda(s \triangleright \eta)
  \ =\ \lambda(\eta \triangleleft \onept)\ =\ c(\eta)\,(\onept)\ \in\ N_\eta \ .} 
\smallskip
The other possibility is that $s = \eta \triangleleft s'$ with
$\eta \in L$ and $s' \in \dom(c(\eta))$, and suppose here 
$\lambda(s) = c(\eta)\,(s')$ is of type $L' \to \sigma$.
Then by Lemma~2.6.1~(2)
\mdisp{s \triangleright \xi
  \ =\ (\eta \triangleleft s') \triangleright \xi
  \ =\ \eta \triangleleft (s' \triangleright \xi)}

and $s' \triangleright \xi \in \dom(c(\eta))$ if and only if 
$\xi \in L'$, since $c(\eta)$ is a labelled tree,
which implies that $s \triangleright \xi \in \dom(\lambda)$  if and only if 
$\xi \in L'$. Moreover, if $\xi \in L'$ then 

\mdisp{\lambda(s \triangleright \xi)
  \ =\ \lambda(\eta \triangleleft (s' \triangleright \xi))
 \ =\ c(\eta)\,(s' \triangleright \xi)}

and $c(\eta)\,(s' \triangleright \xi)$ is an element of $N_\xi$, again 
because $c(\eta)$ is a labelled tree. This shows that
$\lambda$ is a labelled tree. 
\medskip
Next consider $\sigma \in S$ and $\lambda \in {\grave {\cal T}}_\sigma$ with
$\lambda(\onept) = \nu \in N$ of type $L \to \sigma$; it will be shown
that there exists $c \in \ass L {\grave {\cal T}} $ such that 
$\lambda = \amalg_\nu(c)$. By 
Proposition~2.6.1~(3) and Lemma~2.6.2 $\dom(\lambda)$ has a unique representation of 
the form
\mdisp{
\dom(\lambda) \ =\ \{\onept\} \,\cup\, 
\bigcup\limits_{\eta \in B} \eta\,\triangleleft\, A_\eta}
with $B$ a finite subset of $J$ and 
$A_\eta \in {\cal S}$ for each $\eta \in B$.
But (since $\lambda$ is a labelled tree)
$\onept \triangleright \eta  
         = \eta \triangleleft \onept \in \dom(\lambda)$ 
if and only if $\eta \in L$,
and by Lemma~2.6.3 $\onept \in A_\eta$ for each $\eta \in B$;
it therefore follows that $B = L$. 
For each $\eta \in L$ define $\lambda_\eta : A_\eta \to N$ by
$\lambda_\eta(s) = \lambda(\eta \triangleleft s)$, so
$\lambda_\eta \in {\cal L}$ and $\dom(\lambda_\eta) = A_\eta$.
In fact $\lambda_\eta$ is a labelled tree: If 
$s \in \dom(\lambda_\eta)$ then $\eta \triangleleft s \in \dom(\lambda)$ and
$\lambda_\eta(s) = \lambda(\eta \triangleleft s)$; moreover, 
$s \triangleright \xi \in \dom(\lambda_\eta)$ if and only if 
$\eta \triangleleft (s \triangleright \xi) 
= (\eta \triangleleft s ) \triangleright \xi \in \dom(\lambda)$, 
and $\lambda_\eta(s \triangleright \xi) 
= \lambda((\eta \triangleleft s) \triangleright \xi)$.
Hence $\lambda_\eta \in {\grave {\cal T}}_\eta$, since
$\lambda_\eta(\onept) = \lambda(\eta \triangleleft \onept) 
 = \lambda(\onept \triangleright \eta) \in N_\eta$. 
An assignment
$c \in \ass L {\grave {\cal T}} $ can thus be defined 
by letting $c(\eta) = \lambda_\eta$ for 
each $\eta \in L$, and by construction $\lambda = \amalg_\nu(c)$. 
\medskip
Putting these two parts of the proof together now shows that
$\,\bigcup_{\nu \in N_\sigma} \Im({\grave \amalg}_\nu) \,=\, {\grave T}_\sigma\,$
for each $\sigma \in S$. \eop
\bigbreak
Lemma~2.6.4 implies in particular that the family ${\grave {\cal T}}_S$ is 
invariant in $({\cal T}_S,\nabla_N)$, so consider the subalgebra 
$({\grave {\cal T}}_S,{\grave \nabla}_N)$ associated with 
${\grave {\cal T}}_S$. 
Now clearly ${\grave {\cal T}}_S = {\cal {\cal T}}_S$ will hold
if and only if $({\grave {\cal T}}_S,{\grave \nabla}_N)$ is minimal, 
and to establish this Proposition~2.2.5 can be used.
The first requirement in Proposition~2.2.5, namely that 
$\,\bigcup_{\nu \in N_\sigma} \Im({\grave \nabla}_\nu) 
                                             \,=\, {\grave {\cal T}}_\sigma\,$
for each $\sigma \in S$, is just a restatement of Lemma~2.6.4,
and so it remains to satisfy the second requirement, which means 
a suitable family of mappings $\#_S$ must be found.
Let $\# : {\cal L} \to \Nat$ be given by
\mdisp{\#(\lambda) \ =\ \max \lcurl |s| \,:\, s \in \dom(\lambda) \rcurl\ ,}
where $|\cdot| : J^* \to \Nat$ is the mapping given in Lemma~2.6.1~(6),
and for each $\sigma \in S$ put $\#_\sigma = \#$.
Then the requirement in Proposition~2.2.5 is satisfied, since
if $\nu \in N$ is of type $L \to \sigma$ then
for all $c \in \ass L {{\cal L}} $  
\mdisp{\quad  \#_\sigma( \amalg_\nu(c)) 
    \ =\ 1 \,+\, \max \lcurl \#_\eta(c(\eta)) \,:\, 
     \eta \in L \rcurl
    \ >\ \max \lcurl \#_\eta(c(\eta)) \,:\, \eta \in L \rcurl\ .}

Therefore $({\grave {\cal T}}_S,{\grave \nabla}_N)$ is minimal. \eop

\bigbreak

\vfill\eject

\hfline{54}{2.7 \ NOTES}

\sectionhead {2.7} {Notes}
\bigskip\medskip
The material presented in this chapter is all part of standard 
universal algebra. The classical field of universal algebra 
deals with the case of signatures having a single type, and so
the only information required about each operator name is the
number of arguments it takes. In this form the main problems were
stated in Whitehead (1898) and solved in Birkhoff (1935);
standard texts are Cohn (1965) and Gr\"atzer (1968).
(The term `universal algebra' was coined by that great 
mathematical name-giver Sylvester.) 
The generalisation of the theory to 
\indexddef{multi-sorted algebras}
{multi-sorted algebra}{}{algebra}{multi-sorted}
(i.e., to the algebras as they occur here) was made by Higgins (1963)
and Birkhoff and Lipson (1970). 
Birkhoff and Lipson (who speak of 
\indexddef{heterogeneous algebras}
{heterogenous algebra}{}{algebra}{heterogenous})
showed that essentially the whole of the classical theory
carries over to the more general case.  
\medskip
Some of the first uses of multi-sorted algebras in computer science can be
found in Maibaum (1972) and Morris (1973).
Their systematic use has been propagated by the 
ADJ group consisting of J.A.\ Goguen, J.W.\ Thatcher, E.G.\ Wagner 
and J.B.\ Wright, for example in the papers Goguen, Thatcher, Wagner and 
Wright (1977) and Goguen, Thatcher and Wagner (1978). 
\medskip
Minimal algebras are sometimes called 
\indexddef{reachable}{reachable algebra}{}{algebra}{reachable} 
(Wirsing (1990)) or 
\indexddef{term generated}{term generated algebra}{}{algebra}{term generated}
(Bauer, Berghammer {\it et al\/} (1985)). This terminology comes from the fact
that an algebra is minimal if and only if the unique homomorphism to this 
algebra from an initial algebra (and so in particular from the standard 
term algebra) is surjective.
Pervasive signatures are called 
\indexddef{sensible}{sensible signature}{}{signature}{sensible}
in Huet and Oppen (1980).
\medskip
The characterisation of initial algebras in Proposition~2.3.2
is sometimes referred to as stating that initial algebras are exactly
those for which there is {\it no junk\/} and  
{\it no confusion\/}  (i.e., those which are minimal and unambiguous).
Proposition~2.4.2 is just a version of the classical result of
{\L}ukasiewicz ({\L}ukasiewicz (1925)) concerning prefix or (left Polish) 
notation.
\medskip
Bourbaki (1970) thinks that the word {\it algebra\/} is over-used
and suggests the alternative 
\indexdef{magma}{magma}{}
for an algebra in the above sense.
This suggestion has been followed by some
French computer scientists, but it has of course no chance of being 
accepted in the English-speaking world. (What have algebras to do with
volcanic lava?)

\vfill\eject

\hfline{55}{3.1 \ THE BASIC DATA OBJECTS AND THE GROUND TERM ALGEBRA}

\bigskip
{\fourteenbf Chapter 3\quad Data objects}
\bigskip\medskip
With the preparations made in Chapter~2 we can now make a start to building  a
framework for specifying data objects. This involves what will be called a bottomed
extension of an initial algebra. The initial algebra is used to describe the 
basic data objects, and the bottomed extension then specifies which
`undefined' and `partially defined' data objects are to be allowed.
Bottom extensions are defined in Section~3.2. In Sections 3.3 and 3.5 
two conditions, called regularity and monotonicity,
are introduced which the extensions will be required to satisfy.
Meanwhile, in Section~3.4 cored extensions are considered.
These provide simple explicit examples of monotone regular extensions.

\bigskip\bigskip\bigskip

\sectionhead {3.1} {The basic data objects and the ground term algebra}
\bigskip\medskip
The starting point is to choose a signature $\Lambda = (B,K,\Theta,\vartheta)$ and an 
initial \tsym{\Lambda}-algebra $(X_B,p_K)$. The basic data objects are then the elements in 
the family of sets $X_B$. Here $\Lambda$ is called the 
\indexddef{ground signature}{ground signature}{}{signature}{ground}, 
$B$ the set of 
\indexddef{ground types}{ground type}{}{type}{ground} 
and $K$ the set of 
\indexdef{constructor names}{constructor name}{}.
In real applications the set $B$ of ground types will be finite.
\medskip
In all modern functional programming languages the signature $\Lambda$ is
not fixed and can be chosen by the user to fit the application at hand.
Usually there are a few built-in types such as the types
$\,\ftt{int}\,$ and $\,\ftt{bool}\,$ occurring in Example~2.2.1 together with a 
primitive type $\,\ftt{char}\,$ having some multiple of $256$ constructor names. 
(Recall that a type $\beta$ is said to be
primitive if $\kappa$ is of type $\varnothing \to \beta$ for each 
$\kappa \in K_\beta$.)
Once the signature $\Lambda$ has been chosen the initial
\tsym{\Lambda}-algebra $(X_B,p_K)$ is uniquely determined up to isomorphism.
The explicit choice of $(X_B,p_K)$ can be seen as taking place in the 
user's mind.
\medskip
Bearing in mind practical applications we will assume that 
$B$ contains the two primitive types $\,\ftt{int}\,$ and $\,\ftt{bool}\,$ 
as in Example~2.2.1. Thus $\,K_{\fstt{int}} \,=\, \SynInt\,$ (with $\SynInt$ the subset of
$\,\{\ftt{0},\ftt{1},\ftt{2},\ftt{3},\ftt{4},\ftt{5},\ftt{6},
                                  \ftt{7},\ftt{8},\ftt{9},\ftt{-}\}^*$
containing for each integer $n$ its standard representation
${\underline n}$ as a string of characters),
$\,K_{\fstt{bool}} \,=\, \{\ftt{True},\ftt{False}\}\,$ and each of these constructor names
is of type $\varnothing \to \beta$ with $\beta$ either $\ftt{int}$ or $\ftt{bool}$.
Moreover, $X_{\fstt{int}}$ will always taken to be $\Int$ with
$\,p_{\underline n }(\onept) = n\,$ for each $n \in \Int$ and
$X_{\fstt{bool}}$ to be  $\Bool$ with $\,p_{\fstt{True}}(\onept) = T\,$
and $\,p_{\fstt{False}}(\onept) = F\,$.
\medskip
To avoid some trivial complications later it is also convenient
to suppose that $\Lambda$ is pervasive. (Recall this means 
$B$ is the only closed subset of itself, where
a subset $B'$ of $B$ is closed if $\beta \in B'$ whenever there exists 
$\kappa \in K$ of type $L \to \beta$ for some $L$ and $\langle \eta \rangle \in B'$
for each $\eta \in L$.)
By Proposition~2.2.4 this is equivalent to requiring that
$X_\beta \ne \varnothing$ for each $\beta \in B$.
We do  not assume that the sets $K_\beta$, $\beta \in B$, are finite
(since in fact $K_{\fstt{int}}$ is infinite), although
it will almost always be the case that $K_\beta$ is finite whenever $\beta$ 
is not a primitive type. 
\medbreak
In all that follows the signature $\Lambda$ and the initial
\tsym{\Lambda}-algebra $(X_B,p_K)$ are considered to be fixed.
\medskip
As well as the initial \tsym{\Lambda}-algebra $(X_B,p_K)$ we also assume that
a second initial \tsym{\Lambda}-algebra is given. This \tsym{\Lambda}-algebra $(F_B,\fo_K)$
is called the 
\indexddef{ground term algebra}{ground term algebra}{}{algebra}{ground term} 
and is needed in order to be able to 
refer to the basic data objects:
Since $(X_B,p_K)$ and $(F_B,\fo_K)$ are both initial \tsym{\Lambda}-algebras
there exists a  unique isomorphism 
from $(F_B,\fo_K)$ to $(X_B,p_K)$ which will be denoted by 
$\sem{\cdot}_B$.
This means that each basic data object can be denoted by 
a unique element in the ground term algebra, i.e., the data object 
$x \in X_\beta$ can be denoted by the unique element $s \in F_\beta$ 
with $\sem{s}_\beta = x$.
\medskip

Of course, $(F_B,\fo_K)$ could just be taken to be  $(X_B,p_K)$ but, as its name
suggests, it should be thought of rather as some kind 
of term algebra (as defined in Section~2.4).
In fact, we will now introduce such a term algebra which will later be used as the 
basis for a programming language. This is done by 
choosing a family of enumerations $i_K$ for the signature $\Lambda$, i.e.,
by choosing for each $\kappa \in K$ of type $L \to \beta$ a bijective mapping $i_\kappa$ 
from $[m]$ to the set $L$, where $m = |L|$ is the cardinality of $L$.
In the special case when $L \,=\, \list {\beta} m \in B^*$
(and so $[m]$ is the underlying set of the \tsym{B}-typed set $L$)
then $i_\kappa : [m] \to [m]$ will always chosen to be the identity mapping.
Let $(E_B,\blob_K)$ be the standard term \tsym{\Lambda}-algebra defined by the 
family $i_K$. Then by Proposition~2.4.5 
$(E_B,\blob_K)$ is an initial \tsym{\Lambda}-algebra.
Recall that $E_\beta \subset K^*$ for each $\kappa \in B$ and if 
$\kappa \in K$ is of type $L \to \beta$ then 
$\blob_\kappa : \ass L E \to E_\beta$ is the mapping given by

\mdisp{ \blob_\kappa(c)
\ =\ \kappa\ \ c(i_\kappa(1))\,\ \cdots\ \,c(i_\kappa(m))}

for each $c \in \ass L E $, where $m = |L|$. 
Moreover, the family $E_B$ can be regarded as being defined 
by the following rules:
\medskip\smallskip
{\parindent=25pt
\item{(1)} If $\kappa \in K$ is of type $\varnothing \to \beta$ then the list
consisting of the single component $\kappa$ is an 
element of $E_\beta$.
\medskip
\item{(2)} If $\kappa \in K$ is of type $L \to \beta$
with $L \ne \varnothing$
and $e_j \in E_{\beta_j}$ for $\oneto j m $, where 
$\beta_j = \langle i_\kappa(j) \rangle$ and $m = |L|$,
then $\ \kappa\ e_1\ \cdots \ e_m\ $ is an element of $E_\beta$.
\medskip
\item{(3)} The only elements in $E_\beta$ are those which can be obtained
using (1) and (2). 
\bigskip
} 
In this special case the isomorphism $\sem{\cdot}_B$
from $(E_B,\blob_K)$ to $(X_B,p_K)$ is the unique family of mappings satisfying:

\medskip\smallskip
{\parindent=25pt
\item{(1)} If $\kappa \in K$ is of type $\varnothing \to \beta$ then 
$\,\sem{\kappa}_\beta \,=\, p_\kappa(\onept)\,$.
\medskip
\item{(2)} If $\kappa \in K$ is of type $L \to \beta$
with $L \ne \varnothing$
and $e_j \in E_{\beta_j}$ for $\oneto j m $, where 
$\sigma_j = \langle i_\kappa(j) \rangle$ and $m = |L|$, then 
\mdisp{\sem{\,\kappa\ \list e m \,}_\sigma
          \  =\  p_\kappa(\ass L {\sem{c}} )\ ,}
\item{} where $c \in \ass L E $ is the assignment with
$c(i_\kappa(j)) = e_j$ for each $\oneto j m $.
\bigbreak
}  

\vfill\eject

\hfline{57}{3.2 \ BOTTOMED ALGEBRAS AND BOTTOMED EXTENSIONS}

\bigskip\medskip
\sectionhead {3.2} {Bottomed algebras and bottomed extensions}
\bigskip\medskip
The basic data objects were  described in the previous section with the help of the initial
\tsym{\Lambda}-algebra $(X_B,p_K)$. The next step
is to introduce for each $\beta \in B$ a new element $\bot_\beta \notin X_\beta$ 
to play the role
of an `undefined' or `unknown' object of type $\beta$.
(The symbol $\bot$ is referred to as 
\indexdef{bottom}{bottom}{}.)
Thus for each $\beta \in B$ there is now an `extended' set 
$X_\beta \cup \{\bot_\beta\}$ of data objects of type $\beta$ made up
of the `real' objects (i.e., the elements of the set $X_\beta$) together
with the `undefined' object $\bot_\beta$ of type $\beta$. 
\medskip
There are, however, good reasons to go at least one stage further and to also allow 
`partially defined' data objects. For instance, considering the 
\tsym{\Lambda}-algebra introduced in Example~2.2.1, 
if $n \in \Int = X_{\fstt{int}}$ then the pair
$(n,\bot_{\fstt{int}})$ could be regarded as a data object of type $\ftt{pair}$ which is
not completely `undefined' and so which should perhaps be
different from $\bot_{\fstt{pair}}$. (After all, it makes sense to consider 
the first component of $(n,\bot_{\fstt{int}})$.) Similarly, it is often 
useful to allow `partially defined' data objects of type $\ftt{list}$. (For 
example, the data object described as a list containing at least one 
component, with the first component having some specific value.)
\medskip
In order to facilitate the systematic treatment of `partially defined' data 
objects we make the following definition: A \tsym{\Lambda}-algebra $(D_B,f_K)$
is called a 
\indexdef{bottomed extension}{bottomed extension}{}
of $(X_B,p_K)$ if $(D_B,f_K)$ is 
an extension of $(X_B,p_K)$ (i.e., $X_B \subset D_B$ and $p_\kappa$ is the 
restriction of $f_\kappa$ to $\dom(p_\kappa)$ 
for each $\kappa \in K$) such that $\bot_\beta \in D_\beta$ for each 
$\beta \in B$. 
\medskip
The elements in the set $D_\beta \setminus (X_\beta \cup \{\bot_\beta\})$ should be 
thought of for the moment as being the `partially defined' data objects of type $\beta$.
Later we will also consider the possibility of having not only 
`partially defined' but also `infinite' data objects. This more general 
situation can still be described with the help of a bottomed extension. 
The elements in the set $D_\beta \setminus (X_\beta \cup \{\bot_\beta\})$ 
should then, however, be thought of as being made up of both `partially defined' 
and `infinite' data objects.
\medskip
The simplest example of a bottomed extension is obtained by putting
$D_\beta = X_\beta \cup \{\bot_\beta\}$ for each $\beta \in B$ and
for each $\kappa \in K$ of type 
$L \to \beta$ letting $f_\kappa : \ass L D \to D_\beta$ be given by

\mdisp{f_\kappa(c)\ =\ \cases{ p_\kappa(c) & if $\,c \in \ass L X \,$, \cr
                             \bot_\beta & otherwise. \cr}}

Then $(D_B,f_K)$ is a bottomed extension called the 
\indexddef{flat extension}{flat extension}{}{bottomed extension}{flat} 
of $(X_B,p_K)$. 
(The flat extension of the \tsym{\Lambda}-algebra 
defined in Example~2.2.1 is given in Example~3.2.1 on the 
next page.)

\midinsert
\bigskip\medskip
\frame{20pt}{\bigskip 
{\it Example~3.2.1\enspace} The flat extension of the \tsym{\Lambda}-algebra 
$(X_B,p_K)$ first introduced in Example~2.2.1 is the \tsym{\Lambda}-algebra 
$(X^\bot_B,p^\bot_K)$ defined as follows:
\medskip\smallskip
{\leftskip=32pt
$X^\bot_{\fstt{bool}} = \Bool^\bot = \{T,F,\bot_{\fstt{bool}}\}$, 
\quad
$X^\bot_{\fstt{nat}} = \Nat \cup \{\bot_{\fstt{nat}}\}$, 
\smallskip
$X^\bot_{\fstt{int}} = \Int \cup \{\bot_{\fstt{int}}\}$, 
\quad
$X^\bot_{\fstt{pair}} = \Int^2 \cup \{\bot_{\fstt{pair}}\}$,
\quad    
$X^\bot_{\fstt{list}} = \Int^* \cup \{\bot_{\fstt{list}}\}$.
\medskip
$p^\bot_{\fstt{True}} : 
  \Oneptset \to X^\bot_{\fstt{bool}}$\ with\ %
                      $p^\bot_{\fstt{True}}(\onept) = T$, 
\smallskip
$p^\bot_{\fstt{False}} : 
  \Oneptset \to X^\bot_{\fstt{bool}}$\ with\ %
                      $p^\bot_{\fstt{False}}(\onept) = F$, 
\smallskip
$p^\bot_{\fstt{Zero}} : 
    \Oneptset \to X^\bot_{\fstt{nat}}$\ with\ %
                       $p^\bot_{\fstt{Zero}}(\onept) = 0$,
\smallskip
$p^\bot_{\fstt{Succ}} : 
 X^\bot_{\fstt{nat}} \to X^\bot_{\fstt{nat}}$\ %
 with\ %
\mdisp{p^\bot_{\fstt{Succ}}(n)\ =\ \cases{ 
                      n + 1 & if $\,n \in \Nat\,$, \cr
          \bot_{\fstt{nat}} & if $\,n = \bot_{\fstt{nat}}\,$, \cr}} 
\smallskip
$p^\bot_{{\underline n}} : 
 \Oneptset \to X^\bot_{\fstt{int}}$\ with\ %
  $p^\bot_{{\underline n}}(\onept) = n\ $ for each $n \in \Int$,
\smallskip
$p^\bot_{\fstt{Pair}} : 
 X^\bot_{\fstt{int}} \times X^\bot_{\fstt{int}}
 \to X^\bot_{\fstt{pair}}$\ with\ %
\mdisp{p^\bot_{\fstt{Pair}}(m,n)\ =\ \cases{
            	      (m,n) & if $\,m,\, n \in \Int\,$, \cr
         \bot_{\fstt{pair}} & otherwise, \cr}} 
\smallskip
$p^\bot_{\fstt{Nil}} : 
 \Oneptset \to X^\bot_{\fstt{list}}$\ with\ %
                         $p^\bot_{\fstt{Nil}}(\onept) = \onept$, 
\smallskip
$p^\bot_{\fstt{Cons}} : 
X^\bot_{\fstt{int}} \times X^\bot_{\fstt{list}}
 \to X^\bot_{\fstt{list}}$\ with %
\mdisp{p^\bot_{\fstt{Cons}}(n,s)\ =\ \cases{ 
     n \triangleleft s & if $\,n \in \Int\,$ and $\,\ell \in \Int^*\,$, \cr
           \bot_{\fstt{list}} & otherwise. \cr}}
\bigskip
}}
\bigskip
\endinsert

\medskip
The above definition of a bottomed extension will be reformulated somewhat, and for this
it is useful to introduce the following simple notion: A 
\indexddef{bottomed set}{bottomed set}{}{set}{bottomed}
is a pair $(D,\bot)$ consisting of a set $D$ and a distinguished
\indexdef{bottom element}{bottom element}{}
 $\bot \in D$. The bottom element $\bot$ should be 
thought of as representing an `undefined' or `unknown' value which, if it 
were to be `defined' or `become known', would take on the value of one of the
remaining elements of $D$.
\medskip
If $(D,\bot)$ and $(D',\bot')$ are bottomed sets then a mapping
$f : D \to D'$ is said to be 
\indexddef{bottomed}{strict mapping}{}{mapping}{strict}
(or \definition{strict}) if $f(\bot) = \bot'$.
\medskip
If $(D,\bot)$ is a bottomed set then it is usual to write just $D$ instead
of $(D,\bot)$ and to assume that the bottom element $\bot$ can be inferred from
the context.
\medskip
Let $D_B$ be a family of bottomed sets with $\bot_\beta$ the bottom
element of $D_\beta$ for each $\beta \in B$, and let $I$ be a 
\tsym{B}-typed set.
Then $\ass I D $ will be considered as a bottomed set by stipulating that
its bottom element $\bot$ be the assignment given by $\bot(\eta) = \bot_\eta$
for each $\eta \in I$. A special case is that of a cartesian product:
Let $n \ge 2$ and for each $\oneto j n $ let $D_j$ be a bottomed set with
bottom element $\bot_j$; then the product
$D_1 \times \cdots \times D_n$ is considered 
as a bottomed set with bottom element $(\vector {\bot} n )$. 
\medskip
A \tsym{\Lambda}-algebra $(D_B,f_K)$ is said to be
\indexddef{bottomed}{bottomed algebra}{}{algebra}{bottomed}
if $D_\beta$ is a bottomed set for each $\beta \in B$.
If $(D_B,f_K)$ is a bottomed \tsym{\Lambda}-algebra then the bottom element of 
$D_\beta$ will always be denoted by $\bot_\beta$. This means that if
$(D'_B,f'_K)$ is a further bottomed \tsym{\Lambda}-algebra then the bottom 
element of $D'_\beta$ will also be denoted by $\bot_\beta$, although 
it is not to be assumed that the bottom elements of $D_\beta$ and $D'_\beta$
are the same.
\medskip
The definition of a bottomed extension of $(X_B,p_K)$ can now be expressed as 
follows: A 
\indexdef{bottomed extension}{bottomed extension}{}
of $(X_B,p_K)$ is a bottomed 
\tsym{\Lambda}-algebra $(D_B,f_K)$ which is an extension of $(X_B,p_K)$ such that 
$\bot_\beta \notin X_\beta$ for each $\beta \in B$. 
\medskip

If $(D_B,f_K)$ and $(D'_B,f'_K)$ are bottomed 
\tsym{\Lambda}-algebras then a homomorphism 
$\pi_B$ from $(D_B,f_K)$ to $(D'_B,f'_K)$ is said to be 
\indexddef{bottomed}{bottomed homomorphism}{}{homomorphism}{bottomed}
if $\pi_\beta$ is a bottomed mapping 
(i.e., if $\pi_\beta(\bot_\beta) = \bot_\beta$) for each $\beta \in B$. 
\medskip

If $(D_B,f_K)$ and $(D'_B,f'_K)$ are bottomed extensions of $(X_B,p_K)$ then a bottomed 
homomorphism $\pi_B : (D_B,f_K) \to (D'_B,f'_K)$ is said to be an 
\indexddef{extension homomorphism}{extension homomorphism}{}{homomorphism}{extension}
if $\pi_\beta(x) = x$ for all $x \in X_\beta$, 
$\beta \in B$. In fact, the following simple result implies
that a bottomed homomorphism is automatically an extension homomorphism:

\proclaim{Lemma~3.2.1} Let $(Z_B,r_K)$ and $(Z'_B,r'_K)$ be 
extensions of a minimal \tsym{\Lambda}-algebra $(Y_B,q_K)$ and 
$\pi_B : (Z_B,r_K) \to (Z'_B,r'_K)$ be a
homomorphism. Then $\pi_\beta(y) = y$ for all $y \in Y_\beta$, $\beta \in B$.
\endpro

\proof The restriction of $\pi_B$ to $Y_B$ and the family of identity
mappings $\fss{id}_B$ both define homomorphisms from
$(Y_B,q_K)$ to  $(Z'_B,r'_K)$, and so by Proposition~2.2.3~(1)
they must be equal, i.e.,
$\,\pi_\beta(y) = y\,$ for all $y \in Y_\beta$, $\beta \in B$. \eop
\bigbreak
If $(D_B,f_K)$ is a bottomed \tsym{\Lambda}-algebra then the 
family of identity mappings $\fss{id}_B$ defines a
bottomed homomorphism from $(D_B,f_K)$ to itself. 
Also by Proposition~2.2.1 the composition of two bottomed 
homomorphisms is again a bottomed homomorphism.
Moreover, if $\pi_B : (D_B,f_K) \to (D'_B,f'_K)$ is a bottomed homomorphism such that
$\pi_\beta$ is a bijection for each $\beta \in B$
then by Proposition~2.2.2
$\pi^{-1}_B : (D'_B,f'_K) \to (D_B,f_K)$ is again a bottomed homomorphism. 
\medskip
A bottomed homomorphism $\pi_B : (D_B,f_K) \to (D'_B,f'_K)$ is a 
\indexddef{bottomed isomorphism}{bottomed isomorphism}{}{isomorphism}{bottomed}
if the mapping 
$\pi_\beta : D_\beta \to D'_\beta$ is a bijection for each $\beta \in B$;
$(D_B,f_K)$ and $(D'_B,f'_K)$ are then 
\indexddef{isomorphic}{isomorphic bottomed algebras}{}{bottomed algebras}{isomorphic}  
if there 
exists a bottomed isomorphism from $(D_B,f_K)$ to $(D'_B,f'_K)$, and
the property of being isomorphic defines an equivalence relation on the 
class of all bottomed \tsym{\Lambda}-algebras.
\medskip
Bottomed extensions $(D_B,f_K)$ and $(D'_B,f'_K)$ of
$(X_B,p_K)$ are said to be 
\indexddef{conjugate}{conjugate bottomed extensions}{}{bottomed extensions}{conjugate}
if there
exists an extension homomorphism from  $(D_B,f_K)$ to $(D'_B,f'_K)$ which is a 
bottomed isomorphism. Of course, Lemma~3.2.1 implies that
$(D_B,f_K)$ and $(D'_B,f'_K)$ are conjugate if and only if they are
isomorphic as bottomed \tsym{\Lambda}-algebras.
\medskip
We now collect together some elementary results about bottomed \tsym{\Lambda}-algebras 
and then apply them to the case of a bottomed extension of $(X_B,p_K)$.
In what follows let $(D_B,f_K)$ be a bottomed \tsym{\Lambda}-algebra.
A family ${\grave D}_B$ with ${\grave D}_B \subset D_B$ 
is said to be 
\indexddef{bottomed}{bottomed family}{}{family}{bottomed} 
if $\bot_\beta \in {\grave D}_\beta$ for each $\beta \in B$. 
There is clearly a one-to-one correspondence between 
bottomed invariant families and bottomed subalgebras of $(D_B,f_K)$ (i.e., those
subalgebras which are also bottomed \tsym{\Lambda}-algebras with the bottom elements
inherited from $(D_B,f_K)$).

\proclaim{Lemma~3.2.2} The bottomed family ${\breve D}_B$ defined by 
$\,{\breve D}_\beta 
\,=\, \{\bot_\beta\} \cup \bigcup_{\kappa \in K_\beta} \Im(f_\kappa)\,$
for each $\beta \in B$ is invariant. \endpro

\proof This is a special case of Lemma~2.3.2. \eop

\proclaim{Lemma~3.2.3} There exists a minimal bottomed invariant family. 
\endpro

\proof This is a special case of Lemma~2.2.3. \eop
\bigbreak
The subalgebra associated with the minimal bottomed invariant family will be referred 
to as the 
\indexddef{minimal bottomed subalgebra of $(D_B,f_K)$}
{minimal bottomed subalgebra}{}{bottomed subalgebra}{minimal}.
\medskip
If $D_B$ is the only bottomed invariant family then
the bottomed \tsym{\Lambda}-algebra $(D_B,f_K)$  is said to be 
\indexddef{minimal}{minimal bottomed algebra}{}{bottomed algebra}{minimal}.
Note that if 
$(D_B,f_K)$ is a minimal \tsym{\Lambda}-algebra (as defined in Section~2.2)
then it is also a minimal bottomed \tsym{\Lambda}-algebra, but of course in general
the converse does not hold. In fact the
minimal bottomed \tsym{\Lambda}-algebras occurring here will never be minimal 
\tsym{\Lambda}-algebras. The statement that a bottomed 
\tsym{\Lambda}-algebra is minimal will thus mean that it is a minimal bottomed 
\tsym{\Lambda}-algebra; minimality as defined in Section~2.2 will be referred to
by stating explicitly that the algebra is a minimal \tsym{\Lambda}-algebra.
\medskip
It is easy to see that the flat extension of $(X_B,p_K)$  is a minimal bottomed
\tsym{\Lambda}-algebra (since $(X_B,p_K)$  is a minimal \tsym{\Lambda}-algebra). 

\proclaim{Proposition~3.2.1} Let $(D'_B,f'_K)$ be a bottomed \tsym{\Lambda}-algebra.
\medskip
(1)\enskip There exists at most one bottomed
homomorphism from $(D_B,f_K)$ to $(D'_B,f'_K)$ when $(D_B,f_K)$ is minimal. 
\medskip
(2)\enskip Any bottomed homomorphism from $(D_B,f_K)$ to $(D'_B,f'_K)$ is surjective 
when $(D'_B,f'_K)$ is minimal. \endpro

\proof This is the same as the proof of Proposition~2.2.4. \eop

\proclaim{Proposition~3.2.2} If $(D_B,f_K)$ is minimal then 
$\,D_\beta\,=\,\{\bot_\beta\} \cup \bigcup_{\kappa \in K_\beta} \Im(f_\kappa)\,$
for each $\beta \in B$. \endpro
\proof This follows immediately from Lemma~3.2.2. \eop
\bigbreak

\proclaim{Proposition~3.2.3} Suppose that
$\,D_\beta\,=\,\{\bot_\beta\} \cup \bigcup_{\kappa \in K_\beta} \Im(f_\kappa)\,$
for each $\beta \in B$ and there exists a family of mappings
$\#_B$ with $\#_\beta : D_\beta \to \Nat$ for each $\beta$
such that if $\kappa \in K$ is of type $L \to \beta$ 
then for all $c \in \ass L D $, $\eta \in L$
\mdisp{\#_\eta(c(\eta))\ < \ \#_\beta(f_\kappa(c))\ .}
Then $(D_B,f_K)$ is a minimal bottomed \tsym{\Lambda}-algebra. \endpro

\proof This is just a special case of Proposition~2.2.5. \eop
\bigbreak
As expected, $(D_B,f_K)$ is said to be an 
\indexddef{initial bottomed \tsym{\Lambda}-algebra}
{initial bottomed algebra}{}{bottomed algebra}{initial}
if for each bottomed \tsym{\Lambda}-algebra $(D'_B,f'_K)$ there exists a unique bottomed 
homomorphism from $(D_B,f_K)$ to $(D'_B,f'_K)$.
The adjective `initial' will be used in the same way as `minimal':
The statement that a bottomed 
\tsym{\Lambda}-algebra is initial thus means that it is an initial bottomed 
\tsym{\Lambda}-algebra, and initiality as defined in Section~2.3 will be referred
to by stating explicitly that the algebra is an initial \tsym{\Lambda}-algebra.

\proclaim{Lemma~3.2.4} Suppose the bottom elements $\bot_\beta$, $\beta \in B$, are 
all distinct, and let $J$ be the \tsym{B}-typed set consisting of these elements
with $\bot_\beta$ of type $\beta$ for each $\beta \in B$. Then $(D_B,f_K)$ is an 
initial bottomed \tsym{\Lambda}-algebra if and only if it is \tsym{J}-free. \endpro

\proof This is more-or-less clear. (Note that if $(Y_B,q_K)$ is any  
\tsym{\Lambda}-algebra and $c \in \ass J Y $ then 
$(Y_B,q_K)$ can be considered as a bottomed \tsym{\Lambda}-algebra by declaring 
$c(\bot_\beta)$
to be the bottom element of $Y_\beta$ for each $\beta \in B$.) \eop
\bigbreak

\proclaim{Proposition~3.2.4} There exists an initial bottomed \tsym{\Lambda}-algebra, and
any two initial bottomed \tsym{\Lambda}-algebras are isomorphic. \endpro

\proof The existence follows from Lemma~3.2.4 and Proposition~2.3.6. The
rest then follows directly from the definition of being initial. \eop

\bigbreak

Consider the construction given in the proof of Proposition~2.3.6,
tailored somewhat to the special situation occurring here. For each $\beta \in B$ let 
$\flat_\beta$ be some element not in $K$ and such that 
$\flat_{\beta_1} \ne \flat_{\beta_2}$ whenever $\beta_1 \ne \beta_2$. Put 
$K^\flat = K \cup \lcurl \flat_\beta \,:\, \beta \in B \rcurl$ and define 
mappings
$\Theta^\flat : K^\flat \to {\cal F}_B$ and 
$\vartheta^\flat : K^\flat \to B$ by 
\smallskip
\mdisp{\Theta^\flat(\kappa) \ =\ \cases{
      \Theta(\kappa) & if $\,\kappa \in K\,$, \cr
          \varnothing & if $\,\kappa = \flat_\beta\,$ for some 
                              $\,\beta \in B\,$, \cr }}
\vskip-\smallskipamount
\mdisp{\vartheta^\flat(\kappa) \ =\ \cases{
      \vartheta(\kappa) & if $\,\kappa \in K\,$, \cr
          \beta & if $\,\kappa = \flat_\beta\,$ for some 
                              $\,\beta \in B\,$. \cr }}
\smallskip
Then $\Lambda^\flat = (B,K^\flat,\Theta^\flat,\vartheta^\flat)$ is a signature which is 
clearly an extension of $\Lambda$ and in which $\flat_\beta$ is of type 
$\varnothing \to \beta$ for each $\beta \in B$.
\medskip
Now by Proposition~2.3.3 there exists an initial \tsym{\Lambda^\flat}-algebra
$({\breve D}_B,{\breve f}_{K^\flat})$. The
\tsym{\Lambda}-algebra $({\breve D}_B,{\breve f}_K)$ (obtained by omitting the
mappings ${\breve f}_{\flat_\beta}$, $\beta \in B$) can then be considered
as a bottomed \tsym{\Lambda}-algebra by declaring ${\breve f}_{\flat_\beta}(\onept)$ 
to be the bottom element of ${\breve D}_\beta$ for each $\beta \in B$.
\medskip

\proclaim{Proposition~3.2.5} $({\breve D}_B,{\breve f}_K)$ is an initial bottomed
\tsym{\Lambda}-algebra. \endpro

\proof This follows directly from Proposition~2.3.5 and Lemma~3.2.4. \eop
\bigbreak
The bottomed \tsym{\Lambda}-algebra $(D_B,f_K)$ is said to be
\indexddef{unambiguous}{unambiguous bottomed algebra}{}{bottomed algebra}{unambiguous}
if the following hold:
\medskip\smallskip
{\parindent=25pt
\item{(1)} The mapping $f_\kappa$ is injective for each $\kappa \in K$.
\medskip
\item{(2)} For each $\beta \in B$ the sets $\Im(f_\kappa)$, $\kappa \in K_\beta$, are 
disjoint subsets of $D_\beta \setminus \{\bot_\beta\}$.

\bigskip
}
The following analogue of Proposition~2.3.2 then holds:

\proclaim{Proposition~3.2.6} A bottomed \tsym{\Lambda}-algebra is initial 
if and only if it is minimal and unambiguous. \endpro

\proof The general case can clearly be reduced to that in which 
the bottom elements are all distinct. The statement thus follows from 
Lemma~3.2.4 and Proposition~2.3.7.\eop
\bigbreak
Propositions 2.3.2 and 3.2.6 imply that an initial bottomed \tsym{\Lambda}-algebra is 
never an initial \tsym{\Lambda}-algebra and that, conversely, a bottomed 
\tsym{\Lambda}-algebra which is an initial \tsym{\Lambda}-algebra is never an initial 
bottomed \tsym{\Lambda}-algebra. Moreover, Proposition~3.2.6 implies that the flat 
extension of $(X_B,p_K)$ is unlikely to be an initial bottomed \tsym{\Lambda}-algebra.
(In fact this will hold only in the trivial case in which each type in $\Lambda$ is 
primitive.) 
\bigskip
Having said something about bottomed \tsym{\Lambda}-algebras, we next 
look at the special case of bottomed extensions of the 
initial \tsym{\Lambda}-algebra $(X_B,p_K)$.
\medskip
A bottomed extension $(D_B,f_K)$ of $(X_B,p_K)$ is said to be 
\indexddef{minimal}{minimal bottomed extension}{}{bottomed extension}{minimal}
if $(D_B,f_K)$ is a minimal bottomed \tsym{\Lambda}-algebra.
In particular, as was already mentioned, 
the flat extension of $(X_B,p_K)$ is minimal. 
\medskip
Note that for any bottomed extension of $(D_B,f_K)$ of $(X_B,p_K)$ 
the minimal bottomed subalgebra $({\hat D}_B,{\hat f}_K)$ 
of $(D_B,f_K)$ is still an extension, 
and thus a minimal bottomed extension of $(X_B,p_K)$.
\medskip
In the same way as above
a bottomed extension $(D_B,f_K)$ of $(X_B,p_K)$ will be called 
\indexddef{initial}{initial bottomed extension}{}{bottomed extension}{initial}
if $(D_B,f_K)$ is an initial bottomed \tsym{\Lambda}-algebra.
By Proposition~3.2.6 an initial extension is also minimal.

\proclaim{Proposition~3.2.7} There exists an initial extension 
of $(X_B,p_K)$, and any two initial extensions of
$(X_B,p_K)$ are conjugate.
\endpro

\proof The existence follows from Proposition~2.5.3 and Lemma~3.2.4,
and the second statement then follows from Proposition~3.2.4 and Lemma~3.2.1. \eop
\bigbreak

Example~3.2.2 on the 
following page gives an initial extension of the 
\tsym{\Lambda}-algebra $(X_B,p_K)$ introduced in Example~2.2.1. 

\medskip
We defined
a bottomed extension to be initial if it is an initial 
bottomed \tsym{\Lambda}-algebra, and it could be objected that this definition is 
not really correct. However, the next result shows
that it is, even if not correct, equivalent to the correct
definition.

\proclaim{Proposition~3.2.8} Let $(D_B,f_K)$ be a bottomed extension of $(X_B,p_K)$.
Then $(D_B,f_K)$ is initial if and only if for each 
bottomed extension $(D'_B,f'_K)$ of $(X_B,p_K)$ there exists a unique
extension homomorphism from $(D_B,f_K)$ to $(D'_B,f'_K)$. \endpro

\proof Suppose that for each 
bottomed extension $(D'_B,f'_K)$ of $(X_B,p_K)$ there exists a unique
extension homomorphism from $(D_B,f_K)$ to $(D'_B,f'_K)$. Then this holds
in particular with $(D'_B,f'_K)$ an initial extension of 
$(X_B,p_K)$ (and Proposition~3.2.7 states that such an extension exists).
Thus $(D_B,f_K)$ and $(D'_B,f'_K)$ are isomorphic 
(since there also exists a unique bottomed homomorphism from $(D'_B,f'_K)$ to $(D_B,f_K)$)
and hence $(D_B,f_K)$ is initial. The converse is clear. \eop
\bigbreak

\vfill\eject
\bigskip\medskip
\frame{20pt}{\bigskip 
{\it Example 3.2.2\enspace} 
The following notation will be  employed here (and also later): If 
$Z$ is a set then $\Bot{Z}$ denotes a disjoint copy of $Z$; the element in 
$\Bot{Z}$ corresponding to $z \in Z$ will be denoted by $z^\bot$ (so
$\Bot{Z} = \lcurl z^\bot \,:\, z \in Z \rcurl$).
\medskip
For each $\beta \in B$ choose an element $\bot_\beta$ not in $X_\beta$ and
consider the \tsym{\Lambda}-algebra
$(X^\top_B,p^\top_K)$ defined by
\medskip
{\leftskip=32pt
$X^\top_{\fstt{bool}} = \Bool^\bot = \{T,F,\bot_{\fstt{bool}}\}$, 
\smallskip
$X^\top_{\fstt{nat}} = \Nat \cup \Bot{\Nat}$\ with 
                                                $0^\bot = \bot_{\fstt{nat}}$, 
\smallskip
$X^\top_{\fstt{int}} 
  = \Int \cup \{\bot_{\fstt{int}}\}$,
\quad
$X^\top_{\fstt{pair}} 
  = (X^\top_{\fstt{int}})^2 \cup \{\bot_{\fstt{pair}}\}$,
\smallskip    
$X^\top_{\fstt{list}} 
    = (X^\top_{\fstt{int}})^* \cup
       \Bot{(X^\top_{\fstt{int}})^*}$
  \ with $\onept^\bot = \bot_{\fstt{list}}$.
\medskip
$p^\top_{\fstt{True}} : 
    \Oneptset \to X^\top_{\fstt{bool}}$\ with\ %
                      $p^\top_{\fstt{True}}(\onept) = T$, 
\smallskip
$p^\top_{\fstt{False}} : 
    \Oneptset \to X^\top_{\fstt{bool}}$\ with\ %
                      $p^\top_{\fstt{False}}(\onept) = F$, 
\smallskip
$p^\top_{\fstt{Zero}} : 
   \Oneptset \to X^\top_{\fstt{nat}}$\ with\ %
                       $p^\top_{\fstt{Zero}}(\onept) = 0$,
\smallskip
$p^\top_{\fstt{Succ}} : 
 X^\top_{\fstt{nat}} \to X^\top_{\fstt{nat}}$
  \ with\ %
\mdisp{p^\top_{\fstt{Succ}}(x)\ =\ \cases{ n + 1 & 
                           if $\,x = n\,$ for some $\,n \in \Nat\,$, \cr
         (n + 1)^\bot & if $\,x = n^\bot\,$ for some $\,n \in \Nat\,$, \cr
                 }} 
\smallskip
$p^\top_{{\underline n}} : 
   \Oneptset \to X^\top_{\fstt{int}}$\ with\ %
                       $p^\top_{{\underline n}}(\onept) = n\ $
    for each $n \in \Int$,
\smallskip
$p^\top_{\fstt{Pair}} : 
      X^\top_{\fstt{int}} \times X^\top_{\fstt{int}}
 \to X^\top_{\fstt{pair}}$\ with\ %
                   $p^\top_{\fstt{Pair}}(x_1,x_2) = (x_1,x_2)$, 
\smallskip
$p^\top_{\fstt{Nil}} : 
 \Oneptset \to X^\top_{\fstt{list}}$\ with\ %
                         $p^\top_{\fstt{Nil}}(\onept) = \onept$, 
\smallskip
$p^\top_{\fstt{Cons}} : 
 X^\top_{\fstt{int}} \times X^\top_{\fstt{list}}
 \to X^\top_{\fstt{list}}$\ with %
\mdisp{p^\top_{\fstt{Cons}}(x,z) \ =\ \cases{ 
           x \triangleleft s & if $\,z = s\,$ for some
                   $\,s \in  (X^\top_{\fstt{int}})^*\,$, \cr
   \noalign{\smallskip}
        (x \triangleleft s)^\bot & if $\,z = s^\bot\,$ for some
                 $\,s \in  (X^\top_{\fstt{int}})^*\,$. \cr }}
\medskip
}
It is left to the reader to check that $(X^\top_B,p^\top_K)$ is an initial 
extension of the \tsym{\Lambda}-algebra $(X_B,p_K)$ introduced in Example~2.2.1.
\medskip
Note that an element of $X^\top_{\fstt{list}}$ has either the form
$\list x n $ or the form $(\list x n )^\bot$, where $n \ge 0$ and
$x_j \in \Int \cup \{\bot_{\fstt{int}}\}$ for $\oneto j n $.
The element $\list x n $ describes a `real' list with $n$ components 
(although some or all of these components may be `undefined'). The element 
$(\list x n )^\bot$, on the other hand, should be thought of as a `partial' 
list containing at least $n$ components, of which the first $n$ 
components are `known' to be $\lvector x n $.
\bigskip\medskip
}
\bigskip

\vfill\eject

\hfline{64}{3.3 \ REGULAR BOTTOMED ALGEBRAS AND THE TRACE}

\sectionhead {3.3} {Regular bottomed algebras and the trace}
\bigskip\medskip
In Section~3.2 we introduced a framework for dealing with `undefined' and 
`partially defined' data objects. This involved bottomed extensions of the basic
initial \tsym{\Lambda}-algebra $(X_B,p_K)$. However, 
as it stands, this notion is much too general, and 
two conditions will be imposed on the bottomed extensions to be
actually used. In the present section we look at the first
of these conditions, which goes under the name of {\it regularity\/}. 
\medskip
A bottomed \tsym{\Lambda}-algebra $(D_B,f_K)$ is said to be 
\indexddef{regular}{regular bottomed algebra}{}{bottomed algebra}{regular}
if for each $\beta \in B$ and each $u \in D_\beta \setminus \{\bot_\beta\}$ there exists 
a unique $\kappa \in K_\beta$ and a unique element $c \in \dom(f_\kappa)$ such that 
$f_\kappa(c) = u$. 
(This is a possible generalisation of the definition of a regular \tsym{\Lambda}-algebra
given in Section~2.3. However, it is not the generalisation which would
make the analogue of Proposition~2.3.2 hold.)   
\medskip 
A bottomed extension $(D_B,f_K)$ of $(X_B,p_K)$ is now said to be 
\indexddef{regular}{regular bottomed extension}{}{bottomed extension}{regular}
if $(D_B,f_K)$ is a regular bottomed \tsym{\Lambda}-algebra.
In particular, the flat extension of $(X_B,p_K)$ is regular (since
$(X_B,p_K)$ is a regular \tsym{\Lambda}-algebra).
Moreover, by Propositions 3.2.2 and 3.2.6 each initial extension of $(X_B,p_K)$ is regular. 
Note that if $(D_B,f_K)$ is a regular bottomed extension of $(X_B,p_K)$ then 
$D_\beta = X_\beta \cup \{\bot_\beta\}$ must hold for each primitive type $\beta \in B$,
and so in particular $D_{\fstt{int}} = \Int \cup \{\bot_{\fstt{int}}\}$
and $D_{\fstt{bool}} = \Bool \cup \{\bot_{\fstt{bool}}\}$.
\medskip
One reason why regularity is important has to do with the usual `case' constructions
found in all modern functional programming languages. This was already seen 
implicitly in Chapter~1: 
If $(D_B,f_K)$ is a regular extension of $(X_B,p_K)$ then, for example, given a function
$\fss{sqshx} : D_{\fstt{list}} \times D_{\fstt{int}} \times D_{\fstt{list}}
\to D_{\fstt{list}}$, 
it makes sense to define a function
$\fss{sqsh} : D_{\fstt{list}} \times D_{\fstt{list}}\to D_{\fstt{list}}$
by
\smallskip
\mdisp{\fss{sqsh}\,(a,b)\ =\ \cases{
       b & if $\,a = f_{\fstt{Nil}}(\onept)\,$, \cr
       \fss{sqshx}\,(d,m,b)
                   & if $\,a \ne \bot_{\fstt{list}}\,$ and 
		   $\,a = f_{\fstt{Cons}}(m,d)\,$, \cr
		   \bot_{\fstt{list}} & if $\,a = \bot_{\fstt{list}}\,$. \cr} 
}
\smallskip
More generally, consider $\theta \in B$ with $K_\theta$ finite and with
$\kappa$ of type $L_\kappa \to \theta$ for each $\kappa \in K_\theta$.
Let $\beta \in B$ and suppose that for each $\kappa \in K_\theta$ a function
$h^\kappa : \ass {L_\kappa} D \to D_\beta$ has been given. Now if 
$(D_B,f_K)$ is a regular extension of $(X_B,p_K)$ then it makes sense to define
a function $h : D_\theta \to D_\beta$ by
\smallskip
\mdisp{h(u)\ =\ \cases{
                 h^\kappa(b)   & if $\,u \ne \bot_\theta\,$ and
                      $\,u = f_\kappa(b)\,$ with $\,\kappa \in K_\theta\,$
                      and $\,b \in \ass {L_\kappa} D \,$,\cr
\noalign{\smallskip}
       \bot_\beta  & if $\,u = \bot_\theta\,$.\cr}
}
\smallskip
This forms the basis for defining the case operators to be introduced in Section~5.3.
\medskip
We now first present some general facts about regular bottomed \tsym{\Lambda}-algebras.
Then we introduce an initial bottomed extension 
$(F^\flat_B,\fo^\flat_K)$ of the ground term algebra $(F_B,\fo_K)$ called 
the bottomed ground term algebra. 
This extension
$(F^\flat_B,\fo^\flat_K)$ will be used to define what
is called the trace of a bottomed \tsym{\Lambda}-algebra $(D_B,f_K)$. 
For a large class of bottomed algebras the trace has an explicit
finite description, and it will be thought of as providing 
the information about $(D_B,f_K)$ which can be used in practice.
Finally, we show that the trace is essentially a 
complete invariant for minimal regular bottomed \tsym{\Lambda}-algebras.
\medskip
If $(D_B,f_K)$ is a bottomed \tsym{\Lambda}-algebra then the family $C_K$ given by

\mdisp{C_\kappa 
\ =\ \lcurl c \in \dom(f_\kappa) \,:\, f_\kappa(c) \ne \bot_\beta \rcurl}

for each $\kappa \in K_\beta$ will be called the 
\indexdef{core}{core of bottomed algebra}{}
of $(D_B,f_K)$. In particular, 
if $(D_B,f_K)$ is initial then
$C_\kappa = \dom(f_\kappa)$ for each $\kappa \in K$, and
the core of the flat extension of $(X_B,p_K)$ is the family
$C_K$ with $C_\kappa = \dom(p_\kappa)$ for each $\kappa \in K$. 
Regularity can be expressed in terms of the core as follows:
\bigbreak
\proclaim{Lemma~3.3.1} Let $C_K$ be the core of the bottomed \tsym{\Lambda}-algebra
$(D_B,f_K)$. Then $(D_B,f_K)$ is regular if and only if the following
three conditions hold:
\medskip\smallskip
{\parindent=25pt
\item{(1)} The restriction of $f_\kappa$ to $C_\kappa$ is injective for each 
$\kappa \in K$.
\medskip
\item{(2)} For each $\beta \in B$ the sets $f_\kappa(C_\kappa)$, $\kappa \in K_\beta$,
are disjoint subsets of $D_\beta$.
\medskip
\item{(3)}$\bigcup\limits_{\kappa \in K_\beta} f_\kappa(C_\kappa)
                  \,=\, D_\beta \setminus \{\bot_\beta\}\ $ 
for each $\beta \in K$. 
\bigskip
}  \endpro

\proof Assume first that (1), (2) and (3) hold. Let $\beta \in B$ and 
$u \in D_\beta \setminus \{\bot_\beta\}$; by (3) there then exists
$\kappa \in K_\beta$ and $c \in C_\kappa$ such that $f_\kappa(c) = u$.
Suppose also that $u = f_{\kappa'}(c')$ with $\kappa' \in K_\beta$ and
$c' \in \dom(f_{\kappa'})$; then $c' \in C_{\kappa'}$ (since $u \ne \bot_\beta$),
hence $u \in f_\kappa(C_\kappa) \cap f_{\kappa'}(C_{\kappa'})$ and by (2)
this is only possible if $\kappa' = \kappa$. Moreover, by (1) it now follows that 
$c' = c$, and therefore $(D_B,f_K)$ is regular.
\medskip
Assume conversely that $(D_B,f_K)$ is regular. Let $\beta \in B$,
$\kappa_1,\,\kappa_2 \in K_\beta$ and 
$c_1 \in C_{\kappa_1}$, $c_2 \in C_{\kappa_2}$  
with $f_{\kappa_1}(c_1) = f_{\kappa_2}(c_2)$. Then
$f_{\kappa_1}(c_1) \ne \bot_\beta$ and therefore
$\kappa_1 = \kappa_2$ and $c_1 = c_2$. This shows that
(1) and (2) hold. Moreover, 
$\,D_\beta \setminus \{\bot_\beta\}\,\subset\,
               \bigcup_{\kappa \in K_\beta} \Im(f_\kappa)\,$
for each $\beta \in B$ 
and hence by the definition of $C_K$ (3) holds. \eop
\bigbreak
Note that by Proposition~3.2.2 condition (3) in Lemma~3.3.1 holds automatically for a
minimal bottomed \tsym{\Lambda}-algebra, which leads to the following observation:
Let $(D'_B,f'_K)$ be an extension 
of a bottomed \tsym{\Lambda}-algebra $(D_B,f_K)$, considered as a
bottomed \tsym{\Lambda}-algebra in the obvious way, and suppose $(D'_B,f'_K)$ is regular.
Then in general this does not imply that $(D_B,f_K)$ is regular. However,
$(D_B,f_K)$ will be regular if it is minimal (since conditions (1) and (2) 
in Lemma~3.3.1 are inherited by $(D_B,f_K)$).
\medskip
A regular bottomed \tsym{\Lambda}-algebra $(D_B,f_K)$ will be called
\indexddef{fully regular}
{fully regular bottomed algebra}{}{bottomed algebra}{fully regular}
if $\bot_\beta \notin \Im(f_\kappa)$ for each
$\kappa \in K_\beta$, $\beta \in B$. Thus a regular bottomed \tsym{\Lambda}-algebra
$(D_B,f_K)$ is fully regular if and only if its core $C_K$ is maximal, 
i.e., if and only if $C_\kappa = \dom(f_\kappa)$ for each $\kappa \in K$.

\proclaim{Proposition~3.3.1} An initial bottomed \tsym{\Lambda}-algebra is
fully regular. Conversely, a fully regular bottomed \tsym{\Lambda}-algebra is
initial if and only if it is minimal. \endpro

\proof This follows from Propositions 3.2.2 and 3.2.6. \eop
\bigbreak
Recall that in Section~3.1 we introduced the ground term algebra.
This was a further initial \tsym{\Lambda}-algebra $(F_B,\fo_K)$, 
and the unique isomorphism $\sem{\cdot}_B$
from $(F_B,\fo_K)$ to $(X_B,p_K)$ is used to denote the basic
data objects (i.e., the data object 
$x \in X_\beta$ can be denoted by the unique element $s \in F_\beta$ 
with $\sem{s}_\beta = x$).
\medskip
Now by Proposition~3.2.7 there exists an initial bottomed extension
of $(F_B,\fo_K)$ (and, moreover, any two such initial extensions are 
conjugate). Thus in what follows we can suppose that
an initial bottomed extension $(F^\flat_B,\fo^\flat_K)$ of $(F_B,\fo_K)$ has been 
chosen, which will be called the 
\indexddef{bottomed ground term algebra}
{bottomed ground term algebra}{}{ground term algebra}{bottomed}.
The bottom element of $F^\flat_\beta$ will be denoted by $\flat_\beta$ rather than
$\bot_\beta$ (and to be thought of as a name with which 
to refer to the  `undefined' element $\bot_\beta$).
\medskip
Let $(D_B,f_K)$ be a bottomed \tsym{\Lambda}-algebra, which is considered to be
fixed for the rest of the section. Then,
since $(F^\flat_B,\fo^\flat_K)$ is an initial bottomed \tsym{\Lambda}-algebra, there exists
a unique bottomed homomorphism 
$\sem{\cdot}^\bot_B : (F^\flat_B,\fo^\flat_K) \to (D_B,f_K)$, i.e., 
the unique homomorphism from $(F^\flat_B,\fo^\flat_K)$ to $(D_B,f_K)$ with
$\sem{\,\flat_\beta}^\bot_\beta = \bot_\beta$ for each $\beta \in B$.
In particular, if $(D_B,f_K)$ is a bottomed extension of $(X_B,p_K)$ then by
Proposition~2.5.2 $\sem{\cdot}^\bot_B$ is an extension of $\sem{\cdot}_B$,
i.e., $\sem{s}^\bot_\beta = \sem{s}_\beta$ for all $s \in F_\beta$, $\beta \in B$.

\proclaim{Proposition~3.3.2} (1)\enskip The family $\sem{\cdot}^\bot_B$
is surjective (i.e., $\sem{\cdot}^\bot_\beta : F^\flat_\beta \to D_\beta$ is 
surjective for each $\beta \in B$) if and only if $(D_B,f_K)$ is minimal.
\medskip
(2)\enskip If $(D_B,f_K)$ is fully regular then the family $\sem{\cdot}^\bot_B$ is 
injective, i.e., the mapping
$\sem{\cdot}^\bot_\beta : F^\flat_\beta \to D_\beta$ is injective for
each $\beta \in B$.
\medskip
(3)\enskip The family $\sem{\cdot}^\bot_B$ is an isomorphism 
if and only if $(D_B,f_K)$ is initial.
\endpro

\proof (1)\enskip For each $\beta \in B$ let 
$D'_\beta = \sem{F^\flat_\beta}^\bot_\beta$; then $\bot_\beta \in D'_\beta$ 
for each $\beta \in B$ and, as in Lemma~2.2.1~(1), $D'_B$ is  
invariant in $(D_B,f_K)$. Moreover, $D'_B$ is the minimal such family:
Let $D^o_B$ be any family invariant in $(D_B,f_K)$ and
with $\bot_\beta \in D^o_\beta$ for each $\beta \in B$, and for each $\beta \in B$ 
let $F^o_\beta = \lcurl s \in F^\flat_\beta :
\sem{s}^\bot_\beta \in D^o_\beta \rcurl$.
Then $\flat_\beta \in F^o_\beta$ and,
as in Lemma~2.2.1~(2), $F^o_B$ is invariant in $(F^\flat_B,\fo^\flat_K)$;
thus $F^o_B = F^\flat_B$ (since by Proposition~3.2.6 
$(F^\flat_B,\fo^\flat_K)$ is a minimal extension
of $(F_B,\fo_K)$). Hence
$D'_B \subset D^o_B$.
From this it follows that $\sem{\cdot}^\bot_B$
is surjective if and only if $(D_B,f_K)$ is minimal.
\medskip

(3)\enskip This is clear.
\medskip
(2)\enskip Let $({\hat D}_B,{\hat f}_K)$ be the minimal bottomed subalgebra of 
$(D_B,f_K)$. Then $({\hat D}_B,{\hat f}_K)$ is a minimal fully regular and
thus an initial bottomed \tsym{\Lambda}-algebra.
Therefore by (3) $\sem{\cdot}^\bot_B$ must also 
be an isomorphism from $(F^\flat_B,\fo^\flat_K)$ to $({\hat D}_B,{\hat f}_K)$. 
In particular, the mapping
$\sem{\cdot}^\bot_\beta : F^\flat_\beta \to D_\beta$ is injective for
each $\beta \in B$. \eop
\bigbreak

\proclaim{Proposition~3.3.3} Suppose $(D_B,f_K)$ is a regular extension of $(X_B,p_K)$.
Then 
\mdisp{\lcurl s \in F^\flat_\beta\,:\, \sem{s}^\bot_\beta \in X_\beta \rcurl\ =\ F_\beta} 
for each $\beta \in B$. \endpro

\proof It has already been noted that $\sem{\cdot}^\bot_B$ is an extension of
$\sem{\cdot}_B$, which implies that $\sem{s}^\bot_\beta \in X_\beta$ for
each $s \in F_\beta$ (and this holds even if $(D_B,f_K)$ is not regular).
For the converse consider the family ${\grave F}^\flat_B$ with 
${\grave F}^\flat_B \subset F^\flat_B$ given by

\mdisp{{\grave F}^\flat_\beta\ =\ F_\beta \,\cup\,
\lcurl s \in F^\flat_\beta \setminus F_\beta 
  \,:\, \sem{s}^\bot_\beta \in D_\beta \setminus X_\beta\rcurl}

for each $\beta \in B$. If $\kappa \in K$ is of type $\varnothing \to \beta$ then
$\fo^\flat_\kappa(\onept) = \fo_\kappa(\onept )\in F_\beta \subset {\grave F}^\flat_\beta$.
Now let $\kappa \in K$ be of type $L \to \beta$ with $L \ne \varnothing$ and let
$a \in \assb L {{\grave F}^\flat} $, i.e., $a(\eta) \in {\grave F}^\flat_\eta$ for each
$\eta \in L$.  Suppose 
$\sem{\fo^\flat_\kappa(a)}^\bot_\beta = f_\kappa(\assb L {\sem{a}^\bot} ) \in X_\beta$.
Then, since $(D_B,f_K)$ is regular, it follows that
$\assb L {\sem{a}^\bot} \in \ass L X $, i.e., 
$\sem{a(\eta)}_\eta \in X_\eta$ for each $\eta \in L$.
But then by assumption $a(\eta) \in F_\eta$ for each $\eta \in L$ and thus
$\fo^\flat_\kappa(a) = \fo_\kappa(a) \in F_\beta$. This shows that
$\fo^\flat_\kappa(a) \in {\grave F}^\flat_\beta$, and hence that
the family ${\grave F}^\flat_B$ is invariant in $(F^\flat_B,\fo^\flat_K)$.
But clearly $\flat_\beta \in {\grave F}^\flat_\beta$ for each $\beta \in B$
and so by Proposition~3.2.6 ${\grave F}^\flat_B = F^\flat_B$.
From this it follows that $\,\sem{s}^\bot_\beta \in D_\beta \setminus X_\beta\,$ for all
$s \in F^\flat_\beta \setminus F_\beta$. \eop
\bigbreak

\proclaim{Lemma~3.3.2} If $(D_B,f_K)$ is the flat extension of $(X_B,p_K)$
then
\mdisp{\sem{s}^\bot_\beta 
  \ =\ \cases{ \sem{s}_\beta & if $\,s \in F_\beta\,$, \cr
                \bot_\beta &  if 
                   $\,s \in F^\flat_\beta \setminus F_\beta\,$.
         }}
\endpro

\proof This follows immediately from Proposition~3.3.3 and the fact that
$\sem{\cdot}^\bot_B$ is an extension of $\sem{\cdot}_B$. \eop
\bigbreak
We now define a family $R_B \subset F^\flat_B$, called the 
\indexdef{trace}{trace of bottomed algebra}{}
of $(D_B,f_K)$, by putting
\mdisp{R_\beta 
\ =\ \lcurl\, s \in F^\flat_\beta\, :\, 
                           \sem{s}^\bot_\beta \ne \bot_\beta \,\rcurl}

for each $\beta \in B$. Thus in fact 
$\,R_\beta \subset F^\flat_\beta \setminus \{\,\flat_\beta\}\,$
for each $\beta \in B$ and, moreover,
$\,F_B \subset R_B\,$ whenever
$(D_B,f_K)$ is a bottomed extension of $(X_B,p_K)$. 
\medskip
If $(D_B,f_K)$ is the flat extension of $(X_B,p_K)$ then 
by Lemma~3.3.2 $R_B = F_B$, and if $(D_B,f_K)$ is fully regular 
then Proposition~3.3.2~(2) implies that 
$R_\beta = F^\flat_\beta \setminus \{\,\flat_\beta\}$
for each $\beta \in B$. The trace is thus minimal for the flat extension
and maximal for a fully regular algebra.
In the following section a class of bottomed algebras will be introduced (including the 
flat extension and the fully regular algebras) for which the trace can be 
computed explictly.
\medskip
As already indicated, we regard the trace $R_B$ as providing the only information about 
the bottomed extension $(D_B,f_K)$ which can be used in practice (for example, 
in the algorithm for computing values to be introduced in Section~7.3).
This point-of-view may seem somewhat arbitrary; however, 
it turns out that, at least for the purposes of this study, the trace does contains all 
the information we need. Moreover, it emphasises the fact that, in practice, the 
information which can be used about $(D_B,f_K)$ must have some kind of finite description.  
 
\medskip
The final result of this section shows that the trace is essentially a 
complete invariant for minimal regular bottomed \tsym{\Lambda}-algebras.
 
\proclaim{Proposition~3.3.4} Minimal regular bottomed \tsym{\Lambda}-algebras are 
isomorphic if and only if they have the same trace. In particular, 
minimal regular extensions of $(X_B,p_K)$ are 
conjugate if and only if they have the same trace. \endpro

\proof It is clear that isomorphic bottomed \tsym{\Lambda}-algebras 
have the same trace. To prepare for the converse a 
couple of lemmas are needed. 
In the following let $(D_B,f_K)$ be a minimal regular bottomed \tsym{\Lambda}-algebra
with core $C_K$.

\proclaim{Lemma~3.3.3} There exists a unique family of
mappings $\#_B$ with $\#_\beta : D_\beta \to \Nat$ for each 
$\beta \in B$ such that $\,\#_\beta(\bot_\beta) = 0\,$ for each $\beta \in B$, 
$\,\#_\beta(f_\kappa(\onept)) = 0\,$ whenever
$\kappa \in K$ is of type $\varnothing \to \beta$ and such that 
\mdisp{\#_\beta(f_\kappa(c))
  \ =\ 1 \,+\, \max \lcurl \#_\eta(c(\eta))\,:\, \eta \in L \rcurl}
for all $c \in \ass L D $
whenever $\kappa \in K$ is of type $L \to \beta$ 
with $L \ne \varnothing$. \endpro

\proof This is, with the obvious modifications, essentially the same as the proof of 
Lemma~2.3.5. \eop
\bigbreak

\proclaim{Lemma~3.3.4} There exists a unique family of mappings 
$\varrho_B$ with $\varrho_\beta : D_\beta \to F^\flat_\beta$ for each 
$\beta \in B$ such that $\varrho_\beta(\bot_\beta) = \flat_\beta$ 
and if $\kappa \in K$ is of type $L \to \beta$ then
\mdisp{\varrho_\beta(f_\kappa(c)) 
\ =\ \fo^\flat_\kappa(\ass L {\varrho} (c))}
for all $c \in C_\kappa$.
Moreover, $\sem{\varrho_\beta(u)}^\bot_\beta = u$ for all $u \in D_\beta$,
$\beta \in B$. \endpro

\proof Let $\#_B$ be the family of mappings given by Lemma~3.3.3
and for each $\beta \in B$, $m \in \Nat$
let $D^m_\beta = \lcurl u \in D_\beta : \#_\beta(u) = m \rcurl$. Define
$\varrho_\beta$ on $D^m_\beta$ for each $\beta \in B$ using induction on
$m$. If $u \in D^0_\beta$ then either $u = \bot_\beta$, in which case 
put $\varrho_\beta(u) = \flat_\beta$, or there exists a unique $\kappa \in K$
of type $\varnothing \to \beta$ with $u = f_\kappa(\onept)$, and in this case  
put $\varrho_\beta(u) = \fo^\flat_\kappa(\onept)$.
Now let $m > 0$ and suppose $\varrho_\theta$ is already defined on
$D^k_\theta$ for each $\theta \in B$ and for all $k < m$. Let
$u \in D^m_\beta$ (so in particular $u \ne \bot_\beta$); then there exists
a unique $\kappa \in K$ of type $L \to \beta$ for some 
$L \ne \varnothing$ and a unique element $c \in C_\kappa$ such that
$u = f_\kappa(c)$. Moreover, $c(\eta) \in D^{m_\eta}_\eta$
for some $m_\eta$ with $m_\eta < m$, which means that
$\varrho_\eta(c(\eta))$ is already defined for each $\eta \in L$.
It thus makes sense to put

\mdisp{\varrho_\beta(u) \ =\ \fo^\flat_\beta(\ass L {\varrho} (c))}

(where of course $\ass L {\varrho} (c)$ is the element $c' \in \assb L {F^\flat} $ with
$c'(\eta) = \varrho_\eta(c(\eta))$ for each $\eta \in L$).
In this way $\varrho_\beta$ is defined on $D^m_\beta$ for each
$m \in \Nat$ and the family $\varrho_B$ has the required property by
construction. The uniqueness of the family $\varrho_B$ also follows
using induction on $m$.
\medskip
Finally, for each $\beta \in B$ let 
$D'_\beta = \lcurl u \in D_\beta :
                    \sem{\varrho_\beta(u)}^\bot_\beta = u \rcurl$;
then $\bot_\beta \in D'_\beta$ for each $\beta \in B$ and the family
$D'_B$ is invariant. (Let $\kappa \in K$ be of type
$L \to \beta$ and let $c \in \ass L D $ with
$c(\eta) \in D'_\eta $ for each $\eta \in L$; put $u = f_\kappa(c)$. If 
$u = \bot_\beta$
then $u \in D'_\beta$ holds trivially. On the other hand,
if $u \ne \bot_\beta$ then $c \in C_\kappa$; in this case
$\varrho_\beta(u) = \fo^\flat_\kappa(\ass L {\varrho} (c))$
and hence, since 
$\sem{\,\varrho_\eta(c(\eta))\,}^\bot_\eta = c(\eta)$
for each $\eta \in L$ and so by Lemma~2.1.1~(2) 
$\assb L {\sem{\,\ass L {\varrho} (c)\,}^\bot} = c$, it follows that

\mdisp{ \sem{\varrho_\beta(u)}^\bot_\beta 
\ =\  \sem{\,\fo^\flat_\kappa(\ass L {\varrho} (c))\,}^\bot_\beta
\ =\  f_\kappa(\assb L {\sem{\,\ass L {\varrho} (c)\,}^\bot} )
\ =\ f_\kappa(c) \ =\ u\ .}

Therefore in both cases $u \in D'_\beta$.) But $(D_B,f_K)$ is minimal
and thus $D'_B = D_B$, i.e.,
$\sem{\varrho_\beta(u)}^\bot_\beta = u$ for all $u \in D_\beta$, 
$\beta \in B$. \eop
\bigbreak
Now consider a minimal regular bottomed \tsym{\Lambda}-algebra $(D'_B,f'_K)$ having 
the same trace as $(D_B,f_K)$ and denote by $\pi_B$ the unique bottomed homomorphism from 
$(F^\flat_B,\fo^\flat_K)$ to $(D'_B,f'_K)$.

\proclaim{Lemma~3.3.5} Let 
$\varrho_B : D_B \to F^\flat_B$ be the family defined in Lemma~3.3.4.
Then $\omega_B = \pi_B \, \varrho_B$ is a bottomed
homomorphism from $(D_B,f_K)$ to $(D'_B,f'_K)$. \endpro

\proof Clearly $\omega_\beta(\bot_\beta) = \bot_\beta$ for each $\beta \in B$.
Let $\kappa \in K$ be of type $L \to \beta$ and
$c \in \ass L D $. Then by Lemma~2.1.1~(2)

\mdisp{
f'_\kappa(\ass L {\omega} (c))
\ =\ f'_\kappa(\ass L {\pi} (\ass L {\varrho} (c)))
\ =\ \pi_\beta(\fo^\flat_\kappa(\ass L {\varrho} (c)))\ .}

Now if $c \in C_\kappa$ then 
$\,\omega_\beta(f_\kappa(c))
\,  =\,    \pi_\beta(\varrho_\beta(f_\kappa(c)))
\,  =\,  \pi_\beta(\fo^\flat_\kappa(\ass L {\varrho} (c)))\,$,
and hence in this case
$\omega_\beta(f_\kappa(c)) \,=\, f'_\kappa(\ass L {\omega} (c))$.
On the other hand, if $c \notin C_\kappa$ then 
$\,\sem{\,\varrho_\eta(c(\eta))\,}^\bot_\eta = c(\eta)\,$ 
for each $\eta \in L$ and so by Lemma~2.1.1~(2)
$\assb L {\sem{\,\ass L {\varrho} (c)\,}^\bot} = c$, thus

\mdisp{\sem{\, \fo^\flat_\kappa(\ass L {\varrho} (c))\,}^\bot_\beta
\ =\ f_\kappa(\assb L {\sem{ \ass L {\varrho} (c) }^\bot} )
\ =\ f_\kappa(c)\ =\ \bot_\beta\ ,}

which means that
$\fo^\flat_\kappa(\ass L {\varrho} (c)) \in R_\beta$, 
where $R_B$ is the common trace of $(D_B,f_K)$ and $(D'_B,f'_K)$. Therefore
it again follows that
\mdisp{
f'_\kappa(\ass L {\omega} (c))
\ =\ \pi_\beta(\fo^\flat_\kappa(\ass L {\varrho} (c)))
\ =\ \bot_\beta
\ =\ \omega_\beta(\bot_\beta)
\ =\ \omega_\beta(f_\kappa(c))\ .}

This shows that $\omega_B$ is a bottomed homomorphism. \eop
\bigbreak
Let $(D_B,f_K)$ and $(D'_B,f'_K)$ be minimal regular bottomed \tsym{\Lambda}-algebras
having the same trace. By Lemma~3.3.5 there then exists a bottomed
homomorphism $\omega_B$ from $(D_B,f_K)$ to $(D'_B,f'_K)$ and also
a bottomed homomorphism $\omega'_B$ from $(D'_B,f'_K)$ to $(D_B,f_K)$.
By Proposition~2.2.1 $\omega_B \, \omega'_B$ is then a bottomed
homomorphism from $(D_B,f_K)$ to itself and 
the only such homomorphism is $\fss{id}_B$, i.e.,
$\omega_B \, \omega'_B = \fss{id}_B$. In the same way
$\omega'_B \, \omega_B = \fss{id}'_B$, and hence
$\omega_B$ is a bottomed isomorphism, i.e., 
$(D_B,f_K)$ and $(D'_B,f'_K)$ are isomorphic. 
This completes the proof of Proposition~3.3.4. \eop
\bigbreak

\vfill\eject

\hfline{70}{3.4 \ CORED BOTTOMED ALGEBRAS}

\sectionhead {3.4} {Cored bottomed algebras}
\bigskip\medskip
In this section some simple classes of bottomed \tsym{\Lambda}-algebras are introduced
which can be seen as generalising the flat extension and the fully regular algebras.
They provide examples of regular bottomed \tsym{\Lambda}-algebras for which the trace 
has an explicit description.
\medskip
For each $\beta \in B$ let $\natural_\beta$ be an element different from
$\flat_\beta$ and put $H_\beta = \{\flat_\beta,\natural_\beta\}$;
$H_\beta$ is considered as a bottomed set with bottom element $\flat_\beta$.
If $L$ is a \tsym{B}-typed set then $\natural^L$ will denote the element of
$\ass L H $ defined by $\natural^L(\eta) = \natural_\eta$ for each $\eta \in L$.
Let $(H_B,\diamond_K)$ be a \tsym{\Lambda}-algebra; then
$(H_B,\diamond_K)$ is called a 
\indexdef{core type}{core type}{} 
if $\diamond_\kappa(\natural^L) = \natural_\beta$
whenever $\kappa \in K$ is of type $L \to \beta$.
In particular,  
$\diamond_\kappa(\onept) \,=\,\natural_\beta$
whenever $\kappa$ is of type $\varnothing \to \beta$ for some $\beta \in B$.
\medskip
In practice the set of constructor names in $K$ which do not have a
type of the form $\varnothing \to \beta$ will always be finite, and in this case
a core type is essentially a finite object.
\medskip

Now let $(D_K,f_K)$ be a bottomed \tsym{\Lambda}-algebra and for each $\beta \in B$ 
define a mapping $\varepsilon_\beta : D_\beta \to H_\beta$ by putting

\mdisp{\varepsilon_\beta(u) 
\ =\ \cases{\natural_\beta & if $\,u \ne \bot_\beta\,$,\cr
             \flat_\beta & if $\,u = \bot_\beta\,$.}}
\medskip
Let $(H_B,\diamond_K)$ be a core type; then
$(D_B,f_K)$ is said to be 
\indexddef{\tsym{(H_B,\diamond_K)}-cored} 
{cored bottomed algebra}{}{bottomed algebra}{cored}
if the family $\varepsilon_B$ is actually a homomorphism 
(and thus a bottomed homomorphism) from
$(D_B,f_K)$ to $(H_B,\diamond_K)$. In other words, this means that
\mdisp{\varepsilon_\beta(f_\kappa(c))
  \ =\ \diamond_\kappa(\ass L {\varepsilon} (c))}
must hold for all $c \in \ass L D $ whenever
$\kappa \in K$ is of type $L \to \beta$. 
Thus whether or not $f_\kappa(c) \ne \bot_\beta$ holds is completely determined
by $\diamond_\kappa$ and the set 
$\lcurl \eta \in L \,:\, c(\eta) \ne \bot_\eta \rcurl$. 

\medskip
A bottomed extension $(D_B,f_K)$ of $(X_B,p_K)$ is now said to be
\indexddef{\tsym{(H_B,\diamond_K)}-cored} 
{cored bottomed extension}{}{bottomed extension}{cored}
if $(D_B,f_K)$ is an \tsym{(H_B,\diamond_K)}-cored bottomed \tsym{\Lambda}-algebra.
Note that such an extension can be \tsym{(H_B,\diamond_K)}-cored for at most one
core type $(H_B,\diamond_K)$: This follows since the signature
$\Lambda$ is assumed to be pervasive and hence
$X_\beta \ne \varnothing$ for each $\beta \in B$.
\medskip
The class of those bottomed \tsym{\Lambda}-algebras which are 
\tsym{(H_B,\diamond_K)}-cored for some core type $(H_B,\diamond_K)$
includes each fully regular bottomed \tsym{\Lambda}-algebra
and the flat extension of $(X_B,p_K)$:
For each $\kappa \in K$ of type $L \to \beta$ let
$\diamond^\top_\kappa : \ass L H \to H_\beta$
be the mapping with
$\diamond^\top_\kappa(c) = \natural_\beta$ for all
$c \in \ass L H $, and let
$\diamond^\bot_\kappa : \ass L H  \to H_\beta$
be given by
\smallskip
\mdisp{\diamond^\bot_\kappa(c)
\ =\ \cases{
\natural_\beta  & if $\,c = \natural^L\,$, \cr
 \flat_\beta & otherwise. }}
\smallskip
Then $(H_B,\diamond^\top_K)$ and $(H_B,\diamond^\bot_K)$ are both core types. 
Moreover, the flat extension is \tsym{(H_B,\diamond^\bot_K)}-cored and 
each fully regular bottomed \tsym{\Lambda}-algebra is \tsym{(H_B,\diamond^\top_K)}-cored.

\proclaim{Proposition~3.4.1} Let $(H_B,\diamond_K)$ be a core type; 
then there is a unique isomorphism class of minimal regular 
\tsym{(H_B,\diamond_K)}-cored bottomed \tsym{\Lambda}-algebras. 
Moreover, there is a unique conjugacy class of minimal regular 
\tsym{(H_B,\diamond_K)}-cored extensions of $(X_B,p_K)$. \endpro

\proof This is given at the end of the section. \eop
\bigbreak

The trace of an \tsym{(H_B,\diamond_K)}-cored
bottomed \tsym{\Lambda}-algebra will now be considered. This turns out to
only depend on $(H_B,\diamond_K)$ and it
can be computed more-or-less explicitly. In what follows let the core type
$(H_B,\diamond_K)$ be fixed. Since the bottomed ground term algebra 
is initial there exists a unique bottomed homomorphism 
$\delta^\flat_B$ from $(F^\flat_B,\fo^\flat_K)$ to $(H_B,\diamond_K)$, i.e., 
$\delta^\flat_B : (F^\flat_B,\fo^\flat_K) \to (H_B,\diamond_K)$ 
is the unique homomorphism such that 
$\delta^\flat_\beta(\,\flat_\beta) = \flat_\beta$ for each $\beta \in B$. 

\proclaim{Proposition~3.4.2} Suppose that $(D_B,f_K)$ is 
\tsym{(H_B,\diamond_K)}-cored. Then 
\mdisp{\delta^\flat_B\ =\ \varepsilon_B \, \sem{\cdot}^\bot_B\ ,} 
i.e., $\delta^\flat_\beta(s) = \varepsilon_\beta (\sem{s}^\bot_\beta)$ for
all $s \in F^\flat_\beta$, $\beta \in B$. In particular, if $R_B$ is the
trace of $(D_B,f_K)$ then
$\,R_\beta 
 \,=\, \lcurl s \in F^\flat_\beta 
        \,:\, \delta^\flat_\beta(s) = \natural_\beta \rcurl\,$
for each $\beta \in B$.
\endpro

\proof By Proposition~2.2.1 $\varepsilon_B \, \sem{\cdot}^\bot_B$
is a homomorphism from $(F^\flat_B,\fo^\flat_K)$ to
$(H_B,\diamond_K)$ and by definition
$\varepsilon_\beta (\sem{\,\flat_\beta}^\bot_\beta) = \flat_\beta$
for each $\beta \in B$. But $\delta^\flat_B$ is the unique
such homomorphism and therefore
$\delta^\flat_B = \varepsilon_B \, \sem{\cdot}^\bot_B$. \eop
\bigbreak 
Proposition~3.4.2 implies that if $(D_B,f_K)$ is 
\tsym{(H_B,\diamond_K)}-cored then the trace of $(D_B,f_K)$ is 
determined by $(H_B,\diamond_K)$. In fact, more explicitly, the following holds:

\proclaim{Proposition~3.4.3} There is a unique family 
$R_B \subset F^\flat_B$ satisfying the following:
\medskip\smallskip
{\parindent=25pt
\item{(1)} $\flat_\beta \notin R_\beta$ for 
each $\beta \in B$.
\medskip
\item{(2)} If $\kappa \in K$ is of type $\varnothing \to \beta$ for some
$\beta \in B$ then $\fo^\flat_\kappa(\onept) \in R_\beta$.
\medskip
\item{(3)} If $\kappa \in K$ is of type $L \to \beta$ 
with $L \ne \varnothing$ and $c \in \assb L {F^\flat} $ then
$\fo^\flat_\kappa(c) \in R_\beta$ if and only if
$\diamond_\kappa(b) = \natural_\beta$, where $b \in \ass L H $ is given by
\mdisp{b(\eta)
   \ =\ \cases{\natural_{\beta_j} & if $\,c(\eta) \in R_\eta \,$,\cr
                    \flat_\eta  & otherwise.
}}
}
\medskip\smallskip
Moreover, $R_B$ is the trace of each \tsym{(H_B,\diamond_K)}-cored
bottomed \tsym{\Lambda}-algebra. \endpro

\proof For each $\beta \in B$ let 
$R_\beta = \lcurl s \in F^\flat_\beta 
        \,:\, \delta^\flat_\beta(s) = \natural_\beta \rcurl$;
then the definition of the family $\delta^\flat_B$ implies that
$R_B$ satisfies (1), (2) and (3). Moreover, by
Proposition~3.4.2 $R_B$ is the trace of each \tsym{(H_B,\diamond_K)}-cored
bottomed \tsym{\Lambda}-algebra. It thus remains to show that $R_B$ is the unique
family satisfying these conditions and this easily follows from the uniqueness
of the homomorphism $\delta^\flat_B$. \eop
\bigbreak
Proposition~3.4.3 can be used as the basis for an explicit algorithm
for determining whether a given element of $F^\flat_\beta$ is in $R_\beta$ or not.
\medskip
The core types which occur in practice 
(such as $(H_B,\diamond^\bot_K)$ and $(H_B,\diamond^\top_K)$)
all have the additional property of being monotone:
Let $(H_B,\diamond_K)$ be a core type
and for each $\beta \in B$ let $\le_\beta$ be the partial order
defined on the set $H_\beta = \{\flat_\beta,\natural_\beta\}$ by 
requiring that $\flat_\beta \le_\beta \natural_\beta$.
Let $\kappa \in K$ be of type $L \to \beta$; then $\diamond_\kappa : \ass L H \to H_\beta$ 
is {\it monotone\/} if 
$\diamond_\kappa(b) \le_\beta \diamond_\kappa(b')$
whenever $b,\,b' \in \ass L H $ are such that $b \le^L b'$,
where $b \le^L b'$ if and only if $b(\eta) \le_\eta b'(\eta)$
for each $\eta \in L$. The core type
$(H_B,\diamond_K)$ is then said to be 
\indexddef{monotone}{monotone core type}{}{core type}{monotone}
if $\diamond_\kappa$ is monotone for each $\kappa \in K$.
In particular the core types $(H_B,\diamond^\bot_K)$
and $(H_B,\diamond^\top_K)$ are monotone. 
\medskip
In the next section we introduce the second of the two conditions to be imposed on the 
bottomed extensions (in addition to regularity). This is called monotonicity 
and it plays a crucial
role in Chapter~4. The following result shows that if $(D_B,f_K)$ is 
\tsym{(H_B,\diamond_K)}-cored for some monotone core type 
$(H_B,\diamond_K)$ then $(D_B,f_K)$ has a property called structural monotonicity
and, as will be seen in Section~3.5, it turns out that this property is stronger
than that of being monotone.
\medskip  

A bottomed \tsym{\Lambda}-algebra $(D_B,f_K)$ is called
\indexddef{structurally monotone}{structurally monotone bottomed algebra}{}
{bottomed algebra}{structurally monotone}
if whenever $(Y_B,q_K)$ is a \tsym{\Lambda}-algebra 
and $\pi_B$ and $\pi'_B$ are homomorphisms from 
$(Y_B,q_K)$ to $(D_B,f_K)$ then the family $Y'_B$ given by

\mdisp{Y'_\beta \ =\ \lcurl y \in Y_\beta \,:\,
  \,\pi'_\beta(y) = \bot_\beta
   \frm{whenever} \pi_\beta(y) = \bot_\beta \,\rcurl}

for each $\beta \in B$ is invariant in $(Y_B,q_K)$.

\proclaim{Proposition~3.4.4} Suppose that $(D_B,f_K)$ is 
\tsym{(H_B,\diamond_K)}-cored for some monotone core type 
$(H_B,\diamond_K)$. Then $(D_B,f_K)$ is structurally monotone. 
\endpro

\proof Let $(Y_B,q_K)$ be a \tsym{\Lambda}-algebra, let 
$\pi_B$ and $\pi'_B$ be homomorphisms from $(Y_B,q_K)$ to $(D_B,f_K)$, and
define a family $Y'_B$ with $Y'_B \subset Y_B$ by letting

\mdisp{Y'_\beta \ =\ \lcurl y \in Y_\beta \,:\,
  \,\pi'_\beta(y) = \bot_\beta
   \frm{whenever} \pi_\beta(y) = \bot_\beta \,\rcurl}

for each $\beta \in B$. 
Note that if $u \in D_\beta$ then $u = \bot_\beta$ if and only if 
$\varepsilon_\beta(u) = \flat_\beta$, thus

\mdisp{ Y'_\beta \ =\ \lcurl y \in Y_\beta \,:\,
  \varepsilon_\beta(\pi'_\beta(y))
       \le_\beta \varepsilon_\beta(\pi_\beta(y)) \,\rcurl\ .}

If $\kappa \in K$ is of type $\varnothing \to \beta$ then
$\pi_\beta(q_\kappa(\onept)) = f_\kappa(\onept) \ne \bot_\beta$, and so 
$q_\kappa(\onept)$ is trivially an element of $Y'_\beta$. 
Thus consider $\kappa \in K$ of type $L \to \beta$ with 
$L \ne \varnothing$ and let $c \in \ass L Y $ with
$c(\eta) \in Y'_\eta$ for each $\eta \in L$. Then
$\varepsilon_\eta(\pi'_\eta(c(\eta))) \le_\eta \varepsilon_\eta(\pi_\eta(c(\eta)))$
for each $\eta \in L$, which implies that
\smallskip
\ldisp{\quad \varepsilon_\beta(\pi'_\beta(q_\kappa(c)))
 \  =\ \varepsilon_\beta(f_\kappa(\assb L {\pi'} (c)))
  \ =\ \diamond_\kappa(\ass L {\varepsilon} (\assb L {\pi'} (c)))}
\vskip-\medskipamount
\rdisp{
  \ \le_\beta\ \diamond_\kappa(\ass L {\varepsilon} (\ass L {\pi} (c)))
 \  =\ \varepsilon_\beta(f_\kappa(\ass L {\pi} (c)))
\  =\ \varepsilon_\beta(\pi_\beta(q_\kappa(c))) \ ,\quad}
\smallskip
and therefore $q_\kappa(c) \in Y'_\beta$. This shows the family 
$Y'_B$ is invariant in $(Y_B,q_K)$ and thus that $(D_B,f_K)$ is structurally monotone.
\eop
\bigbreak
Proposition~3.4.4 implies in particular that the flat extension of $(X_B,p_K)$
as well as any fully regular bottomed \tsym{\Lambda}-algebra is structurally 
monotone.
\medbreak

Besides $(H_B,\diamond^\top_K)$ and $(H_B,\diamond^\bot_K)$ there is a 
further core type which should be mentioned.
If $\kappa \in K$ is of type $L \to \beta$ with $L \ne \varnothing$ then let
$\diamond^c_\kappa : \ass L H \to H_\beta$
be given by
\smallskip
\mdisp{\diamond^c_\kappa(b)
\ =\ \cases{
\natural_\beta  & if $\,b(\eta) = \natural_\eta \,$
  for at least one $\,\eta \in L\,$, \cr
 \flat_\beta & otherwise, }}
\smallskip
and note that $\diamond^c_\kappa$ is monotone. A type $\beta \in B$ is called a 
\indexdef{product type}{product type}{}
if $K_\beta$ consists of exactly one
constructor name $\kappa$ and $\kappa$ is of type
$L \to \beta$ with $L$ containing at least two elements and $\langle\eta\rangle \ne \beta$
for each $\eta \in L$. (In the signature $\Lambda$ in Example~2.2.1, for instance,
$\ftt{pair}$ is a product type.) Now define a core type
$(H_B,\diamond'_K)$ by letting
\smallskip
\mdisp{ \diamond'_\kappa\ =\ \cases{
   \diamond^c_\kappa & if $\,\kappa\,$ is the single constructor
                 for some product type, \cr
         \noalign{\smallskip}
         \diamond^\top_\kappa & otherwise. \cr}}
\smallskip
Then $(H_B,\diamond'_K)$ is a monotone core type which is sometimes used
instead of $(H_B,\diamond^\top_K)$, and the reason for
perhaps preferring $(H_B,\diamond'_K)$ to $(H_B,\diamond^\top_K)$ will now be
explained.
Let $\beta \in B$ be a product type with
$K_\beta = \{\kappa\}$ and with $\kappa$ of type $L \to \beta$.
Then by Proposition~2.3.2 the mapping $p_\kappa : \ass L X \to X_\beta$
is a bijection, and in most cases $X_\beta$ will just taken to be $\ass L X $
(and hence the name {\it product type}). 
For simplicity suppose that
this has been done. Consider any regular extension $(D_B,f_K)$ of
$(X_B,p_K)$ with core $C_K$, and note that $X_\beta = \ass L X  \subset C_\kappa$.
Then by Lemma~3.3.1 
$f_\kappa$ maps $C_\kappa$ bijectively onto $D_\beta \setminus \{\bot_\beta\}$
and thus, assuming $\bot_\beta \notin C_\kappa$, 
$D_\beta$ can be identified with $C_\kappa \cup \{\bot_\beta\}$.
Now if $(D_B,f_K)$ is \tsym{(H_B,\diamond^\top_K)}-cored then 
$C_\kappa = \dom(f_\kappa) = \ass L D $, which means that
$\,D_\beta \, =\, \ass L D \,\cup \,\{\bot_\beta\}\,$.
On the other hand, if
$(D_B,f_K)$ is \tsym{(H_B,\diamond'_K)}-cored then 
\mdisp{C_\kappa\ =\ \ass L D \,\setminus\, \{\bot^{\!L}\}} 
and hence $\,D_\beta \,=\, \ass L D \,$, provided 
$\bot^{\!L}$ is identified with $\bot_\beta$. A reason for preferring
$(H_B,\diamond'_K)$ to $(H_B,\diamond^\bot_K)$ is 
thus to obtain a more `natural' bottomed product in this case. 
\bigskip

{\it Proof of Proposition~3.4.1\ }\ It is easily checked that a bottomed 
\tsym{\Lambda}-algebra which is isomorphic to a \tsym{(H_B,\diamond_K)}-cored 
bottomed \tsym{\Lambda}-algebra is itself \tsym{(H_B,\diamond_K)}-cored. Thus
by Propositions 3.3.4 and 3.4.3 it is enough to show that there exists a minimal 
regular \tsym{(H_B,\diamond_K)}-cored extension of $(X_B,p_K)$. 
\medskip
To establish this start with any initial bottomed extension $({\breve D}_B,{\breve f}_K)$  
of $(X_B,p_K)$ (whose  existence is guaranteed by Proposition~3.2.7)
and let $\delta_B : ({\breve D}_B,{\breve f}_K) \to (H_B,\diamond_K)$ 
be the unique homomorphism with 
$\delta_\beta(\bot_\beta) = \,\flat_\beta$ for each $\beta \in B$.
For each $\beta \in B$ put
\mdisp{D'_\beta \ =\ \lcurl u \in {\breve D}_\beta \,:\, 
           \delta_\beta(u) = \varepsilon_\beta(u) \rcurl\ ;}
thus
$D'_\beta = \lcurl u \in {\breve D}_\beta \,:\, 
 \delta_\beta(u) = \natural_\beta \rcurl
 \cup \{\bot_\beta\}$, and in particular $X_\beta \subset D'_\beta$. 
For each $\kappa \in K$ of type $L \to \beta$ define a mapping 
$f'_\kappa : \ass L {\breve D} \to {\breve D}_\beta$ by

\mdisp{f'_\kappa(c)\ =\ \cases{
      {\breve f}_\kappa(c) & if $\,{\breve f}_\kappa(c) \in D'_\beta\,$, \cr
       \bot_\beta & otherwise. }}
If $c \in \ass L X $ then
${\breve f}_\kappa(c) = p_\kappa(c) \in X_\beta \subset D'_\beta$ and therefore
$f'_\kappa(c) = {\breve f}_\kappa(c) = p_\kappa(c)$, and thus
$({\breve D}_B,f'_K)$ is a \tsym{\Lambda}-algebra which is a bottomed extension
of $(X_B,p_K)$. Moreover, 
$\varepsilon_\beta(f'_\kappa(c)) = \delta_\beta({\breve f}_\kappa(c))$ for all
$c \in \ass L {\breve D} $ whenever $\kappa \in K$ is of type $L \to \beta$:
If ${\breve f}_\kappa(c) \in D'_\beta$ then 
$f'_\kappa(c) = {\breve f}_\kappa(c)$ and
so 
$\varepsilon_\beta(f'_\kappa(c)) = \varepsilon_\beta({\breve f}_\kappa(c))
 = \delta_\beta({\breve f}_\kappa(c))$; 
on the other hand, if
${\breve f}_\kappa(c) \notin D'_\beta$ 
then $\delta_\beta({\breve f}_\kappa(c)) = \,\flat_\beta$ and hence
$\delta_\beta({\breve f}_\kappa(c)) = \varepsilon_\beta(\bot_\beta)
= \varepsilon_\beta(f'_\beta(v))$.
\medskip
Let $C_K$ be the core of $({\breve D}_B,f'_K)$; then
the restriction of $f'_\kappa$ to $C_\kappa$ is injective for each
$\kappa \in K$ and $f'_{\kappa_1}(C_{\kappa_1})$ and 
$f'_{\kappa_2}(C_{\kappa_2})$ are disjoint subsets of ${\breve D}_\beta$ 
whenever $\kappa_1,\, \kappa_2 \in K_\beta$ with $\kappa_1 \ne \kappa_2$
(since by Lemma~3.3.1 the corresponding statements hold for the family
${\breve f}_K$ if $C_\kappa$ is replaced by $\dom({\breve f}_\kappa)$ for each 
$\kappa \in K$).
In other words, conditions (1) and (2) in Lemma~3.3.1 hold for
$({\breve D}_B,f'_K)$.
Now let $({\hat D}_B,{\hat f}_K)$ be the minimal bottomed subalgebra
of $({\breve D}_B,f'_K)$, so $({\hat D}_B,{\hat f}_K)$ is a minimal bottomed 
extension of $(X_B,p_K)$.
But if ${\hat C}_K$ is the core of 
$({\hat D}_B,{\hat f}_K)$ then ${\hat C}_\kappa \subset C_\kappa$ for
each $\kappa \in K$ and hence conditions (1) and (2)
in Lemma~3.3.1 also hold for $({\hat D}_B,{\hat f}_K)$.
Therefore by Lemma~3.3.1 and Proposition~3.2.2 $({\hat D}_B,{\hat f}_K)$ is regular.
\medskip
Note that ${\hat D}_B \subset D'_B$, since
$D'_B$ is invariant (in $({\breve D}_B,f'_K)$) 
and $\bot_\beta \in D'_\beta$ for each
$\beta \in B$, and hence $\delta_\beta(u) = \varepsilon_\beta(u)$
for all $u \in {\hat D}_\beta$, $\beta \in B$.
Let $\kappa \in K$ be of type $L \to \beta$ and
$c \in \ass L {\hat D} $. Then

\mdisp{
\diamond_\kappa(\ass L {\varepsilon} (c))
     \ =\ \diamond_\kappa(\ass L {\delta} (c))
\ =\ \delta_\beta({\breve f}_\kappa(c))
\ =\ \varepsilon_\beta(f'_\kappa(c))
\ =\ \varepsilon_\beta({\hat f}_\kappa(c)) }

and this implies that
$({\hat D}_B,{\hat f}_K)$ is \tsym{(H_B,\diamond_K)}-cored, i.e.,
$({\hat D}_B,{\hat f}_K)$ is a minimal regular \tsym{(H_B,\diamond_K)}-cored
extension of $(X_B,p_K)$. \eop

\vfill\eject

\hfline{75}{3.5 \ MONOTONE BOTTOMED ALGEBRAS}

\sectionhead {3.5} {Monotone bottomed algebras}
\bigskip\medskip

We now introduce the second of the two conditions which the bottomed 
extensions will be required to satisfy.
This is called {\it monotonicity\/} and, although its definition is somewhat 
technical, it is a simple matter to show (as will be done in Proposition~3.5.2) that 
it is implied by structural monotonicity (as defined in the previous section).
Thus by Proposition~3.4.4 the flat extension of $(X_B,p_K)$ and any fully regular 
bottomed \tsym{\Lambda}-algebra is monotone. 
\medskip
Let $(D_B,f_K)$ be a bottomed \tsym{\Lambda}-algebra. 
If $J$ is a \tsym{B}-typed set, 
$(Z_B,r_K)$ a \tsym{J}-free \tsym{\Lambda}-algebra and $b \in \ass J D $ 
then
$\pi^b_B$ will denote the unique homomorphism from $(Z_B,r_K)$ to $(D_B,f_K)$ such that 
$\pi^b_\eta(\eta) = b(\eta)$ for all $\eta \in J$.   $(D_B,f_K)$ is said to be 
\indexddef{monotone}{monotone bottomed algebra}{}{bottomed algebra}{monotone}
if whenever $J$ is a \tsym{B}-typed set and $(Z_B,r_K)$ 
is a \tsym{J}-free \tsym{\Lambda}-algebra 
and $\pi^{\bot^{\!J}}_\beta(z) \ne \bot_\beta$ for some $z \in Z_\beta$ then
$\pi^b_\beta(z) \ne \bot_\beta$ for all $b \in \ass J D $. 
Here  $\bot^{\!J}$ is the bottom element of $\ass J D $ given by 
$\bot^{\!J}(\eta) = \bot_\eta$ for each $\eta \in J$.
\medskip
A bottomed extension $(D_B,f_K)$ of $(X_B,p_K)$ is now said to be
\indexddef{monotone}{monotone bottomed extension}{}{bottomed extension}{monotone}
if $(D_B,f_K)$ is a monotone bottomed \tsym{\Lambda}-algebra.
\medskip
Monotonicity as defined above is not particularly easy to work with, and so
an equivalent condition (to be called \definition{strong monotonicity})
is now presented which turns out to be more tractable.
\medskip
Let $(D_B,f_K)$ be a bottomed \tsym{\Lambda}-algebra, $(Y_B,q_K)$ be any 
\tsym{\Lambda}-algebra and let 
$\pi_B$ be a homomorphism from $(Y_B,q_K)$ to $(D_B,f_K)$. 
We call the family $U_B \subset Y_B$ with 
\mdisp{U_\beta 
 \ =\ \lcurl y \in Y_\beta \,:\, \pi_\beta(y) = \bot_\beta \rcurl} 

for each $\beta \in B$ the 
\indexdef{kernel}{kernel of a homomorphism}{}
of $\pi_B$, and then say that $\pi_B$ is 
\indexddef{fat bottomed}{fat bottomed homomorphism}{}{homomorphism}{fat bottomed}
if $Y_B$ is the only invariant family containing $U_B$.
\medskip
Now $(D_B,f_K)$ is said to be 
\indexddef{strongly monotone}{strongly monotone bottomed algebra}{}
{bottomed algebra}{strongly monotone}
if whenever $(Y_B,q_K)$ is a \tsym{\Lambda}-algebra 
and ${\breve \pi}_B : (Y_B,q_K) \to (D_B,f_K)$ a fat bottomed homomorphism 
then the kernel ${\breve U}_B$ of
${\breve \pi}_B$ is maximal in the sense that if
$\pi_B : (Y_B,q_K) \to (D_B,f_K)$ is any homomorphism 
with kernel $U_B$ then $U_B \subset {\breve U}_B$.
\medskip

\proclaim{Proposition~3.5.1} A strongly monotone bottomed \tsym{\Lambda}-algebra  
$(D_B,f_K)$ is monotone. \endpro

\proof Let $J$ be a \tsym{B}-typed set and $(Z_B,r_K)$ a \tsym{J}-free 
\tsym{\Lambda}-algebra; denote the kernel of the homomorphism 
$\pi^{\bot^{\!J}}_B : (Z_B,r_K) \to (D_B,f_K)$ by ${\grave U}_B$.
Then $\pi^{\bot^{\!J}}_B$ is fat bottomed, since
$J_\beta = \lcurl \eta \in J : \langle \eta \rangle 
                                   = \beta \rcurl \subset {\grave U}_\beta$
for each $\beta \in B$, and by Proposition~2.3.7 $(Z_B,r_K)$ is \tsym{J}-minimal. 
Thus if $b \in \ass J D $ and $U_B$ is the kernel of $\pi^b_B$ then
$U_B \subset {\grave U}_B$. But this just means that
$\pi^{\bot^{\!J}}_\beta(z) = \bot_\beta$ whenever $\pi^b_\beta(z) = \bot_\beta$, which
shows that $(D_B,f_K)$ is monotone. \eop
\bigbreak
Later it will be seen that the converse of Proposition~3.5.1 holds, i.e., that
a monotone bottomed \tsym{\Lambda}-algebra is also strongly monotone.
\medskip
Recall that a bottomed \tsym{\Lambda}-algebra $(D_B,f_K)$ is said to be
structurally monotone if whenever $(Y_B,q_K)$ is a \tsym{\Lambda}-algebra 
and $\pi_B$ and $\pi'_B$ are homomorphisms from 
$(Y_B,q_K)$ to $(D_B,f_K)$ then the family $Y'_B$ given by

\mdisp{Y'_\beta \ =\ \lcurl y \in Y_\beta \,:\,
  \,\pi'_\beta(y) = \bot_\beta
   \frm{whenever} \pi_\beta(y) = \bot_\beta \,\rcurl}

for each $\beta \in B$ is invariant in $(Y_B,q_K)$.

\proclaim{Proposition~3.5.2} 
If $(D_B,f_K)$ is structurally monotone  then it 
is strongly monotone (and thus also monotone). \endpro

\proof Let $(Y_B,q_K)$ be a \tsym{\Lambda}-algebra, let ${\breve \pi}_B$ be a fat 
bottomed and $\pi_B$ an arbitrary homomorphism from $(Y_B,q_K)$ to $(D_B,f_K)$. Put

\mdisp{Y'_\beta \ =\ \lcurl y \in Y_\beta \,:\,
  \,{\breve \pi}_\beta(y) = \bot_\beta
   \frm{whenever} \pi_\beta(y) = \bot_\beta \,\rcurl}

for each $\beta \in B$. Then $Y'_B$ is invariant, since $(D_B,f_K)$ is structurally
monotone. But if ${\breve U}_B$ is the kernel of ${\breve \pi}_B$ then 
clearly ${\breve U}_B \subset Y'_B$,
and hence $Y'_B = Y_B$, since ${\breve \pi}_B$ is fat bottomed. 
Therefore if $U_B$ is the kernel of $\pi_B$ then 
$U_B \subset {\breve U}_B$. This shows that $(D_B,f_K)$ is strongly monotone. \eop
\bigbreak

Propositions 3.4.4 and 3.5.2 imply that if $(H_B,\diamond_K)$ is a monotone core type
then any \tsym{(H_B,\diamond_K)}-cored bottomed \tsym{\Lambda}-algebra 
is monotone. 
In particular, this means the flat extension of $(X_B,p_K)$
as well as any fully regular bottomed \tsym{\Lambda}-algebra is monotone.
\medskip
On the following page two simple examples are presented which show 
that regularity and monotonicity are independent concepts, i.e., they 
give a regular extension which is not monotone and a monotone extension which is not 
regular. 

\midinsert
\bigskip\bigskip
\frame{20pt}{\bigskip 
{\it Example~3.5.1\enspace} Let
$(X^\bot_B,p^\bot_K)$ be the flat extension
of the \tsym{\Lambda}-algebra $(X_B,p_K)$ from Example~2.2.1 and
define \tsym{\Lambda}-algebras $(Y_B,q_K)$ and $(Y_B,q'_K)$ by letting
\medskip\smallskip
{\leftskip=32pt
$Y_{\fstt{nat}} \,=\, \Nat \cup \{\bot_{\fstt{nat}},\bot^o_{\fstt{nat}}\}$\ with
$\bot^o_{\fstt{nat}} \notin \Nat \cup \{\bot_{\fstt{nat}}\}$,
\smallskip
$Y_\beta = X^\bot_\beta$\ for all 
$\beta \in B \setminus \{\ftt{nat}\}$,
\medskip
$q_{\fstt{Zero}},\,q'_{\fstt{Zero}} : 
   \Oneptset \to Y_{\fstt{nat}}$\ with\ %
                       $q_{\fstt{Zero}}(\onept) \,=\, q'_{\fstt{Zero}}(\onept) \,=\, 0$,
\smallskip
$q_{\fstt{Succ}} : Y_{\fstt{nat}} \to Y_{\fstt{nat}}$
\ with\ %
\mdisp{q_{\fstt{Succ}}(n)\ =\ \cases{ n + 1 & 
                           if $\,n \in \Nat\,$, \cr
            \bot_{\fstt{nat}} & if $\,n = \bot^o_{\fstt{nat}}\,$, \cr
               \bot^o_{\fstt{nat}}    & if $\,n = \bot_{\fstt{nat}}\,$, \cr
                 }} 
\smallskip
$q'_{\fstt{Succ}} : Y_{\fstt{nat}} \to Y_{\fstt{nat}}$
\ with\ %
\mdisp{q'_{\fstt{Succ}}(n)\ =\ \cases{ n + 1 & 
                           if $\,n \in \Nat\,$, \cr
            \bot^o_{\fstt{nat}} & if 
  $\,n \in \{\bot_{\fstt{nat}},\bot^o_{\fstt{nat}}\}\,$, \cr
                 }} 
\smallskip
and with $\,q_\kappa \,=\, q'_\kappa \,=\, p^\bot_\kappa\,$ for all 
$\kappa \in K \setminus \{\ftt{Zero},\ftt{Succ}\}$.
\medskip\smallskip
}
Then $(Y_B,q_K)$ is clearly a minimal regular
extension of $(X_B,p_K)$. However, 
$(Y_B,q_K)$ is not monotone. To see this consider a 
\tsym{B}-typed set $J = \{\ftt{n}\}$ with $\ftt{n}$ of type $\ftt{nat}$
and let $(Z_B,r_K)$ be any \tsym{J}-free \tsym{\Lambda}-algebra. Moreover, let
$z$ be the element $r_{\fstt{Succ}}(\ftt{n})$ of $Z_{\fstt{nat}}$ and $b \in \ass J Y $ 
be such that $b(\ftt{n}) = \bot^o_{\fstt{nat}}$. Then 
$\,\pi^b_{\fstt{nat}}(z) \,=\, q_{\fstt{Succ}}(b(\ftt{n}))
\,=\, q_{\fstt{Succ}}(\bot^o_{\fstt{nat}})\, =\, \bot_{\fstt{nat}}\,$ but
\mdisp{\pi^{\bot^{\!J}}_{\fstt{nat}}(z)\ =\ q_{\fstt{Succ}}(\bot^{\!J}(\ftt{n}))
       \ =\ q_{\fstt{Succ}}(\bot_{\fstt{nat}}) 
                           \ =\ \bot^o_{\fstt{nat}}\ \ne\ \bot_{\fstt{nat}}\ .}
\medskip
On the other hand, $(Y_B,q'_K)$ is a minimal 
extension of $(X_B,p_K)$ which is not regular,
since $\,q'_{\fstt{Succ}}(\bot_{\fstt{nat}})
 \,=\, q'_{\fstt{Succ}}(\bot^o_{\fstt{nat}})
 \,=\, \bot^o_{\fstt{nat}} \,\ne\, \bot_{\fstt{nat}}\,$.
However, the reader is left to show that
$(Y_B,q'_K)$ is monotone. (This will later follow more easily from 
Proposition~4.1.6.)
\bigskip\medskip
}
\bigskip\medskip
\endinsert
\bigskip
We now begin preparing for the proof of the converse of Proposition~3.5.1.
The general construction involving free algebras given below is also needed
several times later.
\medskip 

In what follows let $(D_B,f_K)$ be a bottomed \tsym{\Lambda}-algebra and
$(Y_B,q_K)$ be an arbitrary \tsym{\Lambda}-algebra.
For each $\beta \in B$ let $\prec_\beta$ be a relation on the set
$D_\beta \times Y_\beta$, i.e., $\prec_\beta$ is just a subset of
$D_\beta \times Y_\beta$. As usual, however, infix 
notation will be employed here and 
$u \prec_\beta y$ written to mean that $(u,y)$ belongs to the relation $\prec_\beta$. 
The family of relations $\prec_B$ is said to be 
\indexddef{compatible}{compatible family of relations}{}{family of relations}{compatible} 
if whenever $\kappa \in K$ is of type $L \to \beta$ and $a \in \ass L Y $, 
$u \in D_\beta \setminus \{\bot_\beta\}$ are such that $u \prec_\beta q_\kappa(a)$ then
there exists $c \in \ass L D $ with $u = f_\kappa(c)$ such that
$c \ass L {\prec}  a$, i.e., such that
$c(\eta) \prec_\eta  a(\eta)$ for each $\eta \in L$.
\medskip
Let $\beta \in B$; then $y \in Y_\beta$ is said to be a 
\indexdef{base element}{base element}{}
if the set $\lcurl u \in D_\beta \,:\, u \prec_\beta y \rcurl$ 
contains at most the bottom element $\bot_\beta$, and 
the set of base elements of $Y_\beta$ will be denoted by $U_\beta$. A compatible
family $\prec_B$ is now said to be 
\indexddef{full}{full compatible family of relations}{}
{family of relations}{full compatible} 
if $Y_B$ is the only invariant family containing $U_B$.
\medskip
If $J$ is a \tsym{B}-typed set and $(Z_B,r_K)$ a \tsym{J}-free \tsym{\Lambda}-algebra
then for each $b \in \ass J Y $ let $\varrho^b_B$ denote the unique
homomorphism from $(Z_B,r_K)$ to $(Y_B,q_K)$ such that
$\varrho^b_\eta(\eta) = b(\eta)$ for each $\eta \in L$. Moreover, 
$\omega^J_B$ will denote the unique homomorphism
from $(Z_B,r_K)$ to $(D_B,f_K)$ such that
$\omega^J_\eta(\eta) = \bot_\eta$ for each $\eta \in L$. 
(Thus, with the notation employed above, $\omega^J_B = \pi^{\bot^{\!J}}_B$.)

\proclaim{Proposition~3.5.3} Suppose $\prec_B$ is a full compatible family.
Then for each $y \in Y_\beta$ and each $u \in D_\beta$ with $u \prec_\beta y$
there exists a  finite \tsym{B}-typed set $J$, a
\tsym{J}-free \tsym{\Lambda}-algebra $(Z_B,r_K)$, and elements 
$z \in Z_\beta$ and $b \in \ass J Y $ such that 
$\varrho^b_\beta(z) = y$ and $\omega^J_\beta(z) = u$. \endpro

\proof For each $\beta \in B$ let ${\grave Y}_\beta$ be the set of those elements 
$y \in Y_\beta$ for which the above statement holds (i.e., those $y \in Y_\beta$
such that if $u \in D_\beta$ with $u \prec_\beta y$ then there exists a  finite 
\tsym{B}-typed set $J$, a \tsym{J}-free \tsym{\Lambda}-algebra $(Z_B,r_K)$, and elements 
$z \in Z_\beta$ and $b \in \ass J Y $ such that 
$\varrho^b_\beta(z) = y$ and $\omega^J_\beta(z) = u$). 
The aim is thus to show that ${\grave Y}_B = Y_B$.
\medskip
A basic remark: For each $y \in Y_\beta$ there always
exists a  finite \tsym{B}-typed set $J$, a
\tsym{J}-free \tsym{\Lambda}-algebra $(Z_B,r_K)$ and elements 
$z \in Z_\beta$ and $b \in \ass J Y $ such that 
$\varrho^b_\beta(z) = y$ and $\omega^J_\beta(z) = \bot_\beta$. 
In fact, just let $J = \{\eta\}$ with $\eta$ of type $\beta$, let $(Z_B,r_K)$ be 
any \tsym{J}-free \tsym{\Lambda}-algebra, 
let $b \in \ass J Y $ be the assignment given by
$b(\eta) = y$ and take $z = \eta$.
\medskip
From the basic remark it immediately follows that $U_B \subset Y'_B$
and thus, since the family $\prec_B$ is full, it is enough to
show that the family $Y'_B$ is invariant in $(Y_B,q_K)$.
\medskip
It will be shown first that if $\kappa \in K$ is of type $\varnothing \to \beta$ then 
$q_\kappa(\onept) \in Y'_\beta$. Thus consider
$u \in D_\beta$ with $u \prec_\beta q_\kappa(\onept)$; 
then either $u = f_\kappa(\onept)$ or
$u = \bot_\beta$ (since the family $\prec_B$ is compatible) 
and the latter is dealt with by the basic remark.
But if $u = f_\kappa(\onept)$ then take any 
finite \tsym{B}-typed set $J$, any
\tsym{J}-free \tsym{\Lambda}-algebra $(Z_B,r_K)$, any assignment 
$b \in \ass J Y $ and put $z = r_\kappa(\onept)$; then  
$\,\varrho^b_\beta(z) = \varrho^b_\beta(r_\kappa(\onept)) = q_\kappa(\onept)\,$ and 
$\,\omega^J_\beta(z) = \omega^J_\beta(r_\kappa(\onept)) = f_\kappa(\onept) = u\,$.

\medskip 
Now let $\kappa \in K$ be of type $L \to \beta$ with $L \ne \varnothing$ and let
$a \in \assb L {Y'} $. 
The task on hand is to show that $q_\kappa(a) \in Y'_\beta$ and so, again making use of
the basic remark, it is only necessary to consider
$u \in D_\beta \setminus \{\bot_\beta\}$ with $u \prec_\beta q_\kappa(a)$. 
In this case (since the family $\prec_B$ is compatible) there exists $c \in \ass L D $
with $u = f_\kappa(c)$ such that $c \ass L {\prec} a$.
Then $a(\eta) \in Y'_\eta$ and
$c(\eta) \prec_\eta a(\eta)$ for each $\eta \in L$, 
and hence there exists a finite \tsym{B}-typed set $J_\eta$,
a \tsym{J_\eta}-free \tsym{\Lambda}-algebra $(Z^\eta_B,r^\eta_K)$
and elements $z_\eta \in Z^\eta_\eta$ and  $b_\eta \in \ass {J_\eta} Y $ with 
$\varrho^{b_\eta}_\eta(z_\eta) = a(\eta)$ and 
$\omega^{J_\eta}_\eta(z_\eta) = c(\eta)$.
Moreover, the sets $J_\eta$, $\eta \in L$, can be arranged to be disjoint
(by introducing duplicate variables if necessary).
Put $J = \bigcup_{\eta \in L} J_\eta$, considered as a \tsym{B}-typed set in the obvious
way and choose any \tsym{J}-free \tsym{\Lambda}-algebra $(Z_B,r_K)$; let 
$b \in \ass J Y $ be the sum of the assignments $b_\eta$, $\eta \in L$, i.e.,
$b$ is given by $b(\xi) = b_\eta(\xi)$ for each 
$\xi \in J_\eta$, $\eta \in L$. At this point the following somewhat technical fact
is needed:

\proclaim{Lemma~3.5.1} There exists an assignment $d \in \ass L Z $ such that
$\,\varrho^b_\eta(d(\eta)) \,=\, \varrho^{b_\eta}_\eta(z_\eta)\,$
and $\,\omega^J_\eta(d(\eta))\, =\, \omega^{J_\eta}_\eta(z_\eta)\,$
for each $\eta \in L$. \endpro
\bigbreak
\proof For each $\eta \in L$ let $\lambda^\eta_B : (Z^\eta_B,r^\eta_K) \to (Z_B,r_K)$ 
be the unique homomorphism such that $\lambda^\eta_\xi(\xi) = \xi$ for each 
$\xi \in J_\eta$ (i.e., the homomorphism corresponding to the assignment
$i_\eta \in \ass {J_\eta} Z $ given by $i_\eta(\xi) = \xi$ for each $\xi \in J_\eta$).
Define $d \in \ass L Z $ by putting $\,d(\eta) = \lambda^\eta_\eta(z_\eta)\,$ 
for each $\eta \in L$. Then it is easily checked (using the uniqueness in the definition
of $\varrho^{b_\eta}_B$) that $\,\varrho^{b_\eta}_B = \varrho^b_B\,\lambda^\eta_B\,$, 
and hence that $\,\varrho^b_\eta(d(\eta)) \,=\, \varrho^{b_\eta}_\eta(z_\eta)\,$
for each $\eta \in L$. In the same way 
$\,\omega^{J_\eta}_B = \omega^J_B\,\lambda^\eta_B\,$, thus
$\,\omega^J_\eta(d(\eta)) \,=\, \omega^{J_\eta}_\eta(z_\eta)\,$ for each
$\eta \in L$. \eop
\bigbreak
Let $d \in \ass L Z $ be the assignment in Lemma~3.5.1. Then 
$\,\varrho^b_\eta(d(\eta))\, =\, \varrho^{b_\eta}_\eta(z_\eta)\, =\, a(\eta)\,$
and $\,\omega^J_\eta(d(\eta))\, =\, \omega^{J_\eta}_\eta(z_\eta)\, =\, c(\eta)\,$
for each $\eta \in L$; in other words, 
$\,\assb L {\varrho^b} (d) = a\,$ and $\,\assb L {\omega^J} (d) = c\,$.
Now put $z = r_\kappa(d)$; then
\mdisp{ \varrho^b_\beta(z) \ =\ \varrho^b_\beta(r_\kappa(d))
\ =\ q_\kappa(\assb L {\varrho^b} (d)) \ =\  q_\kappa(a)}

and in the same way $\,\omega^J_\beta(z) = f_\kappa(c) = u\,$, i.e., 
$q_\kappa(a) \in Y'_\beta$. This shows that the family $Y'_B$ is invariant
in $(Y_B,q_K)$, which completes the proof of Proposition~3.5.3. \eop
\bigbreak
The following corollary of Proposition~3.5.3 will be needed in Section~4.1:

\proclaim{Proposition~3.5.4} Suppose $\prec_B$ is a full compatible family.
Let $\kappa \in K$ be of type $L \to \beta$
and let $c \in \ass L D $, $a \in \ass L Y $ 
with $c \ass L {\prec} a$. 
Then there exists a finite \tsym{B}-typed set $J$, a \tsym{J}-free
\tsym{\Lambda}-algebra $(Z_B,r_K)$, and elements $z \in Z_\beta$,
$b \in \ass J D $ such that $f_\kappa(c) = \omega^J_\beta(z)$ and 
$q_\kappa(a) = \varrho^b_\beta(z)$. \endpro

\proof The case $L = \varnothing$ was already dealt with in the
proof of Proposition~3.5.3, and so it can be assumed that $L \ne \varnothing$.
Then $c(\eta) \prec_\eta a(\eta)$ for each $\eta \in L$ 
and hence by Proposition~3.5.3 there exists a finite \tsym{B}-typed set 
$J_\eta$, a \tsym{J_\eta}-free \tsym{\Lambda}-algebra $(Z^\eta_B,r^\eta_K)$,
and elements $z_\eta \in Z^\eta_\eta$ and $b_\eta \in \ass {J_\eta} D $ with 
$\omega^J_\eta(z_\eta) = c(\eta)$ and $\varrho^{b_\eta}_\eta(z_\eta) = a(\eta)$. 
The proof is now exactly the same as in the proof of Proposition~3.5.3:
Again it can be supposed that
the sets $J_\eta$, $\eta \in L$, are disjoint, put $J = \bigcup_{\eta \in L} J_\eta$,
let $(Z_B,r_K)$ be any \tsym{J}-free \tsym{\Lambda}-algebra
and $b \in \ass J D $ be the sum of the assignments $b_\eta$, $\eta \in L$.
Moreover, take the same assignment $d \in \ass L Z $ given by
Lemma~3.5.1 and put $z = r_\kappa(d)$,
and as before it follows that $f_\kappa(c) = \omega^J_\beta(z)$ and 
$q_\kappa(a) = \varrho^b_\beta(z)$. \eop
\bigbreak

The next result is a special case of Proposition~3.5.3 and is the key to
proving the converse of Proposition~3.5.1.

\proclaim{Lemma~3.5.2} Let ${\breve \pi}_B$ and $\pi_B$ be homomorphisms from 
$(Y_B,q_K)$ to $(D_B,f_K)$ with ${\breve \pi}_B$ fat bottomed.
Then for each $y \in Y_\beta$ there exists a  finite \tsym{B}-typed set $J$, a
\tsym{J}-free \tsym{\Lambda}-algebra $(Z_B,r_K)$, and elements 
$z \in Z_\beta$ and $b \in \ass J D $ such that 
$\pi_\beta(y) = \pi^b_\beta(z)$ 
and ${\breve \pi}_\beta(y) = \pi^{\bot^{\!J}}_\beta(z)$. \endpro

\proof For each $\beta \in B$ define a relation $\prec_\beta$ on
$D_\beta \times Y_\beta$ by stipulating that $u \prec_\beta y$ if and only if
$u = {\breve \pi}_\beta(y)$.
Consider $\kappa \in K$ of type $L \to \beta$ and let $a \in \ass L Y $, 
$u \in D_\beta$ be such that $u \prec_\beta q_\kappa(a)$.
Then $u = {\breve \pi}_\beta(q_\kappa(a)) = f_\kappa(c)$ with
$c = \ass L {\breve \pi} (a)$, and so
$c(\eta) = {\breve \pi}_\eta(a(\eta))$ for each $\eta \in L$, i.e.,
$c \ass L {\prec}  a$, which means
the family of relations $\prec_B$ is compatible.
Moreover, $y \in Y_\beta$ is a base element if and only if 
${\breve \pi}_\beta(y) = \bot_\beta$, and hence 
the set of base elements of $Y_\beta$ is just ${\breve U}_\beta$, where
${\breve U}_B$ is the kernel of ${\breve \pi}_B$. This implies that
$\prec_B$ is full, since ${\breve \pi}_B$ is fat bottomed.
Let $y \in Y_\beta$; then by Proposition~3.5.3 there exists
a  finite \tsym{B}-typed set $J$, a
\tsym{J}-free \tsym{\Lambda}-algebra $(Z_B,r_K)$, and elements 
$z \in Z_\beta$ and $b' \in \ass J Y $ such that 
$\varrho^{b'}_\beta(z) = y$ and 
$\pi^{\bot^{\!J}}_\beta(z) = \omega^J_\beta(z) = {\breve \pi}_\beta(y)$. 
Now let $b \in \ass J D $ be the assignment given by
$b(\eta) = \pi_\eta(b'(\eta))$ for each $\eta \in L$.
Then clearly $\pi^b_B = \pi_B \, \varrho^{b'}_B$, and therefore
$\pi^b_\beta(z) = \pi_\beta(\varrho^{b'}_\beta(z)) = \pi_\beta(y)$. \eop
\bigbreak

\proclaim{Proposition~3.5.5} A monotone bottomed \tsym{\Lambda}-algebra $(D_B,f_K)$ is 
strongly monotone. \endpro

\proof This follows immediately from Lemma~3.5.2. \eop
\bigbreak

\bigbreak
The proof of Proposition~3.5.5 shows that in the definition of being monotone it would 
not change anything if the \tsym{B}-typed sets were restricted to being finite. 

\proclaim{Proposition~3.5.6} A bottomed \tsym{\Lambda}-algebra possessing
a monotone extension is itself monotone. \endpro

\proof Note that if $\pi_B : (Y_B,q_K) \to (D_B,f_K)$ is a homomorphism and
$(D'_B,f'_K)$ is an extension of $(D_B,f_K)$, and if $\pi_B$, 
considered as a homomorphism from $(Y_B,q_K)$ to $(D'_B,f'_K)$, is denoted by $\pi'_B$
then $\pi'_B$ and $\pi_B$ have the same kernel (and so in particular
$\pi'_B$ is fat bottomed if and only if $\pi_B$ is). The result thus
follows from the equivalence of monotonicity and strong monotonicity. \eop
\bigbreak

\bigskip\bigskip\bigskip\bigskip
\sectionhead {3.6} {Notes}
\bigskip\medskip

The form of the basic data objects (i.e., given by the family of sets
in an initial algebra) corresponds to that occurring in the
functional programming languages this study is trying to explain. It could be 
objected that what is also required are data types obtained through
equivalence relations defined on initial algebras. For example, this is 
necessary in order to implement the rational numbers. However, the 
implementation of such data types has been left to the user,
who must make do with data types without equivalence relations 
and then check that all the functions he or she defines actually
respect the equivalence relation specifying the intended data type.
\medskip
The concept of a bottomed extension has been introduced to give a 
framework which covers both `lazy' and `eager' programming languages 
(and also everything between these two extremes). The condition of 
regularity is exactly what is needed for the `case' operators to make sense.
\medskip
David Turner's language {\it Miranda\/} 
(see Turner (1985)) uses the monotone core type $(H_B,\diamond'_K)$ 
introduced in Section~3.4, whereas {\it Haskell\/} and most other
`lazy' languages use $(H_B,\diamond^\top_K)$. 
The expression {\it fat bottomed\/} occurring in the definition of
monotonicity is taken of course from Queen (1978).

\vfill\eject

\hfline{81}{}

{\fourteenbf Chapter 4\quad Completion of bottomed extensions}
\bigskip\medskip
In Chapter~3 a framework was introduced for specifying data objects. 
This involved starting with a ground signature $\Lambda$ (whose choice 
depends on the application at hand) and then fixing an initial \tsym{\Lambda}-algebra
$(X_B,p_K)$ to describe the basic data objects. A bottomed extension
$(D_B,f_K)$ of $(X_B,f_K)$ was then chosen in order to specify which 
`undefined' and `partially defined' data objects are to be allowed. 
From now on it will always be assumed that $(D_B,f_K)$ is monotone and
regular (for reasons which should become more apparent in the present and the following 
chapter).
\medskip
Now it is often convenient, and even necessary, to also consider  `infinite' 
data objects. Such an object can be thought of
as representing the limit of some infinite process (for example,
a process which produces a new data element at each step), and so they will be
typically involved in the description of calculations which continue to produce
new data without ever terminating. Moreover, there are also purely mathematical
reasons for including a `correct' set of limiting objects, since this results 
in a set-up which is easier to deal with (in the same way that real analysis,
i.e., analysis done over the real numbers $\Bbb{R}$, is usually much simpler than rational 
analysis, i.e., analysis done over the rational numbers $\Bbb{Q}$).

\medskip
The `infinite' data objects which will occur in this study are described with 
the help of complete posets (i.e., complete partially ordered sets); these are  
studied in Section~4.2. The complete posets which will be used arise as initial (or 
ideal) completions,
and so in Section~4.3 it is shown how to construct the initial completion of a
arbitrary poset.
\medskip
Recall that a 
\indexdef{partially ordered set}{partially ordered set}{}
(or \indexdef{poset}{poset}{}) 
is a pair 
$(Y,\sqle)$ consisting of a 
non-empty set $Y$ and a 
\indexdef{partial order}{partial order}{} 
$\sqle$ on $Y$, i.e., $\sqle$ is a 
binary relation $\sqle$ on $Y$ satisfying:
\medskip\smallskip
{\parindent=25pt
\item{(1)} $y \sqle y$ for all $y \in Y$.
\medskip
\item{(2)} If $y_1 \sqle y_2$ and $y_2 \sqle y_1$ then $y_1 = y_2$.
\medskip
\item{(3)} If $y_1 \sqle y_2$ and $y_2 \sqle y_3$ then $y_1 \sqle y_3$. 
\medskip\smallskip
}  
(In other words, the relation $\sqle$ is reflexive, anti-symmetric and 
transitive.) It is usual, however, to just write $Y$ instead of $(Y,\sqle)$ and 
to assume $\sqle$ can be determined from the context. 
\medskip
How posets come into the picture at all is explained in Section~4.1. 
There it is shown that if $(D_B,f_K)$ is a regular bottomed \tsym{\Lambda}-algebra 
then for each $\beta \in B$ there exists a partial order 
$\sqle_\beta$ on the set $D_\beta$ such that $u \sqle_\beta u'$ can be 
sensibly interpreted as meaning that $u$ is `less-defined' than $u'$. More 
precisely, a family of partial orders 
$\sqle_B$ is said to be  an ordering associated with 
$(D_B,f_K)$ if the following two conditions hold:
\medskip\smallskip
{\parindent=25pt
\item{(1)} $D_\beta$ is a bottomed poset with bottom element $\bot_\beta$,
i.e., with $\bot_\beta \sqle_\beta u$ for each $u \in D_\beta$.
\medskip
\item{(2)} If $\kappa, \, \kappa' \in K_\beta$ and $c \in C_{\kappa}$, 
$c' \in C_{\kappa'}$ then $f_\kappa(c) \sqle_\beta f_{\kappa'}(c')$ if and 
only if $\kappa = \kappa'$ and $c \sqle_\kappa c'$.
\medskip\smallskip
}
Here $C_K$ is the core of $(D_B,f_K)$ and $\sqle_\kappa$ is the 
appropriate `product' partial order on $\dom(f_\kappa)$. (More precisely, if
$\kappa$ is of type $L \to \beta$ then $\sqle_\kappa$ is just the partial order
$\ass L {\sqle} $ on $\ass L D $.)
In Proposition~4.1.2 it is shown that there exists an ordering associated with 
$(D_B,f_K)$. Moreover, if $(D_B,f_K)$ is also minimal 
then it turns out that this associated ordering is unique. 
\medskip
An associated ordering $\sqle_B$ is said to be monotone if the mappings 
$f_\kappa$, $\kappa \in K$, are monotone with respect to $\sqle_B$, 
i.e., if $\kappa \in K$ is of type $L \to \beta$ and $c, \, c' \in \ass L D $ with 
$c \sqle_\kappa c'$ then $f_\kappa(c) \sqle_\beta f_\kappa(c')$. 
Now, although it might be expected that an associated ordering is automatically
monotone, this is not the case.
However, Proposition~4.1.5 states that if $(D_B,f_K)$ is a minimal monotone 
regular bottomed \tsym{\Lambda}-algebra then the (unique) associated ordering 
$\sqle_B$ is monotone.
\medskip
The results of Sections 4.2 and 4.3 are applied in Section~4.4, where the
initial completion of a \tsym{\Lambda}-algebra is defined. More precisely,
if $(Y_B,q_K)$ is a monotone regular extension of $(X_B,p_K)$ and $\sqle_B$ is an 
associated ordering which is monotone then $Y_\beta$ is regarded as a poset with 
the partial order $\sqle_\beta$ for each $\beta \in B$.
The initial completion of $(Y_B,q_K)$ is then a \tsym{\Lambda}-algebra
$(D_B,f_K)$ in which $D_\beta$ is an initial 
completion of the poset $Y_\beta$ for each $\beta \in B$ and such that
$f_\kappa$ is 
the unique continuous extension of $q_\kappa$ for each $\kappa \in K$. 
Proposition~4.3.2 states that an initial completion $(D_B,f_K)$ exists, and 
Proposition~4.4.1 implies it is, in the appropriate sense, unique. Moreover, by 
Proposition~4.4.2 $(D_B,f_K)$ is then a monotone regular extension of $(X_B,p_K)$, 
and the partial orders occurring in the completion constitute
an ordering associated with $(D_B,f_K)$ which is monotone.

\bigskip
The results of this chapter can be applied to give the following `canonical' procedure
for obtaining a complete bottomed extension $(D_B,f_K)$ of $(X_B,p_K)$:
\bigskip
{\parindent=15pt
\item{1.} Start by choosing a minimal monotone regular extension $(Y_B,q_K)$ of $(X_B,p_K)$:
Proposition~3.3.4 implies that such an extension is essentially uniquely determined by its
trace.  
\medskip
\item{2.} By Proposition~4.1.2 there is then a unique ordering $\sqle_B$ 
associated with $(Y_B,q_K)$, which by Proposition~4.1.5 is monotone.
\medskip
\item{3.} Next let $(D_B,f_K)$ be an initial completion of $(Y_B,q_K)$:
The existence of such a completion is guaranteed by Proposition~4.3.2, and 
Proposition~4.4.1 implies that is is essentially unique. 
\medskip
\item{4.} By Proposition~4.4.2 $(D_B,f_K)$ is then a monotone regular 
extension of $(X_B,p_K)$.
Moreover, it follows that $(D_B,f_K)$ has the same trace as
$(Y_B,q_K)$, which implies that $(D_B,f_K)$ is essentially uniquely determined
by this trace.
\bigskip\medskip
}

\vfill\eject

\hfline{83}{4.1 \ ASSOCIATED ORDERINGS FOR REGULAR EXTENSIONS}

\sectionhead {4.1} {Associated orderings for regular extensions}
\bigskip\medskip
In this section 
it is shown that if $(D_B,f_K)$ is a regular bottomed \tsym{\Lambda}-algebra then for 
each $\beta \in B$ there exists a partial order $\sqle_\beta$ on the set 
$D_\beta$ such that $u \sqle_\beta u'$ can be sensibly interpreted as 
meaning that $u$ is `less-defined' than $u'$. Moreover, if $(D_B,f_K)$ is also 
minimal then it turns out that this family $\sqle_B$ of partial orders is unique.
\medskip
First some notation and conventions concerning partial orders must be fixed.
As already mentioned, if $(Y,\sqle)$ is a poset then it is usual to just write
$Y$ instead of $(Y,\sqle)$ and to assume that $\sqle$ can be determined from the context. 
Something like
`$Y$ is a poset with partial order $\sqle$' or `$\sqle$ is the partial order
on $Y$' can be employed when it is 
necessary to refer to the partial order explicitly.
It is useful to consider the set $\Oneptset$ as a poset (naturally
with respect to the unique partial order on $\Oneptset$).
\medskip

Let $Y$ be a poset with partial order $\sqle$; then there can be at most
one element $\bot_Y \in Y$ with $\bot_Y \sqle y$ for all $y \in Y$.
If such an element $\bot_Y$ exists then $Y$ is said to be a
\indexddef{bottomed poset}{bottomed poset}{}{poset}{bottomed} 
with 
\indexdef{bottom element}{bottom element}{of a poset} 
$\bot_Y$.
A bottomed poset $Y$ will always be considered as a bottomed set with $\bot_Y$ as its
bottom element.

\medskip
Let $\lcurl (D_\beta,\sqle_\beta) \,:\, \beta \in B \rcurl$ be a family of 
posets and let $I$ be a \tsym{B}-typed set.
Then $\ass I D $ is considered as a poset with the partial
order $\ass I {\sqle} $ defined by stipulating that $c\ \ass I {\sqle} \ c' $ if and only 
if $c(\eta) \sqle_\eta c'(\eta)$ for each $\eta \in I$. If $\kappa \in K$
is of type $L \to \beta$ then it is convenient to allow $\sqle_\kappa$ as an
alternative notation for the partial order $\ass L {\sqle} $ on
$\ass L D $.
\medskip
Note that if $D_S$ is a family of bottomed posets then the poset $\ass I D $ is also 
bottomed, with the bottom element of the poset $\ass I D $ equal to 
the bottom element of the bottomed set $\ass I D $ (as defined in Section~3.2).
\medskip
Let $(D_B,f_K)$ be a bottomed \tsym{\Lambda}-algebra with core 
$C_K$ (so
\mdisp{C_\kappa \ =\ \lcurl c \in \dom(f_\kappa) \,:\, f_\kappa(c) \ne \bot_\beta
                                                                   \rcurl}
for each $\kappa \in K_\beta$). For each $\beta \in B$ let $\sqle_\beta$ be a 
partial order on the set $D_\beta$. Then the family of partial orders 
$\sqle_B$ will be called an 
\indexddef{ordering associated} 
{associated ordering}{}{ordering}{associated with extension}
with $(D_B,f_K)$ (or just an 
\definition{associated ordering}) if the following 
two conditions hold:
\medskip\smallskip
{\parindent=25pt
\item{(1)} $D_\beta$ is a bottomed poset with bottom element $\bot_\beta$.
\medskip
\item{(2)} If $\kappa, \, \kappa' \in K_\beta$ and $c \in C_{\kappa}$, 
$c' \in C_{\kappa'}$ then $f_\kappa(c) \sqle_\beta f_{\kappa'}(c')$ if and 
only if $\kappa = \kappa'$ and $c \sqle_\kappa c'$.
\bigskip
}
The definition of an ordering associated with $(D_B,f_K)$ is chosen to be 
compatible with the interpretation mentioned above: If $u, \, u' \in D_\beta$
then $u \sqle_\beta u'$ should be thought of as meaning that $u$ is
`less-defined' than $u'$. 
\medskip
Proposition~4.1.2 states that there exists a unique ordering associated with each
minimal regular bottomed \tsym{\Lambda}-algebra. 
In the simple case where $(D_B,f_K)$ is the flat extension of $(X_B,p_K)$ it 
can be seen directly that the unique associated ordering is obtained by 
letting $u \sqle_\beta u'$ if and only if 
$u \in \{\bot_\beta,u'\}$. 
For the initial extension $(X^\top_B,p^\top_K)$ introduced in
in Example~3.2.2 the unique associated ordering 
is given in Example~4.1.1 below.

\bigskip\bigskip
\frame{20pt}{\bigskip 
{\it Example 4.1.1\ } It is straightforward to check that the following 
defines an (and thus the) ordering $\sqle_B$ associated with the initial
extension $(X^\top_B,p^\top_K)$ of the \tsym{\Lambda}-algebra $(X_B,p_K)$ introduced in 
Example~3.2.2.
\medskip\smallskip
{\leftskip=32pt 
If $b, \, b' \in X^{\top}_{\fstt{bool}}$ 
with $b \ne b'$ then $b \sqle_{\fstt{bool}} b'$ if and only if 
$b = \bot_{\fstt{bool}}$.
\medskip
Let $n \in \Nat$ and $x \in X^{\top}_{\fstt{nat}}$. Then
$x \sqle_{\fstt{nat}} n$ if and only if either $x = n$ or $x = m^\bot$ for
some $m \le n$. Moreover, $x \sqle_{\fstt{nat}} n^\bot$ if and only if
$x = m^\bot$ for some $m \le n$. 
\medskip
If $n, \, n' \in X^{\top}_{\fstt{int}}$ 
with $n \ne n'$ then $n \sqle_{\fstt{int}} n'$ if and only if 
$n = \bot_{\fstt{int}}$.
\medskip
If $p, \, p' \in X^{\top}_{\fstt{pair}}$ with 
$p' \ne \bot_{\fstt{pair}}$ then $p' \sqle_{\fstt{pair}} p$ if and only if
$p' = (x',y')$ and $p = (x,y)$ with $x' \sqle_{\fstt{int}} x$ and
$y' \sqle_{\fstt{int}} y$.
\medskip
Let $\ell = \list x n \in (X^{\top}_{\fstt{int}})^*$ and
$z' \in X^{\top}_{\fstt{list}}$. Then 
$z' \sqle_{\fstt{list}} \ell$ if and only if either $z' = \list {x'} n $ with
$x'_j \sqle_{\fstt{int}} x_j$ for $\oneto j n $ or $z' = (\list {x'} m )^\bot$
with $m \le n$ and $x'_j \sqle_{\fstt{int}} x_j$ for $\oneto j m $. Moreover,
$z' \sqle_{\fstt{list}} \ell^\bot$ if and only if $z' = (\list {x'} m )^\bot$
with $m \le n$ and $x'_j \sqle_{\fstt{int}} x_j$ for $\oneto j m $.
\bigskip
}
(Recall here that 
\medskip
{\leftskip=32pt
$X^\top_{\fstt{bool}} = \Bool^\bot = \{T,F,\bot_{\fstt{bool}}\}$, 
\smallskip
$X^\top_{\fstt{nat}} = \Nat \cup \Bot{\Nat}$\ with 
                                                $0^\bot = \bot_{\fstt{nat}}$, 
\smallskip
$X^\top_{\fstt{int}} 
  = \Int \cup \{\bot_{\fstt{int}}\}$,
\quad
$X^\top_{\fstt{pair}} 
  = (X^\top_{\fstt{int}})^2 \cup \{\bot_{\fstt{pair}}\}$,
\smallskip    
$X^\top_{\fstt{list}} 
    = (X^\top_{\fstt{int}})^* \cup
       \Bot{(X^\top_{\fstt{int}})^*}$
  \ with $\onept^\bot = \bot_{\fstt{list}}$.)
\bigskip\medskip
}}
\bigskip\medskip
\bigskip
\medskip
Before coming to Proposition~4.1.2 it will be shown that the converse is 
essentially true.

\proclaim{Proposition~4.1.1} Let $(D_B,f_K)$ be a bottomed \tsym{\Lambda}-algebra
with core $C_K$ for which there exists an associated ordering 
$\sqle_B$. Then:
\medskip\smallskip
{\parindent=25pt
\item{(1)} The restriction of $f_\kappa$ to $C_\kappa$ is injective for each 
$\kappa \in K$.
\medskip
\item{(2)} If $\beta \in B$ and $\kappa_1,\, \kappa_2 \in K_\beta$ with 
$\kappa_1 \ne \kappa_2$ then $f_{\kappa_1}(C_{\kappa_1})$ and
$f_{\kappa_2}(C_{\kappa_2})$ are disjoint subsets of $D_\beta$.
\medskip\smallskip
}  
In particular, any minimal bottomed \tsym{\Lambda}-algebra possessing an associated 
ordering is regular. \endpro

\proof Let $\kappa \in K_\beta$ and $c, \, c' \in C_\kappa$ with
$f_\kappa(c) = f_\kappa(c')$. Then $f_\kappa(c) \sqle_\beta f_\kappa(c')$
and hence $c \sqle_\kappa c'$; in the same way $c' \sqle_\kappa c$. 
Thus $c = c'$, which shows the restriction of $f_\kappa$ to $C_\kappa$ is 
injective.
Next consider $\kappa_1,\, \kappa_2 \in K_\beta$ and let
$u \in f_{\kappa_1}(C_{\kappa_1}) \cap f_{\kappa_2}(C_{\kappa_2})$;
choose $c_1 \in C_{\kappa_1}$, $c_2 \in C_{\kappa_2}$ with
$f_{\kappa_1}(c_1) = u = f_{\kappa_2}(c_2)$. Then
$f_{\kappa_1}(c_1) \sqle_\beta f_{\kappa_2}(c_2)$, which is only possible
if $\kappa_1 = \kappa_2$. Therefore if
$\kappa_1 \ne \kappa_2$ then $f_{\kappa_1}(C_{\kappa_1})$ and
$f_{\kappa_2}(C_{\kappa_2})$ are disjoint subsets of $D_\beta$.
The final statement now follows directly from Proposition~3.2.2 and Lemma~3.3.1. \eop

\proclaim{Proposition~4.1.2} Let $(D_B,f_K)$ be a regular bottomed \tsym{\Lambda}-algebra. 
Then there exists an ordering associated with $(D_B,f_K)$. 
Moreover, if in addition $(D_B,f_K)$ is minimal then this 
associated ordering is unique. \endpro

\proof In order to show the existence of an associated ordering it is useful
to first introduce a somewhat weaker concept. 
If $\sqle_\beta$ is a partial order
on the set $D_\beta$ for each $\beta \in B$ then the family $\sqle_B$ will be called a 
\indexddef{weak ordering}{weak ordering}{}{ordering}{weak}
if the following hold:
\medskip\smallskip
{\parindent=25pt
\item{(1)} $\bot_\beta \sqle_\beta u$ for each $u \in D_\beta$.
\medskip
\item{(2)} If $\kappa, \, \kappa' \in K_\beta$ and $c \in C_{\kappa}$, 
$c' \in C_{\kappa'}$ with $f_\kappa(c) \sqle_\beta f_{\kappa'}(c')$ then 
$\kappa = \kappa'$ and $c \sqle_\kappa c'$.
\medskip\smallskip
}
For each $\beta \in B$ let $\sqle_\beta$ be a partial order on $D_\beta$.
Then (making use of Lemma~3.3.1) a binary relation $\sqle'_\beta$ 
can be defined on $D_\beta$ for each $\beta \in B$ as follows:
\medskip\smallskip
{\parindent=25pt
\item{(1)} $\bot_\beta \sqle'_\beta u$ for each $u \in D_\beta$.
\medskip
\item{(2)} $u \sqle'_\beta \bot_\beta$ holds (if and) only if 
$u = \bot_\beta$.
\medskip
\item{(3)} If $\kappa,\, \kappa' \in K_\beta$ and $c \in C_{\kappa}$, 
$c' \in C_{\kappa'}$ then $f_\kappa(c) \sqle'_\beta f_{\kappa'}(c')$ if 
and only if $\kappa = \kappa'$ and $c \sqle_\kappa c'$.
\medskip\smallskip
}

\proclaim{Lemma~4.1.1} The relation $\sqle'_\beta$ is a partial order for each
$\beta \in B$. \endpro

\proof It is clear that $\sqle'_\beta$ is reflexive. To show that
$\sqle'_\beta$ is anti-symmetric consider $u_1,\, u_2 \in D_\beta$ with 
$u_1 \sqle'_\beta u_2$ and $u_2 \sqle'_\beta u_1$. If 
$\bot_\beta \in \{u_1,u_2\}$ then $u_1 = u_2\ (\,= \bot_\beta)$ by (2). 
Thus suppose that $u_1,\, u_2 \in D_\beta \setminus \{\bot_\beta\}$. Then by 
(3) and Lemma~3.3.1 there exists $\kappa \in K_\beta$ and
$c_1,\, c_2 \in C_\kappa$ such that $u_1 = f_\kappa(c_1)$,
$u_2 = f_\kappa(c_2)$, $c_1 \sqle_\kappa c_2$ and $c_2 \sqle_\kappa c_1$. 
Hence $c_1 = c_2$ and so $u_1 = f_\kappa(c_1) = f_\kappa(c_2) = u_2$. 
A similar argument shows that $\sqle'_\beta$ is transitive. \eop

\proclaim{Lemma~4.1.2} Suppose $\sqle_B$ is a weak ordering.
Then $\sqle'_B$ is also a weak ordering 
and $u_1 \sqle'_\beta u_2$ whenever $u_1 \sqle_\beta u_2$. \endpro

\proof It will be shown first that if $u_1,\, u_2 \in D_\beta$ with 
$u_1 \sqle_\beta u_2$ then $u_1 \sqle'_\beta u_2$. Now since
$\bot_\beta \sqle'_\beta u_2$ holds trivially  it can be assumed here that 
$u_1 \ne \bot_\beta$. Then $u_1,\, u_2 \in D_\beta \setminus \{\bot_\beta\}$,
and hence by Lemma~3.3.1 and the definition of a weak ordering 
there exists $\kappa \in K_\beta$ and $c_1,\, c_2 \in C_\kappa$ such
that $u_1 = f_\kappa(c_1)$, $u_2 = f_\kappa(c_2)$ and $c_1 \sqle_\kappa c_2$.
But then $\,u_1 = f_\kappa(c_1) \sqle'_\beta f_\kappa(c_2) = u_2\,$ holds by
the definition of $\sqle'_\beta$.
\medskip
It remains to show that $\sqle'_B$ is a weak ordering. Thus 
consider $\kappa, \, \kappa' \in K_\beta$
and $c \in C_{\kappa}$, $c' \in C_{\kappa'}$ with 
$f_\kappa(c) \sqle'_\beta f_{\kappa'}(c')$.
Then (by the definition of $\sqle'_\beta$) $\kappa = \kappa'$ and 
$c \sqle_\kappa c'$, and by the first part of the proof 
$c\ \sqle'_\kappa\ c'$. This implies that $\sqle'_B$ is a weak
ordering (since by definition $\bot_\beta \sqle'_\beta u$ 
holds for each $u \in D_\beta$). \eop
\bigbreak
If $\sqle_B$ is a weak ordering then $\sqle'_B$ will be called the 
\indexdef{first refinement}{first refinement of a weak ordering}{}
of $\sqle_B$. Let $\sqle^0_\beta$ be the `trivial' partial order on $D_\beta$ defined so
that $u_1 \sqle^0_\beta u_2$ if and only if $u_1$ is either $\bot_\beta$ or $u_2$.
Then $\sqle^0_B$ is clearly a weak ordering, and thus
by Lemma~4.1.2 a sequence of weak orderings
$\sqle^m_B$, $m \ge 0$, can be defined by 
letting $\sqle^{m+1}_B$ be the first refinement of $\sqle^m_B$ for each
$m \ge 0$.
Now define a partial order $\sqle_\beta$ on $D_\beta$ for each $\beta \in B$ 
by stipulating that $u_1 \sqle_\beta u_2$ if and only if 
$u_1 \sqle^m_\beta u_2$ for some (and thus for all sufficiently large) 
$m \ge 0$. It is easy to see that $\sqle_B$ is a weak ordering. 
\medskip
In fact $\sqle_B$ is an associated ordering: Let $\kappa \in K$ be of type 
$L \to \beta$ and let $c,\, c' \in C_\kappa$ with 
$c \sqle_\kappa c'$. Then $c(\eta) \sqle_\eta c'(\eta)$ for each 
$\eta \in L$, and so for 
each $\eta$ there exists $m_\eta \ge 0$ such that
$c(\eta) \sqle^{m_\eta}_\eta c'(\eta)$. Put
$m = \max \lcurl m_\eta \,:\, \eta \in L \rcurl$; then 
$c(\eta) \sqle^m_\eta c'(\eta)$ for each $\eta$, and this means that
$c\ \sqle^m_\kappa\ c'$. Hence $f_\kappa(c) \sqle^{m+1}_\beta f_\kappa(c')$
by the definition of $\sqle^{m+1}_\beta$, and therefore
$f_\kappa(c) \sqle_\beta f_\kappa(c')$.
\medskip
The uniqueness in the case when 
$(D_B,f_K)$ is minimal still has to be considered. Thus let $\sqle_B$ and $\sqle'_B$ be
two orderings associated with $(D_B,f_K)$ and for each $u \in D_\beta$ define
$L_\beta(u) = \lcurl u' \in D_\beta \,:\, u' \sqle_\beta u \rcurl$ and
$L'_\beta(u) = \lcurl u' \in D_\beta \,:\, u' \sqle'_\beta u \rcurl$; put
$D'_\beta = \lcurl u \in D_\beta \,:\, L'_\beta(u) = L_\beta(u) \rcurl$. Then
$\bot_\beta \in D'_\beta$ for each $\beta \in B$ and it is straightforward to
check that the family $D'_B$ is invariant. Hence if $(D_B,f_K)$ is 
minimal then $D'_B = D_B$, i.e., $\sqle'_B\,=\,\sqle_B$. \eop
\bigbreak

\proclaim{Proposition~4.1.3} Let $(D_B,f_K)$ be a regular bottomed
extension of $(X_B,p_K)$ and $\sqle_B$ be an ordering associated with $(D_B,f_K)$; let 
$\beta \in  B$.
\medskip
(1)\enskip Each element of $X_\beta$ is a maximal 
element of the poset $D_\beta$, i.e., if
$x \in X_\beta$ and $u \in D_\beta$ with $x \sqle_\beta u$ then $u = x$. 
\medskip
(2)\enskip For each $x \in X_\beta$ the set 
$\lcurl u \in D_\beta \,:\, u \sqle_\beta x \rcurl$ is finite. \endpro

\proof (1)\enskip For each $\beta \in B$ let $X'_\beta$ denote the set of
elements in $X_\beta$ which are maximal. Then the family $X'_B$ is clearly
invariant in $(X_B,p_K)$, and thus $X'_B = X_B$, i.e., each element of $X_\beta$
is a maximal element of $D_\beta$. 
\medskip
(2)\enskip For each $\beta \in B$ let $X'_\beta$ denote the set of those
elements $x \in X_\beta$ for which the set
$\lcurl u \in D_\beta \,:\, u \sqle_\beta x \rcurl$ is finite. Then it is 
easily checked that the family $X'_B$ is 
invariant in $(X_B,p_K)$, and thus $X'_B = X_B$, i.e., 
the set $\lcurl u \in D_\beta \,:\, u \sqle_\beta x \rcurl$ is finite
for each $x \in X_\beta$. \eop
\bigbreak
If $(D_B,f_K)$ is a minimal regular extension of $(X_B,p_K)$ then
by Proposition~4.1.2 there exists a unique ordering $\sqle_B$ associated with 
$(D_B,f_K)$; thus in this case it makes sense to speak of {\it the} associated ordering.
Recall the assumption made in Section~3.1 that the signature 
$\Lambda$ is pervasive, from which it follows that $X_\beta \ne \varnothing$ 
for each $\beta \in B$. 

\proclaim{Proposition~4.1.4} Let $(D_B,f_K)$ be a minimal regular
extension of $(X_B,p_K)$ and $\sqle_B$ be the associated ordering; let 
$\beta \in B$.
\medskip
(1)\enskip For each $u \in D_\beta$ there exists $x \in X_\beta$ with 
$u \sqle_\beta x$ (and note here the fact that 
$X_\beta \ne \varnothing$ is needed).
\medskip
(2)\enskip For each $u \in D_\beta$ the set 
$\lcurl u' \in D_\beta \,:\, u' \sqle_\beta u \rcurl$ is finite. 
\medskip
In particular, (1) and Proposition~4.1.3~(1) imply that 
$X_\beta$ is exactly the set of maximal elements of the poset $D_\beta$. 
\endpro

\proof (1)\enskip For each $\beta \in B$ let 
$D'_\beta = \lcurl u \in D_\beta 
    \,:\, u \sqle_\beta x \frm{for some} x \in X_\beta \rcurl$.
Then it easily checked that the family $D'_B$ is invariant in $(D_B,f_K)$ and
$\bot_\beta \in D'_\beta$ for each $\beta \in B$. (Note that
$\bot_\beta \in D'_\beta$ is equivalent to having $X_\beta \ne \varnothing$.) 
Hence $D'_B = D_B$. 
\medskip
(2)\enskip This also follows by considering an appropriate invariant family. \eop
\bigbreak

If $Y_1$ and $Y_2$ are posets then a mapping $h : Y_1 \to Y_2$ is said to be
\indexddef{monotone}{monotone mapping}{}{mapping}{monotone}
if $h(y) \sqle_2 h(y')$ for all $y ,\, y' \in Y_1$ with
$y \sqle_1 y'$ (where $\sqle_j$ is the partial order on $Y_j$ for 
$j = 1,\, 2$ ). 
Let $(D_B,f_K)$ be a regular bottomed \tsym{\Lambda}-algebra and $\sqle_B$ be an 
associated ordering. Then $\sqle_B$ is said to be 
\indexddef{monotone}
{monotone associated ordering}{}{associated ordering}{monotone}
if the mappings $f_\kappa$, $\kappa \in K$, are 
monotone with respect to $\sqle_B$, i.e., if $\kappa \in K_\beta$ and 
$c, \, c' \in \dom(f_\kappa)$ with $c \sqle_\kappa c'$
then $f_\kappa(c) \sqle_\beta f_\kappa(c')$. 
\medskip
Now it might be expected that an associated ordering $\sqle_B$ is automatically
monotone. But, as Example~4.1.2 below shows, this is not true in 
general, even if $(D_B,f_K)$ is a minimal regular bottomed \tsym{\Lambda}-algebra 
and $\sqle_B$ is the (unique) associated ordering.
\medskip
\midinsert
\bigskip\bigskip
\frame{20pt}{\bigskip 
{\it Example~4.1.2\enspace} Let $\Lambda$ and $(X_B,p_K)$ be as in Example~2.2.1 and
let $(Y_B,q_K)$ be the \tsym{\Lambda}-algebra defined (as in Example~3.5.1) by
\medskip\smallskip
{\leftskip=32pt
$Y_{\fstt{nat}} \,=\, \Nat \cup \{\bot_{\fstt{nat}},\bot^o_{\fstt{nat}}\}$\ with
$\bot^o_{\fstt{nat}} \notin \Nat \cup \{\bot_{\fstt{nat}}\}$,
\smallskip
$Y_\beta = X^\bot_\beta$\ for all 
$\beta \in B \setminus \{\ftt{nat}\}$,
\medskip
$q_{\fstt{Zero}} : 
   \Oneptset \to Y_{\fstt{nat}}$\ with\ %
                       $q_{\fstt{Zero}}(\onept) = 0$,
\smallskip
$q_{\fstt{Succ}} : Y_{\fstt{nat}} \to Y_{\fstt{nat}}$
\ with\ %
\mdisp{q_{\fstt{Succ}}(n)\ =\ \cases{ n + 1 & 
                           if $\,n \in \Nat\,$, \cr
            \bot_{\fstt{nat}} & if $\,n = \bot^o_{\fstt{nat}}\,$, \cr
               \bot^o_{\fstt{nat}}    & if $\,n = \bot_{\fstt{nat}}\,$, \cr
                 }} 
\smallskip
and with $q_\kappa = p^\bot_\kappa$ for all 
$\kappa \in K \setminus \{\ftt{Zero},\ftt{Succ}\}$,
\medskip\smallskip
}
where $(X^\bot_B,p^\bot_K)$ is the flat extension
of the \tsym{\Lambda}-algebra 
$(X_B,p_K)$.
Then $(Y_B,q_K)$ is clearly a minimal regular extension of
$(X_B,p_K)$, and so let $\sqle_B$ be
the ordering associated with $(Y_B,q_K)$.
However, $q_{\fstt{Succ}}$ is not monotone, since
$\bot_{\fstt{nat}} \sqle_{\fstt{nat}} \bot^o_{\fstt{nat}}$
but $q_{\fstt{Succ}}(\bot_{\fstt{nat}}) = \bot^o_{\fstt{nat}}$ and
$q_{\fstt{Succ}}(\bot^o_{\fstt{nat}}) = \bot_{\fstt{nat}}$.
\bigskip\medskip
}
\bigskip
\endinsert
However, it will now be shown that if $(D_B,f_K)$ is a minimal monotone regular bottomed
\tsym{\Lambda}-algebra then the associated ordering $\sqle_B$ is monotone.

\proclaim{Proposition~4.1.5} Let $(D_B,f_K)$ be a minimal monotone regular 
bottomed \tsym{\Lambda}-algebra with (unique) associated ordering $\sqle_B$. Then 
$\sqle_B$ is monotone.
\endpro

\proof The family of relations $\sqle_B$ is compatible (as defined in Section~3.5): 
Let $\kappa \in K$ be of type $L \to \beta$ and consider $a \in \ass L D $, 
$u \in D_\beta \setminus \{\bot_\beta\}$ with $u \sqle_\beta f_\kappa(a)$.
Then $u$ has a unique representation of the form $f_{\kappa'}(c)$
(since $u \ne \bot_\beta$ and $(D_B,f_K)$ is regular) and hence
$\kappa' = \kappa$ and $c\ \ass L {\sqle} \ a$
(since $\sqle_B$ is an associated ordering).
Moreover, $\sqle_B$ is full: This follows from the fact that $\bot_\beta$ is a base element
of $D_\beta$ for each $\beta \in B$ and because $(D_B,f_K)$ is minimal.
\medskip
Let $\kappa \in K$ be of type $L \to \beta$ and let
$c,\,a \in \ass L D $ with $c\ \ass L {\sqle} \ a $, and since the task at hand
is to show that
$f_\kappa(c) \sqle_\beta f_\kappa(a)$ it can clearly be assumed that
$f_\kappa(c) \ne \bot_\beta$.
Now by Proposition~3.5.4 there exists a finite \tsym{B}-typed set $J$, a \tsym{J}-free
\tsym{\Lambda}-algebra $(Z_B,r_K)$, and elements $z \in Z_\beta$,
$b \in \ass J D $ such that $f_\kappa(c) = \pi^{\bot^{\!J}}_\beta(z)$ and 
$f_\kappa(a) = \pi^b_\beta(z)$. But $(D_B,f_K)$ is monotone and
$\pi^{\bot^{\!J}}_\beta(z) = f_\kappa(c) \ne \bot_\beta$; hence
$f_\kappa(a) = \pi^b_\beta(z) \ne \bot_\beta$, and so by the definition
of an associated ordering it follows that $f_\kappa(c) \sqle_\beta f_\kappa(a)$.
\eop
\bigbreak

The following result can be considered as a converse to Proposition~4.1.5.

\proclaim{Proposition~4.1.6} Let $(D_B,f_K)$ be a bottomed \tsym{\Lambda}-algebra 
and for each $\beta \in B$ let $\sqle_\beta$ be a partial
order on the set $D_\beta$ such that $\bot_\beta \sqle_\beta u$
for all $u \in D_\beta$. Suppose that the mappings
$f_\kappa$, $\kappa \in K$, are monotone
with respect to the family $\sqle_B$, i.e., if $\kappa \in K_\beta$ and 
$c, \, c' \in \dom(f_\kappa)$ with $c \sqle_\kappa c'$
then $f_\kappa(c) \sqle_\beta f_\kappa(c')$. Then
$(D_B,f_K)$ is monotone. \endpro

\proof By Proposition~3.5.1 it is enough to show that $(D_B,f_K)$ is strongly monotone. 
Let $(Y_B,q_K)$ be a \tsym{\Lambda}-algebra, let ${\breve \pi}_B$ be a fat 
bottomed and and $\pi_B$ be an arbitrary homomorphism from $(Y_B,q_K)$ to $(D_B,f_K)$.
For each $\beta \in B$ put

\mdisp{Y'_\beta  \ =\ \lcurl y \in Y_\beta \,:\, 
   {\breve \pi}_\beta(y) \sqle_\beta \pi_\beta(y) \rcurl\ . }

If $\kappa \in K$ is of type $\varnothing \to \beta$ then
$q_\kappa(\onept) \in Y'_\beta$, because 
${\breve \pi}_\beta(q_\kappa(\onept)) 
        = f_\kappa(\onept) = \pi_\beta(q_\kappa(\onept))$. 
Next consider $\kappa \in K$ of type
$L \to \beta$ with $L \ne \varnothing$ and let $c \in \ass L Y $
be such that  $c(\eta) \in Y'_\eta$ for each $\eta \in L$. Then
$\ass L {\breve \pi} (c) \sqle_\kappa \ass L {\pi} (c)$ and thus

\mdisp{{\breve \pi}_\beta(q_\kappa(c))
  \ =\ f_\kappa(\ass L {\breve \pi} (c))
\ \sqle_\beta\ f_\kappa(\ass L {\pi} (c))\ =\ \pi_\beta(q_\kappa(c))\ ,}

i.e., $q_\kappa(c) \in Y'_\beta$. 
This shows the family $Y'_B$ is invariant in $(Y_B,q_K)$.
\medskip
Let ${\breve U}_B$ (resp.\ $U_B$) be the kernel of ${\breve \pi}_B$
(resp.\ the kernel of $\pi_B$). If $y \in {\breve U}_\beta$ then
${\breve \pi}_\beta(y) = \bot_\beta  \sqle_\beta \pi_\beta(y)$,
and hence ${\breve U}_B \subset Y'_B$.
Thus $Y'_B = Y_B$, since ${\breve \pi}_B$ is fat bottomed. In particular, 
if $y \in U_\beta$ then it follows that
${\breve \pi}_\beta(y) \sqle_\beta \pi_\beta(y) = \bot_\beta$,
i.e., $y \in {\breve U}_\beta$. Therefore $U_B \subset {\breve U}_B$, 
i.e., $(D_B,f_K)$ is strongly monotone. \eop
\bigbreak
\vfill\eject

\hfline{89}{4.2 \ COMPLETE PARTIALLY ORDERED SETS}

\sectionhead {4.2} {Complete partially ordered sets}
\bigskip\medskip
Let $Y_1$ and $Y_2$ be posets and $h : Y_1 \to Y_2$ a mapping; as before $h$ is 
\indexddef{monotone}{monotone mapping}{}{mapping}{monotone} 
if $h(y) \sqle_2 h(y')$ for all $y ,\, y' \in Y_1$ with 
$y \sqle_1 y'$ (where $\sqle_j$ is the partial order on $Y_j$ for 
$j = 1,\, 2$ ). The mapping $h$ is said to be an 
\indexdef{embedding}{embedding}{} 
if $h(y) \sqle_2 h(y')$ holds if and only if $y \sqle_1 y'$. An embedding is, in 
particular, monotone and it is also automatically injective. 
\medskip
A bijective mapping $h : Y_1 \to Y_2$ is called an 
\indexdef{order isomorphism}{order isomorphism}{}
if $h$ and the inverse mapping 
$h^{-1} : Y_2 \to Y_1$ are both monotone. Thus a mapping is an order 
isomorphism if and only if it is a surjective embedding. (Note that if 
$h : Y_1 \to Y_2$ is a monotone mapping which is a bijection then the inverse 
mapping $h^{-1} : Y_2 \to Y_1$ is not necessarily monotone.) The posets $Y_1$ 
and $Y_2$ are said to be 
\indexdef{isomorphic}{isomorphic posets}{}
if there exists an order isomorphism $h : Y_1 \to Y_2$.

\proclaim{Lemma~4.2.1} If $Y_1,\,Y_2,\,Y_3$ are posets and
$f : Y_1 \to Y_2$ and $g : Y_2 \to Y_3$ are monotone mappings then the
composition $g\,f : Y_1 \to Y_3$ is also monotone. \endpro

\proof This is clear. \eop
\bigbreak

Let $Y_S$ be a family of posets (with $\sqle_\sigma$ the partial order on $Y_\sigma$
for each $\sigma \in S$) and let $L$ be an \tsym{S}-typed set. 
Recall then that $\ass L Y $ is considered as a poset with the partial
order $\ass L {\sqle} $ defined by stipulating that $c\ \ass L {\sqle} \ c' $ if and 
only if $c(\eta) \sqle_\eta  c'(\eta)$ for each $\eta \in I$.

\proclaim{Lemma~4.2.2} For each $\eta \in L$ the projection mapping 
$\fss{p}^L_\eta : \ass L Y \to Y_\eta$ defined by $\fss{p}^L_\eta(c) = c(\eta)$
for each $c \in \ass L Y $ is monotone. \endpro

\proof This is also clear. \eop
\bigbreak

\proclaim{Lemma~4.2.3} A mapping $h : Y \to \ass L Y $ from a poset $Y$ to the poset
$\ass L Y $ is monotone if and only if for each $\eta \in L$ the mapping
$\fss{p}^L_\eta\,h$ from $Y$ to $Y_\eta$ is monotone. \endpro

\proof Straightforward. \eop
\bigbreak

If $Y_1$ and $Y_2$ are posets then the set of all monotone mappings from $Y_1$ to $Y_2$
will be denoted by $\mono {Y_1} {Y_2} $.
Note that the set $\mono {Y_1} {Y_2} $ is non-empty, since the
constant mappings are clearly monotone.
\medskip
A partial order $\sqle$ on the set
$\mono {Y_1} {Y_2} $ can be defined by stipulating that $h \sqle h'$ if and only if
$h(y) \sqle_2 h'(y)$ for all $y \in Y_1$ (where $\sqle_2$ is the partial
order on $Y_2$), and
if $\mono {Y_1} {Y_2} $ is considered as a poset then it will always
be with respect to this partial order. Note that if $Y_2$ is a bottomed
poset with bottom element $\bot_2$ then $\mono {Y_1} {Y_2} $ is also a 
bottomed poset: The bottom element $\bot$ of $\mono {Y_1} {Y_2} $ is 
the constant mapping with $\bot(y) = \bot_2$ for all $y \in Y_1$.
\medskip
In what follows let $Y_1$, $Y_2$ and $Y$ be posets.

\proclaim{Lemma~4.2.4} The 
\indexddef{application operator}{application operator}{}{operator}{application} 
$\fss{a} : \mono {Y_1} {Y_2}  \times Y_1 \to Y_2$ given by 
\mdisp{\fss{a}(h,y)\ =\ h(y)}
for all $h \in \mono {Y_1} {Y_2} $, $y \in Y_1$, is monotone. \endpro

\proof This is clear. \eop
\bigbreak

Now consider a monotone mapping $h : Y_1 \times Y_2 \to Y$;
then the mapping $\,y_2 \,\mapsto\, h(y_1,y_2)\,$ from
$Y_2$ to $Y$ is monotone for each $y_1 \in Y_1$, and so 
a mapping $\fss{c}(h) : Y_1 \to \mono Y_2 Y $ can be defined by letting

\mdisp{\fss{c}(h)\,(y_1)\,(y_2) \ =\ h(y_1,y_2)}

for each $y_1 \in Y_1$, $y_2 \in Y_2$. Moreover, it is easy to see that the mapping
$\fss{c}(h)$ is also monotone, which means
that the so-called 
\indexddef{currying operator}{currying operator}{}{operator}{currying} 
can be defined; this is the mapping
$\fss{c} : \mono {Y_1 \times Y_2} {Y} \to \mono Y_1  {\mono Y_2 Y } $  
given by
\mdisp{ \fss{c} (h,y_1)\,(y_2)\ =\ h(y_1,y_2)}
for all $h \in \mono {Y_1 \times Y_2} Y $, $y_1 \in Y_1$, $y_2 \in Y_2$.

\proclaim{Lemma~4.2.5} The currying operator $\fss{c}$
is an order isomorphism whose inverse is the 
\indexddef{uncurrying operator}{uncurrying operator}{}{operator}{uncurrying}
$\,\fss{u} : \mono Y_1 {\mono Y_2 Y } \to \mono {Y_1 \times Y_2} Y \,$
defined by 
\mdisp{\fss{u}(g)\,(y_1,y_2) \,=\, g(y_1)\,(y_2)} 
for all $g \in \mono Y_1 {\mono Y_2 Y } $,
$(y_1,y_2) \in Y_1 \times Y_2$. \endpro

\proof This is straightforward. \eop
\bigbreak

We now define what it means for a poset to be complete and what it means 
for a monotone mapping between complete posets to be continuous. We
then show that the above results hold for complete posets with `monotone' 
replaced by `continuous'.
\medskip 
Let $Y$ be a poset with partial order $\sqle$ and let $A$ be a non-empty
subset of $Y$. An element $y \in Y$ is called an 
\indexdef{upper bound}{upper bound}{}
of $A$ if $y' \sqle y$ for all $y' \in A$; $y$ is called the 
\indexddef{least upper bound}{least upper bound}{}{upper bound}{least}
of $A$ if $y$ is an upper bound of $A$ and
$y \sqle y'$ for each upper bound $y'$ of $A$. (It is clear that there can
be at most one element $y \in Y$ having these properties.)  If the least 
upper bound of $A$ exists then it will be denoted by $\lub A$.
\medskip
A subset $A$ of a poset $Y$ is said to be 
\indexddef{directed}{directed set}{}{set}{directed} 
if it is non-empty and for each $y_1,\, y_2 \in A$ there exists $y \in A$ 
such that $y_1 \sqle y$ and $y_2 \sqle y$. The set of directed 
subsets of $Y$ will be denoted by $\directed{Y}$. If $h : Y_1 \to Y_2$ is monotone then 
clearly $h(A) \in \directed{Y_2}$ for each $A \in \directed{Y_1}$. 
\medskip
A poset $D$ is said to be 
\indexddef{complete}{complete poset}{}{poset}{complete} 
if the least upper bound
$\lub A$ of $A$ exists for each $A \in \directed{D}$. If $D_1$ and $D_2$ are
complete posets then a mapping $h : D_1 \to D_2$ is said to be 
\indexddef{continuous}{continuous mapping}{}{mapping}{continuous}
if $h$ is monotone and $h(\lub A) = \lub h(A)$ for
each $A \in \directed{D_1}$. Note that if $h : D_1 \to D_2$ is monotone then
$\lub h(A) \sqle_2 h(\lub A)$ always holds (with $\sqle_2$ the partial
order on $D_2$). Thus a monotone mapping $h$ is continuous if and only if 
$h(\lub A) \sqle_2 \lub h(A)$ for each $A \in \directed{D_1}$. 

\proclaim{Proposition~4.2.1} If $D_1$,
$D_2$ and $D_3$ are complete posets and $f : D_1 \to D_2$ and
$g : D_2 \to D_3$ are continuous then the composition $g \,f : D_1 \to D_3$
is also continuous. \endpro

\proof This is clear. \eop
\bigbreak

\proclaim{Proposition~4.2.2} Let $D$ be a complete poset and $Y$ a poset
isomorphic to $D$; then $Y$ is also complete. Moreover, if $h : Y \to D$ is
any order isomorphism then $h$ and the inverse mapping $h^{-1} : D \to Y$ 
are both automatically continuous. \endpro

\proof Let $A \in \directed{Y}$; then $B = h(A) \in \directed{D}$, thus let 
$u = \lub B$ and put $y = h^{-1}(u)$. Now $u'$ is an upper bound of $B$ if and 
only if $h^{-1}(u')$ is an upper bound of $A$, and hence $y = \lub A$. This 
shows that $Y$ is complete and that $h$ is continuous (since 
$h(\lub A) = h(y) = u = \lub B = \lub h(A)$. The continuity of $h^{-1}$ 
follows by reversing the roles of $D$ and $Y$. \eop  
\bigbreak
In what follows let $D_S$ be a family of complete posets (with $\sqle_\sigma$ the partial
order on $D_\sigma$ for each $\sigma \in S$) and let $L$ be an \tsym{S}-typed set.

\proclaim{Lemma~4.2.6} Let $A \in \directed{\ass L D }$ and for each $\eta \in L$ put 
$A_\eta = \lcurl c(\eta) : c \in A \rcurl$. Then 
$A_\eta \in \directed{D_\eta}$, and the least upper bound of $A$ is 
the assignment $c \in \ass L D $ defined by $c(\eta) = \lub A_\eta$ for each 
$\eta \in L$. \endpro

\proof Straightforward. \eop
\bigbreak

\proclaim{Proposition~4.2.3} The poset $\ass L D $ is complete. \endpro

\proof This follows immediately from Lemma~4.2.6. \eop
\bigbreak

\proclaim{Proposition~4.2.4} 
Let $\varphi_S : D_S \to D'_S$ be a family of continuous mappings
(with $D'_S$ a further family of complete posets). Then the mapping
$\ass L {\varphi}  : \ass L D \to \assb L {D'} $ is also continuous. \endpro

\proof If $c,\,c' \in \ass L D $ with $c\ \ass L {\sqle} \ c'$ then for each $\eta \in J$ 

\mdisp{\ass L {\varphi} (c)\,(\eta) 
\ =\ \varphi_\eta(c(\eta))\ \sqle_\eta \ \varphi_\eta(c'(\eta))
\ =\ \ass L {\varphi} (c')\,(\eta) \ ,}

since $\varphi_\eta$ is monotone, and hence $\ass L {\varphi} $ is monotone.
Now let $A \in \directed{\ass L D }$; then, with two applications of Lemma~4.2.6 
and since $\varphi_\eta$ is continuous, it follows that
\smallskip
\ldisp{\quad
\ass L {\varphi} (\lub A)\,(\eta)
       \ =\ \varphi_\eta \bigl(\,(\lub A)\,(\eta)\,\bigr)
\ =\ \varphi_\eta \bigl(\,\lub\, \lcurl c(\eta)\,:\,c\in A \rcurl\,\bigr)}
\vskip-\medskipamount
\rdisp{
\ =\ \lub \,\lcurl \varphi_\eta(c(\eta)) \,:\, c \in A \rcurl
 \ =\ \lub \,\lcurl \ass L {\varphi} (c)\,(\eta) \,:\, c \in A \rcurl
 \ =\ \bigl(\,\lub \ass L {\varphi} (A)\,\bigr)\,(\eta)\quad}
\smallskip
for each $\eta \in L$, and therefore 
$\ass L {\varphi} (\lub A) = \lub \ass L {\varphi} (A)$.
Thus $\ass L {\varphi} $ is continuous. \eop
\bigbreak
Note the following special case of Proposition~4.2.4: Let $n \ge 2$ and for $\oneto j n $
let $D_j$ and $D'_j$ be complete posets and 
$h_j : D_j \to D'_j$ be a continuous mapping. Then the mapping 
$\,h : D_1 \times \cdots \times D_n \to D'_1 \times \cdots \times D'_n\,$ defined by
\mdisp{h(\vector u n )\ =\ (h_1(u_1),\ldots,h_n(u_n))}
for each $(\vector u n ) \in D_1 \times \cdots \times D_n$ is continuous.

\proclaim{Lemma~4.2.7} For each $\eta \in L$ the projection mapping 
$\fss{p}^L_\eta : \ass L D \to D_\eta$ defined by $\fss{p}^L_\eta(c) = c(\eta)$
for each $c \in \ass L D $ is continuous. \endpro

\proof This follows immediately from Lemma~4.2.6. \eop
\bigbreak

\proclaim{Proposition~4.2.5} A mapping $h : D \to \ass L D $ from a complete poset $D$ 
to the complete poset
$\ass L D $ is continuous if and only if for each $\eta \in L$ the mapping
$\fss{p}^L_\eta\,h$ from $D$ to $D_\eta$ is continuous. \endpro

\proof If $h$ is continuous then $\fss{p}^L_\eta\,h$ is also continuous for each 
$\eta \in L$ (since it is the composition of two
continuous mappings). Conversely, suppose
$\fss{p}^L_\eta\,h$ is continuous for each $\eta \in L$; in particular, 
$h$ is then monotone (as in Lemma~4.2.2). Let $A \in \directed{D}$; 
then by Lemma~4.2.6 it follows that for each $\eta \in L$
\smallskip
\ldisp{\quad
\bigl(\, \lub h(A) \,\bigr)\,(\eta)
\ =\ \lub \,\lcurl c(\eta)\,:\, c \in h(A) \rcurl}
\vskip-\medskipamount
\vskip-\smallskipamount
\rdisp{
\ =\ \lub \fss{p}^L_\eta(h(A)) \ =\ (\fss{p}^L_\eta\,h)\,\bigl(\,\lub A \,\bigr)
\ =\ h(\lub A)\,(\eta) \quad}
\smallskip
and so $\,\lub h(A) \, =\, h(\lub A)$. This shows $h$ is continuous. \eop 
\bigbreak
If $D_1$ and $D_2$ are complete posets then 
the set of all continuous mappings from $D_1$ to $D_2$ will be denoted by
$\cont D_1 D_2 $. Then
$\cont D_1 D_2 $ is a subset of $\mono D_1 D_2 $, and so $\cont D_1 D_2 $ can be considered 
as a poset with the partial order induced by the partial order on 
$\mono D_1 D_2 $. 
\medskip
Note that if in addition $D_2$ is bottomed with bottom element
$\bot_2$ then the bottom element of $\mono D_1 D_2 $ (i.e., the constant 
mapping $\bot$ with $\bot(u) = \bot_2$ for all $u \in D_1$) is an element
of $\cont D_1 D_2 $, and thus in this case $\cont D_1 D_2 $ is a bottomed poset.

\proclaim{Proposition~4.2.6} If $D_1$, $D_2$ are complete posets then the
poset $D = \cont D_1 D_2 $ is also complete. Moreover, if 
$A \in \directed{D}$ with $h = \lub A$ then 
$h(u) = \lub \,\lcurl f(u) \,:\, f \in A \rcurl$ for each $u \in D_1$. \endpro

\proof Let $A \in \directed{D}$, and for each $u \in D_1$ put
$A_u = \lcurl f(u) \,:\, f \in A \rcurl$. Then $A_u \in \directed{D_2}$ and 
so define a mapping $h : D_1 \to D_2$ by putting $h(u) = \lub A_u$ for each
$u \in D_1$. It will be shown that $h$
is continuous, and, since $h$ is clearly monotone, for this it is enough
to show that if $B \in \directed{D_1}$ with $u' = \lub B$ then
$h(u') \sqle_2 \lub h(B)$. Let $f \in A$; then 
\smallskip
\ldisp{\quad f(u') \ =\ \lub f(B) 
\ =\ \lub\, \bigl(\,\bigcup\limits_{u \in B} f(u) \,\bigr) }
\vskip-\medskipamount
\rdisp{
       \ \sqle_2\ \lub \,\bigl(\,\bigcup\limits_{u \in B} \lub A_u \,\bigr)
          \ =\ \lub\, \bigl(\,\bigcup\limits_{u \in B} h(u) \,\bigr) 
 \ =\ \lub h(B)\ .\quad}
\smallskip
Hence $h(u') = \lub A_{u'} \sqle_2 \lub h(B)$. Now (as in Lemma~4.2.6) it is 
straightforward to check that $h$ is the least upper bound of $A$ in 
$\cont D_1 D_2 $. \eop
\bigbreak

In what follows let $D_1,\, D_2$ and $D$ be complete posets.

\proclaim{Lemma~4.2.8} A monotone mapping $h : D_1 \times D_2 \to D$
is continuous if and only if 
\mdisp{h(\lub A_1,\lub A_2) \ =\ \lub h(A_1 \times A_2)} 
for all $A_1 \in \directed{D_1}$, $A_2 \in \directed{D_2}$. \endpro

\proof If $A \in \directed{D_1 \times D_2}$ then it is easy to see that there exist
$A_1 \in \directed{D_1}$ and $A_2 \in \directed{D_2}$ 
such that $A' = A_1 \times A_2$ is an element of
$\directed{D_1 \times D_2}$ which is equivalent to $A$
in the sense that for each $v \in A$ there exists $v' \in A'$ with $v \sqle v'$
and for each $v' \in A'$ there exists $v \in A$ with $v' \sqle v$.
Then $\lub A = \lub A' = (\lub A_1,\lub A_2)$ and $\lub h(A') = \lub h(A)$,
and this shows that if
$h(\lub A_1,\lub A_2) = \lub h(A_1 \times A_2)$ for all 
$A_1 \in \directed{D_1}$, $A_2 \in \directed{D_2}$ then $h$ is 
continuous. The converse is clear, since 
$A_1 \times A_2 \in \directed{D_1 \times D_2}$
whenever $A_1 \in \directed{D_1}$ and $A_2 \in \directed{D_2}$. \eop 

\proclaim{Lemma~4.2.9}  A mapping $h : D_1 \times D_2 \to D$ 
is continuous if and only if it is separately continuous in each argument, i.e., if and 
only if 
$u_2 \mapsto h(u_1,u_2)$ is a continuous mapping from $D_2$ to $D$ 
for each $u_1 \in D_1$ and
$u_1 \mapsto h(u_1,u_2)$ is a continuous mapping from $D_1$ to $D$ 
for each $u_2 \in D_2$. \endpro

\proof Clearly $h$ is monotone if and only if it is separately monotone in each argument. 
Moreover, if $h$ is continuous then it is separately continuous, since if 
$A \in \directed{D_2}$ then $\{u_1\}\times A \in \directed{D_1\times D_2}$ and
$\,\lub \,(\{u_1\} \times A)\, =\, (u_1,\lub A)\,$, with of course an
analogous statement holding when the roles of the two components are reversed. 
Suppose then that $h$ is separately continuous in each argument. Lemma~4.2.8 will be 
used to show that $h$ is continuous; thus let
$A_1 \in \directed{D_1}$ and $A_2 \in \directed{D_2}$ and put
$u_1 = \lub A_1$ and $u_2 = \lub A_2$. 
Then $A_1 \times \{u_2\} \in \directed{D_1 \times D_2}$ and

\mdisp{h(u_1,u_2) 
  \ =\ \lub \,\lcurl h(v,u_2) \,:\, v \in A_1 \rcurl 
                                               \ =\ \lub h(A_1\times \{u_2\})\ ,}

since $h$ is continuous in its first argument. Moreover,
since $h$ is continuous in its second argument, it follows that
\smallskip
\ldisp{\quad \lub h(A_1 \times \{u_2\})
   \ =\ \lub \,\lcurl h(v_1,\lub A_2) \,:\,
                                       v_1 \in A_1  \rcurl}
\vskip-\smallskipamount
\mdisp{\ =\  \lub \,\lcurl \lub \,\lcurl h(v_1,v_2) 
                           \,:\, v_2 \in A_2 \rcurl \,:\, v_1 \in A_1 \rcurl}
\vskip-\smallskipamount
\rdisp{\ \ =\   \lub \,\lcurl h(v_1,v_2) \,:\, v_1 \in A_1,\, v_2 \in A_2 \rcurl 
                                              \ =\ \lub h(A_1\times A_2)\ .\quad}
\smallskip
Therefore
$\ h(u_1,u_2) \, =\, \lub h(A_1 \times \{u_2\}) \, =\, \lub h(A_1 \times A_2)\ $
and thus by Lemma~4.2.8 $h$ is continuous. \eop

\proclaim{Proposition~4.2.7} The 
\indexddef{application operator}{application operator}{}{operator}{application} 
$\fss{a} : \cont {D_1} {D_2}  \times D_1 \to D_2$ given by 
\mdisp{\fss{a}(h,u)\ =\ h(u)}
for all $h \in \cont {D_1} {D_2} $, $u \in D_1$, is continuous. \endpro

\proof This follows immediately from Lemma~4.2.9, since
$\fss{a}$ is clearly continuous in its second argument, and by
Proposition~4.2.6 it is also continuous in its first argument. 
\eop
\bigbreak
Now consider a continuous mapping $h : D_1 \times D_2 \to D$;
then by Lemma~4.2.9 the mapping $\,u_2 \,\mapsto\, h(u_1,u_2)\,$ from
$D_2$ to $D$ is continuous for each $u_1 \in D_1$, which means 
a mapping $\fss{c}(h) : D_1 \to \cont D_2 D $ can be defined by letting

\mdisp{\fss{c}(h)\,(u_1)\,(u_2) \ =\ h(u_1,u_2)}

for each $u_1 \in D_1$, $u_2 \in D_2$.

\proclaim{Lemma~4.2.10} $\fss{c}(h) : D_1 \to \cont D_2 D $ is continuous for each 
continuous mapping $h : D_1 \times D_2 \to D$. \endpro

\proof It is clear that $\fss{c}(h)$ is monotone. Let $A \in \directed{D_1}$
and put $h = \lub \fss{c}(h) (A)$ (so $h \in \cont D_2 D $). Then by
Proposition~4.2.6 it follows  for each $u_2 \in D_2$ that

\mdisp{h(u_2) \ =\ \lub\, \lcurl \fss{c}(h)(u_1)\,(u_2) \,:\, u_1 \in A \rcurl
               \ =\ \lub \,\lcurl h(u_1,u_2) \,:\, u_1 \in A \rcurl\ ,}

and by Lemma~4.2.9 that

\mdisp{\lub \,\lcurl h(u_1,u_2) \,:\, u_1 \in A \rcurl
                  \ =\ h(\lub A,u_2) \,=\, \fss{c}(h) (\lub A)\,(u_2)\ .}

Hence $\lub \fss{c}(h) (A) = \fss{c}(h)(\lub A)$, and thus $\fss{c}(h)$ is
continuous. \eop
\bigbreak
Lemma~4.2.10 allows the so-called 
\indexddef{currying operator}{currying operator}{}{operator}{currying} 
to be defined;
this is the mapping
$\fss{c} : \cont {D_1 \times D_2} {D} \to \cont D_1  {\cont D_2 D } $  
given by
\mdisp{ \fss{c} (h,u_1)\,(u_2)\ =\ h(u_1,u_2)}
for all $h \in \cont {D_1 \times D_2} D $, $u_1 \in D_1$, $u_2 \in D_2$.

\proclaim{Proposition~4.2.8} The currying operator $\fss{c}$
is an order isomorphism whose inverse is the 
\indexddef{uncurrying operator}{uncurrying operator}{}{operator}{uncurrying}
$\,\fss{u} : \cont D_1 {\cont D_2 D } \to \cont {D_1 \times D_2} D \,$
defined by 
\mdisp{\fss{u}(g)\,(u_1,u_2) \,=\, g(u_1)\,(u_2)} 
for all $g \in \cont D_1 {\cont D_2 D } $,
$(u_1,u_2) \in D_1 \times D_2$. (Proposition~4.2.2 then implies that 
the mappings $\fss{c}$ and $\fss{u}$ are both continuous.) \endpro 

\proof The inverse mapping to $\fss{c}$ will be constructed. For each 
$g \in \cont D_1 {\cont D_2 D } $ define
$\fss{u}(g) : D_1 \times D_2 \to D$ by
$\fss{u}(g) \,(u_1,u_2) = g(u_1)\,(u_2)$. Thus $\fss{u}(g)$ 
is monotone, and if $A_1 \in \directed{D_1},\, A_2 \in \directed{D_2}$
then
\smallskip
\ldisp{\quad \fss{u}(g)\,(\lub A_1,\lub A_2) 
\ =\ g(\lub A_1)\,(\lub A_2)
\ =\ \lub\, \lcurl g(u_1)\,(\lub A_2) \,:\, u_1 \in A_1 \rcurl
}
\vskip-\smallskipamount
\ldisp{\qquad\quad 
\ =\ \lub\, \lcurl \lub \,\lcurl g(u_1)\,(u_2) : u_2 \in A_2 \rcurl \,:\,
                                                       u_1 \in A_1 \rcurl}
\vskip-\smallskipamount
\ldisp{\qquad\qquad\quad
  \ =\ \lub \,\lcurl g(u_1)\,(u_2) \,:\, u_1 \in A_1,\, u_2 \in A_2 \rcurl}
\vskip-\smallskipamount
\rdisp{ 
  \ =\ \lub \,\lcurl \fss{u}(g)\,(u_1,u_2) 
      \,:\, u_1 \in A_1,\, u_2 \in A_2 \rcurl 
                              \ =\ \lub \fss{u}(g)(A_1 \times A_2)\ ;\quad}  
\smallskip
hence by Lemma~4.2.8 $\fss{u}(g)$ is continuous. This gives a mapping

\mdisp{\fss{u} : \cont D_1 {\cont D_2 D } \to \cont {D_1 \times D_2} D }

and it is clear that $\fss{u}$ is monotone. Moreover, it is also clear
that $\fss{c}(\fss{u}(g)) = g$ for each
$g \in \cont D_1 {\cont D_2 D } $ and $\fss{u}(\fss{c}(h)) = h$ for each
$h \in \cont {D_1 \times D_2} D $. Thus $\fss{c}$ is an order isomorphism
and $\fss{u}$ is the corresponding inverse mapping. \eop
\bigbreak

\proclaim{Proposition~4.2.9} Let $D$ be a bottomed complete poset and
$h : D \to D$ be a continuous mapping. Then $h$ possesses a least 
fixed point. More precisely, the set
\mdisp{ \fss{Fix}(h)\ =\ \lcurl u \in D \,:\, h(u) = u \rcurl }
is non-empty and contains a (unique) least element, i.e.,
$\fss{Fix}(h)$ contains an element ${\hat u}$ such that 
${\hat u} \sqle u$ for all $h \in \fss{Fix}(h)$. \endpro

\proof Define a sequence $\{u_n\}_{n \ge 0}$ of elements of $D$ by
$u_0 = \bot$ (with $\bot$ the bottom element of $D$) and
$u_{n+1} = h(u_n)$ for each $n \ge 0$. Then, since $h$ is monotone, it
follows by induction that $u_n \sqle u_{n+1}$ for all $n \ge 0$ 
(since $u_0 = \bot \sqle u_1$ and if $u_n \sqle u_{n+1}$ then
$u_{n+1} = h(u_n) \sqle h(u_{n+1}) = u_{n+2}$). Thus the set

\mdisp{A\ =\ \lcurl u \in D \,:\, u = u_n \frm{for some} n \ge 0 \rcurl} 

is directed and,
because $h(A) = \lcurl u \in D \,:\, u = u_n \frm{for some} n \ge 1 \rcurl$,
it follows that $\lub A = \lub h(A)$.
But $\lub h(A) = h(\lub A)$, since $h$ is continuous, and this implies that
${\hat u} = \lub A \in \fss{Fix}(h)$. Now let $u$ be any element of
$\fss{Fix}(h)$; again by induction it follows that
$u_n \sqle {\hat u}$ for all $n \ge 0$
(since $u_0 = \bot \sqle {\hat u}$ and 
$u_{n+1} = h(u_n) \sqle h({\hat u}) = {\hat u}$ whenever $u_n \sqle {\hat u}$).
Hence $u$ is an upper bound of $A$ and therefore 
${\hat u} = \lub A \sqle u$. \eop
\bigbreak
The proof of Proposition~4.2.9 is just as important as its statement: This shows 
how the minimal fixed point ${\hat u}$ can be constructed `explicitly'.
\bigskip

\vfill\eject

\hfline{96}{4.3 \ INITIAL COMPLETIONS}

\sectionhead {4.3} {Initial completions}
\bigskip\medskip

A poset $Y_1$ is said to be a 
\indexdef{subposet}{subposet}{}
of a poset $Y_2$ if $Y_1 \subset Y_2$ and the partial order on $Y_1$ is obtained by 
restricting the partial order on $Y_2$. 
(This means that if $\sqle_j$ is the partial order on $Y_j$ for 
$j = 1,\, 2$ and  $y,\, y' \in Y_1$ then
$y \sqle_1 y'$ if and only if $y \sqle_2 y'$.) 
The  poset $Y_2$ is then said to be an 
\indexddef{extension}{extension of a poset}{}{poset}{extension of}
of $Y_1$.
\medskip
A poset $D$ is said to be a
\indexddef{completion}{completion of a poset}{}{poset}{completion of}
of a poset $Y$ if $D$ is a complete extension of $Y$
and each element of $D$ is the least upper bound (in $D$) of some element of
$\directed{Y}$. 
\medskip
{\it Warning:\ } Let $D$ be a complete extension of a poset $Y$ and let 
$A \in \directed{Y}$. Then it is possible that $A$ has a least upper bound in
$Y$ which is not equal to the least upper bound of $A$ in $D$. (Note, however,
that if the least upper bound of $A$ in $D$ is an element of $Y$ then in this
case it must also be the least upper bound of $A$ in $Y$.) When speaking
just of the least upper bound of $A$ or writing $\lub A$ then the
least upper bound in $D$ is always meant. 
\medskip
If $Y$ is a poset with partial order $\sqle$ and $D$ is an extension of $Y$
then, unless something explicit to the contrary is stated,  
the partial order on $D$ will also be denoted by $\sqle$.

\proclaim{Lemma~4.3.1} Let $Y$ be a bottomed poset with bottom element $\bot$
and let $D$ be a completion of $Y$. Then $D$ is bottomed and $\bot$ is also 
the bottom element of $D$. \endpro

\proof If $u \in D$ then $u = \lub A$ for some $A \in \directed{Y}$. But
$\bot \sqle y$ for each $y \in A$, and hence also $\bot \sqle u$. \eop
\bigbreak

\proclaim{Proposition 4.3.1} Let $Y_S$ be a family of posets, for each
$\sigma \in S$ let $D_\sigma$ be a completion of $Y_\sigma$, and let $J$ be
an \tsym{S}-typed set. Then $\ass J D $ is a completion of 
$\ass J Y $. \endpro

\proof Let $c \in \ass J D $. Then for each $\eta \in J$ there exists 
$A_\eta \in \directed{Y_\eta}$ with $c(\eta) = \lub A_\eta$
(since $D_\eta$ is a completion of $Y_\eta$).
Put
$A = \lcurl b \in \ass J Y \,:\, b(\eta) \in A_\eta \frm{for each} \eta \in J \rcurl$;
then $A \in \directed{\ass J Y }$: If $c_1,\,c_2 \in A$ then for each $\eta \in J$
there exists $y_\eta \in A_\eta$ with $c_1(\eta) \sqle_\eta  y_\eta$ and
$c_2(\eta) \sqle_\eta y_\eta$; thus if $c \in \ass J Y $ is defined
by letting $c(\eta) = y_\eta$ for each $\eta \in J$ then $c \in A$ and
$c_1\ \ass J {\sqle} \ c$, $c_2\ \ass J {\sqle} \ c$. Moreover, $c = \lub A$:
If $b \in A$ then $b(\eta) \in A_\eta$ and so 
$b(\eta) \sqle_\eta \lub A_\eta = c(\eta)$ for each $\eta \in J$, i.e.,
$b\ \ass J {\sqle} \ c$. On the other hand, if $c' \in \ass J D $ is an upper bound of $A$
then $b(\eta) \sqle_\eta c'(\eta)$ for each $b \in A$ and this implies
that $y \sqle_\eta c'(\eta)$ for all $y \in A_\eta$
(since for each $y \in A_\eta$ there exists $b \in A$ with $b(\eta) = y$)
and thus that $c(\eta) = \lub A_\eta \sqle_\eta c'(\eta)$ for each
$\eta\in J$, i.e., $c\ \ass J \ {\sqle} \ c'$.
This, together with Proposition~4.2.3, shows that $\ass J D $ is a 
completion of $\ass J Y $. \eop
\bigbreak
Extensions $Y_1$ and $Y_2$ of a poset $Y$ are said to be 
\indexdef{conjugate}{conjugate extension of a poset}{}
if there exists an order isomorphism
$h : Y_1 \to Y_2$ with $h(y) = y$ for all $y \in Y$. 
Conjugacy clearly defines an equivalence relation on the class of all 
extensions of $Y$. If $D_1$ and $D_2$ are conjugate extensions of $Y$ then
$D_1$ is a completion of $Y$ if and only if $D_2$ is.
\medskip
A completion $D$ of a poset $Y$ is said to be 
\indexddef{initial}{initial completion of a poset}{}{completion of a poset}{initial}
if for each completion $D'$ of $Y$ there exists a unique continuous mapping
$h : D \to D'$ with $h(y) = y$ for all $y \in Y$. (Note that the mapping $h$
is then automatically surjective.) It is easily checked that if $D_1$ and
$D_2$ are conjugate completions of $Y$ then $D_2$ is initial if and only if
$D_1$ is. Proposition~4.3.2 below states that each poset $Y$ possesses an 
essentially unique initial completion, which is then often called
{\it the\/} initial completion of $Y$. In fact, for reasons to be 
explained at the end of the section, this is also
referred to as the 
\indexddef{ideal completion}{ideal completion of a poset}{}{completion of a poset}{ideal}
of $Y$.
\medskip
A seemingly stronger requirement than being initial on the completion of a 
poset is the following: A completion $D$ of $Y$ is said to have the
\indexdef{continuous extension property}{continuous extension property}{}
if whenever $h : Y \to D'$ is a monotone mapping from $Y$ 
to a complete poset $D'$ then $h$ can be uniquely extended to a continuous 
mapping $h' : D \to D'$ (i.e., there exists a unique continuous mapping 
$h' : D \to D'$ with $h'(y) = h(y)$ for all $y \in Y$). It is clear that a
completion $D$ of $Y$ with the continuous extension property is initial, since
if $D'$ is any completion of $Y$ then the mapping $i : Y \to D'$ with
$i(y) = y$ for all $y \in Y$ is monotone. The next result shows in particular
that the converse is true.

\proclaim{Proposition~4.3.2} Let $Y$ be a poset. Then there is exactly one 
conjugacy class of initial completions of $Y$. Moreover, each initial 
completion of $Y$ has the continuous extension property. \endpro

\proof Let $\sqle$ be the partial order on $Y$. It will be shown first that there is 
at most one conjugacy class of initial completions.

\proclaim{Lemma~4.3.2} Initial completions $D$ and $D'$ of $Y$ are 
conjugate. \endpro

\proof There exists a continuous mapping $h : D \to D'$ with $h(y) = y$ for 
all $y \in Y$ and, reversing the roles of $D$ and $D'$, also a continuous 
mapping $h' : D' \to D$ with $h'(y) = y$ for all $y \in Y$. But if 
$\fss{id}_D : D \to D$ is the identity mapping then 
$h' \, h = \fss{id}_D$, since $h' \, h$ and $\fss{id}_D$ are both 
continuous mappings from $D$ to $D$ with $h' \, h (y) = \fss{id}_D(y) = y$
for all $y \in Y$. In the same way $h \, h' = \fss{id}_{D'}$.
Hence $h$ is a bijection and both $h$ and $h^{-1} = h'$ are continuous. This
shows that $D$ and $D'$ are conjugate. \eop
\bigbreak
If $D_1$ and $D_2$ are conjugate completions of $Y$ then it is easy to see 
that $D_2$ has the continuous extension property if and only if $D_1$ does. 
The proof of Proposition~4.3.2 thus reduces to showing there exists a 
completion of $Y$ having this property. Moreover, the following lemma implies
that in the definition of the continuous
extension property the statement `can be uniquely extended to a continuous 
mapping' can be replaced by just `can be extended to a continuous mapping'. 

\proclaim{Lemma~4.3.3} Let $D$ be a completion of a poset $Y$ and 
$h : Y \to D'$ be a monotone mapping from $Y$ to a complete poset $D'$. Then 
there exists at most one extension of $h$ to a continuous mapping 
$h' : D \to D'$ (i.e., there exists at most one continuous mapping 
$h' : D \to D'$ with $h'(y) = h(y)$ for all $y \in Y$). \endpro

\proof Let $u \in D$; then there exists $A \in \directed{Y}$ with $u = \lub A$
and therefore follows that $h'(u) = \lub h(Y)$. \eop
\bigbreak 
Let $A_1,\, A_2 \in \directed{Y}$; then $A_1$ is said to be 
\indexddef{cofinal}{cofinal directed set}{}{directed set}{cofinal}
in $A_2$ if for each $y_1 \in A_1$ there exists
$y_2 \in A_2$ with $y_1 \sqle y_2$. 
If $D$ is a complete extension of $Y$ 
and $A_1,\, A_2 \in \directed{Y}$ with $A_1$ cofinal in $A_2$ then 
clearly $\lub A_1 \sqle \lub A_2$.

\proclaim{Lemma~4.3.4} Let $D$ be a complete extension of $Y$ (but not
necessarily a completion of $Y$) and suppose that whenever 
$A_1,\, A_2$ are elements of $\directed{Y}$ with 
$\lub A_1 \sqle \lub A_2$ then $A_1$ is cofinal in $A_2$. Let $D'$ denote the 
set of all elements of $D$ having the form $\lub A$ for some 
$A \in \directed{Y}$. Then for each $B \in \directed{D'}$ there exists 
$A \in \directed{Y}$ with $A$ cofinal in $B$ and $\lub A = \lub B$. \endpro

\proof Let $B \in \directed{D'}$ and for each $u \in B$ choose 
$A_u \in \directed{Y}$ with $u = \lub A_u$. Then
$A = \bigcup_{u \in B} A_u \in \directed{Y}$. To see this let 
$y_1,\, y_2 \in A$; then there exist $u_1,\, u_2 \in B$ with
$y_1 \in A_{u_1}$ and $y_2 \in A_{u_2}$ and, since $B \in \directed{D'}$, 
there exists $u \in B$ such that $u_1 \sqle u$ and $u_2 \sqle u$. But then by 
assumption $A_{u_1}$ is cofinal in $A_u$ and $A_{u_2}$ is cofinal in $A_u$. 
Hence 
$y'_1,\, y'_2 \in A_u$ can be found with $y_1 \sqle y'_1$ and $y_2 \sqle y'_2$
and, since $A_u \in \directed{Y}$, it follows that $y_1 \sqle y'_1 \sqle y$ and
$y_2 \sqle y'_2 \sqle y$ for some $y \in A_u \subset A$. 
\medskip
Now if $y \in A$ then $y \in A_u$ for some $u \in B$, and so $y \sqle u $;
this implies $A$ is cofinal in $B$ and thus also that 
$\lub A \sqle \lub B$. On the
other hand, if $u \in B$ then $u \sqle \lub A$, since $u = \lub A_u$ and
$A_u \subset A$, and therefore $\lub B \sqle \lub A$. Hence $\lub A = \lub B$.
\eop

\proclaim{Lemma~4.3.5} Let $D$ be a completion of $Y$ such that whenever 
$A_1,\, A_2 \in \directed{Y}$ with $\lub A_1 \sqle \lub A_2$ then $A_1$ is cofinal in 
$A_2$. Then $D$ has the continuous extension property. \endpro

\proof Let $h : Y \to D'$ be a monotone mapping of $Y$ to a complete poset
$D'$. If $A_1,\, A_2 \in \directed{Y}$ with $\lub A_1 = \lub A_2$ 
then each of $A_1$ and $A_2$ is cofinal in the other and therefore 
$\lub h(A_1) = \lub h(A_2)$. 
Thus, since each element of $D$ has the form
$\lub A$ for some $A \in \directed{Y}$, a mapping 
$h' : D \to D'$ can be defined by putting $h'(u) = \lub h(A)$,
where $A$ is any element of $\directed{Y}$ with $u = \lub A$.
Then $h'$ is monotone and an
extension of $h$, because $\{y\} \in \directed{Y}$ and $\lub \{y\} = y$ for
each $y \in Y$. 
\medskip
It remains to show that the mapping $h'$ is continuous, and for this it is
enough to show that $h'(\lub B) \sqle \lub h'(B)$ for each 
$B \in \directed{D}$ (with $\sqle$ also denoting the partial order on $D'$). 
Let $B \in \directed{D}$; then by Lemma~4.3.4 there exists 
$A \in \directed{Y}$ with $A$ cofinal in $B$ and $\lub A = \lub B$. Hence
$h'(\lub B) = h'(\lub A) = \lub h(A) = \lub h'(A) \sqle \lub h'(B)$. \eop
\bigbreak
A completion of $Y$ will now be constructed and Lemma~4.3.5 then used to show that 
it has the continuous extension property. Elements $A_1,\,A_2 \in \directed{Y}$
are said to be 
\indexddef{mutually cofinal}{mutually cofinal directed sets}{}
{directed sets}{mutually cofinal}
if each is cofinal in the other.
Mutual cofinality clearly defines an equivalence relation on the set
$\directed{Y}$. Let $D$ denote the set of equivalence classes, and if 
$A \in \directed{Y}$ then denote by $[A]$ the equivalence class containing
$A$. If $A_1$ is equivalent to $A'_1$ and $A_2$ equivalent to $A'_2$ then 
clearly $A_1$ is cofinal in $A_2$ if and only if $A'_1$ is cofinal in $A'_2$.
Thus define a relation $\preceq$ on $D$ by letting 
$[A_1] \preceq [A_2]$ if $A_1$ is cofinal in $A_2$; it is immediate that  
$\preceq$ is a partial order. 

\proclaim{Lemma~4.3.6} The poset $D$ (with the partial order $\preceq$) is 
complete. \endpro

\proof Let $B \in \directed{D}$. For each $u \in B$ choose 
$A_u \in \directed{Y}$ with $[A_u] = u$ and put $A = \bigcup_{u \in B} A_u$.
Then $A \in \directed{Y}$. (Let $y_1,\, y_2 \in A$; then there exist
$u_1,\, u_2 \in B$ such that $y_1 \in A_{u_1}$ and $y_2 \in A_{u_2}$
and, since $B \in \directed{D}$, there exists $u \in B$ with 
$u_1 \preceq u$ and $u_2 \preceq u$. But this means that 
$y'_1,\, y'_2 \in A_u$ can be found with $y_1 \sqle y'_1$ and $y_2 \sqle y'_2$
and hence there exists $y \in A_u \subset A$ such that 
$y_1 \sqle y'_1 \sqle y$ and $y_2 \sqle y'_2 \sqle y$, because 
$A_u \in \directed{Y}$.) It is now enough to show that ${\hat u} = [A]$ 
is the least upper bound of $B$ in $D$. If $u \in B$ and $y \in A_u$ 
then $y \sqle y$ and $y \in A$; thus $A_u$ is cofinal in $A$. This implies 
that $u \preceq {\hat u}$, i.e., ${\hat u}$ is an upper bound of $B$. Now let 
$u' = [A']$ be any upper bound of
$B$. If $y \in A$ then $y \in A_u$ for some $u \in B$ and $A_u$ is cofinal in
$A'$, since $u \preceq u'$. There therefore exists $y' \in A'$ with 
$y \sqle y'$ and hence $A$ is cofinal in $A'$, i.e., ${\hat u} \preceq u'$. 
\eop
\bigbreak
Now $\{y\} \in \directed{Y}$ for each $y \in Y$, which means a
mapping $i : Y \to D$ can be defined by putting $i(y) = [\{y\}]$ for each $y \in Y$. 
Then $i$ is an embedding, since
$y_1 \sqle y_2$ if and only if $\{y_1\}$ is cofinal in $\{y_2\}$, and 
$Y$ can thus be considered as a subposet of $D$ by identifying $Y$ with $i(Y)$. But
this just means that $D$ is an extension of $Y$. Moreover, if $u = [A] \in D$
and $B = \bigcup_{y \in A} \{y\}$ then $B$ is a directed subset of 
$Y \subset D$ and, as in the proof of Lemma~4.3.6, $u = \lub B$. This 
shows $D$ is a completion of $Y$.
\medskip
Let $A_1,\, A_2 \in \directed{Y}$ with $\lub A_1 \preceq \lub A_2$.
Then $[A_1] \preceq [A_2]$, since $\lub A = [A]$ for each
$A \in \directed{Y}$, which by definition means that $A_1$ is cofinal in $A_2$.
Therefore by Lemma~4.3.5 $D$ has the continuous extension property. This 
completes the proof of Proposition~4.3.2. \eop
\bigbreak
{\it Warning:\ } If $D$ is an initial completion of a poset $Y$ then in
general $D$ is not an initial completion of itself. In fact, the class of
complete posets which are initial completions of themselves is very special
and will be characterised in Proposition~4.3.7.
\medskip
The next result gives various properties which are equivalent to a completion
being initial.

\proclaim{Proposition 4.3.3} Let $D$ be a completion of a poset $Y$. Then the
following five statements are equivalent:
\medskip\smallskip
{\parindent=25pt
\item{(1)} $D$ is initial. 
\medskip
\item{(2)} $D$ has the continuous extension property.
\medskip
\item{(3)} $A_1$ is cofinal in $A_2$ whenever $A_1,\, A_2$ are 
elements of $\directed{Y}$ with $\lub A_1 \sqle \lub A_2$. 
\medskip
\item{(4)} $y \sqle y'$ for some $y' \in A$ whenever $y \in Y$ and 
$A \in \directed{Y}$ with $y \sqle \lub A$. 
\medskip
\item{(5)} $y \sqle u$ for some $u \in A$ whenever $y \in Y$ and 
$A \in \directed{D}$ with $y \sqle \lub A$. 
\medskip\smallskip
}  \endpro

\proof To show that the first three statements are equivalent it is enough to prove 
the converse of Lemma~4.3.5. Thus assume 
that $D$ has the continuous extension property and let $A_1,\, A_2$ 
be elements of $\directed{Y}$ with $A_1$ not cofinal in $A_2$; then there 
exists $y_1 \in A_1$ such that $y_1 \not\sqle y$ for all $y \in A_2$. Choose 
some complete poset $D'$ containing elements $u,\, u'$ with 
$u \ne u'$ and $u' \sqle u$; then

\mdisp{h(y)\  =\ \cases{u  & if $\,y_1 \sqle y\,$, \cr
                         u' & otherwise, \cr}}

defines a monotone mapping $h : Y \to D'$ which by assumption extends to a 
continuous mapping $h' : D \to D'$. But $h'(\lub A_2) = \lub h'(A_2) = u'$
and $h'(\lub A_1) = \lub h'(A_1) = u$ and hence 
$\lub A_1 \sqle \lub A_2$ cannot hold. Thus if $\lub A_1 \sqle \lub A_2$ then $A_1$ must
be cofinal in $A_2$.
\medskip
This shows that (1), (2) and (3) are equivalent. But it is clear that
(3) and (4) are equivalent and that (5) implies (4); it thus remains to
show that the last statement is implied by the others and this follows 
immediately from Lemma~4.3.4. \eop 

\proclaim{Proposition~4.3.4} Let $D$ be an initial completion of a poset $Y$
and for each $u \in D$ let $A_u = \lcurl y \in Y \,:\, y \sqle u \rcurl$. Then
$A_u \in \directed{Y}$ and $\lub A_u = u$. \endpro

\proof Let $u \in D$; then, since $D$ is a completion of $Y$, there exists 
$A \in \directed{Y}$ with $\lub A = u$. Thus $A \subset A_u$ and so in
particular $A_u \ne \varnothing$. Let $y \in A_u$; then $y \sqle u = \lub A$ 
and hence by Proposition~4.3.3 there exists $y' \in A$ with $y \sqle y'$. 
Therefore if $y_1,\, y_2 \in A_u$ then there exist 
$y'_1,\, y'_2 \in A$ with $y_1 \sqle y'_1$ and $y_2 \sqle y'_2$.
But $A \in \directed{Y}$ and so there exists $y' \in A$ with $y'_1 \sqle y'$
and $y'_1 \sqle y'$, i.e., there exists $y' \in A_u$ with $y_1 \sqle y'$ and
$y_1 \sqle y'$. This shows that $A_u \in \directed{Y}$. It is now clear that
$\lub A_u = u$, since $u = \lub A \sqle \lub A_u$ and by definition $u$ is
an upper bound of $A_u$. \eop

\proclaim{Proposition~4.3.5} Let $D$ be an initial completion of a poset $Y$ 
and $Y'$ be a subposet of $Y$; let $D'$ be the set of all elements of
$D$ having the form $\lub A'$ for some $A' \in \directed{Y'}$. Then $D'$ is an
initial completion of $Y'$. Moreover, if $B \in \directed{D'}$ then $\lub B$
(the least upper bound of $B$ in $D$) is an element of $D'$, and hence
$\lub B$ is also the least  upper bound of $B$ in $D'$. \endpro

\proof If $A'_1,\, A'_2$ are elements of $\directed{Y'}$ (and thus 
also of $\directed{Y}$) with $\lub A'_1 \sqle \lub A'_2$ then by 
Proposition~4.3.3 $A'_1$ is cofinal in $A'_2$. Let 
$B \in \directed{D'}$; then 
Lemma~4.3.4 (applied with $D$ considered as a complete extension of $Y'$) 
implies that there exists $A' \in \directed{Y'}$ with $A'$ cofinal in $B$ and
$\lub A' = \lub B$. Hence $\lub B$ is an element of $D'$ and so $\lub B$ is
also the least upper bound of $B$ in $D'$. In particular, $D'$ is complete,
and therefore a completion of $Y'$. But now Proposition~4.3.3 implies that
$D'$ is an initial completion of $Y'$. (If
$A_1,\, A_2 \in \directed{Y'}$ with $\lub A_1 \sqle \lub A_2$ then
$A_1$ is cofinal in $A_2$, since $A_1,\, A_2 \in \directed{Y}$ and 
$D$ is an initial completion of $Y$.) \eop
\bigbreak

\proclaim{Proposition 4.3.6} Let $Y_S$ be a family of posets, for each
$\sigma \in S$ let $D_\sigma$ be an initial completion of $Y_\sigma$ and let
$J$ be a finite \tsym{S}-typed set. Then $\ass J D $ is an initial completion of 
$\ass J Y $. \endpro

\proof By Proposition~4.3.1 $\ass J D $ is a completion of $\ass J Y $. 
Let $A$ and $A'$ be elements of $\directed{\ass J Y }$ with
$\lub A \ass J {\sqle} \lub A'$. For each $\eta \in J$ let 
$A_\eta = \lcurl b(\eta) : b \in A \rcurl$ and 
$A'_\eta = \lcurl b(\eta) : b \in A' \rcurl$.
Then $A_\eta,\,A'_\eta \in \directed{Y_\eta}$ and 
$\lub A_\eta \sqle_\eta  \lub A'_\eta$ and thus, since 
$D_\eta$ is initial, $A_\eta$ is cofinal in $A'_\eta$ for each $\eta \in J$.
From this it follows that $A$ is cofinal in $A'$:
Let $c \in A$; then for each $\eta \in J$ there exists $c_\eta \in A'$ with
$c(\eta) \sqle_\eta  c_\eta(\eta)$, and therefore, since $J$ is finite
and $A'$ is directed, there exists $c' \in A'$ with $c_\eta\ \ass J {\sqle} \ c'$ for all
$\eta \in J$. But then $c\ \ass J {\sqle} \ c'$.
Hence by Proposition~4.3.3 $\ass J D $ is an initial completion of
$\ass J Y $. \eop
\bigbreak

\proclaim{Proposition~4.3.7} Let $D$ be a complete poset. Then $D$ is an 
initial completion of itself if and only if each directed subset of $D$ 
contains
a maximum element (i.e., if and only if for each $A \in \directed{D}$ then
there exists $u \in A$ with $u' \sqle u$ for all $u' \in A$). \endpro

\proof Suppose first that $D$ is an initial completion of itself and let
$A \in \directed{D}$ with $\lub A = u$. Then $u \sqle \lub A$ and so by
Proposition~4.3.3 $u \sqle u'$ for some $u' \in A$. But this is only possible
if $u = u'$ and hence $u \in A$, i.e., $A$ contains a maximum element (namely
$u$). The converse follows directly from Proposition~4.3.3. \eop 
\bigbreak
If $Y$ is a poset then a sequence of elements $\{y_n\}_{n \ge 0}$ from $Y$ is 
called a 
\indexdef{chain}{chain in a poset}{}
in $Y$ if $y_n \sqle y_{n+1}$ for each $n \ge 0$. 
A chain $\{y_n\}_{n \ge 0}$ is said to be 
\indexdef{finite}{chain in a poset}{finite}
if $y_m = y_n$ 
for all $m \ge n$ for some $n \ge 0$. Now let $D$ be a complete poset; if $D$
is an initial completion of itself then by Proposition~4.3.7 each chain in $D$
must be finite (since the elements in a chain form a directed set). The
converse is in fact also true: If each chain in $D$ is finite then $D$ is an
initial completion of itself. This follows from Proposition~4.3.7 together 
with a standard application of Zorn's lemma.
\medskip
Let $D$ be a complete poset and $D'$ be an initial completion of $D$. Then
the identity mapping $i_D : D \to D$ is monotone and so it extends uniquely
to a continuous mapping $i^*_D : D' \to D$, which is obviously surjective.
However, if $D' \ne D$ then $i^*_D$ cannot be injective. The following simple example
indicates the kind of behaviour which occurs here:
Let $X$ be a set and consider the set $D = {\cal P}(X)$ of all subsets of $X$,
regarded as a poset with the inclusion ordering. Then $D$ is complete,
since if $A \in \directed{D}$ then clearly $\,\lub A \,=\, \bigcup_{B \in A} B\,$.
However, by Proposition~4.3.7 $Y$ is an initial completion of itself if and only
if $X$ is finite. Let $D'$ be an initial completion of $D$ and let
$i^*_D : D' \to D$ be the unique continuous extension of the identity mapping
$i_D$. The proof of Proposition~4.3.2 shows that if $B$ is an infinite subset of $X$ then 
the set $\lcurl u \in D'\,:\,i^*_D(u) = B \rcurl$ is infinite (since there are then
infinitely many non-equivalent directed subsets of $D$ having $B$ as least upper bound).
\medskip
In what follows some remarks will be made about the initial completion of the poset 
$\mono {Y_1} {Y_2} $. The purpose of these remarks is mainly to warn the
reader that such an initial completion is more complicated than perhaps would be expected.
Let $Y_1$ and $Y_2$ be posets and for $j = 1,\,2$ let $D_j$ be an initial completion
of $Y_j$. Then each monotone mapping from $Y_1$ to $Y_2$ extends uniquely to a
continuous mapping from $D_1$ to $D_2$, a mapping
$\psi : \mono {Y_1} {Y_2} \to \cont {D_1} {D_2} $ can thus be defined by letting
$\psi(f)$ be the unique continuous extension of $f \in \mono {Y_1} {Y_2} $.

\proclaim{Lemma~4.3.7} The mapping $\psi : \mono {Y_1} {Y_2} \to \cont {D_1} {D_2} $
is an embedding, i.e.,
$f_1 \sqle f_2$ if and only if $\psi(f_1) \sqle \psi(f_2)$. \endpro

\proof This is straightforward. \eop
\bigbreak
By Lemma~4.3.7 $\mono {Y_1} {Y_2} $ can be considered as a subposet of
$\cont {D_1} {D_2} $ (by identifying $f \in \mono {Y_1} {Y_2} $ with
the continuous mapping $\psi(f)$). However, in general 
$\cont {D_1} {D_2} $ will not be an initial completion of
$\mono {Y_1} {Y_2} $. For example, if the partial order on $Y_1$ is trivial
(i.e., $y \sqle y'$  if and only if $y = y'$) and $Y_2 = \{0,1\}$
with $0 \sqle 1$ then 
$\mono {Y_1} {Y_2} $ can clearly be identified with the poset 
${\cal P}(Y_1)$ (with the inclusion ordering).
But Proposition~4.3.7 implies that $D_1 = Y_1$ and $D_2 = Y_2$, and here 
$\cont {D_1} {D_2} = \mono {Y_1} {Y_2} $. On the other hand, if $Y_1$ is infinite then,
as noted above, ${\cal P}(Y_1)$ is not an initial completion of itself.
\medskip
Now let $D$ be an initial completion of $\mono {Y_1} {Y_2} $. Then, 
the monotone mapping $\psi$ extends uniquely
to a continuous mapping $\psi^* : D \to \cont {D_1} {D_2} $
(since $\cont {D_1} {D_2} $ is complete). As noted above, $\psi^*$ is in general not
injective, and so more then one (and often infinitely many) elements of $D$ will be
associated with the same continuous mapping from $D_1$ to $D_2$.  
\medskip
Consider the application operator $\fss{a} : \mono {Y_1} {Y_2} \times Y_1 \to Y_2$
defined by
\mdisp{ \fss{a}(f,y_1)\ =\ f(y_1)}
for all $f \in \mono {Y_1} {Y_2} $, $y_1 \in Y_1$. This
is clearly monotone and by Proposition~4.3.6 $D \times D_1$ is an initial extension
of $\mono {Y_1} {Y_2} \times Y_1$. Thus $\fss{a}$ extends uniquely to a continuous
mapping $\fss{a}^* : D \times D_1 \to D_2$. Moreover,

\mdisp{\fss{a}^*(u,u_1)\ =\ \psi^*(u)\,(u_1)}
for all $u \in D$, $u_1 \in D_1$, since
$\,\fss{a}(f,y_1)\, =\, f(y_1)\, =\, \psi(f)\,(y_1)\,$
for all $f \in \mono {Y_1} {Y_2} $, $y_1 \in Y_1$, and hence
$\,(u,u_1) \mapsto \fss{a}^*(u,u_1)\,$ and
$\,(u,u_1) \mapsto \psi^*(u)\,(u_1)\,$ are both continuous extensions of the same
monotone mapping. However, $\fss{a}^*$ should not really be regarded as an application
operator, since in general it fails to have the 
\indexdef{extensionality}{extensionality property}{}
property enjoyed by $\fss{a}$, namely that if $f,\,f' \in \mono {Y_1} {Y_2} $ and
$\fss{a}(f,y) = \fss{a}(f',y)$ for all $y \in Y_1$ then $f = f'$. 
\bigbreak
The final topic of this section 
explains why the initial completion also goes under the name of the ideal completion.
In what follows let $Y$ be a fixed poset. A non-empty subset $I$ of $Y$ is called an 
\indexdef{ideal of $Y$}{ideal of a poset}{}
if $y \in I$ whenever $y \sqle y'$ for some $y' \in I$.
An ideal $I$ is said to be 
\indexdef{directed}{ideal of a poset}{directed}
if $I \in \directed{Y}$. In particular, the set

\mdisp{ I(y)\ =\ \lcurl y' \in Y\,:\, y' \sqle y \rcurl}

is a directed ideal for each $y \in Y$ 
(\indexdef{the principal ideal generated by $y$}{ideal of a poset}{principal}).

\proclaim{Lemma~4.3.8} Let $y_1,\,y_2 \in Y$. Then $y_1 \sqle y_2$
if and only if $I(y_1) \subset I(y_2)$. \endpro

\proof This is clear. \eop
\bigbreak
Now denote by ${\cal I}_{\fsss{d}}(Y)$ the set of directed ideals of 
$Y$, regarded as a poset with the inclusion ordering. By Lemma~4.3.8 
${\cal I}_{\fsss{d}}(Y)$ can be considered as an extension of $Y$ (by identifying 
the element $y$ with the principal ideal $I(y)$ for each $y \in Y$).

\proclaim{Proposition~4.3.8} ${\cal I}_{\fsss{d}}(Y)$ is an initial 
completion (called 
\indexddef{the ideal completion}{ideal completion of a poset}{}
{completion of a poset}{ideal}) of $Y$. \endpro

\proof It is clearly enough to show that ${\cal I}_{\fsss{d}}(Y)$ is 
conjugate to 
the initial completion constructed in the proof of Proposition~4.3.2
(since an extension conjugate to an initial completion is also an initial
completion). For each non-empty subset $A$ 
of $Y$ let

\mdisp{ I(A)\ =\ \lcurl y \in Y\,:\,
      y \sqle y' \frm{for some} y' \in A \rcurl\ .}

\proclaim{Lemma~4.3.9} (1)\enskip $I(A)$ is an ideal with $A \subset I(A)$. 
Moreover, 
if $I$ is any ideal with $A \subset I$ then $I(A) \subset I$, i.e., $I(A)$ 
is the smallest ideal containing $A$. 
\medskip
(2)\enskip If $A \in \directed{Y}$ then $I(A)\in \directed{Y}$; moreover, 
$A$ and $I(A)$ are mutually cofinal.
\medskip
(3)\enskip If $A_1,\,A_2 \in \directed{Y}$
then $I(A_1) \subset I(A_2)$ if and only if $A_1$ is cofinal in
$A_2$. \endpro

\proof This is straightforward. \eop
\bigbreak

Now let $D$ be as in the proof of Proposition~4.3.2. Recall that
mutual cofinality defines an equivalence relation 
on the set $\directed{Y}$ and that $D$ is the set of equivalence classes,
considered as a poset with the partial order $\preceq$ defined
by stipulating that $[A_1] \preceq [A_2]$ if $A_1$ is cofinal in $A_2$ 
(and where $[A]$ denotes the equivalence class containing
$A$ for each $A \in \directed{Y}$).
\medskip
If $A_1,\,A_2 \in \directed{Y}$ then by Lemma~4.3.9~(3) 
$I(A_1) = I(A_2)$ if and only if $A_1$ and $A_2$ are equivalent, thus
by Lemma~4.3.9~(2)
a mapping $h : D \to {\cal I}_{\fsss{d}}(Y)$ can be defined by letting 
$h([A]) = I(A)$ for each $A \in \directed{Y}$. Moreover, Lemma~4.3.9~(3)
then implies that $h$ is an embedding.
In fact $h$ is also surjective, since by Lemma~4.3.9~(1) and (2) 
\mdisp{A\ =\ I(A)\ =\ h([A])}
for each $A \in {\cal I}_{\fsss{d}}(Y)$. 
Hence $h$ is an order isomorphism.
But $h(y) = y$ for each $y \in Y$ (because
$y$ is identified with $[\{y\}]$ in $D$, with $I(y)$ in 
${\cal I}_{\fsss{d}}(Y)$, and $I(\{y\}) = I(y)$), and therefore
$D$ and ${\cal I}_{\fsss{d}}(Y)$ are conjugate extensions of $Y$. \eop

\vfill\eject

\hfline{104}{4.4 \ INITIAL COMPLETIONS OF MONOTONE REGULAR EXTENSIONS}

\sectionhead {4.4} {Initial completions of monotone regular extensions}
\bigskip\medskip
For the whole of this section let $(Y_B,q_K)$ be a monotone regular bottomed
\tsym{\Lambda}-algebra and $\sqle_B$ be an associated ordering which is monotone;
$Y_\beta$ will always be considered as a poset with the partial order $\sqle_\beta$ for 
each $\beta \in B$.

\proclaim{Lemma~4.4.1} For each $\beta \in B$ let $D_\beta$ be an initial
completion of the poset $Y_\beta$ and let $\kappa \in K$ be of type 
$L \to \beta$. Then the mapping $q_\kappa : \ass L Y \to Y_\beta$ extends 
uniquely to a continuous mapping $f_\kappa : \ass L D \to D_\beta$. \endpro

\proof By assumption the mapping $q_\kappa : \ass L Y \to Y_\beta$  
is monotone and so it is still monotone considered
as a mapping from $\ass L Y $ to $D_\beta$; moreover, by Proposition~4.3.6 
$\ass L D $ is an initial completion of $\ass L Y $. The result
thus follows from Proposition~4.3.3. \eop 
\bigbreak
Lemma~4.4.1 allows the following definition to be made: A 
\tsym{\Lambda}-algebra $(D_B,f_K)$ is said to be an
\indexddef{initial completion}
{initial completion of algebra}{}{algebra}{initial completion of}
of $(Y_B,q_K)$ if $D_\beta$ is an initial 
completion of the poset $Y_\beta$ for each $\beta \in B$ and $f_\kappa$ is 
the unique continuous extension of $q_\kappa$ for each $\kappa \in K$. 
(Of course, whether or not $(D_B,f_K)$ is an
initial completion of $(Y_B,q_K)$ depends on the associated ordering
$\sqle_B$. However, it will be assumed that this
can be determined from the context.) 
It follows immediately from Proposition~4.3.2 that there exists an initial
completion of $(Y_B,q_K)$.
\medskip

An initial completion $(D_B,f_K)$ of $(Y_B,q_K)$ is, in particular,
an extension of $(Y_B,q_K)$. Moreover,
by Lemma~4.3.1 the bottom element $\bot_\beta$ of $Y_\beta$ is also the
the bottom element of the poset $D_\beta$.
The partial order on $D_\beta$ will always be denoted by 
$\sqle_\beta$.
\medskip
If $(Y_B,q_K)$ is a bottomed extension of $(X_B,p_K)$ then, of course, any
initial completion of $(Y_B,q_K)$ is also 
a bottomed extension of $(X_B,p_K)$.

\proclaim{Proposition~4.4.1} Let $(D_B,f_K)$ and $(D'_B,f'_K)$ be initial
completions of $(Y_B,q_K)$. Then for each $\beta \in B$ there 
exists a unique
order isomorphism $\pi_\beta : D_\beta \to D'_\beta$ with $\pi_\beta(y) = y$
for each $y \in Y_\beta$ (and by Proposition~4.2.2 $\pi_\beta$ and the inverse
mapping $\pi_\beta^{-1}$ are then both continuous). Moreover, the family 
$\pi_B$ is also an isomorphism from $(D_B,f_K)$
to $(D'_B,f'_K)$. \endpro

\proof There exists a unique continuous mapping 
$\pi_\beta : D_\beta \to D'_\beta$ with $\pi_\beta(y) = y$ for all
$y \in Y_\beta$ because $D_\beta$ is an initial completion of $Y_\beta$.
Moreover, exactly as in the proof of Lemma~4.3.2, $\pi_\beta$ is an order
isomorphism. The uniqueness follows from Proposition~4.2.2. and Lemma~4.3.3.
\medskip
It remains to show that $\pi_B : (D_B,f_K) \to (D'_B,f'_K)$ is a 
homomorphism (and thus an isomorphism). Let
$\kappa \in K$ be of type $L \to \beta$; then Proposition~4.2.4 implies that 
$\ass L {\pi} : \ass L D \to \assb L {D'} $ is continuous, and so 
$\pi_\beta \, f_\kappa$ and $f'_\kappa \, \ass L {\pi} $ are both 
continuous mappings from $\ass L D $ to $D'_\beta$ with

\mdisp{\pi_\beta \, f_\kappa\,(c)\ =\ \pi_\beta \, q_\kappa\,(c)
  \ =\ q_\kappa (c)\ =\ q_\kappa \, \ass L {\pi} \,(c)
                                       \ =\ f'_\kappa \, \ass L {\pi} \,(c)}

for all $c \in \ass L Y $. Therefore by Lemma~4.3.3 
$\pi_\beta \, f_\kappa = f'_\kappa \, \ass L {\pi} $, since $\ass L D $ is a
completion of $\ass L Y $. \eop
\bigbreak
Note that in the special case when $(Y_B,q_K)$ is the flat extension of 
$(X_B,p_K)$ (and so there is a  unique associated ordering)
then $(Y_B,q_K)$ itself is the (only possible) initial 
completion of $(Y_B,q_K)$: The poset $Y_\beta$ is clearly 
complete and therefore by
Proposition~4.3.7 $Y_\beta$ is an initial completion of itself, since a 
directed subset of $Y_\beta$ can contain at most two elements and a finite 
directed set always contains a maximum element. Example~4.4.1 on the
next page gives an initial completion of the
\tsym{\Lambda}-algebra $(X^\top_B,p^\top_K)$ 
introduced in Example~3.2.2.

\midinsert
\bigskip\medskip
\frame{20pt}{\bigskip\medskip 
{\it Example 4.4.1\ } Define a \tsym{\Lambda}-algebra 
$(D^\top_B,f^\top_K)$, which is an extension of 
the \tsym{\Lambda}-algebra 
$(X^\top_B,p^\top_K)$ introduced in
Example~3.2.2, as follows (where if $Z$ is a set then
$Z^\infty$ denotes the set of all infinite lists of elements from $Z$):
\bigskip
{\leftskip=32pt
$D^\top_{\fstt{bool}} = X^\top_{\fstt{bool}}$,  
$\ D^\top_{\fstt{nat}} = X^\top_{\fstt{nat}} \cup \{\infty\}$,
where $\infty \notin X^\top_{\fstt{nat}}$,
\smallskip
$D^\top_{\fstt{int}} = X^\top_{\fstt{int}}$,
$\ D^\top_{\fstt{pair}} = X^\top_{\fstt{pair}}$,
\smallskip    
$D^\top_{\fstt{list}} = X^\top_{\fstt{list}} 
             \cup (X^\top_{\fstt{int}})^\infty$,
\medskip
$f^\top_{\fstt{True}} = p^\top_{\fstt{True}}$,
$\ f^\top_{\fstt{False}} = p^\top_{\fstt{False}}$,
\smallskip
$f^\top_{\fstt{Zero}} : \Oneptset \to D^\top_{\fstt{nat}}$ 
with $f^\top_{\fstt{Zero}}(\onept) = 0$,
\smallskip
$f^\top_{\fstt{Succ}} : D^\top_{\fstt{nat}} 
 \to D^\top_{\fstt{nat}}$ with
\mdisp{f^\top_{\fstt{Succ}}(n)\ =\ \cases{
	 p^\top_{\fstt{Succ}}(n) & if 
	   $\,n \in X^\top_{\fstt{nat}}\,$,\cr
            \infty  & if $\,n = \infty\,$, \cr}}
\smallskip
$f^\top_{\underline n} = p^\top_{\underline n}$
for each $n \in \Int$,
$\ f^\top_{\fstt{Pair}} = p^\top_{\fstt{Pair}}$,
\smallskip
$f^\top_{\fstt{Nil}} : \Oneptset \to D^\top_{\fstt{list}}$ 
with $f^\top_{\fstt{Nil}}(\onept) = \onept$,
\smallskip
$f^\top_{\fstt{Cons}} : D^\top_{\fstt{int}}
         \times D^\top_{\fstt{list}} \to D^\top_{\fstt{list}}$ 
with
\mdisp{f^\top_{\fstt{Cons}}(n,s)\ =\ \cases{
	 p^\top_{\fstt{Cons}}(n,s) & if 
	   $\,s \in X^\top_{\fstt{list}}\,$,\cr
 \noalign{\vskip 2pt}
            n\triangleleft s  & if 
            $\,s \in (X^\top_{\fstt{int}})^\infty\,$. \cr}}
\medskip\smallskip
}
Moreover, for each $\beta \in B$ the partial order 
$\sqle_\beta$ on $X^\top_\beta$ can be extended to a partial order $\sqle_\beta$ on
$D^\top_\beta$ (which means that $\sqle_{\fstt{bool}}$, 
$\sqle_{\fstt{int}}$ and $\sqle_{\fstt{pair}}$ are just the corresponding 
partial orders on $X^{\top}_{\fstt{bool}}$, 
$X^{\top}_{\fstt{int}}$ and $X^{\top}_{\fstt{pair}}$): 
\medskip\smallskip
{\leftskip=32pt 
$\sqle_{\fstt{nat}}$ is defined to be the extension of the partial order 
$\sqle_{\fstt{nat}}$ on $X^{\top}_{\fstt{nat}}$ such that 
$\infty \sqle_{\fstt{nat}} n$ if and only if $n = \infty$ and 
$n \sqle_{\fstt{nat}} \infty$ if and only if $n \in \Bot{\Nat}\cup \{\infty\}$.
\medskip
$\sqle_{\fstt{list}}$ is defined to be the extension of the partial order 
$\sqle_{\fstt{list}}$ on $X^{\top}_{\fstt{list}}$ such that if 
$s = \inflist x \in (X^\top_{\fstt{int}})^\infty$
and $z \in X^{\top}_{\fstt{list}}$ then
$s \sqle_{\fstt{list}} z'$ if and only if
$z' = \inflist {x'} \in (X^\top_{\fstt{int}})^\infty$
with $x_n \sqle_{\fstt{int}} x'_n$ for all $n \in \Nat$, and 
$z \sqle_{\fstt{list}} s$ if and only if
$z = (\list {x'} m )^\bot$ for some $m \in \Nat$ and
$x'_n \sqle_{\fstt{int}} x_n$ for each $\nullto n m $.
\bigskip
}
It is left to the reader to check that
$(D^\top_B,f^\top_K)$ (where
$D^\top_\beta$ is considered as a poset with the partial order
$\sqle_\beta$) is an initial completion of
$(X^\top_B,p^\top_K)$.
\bigskip\bigskip
}
\bigskip
\endinsert

\medskip
In what follows let $(D_B,f_K)$ be an initial completion of $(Y_B,q_K)$; 
Proposition~4.4.1 implies that this initial completion is, in the 
appropriate sense, unique. 

\proclaim{Proposition~4.4.2} $(D_B,f_K)$ is a monotone regular bottomed 
\tsym{\Lambda}-algebra and  $\sqle_B$ is an ordering associated with $(D_B,f_K)$
which is monotone.
Moreover, $(Y_B,q_K)$ and $(D_B,f_K)$ have the same trace,
and if $(Y_B,q_K)$ is \tsym{(H_B,\diamond_K)}-cored for some core type
$(H_B,\diamond_K)$ then so is $(D_B,f_K)$. \endpro

\proof Denote the core of $(Y_B,q_K)$ by $C'_K$ and that of 
$(D_B,f_K)$ by $C_K$. The following fact relating the cores $C'_K$ and $C_K$
is needed:

\proclaim{Lemma~4.4.2} Let $\kappa \in K$ be of type $L \to \beta$. Then
\mdisp{C_\kappa\ =\ \lcurl c \in \ass L D  \,:\, c' \sqle_\kappa c 
                                    \frm{for some} c' \in C'_\kappa \rcurl\ .}
Moreover, $C_\kappa$ is also the set of all elements of $\ass L D $ having the 
form $\lub A$ for some $A \in \directed{\ass L Y }$ with $A \subset C'_\kappa$.
\endpro

\proof Let $c \in \ass L D $ and suppose $c' \sqle_\kappa c$ for some 
$c' \in C'_\kappa$; then $\bot_\beta \ne f_\kappa(c') \sqle_\beta f_\kappa(c)$
and hence $f_\kappa(c) \ne \bot_\beta$, i.e., $c \in C_\kappa$. 
Conversely, suppose $c \in \ass L D $ with $f_\kappa(c) \ne \bot_\beta$.
Then, since $\ass L D $ is a completion of $\ass L Y $, there exists 
$A \in \directed{\ass L Y }$ with
$c = \lub A$ and therefore $\lub f_\kappa(A) = f_\kappa(\lub A) \ne \bot_\beta$
(because $f_\kappa$ is continuous). But this implies that
$f_\kappa(c') \ne \bot_\beta$ for some $c' \in A$, and then $c' \in C'_\kappa$
with $c' \sqle_\kappa c$.
\medskip
If $c \in \ass L D $ has the form $\lub A$ for some $A \in \directed{\ass L Y }$
with $A \subset C'_\kappa$ then in particular $c' \sqle_\kappa c$ for all 
$c' \in A$; hence $c \in C_\kappa$. Suppose conversely that 
$c \in C_\kappa$, and let $B \in \directed{\ass L Y }$ with 
$\lub B = c$. Then, putting $A = B \cap C'_\kappa$, it is easily checked that 
also $A \in \directed{\ass L Y }$ and $\lub A = c$. \eop
\bigbreak 
The proof of Proposition~4.4.2 can now be commenced, and
it will be shown first that $\sqle_B$ is an ordering associated with $(D_B,f_K)$.
Clearly $\bot_\beta \sqle_\beta c$ for each $c \in D_\beta$. Moreover,
if $\kappa \in K_\beta$ and $c,\,c' \in C_\kappa$ with $c \sqle_\kappa c'$ 
then $f_\kappa(c) \sqle_\beta f_\kappa(c')$, because $f_\kappa$ is 
monotone.
Thus consider $\kappa,\, \kappa' \in K$ having types $L \to \beta$ and
$L' \to \beta$  respectively, and let $c \in C_\kappa$,
$c' \in C_{\kappa'}$ with
$f_\kappa(c) \sqle_\beta f_{\kappa'}(c')$. By Lemma~4.4.2 there exist 
$A \in \directed{\ass L Y }$ and  $A' \in \directed{\ass {L'} Y }$
with $A \subset C'_\kappa$, $A' \subset C'_{\kappa'}$,
$\lub A = c$ and $\lub A' = c'$, and then 

\mdisp{\lub q_\kappa(A) \ =\ \lub f_\kappa(A') 
   \ =\  f_\kappa(\lub A)\ \sqle_\beta\ f_{\kappa'}(\lub A') 
                     \ =\ \lub f_{\kappa'}(A') \ =\ \lub q_{\kappa'}(A')}

(since $f_\kappa$ is continuous). Therefore by Proposition~4.3.3 
$q_\kappa(A)$ is cofinal in $q_{\kappa'}(A')$, and from the definition of an 
ordering associated with $(Y_B,q_K)$ this is only possible if 
$\kappa = \kappa'$.
Hence in fact $q_\kappa(A)$ is cofinal in $q_\kappa(A')$ which,
again using the definition of an ordering associated with $(Y_B,q_K)$,
implies that $A$ is cofinal in  $A'$, and so in particular 
$c = \lub A \sqle_\kappa \lub A' = c'$.
This shows that $\sqle_B$ is an ordering associated with $(D_B,f_K)$,
and by definition $\sqle_B$ is then monotone (since the mapping $f_\kappa$, 
being continuous, is in particular monotone).
\medskip
It will be shown next that
$\bigcup_{\kappa \in K_\beta} f_\kappa(C_\kappa) 
                = D_\beta \setminus \{\bot_\beta\}$ for each $\beta \in B$.
The regularity of $(D_B,f_K)$ then follows from Proposition~4.1.1 and 
Lemma~3.3.1. Let $u \in D_\beta \setminus \{\bot_\beta\}$ and let 
$A \in \directed{Y_\beta}$ with $\lub A = u$. It can clearly be assumed that
$\bot_\beta \notin A$ and hence for each $y \in A$ there exists a unique
$\kappa \in K_\beta$ and a unique element $u \in C'_\kappa$ such that
$q_\kappa(u) = y$. But from the definition of an ordering associated with 
$(Y_B,q_K)$ this is only possible if $\kappa$ does not depend on $y$. Thus 
there exists $\kappa \in K$ having type $L \to \beta$ for some $L$ and for each 
$y \in A$ a unique element
$c_y \in C'_\kappa$ such that $q_\kappa(c_y) = y$. Let
$B = \lcurl c_y \,:\, y \in A \rcurl$; then, again using the definition 
of an associated ordering, it follows that 
$B \in \directed{\ass L Y }$. Let $c = \lub B$; then by Lemma~4.4.2
$c \in C_\kappa$ and
 
\mdisp{f_\kappa(c) \ =\ f_\kappa(\lub B) \ =\ \lub f_\kappa(B)
                \ =\  \lub q_\kappa(B) \ =\  \lub A \ =  u\ ,}

since $f_\kappa$ is continuous. Therefore $u \in q_\kappa(C_\kappa)$. 
\medskip
The monotonicity of $(D_B,f_K)$ follows directly from Proposition~4.1.6, and 
$(Y_B,q_K)$ and $(D_B,f_K)$ have the same trace because
there is essentially no difference between the homomorphism 
$\sem{\cdot}^\bot_B : (F^\flat_B,\fo^\flat_K) \to (Y_B,q_K)$ 
and the corresponding homomorphism from
$(F^\flat_B,\fo^\flat_K)$ to $(D_B,f_K)$. 
\medskip
It remains to show that if $(Y_B,q_K)$ is \tsym{(H_K,\diamond_K)}-cored 
for some core type $(H_B,\diamond_K)$ then so is $(D_B,f_K)$.
Let $\kappa \in K$ be of type $L \to \beta$ and let $c \in \ass L D $.
Then it is easy to see that there exists $c' \in \ass L Y $ such that
$\varepsilon_\eta(c'(\eta)) = \varepsilon_\eta(c(\eta))$ 
for each $\eta \in L$ and such that
$\varepsilon_{\beta}(f_\kappa(c')) 
                         = \varepsilon_{\beta}(f_\kappa(c))$. 
Thus

\mdisp{\varepsilon_{\beta}(f_\kappa(c))
\ =\ \varepsilon_{\beta}(f_\kappa(c'))
\ =\ \varepsilon_{\beta}(q_\kappa(c'))
\ =\ \diamond_\kappa(\ass L {\varepsilon} (c'))
\ =\ \diamond_\kappa(\ass L {\varepsilon} (c))}

and hence $(D_B,f_K)$ is also \tsym{(H_K,\diamond_K)}-cored. \eop 
\bigbreak
Suppose now that $(Y_B,p_K)$ (and thus $(D_B,f_K)$) is a bottomed extension
of $(X_B,p_K)$.

\proclaim{Proposition~4.4.3} Each element of $X_\beta$ is a maximal element 
of $D_\beta$. \endpro

\proof This follows immediately from Proposition~4.1.3, since by 
Proposition~4.4.2 $(D_B,f_K)$ is a regular extension of $(X_B,p_K)$ and
$\sqle_B$ is an ordering associated with $(D_B,f_K)$.  \eop

\proclaim{Proposition~4.4.4} Suppose in addition that $(Y_B,q_K)$ is
a minimal extension of $(X_B,p_K)$; let $\beta \in B$. Then:
\medskip
(1)\enskip $\lcurl u \in D_\beta \,:\, u \sqle_\beta y \rcurl \subset Y_\beta\,$
for each $y \in Y_\beta$.
\medskip
(2)\enskip For each $u \in D_\beta$ the set
$\lcurl u' \in D_\beta \,:\, u' \sqle_\beta u \rcurl$ is finite if and
only if $u \in Y_\beta$. \endpro

\proof (1)\enskip Let $y \in Y_\beta,\, u \in D_\beta$ with
$u \sqle_\beta y$. Then there exists $A \in \directed{Y_\beta}$ with
$u = \lub A$ and here Proposition~4.3.4 implies that $A \sqle_\beta \{y\}$,
i.e., $y' \sqle_\beta y$ for all $y' \in A$. But then by Proposition~4.1.4~(2)
$A$ is finite and hence $u \in A$, i.e., $u \in Y$.
\medskip
(2)\enskip If $y \in Y_\beta$ then by (1) and Proposition~4.1.4~(2)
$\lcurl u' \in D_\beta \,:\, u' \sqle_\beta y \rcurl$ is finite. The converse 
follows from the fact that if $u \in D_\beta$ with $u = \lub A$ for some 
$A \in \directed{Y_\beta}$ then 
$A \subset \lcurl u' \in D_\beta \,:\, u' \sqle_\beta u \rcurl$. But, 
as in (1), if $A$ is finite then $u \in Y_\beta$. \eop

\bigskip\bigskip\bigskip\bigskip

\sectionhead {4.5} {Notes}
\bigskip\medskip
The special case of an associated ordering for a fully regular extension
can be found in Goguen, Thatcher, Wagner and Wright (1977) and
in Courcelle and Nivat (1976).
\medskip
Section~4.2 contains the elementary results which will be required
from what is called 
\indexdef{domain theory}{domain theory}{}. 
This was originated by Dana Scott 
in the late sixties, and is the main tool for dealing with denotational 
semantics. The reader interested in finding out more about domain theory
should consult Scott and Gunter (1990); an account of its origins can be 
found in Scott (1976).
\medskip
The results in Sections~4.3 concerning the initial completion of a poset 
are well-known and exist in a bewildering variety of forms; see, for example,
Markowsky and Rosen (1976), Markowsky (1977), Wright, Wagner and Thatcher
(1978) and Banaschewski and Nelson (1982). The idea of an ideal completion
goes back to Birkhoff (1976) (which was first published in 1940).

\vfill\eject

\hfline{109}{}
\bigskip
{\fourteenbf Chapter 5\quad Functions}
\bigskip\medskip

Chapters 3 and 4 dealt with the data objects. These were obtained by 
starting with a signature $\Lambda = (B,K,\Theta,\vartheta)$ and fixing an 
initial \tsym{\Lambda}-algebra $(X_B,p_K)$ to describe the basic data objects.
In a further step, a monotone regular extension $(D_B,f_K)$ of $(X_B,p_K)$ was then chosen
to specify which `undefined', `partially defined' and, when appropriate, which
`infinite' data objects are to be allowed. In order to make things more specific 
the following three cases will be concentrated on throughout the study: 
\medskip

\bigskip
\frame{20pt}{\bigskip\medskip
{\it Case 1\/}\enskip $(D_B,f_K)$ is any monotone regular extension of 
$(X_B,p_K)$. Here $D_B$ will be considered just as a family of bottomed sets, and the
mappings in the family $f_K$ have in general no further structure.
\medskip\smallskip
{\it Case 2\/}\enskip $(D_B,f_K)$ is a minimal monotone regular extension of 
$(X_B,p_K)$. Then by Proposition~4.1.2 there exists a unique ordering $\sqle_B$ 
associated with $(D_B,f_K)$ which by Proposition~4.1.5 is monotone.
$D_B$ will be considered as a family of bottomed posets with respect
to this family of partial orders, and hence $f_K$ is a family of monotone mappings.  
\medskip\smallskip
{\it Case 3\/}\enskip $(D_B,f_K)$ is here the initial completion of a minimal monotone 
regular extension $(Y_B,q_K)$ of $(X_B,p_K)$. There is then the ordering
$\sqle_B$ associated with $(D_B,f_K)$ which arises from this completion process
(i.e., from the initial completion of $(Y_B,q_K)$ using its unique associated 
ordering). Here $D_B$ will be considered as a family of bottomed complete posets 
with respect to this family of partial orders, and hence $f_K$ is a family of 
continuous  mappings. (This means that $(D_B,f_K)$ is obtained using the `canonical'
procedure described at the beginning of Chapter~4.)
\bigskip\medskip
}
\bigskip\bigskip
The next task is to define suitable functions over the data objects. 
In Section~5.1 a general framework will be presented involving what are called
concrete cartesian closed categories. These allow a uniform treatment
of the functions which will be associated with the basic cases. 
\medskip
In Section~5.2 the set of ground types is extended
to a set $S$ called the \definition{set of functional types over $B$}.
Somewhat informally, the set $S$ can be thought of as being defined by the following rules:
\medskip\smallskip
{\parindent=25pt
\item{(1)} If $L \in {\cal F}_S$ and $\beta \in B$ then $L \to \beta$ is an element of 
$S$.
\medskip
\item{(2)} Each element of $S$ can be uniquely constructed using rule (1).
\bigskip
}  
(The set $B$ is considered as a subset of $S$ by identifying each ground type 
$\beta \in B$ with the element $\varnothing \to \beta$ of $S$.) 
\medskip
The functions of type $L \to \beta$ will be defined to be
some kind of functions whose arguments are named by the elements of the set $L$, in which 
the values of the argument corresponding to $\eta \in L$ 
range over the set of elements of type $\langle \eta \rangle$, and which take their 
values in the set $D_\beta$. In particular, 
if $\list {\sigma} n \in S^*$ with $n \ge 1$ then the
functions of type $\list {\sigma} n \to \beta$ will be functions of $n$ arguments 
with the values of the 
\tsym{j}-th argument ranging over the set of elements of type $\sigma_j$.
The recursive definition of $S$ (manifested in rule (1)) is necessary
since we want to include higher-order types, i.e., types 
$L \to \beta$ in which $\langle \eta \rangle \notin B$ for at least one $\eta \in L$,
and may need higher-order types 
$L \to \beta$ in which $\langle \eta \rangle$ is itself a higher-order type
for some $\eta \in L$,

\medskip
Having introduced the set $S$, the
family $D_B$ is then extended to a family $D_S$ in such a way 
that for each $\sigma \in S \setminus B$ it is natural to regard $D_\sigma$ 
as a set of functions of type $\sigma$. This family is obtained 
via a construction in an appropriate concrete cartesian closed category, and makes
use of an initiality property enjoyed by $S$.
\medskip
The next step is to extend the signature $\Lambda$ to a signature 
$\Sigma = (S,N,\Delta,\delta)$ by adding names for various
partial application operators;  
the \tsym{\Lambda}-algebra $(D_B,f_K)$ can then be extended to a
\tsym{\Sigma}-algebra $(D_S,f_N)$ by adding these operators to the
family $f_K$.
It is this \tsym{\Sigma}-algebra $(D_S,f_N)$ which forms the basis for what follows.
\medskip
In Section~5.3 the \definition{primitive} or \definition{built-in} functions 
are specified.
These are the functions occurring in the equations which are really 
part of what is given (as opposed to the functions which the equations
are supposed to define). 
There are two kinds of primitive functions which are needed. For practical reasons
it is useful to have a few 
\definition{integer operators}. More importantly,
a suitable family of \definition{case operators} is also required. 
These will facilitate what corresponds
to pattern matching or to the `case' construction in {\it Haskell\/}.
These operators (or something equivalent) are absolutely essential: 
They are the analogue of the $\,\ftt{if then else}\,$ construction found 
in all imperative programming languages and without them it is impossible 
to define any `non-trivial' functions.
\medskip
Finally, in Section~5.4 we take stock of the various constructions made
so far and present the set-up which will be used throughout the rest of the study.

\vfill\eject

\hfline{111}{5.1 \ CONCRETE CARTESIAN CLOSED CATEGORIES}
\bigskip

\sectionhead {5.1} {Concrete cartesian closed categories}
\bigskip\medskip
In this section a general framework will be presented involving what are called
concrete cartesian closed categories. These allow a uniform treatment
of the functions which will be associated with the basic cases introduced at the
beginning of the chapter. 
\medskip
To motivate making use of such a concept let us begin by looking at the 
\tsym{\Lambda}-algebra $(D_B,f_K)$ occurring in one of these three cases. There
is then a natural class ${\cal C}$ of sets equipped with an additional structure 
(depending on the case being considered) such that
$D_\beta \in {\cal C}$ for each $\beta \in B$ and such that $f_K$ is a family of 
mappings which preserve some of this structure. To be more definite:

\bigskip\medskip
\frame{20pt}{\bigskip\medskip
{\it Case 1:\/}\enskip 
${\cal C}$ is the class of bottomed sets and there is no restriction on the family $f_K$.
\medskip
{\it Case 2:\/}\enskip 
${\cal C}$ is the class of bottomed posets and $f_K$ is a family of monotone mappings.
\medskip
{\it Case 3:\/}\enskip 
${\cal C}$ is the class of bottomed complete posets and 
$f_K$ is a family of continuous mappings.
\medskip\smallskip
(Note that the mappings are not required to map the bottom element onto the bottom 
element, and in fact for a fully regular extension such a requirement would never
be met.)
\medskip\bigskip
}
\bigskip\bigskip

Now there are two important properties which these three set-ups 
have in common. The first is the closure of the class ${\cal C}$ under taking 
finite products: Let $L$ be a 
finite \tsym{\cal C}-typed set, i.e., $L$ is a finite set equipped with a mapping
$\langle \cdot \rangle : L \to {\cal C}$, and denote by $\otimes L$ the set 
of all mappings $c : L \to \bigcup_{\eta \in L} \langle \eta\rangle$ such that
$c(\eta) \in \langle\eta \rangle$ for each $\eta \in L$. (In particular, this means
that $\otimes \varnothing = \Oneptset$.)
Then there is a natural way
to equip $\otimes L$ with the appropriate structure  so that it becomes an element
of ${\cal C}$. To be more definite:

\bigskip\medskip
\frame{20pt}{\bigskip\medskip
{\it Case 1:\/}\enskip The bottom element of $\otimes L$ is the mapping
$c$ in which $c(\eta)$ is the bottom element of $\langle \eta \rangle$ for each 
$\eta \in L$. (In particular, this means that $\onept$ is the bottom element of
$\otimes \varnothing = \Oneptset$.)
\medskip
{\it Case 2:\/}\enskip The bottom element is as in Case~1. The set
$\otimes L$ is considered as a poset with the product partial
order $\sqle$ defined by stipulating that $b \sqle c$ if and only if
$b(\eta) \sqle_\eta c(\eta)$ for each $\eta \in L$, with $\sqle_\eta$ the partial
order on $\langle \eta \rangle$. 
\medskip 
{\it Case 3:\/}\enskip 
The bottom element and the partial order are as in Case~2, and the proof of 
Proposition~4.2.3 shows that the poset $\otimes L$ is complete.
\medskip\bigskip
}
\bigskip\medskip

Of course, the existence of finite products
was already implicit in the definition of the family $f_K$, since these mappings
have finite products as their domains. Note that if $L$ is a
finite \tsym{\cal C}-typed set and the family $X_L$ is defined by putting
$X_\eta = \langle \eta \rangle$ for each $\eta \in L$, then 
$\otimes L = \ass L X $, provided that $L$ is regarded as an \tsym{L}-typed set with 
the identity typing in the definition of $\ass L X $. This fact allows
the results in Chapter~4 (which are stated for families) to be applied here.
(For example, it is in this sense that
Proposition~4.2.3 shows that the poset $\otimes L$ is complete in Case~3.)

\medskip

If $X_1,\,X_2 \in {\cal C}$ then the element $\otimes [2]$ of ${\cal C}$
with $\langle 1\rangle = X_1$ and $\langle 2\rangle = X_2$ will be denoted as usual by 
$X_1 \times X_2$.
\medskip
The second common property involves sets of functions: Let $X$ and $Y$ be elements of
${\cal C}$. Then there is a set $\varepsilon(X,Y)$, which should be thought of as
the set of structure preserving mappings from $X$ to $Y$, and there is a natural way 
to equip $\varepsilon(X,Y)$
with the appropriate structure so that it becomes an element of ${\cal C}$. More
precisely:

\bigskip\bigskip
\frame{20pt}{\bigskip\medskip
{\it Case 1:\/}\enskip 
$\varepsilon(X,Y)$ is the set $\total X Y $ of all mappings from $X$ to 
$Y$, and for reasons of legibility it is convenient to allow
$\totale X Y $ as an alternative notation for $\total X Y $.
The bottom element of $\varepsilon(X,Y) = \totale X Y $ is the mapping
$f$ in which $f(x)$ is the bottom element of $Y$ for each $x \in X$. 
\medskip
{\it Case 2:\/}\enskip 
$\varepsilon(X,Y)$ is the set $\mono X Y $ of all monotone mappings from
$X$ to $Y$, with the bottom element as in Case~1. 
The set $\varepsilon(X,Y) = \mono X Y $ is considered as a poset 
with the partial order $\sqle$ defined by stipulating that 
$g \sqle h$ if and only if $g(x) \sqle' h(x)$
for all $x \in X$, with $\sqle'$ the partial order on $Y$.
\medskip
{\it Case 3:\/}\enskip 
$\varepsilon(X,Y)$ is the set $\cont X Y $ of all continuous mappings from
$X$ to $Y$, again with the bottom element as in Case~1.
The set $\varepsilon(X,Y) = \cont X Y $ is here considered as a poset 
with the partial order induced by the partial order on 
$\mono X Y $ (noting that $\cont X Y  \subset \mono X Y $); by Proposition~4.2.6 this 
poset is then complete.
\medskip\bigskip
}
\bigskip\bigskip

In addition, for each $X,\, Y \in {\cal C}$ there is a functional  application (or
evaluation) operator
$\fss{a}_X^Y : \varepsilon(X,Y) \times X \to Y$, which in all three cases is defined
by 
\mdisp{ \fss{a}_X^Y(h,x) \ =\ h(x)}
for all $h \in \varepsilon(X,Y)$, $x \in X$.
(This makes sense, because in all three cases $h$ is a function from $X$ to $Y$.)
Moreover, in Case~2 the mapping $\fss{a}_X^Y$ is monotone by Lemma~4.2.4, and 
Proposition~4.2.7 implies that
in Case~3 the mapping $\fss{a}_X^Y$ is continuous.
\medskip
Finally, if $X,\,Y,\,Z \in {\cal C}$ and $f : X \times Y \to Z$
is a structure preserving mapping (i.e., $f$ is arbitrary, monotone
or continuous depending on which of the cases is being considered) then there
is a unique structure preserving mapping $g : X \to \varepsilon(Y,Z)$ such that
\mdisp{ \fss{a}_Y^Z(g(x),y)\ =\ f(x,y)}
(i.e., such that $g(x)\,(y)\,=\,f(x,y)$)
for all $x \in X$, $y \in Y$. Of course, in all three cases there can be at most one such
mapping, and in Case~1 the requirement that
$g(x)\,(y)\,=\,f(x,y)$ for all $x \in X$, $y \in Y$ actually defines $g$. For the existence
of $g$
in Case~2 (resp.\ Case~3) it must be shown that the mapping $\,y \mapsto f(x,y)\,$
is monotone (resp.\ continuous) for each $x \in X$ and then that the mapping
$\,x \mapsto (y \mapsto f(x,y))\,$ is monotone (resp.\ continuous).
However, this was already established in Chapter~4, and in fact
$g = \fss{c}(f)$, with $\fss{c}$ the appropriate currying operator. More precisely,
in Case~2 $\fss{c}$ is the order isomorphism (given by Lemma~4.2.5) from 
$\mono {X\times Y} Z $ to $\mono X {\mono Y Z } $ 
and in Case~3 the order isomorphism (given by Proposition~4.2.8) from 
$\cont {X\times Y} Z $ to $\cont X {\cont Y Z } $. 
\bigskip\medskip
The above discussion has really shown that there is a concrete cartesian closed category
associated with each of the basic cases. Exactly what this means will now be explained.
It should be noted that not much trouble has been taken up to now to distinguish
between a set with structure (such as a bottomed poset or a complete bottomed poset)
and the underlying set. In the definitions involving  concrete categories which follow
this point will be treated with a bit more care.
\medskip
A 
\indexddef{concrete category}{concrete category}{}{category}{concrete} 
$\fss{C}$ consists of

\bigskip
{\parindent=25pt
\item{---} a class of elements ${\cal C}$ called the
\indexdef{objects}{objects of a category}{} 
of the category,
\medskip
\item{---} for each $X \in {\cal C}$ a set $|X|$ called the 
\indexdef{underlying set}{underlying set}{}
of $X$,
\medskip
\item{---} for each $X,\, Y \in {\cal C}$ a set $\Hom(X,Y)$ which is a subset of the
set of all mappings from $|X|$ to $|Y|$
\bigskip
}

such that the following two conditions hold:
\bigskip
{\parindent=25pt
\item{(C1)} If $f \in \Hom(X,Y)$ and $g \in \Hom(Y,Z)$ then the composition
$g\circ f : |X| \to |Z|$ (i.e., the mapping defined by
$(g\circ f) (x) = g(f(x))$ for all $x \in |X|$) is an element of $\Hom(X,Z)$.
\medskip
\item{(C2)} The identity mapping $\fss{id}_X : |X| \to |X|$
is an element of $\Hom(X,X)$ for each $X \in {\cal C}$.
\bigskip
}

In what follows let $\fss{C}$ be a concrete category; as above the objects of
$\fss{C}$ will be denoted by ${\cal C}$. The elements of $\Hom(X,Y)$ are usually called
\indexdef{morphisms}{morphisms in a category}{}. 
\vfill\eject

\bigskip\bigskip
\frame{20pt}{\bigskip\medskip
The concrete categories associated with the basic cases are, of course, given as
follows:
\medskip\smallskip
{\it Case 1:\/}\enskip 
${\cal C}$ is the class of bottomed sets, for each $X \in {\cal C}$
the underlying set $|X|$ is the set obtained by forgetting the bottom element, and 
$\Hom(X,Y)$ is the set of all mappings from $|X|$ to $|Y|$ for each $X,\, Y \in {\cal C}$.
It is clear that (C1) and (C2) hold.
\medskip
{\it Case 2:\/}\enskip 
${\cal C}$ is the class of bottomed posets,
for each $X \in {\cal C}$
the underlying set $|X|$ is the set obtained by forgetting the bottom element and the
partial order, and 
$\Hom(X,Y)$ is the set of all monotone mappings from $|X|$ to $|Y|$ for each 
$X,\, Y \in {\cal C}$.
It is clear that (C2) holds and (C1) holds by Lemma~4.2.1.
\medskip
{\it Case 3:\/}\enskip 
${\cal C}$ is the class of bottomed complete posets,
for each $X \in {\cal C}$
the underlying set $|X|$ is again the set obtained by forgetting the bottom element and the
partial order, and 
$\Hom(X,Y)$ is the set of all continuous mappings from $|X|$ to $|Y|$ for each 
$X,\, Y \in {\cal C}$.
It is clear that (C2) holds and (C1) holds by Proposition~4.2.1.
\bigskip\medskip
}
\bigskip\bigskip

Let $X,\,Y \in {\cal C}$; a morphism $f \in \Hom(X,Y)$
is called an 
\indexdef{isomorphism}{isomorphism}{in a category}
if there is a morphism $g \in \Hom(Y,X)$
such that $g\circ f = \fss{id}_X$ and $f\circ g = \fss{id}_Y$. In this case $g$
is uniquely determined by $f$, since if also
$g'\circ f = \fss{id}_X$ and $f\circ g' = \fss{id}_Y$ then
\mdisp{g'\ =\ g'\circ \fss{id}_Y\ =\ g' \circ f \circ g\ =\ \fss{id}_X \circ g\ =\ g\ .}
The morphism $g$ is called the 
\indexdef{inverse}{inverse of a morphism}{}
of $f$, and it will be denoted by
$f^{-1}$. 
Note that if $f \in \Hom(X,Y)$ is an isomorphism then the mapping 
$f : |X| \to |Y|$ is a bijection. The converse is, in general, false: If
$f \in \Hom(X,Y)$ is a morphism such that the mapping 
$f : |X| \to |Y|$ is a bijection then $f$ need not be an isomorphism. Case~2 provides
an example of this: If $f$ is a monotone bijection then the set-theoretic inverse
of $f$ need not be monotone. Objects $X,\,Y \in {\cal C}$ are said to be
\definition{isomorphic} if there exists an isomorphism $f \in \Hom(X,Y)$.
\medskip
The notion of a \tsym{{\cal S}}-typed set clearly still makes sense when ${\cal S}$ 
is a class: It is a pair $(L,\langle\cdot\rangle)$ consisting of a set 
$L$ and a mapping $\langle\cdot\rangle : L \to {\cal S}$.
The notation being used for typed sets will also be employed here; 
in particular, the \tsym{\cal S}-typed set $(L,\langle\cdot\rangle)$ will usually be denoted
just by $L$, it being assumed that the typing $\langle\cdot\rangle$ can be inferred
from the context.

\medskip
Let $\fss{Sets}$ denote the class of all sets, and let
$\prod : {\cal F}_{\fsss{Sets}} \to \fss{Sets}$ be the mapping
defined by for each $L \in {\cal F}_{\fsss{Sets}}$ letting
$\prod L$ be the set of all mappings $c : L \to \bigcup_{\eta \in L} \langle\eta\rangle$
such that $c(\eta) \in \langle\eta\rangle$ for each $\eta \in L$.
In particular, $\prod \varnothing = \Oneptset$.
\medskip
If $L\in {\cal F}_{\cal C}$ then $|L|$ will be used to denote the element of
${\cal F}_{\fsss{Sets}}$ having the same underlying set as in $L$ and with the typing
$|\langle\cdot\rangle| : L \to \fss{Sets}$ (with of course
$\langle\cdot\rangle : L \to {\cal C}$ the typing on $L$ as a \tsym{\cal C}-typed set).
\medskip
The concrete category $\fss{C}$ will be said to
\indexdef{possess canonical finite products}{canonical finite products}{}
if there is a mapping 
$\otimes : {\cal F}_{\cal C} \to {\cal C}$ given such that:
\bigbreak
{\parindent=25pt
\item{(P1)} $\,|{\otimes L}| \,=\, \prod |L|\,$ for each 
$L \in {\cal F}_{\cal C}$.
\medskip
\item{(P2)} If $L \in {\cal F}_{\cal C}$ and $\eta \in L$ then the mapping
$p_\eta : |{\otimes L}| \to |\langle\eta \rangle|$ defined by
$p_\eta(c) = c(\eta)$ for each $c \in |{\otimes L}|$
is a morphism, i.e., an element of $\Hom(\otimes L,\langle\eta\rangle)$.
\medskip
\item{(P3)} If $L \in {\cal F}_{\cal C}$, $X \in {\cal C}$ and 
$q_\eta \in \Hom(X,\langle\eta\rangle)$ for each $\eta \in L$ then the mapping
$q : |X| \to |{\otimes L}|$ defined by
$\,q(x)\,(\eta) \,=\, q_\eta(x)\,$ for all $x \in |X|$, $\eta \in L$,
is a morphism, i.e., an element of $\Hom(X,\otimes L)$.
If $L = \varnothing$ then this is to be interpreted as requiring that the
unique mapping $q : |X| \to \Oneptset$ (i.e., with $q(x) = \onept$  for all
$x \in |X|$) be a morphism, i.e., an element of $\Hom(X,\otimes \varnothing)$.
\bigskip
}
\bigskip
\frame{20pt}{\bigskip\smallskip 
The concrete categories associated with the basic cases possess canonical 
finite products: In all three cases condition (P1) holds by definition, and (P2) 
and (P3) hold trivially in Case~1.
\medskip
In Case~2 (resp.\ in Case~3) Lemma~4.2.2 (resp.\  Lemma~4.2.7) implies that (P2) holds,
and then Lemma~4.2.3 (resp.\ Proposition~4.2.5) implies, making use of 
Lemma~5.1.1 below, that (P3) holds.
\bigskip\smallskip
}
\bigskip\bigskip

In what follows assume that $\fss{C}$ possess canonical finite products. By (P1) (and
as already noted in (P3)) $|{\otimes \varnothing}| = \prod |\varnothing| = \Oneptset$,
and in fact $\Oneptset$ will also be used to denote the object $\otimes \varnothing$.
(The object and its
underlying set having the same denotation should cause no problem in this  particular
case.) Note (P3) implies that $\Hom(X,\Oneptset)$ consists of a unique
morphism for each $X \in {\cal C}$; in the language of category theory this means that
$\Oneptset$ is a 
\indexdef{terminal object}{terminal object of a category}{}
of the category $\fss{C}$.

\proclaim{Lemma~5.1.1} Let $L \in {\cal F}_{\cal C}$, $X \in {\cal C}$ and 
for each $\eta \in L$ let $q_\eta \in \Hom(X,\langle\eta\rangle)$. Then the 
morphism $q \in \Hom(X,\otimes L)$ given by (P3) 
is the unique mapping $q : |X| \to |{\otimes L}|$ such that
$p_\eta \circ q = q_\eta$ for each $\eta \in L$. 
\endpro

\proof This holds trivially if $L = \varnothing$, so assume that
$L \ne \varnothing$. If $\,q(x)\,(\eta) \,=\, q_\eta(x)\,$ for all $x \in |X|$
then $\,(p_\eta \circ q)\,(x)\,=\, p_\eta(q(x))\, =\, q_\eta(x)\,$ for all
$x \in |X|$, and thus $p_\eta \circ q = q_\eta$. Conversely, if
$p_\eta \circ q = q_\eta$ then for all $x \in |X|$
\mdisp{q(x)\,(\eta)\ =\ p_\eta(q(x))\ =\ (p_\eta \circ q)\,(x)\ =\ q_\eta(x)\ .\quad\eop}
\bigbreak
\proclaim{Lemma~5.1.2} Let $L \in {\cal F}_{\cal C}$, $X \in {\cal C}$ and
let $q : |X| \to |{\otimes L}|$ be a mapping. Then $q$ is a morphism, i.e., an
element of $\Hom(X,\otimes L)$, if and only if
$p_\eta \circ q \in \Hom(X,\langle\eta\rangle)$
for all $\eta \in L$.
\endpro

\proof This follows immediately from Lemma~5.1.1. \eop
\bigbreak
Let $X_1,\,X_2 \in {\cal C}$; then the object $\otimes [2]$ of ${\cal C}$
with $\langle 1\rangle = X_1$ and $\langle 2\rangle = X_2$ will be denoted by 
$X_1 \times X_2$.
It is easily checked that $|X_1 \times X_2| = |X_1|\times |X_2|$ and that the morphisms
$p_1 \in \Hom(X_1\times X_2,X_1)$ and $p_2 \in \Hom(X_1\times X_2,X_2)$
occurring in (P2) are the mappings 
$p_1 : |X_1|\times |X_2| \to |X_1|$ and $p_2 : |X_1|\times |X_2| \to |X_2|$ 
given by $p_1(x_1,x_2) = x_1$ and $p_2(x_1,x_2) = x_2$ for all 
$x_1 \in |X_1|$, $x_2 \in |X_2|$. Moreover, if $X \in {\cal C}$ and
$q_1 \in \Hom(X,X_1)$, $q_2 \in \Hom(X,X_2)$ then the morphism 
$q \in \Hom(X,X_1\times X_2)$ given by (P3) is the mapping
$q : |X| \to |X_1|\times |X_2|$ given by 
$q(x) = (q_1(x),q_2(x))$ for each $x \in |X|$.
\medskip

\proclaim{Lemma~5.1.3} Let $(L,\langle\cdot\rangle)$ and $(L,\langle\cdot\rangle')$ 
be two finite
\tsym{\cal C}-typed sets (with the same underlying set $L$) and for each
$\eta \in L$ let $f_\eta \in \Hom(\langle\eta\rangle,\langle\eta\rangle')$. Then the mapping
$f : |{\otimes(L,\langle\cdot\rangle)}| \to |{\otimes(L,\langle\cdot\rangle')}|$ defined
for all $c \in |{\otimes(L,\langle\cdot\rangle)}|$, $\eta \in L$ by
\mdisp{ f(c)\,(\eta)\ =\ f_\eta(c(\eta))}
is a morphism, i.e., an element of
$\Hom(\otimes(L,\langle\cdot\rangle),\otimes(L,\langle\cdot\rangle'))$. \endpro

\proof For each $\eta \in L$ let 
$p_\eta \in \Hom(\otimes (L,\langle\cdot\rangle),\langle\eta\rangle)$
be the morphism occurring in (P2) in the definition of 
$\otimes (L,\langle\cdot\rangle)$ and let
$p'_\eta \in \Hom(\otimes (L,\langle\cdot\rangle'),\langle\eta\rangle')$
be the corresponding morphism in the definition of 
$\otimes (L,\langle\cdot\rangle')$. Then for all $\eta \in L$, 
$c \in \otimes(L,\langle\cdot\rangle)$
\mdisp{
(p'_\eta \circ f)\, (c)\ =\ f(c)\,(\eta)\ =\ f_\eta(c(\eta))
\ =\ (f_\eta \circ p_\eta)\,(c)\ ,}
i.e., $p'_\eta \circ f = f_\eta \circ p_\eta$ 
and hence $p'_\eta \circ f \in \Hom(\otimes (L,\langle\cdot\rangle),\langle\eta\rangle')$
for each $\eta \in L$. Thus by Lemma~5.1.2
$f \in \Hom(\otimes(L,\langle\cdot\rangle),\otimes(L,\langle\cdot\rangle'))$. \eop
\bigbreak
If $X_1,\,X_2,\,Y_1,\,Y_2 \in {\cal C}$ and $f_1 \in \Hom(X_1,Y_1)$,
$f_2 \in \Hom(X_2,Y_2)$ then the morphism
$f \in \Hom(X_1\times X_2,Y_1\times Y_2)$ given by Lemma~5.1.3 will be denoted by
$f_1 \times f_2$. It is easily checked that
$f_1 \times f_2 : |X_1|\times |X_2| \to |Y_1|\times |Y_2|$ is the mapping defined by
$\,(f_1\times f_2)(x_1,x_2)\, =\, (f_1(x_1),f_2(x_2))\,$ for all 
$x_1 \in |X_1|$, $x_2 \in |X_2|$.

\bigskip
The concrete category $\fss{C}$ is said to be 
\indexddef{cartesian closed}{cartesian closed category}{}{category}{cartesian closed}
if (in addition to possessing canonical finite products) there is given
a mapping $\varepsilon : {\cal C} \times {\cal C} \to {\cal C}$
such that the following conditions holds:
\medskip\smallskip
{\parindent=27pt
\item{(E1)} For each $X,\, Y \in {\cal C}$ the underlying set of the object
$\varepsilon(X,Y)$ is just $\Hom(X,Y)$, i.e., 
$|{\varepsilon(X,Y)}| = \Hom(X,Y)$. 
\medskip
\item{(E2)} For each $X,\, Y \in {\cal C}$ the mapping
$\fss{a}^Y_X : |{\varepsilon(X,Y)}| \times |X| \to |Y|$ given by
\mdisp{\fss{a}^Y_X (h,x) \ =\ h(x)}
\item{} for all $h \in |{\varepsilon(X,Y)}| = \Hom(X,Y)$, 
$x \in |X|$, is a morphism, i.e., $\fss{a}^Y_X$ is an element
of $\Hom(\varepsilon(X,Y)\times X,Y)$.
\medskip
\item{(E3)}
If $X,\, Y,\,Z \in {\cal C}$ then for each 
$f \in \Hom(X\times Y,Z)$ there is a unique morphism 
$g \in \Hom(X,\varepsilon(Y,Z))$ such that
\mdisp{\fss{a}_Y^Z \,\circ\, (g \times \fss{id}_Y )\ =\ f\ ,}
\item{} i.e., such that $\,\fss{a}_Y^Z(g(x),y) \,=\, f(x,y)\,$
for all $x \in |X|$, $y \in |Y|$.
\medskip\smallskip
}

\bigbreak

The object $\varepsilon(X,Y)$ is often called the 
\indexdef{exponential object}{exponential object}{}
of $X$ and $Y$; the morphism $\fss{a}_X^Y$ is called a 
\indexdef{functional application operator}{functional application operator}{} 
or just an 
\indexddef{application operator}{application operator}{}{operator}{application}.
\medskip
Note that (E3) is really the assertion that for each $f \in \Hom(X\times Y,Z)$ 
the mapping $\,y \mapsto f(x,y)\,$ is an element of $\Hom(Y,Z)$ for each $x \in |X|$, 
and that the mapping $\,x \mapsto (y \mapsto f(x,y))\,$ is then an element of
$\Hom(X,\varepsilon(Y,Z))$.

\bigskip\bigskip
\frame{20pt}{\bigskip\medskip 
In the concrete categories associated with the basic cases 
the object $\varepsilon(X,Y)$ was defined to be
respectively the bottomed set $\totale X Y $ of all mappings from $|X|$ to 
$|Y|$, the bottomed poset $\mono X Y $ of all monotone mappings from
$|X|$ to $|Y|$, and the bottomed complete poset $\cont X Y $ of all continuous mappings 
from $|X|$ to $|Y|$. (Note that $\totale X Y $, $\mono X Y $ and $\cont X Y $ are now
being used to denote the objects and not the underlying sets.)
\medskip
In each of these cases $|\varepsilon(X,Y)|$ 
is just the corresponding set of mappings without  the additional structure
which  makes $\varepsilon(X,Y)$ respectively a 
bottomed set, a bottomed poset and a bottomed complete poset. 
In particular, this means that $|\varepsilon(X,Y)| = \Hom(X,Y)$. 
\medbreak
In all three cases it was shown that
$\fss{a}^Y_X \in \Hom(\varepsilon(X,Y)\times X,Y)$.
Moreover, it was also shown that (E3) holds, with $g = \fss{c}(f)$ and with
$\fss{c}$ the appropriate currying operator. This implies that the
concrete categories associated with the basic cases are all cartesian closed.
\bigskip\medskip
}
\bigskip\bigskip

In what follows assume that $\fss{C}$ is cartesian closed.
One of the main reasons for working in the present generality is that it allows
a uniform treatment of partial application operators. These operators, which are a 
generalisation of the application operators, will play an important role in the remainder
of the study.
 
\medskip
In order to be able to define the partial application operators some preparation
is needed. Note that if $X,\,Y,\,Z \in {\cal C}$ then the objects
$(X \times Y) \times Z$ and $X \times (Y \times Z)$ cannot be equal, since their
underlying sets are not even the same. However, they are isomorphic:

\proclaim{Lemma~5.1.4} The mapping
$\vartheta : (|X| \times |Y|) \times |Z| \to |X| \times (|Y| \times |Z|)$
defined by
\mdisp{ \vartheta((x,y),z)\ =\ (x,(y,z))}
for all $x \in |X|$, $y \in |Y|$ and $z \in |Z|$ is an element of
$\Hom( (X \times Y) \times Z, X \times (Y \times Z))$.
Moreover, $\vartheta$ is an isomorphism with inverse the mapping $\vartheta^{-1}$
given by 
\mdisp{ \vartheta^{-1}(x,(y,z))\ =\ ((x,y),z)}
for all $x \in |X|$, $y \in |Y|$ and $z \in |Z|$.
\endpro

\proof Put $U = (X\times Y)\times Z$; thus
$|U| = |X\times Y|\times |Z| = (|X|\times|Y|)\times |Z|$.
By (P2) the mappings 
$q_1 : |X\times Y| \to |X|$ and $q_2 : |X\times Y| \to |Y|$ 
defined by $q_1(x,y) = x$ and $q_2(x,y) = y$ are 
morphisms, i.e., $q_1 \in \Hom(X\times Y,X)$ and $q_2 \in \Hom(X\times Y,Y)$. In the same
way, by (P2) the mappings $r_1 : |U| \to |X\times Y|$ and 
$r_2 : |U| \to |Z|$ defined
by $r_1(w,z) = w$ and $r_2(w,z) = z$ are 
also morphisms, i.e., $r_1 \in \Hom(U,X\times Y)$ and $r_2 \in \Hom(U,Z)$. 
Put $q'_1 = q_1 \circ r_1$, $q'_2 = q_2 \circ r_1$ and $q'_3 = r_2$, hence
$q'_1 \in \Hom(U,X)$, $q'_2 \in \Hom(U,Y)$ and $q'_3 \in \Hom(U,Z)$, and let
$\varphi : U \to X \times Y \times Z$ be the mapping given by
$\varphi((x,y),z) = (x,y,z)$ for all $x \in |X|$, $y \in |Y|$, $z \in |Z|$.
Then $p_j \circ \varphi = q'_j$ for each $j = 1,\,2,\,3$, where
$p_1 \in \Hom(X\times Y\times Z,X)$, $p_2 \in \Hom(X\times Y\times Z,Y)$ and 
$p_3 \in \Hom(X\times Y\times Z,Z)$ are the morphisms occurring in (P2) in the definition
of $X\times Y \times Z$. Therefore by Lemma~5.1.2 $\varphi$ is a morphism, i.e., 
$\varphi$ is an element of $\Hom(U,X\times Y \times Z)$. 
\medskip
Now let $p'_1 : |X\times Y\times Z| \to |X\times Y|$ be the mapping
given by $p'_1(x,y,z) = (x,y)$;
then by Lemma~5.1.2 $p'_1 \in \Hom(X\times Y\times Z,X\times Y)$,
since $q_1 \circ p'_1 = p_1$ and $q_2 \circ p'_1 = p_2$. Thus, again making use of 
Lemma~5.1.2,
the mapping $\varphi' : X \times Y \times Z \to U$ given by
$\varphi'(x,y,z) = ((x,y),z)$ for all $x \in |X|$, $y \in |Y|$, $z \in |Z|$,
is a morphism, since $r_1 \circ \varphi' = p'_1$ and $r_2 \circ \varphi' = p_3$.
But clearly $\varphi' \circ \varphi = \fss{id}_U$ and 
$\varphi \circ \varphi' = \fss{id}_{X\times Y\times Z}$, and this shows that
$\varphi$ is an isomorphism with inverse $\varphi'$.
\medskip
Finally, let $\psi : |X| \times |Y\times Z| \to |X\times Y \times Z|$ be
given by $\psi(x,(y,z)) = (x,y,z)$. The same proof as above shows that
$\psi \in \Hom(X\times (Y\times Z), X\times Y \times Z)$ and that
$\psi$ is an isomorphism with inverse 
$\psi' \in \Hom(X\times Y\times Z, X\times (Y \times Z))$ the mapping
given by $\psi'(x,y,z) = (x,(y,z))$ for all $x \in |X|$, $y \in |Y|$, $z \in |Z|$.
But clearly $\vartheta = \psi' \circ \varphi$ and therefore 
$\Hom( (X \times Y) \times Z, X \times (Y \times Z))$; moreover,
$\varphi$, as the composition of two isomorphisms, is itself an isomorphism
with inverse $\vartheta^{-1} = \varphi'\circ \psi$, which in particular implies that
$\,\vartheta^{-1}(x,(y,z))\, =\, ((x,y),z)\,$
for all $x \in |X|$, $y \in |Y|$ and $z \in |Z|$.
\eop
\bigbreak
The mapping in Lemma~5.1.4 will always be denoted by $\vartheta$; the 
objects $X$, $Y$ and $Z$ can always be determined from the context.
\medskip
If $L \in {\cal F}_{\cal C}$ and $J$ is a subset of the underlying
set $L$ then $J$ will be considered as a \tsym{\cal C}-typed set with the typing
induced from $L$. 

\proclaim{Lemma~5.1.5} Let $L \in {\cal F}_{\cal C}$ and $J$ be a non-empty proper
subset of $L$. Then the mapping 
$\fss{p}_{L,J} : |{\otimes L}| \to |{\otimes J}|$ given by
\smallskip
\mdisp{ \fss{p}_{L,J}(c)\,(\eta)\ =\ c(\eta)}
for all $c \in |{\otimes L}|$, $\eta \in J$, is a morphism, i.e., an element of
$\Hom(\otimes L,\otimes J)$. \endpro

\proof For each $\eta \in L$ let $p_\eta \in \Hom(\otimes L,\langle\eta\rangle)$
be the morphism occurring in (P2) in the definition of $\otimes L$, and for each
$\eta \in J$ let $p'_\eta \in \Hom(\otimes J,\langle\eta\rangle)$ be the corresponding
morphism in the definition of $\otimes J$.
Then clearly $p'_\eta \circ \fss{p}_{L,J} = p_\eta$ for each $\eta \in J$, and hence by
Lemma~5.1.2 $\fss{p}_{L,J}$ is a morphism. \eop 
\eject

\proclaim{Lemma~5.1.6} Let $L \in {\cal F}_{\cal C}$ and $J$ be a non-empty proper
subset of $L$. Then the mapping 
$\fss{s}_{L,J} : |{\otimes J}| \times |{\otimes (L{\setminus} J)}| \to |{\otimes L}|$ 
given by
\smallskip
\mdisp{ \fss{s}_{L,J}(a,b)\,(\eta)
\ =\ \cases{ a(\eta) & if $\,\eta \in J\,$,\cr
\noalign{\smallskip}
             b(\eta) & if $\,\eta \in L{\setminus} J\,$,\cr}}
\smallskip
is a morphism, i.e., an element of 
$\Hom(\otimes J \times \otimes (L{\setminus} J), \otimes L)$. Moreover, $\fss{s}_{L,J}$
is an isomorphism with inverse the mapping from $|{\otimes L}|$ to
$|{\otimes J}| \times |{\otimes (L{\setminus} J)}|$ given by
\mdisp{ \fss{s}_{L,J}^{-1}(c)\ =\ (\fss{p}_{L,J}(c),\fss{p}_{L,L{\setminus} J}(c))}
for all $c \in |{\otimes J}|$. \endpro

\proof This is very similar to the proof of Lemma~5.1.4 and is left to the reader. \eop
\bigbreak
\proclaim{Lemma~5.1.7} Let $L \in {\cal F}_{\cal C}$ and let $J_1,\,J_2,\,J_3$ 
be non-empty disjoint subsets of $L$ with $L = J_1 \cup J_2 \cup J_3$. Then
\mdisp{\fss{s}_{L,J_1}\,\circ\,(\fss{id}_{\otimes J_1} \times \fss{s}_{J_2\cup J_3,J_2})
\,\circ\, \vartheta
\ =\ \fss{s}_{L,J_1\cup J_2}\,\circ\,
(\fss{s}_{J_1\cup J_2,J_1}\times \fss{id}_{\otimes J_3})}
with $\vartheta \in \Hom(\,(\otimes J_1\times \otimes J_2)\times \otimes J_3\,,\,
\otimes J_1\times (\otimes J_2 \times \otimes J_3)\,)$. In other words,
\mdisp{\fss{s}_{L,J_1}(c_1,\fss{s}_{J_2\cup J_3,J_2}(c_2,c_3))
\ =\ \fss{s}_{L,J_1\cup J_2}(\fss{s}_{J_1\cup J_2,J_1}(c_1,c_2),c_3)}
for all $c_1 \in |{\otimes J_1}|$, $c_2 \in |{\otimes J_2}|$ and $c_3 \in |{\otimes J_3}|$.
\endpro
\proof This is straightforward; for example, if
$\eta \in J_2$ then
\smallskip
\ldisp{\quad\fss{s}_{L,J_1}(c_1,\fss{s}_{J_2\cup J_3,J_2}(c_2,c_3))\,(\eta)
\ =\ \fss{s}_{J_2\cup J_3,J_2}(c_2,c_3))\,(\eta)
\ =\ c_2(\eta)}
\vskip-\medskipamount
\rdisp{ 
\ =\ \fss{s}_{J_1\cup J_2,J_1}(c_1,c_2)\,(\eta)
\ =\ \fss{s}_{L,J_1\cup J_2}(\fss{s}_{J_1\cup J_2,J_1}(c_1,c_2),c_3)\,(\eta)\ ,\quad}
\smallskip
with similar calculations in the other two cases. \eop
\bigbreak

Let $L \in {\cal F}_{\cal C}$ and $Y \in {\cal C}$ and, to simplify the notation, it 
is convenient here to just write $\varepsilon(L,Y)$ instead of $\varepsilon(\otimes L,Y)$
and $\fss{a}^Y_L$ instead of $\fss{a}^Y_{\otimes L}$. Now the application operator 
$\fss{a}^Y_L$ is a mapping
from $|{\varepsilon(L,Y)}| \times |{\otimes L}|$ to $|Y|$, in which the element
$\fss{a}^Y_L(h,c)$ of $|Y|$ is the value obtained
by applying the function $h \in |{\varepsilon(L,Y)}| = \Hom(\otimes L,Y)$ to its arguments
$c \in |{\otimes L}|$.
\medskip 
The partial application operators will generalise this, but with 
the function applied to only some of its arguments. To be more precise,
let $J$ be a non-empty proper subset of $L$ (which means in particular that $L$ must 
contain at least two elements). Then there will be a
partial application operator $\fss{pa}^Y_{L,J}$, which will be a mapping from
$|{\varepsilon(L,Y)}| \times |{\otimes J}|$ to $|{\varepsilon(L{\setminus} J,Y)}|$,
in which the element $\fss{pa}^Y_{L,J}(h,b)$ of 
$|{\varepsilon(L{\setminus} J,Y)}|$ 
is the value obtained
by applying the function $h \in |{\varepsilon(L,Y)}| = \Hom(\otimes L,Y)$ just to the 
arguments $b \in |{\otimes J}|$. This means that the value is now an element of
$|{\varepsilon(L{\setminus} J,Y)}| = \Hom(\otimes (L\setminus J),Y)$, and 
is thus a function from $|{\otimes (L{\setminus} J)}|$ to $|Y|$.
In fact, it should be the function which, when the remaining arguments are
applied to it, gives the same value (i.e., the same element of $|Y|$) which would
have been obtained if all the arguments had been applied directly to 
$h$. In other words, 
$\fss{pa}^Y_{L,J}(h,b)\,(c)\, =\, h(\fss{s}_{L,J}(b,c))\,$, i.e., 

\mdisp{ \fss{a}^Y_{L\setminus J}(\fss{pa}^Y_{L,J}(h,b),c)
\ =\ \fss{a}^Y_{L}(h,\fss{s}_{L,J}(b,c))}
should hold for all $h \in |{\varepsilon(L,Y)}|$, $b \in |{\otimes J}|$ and all
$c \in |{\otimes (L{\setminus} J)}|$. The morphism $\fss{s}_{L,J}$ is needed
here to combine the two partial sets of arguments $b$ and $c$ into a complete set
of arguments $\fss{s}_{L,J}(b,c) \in |{\otimes L}|$.
\medskip
The above equation describes how $\fss{pa}^Y_{L,J}$ should behave as a mapping, and
clearly there is at most one such mapping from
$|{\varepsilon(L,Y)}| \times |{\otimes J}|$ to $|{\varepsilon(L{\setminus} J,Y)}|$. 
The existence is not so clear: This requires
that the mapping $\,c \mapsto h(\fss{s}_{L,J}(b,c))\,$ be a morphism 
(i.e., an element of $\Hom(\otimes (L{\setminus} J),Y)$)
for each $h \in |{\varepsilon(L,Y)}|$, $b \in |{\otimes J}|$.
However,
what is really needed is that $\fss{pa}^Y_{L,J}$ itself should actually be a morphism,
i.e., an element of
$\Hom(\varepsilon(L,Y)\times \otimes J, \varepsilon(L{\setminus} J,Y))$. 
It will be shown below that there is, in fact, a morphism (and thus a unique morphism) 
$\fss{pa}^Y_{L,J}$ satisfying this equation.
\medskip
In order to establish the existence of the morphism $\fss{pa}_{L,J}^Y$
it is useful to rewrite the above equation in a form which does
not involve the arguments of the mappings. 

\proclaim{Lemma~5.1.8} Let $L \in {\cal F}_{\cal C}$, $J$ be a non-empty proper subset of 
$L$ and $Y \in {\cal C}$. Let 
$\psi \in \Hom(\varepsilon(L,Y)\times \otimes J, \varepsilon(L{\setminus} J,Y))$; then
\mdisp{ \fss{a}^Y_{L\setminus J}(\psi(h,b),c)
\ =\ \fss{a}^Y_{L}(h,\fss{s}_{L,J}(b,c))}
holds for all $h \in |{\varepsilon(L,Y)}|$, $b \in |{\otimes J}|$ and all
$c \in |{\otimes (L{\setminus} J)}|$ if and only if
\mdisp{\fss{a}_{L\setminus J}^Y \,\circ\,
(\psi \times \fss{id}_{\otimes (L{\setminus} J)})
\ =\ \fss{a}_{L}^Y\,\circ\,(\fss{id}_{\varepsilon(L,Y)} \times \fss{s}_{L,J})
\,\circ\, \vartheta
}
with
$\vartheta \in \Hom(\,(\varepsilon(L,Y)\times \otimes J)\times \otimes (L{\setminus} J)\,,\,
\varepsilon(L,Y)\times (\otimes J \times \otimes (L{\setminus} J))\,)$.
\endpro

\proof Let $h \in |{\varepsilon(L,Y)}|$, $b \in |{\otimes J}|$ and
$c \in |{\otimes (L{\setminus} J)}|$; then
\smallskip
\ldisp{\quad
\fss{a}^Y_{L}(h,\fss{s}_{L,J}(b,c))
\ =\ \bigl(\,\fss{a}^Y_{L}
\circ (\fss{id}_{\varepsilon(L,Y)} \times \fss{s}_{L,J})\,\bigr)\,(h,(b,c))
}
\vskip-\medskipamount
\rdisp{
\ =\ \bigl(\,\fss{a}^Y_{L}
\circ (\fss{id}_{\varepsilon(L,Y)} \times \fss{s}_{L,J})\,\circ\,
 \vartheta\,\bigr)\,((h,b),c))\ ,\quad
}
\smallskip
and on the other hand
\mdisp{ \fss{a}^Y_{L\setminus J}(\psi(h,b),c)
\ =\ \bigl(\,\fss{a}^Y_{L\setminus J}
 \circ (\psi \times \fss{id}_{\otimes (L\setminus J)})\,\bigr)\, ((h,b),c)\ .}
The result follows immediately from these two facts. \eop
\bigbreak

\proclaim{Proposition~5.1.1}
Let $L \in {\cal F}_{\cal C}$, $J$ be a non-empty proper subset of $L$ and
$Y \in {\cal C}$. Then there is a  unique morphism
$\fss{pa}_{L,J}^Y \in \Hom(\varepsilon(L,Y)\times \otimes J, 
\varepsilon(L{\setminus} J,Y))$ 
such that
\mdisp{\fss{a}_{L\setminus J}^Y \,\circ\,
(\fss{pa}_{L,J}^Y \times \fss{id}_{\otimes (L\setminus J)})
\ =\ \fss{a}_{L}^Y\,\circ\,(\fss{id}_{\varepsilon(L,Y)} \times \fss{s}_{L,J})
\,\circ\, \vartheta\ ,
}
again with
$\vartheta \in \Hom(\,(\varepsilon(L,Y)\times \otimes J)\times \otimes (L{\setminus} J)\,,\,
\varepsilon(L,Y)\times (\otimes J \times \otimes (L{\setminus} J))\,)$.
\endpro

\proof Just apply (E3) with
$\,f\,=\,\fss{a}_{L}^Y\,\circ\,(\fss{id}_{\varepsilon(L,Y)} \times \fss{s}_{L,J})
\,\circ\, \vartheta\,$
(and note that the objects $X$, $Y$ and $Z$ occurring in (E3) are here equal to
$\varepsilon(L,Y)\times \otimes J$, $\otimes (L{\setminus} J)$ and $Y$ respectively). \eop
\bigbreak
Let $L \in {\cal F}_{\cal C}$, $J$ be a non-empty proper subset of $L$ and 
$Y \in {\cal C}$. Then by Lemma~5.1.8 
$\fss{pa}_{L,J}^Y \in \Hom(\varepsilon(L,Y)\times \otimes J, 
\varepsilon(L{\setminus} J,Y))$ 
is the unique morphism such that
\mdisp{ \fss{a}^Y_{L\setminus J}(\fss{pa}_{L,J}^Y(h,b),c)
\ =\ \fss{a}^Y_{L}(h,\fss{s}_{L,J}(b,c))}
for all $h \in |{\varepsilon(L,Y)}|$, $b \in |{\otimes J}|$ and all
$c \in |{\otimes (L{\setminus} J)}|$. In fact, as already noted,
$\fss{pa}_{L,J}^Y$ is the unique mapping from $\varepsilon(L,Y)\times \otimes J$
to $\varepsilon(L{\setminus} J,Y))$ satisfying this equation.

\proclaim{Proposition~5.1.2} Let $L \in {\cal F}_{\cal C}$, let
$J$ and $J'$ be non-empty disjoint subsets of $L$ with $J\cup J' \ne L$ 
and let $Y \in {\cal C}$; then
\mdisp{\fss{pa}_{L,J\cup J'}^Y \,\circ\,(\fss{id}_{\varepsilon(L,Y)} 
\times \fss{s}_{J\cup J',J})\,\circ\, \vartheta
\ =\ \fss{pa}^Y_{L\setminus J,J'}\,\circ\,
(\fss{pa}_{L,J}^Y \times \fss{id}_{\otimes J'})}
with 
$\vartheta \in \Hom(\,(\varepsilon(L,Y)\times \otimes J)\times \otimes J'\,,\,
\varepsilon(L,Y)\times (\otimes J \times \otimes J')\,)$. 
Equivalently, 
\mdisp{\fss{pa}_{L,J\cup J'}^Y(h,\fss{s}_{J\cup J',J}(b,b'))
\ =\ \fss{pa}_{L\setminus J,J'}^Y (\fss{pa}_{L,J}^Y(h,b),b')}
holds for all $h \in |{\varepsilon(L,Y)}|$, $b \in |{\otimes J}|$ and
$b' \in |{\otimes J'}|$. \endpro

\proof Let $h \in |{\varepsilon(L,Y)}|$, $b \in |{\otimes J}|$ and 
$b' \in |{\otimes J'}|$; then by Lemma~5.1.7

\mdisp{
\fss{s}_{L,J}(b,\fss{s}_{L\setminus J,J'}(b',a))
\ =\ \fss{s}_{L,J\cup J'}(\fss{s}_{J\cup J',J}(b,b'),a)
}
for all $a \in |{\otimes (L\setminus (J\cup J'))}|$. Thus
\smallskip
\ldisp{\quad \fss{pa}_{L\setminus J,J'}^Y (\fss{pa}_{L,J}^Y(h,b),b')\,(a)
\ =\ \fss{pa}_{L,J}^Y(h,b)\,(\fss{s}_{L\setminus J,J'}(b',a))}

\vskip-\medskipamount
\mdisp{
\ =\ h(\fss{s}_{L,J}(b,\fss{s}_{L\setminus J,J'}(b',a)))
\ =\ h(\fss{s}_{L,J\cup J'}(\fss{s}_{J\cup J',J}(b,b'),a))}
\vskip-\medskipamount
\rdisp{
\ =\ \fss{pa}_{L\setminus J,J'}^Y (\fss{pa}_{L,J}^Y(h,b),b')\,(a) \quad
}
for all $a \in |{\otimes (L\setminus (J\cup J'))}|$, and therefore
\mdisp{\fss{pa}_{L,J\cup J'}^Y(h,\fss{s}_{J\cup J',J}(b,b'))
\ =\ \fss{pa}_{L\setminus J,J'}^Y (\fss{pa}_{L,J}^Y(h,b),b')\ .\quad\eop}
\bigbreak
\bigbreak
It is useful to extend the class of partial application operators by putting
$\fss{pa}^Y_{L,L} = \fss{a}^Y_L$ for each non-empty $L \in {\cal F}_{\cal C}$.
Note, however, that the morphism $\fss{pa}^Y_{L,L}$ is defined to be an element of
$\Hom(\varepsilon(L,Y)\times \otimes L, Y)$ (and not an element of
$\Hom(\varepsilon(L,Y)\times \otimes L,\varepsilon(\Oneptset,Y)$).
The reason for making this extension
is that the partial application operators then include the
application operators, and it is justified by the fact that Proposition~5.1.2
still holds:

\proclaim{Proposition~5.1.3} Let $L \in {\cal F}_{\cal C}$, let
$J$ and $J'$ be non-empty disjoint subsets of $L$  
and let $Y \in {\cal C}$; then
\mdisp{\fss{pa}_{L,J\cup J'}^Y \,\circ\,(\fss{id}_{\varepsilon(L,Y)} 
\times \fss{s}_{J\cup J',J})\,\circ\, \vartheta
\ =\ \fss{pa}^Y_{L\setminus J,J'}\,\circ\,
(\fss{pa}_{L,J}^Y \times \fss{id}_{\otimes J'})}
with 
$\vartheta \in \Hom(\,(\varepsilon(L,Y)\times \otimes J)\times \otimes J'\,,\,
\varepsilon(L,Y)\times (\otimes J \times \otimes J')\,)$.
Equivalently,
\mdisp{\fss{pa}_{L,J\cup J'}^Y(h,\fss{s}_{J\cup J',J}(b,b'))
\ =\ \fss{pa}_{L\setminus J,J'}^Y (\fss{pa}_{L,J}^Y(h,b),b')}
holds for all $h \in |{\varepsilon(L,Y)}|$, $b \in |{\otimes J}|$ and
$b' \in |{\otimes J'}|$. \endpro

\proof If $J \cup J' \ne L$ then this is just Proposition~5.1.2, and if
$J \cup J' = L$ then it follows immediately from
Proposition~5.1.1 and Lemma~5.1.8. \eop
\bigbreak

The concrete cartesian closed category $\fss{C}$ will be called 
\indexdef{bottomed}{bottomed category}{} 
if for each $X \in {\cal C}$ the underlying set
$|X|$ contains a bottom element, to be denoted by $\bot_X$, 
such that the following two conditions hold:
\medskip\smallskip
{\parindent=25pt
\item{(B1)} For each $L \in {\cal F}_{\cal C}$ the bottom element
$\bot_{\otimes L}$ of $|{\otimes L}| = \prod |L|$ is the mapping defined by
$\bot_{\otimes L}(\eta) = \bot_{\langle \eta\rangle}$ for each $\eta \in L$.
\medskip
\item{(B2)} For each $X,\,Y \in {\cal C}$ the bottom element $\bot_{\varepsilon(X,Y)}$
of $|{\varepsilon(X,Y)}|$ is the unique element $h \in |{\varepsilon(X,Y)}|$  such that
$\fss{a}_X^Y(h,x) = \bot_{Y}$ for all $x \in |X|$. (In other words,
$\fss{a}_X^Y(\bot_{\varepsilon(X,Y)},x) = \bot_{Y}$ for all $x \in |X|$, and if
$\fss{a}_X^Y(h,x) = \bot_{Y}$ for all $x \in |X|$ then $h = \bot_\varepsilon(X,Y)$.)
\medskip\smallskip
}
The concrete categories associated with the basic cases are clearly bottomed.
In what follows assume that $\fss{C}$ is bottomed.

\proclaim{Proposition~5.1.4} Let $L \in {\cal F}_{\cal C}$, $J$ be a non-empty subset 
of $L$ and let $Y \in {\cal C}$; then
\mdisp{ \fss{pa}^Y_{L,J}(\bot_{\varepsilon(L,Y)},b)\ =\ \bot_Z}
for all $b \in |{\otimes J}|$, where of course $Z = Y$ if $J = L$ and
$Z = \varepsilon(L{\setminus}J,Y)$ if $J$ is a proper subset of $L$. \endpro

\proof This follows directly from (B2) if $J = L$, and so it can assumed that
$J$ is a proper subset of $L$. Then by Proposition~5.1.1, Lemma~5.1.8 and (B2)

\mdisp{ \fss{a}^Y_{L\setminus J}(\fss{pa}_{L,J}^Y(\bot_{\varepsilon(L,Y)},b),c)
\ =\ \fss{a}^Y_{L}(\bot_{\varepsilon(L,Y)},\fss{s}_{L,J}(b,c))\ =\ \bot_Y}
for all $b \in |{\otimes J}|$, $c \in |{\otimes (L{\setminus} J)}|$, and thus
by the uniqueness in (B2) it follows that
\mdisp{ \fss{pa}^Y_{L,J}(\bot_{\varepsilon(L,Y)},b)
  \ =\ \bot_{\varepsilon(L\setminus J,Y)}}
for all $b \in \otimes J$. \eop
\bigbreak
\vfill\eject

\hfline{123}{5.2 \ FUNCTIONAL TYPES AND FUNCTIONAL ALGEBRAS}
\bigskip

\sectionhead {5.2} {Functional types and functional algebras}
\bigskip\medskip
This chapter deals with the functions to be defined over the data objects.
These will be obtained by first 
introducing a set $S$ called the set of functional types over $B$
and then extending the family $D_B$ to a family $D_S$ in such a way 
that it is natural to regard $D_\sigma$ 
as a set of functions of type $\sigma$ for each $\sigma \in S \setminus B$. 
\medskip
For any set $S$ denote by ${\cal F}^o_S$ the set of non-empty finite 
\tsym{S}-typed sets, in other words 
${\cal F}^o_S \,=\, {\cal F}_S \setminus \{\varnothing\}$.
Somewhat informally, the set $S$ of functional types over $B$ can be thought 
of as being defined by the following rules:
\medskip\smallskip
{\parindent=25pt
\item{(1)} Each ground type $\beta \in B$ is an element of $S$.
\medskip
\item{(2)} If $L \in {\cal F}^o_S$ and $\beta \in B$
then $L \to \beta$ is an element of $S$.
\medskip
\item{(3)} Each element of $S$ can be uniquely constructed using rules (1)
and (2).
\bigskip
}  
The functions of type $L \to \beta$ will be defined to be
some kind of functions whose arguments are named by the elements of the set $L$, in which 
the values of the argument corresponding to $\eta \in L$ 
range over the set of elements of type $\langle \eta \rangle$, and which take their 
values in the set $D_\beta$. In particular, 
if $\list {\sigma} n \in S^*$ with $n \ge 1$ then the
functions of type $\list {\sigma} n \to \beta$ will be functions of $n$ arguments 
with the values of the 
\tsym{j}-th argument ranging over the set of elements of type $\sigma_j$.
As mentioned at the beginning of the chapter, this recursive definition is 
necessary in order to obtain higher-order types. 
\medskip
Now it is convenient to identify the ground type
$\beta \in B$ with a functional type having no arguments, i.e., with the type
$\varnothing \to \beta$, and this reduces the informal definition to just the
following two rules:
\medskip\smallskip
{\parindent=25pt
\item{(1)} If $L \in {\cal F}_S$ and $\beta \in B$ then $L \to \beta$ is an element of 
$S$.
\medskip
\item{(2)} Each element of $S$ can be uniquely constructed using rule (1).
\bigskip
}  

This must be made precise. Thus suppose $S$ is a set which is going
to be a candidate for the set of functional types over $B$. Then
a mapping $\# : {\cal F}_S \times B \to S$ must also be given
which `constructs' the new types, i.e., so that $\#(L,\beta)$ is the new type
$\,L \to \beta\,$ occurring in (1)
for each  $L \in {\cal F}_S$, $\beta \in B$.
Such a pair $(S,\#)$ will be called a 
\indexdef{set of functional types over}{set}{of functional types}
$B$.
If $(S,\#)$ and $(S',\#')$ are sets of functional types over $B$ then a 
\indexdef{homomorphism}{homomorphism}{}
from $(S,\#)$ to $(S',\#')$ is any mapping
$\pi : S \to S'$ such that
\mdisp{\pi(\#(L , \beta))
\ =\ \#'(\pi\,L,\beta)}
for all $L \in {\cal F}_S$, $\beta \in B$, where
$\pi\,L$ is the \tsym{S'}-typed set having the same underlying set as $L$ together
with the typing $\langle\cdot\rangle' : L \to S'$ 
defined by $\langle \eta\rangle' = \pi(\langle\eta\rangle)$ for each $\eta \in L$ 
(with $\langle\cdot\rangle$ the typing on $L$). 
As is to be expected, a set of functional types $(S,\#)$ over $B$  is said 
to be 
\indexddef{initial}{initial set of functional types}{}{set of functional types}{initial} 
if for each set of functional types $(S',\#')$ over $B$ there exists a unique
homomorphism from $(S,\#)$ to $(S',\#')$. If such an initial 
$(S,\#)$ exists then it is, of course, up to isomorphism unique.

\proclaim{Proposition~5.2.1} There exists an initial set of functional types over $B$.
Moreover, a set of functional types $(S,\#)$ over $B$ is initial if and only
if the following hold:
\medskip\smallskip
{\parindent=25pt
\item{(1)} The mapping $\# : {\cal F}_S \times B \to S$ is bijective.
\medskip
\item{(2)} If $T \subset S$ with  $\#({\cal F}_T \times B) \subset T$ then $T = S$.
\medskip\smallskip
}  
(Note that by (2) `bijective' could be replaced in (1) by `injective',
since if $T = \Im(\#)$ then clearly
$\,\#({\cal F}_T \times B) \,\subset \,\#({\cal F}_S \times B) \,=\, T\,$.)
\endpro

\proof Consider the single-sorted signature
$\Xi = ({\cal F}_\Oneptset \times B,\Phi)$ in which
the mapping $\Phi: {\cal F}_\Oneptset \times B \to {\cal F}_\Oneptset$ is just the
projection onto the first factor (i.e., $\Phi(A,\beta) = A$ for all
$A \in {\cal F}_\Oneptset$, $\beta \in B$).
Thus $\Xi$ contains a single type $\onept$ and for each $A \in {\cal F}_\Oneptset$ and 
each $\beta \in B$ an operator name $(A,\beta)$ of type 
$A \to \onept$. 
Let $(S,h_{{\cal F}_\Oneptset\times B})$ be a
\tsym{\Xi}-algebra. Note that $h_{(A,\beta)}$ is then
a mapping from $\total A S $ to $S$ for each $A \in {\cal F}_\Oneptset$, $\beta \in B$,
and that the elements of the set $\total A S $ can be considered as typings on the set $A$.
A mapping $\# : {\cal F}_S \times B \to S$ can thus be defined by letting
\mdisp{\#(L , \beta) \ =\ h_{(A,\beta)}(\langle\cdot\rangle)}
for each $L = (A,\langle\cdot\rangle) \in {\cal F}_S$, $\beta \in B$;
$(S,\#)$ will be called the 
set of functional types \definition{associated} with 
$(S,h_{{\cal F}_\Oneptset\times B})$. Conversely, if
$(S,\#)$ is any set of functional types over $B$ then
for each $A \in {\cal F}_\Oneptset$, $\beta \in B$,
a mapping from $h_{(A,\beta)} : \total A S \to S$ can be defined by
\mdisp{h_{(A,\beta)}(\langle\cdot\rangle)\ =\ \#((A,\langle\cdot\rangle) , \beta)}

for each $\langle\cdot\rangle \in \total A S $;
$(S,h_{{\cal F}_\Oneptset\times B})$ will be called the
\tsym{\Xi}-algebra \definition{associated} with $(S,\#)$.
\medskip
Let $(S,\#)$ be a set of functional types over $B$ and
$(S,h_{{\cal F}_\Oneptset\times B})$ be a \tsym{\Xi}-algebra (with the same
set $S$); then it is clear that $(S,\#)$ is the
set of functional types associated with $(S,h_{{\cal F}_\Oneptset\times B})$
if and only of $(S,h_{{\cal F}_\Oneptset\times B})$ is the
\tsym{\Xi}-algebra associated with $(S,\#)$.
Moreover, if $(S',\#')$ is a further
set of functional types over $B$ and $\pi : S \to S'$ is a mapping then it
is easily checked that $\pi$ is a homomorphism from $(S,\#)$ to $(S',\#')$
if and only if $\pi$ is a \tsym{\Xi}-homomorphism between the associated
\tsym{\Xi}-algebras. From this it follows that a
\tsym{\Xi}-algebra $(S,h_{{\cal F}_\Oneptset\times B})$ is 
initial if and only if the set of functional types associated with 
$(S,h_{{\cal F}_\Oneptset\times B})$ is initial.
Thus by Proposition~2.3.3 there exists an initial set of functional types over $B$.
\medskip
The second part of Proposition~5.2.1 now follows from the above correspondence
between \tsym{\Xi}-algebras and sets of functional types over $B$, together
with the characterisation
of initial \tsym{\Xi}-algebras given in Proposition~2.3.2. \eop
\bigbreak

In what follows let $(S,\#)$ be an initial set of functional types over $B$
(whose existence is guaranteed by Proposition~5.2.1). In the sense that
initial objects are unique, $(S,\#)$ will be referred to as
{\it the} set of functional types over $B$. Moreover, this pair will usually
be denoted just by $S$, and the notation 
$L \to \beta$ will almost always be employed instead of $\#(L ,\beta)$. The symbol $\to$ 
thus has two different usages; it will be seen, however, that these two usages are, in a
certain sense, compatible. 
\medskip
By (1) in Proposition~5.2.1 each element of $S$ has a unique representation of the form 
$L \to \beta$ with $L \in {\cal F}_S$ and $\beta \in B$, which implies, in particular, that 
$\varnothing \to \beta$ and $\varnothing \to \beta'$ 
are different elements of $S$ whenever $\beta \ne \beta'$. 
This means that $\varnothing \to \beta$ can (and will) be identified 
with $\beta$ for each $\beta \in B$; in this way $B$ is considered to be a subset 
of $S$.
\medskip
Finally, although $S$ is referred to as the set of functional types,
the statement that $\sigma \in S$ is a 
\indexddef{functional type}{functional type}{}{type}{functional}
will mean that $\sigma \in S \setminus B$, i.e., that 
$\sigma$ has the form $L \to \beta$ with $L \in {\cal F}^o_S$.
The set $S$ is therefore
divided up into functional types (the elements of $S \setminus B$) and
ground types (the elements of $B$).
\medskip
If $\sigma = L\to \beta$ is a functional type and $J$ is a subset of $L$
then $\sigma_J$ will be used to denote the type $J\to \beta$, where $J$ is considered as 
an \tsym{S}-typed set with the typing induced from $L$. In particular, 
$\sigma_\varnothing = \beta$.
\bigskip\medskip
\frame{20pt}{\bigskip 
{\it Example~5.2.1\enspace} In the signature $\Lambda$ in Example~2.2.1
the set of ground types is the set
$B = \{\ftt{bool},\ftt{nat},\ftt{int},\ftt{pair},\ftt{list}\}$. 
The set $S$ thus contains the functional types
$\ftt{int} \to \ftt{int}$, 
$\ftt{list}\ \ftt{list} \to \ftt{list}$,  
$\ftt{list}\ \ftt{int}\ \ftt{list} \to \ftt{list}$,
$\ftt{int} \to \ftt{list}$ and
$(\ftt{int}\to\ftt{int})\ \ftt{int} \to \ftt{list}$
which already occurred implicitly in Example~1.2.1.
\bigskip
}
\bigskip\bigskip

\proclaim{Lemma~5.2.1} There is a unique mapping $|\cdot| : S \to \Nat$ 
with $|\beta| = 0$ for each ground type $\beta \in B$ and such that 
\mdisp{|L \to \beta| \ =\ 1 \,+ \,\sum\limits_{\eta \in L} |\langle \eta \rangle|}
for each functional type $L \to \beta$.
\endpro

\proof Let $g : {\cal F}_\Nat \times B \to \Nat$ be the mapping given by
$g(\varnothing,\beta) = 0$ and
\mdisp{g(L,\beta)\ =\ 1 \,+\,\sum\limits_{\eta \in L} \langle\eta\rangle }

for all $L \in {\cal F}^o_\Nat$, $\beta \in B$.
Then $(\Nat,g)$ is a set of functional types over $B$ and so there exists a
unique homomorphism $|\cdot|$ from $(S,\#)$ to $(\Nat,g)$. Thus
$|\cdot| : S \to \Nat$ is such that
$\,|\beta| \,=\, |\varnothing \to \beta| \,=\, |\#(\varnothing,\beta)| \,=\,
  g(\varnothing,\beta) \,=\, 0\,$ 
for all $\beta \in B$ and
\mdisp{|L \to \beta|\ =\ |\#(L , \beta)|
\ =\ g(|\cdot|\,L,\beta)\ =\ 1 \,+ \,\sum\limits_{\eta \in L} |\langle \eta \rangle|}
for all $L \in {\cal F}^o_S$, $\beta \in B$. The uniqueness of the mapping $|\cdot|$ 
follows either from the uniqueness of the homomorphism from $(S,\#)$ to $(\Nat,g)$, or 
directly from condition (2) in Proposition~5.2.1. \eop
\bigbreak
The elements $\sigma$ of $S$ with $|\sigma| = 1$ are called 
\indexddef{first order functional types}
{first order functional type}{}{functional type}{first order}; 
these are exactly the types having the
form $L \to \beta$ with $L \in {\cal F}^o_S$ and $\langle \eta \rangle \in B$ for each 
$\eta \in L$ (i.e., each of the `parameters' of a first order functional type is of 
ground type). The elements $\sigma$ of $S$ with $|\sigma| > 1$ are then called
\indexddef{higher order functional types}
{higher order functional type}{}{functional type}{higher order}: 
These are therefore the types having the
form $L \to \beta$ with $L \in {\cal F}^o_S$ and $\langle \eta \rangle \in S \setminus B$
for at least one $\eta \in L$ (i.e., a higher order functional type has at least one 
`parameter' which is of functional type). 

\bigskip
Having constructed the set $S$ of functional types, the next step is to
extend the family $D_B$ in the \tsym{\Lambda}-algebra $(D_B,f_K)$ to a family $D_S$ 
in such a way that it is natural to regard $D_\sigma$ 
as a set of functions of type $\sigma$ for each $\sigma \in S \setminus B$.
This will be done using the concepts introduced in the previous section, i.e., 
within the framework of a bottomed concrete cartesian closed category.

\medskip
Thus in what follows let $\fss{C}$ be a bottomed concrete cartesian closed category.
The notation employed here is that introduced in Section~5.1; in particlar,
the objects of $\fss{C}$ will be denoted by ${\cal C}$.
\medskip

Let ${\bf X}_T$ be a family from ${\cal C}$, i.e., a family such that 
${\bf X}_\tau \in {\cal C}$ for each $\tau \in T$
and let 
$L \in {\cal F}_T$ be a finite \tsym{T}-typed set with typing 
$\langle \cdot \rangle : L \to T$. Then $\ass L {\bf X} $ will be used to denote the object
$\otimes L'$ of ${\cal C}$, 
where $L' \in {\cal F}_{\cal C}$ is the
\tsym{\cal C}-typed set having the same underlying set as $L$ together with the
typing $\langle\cdot\rangle' : L' \to {\cal C}$ given by
$\langle \eta \rangle' = {\bf X}_{\langle\eta\rangle}$ for each $\eta \in L' = L$.
In particular, this means that $|\ass L {\bf X} | = \ass L X $, with 
$\ass L X $ as defined in Section~2.1, and where $X_T$ is the family of sets
given by $X_\tau = |{\bf X}_\tau|$ for each $\tau \in T$.
The element
$L' \in {\cal F}_{\cal C}$ defined above will be referred to as the 
\definition{\tsym{\cal C}-typed set associated with $L$ and ${\bf X}_T$}.

\bigskip\bigskip
\frame{20pt}{\bigskip 
In the concrete categories associated with the basic cases the above definition
of the object $\ass L {\bf X} $ is correct in that it equips the underlying set
$\ass L X $ with the proper structure.
\bigskip
}
\bigskip\bigskip
Consider now the \tsym{\Lambda}-algebra $(D_B,f_K)$ which describes the data objects.
Then $D_\beta$ is at least a bottomed set for each $\beta \in B$, and 
(as in Case~2 and Case~3) it is a often bottomed set endowed with additional structure. 
Being somewhat pedantic, ${\bf D}_\beta$ will now be used to denote 
this structured set and $D_\beta$ will be reserved to denote just the underlying
set without the structure. In other words, ${\bf D}_\beta$ is the object in the 
appropriate concrete category and $D_\beta = |{\bf D}_\beta|$. Note that it is 
$(D_B,f_K)$, and not $({\bf D}_B,f_K)$, which is really a \tsym{\Lambda}-algebra.
\medskip
This point of view leads to the following 
definition: A pair $({\bf Y}_B,q_K)$ will be called a
\indexddef{\tsym{\fss{C}}-based \tsym{\Lambda}-algebra}
{category-based algebra}{}{algebra}{category-based}
if ${\bf Y}_B$ is a family from ${\cal C}$ (i.e., ${\bf Y}_\beta \in {\cal C}$ for each 
$\beta \in B$)
and $q_K$  is a family of morphisms such that 
$q_\kappa \in \Hom(\ass L {{\bf Y}} ,{\bf Y}_\beta)$ 
whenever $\kappa \in K$ is of type $L \to \beta$. 
Of course, if $({\bf Y}_B,q_K)$ is a \tsym{\fss{C}}-based \tsym{\Lambda}-algebra 
and $Y_\beta = |{\bf Y}_\beta|$ for each $\beta \in B$ then
$(Y_B,q_K)$ is a \tsym{\Lambda}-algebra as defined in Section~2.2.
\medskip
Clearly each of the pairs $({\bf D}_B,f_K)$ occurring in the basic cases
is a \tsym{\fss{C}}-based \tsym{\Lambda}-algebra (with $\fss{C}$ the
associated category).

\proclaim{Proposition~5.2.2} If $({\bf Y}_B,q_K)$ is a \tsym{\fss{C}}-based 
\tsym{\Lambda}-algebra then ${\bf Y}_B$ extends uniquely 
to a family ${\bf Y}_S$ from ${\cal C}$ such that
\mdisp{ {\bf Y}_{L\to \beta}\ =\  \varepsilon(\ass L {\bf Y} ,{\bf Y}_\beta) } 
for each functional type $L \to \beta$. \endpro

\proof Define a mapping $\#' : {\cal F}_{\cal C} \times B \to {\cal C}$ by
\smallskip
\mdisp{ \#'(L,\beta)\ =\ \cases{
                 \varepsilon(\otimes L,{\bf Y}_\beta) & if $\,L \ne \varnothing\,$,\cr
\noalign{\smallskip}
                      {\bf Y}_\beta & if $\,L = \varnothing\,$.}}
\smallskip
Then $({\cal C},\#')$ is a set of functional types over $B$, except that
${\cal C}$ is a class rather than a set. However, there is still a unique
homomorphism $\pi$ from $(S,\#)$ to $({\cal C},\#')$, i.e., a unique mapping
$\pi : S \to {\cal C}$ such that
\mdisp{\pi(\#(L , \beta))
\ =\ \#'(\pi\,L,\beta)}
for all $L \in {\cal F}_S$, $\beta \in B$, where
$\pi\,L$ is the \tsym{\cal C}-typed set having the same underlying set as $L$ together
with the typing $\langle\cdot\rangle' : L \to {\cal C}$ 
defined by $\langle \eta\rangle' = \pi(\langle\eta\rangle)$ for each $\eta \in L$. 
(This can be seen by examining the construction of the homomorphism in the
proof of Proposition~2.3.2. Alternatively, $\pi$ can be defined directly with the help
the mapping $|\cdot| : S \to \Nat$ given in Lemma~5.2.1.) 
Now define ${\bf Y}_\sigma = \pi(\sigma)$ for each 
$\sigma \in S\setminus B$; then ${\bf Y}_S$ is a family  of objects from ${\cal C}$ 
extending ${\bf Y}_B$, and ${\bf Y}_\sigma = \pi(\sigma)$ for all $\sigma \in S$, since

\mdisp{ \pi(\beta)\ =\ \pi(\#(\varnothing,\beta))
\ =\ \#'(\varnothing,\beta)\ =\ {\bf Y}_\beta}
for each $\beta \in B$. Note that if $L \in {\cal F}_S$ then
$\otimes\,\pi L = \ass L {\bf Y} $, since $\pi(\langle \eta\rangle) = {\bf Y}_\eta$ for 
each $\eta \in L$. Thus if $L \to \beta$ is a functional type then
\mdisp{ {\bf Y}_{L\to \beta}\ =\ \pi(\#(L,\beta))
\ =\ \#'(\pi\,L,\beta)\ =\ \varepsilon(\otimes\, \pi L,{\bf Y}_\beta) 
\ =\ \varepsilon(\ass L {\bf Y} ,{\bf Y}_\beta)\ . }
The uniqueness of the family ${\bf Y}_S$ follows from the uniqueness of the
homomorphism $\pi$ (or by induction using the mapping
$|\cdot|$). \eop

\bigskip\bigskip
\frame{20pt}{\bigskip\medskip
Let $\fss{C}$ be the concrete category associated with one of the basic cases,
let $({\bf D}_B,f_K)$ be a \tsym{\fss{C}}-based 
\tsym{\Lambda}-algebra and let ${\bf D}_S$ be the family of objects given 
by Proposition~5.2.2.
Thus in Case~1 ${\bf D}_S$ is a family of bottomed sets and
\mdisp{ {\bf D}_{L\to \beta}\ =\  \totale {\ass L {\bf D} } {{\bf D}_\beta} } 
for each functional type $L \to \beta$. 
In Case~2 ${\bf D}_S$ is a family of bottomed posets and
for each functional type $L \to \beta$
\mdisp{ {\bf D}_{L\to \beta}\ =\  \mono {\ass L {\bf D} } {{\bf D}_\beta} \ .} 
\medskip
Finally, in Case~3 ${\bf D}_S$ is a family of bottomed complete posets and
\mdisp{ {\bf D}_{L\to \beta}\ =\  \cont {\ass L {\bf D} } {{\bf D}_\beta} } 
for each functional type $L \to \beta$. 
\bigskip\medskip
}
\bigskip\bigskip

\medskip
In what follows assume that $({\bf D}_B,f_K)$ is a \tsym{\fss{C}}-based 
\tsym{\Lambda}-algebra 
and let ${\bf D}_S$ be the family from ${\cal C}$ given by Proposition~5.2.2.
As above put $D_\sigma = |{\bf D}_\sigma|$ for each $\sigma \in S$, and note that 
$|{\ass L {\bf D} }| = \ass L D $ for each $L \in {\cal F}_S$.
\medskip
The application and partial application operators introduced in 
Section~5.1 will now be specialised to the objects in the family ${\bf D}_S$.
\medskip
Let $\sigma = L \to \beta$ be a functional type and let $L' \in {\cal F}_{\cal C}$
be the \tsym{\cal C}-typed set associated with $L$ and ${\bf D}_S$;
then the application operator $\fss{a}_{L'}^{{\bf D}_\beta}$ will be denoted by 
$\fss{a}_\sigma$. Thus 
$\fss{a}_\sigma \in \Hom({\bf D}_\sigma \times \ass L {\bf D} , {\bf D}_\beta)$, since
$\ass L {\bf D} = \otimes L'$ and
${\bf D}_\sigma = \varepsilon(\ass L {\bf D} ,{\bf D}_\beta) 
=  \varepsilon(\otimes L',{\bf D}_\beta)$. Moreover,
$\fss{a}_\sigma : D_\sigma \times \ass L D \to D_\beta$ is the mapping 
given by
\mdisp{ \fss{a}_\sigma(h,c) \ =\ h(c)}
for all $h \in D_\sigma$, $c \in \ass L D $.
\medskip

Now let $J$ be a non-empty
proper subset of $L$ and let $J'$ be the corresponding subset of $L'$
(i.e., $J'$ has the same underlying set as $J$ but is considered as a
\tsym{\cal C}-typed set  with the typing induced from $L'$);
then the partial application operator
$\fss{pa}^{{\bf D}_\beta}_{L',J'}$ will be denoted by $\fss{pa}^J_\sigma$.
The morphism $\fss{pa}^J_\sigma$ is thus an element of
$\Hom({\bf D}_\sigma \times \ass J {\bf D} , {\bf D}_{\sigma_{L\setminus J}})$
(recalling that $\sigma_{L\setminus J}$ is the type $L\setminus J \to \beta$), since
${\bf D}_\sigma = \varepsilon(\otimes L',{\bf D}_\beta)$, $\ass J {\bf D} = \otimes J'$ and
${\bf D}_{\sigma_{L\setminus J}} = \varepsilon(\ass {L\setminus J} {\bf D} ,{\bf D}_\beta) 
=  \varepsilon(\otimes (L'{\setminus}J'),{\bf D}_\beta)$.

\medskip
By Lemma~5.1.8 
$\fss{pa}^J_\sigma \in \Hom({\bf D}_\sigma \times \ass J {\bf D} , 
                                               {\bf D}_{\sigma_{L\setminus J}})$
is the unique morphism such that
\mdisp{ \fss{a}_{\sigma_{L\setminus J}}(\fss{pa}^J_\sigma(h,b),c)
\ =\ \fss{a}_\sigma(h,\fss{s}_{L,J}(b,c))}
(i.e., such that 
$\fss{pa}^J_\sigma(h,b)\,(c)
\, =\, h(\fss{s}_{L,J}(b,c))$)
for all $h \in D_\sigma$, $b \in \ass J D $ and all
$c \in \ass {L\setminus J} D $.
\medskip
To increase the legibility $b\oplus c$ will now be written instead of
$\fss{s}_{L,J}(b,c)$ (with the sets $L$ and $J$ being determined from the context).
Thus
\mdisp{ \fss{pa}^J_\sigma(h,b)\,(c) \ =\ h(b\oplus c)}
for all $h \in D_\sigma$, $b \in \ass J D $, $c \in \ass {L\setminus J} D $.
\medskip
As in Section~5.1 it is useful to put $\fss{pa}^L_\sigma = \fss{a}_\sigma$, in order
to include $\fss{a}_\sigma$ in with the partial application 
operators. Note then that
$\fss{pa}^J_\sigma \in \Hom({\bf D}_\sigma 
                         \times \ass J {\bf D} , {\bf D}_{\sigma_{L\setminus J}})$
for each non-empty subset $J$ of $L$, since $\sigma_\varnothing = \beta$.
If it is deemed necessary to emphasise that 
$\fss{pa}^L_\sigma$ is not really a partial application operator then it will be
referred to as a 
\indexddef{total application operator}{total application operator}{}
{application operator}{total}.

\proclaim{Proposition~5.2.3} Let $\sigma = L\to \beta$ be a functional type and let
$J$ and $J'$ be non-empty disjoint subsets of $L$; then
\mdisp{\fss{pa}^{J\cup J'}_\sigma(h,b\oplus b')
\ =\ \fss{pa}^{J'}_{\sigma_{L\setminus J}} (\fss{pa}_\sigma^J(h,b),b')}
for all $h \in D_\sigma$, $b \in \ass J D $ and
$b' \in \ass {J'} D $. \endpro

\proof This is a special case of Proposition~5.1.3. \eop
\bigbreak

The following situation has been arrived at: 
\medskip\smallskip
{\parindent=20pt
\item{1.} Assuming that $({\bf D}_B,f_K)$ is a \tsym{\fss{C}}-based \tsym{\Lambda}-algebra, 
Proposition~5.2.2 produces a family ${\bf D}_S$ from ${\cal C}$ which extends the family ${\bf D}_B$. 
\medskip
\item{2.} For each functional type $\sigma = L \to \beta$
the object ${\bf D}_\sigma$ can be thought of as a set of functions endowed with an
appropriate structure. In fact, ${\bf D}_\sigma$ should be regarded as
consisting of functions from the object $\ass L {\bf D} $ to the object ${\bf D}_\beta$.
\medskip
\item{3.} For each functional type $\sigma = L \to \beta$ and each non-empty subset
$J$ of $L$ there is a partial application operator
$\fss{pa}^J_\sigma \in \Hom({\bf D}_\sigma \times \ass J {\bf D} , {\bf D}_{\sigma_{L\setminus J}})$.
\bigskip
}
\midinsert
\bigskip\medskip
\frame{20pt}{\bigskip 
{\it Example~5.2.2\enspace} Let $({\bf D}_B,f_K)$ be the flat extension
of the \tsym{\Lambda}-algebra $(X_B,p_K)$ introduced in Example~2.2.1,
so $({\bf D}_B,f_K)$ is a \tsym{\fss{C}}-based \tsym{\Lambda}-algebra with
$\fss{C}$ the concrete category associated with Case~1. Let ${\bf D}_S$ be the family 
of objects given by Proposition~5.2.2.
Then in particular 
\medskip\smallskip
{\leftskip=32pt
${\bf D}_{\fstt{int}\to\fstt{int}} = \totale {\Int^\bot} {\Int^\bot} $,
${\bf D}_{\fstt{list}\ \fstt{list}\to\fstt{list}} 
  = \totale {{\Bbb L}^\bot\times {\Bbb L}^\bot}
       {{\Bbb L}^\bot} $, \smallskip
${\bf D}_{\fstt{list}\ \fstt{int}\ \fstt{list}\to\fstt{list}} 
  = \totale {{\Bbb L}^\bot\times \Int^\bot \times {\Bbb L}^\bot}
       {{\Bbb L}^\bot} $, \smallskip
${\bf D}_{\fstt{int}\to\fstt{list}} 
  = \totale {\Int^\bot} {{\Bbb L}^\bot} $, \smallskip
${\bf D}_{(\fstt{int}\to\fstt{int})\ \fstt{int}\to\fstt{list}} 
  = \totale {\totale {\Int^\bot} {\Int^\bot} \times \Int^\bot} {{\Bbb L}^\bot} $, 
\medskip\smallskip  
}
with ${\Bbb L} = \Int^*$ (and where $\Int^\bot$ and ${\Bbb L}^\bot$ are here considered
as objects in $\fss{C}$, i.e., as bottomed sets).
This is what is needed to apply the second interpretation in Section~1.1 to
the equations in Example~1.2.1.
\bigskip\medskip
}
\bigskip
\endinsert

The next step is to extend the signature $\Lambda$ to a signature
$\Sigma = (S,N,\Delta,\delta)$ by adding names for the partial application
operators, and then to extend 
$({\bf D}_B,f_K)$ to a \tsym{\fss{C}}-based
\tsym{\Sigma}-algebra $({\bf D}_S,f_N)$ by adding these operators to the family $f_K$.
\medskip
Before doing this, though, there is one small point which has to be dealt with.
The partial application operators occurring above have
two arguments, the first being a function and the remainder the function's arguments. 
However, to conform with the set-up being employed for the mappings
occurring in algebras, they should only have a 
single argument, and to achieve this it is convenient to introduce the 
following simple device:
For each functional type $\sigma = L \to \beta \in S$ fix some element
$\diamond_\sigma$ not in $L$, and for each non-empty subset $J$ of $L$ put 
$\sigma \cdot J = J \cup \{\diamond_\sigma\}$,
considered as an \tsym{S}-typed set with the typing induced from $L$ and with
$\diamond_\sigma$ of type $\sigma$. Now define a mapping 
$\fss{t}_\sigma^J : D_\sigma \times \ass J D \to \ass {\sigma \cdot J} D $
by

\mdisp{\fss{t}_\sigma^J(h,c)\,(\eta)\ =\ \cases{c(\eta) & if $\,\eta \in J\,$,\cr
                                h & if $\,\eta = \diamond_\sigma\,$.}}

\proclaim{Lemma~5.2.2} The mapping 
$\fss{t}_\sigma^J : D_\sigma \times \ass J D \to \ass {\sigma \cdot J} D $
is a morphism, i.e., an element of
$\Hom({\bf D}_\sigma \times \ass J {\bf D} ,\ass {\sigma \cdot J} {\bf D} )$.
Moreover $\fss{t}_\sigma^J$ is an isomorphism. \endpro 

\proof This is similar to the proof of Lemma~5.1.4 and is left the reader. \eop
\bigbreak
In order to increase the legibility $h\triangleleft c$ will be written instead of
$\fss{t}_\sigma^J(h,c)$ (with $\sigma$ and $J$ being determined from the context).
Thus, since an isomorphism is in particular a bijection, 
each element of $\ass {\sigma \cdot J} D $ has a unique 
representation of the form $h \triangleleft c$ with $h \in D_\sigma$ and 
$c \in \ass J D $.
\medskip
The morphism $\fss{t}_\sigma^J$ will allow the partial application operator
$\fss{pa}_\sigma^J$, which is a mapping with domain $D_\sigma \times \ass J D $, 
to be converted into a mapping with domain $\ass {\sigma\cdot J} D $.
\medskip
For each functional type $\sigma = L \to \beta$
and each non-empty subset $J$ of $L$ let $\triangleleft\{\sigma,J\}$ be some element not 
in $K$ and such that
$\triangleleft\{\sigma,J\} \ne \triangleleft\{\tau,J'\}$ whenever
$(\sigma,J) \ne (\tau,J')$.
Put $N = K \cup A$, where $A$ is the set consisting of these new elements
and define mappings $\Delta : N \to {\cal F}_S$ and 
$\delta : N \to S$ by
\smallskip
\mdisp{\Delta(\nu)\ =\ \cases{ \Theta(\nu) & if $\,\nu \in K\,$,\cr
           \sigma \cdot J 
             & if $\,\nu = \triangleleft\{\sigma,J\}\,$,\cr
                     }}
\mdisp{\delta(\nu)\ =\ \cases{ \vartheta(\nu) & if $\,\nu \in K\,$,\cr
           \sigma_{L\setminus J}
             & if $\,\nu = \triangleleft\{\sigma,J\}\,$
              with $\,\sigma = L \to \beta\,$.\cr
                     }}
\smallskip
Then $\Sigma = (S,N,\Delta,\delta)$ is a signature which is an extension of $\Lambda$, 
and in which $\triangleleft\{\sigma,J\}$ is of type 
$\sigma \cdot J \to \sigma_{L\setminus J}$ whenever
$\sigma = L \to \beta \in S \setminus B$ and $J$ is a non-empty subset of $L$. 
$\Sigma$ will be  called the 
\indexddef{functional signature associated with} 
{functional signature}{}{signature}{functional}
$\Lambda$. 
\medskip
Now if $\sigma = L \to \beta $ is a functional type and $J$ is a non-empty subset of $L$
then put
\mdisp{f_{\triangleleft\{\sigma,J\}}
                        \ =\ \fss{pa}^J_\sigma\,\circ\,(\fss{t}_\sigma^J)^{-1}\ .}
Thus 
$f_{\triangleleft\{\sigma,J\}} 
\in \Hom(\ass {\sigma\cdot J} {\bf D} ,{\bf D}_{\sigma_{L\setminus J}})$
and the mapping
$f_{\triangleleft\{\sigma,J\}} : \ass {\sigma\cdot J} D  \to
D_{\sigma_{L\setminus J}}$
satisfies

\mdisp{f_{\triangleleft\{\sigma,J\}}(h \triangleleft b)\ =\ \fss{pa}^J_\sigma(h,b)}

for all $h \in D_\sigma$, $b \in \ass J D $. In other words, 
the mapping $f_{\triangleleft\{\sigma,J\}}$ satisfies
\mdisp{f_{\triangleleft\{\sigma,J\}}(h \triangleleft b)\,(c)
\ =\ h(b\oplus c)}
for all $h \in D_\sigma$, $b \in \ass J D $,
$c \in \ass {L\setminus J} D $.
This results in a \tsym{\fss{C}}-based
\tsym{\Sigma}-algebra $({\bf D}_S,f_N)$ which will be called the 
\indexddef{functional \tsym{\fss{C}}-based \tsym{\Sigma}-algebra
derived from $({\bf D}_B,f_K)$ in $\fss{C}$} 
{functional category-based algebra}{}{category-based algebra}{functional}.
The \tsym{\Sigma}-algebra $(D_S,f_N)$ will be called the  
\indexdef{functional \tsym{\Sigma}-algebra derived from $({\bf D}_B,f_K)$ in $\fss{C}$}
{derived functional algebra}{}.
\medskip

The functional \tsym{\Sigma}-algebra $(D_S,f_N)$ 
possesses a couple of properties which will play an important role in what follows.
These are given in the next two results. For each $\sigma \in S$ denote by $\bot_\sigma$
the bottom element of $D_\sigma$ (in the bottomed category $\fss{C}$).

\proclaim{Proposition~5.2.4} Let $\sigma = L \to \beta $ be a functional type and $J$ be a 
non-empty subset of $L$. Then
$\,f_{\triangleleft\{\sigma,J\}}(\bot_\sigma \triangleleft c) 
\,=\, \bot_{\sigma_{L\setminus J}}\,$
for all $c \in \ass J D $. \endpro

\proof This is a special case of Proposition~5.1.4. \eop
\bigbreak

\proclaim{Proposition~5.2.5} Let $\sigma = L \to \beta $ be a functional type 
and $J$ and $J'$ be 
non-empty disjoint subsets of $L$ (and so in fact $L$ must contain at least two elements).
Then
\mdisp{f_{\triangleleft\{\sigma,J\cup J'\}}\bigl(h \triangleleft (c \oplus c')\bigr)
      \ =\ f_{\triangleleft\{\sigma_{L\setminus J},J'\}}
      \bigl(f_{\triangleleft\{\sigma,J\}}(h \triangleleft c) \,\triangleleft\, c'\bigr)}
for all $h \in D_\sigma$, $c \in \ass {J} D $ and all
$c' \in \ass {J'} D $. \endpro

\proof This is a special case of Proposition~5.1.3. \eop
\bigbreak

Finally, the following fact will be needed in Chapter~6.

\proclaim{Lemma~5.2.3} If $(Y_S,q_N)$ is a minimal \tsym{\Sigma}-algebra 
(and in particular if $(Y_S,q_N)$ is an initial \tsym{\Sigma}-algebra)
then $Y_\sigma = \varnothing$ for each functional type $\sigma \in S \setminus B$.
\endpro

\proof Let $Z_S$ be the family with $Z_\beta = Y_\beta$ for each $\beta \in B$
and $Z_\sigma = \varnothing$ for each $\sigma \in S \setminus B$.
Then it is easily checked that $Z_S$ is invariant in $(Y_S,q_N)$.
(Note that if $\sigma = L \to \beta$ is a functional type and $J$ a non-empty
subset of $L$ then $\ass {\sigma\cdot J} {Z} = \varnothing$, since
$\sigma\cdot J$ contains the element $\diamond_\sigma$ of type $\sigma$).
Thus $Y_\sigma = \varnothing$ for each functional type 
$\sigma \in S \setminus B$. \eop
\bigbreak
\vfill\eject

\hfline{132}{5.3 \ PRIMITIVE FUNCTIONS}

\sectionhead {5.3} {Primitive functions}
\bigskip\medskip
For the  whole of the section let $\fss{C}$ be a bottomed concrete cartesian closed
category and $({\bf D}_B,f_K)$ be a \tsym{\fss{C}}-based \tsym{\Lambda}-algebra.
As in Section~5.2 put $D_\beta = |{\bf D}_\beta|$ for each $\beta \in B$.
It will be assumed that the \tsym{\Lambda}-algebra $(D_B,f_K)$ is a monotone
regular extension of the initial \tsym{\Lambda}-algebra $(X_B,p_K)$ (with
the bottom element $\bot_\beta$ of the bottomed set $D_\beta$ the bottom element
of the object ${\bf D}_\beta$ for each $\beta \in B$).

\medskip
Before introducing the equations in Chapter~7 there is a further point that 
has to be considered which concerns the so-called 
\indexddef{primitive}{primitive function}{}{function}{primitive}
or 
\indexddef{built-in}{built-in function}{}{function}{built-in} 
functions. 
These are the functions occurring in the equations which are really 
part of what is given (as opposed to the functions which 
the equations are supposed to define). 
For instance, the symbols $+$ and $-$ occur in the original equations in 
Chapter~1 and in the discussion there it was implicitly 
assumed that $+$ and $-$ refer to the usual arithmetical 
operations of addition and subtraction. Somewhat more importantly, the 
right-hand sides of the equations for $\fss{it}$, $\fss{fst}$ and $\fss{pfb}$ 
clearly also involve some kind of built-in `case' operation.
\medskip
As in these original equations, in general there are two kinds of primitive functions 
which are needed.
On the one hand it is useful to have a few 
\indexddef{integer operators}{integer operator}{}{operator}{integer}, 
and on the other hand nothing non-trivial can be achieved without a suitable family of 
\indexddef{case operators}{case operator}{}{operator}{case}. 
In this section we explain exactly what these operators are
and introduce the assumptions needed for them to exists.

\bigskip\medskip

{\it Integer operators:\/}\enskip
Recall that $X_{\fstt{int}} = \Int$, $X_{\fstt{bool}} = \Bool$ and that
$D_{\fstt{int}} = X_{\fstt{int}} \cup \{\bot_{\fstt{int}}\}$ and
$D_{\fstt{bool}} = X_{\fstt{bool}} \cup \{\bot_{\fstt{bool}}\}$,
since $(D_B,f_K)$ is a regular extension of $(X_B,p_K)$
and $\ftt{int}$ and $\ftt{bool}$ 
are primitive types.
\medskip
Let $\theta$ be either $\ftt{int}$ or $\ftt{bool}$ and 
$f : X_{\fstt{int}} \times X_{\fstt{int}} \to X_\theta$ 
be a mapping; then the mapping 
$f^\bot : D_{\fstt{int}} \times D_{\fstt{int}} \to D_\theta$ given by
\smallskip 
\mdisp{ f^\bot(m,n)\ =\ \cases{ f(m,n) & if $\,m,\,n \in X_{\fstt{int}}\,$,\cr
\noalign{\smallskip}
                          \bot_\theta  & otherwise,}}
\smallskip
will be called the 
\indexddef{strict extension}{strict extension}{}{extension}{strict} 
of $f$.
$({\bf D}_B,f_K)$ is said to 
\indexdef{support integer operators}{supports integer operators}{}
if for each mapping
$f : X_{\fstt{int}} \times X_{\fstt{int}} \to X_\theta$ the strict extension
of $f$ is a morphism, i.e., an element of
$\Hom({\bf D}_{\fstt{int}}\times {\bf D}_{\fstt{int}}, {\bf D}_\theta)$.

\bigskip\bigskip
\frame{20pt}{\bigskip\medskip
In each of the three basic cases $({\bf D}_B,f_K)$
supports integer operators. This holds trivially in Case~1, and it holds in Cases 2 and 3
since if $n \in X_{\fstt{int}}$ then $n' \sqle_{\fstt{int}} n$ if and only if
$n'$ is either $n$ or $\bot_{\fstt{int}}$; moreover, the analogous statement
holds for $D_{\fstt{bool}}$.
\medskip\bigskip
}
\bigskip\bigskip

In what follows assume that $({\bf D}_B,f_K)$ supports integer operators.
Let $\fss{Add}$, $\fss{Sub}$ and $\fss{Mul}$ be the mappings from
$D_{\fstt{int}} \times D_{\fstt{int}}$ to $D_{\fstt{int}}$
which are respectively the strict extensions of the
arithmetical operators $+$, $-$ and $\times$. Thus, for instance
\smallskip
\vbox{
\medskip
\mdisp{\fss{Add}\,(m,n)\ =\ \cases{ m + n & if 
                                $\,m,\,n \in X_{\fstt{int}}\,$,\cr
                        	 \bot_{\fstt{int}} & otherwise. \cr}}
\smallskip
}
Moreover, let
$\fss{Eq}$, $\fss{Neq}$, $\fss{Le}$ and $\fss{Ge}$ be the mappings
from $D_{\fstt{int}} \times D_{\fstt{int}}$ to $D_{\fstt{bool}}$
which are respectively the strict extensions of the
relational operators $=$, $\neq$, $\le$ and $\ge$. For instance, this means that
\smallskip
\mdisp{\fss{Eq}\,(m,n) \ =\ \cases{ T & if $\,m,\,n \in X_{\fstt{int}}\,$ and
                                        $\,m = n\,$,\cr
                                 F & if $\,m,\,n \in X_{\fstt{int}}\,$ and
                                        $\,m \ne n\,$,\cr
                        	 \bot_{\fstt{bool}} & otherwise. \cr}}
\smallskip
From now on $\ftt{int}_2$ will be
used to denote the list $\,\ftt{int}\ \ftt{int}\,$ (this list being considered as 
a \tsym{B}-typed set). 

\proclaim{Proposition~5.3.1} (1)\enskip There exists a unique element
$\fss{add} \in D_{\fstt{int}_2\to \fstt{int}}$ such that
\mdisp{ f_{\triangleleft\{\fstt{int}_2\to \fstt{int},\fstt{int}_2\}}(
  \fss{add} \triangleleft (m,n))\ =\ \fss{Add}(m,n)}
for all $m,\,n \in D_{\fstt{int}}$. In the same way there exist unique elements
$\fss{sub}$ and $\fss{mul}$ of $D_{\fstt{int}_2\to \fstt{int}}$
corresponding to $\fss{Sub}$ and $\fss{Mul}$.
\medskip
(2)\enskip There exists a unique element
$\fss{eq} \in D_{\fstt{int}_2\to \fstt{bool}}$ such that
\mdisp{ f_{\triangleleft\{\fstt{int}_2\to \fstt{int},\fstt{int}_2\}}(
  \fss{eq} \triangleleft (m,n))\ =\ \fss{Eq}(m,n)}
for all $m,\,n \in D_{\fstt{int}}$. In the same way there exist unique elements
$\fss{neq}$, $\fss{le}$ and $\fss{ge}$ of $D_{\fstt{int}_2\to \fstt{bool}}$
corresponding to $\fss{Neq}$, $\fss{Le}$ and $\fss{Ge}$.
\endpro

\proof Note first that for any $\beta \in B$
\smallskip
\ldisp{\quad\Hom({\bf D}_{\fstt{int}}\times {\bf D}_{\fstt{int}}, {\bf D}_\beta)
\ =\ |{\varepsilon({\bf D}_{\fstt{int}}\times {\bf D}_{\fstt{int}}, {\bf D}_\beta)}|}
\vskip-\medskipamount
\rdisp{
\ =\ |{\varepsilon({\bf D}_{\fstt{int}_2}, {\bf D}_\beta)}|
\ =\ |{\bf D}_{\fstt{int}_2 \to \beta}| \ =\ D_{\fstt{int}_2 \to \beta\ .}
\quad}
\smallskip
(1)\enskip By assumption $\fss{Add}$ is an element of
$\Hom({\bf D}_{\fstt{int}}\times {\bf D}_{\fstt{int}}, {\bf D}_{\fstt{int}})$,
and thus an element of $D_{\fstt{int}_2\to \fstt{int}}$. Thus $\fss{add}$ can just
taken to be $\fss{Add}$, and it then clearly the unique element of
$D_{\fstt{int}_2\to \fstt{int}}$ such that
\mdisp{ f_{\triangleleft\{\fstt{int}_2\to \fstt{int},\fstt{int}_2\}}(
  \fss{add} \triangleleft (m,n))\ =\ \fss{Add}(m,n)}
for all $m,\,n \in D_{\fstt{int}}$. The other two cases are the same.
\medskip
(2)\enskip This is the same as (1). \eop
\bigbreak
As the proof of Proposition~5.3.1 shows, there is really no difference
between each entity starting with an upper case letter and the corresponding entity
starting with a lower case letter. However, the former will be considered as
mappings and the latter as elements of
$D_{\fstt{int}_2\to \beta}$ with $\beta$ either $\ftt{int}$ or 
$\ftt{bool}$. For instance, $\ftt{Add}$ is considered simply as a mapping
from $D_{\fstt{int}} \times D_{\fstt{int}}$ to $D_{\fstt{int}}$, whereas
$\fss{add}$ is considered as an element of
$D_{\fstt{int}_2\to \fstt{int}}$.
\eject
{\it Case operators:\/}\enskip 
Let $\Bf$ denote the set of ground types  $\theta \in B$ with $K_\theta$ 
finite. 
For each $\kappa \in K$ let $\bdom(f_\kappa) \in {\cal C}$ be  such that 
$f_\kappa \in \Hom(\bdom(f_\kappa),{\bf D}_\beta)$, thus if
$\kappa \in K$ is of type $L \to \beta$ then 
$\bdom(f_\kappa) = \ass L {\bf D} $ and
$\,\dom(f_\kappa) \,=\, |\bdom(f_\kappa)| \,=\, \ass L D \,$ is the domain of the mapping
$f_\kappa$.
\medskip
Let $\theta \in B_f$, $X \in {\cal C}$ and for each 
$\kappa \in K_\theta$ let $q_\kappa \in \Hom(\bdom(f_\kappa),X)$. 
If $u \ne \bot_\theta$ then, since $(D_B,f_K)$ is a regular
extension of $(X_B,p_K)$, there exists a unique $\kappa \in K$ having
type $L \to \theta$ for some $L \in {\cal F}_B$ and a unique
$b \in \ass L D $ such that $u = f_\kappa(b)$. This means that
a mapping $\varphi : D_\theta \to |X|$  can be defined by putting
\smallskip
\mdisp{  \varphi(u) \ =\ \cases{
    q_\kappa(b) & if $\,u\ne \bot_\theta\,$ and 
             $\,u = f_\kappa(b)\,$\cr
         & \qquad\qquad\qquad 
           for some $\,b \in \dom(f_\kappa)\,$, $\,\kappa \in K_\theta\,$,\cr
\noalign{\smallskip}
          \bot_X & if $\,u = \bot_\theta\,$\ ,}
}
\smallskip
with $\bot_X$ the bottom element of $X$. Now $({\bf D}_B,f_K)$ is said to 
\indexdef{support case operators}{supports case operators}{}
if for each $\theta \in B_f$, each $X \in {\cal C}$ and for each family of morphisms
$q_{K_\theta}$, this mapping $\varphi$ is actually a morphism, i.e., an element of
$\Hom({\bf D}_\theta,X)$. 
\medskip
{\it Remark\enskip} For those who know what a coproduct is, there might seem
to be one involved here. However, this is not so, since 
in general it will not be the case that
$\varphi\circ f_\kappa = q_\kappa$ holds for each $\kappa \in K_\theta$. 
\medskip

In what follows assume that $({\bf D}_B,f_K)$ supports case operators.
In particular, and as is shown on the next page, this is true
for the three basic cases.

\medskip

\midinsert
\bigskip
\frame{20pt}{\bigskip\smallskip
In the three basic cases $({\bf D}_B,f_K)$ supports 
case operators:
\medskip
This is trivially true in Case~1, so only Cases 2 and 3 need to be dealt with.
Let $\theta \in B_f$, let $X \in {\cal C}$ with bottom element $\bot_X$, 
let $q_\kappa \in \Hom(\bdom(f_\kappa),X)$ for each $\kappa \in K_\theta$ 
and let
$\varphi : D_\theta \to |X|$ be defined by putting
\smallskip
\mdisp{  \varphi(u) \ =\ \cases{
    q_\kappa(b) & if $\,u\ne \bot_\theta\,$ and 
             $\,u = f_\kappa(b)\,$\cr
         & \qquad\qquad\qquad 
           for some $\,b \in \dom(f_\kappa)\,$, $\,\kappa \in K_\theta\,$,\cr
\noalign{\smallskip}
          \bot_X & if $\,u = \bot_\theta\,$\ .}
}
\smallskip
Therefore in Case~2 it must be shown that $\varphi$ is monotone and in Case~3 
that it is continuous. 
\medskip
Consider first Case~2.
Let $u,\,u' \in D_\theta$ with $u \sqle_\theta u'$. If $u = \bot_\theta$ then 
\mdisp{\varphi(u) \ =\ \bot_X
           \ \sqle\  \varphi(u')\ ,}
with $\sqle$ the partial order on $|X|$, 
and so it can be assumed that $u \ne \bot_\theta$ (and hence that
$u' \ne \bot_\theta$). But by the definition of an associated ordering there
then exists a unique $\kappa$ of type $L \to \theta$ for some $L \in {\cal F}_B$
and unique elements
$b,\, b' \in \ass L D $ with $b \ \ass L {\sqle} \ b'$ such that
$u = f_\kappa(b)$ and $b' = f_\kappa(b')$. Thus 
\mdisp{\varphi(u)
 \ =\ q_\kappa(b) \ \sqle\ q_\kappa(b')
\ =\ \varphi(u')\ ,}
since $q_\kappa : \dom(f_\kappa) \to |X|$ is monotone, and this shows that
$\varphi$ is monotone.
\medskip
Consider now Case~3. As above the mapping $\varphi$ is monotone, hence it must be
shown that if $A \in \directed{D_\theta}$ with $v = \lub A$ then
\mdisp{\varphi(v)
\ =\ \lub \,\lcurl \varphi(u) \,:\, u \in A \rcurl\ , }
and clearly it can be assumed here that
$v \ne \bot_\theta$. There thus exists a unique $\kappa \in K$ having type
$L \to \theta$ for some $L \in {\cal F}_B$ and a unique element
$b \in \ass L D $ such that $v = f_\kappa(b)$, and by the 
definition of an associated ordering it then follows that each element
in $A \setminus \{\bot_\theta\}$ has a unique representation of the form
$f_\kappa(b')$ with $b' \in \ass L D $. Let 
\mdisp{A'\ =\ \lcurl b' \in \ass L D \,:\, 
                  f_\kappa(b') \in A \setminus \{\bot_\theta\}\rcurl\ ;}
then it is easily checked that $A' \in \directed{ \ass L D }$ and hence
\mdisp{
f_\kappa(\lub A') \ =\ \lub \,\lcurl f_\kappa(b') \,:\, b' \in A' \rcurl
\ =\ \lub \,(A \setminus \{\bot_\theta\}) \ =\ \lub A \ =\ v \ ,}
from which it follows that $b = \lub A'$. Therefore
\smallskip
\ldisp{\qquad
 \lub \,\lcurl \varphi(u) \,:\, u \in A \rcurl 
\ =\ \lub \,\lcurl \varphi(u) 
           \,:\, u \in A \setminus \{\bot_\theta\} \rcurl } 
\vskip-\medskipamount
\rdisp{\ =\ \lub\, \lcurl \varphi(f_\kappa(b')) 
           \,:\, b' \in A' \rcurl 
\ =\ \lub \,\lcurl g_\kappa(b') \,:\, b' \in A' \rcurl 
\ =\ g_\kappa(b) 
\ =\  \varphi(v)\ , \qquad} 
\smallskip 
since $q_\kappa : \dom(f_\kappa) \to |X|$ is continuous, and this shows that
$\varphi$ is continuous.
\bigskip\smallskip
}
\bigskip
\endinsert

For each $\theta \in \Bf$, $\beta \in B$ let $K_{\theta,\beta}$ be the 
\tsym{S}-typed set with underlying set $K_\theta$
in which $\kappa$ has type $L \to \beta$ in $K_{\theta,\beta}$ whenever
$\kappa$ is of type $L \to \theta$ in $\Lambda$.
\medskip
Let $\theta \in \Bf$, $\beta \in B$; then, again using the fact that
$(D_B,f_K)$ is a regular
extension of $(X_B,p_K)$, a mapping 
$\fss{Case}^\theta_\beta : D_\theta \times \ass {K_{\theta,\beta}} D \to D_\beta$
can be defined by letting
\smallskip
\mdisp{\fss{Case}^\theta_\beta(u,c)
\ =\ \cases{f_{\triangleleft\{J\to \beta,J\}}(c(\kappa) \triangleleft b) 
            & if $\,u \ne \bot_\theta\,$ and 
          $\,u = f_{\kappa}(b)\,$ with\cr
         &  \qquad $\,\kappa\,$ of type $\,J \to \theta\,$ and 
             $\,b \in \ass J D \,$, \cr
\noalign{\smallskip}
          \bot_\beta  & if $\,u = \bot_\theta\,$, \cr}} 
\smallskip
noting that in fact
$\,f_{\triangleleft\{J\to\beta,J\}}(c(\kappa) \triangleleft b)\,$ is just equal to 
$\,c(\kappa)\,(b)\,$, i.e., the value obtained by applying the function
$c(\kappa)$ to the argument $b$.

\proclaim{Lemma~5.3.1} For all $\theta \in \Bf$, $\beta \in B$ 
the mapping $\fss{Case}^\theta_\beta$ is a morphism, 
i.e., an element of
$\Hom({\bf D}_\theta \times \ass {K_{\theta,\beta}} {\bf D} ,{\bf D}_\beta)$. \endpro

\proof Consider $\kappa \in K_\theta$ of type $J \to \theta$ (so
$\bdom(f_\kappa) = \ass J {\bf D} $) and define a mapping
$r_\kappa : \dom(f_\kappa) \times \ass {K_{\theta,\beta}} D \to D_\beta$ by letting
$\,r_\kappa(b,c) \,=\, f_{\triangleleft\{\tau,J\}}(c(\kappa)\triangleleft b)\,$
for all $b \in \dom(f_\kappa)$, $c \in \ass {K_{\theta,\beta}} D $, where
$\tau = J \to \beta$. Then
\mdisp{r_\kappa\ =\ f_{\triangleleft\{\tau,J\}}\,\circ\,t^J_\tau\,\circ\,
  \psi\,\circ\,(\fss{id}_Z \times p_\kappa)\ ,}
where $Z = \bdom{f_\kappa}$, $p_\kappa : \ass {K_{\theta,\beta}} D \to D_\tau$ is the 
morphism  
occurring in (P2) in the definition of $\ass {K_{\theta,\beta}} {\bf D} $ and
$\psi : \dom(f_\kappa) \times D_\tau \to D_\tau \times \dom(f_\kappa)$
is the mapping given by $\psi(b,a) = (a,b)$ for all $b \in \dom(f_\kappa)$,
$a \in D_\tau$. Now it is easily checked that $\psi$ is a morphism (in fact an
isomorphism); thus by Lemmas 5.1.3 and 5.2.2 $r_\kappa$ is a morphism, i.e., an element
of $\Hom(\bdom(f_\kappa) \times \ass {K_{\theta,\beta}} {\bf D} , {\bf D}_\beta)$.
Therefore by (E3) there exists a unique morphism
$q_\kappa \in \Hom(\bdom(f_\kappa), 
             \varepsilon( \ass {K_{\theta,\beta}} {\bf D} , {\bf D}_\beta))$
such that $q_\kappa(b)\,(c)\,=\,r_\kappa(b,c)$ for all
$b \in \dom(f_\kappa)$, $c \in \ass {K_{\theta,\beta}} D $.
Carrying out this procedure for each $\kappa \in K_\theta$ results in a family of 
morphisms $q_{K_\theta}$ with $q_\kappa \in \Hom(\bdom(f_\kappa),X)$ for each
$\kappa \in K_\theta$, and where
$X = \varepsilon( \ass {K_{\theta,\beta}} {\bf D} , {\bf D}_\beta)$.
Hence, since $({\bf D}_B,f_K)$ supports case operators, the mapping
$\varphi : D_\theta \to |X|$  given by
\smallskip
\mdisp{  \varphi(u) \ =\ \cases{
    q_\kappa(b) & if $\,u\ne \bot_\theta\,$ and 
             $\,u = f_\kappa(b)\,$\cr
         & \qquad\qquad\qquad 
           for some $\,b \in \dom(f_\kappa)\,$, $\,\kappa \in K_\theta\,$,\cr
\noalign{\smallskip}
          \bot_X & if $\,u = \bot_\theta\,$\ ,}
}
\smallskip
is an element of $\Hom({\bf D}_\theta,X)$.
Now put $\varphi' = \fss{a}_Y^{{\bf D}_\beta} \circ (\varphi \times \fss{id}_Y)$, where
$Y = \ass {K_{\theta,\beta}} {\bf D} $, and so
$\varphi' \in \Hom({\bf D}_\theta \times \ass {K_{\theta,\beta}} {\bf D} , {\bf D}_\beta)$.
But for all $u \in D_\theta$, $c \in \ass {K_{\theta,\beta}} D $ it follows that
\medskip
\ldisp{\quad \varphi'(u,c)\ =\ \varphi(u)\,(c)}
\ldisp{\quad\qquad  \ =\ \cases{
    q_\kappa(b)\,(c) & if $\,u\ne \bot_\theta\,$ and 
             $\,u = f_\kappa(b)\,$\cr
         & \qquad\qquad\qquad 
           for some $\,b \in \dom(f_\kappa)\,$, $\,\kappa \in K_\theta\,$,\cr
\noalign{\smallskip}
          \bot_X(c) & if $\,u = \bot_\theta\,$\ ,}
}
\smallskip
\rdisp{
\ =\ \cases{f_{\triangleleft\{J\to \beta,J\}}(c(\kappa) \triangleleft b) 
            & if $\,u \ne \bot_\theta\,$ and 
          $\,u = f_{\kappa}(b)\,$ with\cr
         &  \qquad $\,\kappa\,$ of type $\,J \to \theta\,$ and 
             $\,b \in \ass J D \,$, \cr
\noalign{\smallskip}
          \bot_\beta  & if $\,u = \bot_\theta\,$, \cr}\qquad\quad} 
\vskip-\smallskipamount
\rdisp{\ =\ \fss{Case}^\theta_\beta(u,c)\qquad}
\medskip
and hence $\fss{Case}^\theta_\beta = \varphi'$, which shows that
$\fss{Case}^\theta_\beta 
  \in \Hom({\bf D}_\theta \times \ass {K_{\theta,\beta}} {\bf D} , {\bf D}_\beta)$.
\eop
\bigbreak
The device introduced in Section~5.2 will also be employed here to regard
$\fss{Case}^\theta_\beta$ as a function of a single argument:
Put $\theta \cdot K_{\theta,\beta} = K_{\theta,\beta} \cup \{\diamond_\theta\}$, where
$\diamond_\theta$ is some element not in $K_\theta$; 
consider $\theta \cdot K_{\theta,\beta}$ as an \tsym{S}-typed set with the typing 
induced from $K_{\theta,\beta}$ and 
with $\diamond_\theta$ of type $\theta$. 
Now define a mapping 
$\fss{t}_\theta^\beta : D_\theta \times \ass {K_{\theta,\beta}} D 
\to \ass {\theta \cdot K_{\theta,\beta}} D $
by
\smallskip
\mdisp{\fss{t}_\theta^\beta(u,c)\,(\eta)\ =\ \cases{c(\eta) & 
                                     if $\,\eta \in K_{\theta,\beta}\,$,\cr
                                u & if $\,\eta = \diamond_\sigma\,$.}}
\smallskip

\proclaim{Lemma~5.3.2} The mapping 
$\fss{t}_\theta^\beta : D_\theta \times \ass {K_{\theta,\beta}} D 
\to \ass {\theta \cdot K_{\theta,\beta}} D $
is a morphism, i.e., an element of
$\Hom({\bf D}_\theta \times \ass {K_{\theta,\beta}} {\bf D} ,
\ass {\sigma \cdot {K_{\theta,\beta}}} {\bf D} )$.
Moreover $\fss{t}_\theta^\beta$ is an isomorphism. \endpro 

\proof This is the same as Lemma~5.2.2. \eop
\bigbreak
In order to increase the legibility
$u\diamond c$ will be written instead of
$\fss{t}_\theta^\beta(u,c)$ (with $\theta$ and $\beta$ being determined from the context).
Thus, since an isomorphism is in particular a bijection, 
each element of $\ass {\theta \cdot {K_{\theta,\beta}}} D $ has a unique 
representation of the form $u \diamond c$ with $u \in D_\theta$ and 
$c \in \ass {K_{\theta,\beta}} D $.
\medskip

\proclaim{Proposition~5.3.2} Let $\theta \in \Bf$, $\beta \in B$ and put
$L = \theta\cdot K_{\theta,\beta}$. Then there exists a unique element
$\fss{case}^\theta_\beta \in D_{L\to\beta}$ such that
\mdisp{f_{\triangleleft\{L\to \beta,L\}}( \fss{case}^\theta_\beta 
   \triangleleft (u \diamond c))
\ =\ \fss{Case}^\theta_\beta (u,c)}
for all $u \in D_\theta$, $c \in \ass {\theta\cdot K_{\theta,\beta}} D $.
\endpro

\proof Put
$\,\fss{case}^\theta_\beta
\, =\, \fss{Case}^\theta_\beta \,\circ\, (\fss{t}_\theta^\beta)^{-1}\,$; 
thus
$\fss{case}^\theta_\beta \in 
\Hom(\ass {\theta\cdot K_{\theta,\beta}} {\bf D} ,{\bf D}_\beta)$ and the mapping
$\fss{case}^\theta_\beta :
\ass {\theta\cdot K_{\theta,\beta}} D \to D_\beta$ satisfies
\mdisp{\fss{case}^\theta_\beta (u \diamond c)
\ =\ \fss{Case}^\theta_\beta (u,c)}
for all $u \in D_\theta$, $c \in \ass {\theta\cdot K_{\theta,\beta}} D $.
Note that
\mdisp{\Hom(\ass {\theta\cdot K_{\theta,\beta}} {\bf D} ,{\bf D}_\beta)
\ =\ |{\varepsilon(\ass {\theta\cdot K_{\theta,\beta}} {\bf D} ,{\bf D}_\beta)}|
\ =\ |{\bf D}_{L\to \beta}|\ =\ D_{L\to \beta}}
and hence $\fss{case}^\theta_\beta \in D_{L\to\beta}$. Thus
$\fss{case}^\theta_\beta$ is the unique
element of $D_{L\to\beta}$ such that
\mdisp{f_{\triangleleft\{L\to \beta,L\}}( \fss{case}^\theta_\beta 
   \triangleleft (u \diamond c))
\ =\ \fss{Case}^\theta_\beta (u,c)}
for all $u \in D_\theta$, $c \in \ass {\theta\cdot K_{\theta,\beta}} D $. \eop
\bigbreak

The requirement that certain integer operators be 
`built-in' is important if the equations are to be used as the basis for a 
practical programming language. In fact, since $\ftt{int}$ is a
primitive type with infinitely many constructor names, the only way to
make use of this type is via the integer operators.
However, it is essentially a requirement about one specific data type. 
\medskip
On the other hand, the inclusion of the `case' operators is of a much more 
fundamental character and will facilitate in the present set-up what corresponds
to pattern matching or to the `case' construction in {\it Haskell\/}.
These `case' operators (or something equivalent) are absolutely essential: 
They are the analogue of the $\,\ftt{if then else}\,$ construction occurring 
in all imperative programming languages and without them it is impossible 
to define any `non-trivial' functions.
\medskip
The choice of three arithmetical and four relational operators is, of
course, somewhat arbitrary. 
There is no reason to prevent further 
operators being added, provided that each of these is
the strict extension of a corresponding (total) operator with domain
$X_{\fstt{int}} \times X_{\fstt{int}}$ and codomain either 
$X_{\fstt{int}}$ or $X_{\fstt{bool}}$.
\medskip
Now choose a set $P$ to consist of the distinct elements 
\smallskip
\mdisp{\lcurl \ftt{add},\,\ftt{sub},\,\ftt{mul},\,
\ftt{eq},\, \ftt{neq},\, \ftt{le},\, \ftt{ge}\rcurl
\,\cup\, \lcurl \ftt{case}^\theta_\beta \,:\,
                            \theta \in \Bf,\  \beta \in B \rcurl
}
\smallskip
and which is disjoint from $K$. Regard $P$ as an \tsym{S}-typed set 
by stipulating that
$\ftt{add},\,\ftt{sub}$ and $\ftt{mul}$ be of type $\ftt{int}_2\to \ftt{int}$,
that
$\ftt{eq},\, \ftt{neq}, \,\ftt{le}$ and $\ftt{ge}$ be of
type $\ftt{int}_2 \to \ftt{bool}$ and that
$\ftt{case}^\theta_\beta$ be of type $\theta\cdot K_{\theta,\beta}\to\beta$ for each
$\theta \in \Bf$, $\beta \in B$.
\medskip

Finally, let $p \in \ass P D $ be the assignment which assigns 
the corresponding `operator' to each name of an integer operator 
(i.e., $p(\ftt{add}) = \fss{add}$, $p(\ftt{sub}) = \fss{sub}$ and so on) 
and for which $p(\ftt{case}^\theta_\beta) = \fss{case}^\theta_\beta$
for each $\theta \in \Bf$, $\beta \in B$.
\medskip
Example~5.3.1 below indicates how some of the primitive functions
are involved in the equations introduced in Chapter~1.

\bigskip\bigskip
\frame{20pt}{\bigskip 
{\it Example~5.3.1\enspace} Let $\fss{C}$ be as in one of the basic cases with
$(D_B,f_K)$ a regular bottomed
extension of the \tsym{\Lambda}-algebra $(X_B,p_K)$ introduced in Example~2.2.1.
\medskip
Let
$\fss{shunx} \in D_{\fstt{list}\ \fstt{int}\ \fstt{list} \to \fstt{list}}$; then
for all $a,\, b \in D_{\fstt{list}}$ 
\smallskip
\ldisp{\qquad\quad \fss{Case}^{\fstt{list}}_{\fstt{list}}
     (\,a\,,\,\{\,\ftt{Nil}\to b\,,\,\ftt{Cons}\to \fss{shunx}(b,\cdot,\cdot)\,\}\,)}
\vskip-\smallskipamount
\rdisp{
\ =\ \cases{
       \bot_{\fstt{list}} & if $\,a = \bot_{\fstt{list}}\,$,\cr
       b  & if $\,a = f_{\fstt{Nil}}\,$,\cr
    \fss{sqshx}(b,m,d) 
       & if $\,a \ne \bot_{\fstt{list}}\,$ and $\,a = f_{\fstt{Cons}}(m,d)\,$.\cr
}\qquad\quad}
\smallskip
This should be compared with the right-hand side of the equation for 
$\fss{shunt}$ in Example~2.1.1.  
Now if
$b \in D_{\fstt{bool}}$ and $a_1,\,a_2 \in D_{\fstt{list}} $ then
\smallskip
\mdisp{\fss{Case}^{\fstt{bool}}_{\fstt{list}}(\,b\,,
                \,\{\,\ftt{True}\to a_1\,,\,\ftt{False}\to a_2\,\}\,)
      \ =\ \cases{
         a_1 & if $\,b = T\,$, \cr
         a_2 & if $\,b = F\,$, \cr
         \bot_{\fstt{list}} & if $\,b = \bot_{\fstt{bool}}\,$. \cr}}
\smallskip         
Thus if $\fss{revs} \in D_{(\fstt{int}\to\fstt{int})\ \fstt{int} \to \fstt{list}}$ then
\smallskip
\ldisp{\qquad\quad  \fss{Case}^{\fstt{bool}}_{\fstt{list}}
 (\,\fss{Eq}(n,0)\,,\,\{\,\ftt{True} \to \onept\,,\,
     \ftt{False} \to f_{\fstt{Cons}}(p(n),\fss{revs}(\fss{Sub}(n,1)))\,\}\,)}
\vskip-\smallskipamount
\mdisp{
\ =\ \cases{
       \bot_{\fstt{list}} & if $\,\fss{Eq}(n,0) = \bot_{\fstt{bool}}\,$,\cr
      f_{\fstt{Nil}}  & if $\,\fss{Eq}(n,0) = T\,$,\cr
      f_{\fstt{Cons}}(p(n),\fss{revs}(p,\fss{Sub}(n,1))
       & if $\,\fss{Eq}(n,0) = F\,$, \cr} \qquad\quad}
\rdisp{
\ =\ \cases{
       \bot_{\fstt{list}} & if $\,n = \bot_{\fstt{int}}\,$,\cr
      f_{\fstt{Nil}}  & if $\,n = 0\,$,\cr
      f_{\fstt{Cons}}(p(n),\fss{revs}(p,n-1))
       & if $\,n \in \Int \setminus \{0\}\,$, \cr} \qquad\quad}
\smallskip
for all $n \in D_{\fstt{int}}$, $p \in D_{\fstt{int}\to\fstt{int}}$.
This should be 
compared with the right-hand side of the equation for $\fss{revs}$ in 
Example~2.1.1.  
\bigskip\medskip
}
\bigskip

\vfill\eject

\hfline{139}{5.4 \ THE BASIC SET-UP}

\sectionhead {5.4} {The basic set-up}
\bigskip\medskip
Let us take stock of the various constructions which have been introduced so far.
Putting these together results in the set-up described below, and it is this
set-up which forms the starting point for the rest of the study.
\medskip

We assume that the following objects are given:
\bigskip
{\parindent=20pt
\item{1.} A signature $\Lambda = (B,K,\Theta,\vartheta)$ to specify the basic data types.
\medskip
\item{2.} An initial \tsym{\Lambda}-algebra $(X_B,p_K)$ consisting of the
basic data objects together with their constructors.
\medskip
\item{3.} A bottomed concrete cartesian closed category $\fss{C}$ which specifies
the mathematical framework within which everything takes place.
\medskip
\item{4.} A \tsym{\fss{C}}-based \tsym{\Lambda}-algebra $({\bf D}_B,f_K)$ 
\medskip
{\parindent=30pt
\item{---} such that $(D_B,f_K)$ is a monotone regular bottomed extension
of $(X_B,p_K)$ (with $D_\beta = |{\bf D}_\beta|$ for each $\beta \in B$),
where the bottom element $\bot_\beta$ of $D_\beta$ in $(D_B,f_K)$
is the bottom element of ${\bf D}_\beta$ in the bottomed category
$\fss{C}$.
\medskip
\item{---} and which supports both integer and case operators.
\medskip
}
\medskip
}
Given these objects, the following steps are then carried out:
\bigskip
{\parindent=20pt
\item{5.} The signature $\Lambda$ is extended to a signature 
$\Sigma = (S,N,\Delta,\delta)$ by adding to $B$ the functional types and to
$K$ the names for the partial application operators.
\medskip
\item{6.} Using Proposition~5.2.2 and the results in Section~5.1 concerning the existence
of partial application operators, $({\bf D}_B,f_K)$ is extended to a
\tsym{\fss{C}}-based \tsym{\Sigma}-algebra $({\bf D}_S,f_N)$, called
the functional \tsym{\fss{C}}-based \tsym{\Sigma}-algebra derived from 
$({\bf D}_B,f_K)$ in $\fss{C}$. 
\medskip
\item{7.} Putting $D_\sigma = |{\bf D}_\sigma|$ for all $\sigma \in S$ results in a
\tsym{\Sigma}-algebra $(D_S,f_N)$, called the
functional \tsym{\Sigma}-algebra derived from 
$({\bf D}_B,f_K)$ in $\fss{C}$.
\bigskip
}

For the three basic cases introduced at the beginning of Chapter~5 the 
objects presented above are described on the following page. Note that the choice
of the signature $\Lambda$ and the initial \tsym{\Lambda}-algebra $(X_B,p_K)$
has nothing to do with which case is being considered.

\bigskip 
However, we are not going to work directly with the set-up described above,  but rather
with another set-up, to be introduced below, which is a bit more general.
In this new approach we do not suppose explicitly that
the elements in the sets $D_\sigma$, $\sigma \in D \setminus B$, are functions
and thus cannot suppose that the mappings $f_{\triangleleft\{\sigma,J\}}$ are
partial application operators. These assumptions in the original set-up will be
replaced by a number of assumptions concerning the
mappings $f_{\triangleleft\{\sigma,J\}}$ which are exactly the properties
which we will need in what follows.

\eject
\frame{20pt}{\bigskip\medskip
{\it Case 1:\/}\enskip 
${\cal C}$ is the class of bottomed sets, for each $X \in {\cal C}$
the underlying set $|X|$ is the set obtained by forgetting the bottom element, and 
$\Hom(X,Y)$ is the set of all mappings from $|X|$ to $|Y|$ for each $X,\, Y \in {\cal C}$.
\medskip
The \tsym{\fss{C}}-based \tsym{\Lambda}-algebra $({\bf D}_B,f_K)$ is obtained by
starting with any monotone regular extension $(D_B,f_K)$ of $(X_B,p_K)$ and letting
${\bf D}_B$ be the corresponding family of bottomed sets.
\medskip
The family ${\bf D}_S$ given by Proposition~5.2.2 is a family of bottomed sets and
\mdisp{ {\bf D}_{L\to \beta}\ =\  \totale {\ass L {\bf D} } {{\bf D}_\beta} } 
for each functional type $L \to \beta$. 
\medskip
{\it Case 2:\/}\enskip 
${\cal C}$ is the class of bottomed posets,
for each $X \in {\cal C}$
the underlying set $|X|$ is the set obtained by forgetting the bottom element and the
partial order, and 
$\Hom(X,Y)$ is the set of all monotone mappings from $|X|$ to $|Y|$ for each 
$X,\, Y \in {\cal C}$.
\medskip
$({\bf D}_B,f_K)$ is obtained
by starting with
a minimal monotone regular extension $(D_B,f_K)$ 
of $(X_B,p_K)$. By Proposition~4.1.2 there exists a unique ordering $\sqle_B$ 
associated with $(D_B,f_K)$, which by Proposition~4.1.5 is monotone,
and ${\bf D}_B$ is then the corresponding family of bottomed posets.
\medskip
The family ${\bf D}_S$ given by Proposition~5.2.2 is a family of bottomed posets and
\mdisp{ {\bf D}_{L\to \beta}\ =\  \mono {\ass L {\bf D} } {{\bf D}_\beta} } 
for each functional type $L \to \beta$. 
\medskip
{\it Case 3:\/}\enskip 
${\cal C}$ is the class of bottomed complete posets,
for each $X \in {\cal C}$
the underlying set $|X|$ is again the set obtained by forgetting the bottom element and the
partial order, and 
$\Hom(X,Y)$ is the set of all continuous mappings from $|X|$ to $|Y|$ for each 
$X,\, Y \in {\cal C}$.
\medskip
$({\bf D}_B,f_K)$ is obtained
by starting with a minimal monotone regular extension $(Y_B,q_K)$ of $(X_B,p_K)$ 
and then letting $(D_B,f_K)$ be the initial completion of $(Y_B,q_K)$.
${\bf D}_B$ is the corresponding family of bottomed complete posets.
\medskip
The family ${\bf D}_S$ given by Proposition~5.2.2 is a family of bottomed complete 
posets and for each functional type $L \to \beta$
\mdisp{ {\bf D}_{L\to \beta}\ =\  \cont {\ass L {\bf D} } {{\bf D}_\beta} \ . } 
\smallskip
If $\sigma = L \to \beta \in S\setminus B$  and $J$ is a non-empty subset of $L$
then in all three cases
$f_{\triangleleft\{\sigma,J\}} 
\in \Hom(\ass {\sigma\cdot J} {\bf D} ,{\bf D}_{\sigma_{L\setminus J}})$
is the mapping
$f_{\triangleleft\{\sigma,J\}} : \ass {\sigma\cdot J} D  \to
D_{\sigma_{L\setminus J}}$
given for all $h \in D_\sigma$, $b \in \ass J D $, $c \in \ass {L\setminus J} D $ by
\mdisp{f_{\triangleleft\{\sigma,J\}}(h \triangleleft b)\,(c)
\ =\ h(b\oplus c)\ .}
\smallskip\bigskip
}
\vfill\eject

Note that if $(Y_S,q_N)$ is any \tsym{\Sigma}-algebra then 
$(Y_B,q_K)$ is a \tsym{\Lambda}-algebra (where $(Y_B,q_K)$ is obtained from $(Y_S,q_N)$ 
by omitting the sets $\lcurl Y_\sigma\,:\, \sigma \in Y\setminus B \rcurl$ from the family
$Y_S$ and the mappings $\lcurl q_\nu\,:\, \nu \in N\setminus K \rcurl$ from the family
$q_N$).
\medskip

In the new set-up we assume that the following objects are given:
\bigskip
{\parindent=20pt
\item{1.} A bottomed concrete category $\fss{C}$ with finite products. 
More precisely, this means that $\fss{C}$ is a concrete
category possessing canonical finite products (there is thus a mapping
$\otimes : {\cal F}_{\cal C} \to {\cal C}$ such that (P1), (P2) and (P3) hold)
which is bottomed in the sense that for each $X \in {\cal C}$ the underlying set
$|X|$ contains a bottom element, denoted by $\bot_X$, 
such that (B1) holds.
\medskip
\item{2.} A \tsym{\fss{C}}-based \tsym{\Sigma}-algebra $({\bf D}_S,f_N)$ such that
the \tsym{\Lambda}-algebra $(D_B,f_K)$ is a monotone regular bottomed extension of 
$(X_B,p_K)$ (where the bottom
element $\bot_\beta$ of $D_\beta$ in $(D_B,f_K)$ is
the bottom element of ${\bf D}_\beta$ in the bottomed category
$\fss{C}$),
\medskip
\medskip
}
where as usual $D_\sigma = |{\bf D}_\sigma|$ for each $\sigma \in S$.
Note that the mappings $\fss{s}_{L,J}$, $\fss{t}^J_\sigma$ and $\fss{t}^\beta_\theta$
introduced in Sections 5.2 and 5.3 (and denoted as usual by $\oplus$, $\triangleleft$ 
and $\diamond$ respectively) are still
morphisms (and in fact isomorphisms) in this set-up.
\medskip\smallskip
We then assume that the \tsym{\Sigma}-algebra $(D_S,f_N)$ has the following properties:
\bigskip
{\parindent=25pt
\item{(A)} If $\sigma = L \to \beta $ is a functional type and $J$ is a 
non-empty subset of $L$ then 
\mdisp{f_{\triangleleft\{\sigma,J\}}(\bot_\sigma \triangleleft c) 
\ =\ \bot_{\sigma_{L\setminus J}}}
\item{} for all $c \in \ass J D $ (with $\bot_\tau$ the bottom element of
${\bf D}_\tau$ for each $\tau \in S$).
\medskip\smallskip
\item{(B)} If $\sigma = L \to \beta $ is a functional type 
and $J$ and $J'$ are non-empty disjoint subsets of $L$ then
for all $h \in D_\sigma$, $c \in \ass {J} D $ and all
$c' \in \ass {J'} D $
\mdisp{f_{\triangleleft\{\sigma,J\cup J'\}}\bigl(h \triangleleft (c \oplus c')\bigr)
      \ =\ f_{\triangleleft\{\sigma_{L\setminus J},J'\}}
      \bigl(f_{\triangleleft\{\sigma,J\}}(h \triangleleft c) \,\triangleleft\, c'\bigr)}
\item{} for all $h \in D_\sigma$, $c \in \ass {J} D $ and all
$c' \in \ass {J'} D $. 
\medskip\smallskip
\item{(C)} For each $\theta \in \Bf$ and each $\beta \in B$ there exists an element
$\fss{case}^\theta_\beta$ of $D_{L\to\beta}$ with $L = \theta\cdot K_{\theta,\beta}$
such that
\mdisp{f_{\triangleleft\{L\to \beta,L\}}( \fss{case}^\theta_\beta 
   \triangleleft (u \diamond c))
\ =\ \fss{Case}^\theta_\beta (u,c)}
\item{} for all $u \in D_\theta$, $c \in \ass {K_{\theta,\beta}} D $.
\medskip\smallskip
\item{(D)} There exists an element
$\fss{add}$ of $D_{\fstt{int}_2\to \fstt{int}}$ such that
\mdisp{ f_{\triangleleft\{\fstt{int}_2\to \fstt{int},\fstt{int}_2\}}(
  \fss{add} \triangleleft (m,n))\ =\ \fss{Add}(m,n)}
\item{}for all $m,\,n \in D_{\fstt{int}}$ (with $\ftt{int}_2$ 
the list $\,\ftt{int}\ \ftt{int}\,$). In the same way there exist elements
$\fss{sub}$ and $\fss{mul}$ of $D_{\fstt{int}_2\to \fstt{int}}$
corresponding to $\fss{Sub}$ and $\fss{Mul}$, as well as elements
$\fss{eq}$, $\fss{neq}$, $\fss{le}$ and $\fss{ge}$ of $D_{\fstt{int}_2\to \fstt{bool}}$
corresponding to $\fss{Eq}$, $\fss{Neq}$, $\fss{Le}$ and $\fss{Ge}$.
\bigskip
}
\eject
Recall that 
$\fss{Case}^\theta_\beta : D_\theta \times \ass {K_{\theta,\beta}} D \to D_\beta$
is the mapping defined by letting
\smallskip
\mdisp{\fss{Case}^\theta_\beta(u,c)
\ =\ \cases{f_{\triangleleft\{J\to \beta,J\}}(c(\kappa) \triangleleft b) 
            & if $\,u \ne \bot_\theta\,$ and 
          $\,u = f_{\kappa}(b)\,$ with\cr
         &  \qquad $\,\kappa\,$ of type $\,J \to \theta\,$ and 
             $\,b \in \ass L D \,$, \cr
\noalign{\smallskip}
          \bot_\beta  & if $\,u = \bot_\theta\,$, \cr}} 
\smallskip
which still makes sense here because $(D_B,f_K)$ is a regular 
extension of $(X_B,p_K)$.
\medbreak
Recall also that $\fss{Add}$, $\fss{Sub}$ and $\fss{Mul}$ are the mappings from
$D_{\fstt{int}} \times D_{\fstt{int}}$ to $D_{\fstt{int}}$
which are respectively the strict extensions of the
arithmetical operators $+$, $-$ and $\times$, and 
$\fss{Eq}$, $\fss{Neq}$, $\fss{Le}$ and $\fss{Ge}$ are the mappings
from $D_{\fstt{int}} \times D_{\fstt{int}}$ to $D_{\fstt{bool}}$
which are respectively the strict extensions of the
relational operators $=$, $\neq$, $\le$ and $\ge$. 
\medskip
As in Section~5.3 let $P$ be the \tsym{S}-typed set
specifying the names and types of the primitive functions 
and let $p \in \ass P D $ be the assignment which determines the meaning 
of each such function, where now, of course, the meanings are the
elements given in (C) and (D).

\medskip
This new set-up includes the original one:
If $\fss{C}$ is a bottomed concrete cartesian closed category then
$\fss{C}$ is clearly a bottomed concrete category with finite products and
the functional \tsym{\Sigma}-algebra $(D_S,f_N)$ derived from $({\bf D}_B,f_K)$ in 
$\fss{C}$ possesses all the above
properties: Property (A) follows from Proposition~5.2.4, (B) from 
Proposition~5.2.5, (C) from Proposition~5.3.2 and (D) from Proposition~5.3.1.
\medskip

One reason for taking this new approach is that it allows some parts of Case~3 to be
dealt with within the framework of Case~2. More precisely, let $\fss{C}'$ be as in Case~3
and let $({\bf D}_S,f_N)$ be the \tsym{\fss{C}'}-based functional \tsym{\Sigma}-algebra 
derived from $({\bf D}_B,f_K)$ in $\fss{C}'$.
Now let $\fss{C}$ be as in Case~2. Then
$({\bf D}_S,f_N)$ is also a \tsym{\fss{C}}-based \tsym{\Sigma}-algebra 
and $(D_S,f_N)$ has properties (A), (B), (C) and (D) in the category $\fss{C}$.

\bigskip
We end the section by extending the notation involving $\oplus$, $\triangleleft$ and 
$\diamond$ to an arbitrary \tsym{\Sigma}-algebra.
If $L$ and $L'$ are disjoint \tsym{S}-typed sets then consider $L\cup L'$ as 
an \tsym{S}-typed set with the typing induced by the typings on $L$ and $L'$.
Let $Y_S$ be a family of sets; then for each
$c \in \ass L Y $, $c' \in \ass {L'} Y $ denote by
$c \oplus c'$ the element of $\ass {L\cup L'} Y $ defined by
\smallskip
\mdisp{(c \oplus c')\,(\eta)\ =\ \cases{c(\eta) & if $\,\eta \in L\,$,\cr
                                c'(\eta) & if $\,\eta \in L'\,$.}}
\smallskip

\proclaim{Lemma~5.4.1} Each element of $\ass {L\cup L'} Y $ has a unique representation 
of the form $c \oplus c'$ with $c \in \ass L Y $ and  $c' \in \ass {L'} Y $.
The mapping $\,(c,c') \mapsto c \oplus c'\,$ thus defines a 
bijection between $\ass J Y \times \ass {J'} Y $ and $\ass {J\cup J'} Y $.
\endpro

\proof This is clear. \eop
\bigbreak
Now let $\sigma = L \to \beta$ be a functional type and $J$ be a non-empty subset of $L$,
and for each
$y \in Y_\sigma$, $c \in \ass J Y $ denote by $y \triangleleft c$ 
the element of $\ass {\sigma \cdot J} Y $ given by
\mdisp{(y \triangleleft c)\,(\eta)\ =\ \cases{c(\eta) & if $\,\eta \in J\,$,\cr
                                y & if $\,\eta = \diamond_\sigma\,$.}}

(As before $\sigma \cdot J = J \cup \{\diamond_\sigma\}$, with
$\diamond_\sigma$ some element not in $L$, 
considered as an \tsym{S}-typed set with the typing induced from $L$ and with
$\diamond_\sigma$ of type $\sigma$.)

\proclaim{Lemma~5.4.2} Each element of $\ass {\sigma \cdot J} Y $ has a unique 
representation of the form $y \triangleleft c$ with $y \in Y_\sigma$ and $c \in \ass J Y $.
The mapping $\,(y,c) \mapsto y \triangleleft c\,$ thus defines a 
bijection between $Y_\sigma \times \ass J Y $ and $\ass {\sigma\cdot J} Y $.
\endpro

\proof This is clear. \eop
\bigbreak
Finally, let $\theta \in B_f$, $\beta \in B$, and for each
$s \in Y_\theta$, $a \in \ass {K_{\theta,\beta}} Y $ denote by $s \diamond a$ 
the element of $\ass {\theta \cdot {K_{\theta,\beta}}} Y $ given by
\smallskip
\mdisp{(s \diamond a)\,(\eta)\ =\ \cases{a(\eta) & 
                                     if $\,\eta \in K_{\theta,\beta}\,$,\cr
                                s & if $\,\eta = \diamond_\sigma\,$.}}
\smallskip
(As before $\theta \cdot K_{\theta,\beta} = K_{\theta,\beta} \cup \{\diamond_\theta\}$, 
with $\diamond_\theta$ some element not in $K_\theta$, considered 
as an \tsym{S}-typed set with the typing induced from $K_{\theta,\beta}$ and 
with $\diamond_\theta$ of type $\theta$. Recall that
$K_{\theta,\beta}$ is the 
\tsym{S}-typed set with underlying set $K_\theta$
in which $\kappa$ has type $L \to \beta$ in $K_{\theta,\beta}$ whenever
$\kappa$ is of type $L \to \theta$ in $\Lambda$.)

\proclaim{Lemma~5.4.3} Each element of $\ass {\theta \cdot K_{\theta,\beta}} Y $ 
has a unique representation of the form $s \diamond a$ with $s \in Y_\theta$ and 
$a \in \ass {K_{\theta,\beta}} Y $.
The mapping $\,(s,a) \mapsto s \diamond a\,$ thus defines a 
bijection between $Y_\theta \times \ass {K_{\theta,\beta}} Y $ 
and $\ass {\theta\cdot K_{\theta,\beta}} Y $.
\endpro

\proof This is the same as Lemma~5.4.2. \eop
\bigbreak
This usage of the symbols $\oplus$, $\triangleleft$ and $\diamond$ 
is, of course, compatible with that introduced
in Sections 5.2 and 5.3.
\medskip
The following simple facts will be needed continually 
in connection with the operations $\oplus$, $\triangleleft$ and $\diamond$:
Let $Y_S$ and $D_S$ be families of sets and let $\pi_S : Y_S \to D_S$ be a family
of mappings, i.e., $\pi_\sigma : Y_\sigma \to D_\sigma$ for each $\sigma \in S$.
For each \tsym{I}-typed set $I$ there is then the mapping
$\ass I {\pi} : \ass I Y \to \ass I D $ 
defined for each $c \in \ass I Y $, $\eta \in I$ by

\mdisp{\ass I {\pi} (c)\,(\eta)\ =\ \pi_\eta (c(\eta))\ .}

\proclaim{Proposition~5.4.1} (1)\enskip If $L$ and $L'$ are disjoint \tsym{S}-typed sets
then
\mdisp{ \ass {J\cup J'} {\pi} (c \oplus c')
   \ =\ \ass {J} {\pi} (c) \,\oplus\,  \ass {J'} {\pi} (c')}
for all $c \in \ass J Y $, $c' \in \ass {J'} Y $.
\medskip
(2)\enskip If $\sigma = L \to \beta$ is a functional type and $J$ a 
non-empty subset of $L$ then
\mdisp{ \ass {\sigma\cdot J} {\pi} (y \triangleleft c)
   \ =\ \pi_\sigma(y) \,\triangleleft\,  \ass {J} {\pi} (c)}
for all $y \in Y_\sigma$, $c \in \ass {J} Y $.
\medskip
(3)\enskip If $\theta \in B_f$ and $\beta \in B$ then
\mdisp{ \ass {\theta\cdot K_{\theta,\beta}} {\pi} (s \diamond a)
   \ =\ \pi_\theta(s) \,\diamond\,  \ass {K_{\theta,\beta}} {\pi} (a)}
for all $s \in Y_\theta$, $a \in \ass {K_{\theta,\beta}} Y $.
\endpro

\proof Straightforward. \eop
\bigbreak

\bigskip\bigskip\bigskip\bigskip

\sectionhead {5.5} {Notes}
\bigskip\medskip
The concept of a concrete category in the form used here is taken from
Mac Lane and Birkhoff (1967) (Chapter II, Section 10). Cartesian closed categories
are often used in Theoretical Computer Science, see, for example, 
Barr and Wells (1990) and the references therein. Purists are probably appalled
by some of the proofs in Section~5.1: Identities involving morphisms such as that 
occurring in Proposition~5.1.2 have been established by just showing that the 
left- and right-hand sides define the same mapping.
However, in a concrete category this is the easiest way to do things.
\medskip
Each functional type has the form
$L \to \beta$ with $\beta \in B$, i.e., the result is always
a ground type. It is possible to consider types of the form
$L \to \tau$ with $\tau \in S\setminus B$ a functional type. However, 
there is then a problem of non-unique representation since, for example, if
$\tau = \list {\tau} n \to \beta$ then $\list {\sigma} m \to \tau$ 
and $\list {\sigma} m \ \list {\tau} n \to \beta$ 
are (in any sensible interpretation) the same type.
 
\medskip
The construction made in Section~5.2 more-or-less corresponds to
what in Mitchell (1990) is referred to as a 
\indexdef{type frame}{type frame}{}.
The family ${\bf D}_S$ then corresponds in Case~1 to the 
\indexdef{full set-theoretic function hierarchy}{function hierarchy}{full set-theoretic} 
and in Case~3 to the 
\indexdef{full continuous function hierarchy}{function hierarchy}{full continuous}.

\vfill\eject

\hfline{145}{}
\bigskip
{\fourteenbf Chapter 6\quad Functionally free term algebras}
\bigskip\medskip

For the whole of the chapter let $\fss{C}$ be a bottomed concrete category with finite 
products and let $({\bf D}_S,f_N)$ be a \tsym{\fss{C}}-based \tsym{\Sigma}-algebra 
such that $(D_B,f_K)$ is a monotone regular bottomed extension
of $(X_B,p_K)$. We assume that the \tsym{\Sigma}-algebra $(D_S,f_N)$ has
the properties (A), (B), (C) and (D) listed in Section~5.4.
\medskip
Moreover, let $P$ be the \tsym{S}-typed set specifying the names and types of the 
primitive functions and let $p \in \ass P D $ be the assignment which determines the 
meaning of each such function.

\medskip
The aim of the present chapter is to  find some kind of generalisation 
of the ground term algebra $(F_B,\fo_K)$ which will allow references to be made
to the functions occurring in the sets in the family $D_{S\setminus B}$. 
(We call the elements in the sets in this family `functions' because that is what 
they are in the three basic cases.
In general, however, this will be a misnomer.)
\medskip
Now Lemma~5.2.3 shows that there is no simple direct generalisation of the 
ground term algebra. In fact the only obvious way of referring to functions 
is the trivial one
of simply giving some of them names: This at least provides a 
meaning to expressions involving these names, and such expressions or terms provide
the basic components of any functional programming language.
\medskip
The set of names of the functions of interest (together with their types)
specifies an \tsym{S}-typed set $I$. What is then required is a 
\tsym{\Sigma}-algebra $(Y_S,q_N)$ containing $I$ (i.e., such that $\eta \in Y_\sigma$
for each $\eta \in I$ of type $\sigma$) with the property
that for each $c \in \ass I D $ there exists a unique \tsym{\Sigma}-homomorphism 
$\pi^c_S : (Y_S,q_N) \to (D_S,f_N)$ such that $\pi^c_\xi(\xi) = c(\xi)$ for each 
$\xi \in I$. Then $\pi^c_S$ can be regarded as giving a meaning, 
relative to the assignment $c$, to the elements in the sets in the family $Y_S$:
For each $y \in Y_\sigma$ the `value' $\pi^c_\sigma(y)$ is interpreted as the meaning of
the `expression' $y$, given that $\xi$ has already been assigned the meaning 
$c(\xi)$ for each $\xi \in I$.
\medskip
Of course, an \tsym{I}-free \tsym{\Sigma}-algebra $(Y_S,q_N)$ satisfies the
above requirement, but we have decided to work instead with another kind of
free algebra. 
The reason is that there are advantages in only considering \tsym{\Sigma}-algebras 
having property (B), and such \tsym{\Sigma}-algebras will be called 
\indexddef{functional}{functional algebra}{}{algebra}{functional}. 
Now the class of functional \tsym{\Sigma}-algebras has its own notion of being free.
To be more precise, 
if $I$ is an \tsym{S}-typed set then a functional \tsym{\Sigma}-algebra $(Y_S,q_N)$ 
is said to be 
\indexddef{functionally \tsym{I}-free}{functionally free algebra}{}
{algebra}{functionally free} 
if $(Y_S,q_N)$ contains $I$ 
and for each functional \tsym{\Sigma}-algebra $(Z_S,r_N)$ and each
$c \in \ass I Z $ there exists a unique \tsym{\Sigma}-homomorphism $\pi^c_S$ 
from $(Y_S,q_N)$ to $(Z_S,r_N)$ such that $\pi^c_\eta(\eta) = c(\eta)$ for each 
$\eta \in I$. 
\medskip
The class of functionally free 
\tsym{\Sigma}-algebras includes the term algebras 
which will provide the basic building blocks for the rudimentary programming language to be
defined in Chapter~7.
Most of the properties of these algebra which are needed in
Chapters 7 and 8 are worked out in Section~6.1.

\medskip
In Section~6.2 the basic terms algebras are introduced: This involves choosing
for each \tsym{S}-typed set $I$ disjoint from $P$ 
a functionally \tsym{I\cup P}-free \tsym{\Sigma}-algebra $(F^I_S,\fo^I_N)$ which is an 
extension of the ground term algebra $(F_B,\fo_K)$.
If $c \in \ass I D $ then $\sem{\cdot}^c_S$ will be used to denote 
 the unique \tsym{\Sigma}-homomorphism from $(F^I_S,\fo^I_N)$ to $(D_S,f_N)$ such that 
$\sem{\,\xi\,}^c_\xi = c(\xi)$ for each 
$\xi \in I$ and $\sem{\,\zeta\,}^c_\zeta = p(\zeta)$ for each $\zeta \in P$.
As indicated above, the homomorphism
$\sem{\cdot}^c_S$ is regarded as 
giving a meaning, relative to the assignment $c$, to the terms in the family $F^I_S$.
In Lemma~6.2.1 it is shown that this \tsym{\Sigma}-homomorphism is an extension of the 
\tsym{\Lambda}-homomorphism $\sem{\cdot}_B$, i.e.,
$\sem{s}^c_\beta =  \sem{s}_\beta$ for all $s \in F_\beta$, $\beta \in B$
\medskip
At the end of Section~6.2 a functionally \tsym{J}-free \tsym{\Sigma}-algebra 
is explicitly constructed for each \tsym{S}-typed set $J$; 
this is essentially a term algebra as defined in Section~2.4 and is an extension
of the explicit term algebra introduced in Section~3.1.
These algebras (with sets $J$ of the form $I \cup P$) then provide the basis for 
a rudimentary functional programming language. 
\medskip
Let $I$ be an \tsym{S}-typed set disjoint from $P$.
A family $N^I_B$ with $F_\beta \subset N^I_\beta \subset \kterm{F}^I_\beta$ for each 
$\beta \in B$ will be called a 
\indexdef{support system}{support system}{} 
for $(D_S,f_N)$ if
$\sem{s}^c_\beta \ne \bot_\beta$ for all $s \in N^I_\beta$, $\beta \in B$.
In Section~6.3 we show that there is a `natural' support system $\bterm{F}^I_B$ 
which is determined entirely by the trace $R_B$ of $(D_B,f_K)$. This family will
play an important role in Chapters 7 and 8.
\medskip

\vfill\eject

\hfline{147}{6.1 \ FUNCTIONAL AND FUNCTIONALLY FREE ALGEBRAS}

\sectionhead {6.1} {Functional and functionally free algebras}
\bigskip\medskip

The present  section introduces what will be called functionally free 
\tsym{\Sigma}-algebras. This class of algebras includes the term algebras 
which will provide the basic building blocks for the rudimentary programming 
language to be defined in Chapter~7.
\medskip
Let $I$ be an \tsym{S}-typed set, which is considered to be fixed in what follows.
As already indicated, $I$ should be thought of as
the names (together with their types) of the functions in which we are interested.
What we are looking for is a suitable
\tsym{\Sigma}-algebra $(Y_S,q_N)$ containing $I$  
with the property
that for each $c \in \ass I D $ there exists a unique \tsym{\Sigma}-homomorphism 
$\pi^c_S : (Y_S,q_N) \to (D_S,f_N)$ such that $\pi^c_\xi(\xi) = c(\xi)$ for each 
$\xi \in I$. Then $\pi^c_S$ will be regarded as giving a meaning, 
relative to the assignment $c$, to the elements in the sets in the family $Y_S$.
\medskip

The most obvious candidate for $(Y_S,q_N)$ is an \tsym{I}-free \tsym{\Sigma}-algebra.
However, this is not what we will use because we have chosen, 
for reasons to be given later, to consider only \tsym{\Sigma}-algebras having property (B)
(as introduced in Section~5.4). Having made this choice,
there is then a more appropriate notion of being free, namely that of being
functionally free.
\medskip
Define a family $I_S$ by putting
$\,I_\sigma\,  =\, \lcurl \eta \in I \,:\, \langle \eta \rangle = \sigma \rcurl\,$
for each $\sigma \in S$. Thus a \tsym{\Sigma}-algebra $(Y_S,q_N)$ contains $I$  
if and only if $I_S \subset Y_S$. 
\medskip
If $(Y_S,q_N)$ is a \tsym{\Sigma}-algebra then, to increase the legibility,
it is convenient to allow $q_{\sigma,J}$ as an alternative notation for
$q_{\triangleleft\{\sigma,J\}}$. Thus the assumption that $(D_S,f_N)$ has
Property (B) states, in this new notation, that
if $\sigma = L \to \beta $ is a functional type 
and $J$ and $J'$ are
non-empty disjoint subsets of $L$ then

\mdisp{f_{\sigma,J\cup J'}\bigl(h \triangleleft (c \oplus c')\bigr)
      \ =\ f_{\sigma_{L\setminus J},J'}
      \bigl(f_{\sigma,J}(h \triangleleft c) \,\triangleleft\, c'\bigr)}
for all $h \in D_\sigma$, $c \in \ass {J} D $ and all
$c' \in \ass {J'} D $.
\medskip
In general, a \tsym{\Sigma}-algebra $(Y_S,q_N)$ having Property (B) 
will be called 
\indexddef{functional}{functional algebra}{}{algebra}{functional}.
In other words, a \tsym{\Sigma}-algebra $(Y_S,q_N)$ is functional if
whenever $\sigma = L \to \beta $ is a functional type, and $J$ and $J'$ are 
non-empty disjoint subsets of $L$ then 
\mdisp{q_{\sigma,J\cup J'}\bigl(y \triangleleft (c \oplus c')\bigr)
      \ =\ q_{\sigma_{L\setminus J},J'}
     \bigl(q_{\sigma,J}(y \triangleleft c) \,\triangleleft\, c'\bigr)}
for all $y \in Y_\sigma$, $c \in \ass {J} Y $ and all
$c' \in \ass {J'} Y $. 

\medskip
Note that
an \tsym{I}-free \tsym{\Sigma}-algebra $(Y_S,q_N)$ cannot be functional 
(since Proposition~2.3.8 implies that the sets $\Im(q_\nu)$, 
$\nu \in N_\sigma$, form a partition of $Y_\sigma \setminus I_\sigma$ for each 
$\sigma \in S$).
This would later cause some difficulties, and is one of the reasons for not
using \tsym{I}-free \tsym{\Sigma}-algebras. 
\medskip

For the class of functional \tsym{\Sigma}-algebras the following is the
appropriate notion of being free:
A functional \tsym{\Sigma}-algebra $(Y_S,q_N)$ will be called
\indexddef{functionally \tsym{I}-free}{functionally free algebra}{}
{algebra}{functionally free} 
if it contains $I$ and 
if for each functional \tsym{\Sigma}-algebra $(Z_S,r_N)$ and each $c \in \ass I Z $ there
exists a unique \tsym{\Sigma}-homomorphism $\pi^c_S$ from $(Y_S,q_N)$ to $(Z_S,r_N)$ 
such that $\pi^c_\xi(\xi) = c(\xi)$ for all $\xi \in I$.
\medskip
This modified definition provides a reasonable framework to work with, at least in as 
much as functionally free \tsym{\Sigma}-algebras exist:

\proclaim{Proposition~6.1.1} (1)\enskip There exists a functionally \tsym{I}-free 
\tsym{\Sigma}-algebra $(Y_S,q_N)$ which is, in the usual sense, unique:
For each functionally \tsym{I}-free \tsym{\Sigma}-algebra $(Y'_S,q'_N)$ 
there exists a unique isomorphism $\pi_S : (Y_S,q_N) \to (Y'_S,q'_N)$ such that 
$\pi_\eta(\xi) = \xi$ for all $\xi \in I$. 
\medskip
(2)\enskip Let $(Z_B,r_K)$ be an initial \tsym{\Lambda}-algebra 
disjoint from $I$ (i.e., such that $I_\beta \cap Z_\beta = \varnothing$ for each 
$\beta \in B$). Then there exists a functionally \tsym{I}-free \tsym{\Sigma}-algebra 
$(Y_S,q_N)$ which is an extension of $(Z_B,r_K)$. 
Moreover, if $(Y'_S,q'_N)$ is any such algebra and  $\pi_S$ is the unique isomorphism 
from $(Y_S,q_N)$ to $(Y'_S,q'_N)$ such that $\pi_\eta(\xi) = \xi$ for all $\xi \in I$
then $\pi_\beta(z) = z$ for all $z \in Z_\beta$, $\beta \in B$. \endpro

\proof Later in this section. \eop
\bigbreak
A functionally \tsym{I}-free \tsym{\Sigma}-algebra $(Y_S,q_N)$ clearly meets the 
requirements stated above: If $c \in \ass I D $ then, since the \tsym{\Sigma}-algebra
$(D_S,f_N)$ is functional, there exists a unique \tsym{\Sigma}-homomorphism 
$\pi^c_S : (Y_S,q_N) \to (D_S,f_N)$ such that $\pi^c_\xi(\xi) = c(\xi)$ for each 
$\xi \in I$. 
\medskip
At this point the reader is justified in being somewhat sceptical about the need for 
functionally free rather than just free algebras. However, it will now be shown 
that functionally 
free algebras also arise when taking a different approach, which
gives a much more convincing reason for employing them. 
\medskip
This alternative approach begins by noting that in all modern functional
programming languages the functional application operators do not have to be named
explicitly. The reason is essentially that the names of the functions involved
(i.e., the elements of $I$) are themselves used as names for operators.
More precisely, $\xi \in I$ is also used as the name of the operation
of applying the function which $\xi$ denotes. This turns out to be enough,
since any expression involving general functional application
operators can be expressed equivalently using only these more elementary 
operations.
\medskip
To implement this approach formally names have to be introduced for the 
operations of applying elements of $I$ to some or all of their arguments.
Thus for each $\xi \in I$ of type $L \to \beta$
and each subset $J$ of $L$ let $\xi\{J\}$ be some element not 
in $K$ and such that
$\xi\{J\} \ne \xi'\{J'\}$ whenever
$(\xi,J) \ne (\xi',J')$.
The `name' $\xi\{J\}$ will be used for the operation
of applying the function which $\xi$ denotes just to the arguments in $J$. This means
that if $\xi$ is of type $L \to \beta$ then
$\xi\{L\}$ refers to a total application operation, whereas
$\xi\{J\}$ refers to a partial application operation if $J$ is a non-empty
proper subset of $L$, and $\xi\{\varnothing\}$ will be essentially a synonym
for $\xi$. 
\medskip
In line with this interpretation there is then the following new
signature $\Sigma^I$: Put $N^I = K \cup A^I$, where  
$A^I$ is the set consisting of the new elements introduced above,
and define mappings $\Delta^I : N^I \to {\cal F}_S$ and 
$\delta^I : N^I \to S$ by

\vbox{\medskip
\mdisp{\Delta^I(\nu)\ =\ \cases{ \Delta(\nu) & if $\,\nu \in K\,$,\cr
     J    & if $\,\nu = \xi\{J\}\,$,
                     }}
\mdisp{\delta^I(\nu)\ =\ \cases{ \vartheta(\nu) & if $\,\nu \in K\,$,\cr
     \sigma_{L\setminus J}
             & if $\,\nu = \xi\{J\}\,$ with $\,\xi\,$ of type $\,\sigma\,$.
                     }}
\smallskip
}
Then $\Sigma^I = (S,N^I,\Delta^I,\delta^I)$ is a signature which is an extension of 
$\Lambda$. If $\xi \in I$ is of type $\sigma = L \to \beta$ and $J \subset L$ then
$\xi\{J\}$ is of type $J \to \sigma_{L\setminus J}$ in $\Sigma^I$.
In particular, this means that $\xi\{L\}$ is of type $L \to \beta$ and
$\xi\{\varnothing\}$ is of type 
$\varnothing \to \sigma$  (i.e., $\xi\{\varnothing\}$ is of type 
$\varnothing \to \langle \xi \rangle$).  
\medskip
If $(Y_S,q_{N^I})$ is a \tsym{\Sigma^I}-algebra then, to increase the legibility,
it is convenient to allow $q_{\xi,J}$ as an alternative notation for
$q_{\{\xi,J\}}$. 
\medskip
{\it Warning:\enskip} In what follows it will often be the case that 
expressions of the form $q_{\xi,J}$ and $q_{\sigma,J}$ (as alternative notation for
$q_{\{\xi,J\}}$ and $q_{\triangleleft\{\sigma,J\}}$ respectively) occur in the
same context.

\proclaim{Proposition~6.1.2} Let $(Y_S,q_{N^I})$ be an initial 
\tsym{\Sigma^I}-algebra and $c \in \ass I D $. Then 
there exists a unique family of mappings $\pi^c_S$ with
$\pi^c_\sigma : Y_\sigma \to D_\sigma$ for each $\sigma \in S$ such that
\medskip\smallskip
{\parindent=25pt
\item{(1)} $\pi^c_B$ is a \tsym{\Lambda}-homomorphism from
$(Y_B,q_K)$ to $(D_B,f_K)$.
\medskip
\item{(2)} $\pi^c_\xi\bigl(\,q_{\xi,\varnothing}(\onept)\,\bigr)
                                                                  \, =\, c(\xi)\,$ 
for each $\xi \in I$.
\medskip
\item{(3)} 
If $\xi \in I$ is of type $\sigma = L \to \beta$ and $J$ is a non-empty subset of $L$ then
\mdisp{\pi^c_{\sigma_{L\setminus J}}\bigl(\,q_{\xi,J}(b)\,\bigr)
 \ =\ f_{\sigma,J}\bigl(\,c(\xi)\,\triangleleft
                           \,\ass {J} {(\pi^c)} (b)\,\bigr)}
\item{} for all $b \in \ass J Y $.
\bigskip
} \endpro

\medskip

(Note in (1) that $(Y_B,q_K)$, obtained by omitting the sets 
$\lcurl Y_\sigma \,:\, \sigma \in S \setminus B \rcurl$ from the family $Y_S$ and 
the mappings $\lcurl q_\nu \,:\, \nu \in N^I \setminus K \rcurl$ 
from the family $q_{N^I}$, is a \tsym{\Lambda}-algebra.)
\bigbreak
\proof Define a \tsym{\Sigma^I}-algebra
$(D_S,f^c_{N^I})$ with $f^c_\kappa = f_\kappa$ for each $\kappa \in K$ 
as follows:
If $\xi \in I$ is of type $\sigma = L \to \beta$ and $J$ is a non-empty subset of $L$
then $f^c_{\xi\{J\}} : \ass J D \to D_{\sigma_{L\setminus J}}$
is the mapping given by 
$\,f^c_{\xi\{J\}}(b)\, =\, f_{\sigma,J}(c(\xi)\triangleleft b)\,$
for all $b \in \ass J D $, and for each $\xi \in I$ the mapping
$f^c_{\xi\{\varnothing\}} : \Oneptset \to D_\xi$ is given by
$f^c_{\xi\{\varnothing\}}(\onept) = c(\xi)$.
By construction $(D_S,f^c_{N^I})$ is then an extension of 
$(D_B,f_K)$. Let $\pi^c_S$ be the unique \tsym{\Sigma^I}-homomorphism from 
$(Y_S,q_{N^I})$ to $(D_S,f^c_{N^I})$. Then it is easy to see that the family
$\pi^c_S$ satisfies (1), (2) and (3).
The uniqueness follows since, conversely, any family satisfying (1), (2) and (3) is 
actually a \tsym{\Sigma^I}-homomorphism from $(Y_S,q_{N^I})$ to $(D_S,f^c_{N^I})$. \eop
\bigbreak
If $(Y_S,q_{N^I})$ is an initial \tsym{\Sigma^I}-algebra and $c \in \ass I D $ then 
the unique family of mappings $\pi^c_S$ given in Proposition~6.1.2 can be used
in the same way as the corresponding families defined in terms of either an
\tsym{I}-free or a functionally \tsym{I}-free \tsym{\Sigma}-algebra:
$\pi^c_S$ is considered as giving a meaning, 
relative to the assignment $c$, to the elements in the sets in the family $Y_S$,
i.e., for each $y \in Y_\sigma$ the `value' $\pi_\sigma^c(y)$ is interpreted
as the meaning of
the `expression' $y$, given that $\xi$ has already been assigned the meaning 
$c(\xi)$ for each $\xi \in I$.
\medskip
There are now at least two seemingly different approaches to constructing `expressions' 
involving the names in $I$. The first involves functionally \tsym{I}-free
\tsym{\Sigma}-algebras (or even just \tsym{I}-free
\tsym{\Sigma}-algebras) and the second initial \tsym{\Sigma^I}-algebras.
However, as will be seen below, 
initial \tsym{\Sigma^I}-algebras
and functionally \tsym{I}-free \tsym{\Sigma}-algebras are essentially the same
objects, and so the choice of which to use is really just a matter of taste.
This is the real reason for using functionally \tsym{I}-free rather than just 
\tsym{I}-free \tsym{\Sigma}-algebras.
\medskip
A procedure will be developed for switching between functional \tsym{\Sigma}-algebras 
containing $I$ and \tsym{\Sigma^I}-algebras, and it will be shown that under this procedure 
the initial \tsym{\Sigma^I}-algebras correspond exactly to the functionally \tsym{I}-free
\tsym{\Sigma}-algebras. These transformations only change the families of
mappings in the algebras and not the family of sets, i.e., there is a family of sets
$Y_S$ which occurs unchanged in all the algebras which will be constructed.

\medskip
We start by
considering a \tsym{\Sigma}-algebra $(Y_S,q_N)$ containing $I$; then a
\tsym{\Sigma^I}-algebra can be defined as follows:
Let $\xi \in I$ be of type $\sigma = L \to \beta$.
If $J$ is a non-empty subset of $L$ then let
$q_{\xi\{J\}} : \ass {J} Y \to Y_{\sigma_{L\setminus J}}$
be the mapping given by 

\mdisp{q_{\xi\{J\}}(b)\ =\ q_{\sigma,J}(\xi \triangleleft b )}

for each $b \in \ass J Y $. Moreover, let
$q_{\xi\{\varnothing\}} : \Oneptset \to Y_{L\to \beta}$ be the mapping defined by
\mdisp{q_{\xi\{\varnothing\}}(\onept)\ =\ c(\xi)\ .}
In this way a \tsym{\Sigma^I}-algebra
$(Y_S,q_{N^I})$ is obtained, which will be called the \tsym{\Sigma^I}-algebra 
\indexddef{associated with $(Y_S,q_N)$}{associated algebra}{}{algebra}{associated}: 
The family $q_{N^I}$ is derived from the family
$q_N$ by omitting the mappings in $\lcurl q_\nu \,:\, \nu \in A \rcurl$
and replacing them with the mappings $\lcurl q_\mu \,:\, \mu \in A^I \rcurl$.
\medskip
Let us say that a \tsym{\Sigma^I}-algebra $(Y_S,q_{N^I})$ 
\indexdef{is adapted to}{algebra}{adapted to typed set} 
$I$ if $q_{\xi,\varnothing}(\onept) = \xi$ for each
$\xi \in I$ (which in particular implies that $(Y_S,q_{N^I})$ must contain $I$).
Note that if $(Y_S,q_N)$ is a \tsym{\Sigma}-algebra 
containing $I$ then by definition the \tsym{\Sigma^I}-algebra associated with 
$(Y_S,q_N)$ is adapted to $I$. 
\medskip
The following result is really the key to understanding the relationship 
between initial \tsym{\Sigma^I}-algebras and functionally \tsym{I}-free 
\tsym{\Sigma}-algebras.

\proclaim{Proposition~6.1.3} (1)\enskip
A functional \tsym{\Sigma}-algebra $(Y_S,q_N)$ containing 
$I$ is functionally \tsym{I}-free if and only if the 
\tsym{\Sigma^I}-algebra 
$(Y_S,q_{N^I})$ associated with $(Y_S,q_N)$ is initial. 
\medskip
(2)\enskip Each initial \tsym{\Sigma^I}-algebra 
$(Y_S,q_{N^I})$ adapted to $I$
is the \tsym{\Sigma^I}-algebra associated with 
a unique functionally \tsym{I}-free 
\tsym{\Sigma}-algebra $(Y_S,q_N)$.
\endpro

\proof Later. \eop
\bigbreak
Proposition~6.1.3 implies that there 
is a natural one-to-one correspondence between functionally \tsym{I}-free 
\tsym{\Sigma}-algebras and initial \tsym{\Sigma^I}-algebras adapted to $I$.
Moreover, the following result shows that for each $c \in \ass I D $ the
meaning, relative to the assignment $c$, of the elements in the sets in the family $Y_S$
does not depend on which approach is taken.

\proclaim{Proposition~6.1.4} Let $(Y_S,q_N)$ be a functionally \tsym{I}-free 
\tsym{\Sigma}-algebra, let $c \in \ass I D $ and $\pi^c_S :(Y_S,q_N) \to (D_S,f_N)$ 
be the unique \tsym{\Sigma}-homomorphism 
such that $\pi^c_\eta(\eta) = c(\eta)$ for each
$\eta \in I$. Then $\pi^c_S$ is also the unique family of mappings  
given by Proposition~6.1.2 in terms of the \tsym{\Sigma^I}-algebra 
$(Y_S,q_{N^I})$ associated with $(Y_S,q_N)$.
\endpro

\proof Condition (3) in Proposition~6.1.2 
holds because if $\xi \in I$ is of type $\sigma = L \to \beta$ and 
$J$ is a non-empty subset of $L$ then 
\mdisp{\pi^c_{\sigma_{L\setminus J}}\bigl(\,q_{\xi\{J\}}(b)\,\bigr)
\ =\ \pi^c_{\sigma_{L\setminus J}}\bigl(
\,q_{\sigma,J}(\xi \triangleleft b)\,\bigr)
    \ =\ f_{\sigma,J}\bigl(\,c(\xi)\,
             \triangleleft\,\ass {J} {(\pi^c)} (b)\,\bigr)}
for all $b \in \ass J Y $. Conditions (1) and (2) hold trivially. \eop
\bigbreak
The above results imply that there is very little difference between a
functionally \tsym{I}-free \tsym{\Sigma}-algebra and an initial \tsym{\Sigma^I}-algebra
adapted to $I$. The choice has been made to work formally with functionally \tsym{I}-free 
\tsym{\Sigma}-algebras, but continual use will be made of the fact that the 
associated \tsym{\Sigma^I}-algebras are initial.
One advantage of making definitions and stating results in terms of
functionally \tsym{I}-free \tsym{\Sigma}-algebras rather than initial
\tsym{\Sigma^I}-algebras is that there is only one signature $\Sigma$ involved
(independent of $I$) and that the family $\pi^c_S$ is in this case
explicitly a homomorphism. 
\medskip
Finally, the following fact will be needed several times:

\proclaim{Proposition~6.1.5} Let $(Y_S,q_N)$ be a functionally \tsym{I}-free
and for each $\sigma \in S$ let ${\breve Y}_\sigma \subset Y_\sigma$. Suppose that:
\medskip\smallskip
{\parindent=25pt
\item{(1)} The family ${\breve Y}_B$ is invariant in the \tsym{\Lambda}-algebra
$(Y_B,q_K)$.
\medskip
\item{(2)} $I_\sigma \subset {\breve Y}_\sigma$ for each $\sigma \in S$.
\medskip
\item{(3)} If $\xi \in I$ is of type $\sigma = L \to \beta$ and $J$ is a non-empty
subset of $L$ 
then $q_{\sigma,J}(\xi \triangleleft b) \in {\breve Y}_{\sigma_{L\setminus J}}$ for all
$b \in \ass J {\breve Y} $.
\medskip\smallskip
}
Then ${\breve Y}_S = Y_S$. \endpro 

\proof The three conditions just say that the family ${\breve Y}_S$ is invariant
in the \tsym{\Sigma^I}-algebra 
$(Y_S,q_{N^I})$ associated with $(Y_S,q_N)$. But by Proposition~6.1.3~(1)
$(Y_S,q_{N^I})$ is initial, thus by Proposition~2.3.2 it is minimal, and therefore
${\breve Y}_S = Y_S$. \eop

\bigbreak

The rest of the section is taken up with proving Propositions 6.1.1 and 6.1.3.
\medskip
Until further notice let $(Y_S,q_{N^I})$ now be an initial \tsym{\Sigma^I}-algebra 
adapted to $I$. It will be shown that in this case the above construction can be reversed 
to obtain a functional \tsym{\Sigma}-algebra $(Y_S,q_N)$, which in fact turns out to
be functionally \tsym{I}-free.
\medskip
\proclaim{Lemma~6.1.1} Let $\sigma = L \to \beta $ be a functional type 
and $y \in Y_\sigma$; then there exists a unique 
$\xi \in I$ having type $L' \cup L \to \beta$ for some $L' \in {\cal F}_S$ disjoint
from $L$ and a unique assignment $b \in \ass {L'} Y $ such that
$y = q_{\xi,L'}(b)$. \endpro

\proof This follows since by Proposition~2.3.2 $(Y_S,q_{N^I})$ is a regular 
\tsym{\Sigma^I}-algebra (and because $\sigma \notin B$). \eop
\bigbreak
Note that if $\xi \in I_\sigma$ then the unique representation of $\xi$ given in
Lemma~6.1.1 is just $q_{\xi,\varnothing}(\onept)$. Conversely, if
$y \in Y_\sigma \setminus I_\sigma$ then the
unique representation of $y$ has the form
$y = q_{\xi,L'}(b)$ with $L' \in {\cal F}^o_S$.
\medskip
Let $\sigma = L \to \beta $ be a functional type and $J$ be a non-empty subset of $L$; 
consider $y \in Y_\sigma$ and $c \in \ass {J} Y $. By Lemma~6.1.1 
there then exists a unique $\xi \in I$ having type $L'\cup L \to \beta$ for some 
$L' \in {\cal F}_S$ disjoint 
from $L$ and a unique assignment $b \in \ass {L'} Y $ such that
$y = q_{\xi,L'}(b)$. There is thus an element 
$q_{\xi,L'\cup J}(b \oplus c)$ of 
$Y_{\sigma_{L\setminus J}}$ which is it convenient to denote simply by $y(c)$.
In particular, if $y = \xi \in I_\sigma$ then 
$y(c) = q_{\xi,J}(c)$.

\proclaim{Lemma~6.1.2} Let $\sigma = L \to \beta $ be a functional type and $J$ and $J'$ be 
non-empty disjoint subsets of $L$.
Then for all $y \in Y_\sigma$, $c \in \ass {J} D $ and all
$c' \in \ass {J'} D $ 
\mdisp{y(c\oplus c')\ =\ y(c)\,(c')\ .}
\endpro

\proof Let $y \in Y_\sigma$ and let $q_{\xi,L'}(b)$ be the unique 
representation of $y$ given in Lemma~6.1.1. Then it follows that
$y(c) = q_{\xi,L'\cup J}(b \oplus c)$, which means 
$q_{\xi,L'\cup J}(b \oplus c)$ is the unique representation of the element
$y(c)$, and hence that
\smallskip
\ldisp{\qquad y(c)\,(c')\ =\ q_{\xi,(L'\cup J)\cup J'}((b \oplus c)\oplus c')}
\vskip-\medskipamount
\vskip-\smallskipamount
\rdisp{
\ =\ q_{\xi,L'\cup (J\cup J')}(b \oplus (c\oplus c'))\ =\ y(c\oplus c') \qquad}
\smallskip
for all $c \in \ass {J} Y $, $c' \in \ass {J'} Y $. \eop
\bigbreak

Let $\sigma \,=\, L \to \beta$ be a functional type and $J$ be a non-empty subset of
$L$. Then a mapping
$q_{\triangleleft\{\sigma,J\}} : \ass {\sigma\cdot J} Y \to Y_{\sigma_{L\setminus J}}$ 
can be defined by letting

\mdisp{q_{\triangleleft\{\sigma,J\}}(y \triangleleft c ) \ =\ y(c)} 

for all $y \in Y_\sigma$, $c \in \ass J Y $.
(In particular, if $\xi \in I_\sigma$ then
$\,q_{\triangleleft\{\sigma,J\}}(\xi \triangleleft c ) \, =\, q_{\xi,J}(c)\,$ 
for each $c \in \ass J Y $.)
This defines a \tsym{\Sigma}-algebra $(Y_S,q_N)$: The family $q_N$ is 
obtained from the family $q_{N^I}$ by omitting the mappings in 
$\lcurl q_\mu \,:\, \mu \in A^I \rcurl$ and replacing them with the mappings 
$\lcurl q_\nu \,:\, \nu \in A \rcurl$. Of course, $(Y_S,q_N)$ contains $I$ because
$(Y_S,q_{N^I})$ is adapted to $I$.

\proclaim{Lemma~6.1.3} $(Y_S,q_N)$ is a functional \tsym{\Sigma}-algebra. \endpro

\proof Let $\sigma = L \to \beta $ be a functional type and $J$ and $J'$ be 
non-empty disjoint subsets of $L$. Then by Lemma~6.1.2 
$\,y(c\oplus c')\, =\, y(c)\,(c')\,$, which just means that
\mdisp{q_{\sigma,J\cup J'}\bigl(y \triangleleft (c \oplus c')\bigr)
      \ =\ q_{\sigma_{L\setminus J},J'}
      \bigl(q_{\sigma,J}(y \triangleleft c) \,\triangleleft\, c'\bigr)}

for all $y \in Y_\sigma$, $c \in \ass {J} Y $ and $c' \in \ass {J'} Y $. \eop

\bigbreak

Let us call $(Y_S,q_N)$ the 
\indexddef{functional \tsym{\Sigma}-algebra associated with $(Y_S,q_{N^I})$}
{associated algebra}{}{algebra}{associated}, 
and note that the first construction introduced above 
is in the following sense the inverse of the second:
 
\proclaim{Lemma~6.1.4} If $(Y_S,q_N)$ is the functional \tsym{\Sigma}-algebra 
associated with $(Y_S,q_{N^I})$ then $(Y_S,q_{N^I})$ is the \tsym{\Sigma^I}-algebra 
associated with $(Y_S,q_N)$. \endpro

\proof This follows because if $\xi \in I$ is of type $\sigma = L \to \beta$ and
$J$ is a non-empty subset of $L$ then by definition 
$\,q_{\sigma,J}(\xi \triangleleft c ) \, =\, q_{\xi,J}(c)\,$ 
for each $c \in \ass J Y $. \eop
\bigbreak
At this point the assumption that $(Y_S,q_{N^I})$ is an initial \tsym{\Sigma^I}-algebra 
will be dropped. 

\proclaim{Lemma~6.1.5} Let $(Y_S,q_N)$ be a functionally \tsym{I}-free 
\tsym{\Sigma}-algebra and let $(Y_S,q_{N^I})$ be the \tsym{\Sigma^I}-algebra 
associated with $(Y_S,q_N)$. Then $(Y_S,q_N)$ is the functional 
\tsym{\Sigma}-algebra associated with $(Y_S,q_{N^I})$.
\endpro

\proof  Let $\sigma = L \to \beta$ be a functional type and $J$ be a non-empty
subset of $L$; let $y \in Y_\sigma$ and $c \in \ass J Y $. 
Then $y$ has a unique representation of the form
$y = q_{\xi,L'}(b)$ with $\xi \in I$ of type $\tau = L' \cup L \to \beta$ for some
$L'$ disjoint from $L$ and with $b \in \ass {L'} Y $.
But then by definition
$\,q_{\xi,L'}(b) 
      \,=\, q_{\tau,L'}(\xi \triangleleft b)\,$
and
\mdisp{q_{\xi,L'\cup J }(b\oplus c) 
    \ =\ q_{\tau,L'\cup J}(\xi \triangleleft (b\oplus c))}

and therefore, since $(Y_S,q_N)$ is a functional \tsym{\Sigma}-algebra, 
it follows that
\mdisp{\quad q_{\sigma,J}(y \triangleleft c) 
\ =\ q_{\sigma,J}\bigl(\,
q_{\tau,L'}(\xi \triangleleft b)\,
 \triangleleft\, c \,\bigr)\ =\ q_{\tau,L'\cup J}(\xi \triangleleft (b\oplus c))
                            \ =\ q_{\xi,L'\cup J}(b \oplus c)\ ,}

i.e., 
$\,q_{\sigma,J}(y \triangleleft c) 
                            \,=\, q_{\xi,L'\cup J}(b \oplus c)\,$. \eop

\bigbreak
The construction made before the statement of Proposition~6.1.3 needs to be generalised.
Let $(Y_S,q_N)$ be any \tsym{\Sigma}-algebra and $c \in \ass I Y $ be an assignment.
Then a \tsym{\Sigma^I}-algebra can be defined as follows:
Let $\xi \in I$ be of type $\sigma = L \to \beta$.
If $J$ is a non-empty subset of $L$ then let
$q_{\xi\{J\}} : \ass {J} Y \to Y_{\sigma_{L\setminus J}}$
be the mapping given by 

\mdisp{q_{\xi\{J\}}(b)\ =\ q_{\sigma,J}(c(\xi) \triangleleft b )}

for each $b \in \ass J Y $. Moreover, let
$q_{\xi\{\varnothing\}} : \Oneptset \to Y_{L\to \beta}$ be the mapping defined by
\mdisp{q_{\xi\{\varnothing\}}(\onept)\ =\ c(\xi)\ .}
The resulting \tsym{\Sigma^I}-algebra
$(Y_S,q_{N^I})$ is called the \tsym{\Sigma^I}-algebra 
\definition{associated with $(Y_S,q_N)$ and the assignment $c$}.
Of course, if $(Y_S,q_N)$ is a \tsym{\Sigma}-algebra 
containing $I$ and $i \in \ass I Y $ is the assignment given by $i(\xi) = \xi$ 
for each $\xi \in I$ then this construction just gives the \tsym{\Sigma^I}-algebra 
associated with $(Y_S,q_N)$. 

\proclaim{Proposition~6.1.6} Let $(Y_S,q_{N^I})$  be an initial \tsym{\Sigma^I}-algebra 
adapted to $I$. Then the functional \tsym{\Sigma}-algebra $(Y_S,q_N)$ associated with 
$(Y_S,q_{N^I})$ is functionally \tsym{I}-free. 
\endpro

\proof Let $(Z_S,r_N)$ be a functional \tsym{\Sigma}-algebra and 
$c \in \ass I Z $, and let $(Z_S,r_{N^I})$ be the \tsym{\Sigma^I}-algebra
associated with $(Z_S,r_N)$ and $c$. Then, since $(Y_S,q_{N^I})$ is initial, 
there exists a unique \tsym{\Sigma^I}-homomorphism 
$\pi_S : (Y_S,q_{N^I}) \to (Z_S,r_{N^I})$, and it will be shown that $\pi_S$ is
actually the unique \tsym{\Sigma}-homomorphism from
$(Y_S,q_N)$ to $(Z_S,r_N)$ with $\pi_\eta(\eta) = c(\eta)$ for each
$\eta \in I$.
\medskip
Thus consider a functional type $\sigma\,=\,L \to \beta$, let
$J$ be a non-empty subset of $L$, and let $y \in Y_\sigma$ and 
$b \in \ass J Y $. Let $q_{\xi,L'}(b')$ be the unique representation of $y$
given in Lemma~6.1.1 (so $L' \in {\cal F}_S$ is disjoint from $L$,
$\xi \in I$ is of type $\tau = L'\cup L\to \beta$ and $b' \in \ass {L'} Y $).
Assume first that $L' \ne \varnothing$. Then, 
since
$\pi_S$ is a \tsym{\Sigma^I}-homomorphism, it follows that
\smallskip
\ldisp{\quad \pi_{\sigma_{L\setminus J}}
  \bigl(\,q_{\sigma ,J}(y\triangleleft b)\,\bigr) 
    \ =\ \pi_{\sigma_{L\setminus J}}\bigl(\,q_{\xi,L'\cup J}(b' \oplus b )\,\bigr)} 
\vskip-\medskipamount
\rdisp{
 \ =\ r_{\xi,L'\cup J}\bigl(\,\ass {L'\cup J} {\pi} (b'\oplus b)\,\bigr)
    \ =\ r_{\tau,L'\cup J}
      \bigl(\,c(\xi) \,\triangleleft \,\ass {L'\cup J} {\pi} (b'\oplus b)\,\bigr)\ .\quad}
\smallskip
But
$\,\ass {L'\cup J} {\pi} (b'\oplus b) 
                      \,=\, \ass {L'} {\pi} (b') \oplus \ass {J} {\pi} (b)\,$
and $(Z_S,q_N)$ is a functional \tsym{\Sigma}-algebra, thus
\smallskip
\ldisp{\quad r_{\tau,L'\cup J}
      \bigl(\,c(\xi) \,\triangleleft \,\ass {L'\cup J} {\pi} (b'\oplus b)\,\bigr)
\ =\ r_{\tau,L'\cup J}
      \bigl(\,c(\xi) \,\triangleleft 
              \,(\ass {L'} {\pi} (b') \oplus \ass {J} {\pi} (b))\,\bigr)}
\vskip-\medskipamount
\mdisp{\ =\ r_{\tau,L'\cup J}
 \bigl(\,(c(\xi) \triangleleft \ass {L'} {\pi} (b')) \,\oplus\, \ass {J} {\pi} (b)\,\bigr)}
\vskip-\medskipamount
\rdisp{\ =\ r_{\sigma,J}
      \bigl(\,  r_{\tau,L'}(
       c(\xi)\triangleleft \ass {L'} {\pi} (b') )
           \,\triangleleft\, \ass {J} {\pi} (b)\, \bigr)\ .\quad} 
\smallskip
Moreover, again using that $\pi_S$ is a \tsym{\Sigma^I}-homomorphism, 
\smallskip

\mdisp{  
 r_{\tau,L'}(
       c(\xi)\triangleleft \ass {L'} {\pi} (b') )
\ = \ r_{\xi,L'}(\ass {L'} {\pi} (b')) 
      \ =\  \pi_{\sigma}(q_{\xi,L'}(b'))  
\ =\  \pi_\sigma(y)} 
\smallskip
and therefore

\mdisp{r_{\sigma,J}
      \bigl(\,  r_{\tau,L'}(
       c(\xi)\triangleleft \ass {L'} {\pi} (b') )
           \,\triangleleft\, \ass {J} {\pi} (b)\, \bigr)
\ =\ r_{\sigma,J}\bigl(  
   \pi_\sigma(y) \triangleleft \ass {J} {\pi} (b) \bigr)
\ =\ r_{\sigma,J}\bigl(  
   \ass {\sigma \cdot J} {\pi} (y \triangleleft b) \bigr)\ .} 

Putting these steps together then gives the equality

\mdisp{ \pi_{\sigma_{L\setminus J}}
  \bigl(\,q_{\sigma ,J}(y\triangleleft b)\,\bigr) 
\ =\ r_{\sigma,J}\bigl(  
   \ass {\sigma \cdot J} {\pi} (y \triangleleft b) \bigr)\ ,} 
and a somewhat easier calculation shows that this also holds when $L' = \varnothing$.
Therefore $\pi_S$ is a \tsym{\Sigma}-homomorphism from $(Y_S,q_N)$ to $(Z_S,r_N)$ (since
$\pi_B$ is automatically a \tsym{\Lambda}-homomorphism from $(Y_B,q_K)$ to $(Z_B,r_K)$).
Moreover,
\mdisp{\pi_\eta(\eta) \ =\ \pi_\eta(q_{\eta,\varnothing}(\onept)) 
         \ =\ r_{\eta,\varnothing}(\onept) \ =\ c(\eta)}
for each $\eta \in I$, since $(Y_S,q_{N^I})$ is adapted to $I$.
\medskip
To show the uniqueness, consider any invariant family $Y'_S$ in 
$(Y_S,q_N)$ containing $I_S$. Then $\,q_{\xi,\varnothing}(\onept) = \xi$ for each 
$\xi \in I$ and 
$q_{\xi,J}(b)\, =\, q_{\sigma,J}(\xi \triangleleft b )$
for each $b \in \ass {J} Y $
whenever $\xi \in I$ is of type $\sigma = L \to \beta$ and
$J$ is a non-empty subset of $L$;  it thus follows that
the family $Y'_S$ is also invariant in $(Y_S,q_{N^I})$, and so
$Y'_S = Y_S$, since $(Y_S,q_{N^I})$ is a minimal \tsym{\Sigma^I}-algebra.
But if $\pi_S$ and $\pi'_S$ are \tsym{\Sigma}-homomorphisms from $(Y_S,q_N)$ to 
$(Z_S,r_N)$ with $\pi_\eta(\eta) = c(\eta) = \pi'_\eta(\eta)$ for each 
$\eta \in I$ then the family $Y'_S$ defined by
$\,Y'_\sigma 
\, =\, \lcurl y \in Y_\sigma \,:\, \pi'_\sigma(y) = \pi_\sigma(y) \rcurl\,$
for each $\sigma \in S$ contains $I_S$ 
and is invariant in $(Y_S,q_N)$. 
Thus from the above $Y'_S = Y_S$, and therefore $\pi'_S = \pi_S$. \eop
\bigbreak

{\it Proof of Proposition~6.1.1\enskip} (1)\enskip
By Proposition~2.3.2 there exists an initial 
\tsym{\Sigma^I}-algebra and there then also exists such a \tsym{\Sigma^I}-algebra 
$(Y_S,q_{N^I})$ which is adapted to $I$. (An initial \tsym{\Sigma^I}-algebra 
$(Y_S,q_{N^I})$ is unambiguous, and hence 
$q_{\xi,\varnothing}(\onept) \ne q_{\xi',\varnothing}(\onept)$
whenever $\xi$ and $\xi'$ are different elements of $I_\sigma$ for some $\sigma \in S$.
It is thus easy to arrange that 
$q_{\xi,\varnothing}(\onept) \ne q_{\xi',\varnothing}(\onept)$
whenever $\xi$ and $\xi'$ are different elements of $I$, and in this case
$q_{\xi,\varnothing}(\onept)$ can be identified with $\xi$ for each $\xi \in I$.)
By Proposition~6.1.6 the functional 
\tsym{\Sigma}-algebra associated with $(Y_S,q_{N^I})$ is then functionally \tsym{I}-free.
The uniqueness follows in the usual way. 
\medskip
(2)\enskip By Proposition~2.5.1 there exists an initial \tsym{\Sigma^I}-algebra which
is an extension of $(Z_B,r_K)$, and exactly as in the proof of
(1) there then also exists such a \tsym{\Sigma^I}-algebra $(Y_S,q_{N^I})$ 
which is adapted to $I$. Let $(Y_S,q_N)$ be the functional 
\tsym{\Sigma}-algebra associated with $(Y_S,q_{N^I})$.
Proposition~6.1.6 then implies that $(Y_S,q_N)$ is functionally \tsym{I}-free
and by definition $(Y_S,q_N)$ is an extension of $(Z_B,r_K)$.
The final statement follows by considering the appropriate
invariant family in $(Z_B,r_K)$. \eop
\bigbreak

{\it Proof of Proposition~6.1.3\enskip} (1)\enskip
As in the proof of Proposition~6.1.1 there exists an initial 
\tsym{\Sigma^I}-algebra $(Y'_S,q'_{N^I})$ which is adapted to $I$. Let $\pi_S$ be the 
unique \tsym{\Sigma^I}-homomorphism from $(Y'_S,q'_{N^I})$ to 
$(Y_S,q_{N^I})$; then $(Y_S,q_{N^I})$ is initial if and only if 
$\pi_S$ is a \tsym{\Sigma^I}-isomorphism, which is the case if and only if
the mapping $\pi_\sigma$ is a bijection for each $\sigma \in S$.
\medskip
Consider also the functional \tsym{\Sigma}-algebra 
$(Y'_S,q'_N)$ associated with $(Y'_S,q'_{N^I})$. Then by Proposition~6.1.3 
$(Y'_S,q'_N)$ is functionally \tsym{I}-free and therefore there 
exists a unique \tsym{\Sigma}-homomorphism $\pi'_S$ from $(Y'_S,q'_N)$ to 
$(Y_S,q_N)$ such that $\pi'_\xi(\xi) = \xi$ for all $\xi \in I$. 
Moreover, it is easy to see that
$(Y_S,q_N)$ is functionally \tsym{I}-free if and only if $\pi'_S$ is
a \tsym{\Sigma}-isomorphism, which is again the case if and only if
the mapping $\pi'_\sigma$ is a bijection for each $\sigma \in S$.
\medskip
This means it is enough to to show that $\pi_S = \pi'_S$ or, equivalently, that
$Y^o_S = Y_S$, where 
$Y^o_\sigma = \lcurl y \in Y'_\sigma 
               \,:\, \pi_\sigma(y) = \pi'_\sigma(y) \rcurl$
for each $\sigma \in S$.
Consider $\xi \in I$ having functional type 
$\sigma = L \to \beta$ and a non-empty subset $J$ of $L$, and let $c \in \assb J {Y^o} $.
Then it follows that
\smallskip
\ldisp{\quad
\pi_{\sigma_{L\setminus J}}(q'_{\xi,J}(c))
\ =\ q_{\xi,J}(\ass {J} {\pi} (c))
\ =\ q_{\xi,J}(\assb {J} {\pi'} (c))
}
\vskip-\medskipamount
\mdisp{
\ =\ q_{\sigma,J}
         \bigl(\,\xi \,\triangleleft \,\assb {J} {\pi'} (c)\,\bigr)
\ =\ q_{\sigma,J}
    \bigl(\,\assb {\sigma\cdot J} {\pi'} (\xi \triangleleft c)\,\bigr)}
\vskip-\medskipamount
\rdisp{
\ =\ \pi'_{\sigma_{L\setminus J}}(q'_{\sigma,J}
         ( \xi \triangleleft c ))
\ =\ \pi'_{\sigma_{L\setminus J}}(q'_{\xi,J}(c)) \quad}
\smallskip
and hence $q'_{\xi,J}(c) \in Y^o_{\sigma_{L\setminus J}}$. The family $Y^o_S$ is 
therefore invariant in 
$(Y'_S,q'_{N^I})$ (noting that the other cases which should be checked hold trivially 
because $\pi_B$ and $\pi'_B$ are both \tsym{\Lambda}-homomorphisms from $(Y'_B,q'_K)$ to 
$(Y_B,q_K)$). Thus $Y^o_S = Y'_S$, since $(Y'_S,q'_{N^I})$ is a minimal 
\tsym{\Sigma^I}-algebra. 

\medskip
(2)\enskip This follows from (1), Proposition~6.1.6 and Lemmas 6.1.4 and 6.1.5.
\eop
\bigbreak
\vfill\eject

\hfline{156}{6.2 \ THE BASIC TERM ALGEBRAS}

\sectionhead {6.2} {The basic term algebras}
\bigskip\medskip
In this section we introduce the basic terms algebras. More precisely,
this will mean that for each \tsym{S}-typed set $I$ disjoint from $P$ 
a suitable functionally 
\tsym{I\cup P}-free \tsym{\Sigma}-algebra $(F^I_S,\fo^I_N)$ is chosen
in order to give a meaning
to `expressions' or `terms' involving the names in $I$.  
(Recall that $P$ is the
\tsym{S}-typed set specifying the names and types of the primitive functions).
\medskip
Recall that in Section~3.1 we introduced an initial \tsym{\Lambda}-algebra 
$(F_B,\fo_K)$ called the ground term algebra, and that 
the unique \tsym{\Lambda}-isomorphism from $(F_B,\fo_K)$ 
to $(X_B,p_K)$ is denoted by $\sem{\cdot}_B$. Note that $\sem{\cdot}_B$ is then also
the unique \tsym{\Lambda}-homomorphism from $(F_B,\fo_K)$ to $(D_B,f_K)$.
\medskip
Let us assume (without any real loss of generality) that 
$(F_B,\fo_K)$ is disjoint from every \tsym{S}-typed set. 
Then by Proposition~6.1.1~(2)
we can choose for each \tsym{S}-typed set $I$
disjoint from $P$ 
a functionally \tsym{I\cup P}-free \tsym{\Sigma}-algebra 
$(F^I_S,\fo^I_N)$ which 
is an extension of $(F_B,\fo_K)$. These \tsym{\Sigma}-algebras are referred to as the
\indexddef{basic term algebras}{basic term algebra}{}{term algebra}{basic}. 
Later in the section a specific choice of these algebras is made
to serve as a basis for an explicit functional programming language. 
\medskip
For each $c \in \ass I D $ denote by $\sem{\cdot}^c_S$ the unique 
\tsym{\Sigma}-homomorphism from $(F^I_S,\fo^I_N)$ to the functional 
\tsym{\Sigma}-algebra $(D_S,f_N)$ such that $\sem{\,\xi\,}^c_\xi = c(\xi)$ for each 
$\xi \in I$ and $\sem{\,\zeta\,}^c_\zeta = p(\zeta)$ for each $\zeta \in P$
(with $p \in \ass P D $ the assignment
which determines the meaning of the primitive functions). 	
As already stated several times before, $\sem{\cdot}^c_S$ is considered as 
giving a meaning, relative to the assignment $c$, to the terms in the family $F^I_S$:
In other words, for each $s \in F^I_\sigma$ the `value' $\sem{s}^c_\sigma$ is interpreted
as being the  meaning of the term $s$, given that $\xi$ has already been assigned the 
meaning $c(\xi)$ for each $\xi \in I$.

\proclaim{Lemma~6.2.1} For each $c \in \ass I D $ the \tsym{\Sigma}-homomorphism 
$\sem{\cdot}^c_S$ is an extension of the \tsym{\Lambda}-homomorphism $\sem{\cdot}_B$, i.e.,
$\sem{s}^c_\beta =  \sem{s}_\beta$ for all $s \in F_\beta$, $\beta \in B$. \endpro

\proof For each $\beta \in B$ let $\sem{\cdot}'_\beta$ be the restriction of 
$\sem{\cdot}^c_\beta$ to $F_\beta$. Then $\sem{\cdot}'_B$ is also a 
\tsym{\Lambda}-homomorphism from $(F_B,\fo_K)$ to $(D_B,f_K)$ and thus
$\sem{\cdot}'_B = \sem{\cdot}_B$, since 
$\sem{\cdot}_B$ is the unique such homomorphism, 
i.e., $\sem{\cdot}^c_S$ is an extension of $\sem{\cdot}_B$. \eop
\bigbreak

Let $I$ be an \tsym{S}-typed set disjoint from $P$ which is considered to be fixed 
in what follows. It is useful to make the following classification: 
\bigskip
{\parindent=22pt
\item{---} For each $\beta \in B$ denote by $\kterm{F^I}_\beta$ the set of all 
elements in $F^I_\beta$ having the form $\fo^I_\kappa(b)$ with  $\kappa \in K$ of 
type $L \to \beta$ for some $L \in {\cal F}_S$ and with $b \in \assb L {F^I} $. 
\medskip
\item{---} For each $\sigma = L \to \beta \in S$ denote by $\aterm{F^I}_\sigma$ 
the set of all elements in $F^I_\sigma$ having the form
$\fo^I_{\tau,J}(\xi \triangleleft b)$ 
with $\xi \in I\cup P$ having functional type $\tau = J \cup L \to \beta$ 
for some $J \in {\cal F}^o_S$ disjoint from $L$ and with $b \in \assb J {F^I} $.
\bigskip
}
If $J$ is an \tsym{S}-typed set then as usual put 
$\,J_\sigma \,=\, \lcurl \eta \in J \,:\, \langle \eta \rangle = \sigma \rcurl\,$
for each $\sigma \in S$.
The unique representations appearing in the next result will play a decisive role.

\proclaim{Proposition~6.2.1} (1)\enskip For each $\beta \in B$ 
the sets $\kterm{F^I}_\beta$, $\aterm{F^I}_\beta$ and $I_\beta$ form a partition of
$F^I_\beta$. Furthermore, for each $\beta \in B$ the following statements hold:
\medskip\smallskip
{\parindent=22pt
\item{---} If $s \in \kterm{F^I}_\beta$ then
there exists a unique $\kappa \in K$ having type $L \to \beta$ 
for some $L \in {\cal F}_S$ and a unique element $b \in \assb L {F^I} $
such that $s = \fo^I_\kappa(b)$.
\medskip
\item{---} If $s \in \aterm{F^I}_\beta$ then there exists a unique
$\xi \in I\cup P$ having functional type $\sigma = L \to \beta$ 
for some $L \in {\cal F}^o_S$ and a unique element $b \in \assb L {F^I} $ such that
$s = \fo^I_{\sigma,L}(\xi \triangleleft b)$.  
\bigskip
} 
(2)\enskip For each $\sigma \in S \setminus B$ the sets $\aterm{F^I}_\sigma$,
$I_\sigma$ and $P_\sigma$ form a  partition of $F^I_\sigma$. Moreover, if
$\sigma = L \to \beta \in S \setminus B$ and
$s \in \aterm{F^I}_\sigma$ then there exists a unique 
$\xi \in I\cup P$ having type $\tau = J \cup L \to \beta$ for some  
$J \in {\cal F}^o_S$ disjoint
from $L$ and a unique assignment $b \in \assb {J} {F^I} $ such that
$s = \fo^I_{\tau,J}(\xi \triangleleft b)$. \endpro

\proof This follows since Proposition~6.1.3~(1) states that the 
\tsym{\Sigma^{I\cup P}}-algebra associated with $(F^I_S,\fo^I_N)$ is initial, and so in 
particular regular. (Note also that $P_\beta = \varnothing$ for each $\beta \in B$.)
\eop
\bigbreak

If $\xi \in I\cup P$ is of type $\sigma = L \to \beta$ then it is 
convenient introduce the mapping $\fo^I_\xi : \assb L {F^I} \to F^I_\beta$ 
defined by $\,\fo^I_\xi(\onept)\, =\, \xi\,$ when $L = \varnothing$ and by

\mdisp{\fo^I_\xi(b)\ =\ \fo^I_{\sigma,L}(\xi \triangleleft b)}

for each $b \in \assb L {F^I} $ when $L \ne \varnothing$. One reason for introducing
this last notation is because if $\beta \in B$ and $s \in F^I_\beta$ then 
Proposition~6.2.1~(1) implies that
there exists a unique element $\lambda \in K \cup I \cup P$ having type $L \to \beta$ 
for some $L \in {\cal F}_S$ and a unique assignment $b \in \assb L {F^I} $
such that $s = \fo^I_\lambda(b)$.
\medskip
\proclaim{Proposition~6.2.2} (1)\enskip
Let $\xi \in I$ be of type $\sigma = L \to \beta$ with $L \ne \varnothing$
and let $b \in \assb L {F^I} $. Then for each $c \in \ass I D $
\mdisp{
\sem{\,\fo^I_\xi(b)\,}^c_\beta
  \ =\ f_{\sigma,L}\bigl(c(\xi)\triangleleft \assb L {\sem{b}^c} \bigr)\ .} 
\medskip
(2)\enskip Let $\zeta = \ftt{case}^\theta_\beta$ with
$\theta \in \Bf$, $\beta \in B$, and let
$s \in F^I_\theta$, $b \in \assb {K_{\theta,\beta}} {F^I} $.
Then
\mdisp{\sem{\,\fo^I_\zeta(s \diamond b)\,}^c_\beta
\ =\ \fss{Case}^\theta_\beta 
            \bigl(\, \sem{s}^c_\theta\,,\,\assb {K_{\theta,\beta}} {\sem{b}^c} \bigr)}  
for all $c \in \ass I D $.
\medskip
(3)\enskip 
Let $\zeta \in P$ be the name of an integer operator (and so $\zeta$ is of type 
$\ftt{int}_2 \to \beta$ with $\beta$ either $\ftt{int}$ or $\ftt{bool}$)
and let $s_1,\, s_2 \in F^I_{\fstt{int}}$. Then
\mdisp{\sem{\,\fo^I_\zeta(s_1,s_2)\,}^c_\beta
\ =\ {\hat p}(\zeta)(\sem{s_1}^c_{\fstt{int}},\sem{s_2}^c_{\fstt{int}})}
for all $c \in \ass I D $, where 
${\hat p}(\zeta) : D_{\fstt{int}}\times D_{\fstt{int}} \to D_{\beta}$ is the mapping
corresponding to $\zeta$ (i.e., ${\hat p}(\ftt{add}) = \fss{Add}$,
${\hat p}(\ftt{sub}) = \fss{Sub}$ and so on).
\endpro

\proof (1)\enskip By Proposition~5.4.1~(2) it follows that each $c \in \ass I D $ 
\smallskip
\ldisp{\quad
\sem{\,\fo^I_\xi(b)\,}^c_\beta
\ =\ \sem{\,\fo^I_{\sigma,L}(\xi \triangleleft b)\,}^c_\beta
  \ =\ f_{\sigma,L}\bigl( 
  \assb {\sigma\cdot L} {\sem{\xi \triangleleft b}^c} \bigr)}
\vskip-\medskipamount
\rdisp{
  \ =\ f_{\sigma,L}\bigl( \sem{\xi}^c_\sigma \triangleleft
  \assb {L} {\sem{b}^c} \bigr)
  \ =\ f_{\sigma,L}\bigl(c(\xi)\triangleleft \assb L {\sem{b}^c} \bigr)\ .\quad} 
\medskip
(2) \enskip As in (1), and then making use of Proposition~5.4.1~(3) and (C), 
it follows that 
\smallskip
\ldisp{\quad
\sem{\,\fo^I_\zeta(s \diamond b)\,}^c_\beta
  \ =\ f_{L\to \beta,L}\bigl( \sem{\zeta}^c_{L\to \beta} \triangleleft
  \assb {L} {\sem{s \diamond b}^c} \bigr)
}
\vskip-\medskipamount
\rdisp{
  \ =\ f_{L\to \beta,L}\bigl( \fss{case}^\theta_\beta \triangleleft
( \sem{s}^c_\theta \diamond \assb {K_{\theta,\beta}} {\sem{b}^c} )
\bigr)
\ =\ \fss{Case}^\theta_\beta 
            \bigl(\, \sem{s}^c_\theta\,,\,\assb {K_{\theta,\beta}} {\sem{b}^c} \bigr)
\quad} 
\smallskip
for each $c \in \ass I D $, where $L =  \theta\cdot K_{\theta,\beta}$.
\medskip 
(3)\enskip This is the same as (2), but using (D) instead of (C).
\eop
\bigbreak

In what follows the \tsym{\Lambda}-algebra $(F^I_B,\fo^I_K)$ 
will usually play a more important role than the \tsym{\Sigma}-algebra 
$(F^I_S,\fo^I_N)$ itself.
In this regard the following two facts will be needed:

\proclaim{Lemma~6.2.2} $F^I_B$ is the only invariant family ${\grave F}^I_B$ in
the \tsym{\Lambda}-algebra $(F^I_B,\fo^I_K)$ such that 
$\,\aterm{F^I}_\beta \cup I_\beta \,\subset\, {\grave F^I}_\beta\,$ for each 
$\beta \in B$. \endpro

\proof Let ${\grave F}^I_B$ be invariant in $(F^I_B,\fo^I_K)$ and such that 
$\,\aterm{F^I}_\beta \cup I_\beta \,\subset\, {\grave F^I}_\beta\,$ for each 
$\beta \in B$. For each $\sigma \in S \setminus B$ put 
${\grave F^I}_\sigma = F^I_\sigma$; then the family 
${\grave F}^I_S$ satisfies the three conditions in Proposition~6.1.5.
Hence ${\grave F}^I_S = F^I_S$, and so in particular
${\grave F}^I_B = F^I_B$. \eop
\bigskip

\proclaim{Lemma~6.2.3} There exists a family of mappings
$\ell_B$ with $\ell_\beta : F^I_\beta \to \Nat$ for each $\beta \in B$
such that:
\medskip\smallskip
{\parindent=25pt
\item{(1)} If $\kappa \in K$ is of type $\varnothing \to \beta$ then
$\,\ell_\beta(\fo^I_\kappa(\onept)) = 0\,$.
\medskip
\item{(2)} If $\kappa \in K$ is of type $L \to \beta$ with $L \ne \varnothing$ 
and $a \in \assb L {F^I} $ then
\mdisp{\ell_\beta(\fo^I_\kappa(a))
  \ >\  \max \lcurl \ell_\eta(a(\eta))\,:\, \eta \in L \rcurl\ .}
\medskip
\item{(3)} If $\xi \in I_\beta$ for some $\beta \in B$ then
$\ell_\beta(\xi) = 0\,$.
\medskip
\item{(4)} If $\xi \in I_\sigma$ with $\sigma = L \to \beta$ a functional type then
$\ell_\beta(\fo^I_\xi(a)) > 0$ for all $a \in \assb L {F^I} $.
\medskip
\item{(5)} If $\zeta = \ftt{case}^\theta_\beta$ for some
$\beta \in B$, $\theta \in \Bf$ then 
\mdisp{\ell_\beta( \fo^I_\zeta(s \triangleleft a )) 
\ >\ \ell_\theta(s)}
\item{} 
for all $s \in F^I_\theta$, $a \in \assb {K_{\theta,\beta}} {F^I} $. 
\medskip
\item{(6)} If $\zeta \in P$ is the name of an integer operator
(so $\zeta$ is of type $\ftt{int}_2 \to \beta$ with
$\beta$ either $\ftt{int}$ or $\ftt{bool}$) then
\mdisp{\ell_\beta(\,\fo^I_\zeta(s_1,s_2)\,)
   \ > \    \max \{\ell_{\fstt{int}}(s_1),\ell_{\fstt{int}}(s_2)\}}
\item{} for all $s_1,\,s_2 \in F^I_{\fstt{int}}$. 
\medskip\smallskip
} \endpro

\proof Let $\ell_S$ be the family of mappings obtained by applying
Lemma~2.3.4 to the \tsym{\Sigma^{I\cup P}}-algebra $(F^I_S,\fo^I_{N^{I\cup P}})$
associated with $(F^I_S,\fo^I_N)$.
(This \tsym{\Sigma^{I\cup P}}-algebra is initial and so minimal and regular.)
Then it is easily checked that the restricted family $\ell_B$ satisfies (1), (2),
(3), (4), (5) and (6). \eop
\bigbreak

The final topic considered in this section is the explicit construction of a 
functionally \tsym{J}-free \tsym{\Sigma}-algebra $(G^J_S,\nabla^J_N)$
for each \tsym{S}-typed set $J$. Each of these algebras 
involves a term algebra as defined in Section~2.4 and is an extension
of the term algebra $(E_B,\blob_K)$ introduced in Section~3.1.
If $I$ is an \tsym{S}-typed set disjoint from $P$ then the functionally 
\tsym{I \cup P}-free \tsym{\Sigma}-algebra
$(G^{I\cup P}_S,\nabla^{I\cup P}_N)$ will be denoted just by
$(E^I_S,\blob^I_N)$.
These algebras will be used as the basis for an explicit functional programming language. 
\medskip
Consider the \tsym{S}-typed set $J$ to be fixed. The functionally \tsym{J}-free
\tsym{\Sigma}-algebra $(G^J_S,\nabla^J_N)$ will be obtained 
with the help of Proposition~6.1.6, which means the first task
is to construct an initial \tsym{\Sigma^J}-algebra 
$(G^J_S,\nabla^J_{N^J})$ adapted to $J$.

\medskip
Recall that in Section~3.1 $(E_B,{\blob}_K)$ was defined to be the standard term 
\tsym{\Lambda}-algebra defined by some family of enumerations $i_K$
(which means that if $\kappa$ is of type $L \to \beta$ then
$i_\kappa$ is a bijective mapping from $[m]$ to the 
set $L$, where $m = |L|$ is the cardinality of $L$).
This family $i_K$ must now be extended to a family of enumerations
$i^J_{N^J}$ for the signature $\Sigma^J$:
For each $\kappa \in K$ let $i^J_\kappa = i_\kappa$, and for each
$\xi \in J$ of type $L \to \beta$ and each subset $V$ of $L$ choose
a bijective mapping $i^J_{\xi\{V\}} : [m] \to V$, where $m = |V|$.
Without loss of generality it can be assumed that $K$ is disjoint from $J$, and 
hence a term algebra specifier $\Gamma^J : N^J \to K \cup J$ can then be defined 
by letting 
\smallskip
\mdisp{\Gamma^J(\nu)\ =\ \cases{ 
     \kappa  & if $\,\nu = \kappa \in K\,$,\cr
\noalign{\smallskip}
     \xi     & if $\,\nu = \xi\{V\}\in N^J \setminus K\,$.}}
\smallskip
Let $(G^J_S,\nabla^J_{N^J})$ be the term
\tsym{\Sigma^J}-algebra specified by
$\Gamma^J$ and the family of enumerations $i^J_{N^J}$.
Thus $G^J_\sigma \subset (K \cup J)^*$ for each $\sigma \in S$ 
and the family $G^J_S$ can be regarded as being defined by the following rules:

\bigbreak
{\parindent=30pt
\item{(1)} If $\kappa \in K$ is of type $\varnothing \to \beta$ then the list
consisting of the single component $\kappa$ is an 
element of $G^J_\beta$.
\medskip
\item{(2)} If $\kappa \in K$ is of type $L \to \beta$
with $L \ne \varnothing$
and $e_j \in G^J_{\beta_j}$ for $\oneto j m $, where 
$\beta_j = \langle i_\kappa(j) \rangle$ and $m = |L|$,
then $\ \kappa\ e_1\ \cdots \ e_m\ $ is an element of $G^J_\beta$.
\medskip
\item{(3)} If $\xi \in J$ is of type $\sigma$ then the list
consisting of the single component $\xi$ is an 
element of $G^J_\sigma$.
\medskip
\item{(4)} If $\xi \in J$ is of type $L \to \beta$, $V$ is a non-empty subset
of $L$ and $e_j \in G^J_{\sigma_j}$ for $\oneto j m $, where 
$\sigma_j = \langle \,i^J_{\xi\{V\}}(j) \,\rangle$ and $m = |V|$,
then $\ \xi\ e_1\ \cdots \ e_m\ $ is an element of $G^J_\sigma$
with $\sigma = L\setminus V \to \beta$.
\medskip
\item{(5)} The only elements in $G^J_\sigma$ are those obtained
using (1), (2), (3) and  (4). 
\bigskip\medskip
}

\proclaim{Proposition~6.2.3} $(G^J_S,\nabla^J_{N^J})$ is an initial
\tsym{\Sigma^J}-algebra
adapted to $J$. Moreover, it is an extension of $(E_B,{\blob}_K)$. \endpro

\proof If $\sigma = L \to \beta$ and $\xi\{V\} \in N^J_\sigma$ then the sets $V$ and 
$L$ are 
disjoint and $\xi$ is of type $V \cup L \to \beta$. This implies that the restriction 
of $\Gamma^J$ to $N^J_\sigma$ is injective for each $\sigma \in S$ and hence by 
Proposition~2.4.5 $(G^J_S,\nabla^J_{N^J})$ is an initial 
\tsym{\Sigma^J}-algebra.
Moreover, by definition $\nabla^J_{\xi\{\varnothing\}}(\onept) = \xi$
for each $\xi \in J$ (recalling that each set $Z$ is considered to be a subset
of $Z^*$ by identifying $z \in Z$ with the list whose single component is equal to $z$)
and therefore $(G^J_S,\nabla^J_{N^J})$ is adapted to $J$. Finally, 
Proposition~2.5.5 implies  that $(G^J_S,\nabla^J_{N^J})$ is an 
extension of $(E_B,{\blob}_K)$, since
$\Sigma^J$ is an extension of the signature $\Lambda$ and the family of enumerations
$i^J_{N^J}$ is an extension of the family $i_K$. \eop
\bigbreak
Now let $(G^J_S,\nabla^J_N)$ be the functional \tsym{\Sigma}-algebra 
associated with
$(G^J_S,\nabla^J_N)$. Then by Proposition~6.1.6  
$(G^J_S,\nabla^J_N)$ is functionally \tsym{J}-free, and clearly it is 
also an extension of $(E_B,{\blob}_K)$. 
\medskip
As already stated,
these algebras (at least for sets $J$ of the form $I\cup P$)
will provide the basis for a rudimentary functional programming language. 
\medskip
Consider $\xi \in J$ of type $\sigma = L \to \beta$, let $V$ be a non-empty subset of $L$
and $b \in \assb {V} {G^J} $. Then the term 
$\nabla^J_{\sigma,V}(\xi \triangleleft b)$ is by definition equal to
$\nabla^J_{\xi,V}(b)$ and this is a list whose first component is equal to
$\xi$. In the same way, if $\kappa \in K$ is of type $L \to \beta$ and
$b \in \assb {L} {G^J} $ then the term 
$\nabla^J_{\kappa}(b)$ is a list whose first component is equal to $\kappa$.

\medskip
Some care must be taken with the operations in the
\tsym{\Sigma}-algebra $(G^J_S,\nabla^J_N)$: 
Suppose, for example, that $\xi \in J$ is of type $L \to \beta$,
where $L = \{\eta_1,\,\eta_2,\,\eta_3,\,\eta_4\}$
and for $j = 1,\,2,\,3,\,4$ let $e_j \in G^J_{\eta_j}$.
Consider $V = \{\eta_2,\,\eta_4\}$  and suppose that $i^J_{\xi\{L\}}(j) = \eta_j$ for each
$j$ and that $i^J_{\xi\{V\}}(1) = \eta_2$ and  $i^J_{\xi\{V\}}(2) = \eta_4$.
Then $\xi\ e_2\ e_4$ is an element of $G^J_\tau$ with
$\tau = L\setminus V \to \beta$ and so there is the following element
\mdisp{\nabla^J_{\tau,L\setminus V}
 (\,\xi\ e_2\ e_4\, \triangleleft\, \{\eta_1 \to e_1 ,\,\eta_3 \to e_3\}\,)}
of $G^J_\beta$. However, this term has the unique representation
$\,\xi\ e_1\ e_2\ e_3\ e_4\,$ (and not $\,\xi\ e_2\ e_4\ e_1\ e_3\,$).
Moreover, this problem cannot be avoided by a `better' choice of the enumerations,
since for any choice of $i^J_{\xi\{L\}}$ there will be subsets $V$ of $L$ for which
the order is then not `correct'. Of course, the problem does not arise
in the usual functional programming languages, because they restrict partial application
to subsets $V$ which are compatible with the enumeration on $L$ (more precisely,
to initial segments of $L$).
\medskip
If $\xi \in J$ is of type $L \to \beta$ with $L \,=\, \list {\sigma} m $
(and so $[m]$ is the underlying set of the \tsym{S}-typed set $L$)
then $i^J_{\xi\{L\}} : [m] \to [m]$ will always chosen to be the identity mapping.
Moreover, if $V$ is a non-empty subset of $L$ then
$i^J_{\xi\{V\}}$ will taken to be the induced enumeration (i.e., so that the 
elements in $V$ occur in the same relative order as they occur in $L$). 
\medskip
The \tsym{S}-typed sets $J$ which occur in what follows are usually of the form $I \cup P$,
and so the enumerations corresponding to each name of a primitive function
have to be decided upon. The names of the integer operators were already dealt with 
above (since each is of type $\ftt{int}\ \ftt{int} \to \beta$ with 
$\beta$ either $\ftt{int}$ or $\ftt{bool}$). It thus remains to
consider a name $\zeta = \ftt{case}^\theta_\beta$ with
$\beta \in B$ and $\theta \in \Bf$, and here an enumeration of the 
elements of $V$ for each
non-empty subset $V$ of $\theta\cdot K_{\theta,\beta}$ has to be chosen.
However, in all the examples to be given the only terms which occur involving $\zeta$ 
are of the form $\nabla^J_{\zeta,V}(s \triangleleft a )$ with
$V = \theta\cdot K_{\theta,\beta}$ (i.e., total applications), and instead of 
fixing an enumeration of the set $\theta\cdot K_{\theta,\beta}$ once and for all,
it is preferable rather to write such a term
more `legibly' as

\mdisp{ \ftt{case } s \ftt{ of \lcb} \kappa_1 \ftt{ -> } a(\kappa_1)\, \ftt{, }
     \cdots\, \ftt{, } \kappa_m \ftt{ -> } a(\kappa_m) \ftt{\rcb}}

where $\lvector {\kappa} m $ is any enumeration of the elements
of the set $K_{\theta,\beta} = K_\theta$ 
(which can be chosen arbitrarily by the author of the term).
It is clear that $\theta$ and $\beta$, which have been omitted here
from the name $\ftt{case}^\theta_\beta$, can always be 
uniquely determined from the context.
\medskip
Finally, when working with explicit examples of terms it is sensible to 
allow the use of brackets if this increases the legibility. 
The bracketing that will be employed here is essentially that used in {\it Haskell\/}. 
\medskip
As already stated, if $I$ is an \tsym{S}-typed set disjoint from $P$ then 
$(G^{I\cup P}_S,\nabla^{I\cup P}_N)$ will be denoted just by
$(E^I_S,\blob^I_N)$, so $(E^I_S,\blob^I_N)$ is the
functionally \tsym{I \cup P}-free \tsym{\Sigma}-algebra corresponding to
$(F^I_S,\fo^I_N)$.
Example~6.2.1 below gives a first indication of how these 
term algebras will be involved in the equations from Chapter~1. 
The meanings of the terms occurring in this example should be compared 
with the right-hand sides of the equations in Example~2.1.1.  
\bigskip\bigskip

\frame{20pt}{\bigskip {\it Example~6.2.1\enspace} Let $(X_B,p_K)$ be as in 
Example~2.2.1 and $(D_S,f_N)$ be as in one of the three basic Cases.
Let
\mdisp{I\  =\  \{\,\ftt{revs},\,\ftt{sq},\,\ftt{shunt},\,
                        \ftt{shunx},\, \ftt{sqs}\,\}}
be the \tsym{S}-typed set with 
\medskip
{\leftskip=25pt
$\ftt{revs}$ of type $(\ftt{int}\to\ftt{int})\ \ftt{int} \to \ftt{list}$,
$\ftt{sq}$ of type $\ftt{int} \to \ftt{int}$,
\smallskip
$\ftt{shunt}$ of type $\ftt{list}\ \ftt{list} \to \ftt{list}$, \smallskip
$\ftt{shunx}$ of type $\ftt{list}\ \ftt{int}\ \ftt{list} \to \ftt{list}$
and $\ftt{sqs}$ of type $\ftt{int} \to \ftt{list}$.
\medskip
}
Let $L$ be the \tsym{S}-typed set consisting of the element
$\ftt{n}$ of type $\ftt{int}$.   Then 
\mdisp{\ftt{shunt (revs sq n) Nil}}
is an element of $E^{I\cup L}_{\fstt{list}}$
which will be denoted by $e_{\fstt{sqs}}$.
This corresponds to the right-hand side of the equation for $\ftt{sqs}$ in 
Example~1.2.1.
\medskip
{\it (Example~6.2.1 is continued on the next page.)}
\bigskip\medskip
}

\vfill\eject
\frame{20pt}{\bigskip {\it Example~6.2.1 (continued)\enspace} 
Consider $n \in D_{\fstt{int}}$ and an assignment 
$c \in \ass {I\cup L} D $ with $c(\ftt{n}) = n$. Then by Proposition~6.2.2~(1)
\smallskip
\ldisp{\qquad\quad \sem{e_{\fstt{sqs}}}^c_{\fstt{list}}
   \ =\  \sem{\,\ftt{shunt (revs sq n) Nil}\,}^c_{\fstt{list}}}
\vskip-\medskipamount
\ldisp{\qquad\qquad\quad
   \ =\  c_{\fstt{shunt}}(\sem{\,\ftt{revs sq n}\,}^c_{\fstt{list}},
                                         \sem{\ftt{Nil}}^c_{\fstt{list}})}
\vskip-\medskipamount
\ldisp{\qquad\qquad\qquad\qquad
\ =\ c_{\fstt{shunt}}(c_{\fstt{revs}}(\sem{\,\ftt{sq}\,}^c_{\fstt{int}\to\fstt{int}},
\sem{\,\ftt{n}\,}^c_{\fstt{int}}),f_{\fstt{Nil}})}
\vskip-\medskipamount
\rdisp{
\ =\ c_{\fstt{shunt}}(c_{\fstt{revs}}(c_{\fstt{sq}},n),f_{\fstt{Nil}})\ ,\qquad\quad}
\smallskip
where, to increase the legibility, $c_\eta$ is used as an 
alternative notation for $c(\eta)$. 
\medskip
If $L$ is a \tsym{S}-typed set consisting of the element
$\ftt{n}$ of type $\ftt{int}$ then $\,\ftt{mul n n}\,$
is an element of $E^{I\cup L}_{\fstt{int}}$ which will be denoted by
$e_{\fstt{sq}}$ and
corresponds to the right-hand side of the equation for $\ftt{sq}$ in Example~1.2.1.
\medskip
Let $n \in D_{\fstt{int}}$
and consider an assignment $c \in \ass {I\cup L} D $ with 
$c(\ftt{n}) = n$. Then by Proposition~6.2.2~(3) it follows that
\mdisp{  \sem{e_{\fstt{sq}}}^c_{\fstt{int}}
\ =\ \fss{Mul}(\sem{\ftt{n}}^c_{\fstt{int}},\sem{\ftt{n}}^c_{\fstt{int}})
\ =\ \fss{Mul}(n,n)
\ .}
\medskip
Now let $L$ be the \tsym{S}-typed set consisting of the element
$\ftt{n}$ of type $\ftt{int}$ and elements $\ftt{a}$ and $\ftt{b}$ 
of type $\ftt{list}$. Then the following element $e_{\fstt{shunx}}$
of $E^{I\cup L}_{\fstt{list}}$
\mdisp{\ftt{shunt b (Cons n a)}}
corresponds to the right-hand side of the equation for $\ftt{shunx}$ in Example~1.2.1.
\medskip
Let $n \in D_{\fstt{int}}$, $a,\,b \in D_{\fstt{list}}$ 
and consider an assignment $c \in \ass {I\cup L} D $ with 
$c(\ftt{n}) = n$, $c(\ftt{a}) = a$ and $c(\ftt{b}) = b$. Then by Proposition~6.2.2~(1)
\smallskip
\ldisp{\qquad\quad
 \sem{e_{\fstt{shunx}}}^c_{\fstt{list}}
   \ =\  c_{\fstt{shunt}}(
         \sem{\,\ftt{b}\,}^c_{\fstt{list}},
         \sem{\,\ftt{Cons n a}\,}^c_{\fstt{list}})}
\vskip-\medskipamount
\rdisp{
   \ =\  c_{\fstt{shunt}}(b,
         f_{\fstt{Cons}}(
         \sem{\,\ftt{n}\,}^c_{\fstt{int}},
         \sem{\,\ftt{a}\,}^c_{\fstt{list}}))
   \ =\  c_{\fstt{shunt}}(b,
         f_{\fstt{Cons}}(n,a))\ .\qquad\quad}
\medskip
Next let $L$ be the \tsym{S}-typed set consisting of the elements
$\ftt{a}$ and $\ftt{b}$ 
of type $\ftt{list}$. Then the following element $e_{\fstt{shunt}}$
of $E^{I\cup L}_{\fstt{list}}$
\mdisp{\ftt{case a of \lcb Nil -> b, Cons -> shunx b\rcb}}
corresponds to the right-hand side of the equation for $\ftt{shunt}$ in Exampls~1.2.1.
\medskip
{\it (Example~6.2.1 is continued on the next page.)}
\bigskip\medskip
}

\vfill\eject

\frame{20pt}{\bigskip 
{\it Example~6.2.1 (continued)\enskip} 
Let $a,\,b \in D_{\fstt{list}}$ 
and consider an assignment $c \in \ass {I\cup L} D $ with 
$c(\ftt{a}) = a$ and $c(\ftt{b}) = b$. Then by Proposition~6.2.2~(2)
\smallskip
\ldisp{ \qquad \sem{e_{\fstt{shunt}}}^c_{\fstt{list}}}
\vskip-\medskipamount
\mdisp{  \ =\ \fss{Case}^{\fstt{list}}_{\fstt{list}}
     (\,\sem{\,\ftt{a}\,}^c_{\fstt{list}}\,,
      \,\{\,\ftt{Nil}\to\sem{\,\ftt{b}\,}^c_{\fstt{list}}\,,\,
     \ftt{Cons}\to\sem{\,\ftt{shunx b}\,}^c_{\fstt{list}}\,\}\,)}
\vskip-\smallskipamount
\mdisp{  \ =\ \fss{Case}^{\fstt{list}}_{\fstt{list}}
     (\,a\,,\,\{\,\ftt{Nil}\to b\,,\,\ftt{Cons}\to c_{\fstt{shunx}}(b,\cdot,\cdot)\,\}\,)}
\vskip-\smallskipamount
\rdisp{
\ =\ \cases{
    \bot_{\fstt{list}} & if $\,a = \bot_{\fstt{list}}\,$,\cr
    b  & if $\,a = f_{\fstt{Nil}}\,$,\cr
    c_{\fstt{shunx}}(b,m,d) 
       & if $\,a \ne \bot_{\fstt{list}}\,$ and $\,a = f_{\fstt{Cons}}(m,d)\,$.\cr
}\qquad}
\medskip
Finally, consider the \tsym{S}-typed set $L$ consisting of the element
$\ftt{n}$ of type $\ftt{int}$ and $\ftt{p}$ of type
$\ftt{int}\to\ftt{int}$. Then
the following element $e_{\fstt{revs}}$ of $E^{I\cup L}_{\fstt{list}}$ 
\smallskip
\ldisp{\qquad\ftt{case (eq n 0) of \lcb True -> Nil,}}
\vskip-\medskipamount
\vskip-\smallskipamount
   \rdisp{ \ftt{False -> Cons (p n) (revs (sub n 1))\rcb}\qquad}
\smallskip
corresponds to the right-hand side of the equation for $\ftt{revs}$ in 
Example~1.2.1.
\medskip
Let $n \in D_{\fstt{int}}$, $p \in D_{\fstt{int}\to\fstt{int}}$ and 
$c \in \ass {I\cup L} D $ be an assignment with 
$c(\ftt{n}) = n$ and $c(\ftt{p}) = p$. Then by Proposition~6.2.2~(3)
\smallskip
\mdisp{\sem{\,\ftt{eq n 0}\,}^c_{\fstt{bool}}
   \ =\ \fss{Eq}(n,0)
   \ =\ \cases{
               \bot_{\fstt{bool}} & if $\,n = \bot_{\fstt{int}}\,$,\cr 
               T & if $\,n = 0\,$,\cr
               F & if $\,n \in \Int \setminus \{0\}\,$.\cr}}
\medskip
Thus, putting $z = \sem{\,\ftt{eq n 0}\,}^c_{\fstt{bool}}$, it follows from
Proposition~6.2.2 that
\medskip
\ldisp{ \qquad \sem{e_{\fstt{revs}}}^c_{\fstt{list}}
 \ =\ \fss{Case}^{\fstt{bool}}_{\fstt{list}}
     (\,z\,,\,\{\,\ftt{True}\to\sem{\,\ftt{Nil}\,}^c_{\fstt{list}}\,,\,}
\vskip-\medskipamount
\rdisp{
  \ftt{False}\to\sem{\,\ftt{Cons (p n) (revs (sub n 1))}\,}^c_{\fstt{list}}\,\}\qquad\qquad}
\vskip-\smallskipamount
\ldisp{\qquad\qquad \ =\ \fss{Case}^{\fstt{bool}}_{\fstt{list}}
     (\,\fss{Eq}(n,0)\,,\,
            \{\,\ftt{True}\to \onept\,,}
\vskip-\medskipamount
\rdisp{
        \,\ftt{False}\to 
       f_{\fstt{Cons}}(p(n),c_{\fstt{revs}}(\fss{Sub}(n,1)))\,\}\,)\qquad\qquad}
\vskip-\smallskipamount
\rdisp{
\ =\ \cases{
       \bot_{\fstt{list}} & if $\,n = \bot_{\fstt{int}}\,$, \cr
        f_{\fstt{Nil}}  & if $\,n = 0\,$,\cr
        f_{\fstt{Cons}}(p(n),c_{\fstt{revs}}(n-1))
            & if $\,n \in \Int \setminus \{0\}\,$. \cr}\qquad}
\bigskip\medskip
}

\vfill\eject

\hfline{164}{6.3 \ SUPPORT SYSTEMS AND THE TRACE}

\sectionhead {6.3} {Support systems and the trace}
\bigskip\medskip
In the previous section the basic term algebras were introduced:
For each \tsym{S}-typed set $I$ disjoint from $P$ we have chosen
a functionally \tsym{I\cup P}-free \tsym{\Sigma}-algebra 
$(F^I_S,\fo^I_N)$ which is an extension of the ground term algebra $(F_B,\fo_K)$. 
\medskip
For the whole of the section let $I$ be some fixed \tsym{S}-typed set disjoint from $P$.
The problem which will be addressed here is to find reasonable sufficient 
conditions on a term $s \in \kterm{F}^I_\beta$ (with $\beta \in B$) which will ensure that 
$\sem{s}^c_\beta \ne \bot_\beta$ for all $c \in \ass I D $.
\medskip
A typical situation in which such conditions will be needed is the following:
Let $\beta \in B$, $\theta \in \Bf$, and suppose we are interested in
evaluating $\fss{Case}^\theta_\beta(\sem{s}^c_\theta,b)$
for some $c \in \ass I D $, where $s \in \kterm{F}^I_\theta$ and
$b \in \assb {K_{\theta,\beta}} D $ (and where $b$ will be of the form
$\assb {K_{\theta,\beta}} {\sem{a}^c} $
for some $a \in \assb {K_{\theta,\beta}} {F^I} $).
Now by Proposition~6.2.1~(1) 
there exists a unique $\kappa \in K_\theta$ having type $J \to \theta$ 
for some $J \in {\cal F}_S$ and a unique element $a' \in \assb J {F^I} $
such that $s = \fo^I_\kappa(a')$. Therefore,
if we knew that $\sem{s}^c_\theta \ne \bot_\theta$ then we could conclude that
\smallskip
\ldisp{\quad
\fss{Case}^\theta_\beta\bigl(\,\sem{s}^c_\theta,\, b\,\bigr)
\ =\ \fss{Case}^\theta_\beta\bigl(\,\sem{\,\fo^I_\kappa(a')\,}^c_\theta, \,b\,\bigr)}
\vskip-\medskipamount
\rdisp{
\ =\ \fss{Case}^\theta_\beta\bigl(\,   f_\kappa(\,\assb {J} {\sem{a'}^c} ),\, b\,\bigr)
\ =\ f_{\triangleleft\{J\to \beta,J\}}\bigl(\,b(\kappa) \triangleleft 
\assb {J} {\sem{a'}^c} \,\bigr)
\ .\quad}
\smallskip
But we will not usually have any detailed information about $c$, and so the only 
practical way to be sure that this conclusion is valid is to know that $s$ is such that
$\sem{s}^c_\theta \ne \bot_\theta$ for {\it all\/} $c \in \ass I D $.

\medskip
Let us say that a family $N^I_B$ is a 
\indexdef{support system}{support system}{} 
for $(D_S,f_N)$ if
$F_\beta \subset N^I_\beta \subset \kterm{F}^I_\beta$ for each 
$\beta \in B$ and
$\sem{s}^c_\beta \ne \bot_\beta$ for all $s \in N^I_\beta$, $\beta \in B$.
The above problem is thus that of finding suitable support systems for $(D_S,f_N)$. 
To be more precise, what we are looking for is a support system $N^I_B$ which is,
on the one hand, not too small and for which, on the other hand, it is possible
to effectively determine whether a given term
$s \in \kterm{F}^I_\beta$ is an element of $N^I_\beta$. 
\medskip

For example, the family $F_B$ 
provides a somewhat trivial example of a support system 
(since by Lemma~6.2.1 
$\sem{s}^c_\beta = \sem{s}_\beta \in X_\beta$ for all $s \in F_\beta$, $c \in \ass I D $). 
However, except when $(D_B,f_K)$ is the flat extension of $(X_B,p_K)$ this family
is too small to be of any use.
At the other extreme, the family $N^I_B$ defined for each $\beta \in B$ by
\mdisp{ N^I_\beta \ =\ \lcurl s \in \kterm{F}^I_\beta \,:\, \sem{s}^c_\beta \ne \bot_\beta
\frm{for all} c \in \ass I D \rcurl}
is trivially the maximal support system, but in general this will certainly not satisfy the
second requirement. 
\medskip
As we will see,
there is, however, a `natural' support system $\bterm{F}^I_B$ for $(D_S,f_N)$ which
satisfies both of the requirements. In fact this family also complies with the general
principle formulated in Section~3.3, since it is determined entirely by the
trace $R_B$ of $(D_B,f_K)$.
\medskip
Before coming to the definition of the support system $\bterm{F}^I_B$ it is first necessary
to extend the bottomed ground term algebra to a \tsym{\Sigma}-algebra and then
to look at the relationship between this \tsym{\Sigma}-algebra and the 
\tsym{\Sigma}-algebra $(F^I_S,\fo^I_N)$.
\medskip
Recall from Chapter~3 that the bottomed ground term algebra
$(F^\flat_B,\fo^\flat_K)$ is a \tsym{\Lambda}-algebra which
is an initial bottomed extension of $(F_B,\fo_K)$
(but with the bottom element of $F^\flat_\beta$ being denoted by $\flat_\beta$ 
rather than $\bot_\beta$).
The unique bottomed \tsym{\Lambda}-homomorphism from 
$(F^\flat_B,\fo^\flat_K)$ to $(D_B,f_K)$ 
(i.e., the unique \tsym{\Lambda}-homomorphism with 
$\sem{\,\flat_\beta}^\bot_\beta = \bot_\beta$ for 
each $\beta \in B$) is denoted by $\sem{\cdot}^\bot_B$.
By Proposition~2.5.3 $\sem{\cdot}^\bot_B$ is an extension of $\sem{\cdot}_B$,
i.e., $\sem{s}^\bot_\beta = \sem{s}_\beta$ for all $s \in F_\beta$, $\beta \in B$.
Recall also that the trace $R_B$ of $(D_B,f_K)$ is the  family with 
$R_B \subset F^\flat_B$ defined by putting
\mdisp{R_\beta 
\ =\ \lcurl\, s \in F^\flat_\beta\, :\, 
                           \sem{s}^\bot_\beta \ne \bot_\beta \,\rcurl}
for each $\beta \in B$.
\medskip
The bottomed ground term algebra
$(F^\flat_B,\fo^\flat_K)$ will now be extended in a trivial way to obtain a bottomed
\tsym{\Sigma}-algebra $(F^\flat_S,\fo^\flat_N)$:
For each $\sigma \in S \setminus B$ let $F^\flat_\sigma$ consist of a single element
$\flat_\sigma$ (which is considered to be the bottom
element of $F^\flat_\sigma$), and if $\sigma = L \to \beta$ is a functional type 
and $J$ is a non-empty subset of $L$ then define 
$\,\fo^\flat_{\triangleleft\{\sigma,J\}} : \assb J {F^\flat} 
                                    \to F^\flat_{\sigma_{L\setminus J}}\,$
simply to be the mapping given by 
\mdisp{\fo^\flat_{\triangleleft\{\sigma,J\}}(c) \ =\ \flat_{\sigma_{L\setminus J}}}
for all $c \in \assb J {F^\flat} $. 

\proclaim{Lemma~6.3.1} $(F^\flat_S,\fo^\flat_N)$ is a functional \tsym{\Sigma}-algebra. 
\endpro

\proof If
$\sigma = L \to \beta $ is a functional type, $J$ and $J'$ are non-empty disjoint  
subsets of $L$ and $s \in F^\flat_{\sigma}$ then the terms
$\,\fo^\flat_{\sigma,J\cup J'}\bigl(s \triangleleft (c \oplus c')\bigr)\,$
and
$\,\fo^\flat_{\sigma_{L\setminus J},J'}
  \bigl(\fo^\flat_{\sigma,J}(s \triangleleft c) \,\triangleleft\, c'\bigr)\,$
are both equal to $\,\flat_{\sigma_{L\setminus (J\cup J')}}\,$
for each $c \in \assb {J} {F^\flat} $, $c' \in \assb {J'} {F^\flat} $. 
This implies that
$(F^\flat_S,\fo^\flat_N)$ is a functional \tsym{\Sigma}-algebra. 
\eop
\bigbreak
The \tsym{\Lambda}-homomorphism $\sem{\cdot}^\bot_B$ will also be extended to a family
$\sem{\cdot}^\bot_S$ by letting $\sem{\cdot}^\bot_\sigma : F^\flat_\sigma \to D_\sigma$
be the mapping with $\sem{\,\flat_\sigma}^\bot_\sigma = \bot_\sigma$ for each
$\sigma \in S \setminus B$.

\proclaim{Lemma~6.3.2} $\sem{\cdot}^\bot_S$ is a \tsym{\Sigma}-homomorphism,
and in fact it is the unique bottomed 
\tsym{\Sigma}-homomorphism from $(F^\flat_S,\fo^\flat_N)$ to $(D_S,f_N)$. \endpro

\proof Consider a functional type $\sigma = L \to \beta$ and a non-empty subset $J$
of $L$. Then from Property (A) and Proposition~5.4.1~(1) it follows 
for all $s \in F^\flat_\sigma$, $c \in \assb J {F^\flat} $ that
\smallskip
\ldisp{\quad f_{\sigma,J}
  \bigl(\assb {\sigma\cdot J} {\sem{\,s \triangleleft c\,}^\bot} \bigr)
\ =\ f_{\sigma,J}
  \bigl(\sem{s}^\bot_\sigma \triangleleft \assb  J {\sem{c}^\bot} \bigr)
}
\vskip-\smallskipamount
\rdisp{
\ =\ f_{\sigma,J}
  \bigl(\bot_\sigma \triangleleft \assb  J {\sem{c}^\bot} \bigr)
\ =\ \bot_{\sigma_{L\setminus J}} 
\ =\ \sem{\,\flat_{\sigma_{L\setminus J}}}^\bot_{\sigma_{L\setminus J}}
\ =\ \sem{\,\fo^\flat_{\sigma,J} (s \triangleleft c)\,}^\bot_{\sigma_{L\setminus J}} 
\ .\quad}
\smallskip
On the other hand, if
$\kappa \in K$ is of type $L \to \beta$ then
$\,\sem{\,\fo^\flat_\kappa(c)\,}^\bot_\beta 
                  \,=\, f_\kappa(\assb L {\sem{c}^\bot} )\,$ 
holds for all $c \in \assb L {F^\flat} $  
because  $\sem{\cdot}^\bot_B$ is a \tsym{\Lambda}-homomorphism. 
This implies that 
$\sem{\cdot}^\bot_S$ is a \tsym{\Sigma}-homomorphism, which by definition
is bottomed. The uniqueness now follows from the fact that
$\sem{\cdot}^\bot_B$  is the unique bottomed \tsym{\Lambda}-homomorphism  from
$(F^\flat_B,\fo^\flat_K)$ to $(D_B,f_K)$. \eop
\bigbreak
By Lemma~6.3.1 $(F^\flat_S,\fo^\flat_N)$ is a functional \tsym{\Sigma}-algebra and
$(F^I_S,\fo^I_N)$ is functionally \tsym{I\cup P}-free;
there thus exists a unique
\tsym{\Sigma}-homomorphism 
$\Omega^I_S : (F^I_S,\fo^I_N) \to (F^\flat_S,\fo^\flat_N)$ such that
$\,\Omega^I_\xi(\xi) \,=\, \flat_\xi\,$ for each $\xi \in I\cup P$. 
\medskip
It is important to note that $\Omega^I_S$ has nothing to to with the
\tsym{\Sigma}-algebra $(D_S,f_N)$.
In fact, in the sense in which a functionally 
\tsym{I\cup P}-free \tsym{\Sigma}-algebra is unique, it can be said that $\Omega^I_S$ is 
uniquely determined by the signature $\Sigma$ and the \tsym{S}-typed sets $I$
and $P$.
Now define a family $\bterm{F}^I_B$
with $\bterm{F}^I_B \subset F^I_B$ by putting
\mdisp{ \bterm{F}^I_\beta\ =\ \lcurl s \in F^I_\beta \,:\,
           \Omega^I_\beta(s) \in R_\beta \rcurl}
for each $\beta \in B$, with $R_B$ again the trace of $(D_B,f_K)$.
Thus $\bterm{F}^I_B$ depends on 
$(D_B,f_K)$ (but not on the whole of $(D_S,f_N)$), and it only depends on
$(D_B,f_K)$ via the trace $R_B$.

\proclaim{Proposition~6.3.1} The family $\bterm{F}^I_B$ is a support system for
$(D_B,f_K)$.
Moreover, if $(D_B,f_K)$ is the flat extension of $(X_B,p_K)$ then
$\bterm{F}^I_B = F_B$, and if $(D_B,f_K)$ is a fully regular bottomed extension of 
$(X_B,p_K)$ then $\bterm{F}^I_B = \kterm{F}^I_B$. \endpro 

\proof Later. \eop
\bigbreak

In what follows the \tsym{\Lambda}-algebra $(F^I_B,\fo^I_K)$ 
will usually play a more important role than the \tsym{\Sigma}-algebra 
$(F^I_S,\fo^I_N)$ itself.
Note that the family $\Omega^I_B$ (obtained by omitting the mappings 
$\lcurl \Omega^I_\sigma \,:\, \sigma \in S \setminus B \rcurl$ 
from the family $\Omega^I_S$) is then a \tsym{\Lambda}-homomorphism
from $(F^I_B,\fo^I_K)$ to $(F^\flat_B,\fo^\flat_K)$. In this regard
the following facts will be needed:

\proclaim{Lemma~6.3.3}
(1)\enskip $\,\Omega^I_\beta(s) =\, \flat_\beta\,$
for all $s \in \aterm{F}^I_\beta$, $\beta \in B$.
\medskip
(2)\enskip  $\,\Omega^I_\beta(s) = s\,$ for all ground terms $s \in F_\beta$, 
$\beta \in B$. 
\medskip
(3)\enskip  $\,\lcurl s \in F^I_\beta \,:\,
              \Omega^I_\beta(s) \in F_\beta \rcurl\,=\, F_\beta\,$
for each $\beta \in B$.
\endpro

\proof (1) and (2) are special cases of Lemma~6.3.4 below.
\medskip
(3)\enskip Part (2) implies that $\Omega^I_\beta(s) \in F_\beta$
for all $s \in F_\beta$, $\beta \in B$. The proof of the converse is essentially the
same as the proof of Proposition~3.3.3. Consider the family 
${\grave F}^I_B$ with ${\grave F}^I_B \subset F^I_B$ given by

\mdisp{{\grave F}^I_\beta\ =\ F_\beta \,\cup\,
\lcurl s \in F^I_\beta \setminus F_\beta 
          \,:\, \Omega^I_\beta(s) \in F^\flat_\beta \setminus F_\beta\rcurl}

for each $\beta \in B$. 
If $\kappa \in K$ is of type $\varnothing \to \beta$ then
$\fo^I_\kappa(\onept) = \fo_\kappa(\onept )\in F_\beta \subset {\grave F}^I_\beta$.
Now let $\kappa \in K$ be of type $L \to \beta$ with $L \ne \varnothing$ and let
$a \in \assb L {{\grave F}^I} $, i.e., $a(\eta) \in {\grave F}^I_\eta$ for each
$\eta \in L$.  
Suppose 
$\Omega^I_\beta(\fo^I_\kappa(a)) = \fo^\flat_\kappa(\assb L {\Omega^I} (a)) \in F_\beta$.
Then by Proposition~3.2.6 $\assb L {\Omega^I} (a) \in \ass L F $
(since $(F^\flat_B,\fo^\flat_K)$ is an initial bottomed
extension of $(F_B,\fo_K)$), i.e., 
$\Omega^I_\eta(a(\eta)) \in F_\eta$ for each $\eta \in L$.
Then by assumption $a(\eta) \in F_\eta$ for each $\eta \in L$ and thus
$\fo^I_\kappa(a) = \fo_\kappa(a) \in F_\beta$. This shows that
$\fo^I_\kappa(a) \in {\grave F}^I_\beta$, and hence that
the family ${\grave F}^I_B$ is invariant in $(F^I_B,\fo^I_K)$.
But by definition $I_\beta \subset {\grave F}^I_\beta$
and by part (1) $\aterm{F}^I_\beta \subset {\grave F}^I_\beta$ for each
$\beta \in B$. Therefore by Lemma~6.2.2 ${\grave F}^I_B = F^I_B$, which
of course implies that $\Omega^I_\beta(s) \in F^\flat_\beta \setminus F_\beta$
for all $s \in F^I_\beta \setminus F_\beta$, $\beta \in B$. \eop
\bigbreak
 
\proclaim{Lemma~6.3.4} Let $\pi_S : (F^I_S,\fo^I_N) \to (F^\flat_S,\fo^\flat_N)$ 
be any \tsym{\Sigma}-homomorphism. Then:
\medskip
(1)\enskip $\,\pi_\beta(s) =\, \flat_\beta\,$
for all $s \in \aterm{F}^I_\beta$, $\beta \in B$.
\medskip
(2)\enskip  $\,\pi_\beta(s) = s\,$ for all ground terms $s \in F_\beta$, 
$\beta \in B$. \endpro

\proof (1)\enskip This follows since if $\xi \in I\cup P$ is of 
functional type 
$\sigma = L \to \beta$ and $b \in \assb L {F^I} $ then
$\,\pi_\beta\bigl(\,\fo^I_{\sigma,L}(\xi \triangleleft b)\,\bigr)\,$ has 
the form $\fo^\flat_{\sigma,L}(a)$ for some
$a \in \assb {\sigma\cdot L} {F^\flat} $, which is by definition is equal to
$\,\flat_\beta$.
\medskip
(2)\enskip This is a special case of Lemma~3.2.1. \eop
\bigbreak

\proclaim{Proposition~6.3.2} 
$F_B \subset \bterm{F}^I_B \subset \kterm{F}^I_B$.
Moreover, if $(D_B,f_K)$ is the flat extension of $(X_B,p_K)$ then
$\bterm{F}^I_B = F_B$, and if $(D_B,f_K)$ is a fully regular bottomed extension of 
$(X_B,p_K)$ then $\bterm{F}^I_B = \kterm{F}^I_B$. \endpro 

\proof Since $F_\beta \subset R_\beta \subset F^\flat_\beta \setminus \{\,\flat_\beta\}$
for each $\beta \in B$,
the first statement follows from parts (1) and (2) of Lemma~6.3.3.
The rest then follows from Lemma~6.3.3~(3), since
if  $(D_B,f_K)$ is the flat extension of $(X_B,p_K)$ then 
by Lemma~3.3.2 $R_B = F_B$, and if $(D_B,f_K)$ is a fully regular bottomed extension 
then Proposition~3.3.2~(2) implies that 
$R_\beta = F^\flat_\beta \setminus \{\,\flat_\beta\}$
for each $\beta \in B$. \eop
\bigbreak
The first statement in Proposition~6.3.2 implies that
$\bterm{F}^I_\beta = F_\beta$ for each primitive type $\beta \in B$
(since then $\kterm{F}^I_\beta = F_\beta$). Thus in particular
$\bterm{F}^I_{\fstt{int}} = F_{\fstt{int}}$.
\medskip

Denote by $\varrho^I_S$ the unique 
\tsym{\Sigma}-homomorphism from $(F^I_S,\fo^I_N)$ to  $(D_S,f_N)$
such that $\varrho^I_\xi(\xi) = \bot_\xi$ for each $\xi \in I\cup P$.

\proclaim{Lemma~6.3.5} $\,\sem{\cdot}^\bot_S \,\Omega^I_S\, =\, \varrho^I_S\,$,
i.e., $\,\sem{\Omega^I(s)}^\bot_\sigma\, =\, \varrho^I_\sigma(s)\,$ for all
$s \in F^I_\sigma$, $\sigma \in S$. \endpro

\proof By Proposition~2.2.1 $\,\omega^I_S \,=\, \sem{\cdot}^\bot_S \, \Omega^I_S\,$ 
a \tsym{\Sigma}-homomorphism from $(F^I_S,\fo^I_N)$  to $(D_S,f_N)$ and 
$\,\omega^I_\xi(\xi) \,=\, \sem{\Omega^I(\xi)}^\bot_\xi 
\,=\, \sem{\,\flat_\xi}^\bot_\xi \,=\, \bot_\xi\,$
for each $\xi \in I\cup P$. Thus by the uniqueness of $\varrho^I_S$ 
it follows that $\omega^I_S = \varrho^I_S$. \eop
\bigbreak  
Of course, Lemma~6.3.5 implies in particular that
$\,\bterm{F}^I_\beta\, =\, \lcurl s \in F^I_\beta \,:\,
           \varrho^I_\beta(s) \ne \bot_\beta \rcurl\,$
for each $\beta \in B$.
Note that the family $\varrho^I_B$ is a \tsym{\Lambda}-homomorphism from $(F^I_B,\fo^I_K)$
to $(D_B,f_K)$. 

\proclaim{Lemma~6.3.6} 
$F^I_B$ is the only invariant family in
$(F^I_B,\fo^I_K)$ containing $U_B$, where
$\,U_\beta\, =\, \lcurl s \in F^I_\beta \,:\,
                            \varrho^I_\beta(s) = \bot_\beta \rcurl\,$
for each $\beta \in B$; in other words, in the terminology of Section~3.5, the 
\tsym{\Lambda}-homomorphism $\varrho^I_B$ is fat bottomed. \endpro

\proof If $\xi \in I\cup P$ has functional type $\sigma = L \to \beta$ then
by Proposition~5.4.1~(1) and Property (A) it follows that
\smallskip
\ldisp{\quad
\varrho^I_\beta\bigl(\fo^I_\xi(b)\bigr)
\ =\ \varrho^I_\beta\bigl(\fo^I_{\sigma,L}(\xi \triangleleft b)\bigr)
  \ = \ f_{\sigma,L}\bigl(
      \assb {\sigma\cdot L} {\varrho^I} (\xi \triangleleft b)  \bigr)}
\vskip-\medskipamount
\rdisp{
  \ = \ f_{\sigma,L}\bigl(
    \varrho^I_\sigma(\xi) \triangleleft \assb L {\varrho^I} (b)  \bigr)
  \ = \ f_{\sigma,L}\bigl(
    \bot_\sigma \triangleleft \assb L {\varrho^I} (b)  \bigr)
\ =\ \bot_\beta\quad}
\smallskip
for all $b \in \assb L {F^I} $. Moreover, by definition
$\varrho^I_\beta(\xi) = \bot_\beta$ for each $\xi \in I_\beta$, and hence
$\aterm{F^I}_\beta \cup I_\beta \subset U_\beta$ for each $\beta \in B$.
Thus by Lemma~6.2.2 $F^I_B$ is the only invariant family in
$(F^I_B,\fo^I_K)$ containing $U_B$. \eop
\bigbreak
{\it Proof of Proposition~6.3.1\ }\enskip By Proposition~6.3.2 it is enough to show
that $\sem{s}^c_\beta \ne \bot_\beta$ for all $s \in \bterm{F}^I_\beta$, $c \in \ass I D $.
Consider $s \in \bterm{F}^I_\beta$, so by Lemma~6.3.5 
$\,\varrho^I_\beta(s) \,\ne\, \bot_\beta\,$.
But by Lemma~6.3.6 the \tsym{\Lambda}-homomorphism $\varrho^I_B$ is 
fat bottomed and by Proposition~3.5.5 $(D_B,f_K)$ is strongly monotone. Therefore
$\pi_\beta(s) \ne \bot_\beta$ for every \tsym{\Lambda}-homomorphism
$\pi_B : (F^I_B,\fo^I_K) \to (D_B,f_K)$, and so in particular
$\sem{s}^c_\beta \ne \bot_\beta$ for each $c \in \ass I D $. \eop

\bigskip\bigskip\bigskip\bigskip

\sectionhead {6.4} {Notes}
\bigskip\medskip
Proposition~6.1.3, which gives the exact relationship between initial 
\tsym{\Sigma^I}-algebras and functionally \tsym{I}-free \tsym{\Sigma}-algebras, should 
be well-known, at least at an intuitive level, to any experienced functional programmer.

\vfill\eject

\hfline{169}{}
\bigskip
{\fourteenbf Chapter 7\quad Equations}
\bigskip\medskip
In the present chapter we explain what is meant by a system of equations and what 
the solutions of such a system are. The problem of computing values
will then be considered.
\medskip
As before let $\fss{C}$ be a bottomed concrete category with finite 
products and $({\bf D}_S,f_N)$ be a \tsym{\fss{C}}-based \tsym{\Sigma}-algebra 
such that $(D_B,f_K)$ is a monotone regular bottomed extension
of $(X_B,p_K)$. It is again assumed that the \tsym{\Sigma}-algebra $(D_S,f_N)$ has
the properties (A), (B), (C) and (D) listed in Section~5.4.
Moreover, we assume that for each \tsym{S}-typed set $I$ disjoint from $P$ 
a functionally \tsym{I\cup P}-free \tsym{\Sigma}-algebra $(F^I_S,\fo^I_N)$ has been
chosen which is an 
extension of the ground term algebra $(F_B,\fo_K)$
(with $P$ the \tsym{S}-typed set specifying the names and types of the primitive
functions). 

\medskip
Equations are introduced in Section~7.1.
Let $I$ be a non-empty finite \tsym{S}-typed set disjoint from $P$.
Consider $\xi \in I$ of type 
$\sigma = L \to \beta$ and assume that $L$ is disjoint from $I$ and $P$.
Then $F^{I \cup L}_\beta$ is regarded as the set of possible (right-hand sides of)
equations for $\xi$ and this set will be denoted
by $\fss{Eq}(\xi,I)$.
The set of solutions $\fss{Sol}(s)$ of an equation $s \in \fss{Eq}(\xi,I)$ is
the set of assignments $c \in \ass I D $ for which

\mdisp{f_{\sigma,L}(c(\xi) \triangleleft b) 
\  =\ \sem{s}_\beta^{\,c \oplus b}}  

for all $b \in \ass L D $ if $L \ne \varnothing$, and for which
$\,c(\xi) = \sem{s}_\beta^c\,$ if $L = \varnothing$.
\medskip
A system of equations then consists of a family of equations, one for each name in 
$I$, i.e., a family of the form $s_I$, 
where $s_\xi \in \fss{Eq}(\xi,I)$ for each $\xi \in I$, and
the set of all such families is denoted by $\fss{Eq}(I)$. 
If $\alpha = s_I\in \fss{Eq}(I)$ is a system of equations then an assignment 
$c \in \ass I D $ is said to be a solution of the equations $\alpha$ if 
$c \in \fss{Sol}(s_\xi)$ for each $\xi \in I$; the set of  
solutions will be denoted by $\fss{Sol}(\alpha)$. 
\medskip

In Section~7.2 the problem of computing values is looked at, which means something like 
the following: Let $\alpha \in \fss{Eq}(I)$ be a system of equations for which 
$\fss{Sol}(\alpha) \ne \varnothing$ 
and consider the family $F^\alpha_B$ with $F^\alpha_B \subset F^I_B$ defined by

\mdisp{F^\alpha_\beta\ =\ \lcurl s \in F^I_\beta
    \,: \frm{there exists} x \in X_\beta \frm{such that}
   \sem{s}^{c}_\beta  = x \frm{for all} 
                          c \in \fss{Sol}(\alpha) \rcurl}
for each $\beta \in B$. 
(In particular, $F_B \subset F^\alpha_B$, since if $s \in F_\beta$ then Lemma~6.2.1 
implies that $\sem{s}^{c}_\beta  = \sem{s}_\beta$ for all $c \in \ass I D $ 
and so in particular for $c \in \fss{Sol}(\alpha)$.) 
For which elements $s \in F^\alpha_\beta$ is it then possible to compute
the value of $s$, i.e., the unique element $x \in X_\beta$ such that 
$\sem{s}^{c}_\beta = x$ for all $c \in \fss{Sol}(\alpha)$?
\medskip

To answer this question some kind of algorithm is needed which will
manipulate the elements from the sets in the family $F^I_B$: In response to the `input' 
$s$ it should either produce the `output' $s'$ (whenever possible) or produce 
no output at all. We will look for such an algorithm defined in terms of 
a \tsym{\Lambda}-homomorphism
$\Phi_B : (F^I_B,\fo^I_K) \to (F^I_B,\fo^I_K)$ as follows:
For each $\beta \in B$ let
\smallskip
\mdisp{F^\Phi_\beta \ =\ \lcurl s \in F^I_\beta \,:\,
        \Phi^n_\beta(s) \in F_\beta \frm{for some} n \ge 0 \rcurl\ ,}
where $\{\Phi^n_B\}_{n \ge 0}$ is the sequence of iterates of $\Phi_B$.
Now it is easy to see that $\Phi_\beta(s) = s$ for each $s \in F_\beta$ and
hence a mapping $\Phi^\infty_\beta : F^\Phi_\beta \to F_\beta$ can 
be defined by putting $\Phi^\infty_\beta(s) = \Phi^n_\beta(s)$, where 
$n \ge 0$ is chosen so that $\Phi^n_\beta(s) \in F_\beta$.
Then $\Phi_B$ can be considered as defining
an algorithm which produces the 
`output' $\Phi^\infty_\beta(s)$ if $s \in F^\Phi_\beta$  
and which produces no output if $s \in F^I_\beta \setminus F^\Phi_\beta$.  
\medskip
Of course, if the algorithm should compute values then there must be some relationship
between the equations $\alpha$ and the homomorphism $\Phi_B$, and this leads to the
following definition: The homomorphism  $\Phi_B$ is said to be 
\indexddef{\tsym{\alpha}-consistent}{consistent homomorphism}{} 
{homomorphism}{consistent}
if 

\mdisp{\sem{\,\Phi_\beta(s)\,}^c_\beta \ =\ \sem{s}^c_\beta}

for all $c \in \fss{Sol}(\alpha)$ for each $s \in F^I_\beta$, $\beta \in B$.
Proposition~7.2.1 will show that if $\Phi_B$ is 
\tsym{\alpha}-consistent then $F^\Phi_B \subset F^\alpha_B$ and moreover, that
\mdisp{\sem{\,\Phi^\infty_\beta(s)\,}_\beta\ =\ \sem{s}^c_\beta}
for all $c \in \fss{Sol}(\alpha)$ for each $s \in F^\Phi_\beta$, $\beta \in B$. 
This says that the algorithm defined by an \tsym{\alpha}-consistent 
homomorphism only outputs correct values. However, this is of limited use if the family  
$F^\Phi_B$ is much smaller than $F^\alpha_B$. For example, the identity 
\tsym{\Lambda}-homomorphism 
$\fss{id}_B : (F^I_B,\fo^I_K) \to (F^I_B,\fo^I_K)$ is always \tsym{\alpha}-consistent,
but, as must be expected, it does not really provide any information at all, since 
$F^{\fsss{id}}_B = F_B$.
\medskip
The best to be hoped for is an \tsym{\alpha}-consistent \tsym{\Lambda}-homomorphism 
$\Phi_B$ with  $F^\Phi_B = F^\alpha_B$ and such a homomorphism $\Phi_B$ will be called 
\indexddef{\tsym{\alpha}-complete}{complete homomorphism}{}{homomorphism}{complete}. 
Thus if $\Phi_B$ is \tsym{\alpha}-complete then the corresponding algorithm computes 
the value whenever this exists.
\medskip
In Section~7.3 an \tsym{\alpha}-consistent \tsym{\Lambda}-homomorphism $\Phi_B$
is constructed for each system of equations $\alpha$. 
$\Phi_B$ depends, of course, on the system of equations under consideration.
However, apart from this it only depends on the trace $R_B$
of the bottomed extension $(D_B,f_K)$.
In Chapter~8 the main results of the study are presented. They imply that
in Case~3 the equations $\alpha$ always have a solution and that
$\Phi_B$ is \tsym{\alpha}-complete.
\medskip
Finally, in Section~7.4 replacement rules are introduced
which allow valid assertions to be made about the solutions of equations. 
These are rules which should be thought of as tools which a human being
(rather than a machine)
can apply to prove that the solutions have certain properties.

\vfill\eject

\hfline{171}{7.1 \ EQUATIONS AND THEIR SOLUTIONS}

\sectionhead {7.1} {Equations and their solutions}
\bigskip\medskip
This section explains what is meant by equations and their solutions.
An \tsym{S}-typed set $I$ will be called 
\indexddef{global}{global typed set}{}{typed set}{global}
if $I$ is disjoint from $P$ and if $L$ is disjoint from  $I \cup P$
whenever $\xi \in I$ is of type $L \to \beta$.
For the whole of the section let $I$ be a non-empty finite global \tsym{S}-typed set.

\medskip
In general, what is meant by an equation for the names in $I$ is a pair 
$(s_1,s_2)$, where $s_1,\, s_2 \in F^{I\cup V}_\beta$ for some 
$\beta \in B$ and where $V$ is some \tsym{S}-typed set disjoint from
$I$ and $P$. Here $s_1$ should be thought of 
as the left-hand and $s_2$ as the right-hand side 
of the equation. A solution of this equation is then any assignment 
$c \in \ass I D $ for which

\mdisp{\sem{s_1}_\beta^{c \oplus b}\ =\ \sem{s_2}_\beta^{c \oplus b}}

for all $b \in \ass V D $. Given a system (i.e., a family) of such equations, 
the solutions of the system are those assignments which are solutions
of each of the equations in the family.
\medskip
However, this notion of an equation is much too general. Note that in these
equations there is really no difference in the role played by the left-hand 
and the right-hand side, whereas in the equations in 
Chapter~1 this is not the case, since there the left-hand sides of the 
equations clearly have a very special
form. Thus the elements which can occur as the left-hand side of an equation 
will now be restricted, and this involves what will be called an abstractor. 
\medskip 
Until further notice let $\xi$ be an element of $I$ of type 
$\sigma = L \to \beta$ (and so $L$ is disjoint from $I$ and $P$). 
Define an element 
$a_\xi \in F^{I\cup L}_\beta$, called the 
\indexdef{\tsym{\xi}-abstractor}{abstractor}{}, 
by putting 

\mdisp{a_\xi \ =\ \fo^{I\cup L}_\xi(a_L)\ ,} 
 
where $a_L \in \assb L {F^{I\cup L}} $ is given by
$a_L(\eta) = \eta$ for each $\eta \in L$
(with the $\eta$ on the right-hand side of this definition considered as
an element of $F^{I\cup L}_\eta$). 
Thus 

\mdisp{a_\xi \ =\ \fo^{I\cup L}_{\sigma,L}(\xi \triangleleft a_L )} 

when $L \ne \varnothing$, and $a_\xi = \xi$ if $L = \varnothing$.
Of course, if the `genuine' term algebra  
$(E^{I\cup L}_S,\blob^{I\cup L}_N)$ is being used then $a_\xi$ is simply the list
\mdisp{\xi\ \list {\eta} m }
where $m = |L|$ and $\lvector {\eta} m $ is the
enumeration of the elements of $L$ which comes with the definition of $E^{I\cup L}_S$.

\proclaim{Lemma~7.1.1} If  $L = \varnothing$ then
$\sem{a_\xi}_\beta^c = c(\xi)$ for all $c \in \ass I D $, and if
$L \ne \varnothing$ then
\mdisp{\sem{a_\xi}_\beta^{c \oplus b}\ =\ f_{\sigma,L}(c(\xi)\triangleleft b)}
for all $c \in \ass I D $ and all $b \in \ass L D $.
\endpro
\proof This holds trivially if $L = \varnothing$, and if $L \ne \varnothing$ then
by Proposition~5.4.1~(1)
\smallskip
\ldisp{\quad
\sem{a_\xi}_\beta^{c \oplus b}
\ =\ \sem{\,\fo^{I\cup L}_{\sigma,L}(\xi \triangleleft a_L )\,}_\beta^{c \oplus b}}
\vskip-\medskipamount
\mdisp{
\ =\ f_{\sigma,L}\bigl(
  \assb {\sigma \cdot L} {\sem{\xi \triangleleft a_L}^{c \oplus b}} \,\bigr)
\ =\ f_{\sigma,L}\bigl(\sem{\xi}^{c \oplus b}_\sigma
    \triangleleft \assb L {\sem{a_L}^{c \oplus b}} \,\bigr)
}
\vskip-\medskipamount
\rdisp{
\ =\ f_{\sigma,L}\bigl(c(\xi)
    \triangleleft \assb L {\sem{a_L}^{c \oplus b}} \,\bigr)
\ =\ f_{\sigma,L}(c(\xi) \triangleleft b) \quad}
\smallskip
since
$\,\assb L {\sem{a_L}^{c \oplus b}} \,(\eta) \,=\, \sem{a_L(\eta)}^{c\oplus b}_\eta 
\,=\,  \sem{\eta}^{c\oplus b}_\eta \,=\,  b(\eta)\,$
for each $\eta \in L$, which implies that $\,\assb L {\sem{a_L}^{c \oplus b}} \,=\,b\,$.
\eop
\bigbreak
The set $F^{I \cup L}_\beta$ will be denoted by $\fss{Eq}(\xi,I)$. 
Thus to each element $s \in \fss{Eq}(\xi,I)$ there corresponds the equation $(a_\xi,s)$, 
and with this correspondence $s$ can also be regarded as being an 
\indexdef{equation}{equation}{}
for $\xi$. 
\medskip
If $s \in \fss{Eq}(\xi,I)$ is an equation for $\xi$
then $\fss{Sol}(s)$ will denote the set of solutions, i.e., 
$\fss{Sol}(s)$ is the set of assignments 
$c \in \ass I D $ for which

\mdisp{\sem{a_\xi}_\beta^{c \oplus b}\ =\ \sem{s}_\beta^{c \oplus b}}

for all $b \in \ass L D $.
Because of the special form of the left-hand side $a_\xi$, 
Lemma~7.1.1 implies that if $L \ne \varnothing$ then $\fss{Sol}(s)$ is the set of 
assignments $c \in \ass I D $ for which
\mdisp{f_{\sigma,L}(c(\xi)\triangleleft b) 
                                   \ =\ \sem{s}_\beta^{c \oplus b}} 
for all $b \in \ass L D $, while if $L = \varnothing$ then
then $\fss{Sol}(s)$ is just the set of 
assignments $c \in \ass I D $ for which $\,c(\xi) = \sem{s}^{c}_\beta\,$.

\medskip
Now a system of equations is just a family, made up of one equation 
 for each name in $I$. More precisely, denote by $\fss{Eq}(I)$ 
the set of all families $s_I$, 
where $s_\xi \in \fss{Eq}(\xi,I)$ for each $\xi \in I$. 
Each element of $\fss{Eq}(I)$ will be regarded as a 
\indexddef{system of equations}{system of equations}{}{equations}{system of}
for the names in $I$. Note that the definition of the set $\fss{Eq}(I)$ 
has nothing to do with the \tsym{\Sigma}-algebra $(D_S,f_N)$. It really only depends 
on the the signature $\Sigma$ and the \tsym{S}-typed set $I$.

\medskip
If $\alpha = s_I\in \fss{Eq}(I)$ is a system of equations then an assignment 
$c \in \ass I D $ is said to be a 
\indexdef{solution}{solution}{}
of the equations $\alpha$ if 
$c \in \fss{Sol}(s_\xi)$ for each $\xi \in I$; the set of such 
solutions will be denoted by $\fss{Sol}(\alpha)$. 
An assignment $c \in \ass I D $ 
being a solution thus means that

\mdisp{f_{\sigma,L}(c(\xi) \triangleleft b) 
       \  =\ \sem{s_\xi}_\beta^{\,c \oplus b}}  

must hold for all $b \in \ass L D $ whenever $\xi \in I$ is of type $\sigma = L \to \beta$
with $L \ne \varnothing$ and that $\,c(\xi) = \sem{s_\xi}^{c}_\beta\,$ whenever
$\xi$ is of type $\beta$. (Of course, if $(D_S,f_N)$ is as in one of the three basic cases
then
$\,f_{\sigma,L}(c(\xi) \triangleleft b)\,$ is just equal to
$\,c(\xi)\,(b)\,$, i.e., equal to the value obtained by applying the function
$c(\xi)$ to the argument $b$.)

\medskip
Although the definition of $\fss{Eq}(I)$ has 
nothing to do with $(D_S,f_N)$, the set $\fss{Sol}(\alpha)$ is a subset of 
$\ass I D $, and for this trivial reason it thus depends on $(D_S,f_N)$. 
In fact, the dependence is 
perhaps more subtle than would at first be expected. Examples 7.1.2 and 7.1.3 towards
the end of the section indicate the typical kind of behaviour which occurs.
\medskip
In what follows consider the explicit programming language obtained using the `genuine' 
term algebras. An equation $e \in \fss{Eq}(I)$ for $\xi$ will here be written as

\mdisp{a_\xi \ =\ {\hat e}} 

where $a_\xi$ is the \tsym{\xi}-abstractor (and, as already noted, $a_\xi$ just has 
the form $\xi\ \list {\eta} m $), and ${\hat e}$ is the `legible' notation for 
the term $e$ introduced in Section~6.2 (i.e., with the particular notation for the
case operators and permitting the use of brackets). For a  system of equations this 
notation will be used for each single equation. Moreover, such a system will be 
represented as 
a list, with each single equation written on a separate line. (This involves, of course,
choosing some enumeration of the elements of $I$.) 
\medskip
To prepare for the examples there a further point which has to be dealt with.
In all of these examples each element $\xi \in I$ has a type of the form
$\list {\sigma} m \to \beta$. Now the underlying set of the \tsym{S}-typed
set $\list {\sigma} m $ is just $[m]$, which means that the left-hand side of the equation
for $\xi$ should always have the form 
\mdisp{\xi\,\ 1\ 2\ \cdots\ m}
with the `names' $1\,\,2,\, \ldots,\,m$ also occurring as the 
`local variables' on the right-hand side of the equation. This is not very
satisfactory, in particular since these `names' give no information about their types.
Moreover, the problem then arises of representing the `name' $k \in [m]$ 
in a way which distinguishes it 
from the constructor name ${\underline k}$ of type $\ftt{int}$
(i.e., from the element ${\underline k}$ of $\SynInt$).

\medskip
To avoid these problems it is convenient to replace each $j \in [m]$ in the 
equation for $\xi$ by some more 
suggestive alias $\eta_j$. The left-hand side of the equation then becomes
\mdisp{\xi\,\ \eta_1\ \eta_2\ \cdots\ \eta_m}
with the right-hand side changed accordingly. In any example the choice of the
aliases used in an equation can of course be determined from the modified left-hand side.
\medskip

It should be emphasised that this aliasing device is simply a method of making
explicit examples more readable, and there is no reason compelling its use. 
(The procedure also has a more elegant interpretation, which the interested 
reader is left to work out.) 
\medskip

The system of equations introduced at the end of Chapter~1 is treated in Example~7.1.1. 
These equations are first presented without aliasing, just to show how unreadable 
they are. In this version it is  necessary to distinguish each element of $[m]$ 
from the corresponding element of $\SynInt$, which is done by enclosing the
elements of $[m]$ in single quotes 
(so $1$ will be represented as $\ftt{`1'}$, $2$ as $\ftt{`2'}$, and so on). 
In all the remaining examples in these notes aliasing is employed without comment.  

\vfill\eject
\bigskip\bigskip
\frame{20pt}{\bigskip 
{\it Example~7.1.1\enspace} Let $(X_B,p_K)$ be as Example~2.2.1. As in Example~6.2.1 let
\mdisp{I\  =\  \{\,\ftt{revs},\,\ftt{sq},\,\ftt{shunt},\,
                        \ftt{shunx},\, \ftt{sqs}\,\}}
be the \tsym{S}-typed set with 
\medskip
{\leftskip=35pt
$\ftt{revs}$ of type $(\ftt{int}\to\ftt{int})\ \ftt{int} \to \ftt{list}$,
$\ftt{sq}$ of type $\ftt{int}\to \ftt{int}$,
\smallskip
$\ftt{shunt}$ of type $\ftt{list}\ \ftt{list} \to \ftt{list}$, \smallskip
$\ftt{shunx}$ of type $\ftt{list}\ \ftt{int}\ \ftt{list} \to \ftt{list}$
and $\ftt{sqs}$ of type $\ftt{int} \to \ftt{list}$.
\bigskip
}
Consider that following terms (each being an element of $E^{I\cup \gamma}_\beta$ for
some list of types $\gamma \in S^*$ and some $\beta \in B$):
\medskip\smallskip
{\leftskip=35pt
$e_{\fstt{revs}}\ =\ \ftt{case (eq `2' 0) of \lcb True -> Nil,}$ 
\vskip 2.5pt
$\qquad\qquad\quad
 \ftt{False -> Cons (`1' `2') (revs `1' (sub `2' 1))\rcb}$
\vskip 2.5pt
$\qquad\qquad\qquad\qquad
\in\ E^{I\,\cup \,\fstt{int}}_{\fstt{list}}\,$,
\vskip 5.0pt
$e_{\fstt{sq}}\ =\ \ftt{mul `1' `1'}
       \ \in\ E^{I\,\cup \,\fstt{int}}_{\fstt{int}}\,$,
\vskip 5.0pt
$e_{\fstt{shunt}}\ = \ \ftt{case `1' of \lcb Nil -> `2',}$ 
\vskip 2.5pt
$\qquad\qquad\qquad\qquad\qquad\qquad\quad
\ftt{Cons -> shunx `2'\rcb}
                        \ \in\ E^{I\,\cup \,\fstt{list}\ \fstt{list}}_{\fstt{list}}\,$,
\vskip 5.0pt
$e_{\fstt{shunx}}\ =\ \ftt{shunt `3' (Cons `2' `1')}
       \ \in\ E^{I\,\cup \,\fstt{list}\ \fstt{int}\ \fstt{list}}_{\fstt{list}}\,$,
\vskip 5.0pt
$e_{\fstt{sqs}}\ =\ \ftt{shunt (revs sq `1') Nil}
       \ \in\ E^{I\,\cup \,\fstt{int}}_{\fstt{list}}\,$,
\bigskip
}
Then $\alpha = e_I$ is an element of $\fss{Eq}(I)$ which should
be written as
\bigskip
{\leftskip=35pt
$\ftt{revs `1' `2'}$
\vskip 2.5pt
$\qquad
\ =\ \ftt{ case (eq `2' 0) of \lcb True -> Nil,}$ 
\vskip 2.5pt
$\qquad\qquad\qquad
 \ftt{False -> Cons (`1' `2') (revs `1' (sub `2' 1))\rcb}$
\vskip 5.0pt
$\ftt{sq `1' }\ =\ \ftt{ mul `1' `1'}$
\vskip 5.0pt
$\ftt{shunt `1' `2' }\ = \ \ftt{ case `1' of \lcb Nil -> `2',}$ 
\vskip 2.5pt
$\qquad\qquad\qquad\qquad\qquad\qquad\qquad\qquad\qquad\qquad
\ftt{Cons -> shunx `2'\rcb}$
\vskip 5.0pt
$\ftt{shunx `1' `2' `3' }\ =\ \ftt{ shunt `3' (Cons `2' `1')}$
\vskip 5.0pt
$\ftt{sqs `1' }\ =\ \ftt{ shunt (revs sq `1') Nil}$
\bigskip\smallskip
}
{\it (Example~7.1.1 is continued on the next page.)}
\bigskip\medskip
}

\vfill\eject

\bigskip\bigskip
\frame{20pt}{\bigskip 
{\it Example~7.1.1 (continued)\enspace}
However, with an appropriate choice of aliases, $\alpha$ can be written more 
`legibly' as
\bigskip
{\leftskip=35pt
$\ftt{revs p n }$
\vskip 2.5pt
$\qquad \ =\ \ftt{ case (eq n 0) of \lcb True -> Nil,}$ 
\vskip 2.5pt
$\qquad\qquad\qquad\qquad\qquad
 \ftt{False -> Cons (p n) (revs p (sub n 1))\rcb}$
\vskip 5.0pt
$\ftt{sq n }\ =\ \ftt{ mul n n}$
\vskip 5.0pt
$\ftt{shunt a b }\ = \ \ftt{ case a of \lcb Nil -> b, Cons -> shunx b\rcb}$
\vskip 5.0pt
$\ftt{shunx a n b }\ =\ \ftt{ shunt b (Cons n a)}$
\vskip 5.0pt
$\ftt{sqs n }\ =\ \ftt{ shunt (revs sq n) Nil}$
\bigskip
}
Again let $(D_S,f_N)$ be as in one of the three basic cases.
Exactly the same calculations as in Example~6.2.1 show that an assignment
$c \in \ass I D $ is in
$\fss{Sol}(\alpha)$ if and only if 
for all $a, \,b \in D_{\fstt{list}}$, $n \in D_{\fstt{int}}$,
$p \in D_{\fstt{int}\ \fstt{int}}$
\medskip
\ldisp{\qquad\quad c_{\fstt{revs}}(p,n)
\ =\ \cases{
 \bot_{\fstt{list}} & if $\,n = \bot_{\fstt{int}}\,$,\cr
 f_{\fstt{Nil}}  & if $\,n = 0\,$,\cr
 f_{\fstt{Cons}}(p(n),c_{\fstt{rev}}(p,n-1))
       & if $\,n \in \Int \setminus \{0\}\,$, \cr
}}
\ldisp{\qquad\quad c_{\fstt{sq}}(n)
   \ =\  \fss{mul}(n,n)\, ,}
\ldisp{\qquad\quad c_{\fstt{shunt}}(a,b)
\ =\ \cases{
       \bot_{\fstt{list}} & if $\,a = \bot_{\fstt{list}}\,$,\cr
    b  & if $\,a = f_{\fstt{Nil}}\,$,\cr
    c_{\fstt{shunx}}(b,m,d) 
       & if $\,a \ne \bot_{\fstt{list}}\,$ and $\,a = f_{\fstt{Cons}}(m,d)\,$,\cr
}}
\ldisp{\qquad\quad c_{\fstt{shunx}}(a,n,b)
   \ =\  c_{\fstt{shunt}}(b,
         f_{\fstt{Cons}}(n,a))\,,}
\ldisp{\qquad\quad c_{\fstt{sqs}}(n)
   \ =\  c_{\fstt{shunt}}(c_{\fstt{revs}}(c_{\fstt{sq}},n),f_{\fstt{Nil}})\, .}
\bigskip\medskip
}
\bigskip\bigskip\bigskip

\vfill\eject

As already mentioned, Examples 7.1.2 and 7.1.3 below
show that the dependence of the solutions on $(D_S,f_N)$ is 
perhaps more subtle than would at first be expected. 

\bigskip\bigskip\bigskip
\frame{20pt}{\bigskip 
{\it Example~7.1.2\enspace} Let $(X_B,p_K)$ be as in Example~2.2.1 and let
\mdisp{I\ =\ \{\,\ftt{fst},\,\ftt{fstx},\,\ftt{inf},\,\ftt{one}\,\}}
be the \tsym{S}-typed set with 
\medskip\smallskip
{\leftskip=40pt
$\,\ftt{fst}\,$ of type $\ftt{pair} \to \ftt{int}$,
$\,\ftt{fstx}\,$ of type $\ftt{int}\ \ftt{int} \to \ftt{int}$,
\smallskip
$\,\ftt{inf}\,$  and $\,\ftt{one}\,$ of type $\ftt{int}$.
\bigskip
}
Consider the system of equations $\alpha \in \fss{Eq}(I)$ given by
\medskip\smallskip
{\leftskip=35pt
$\ftt{fst p } = \ftt{ case p of \lcb Pair -> fstx\rcb}$
\vskip 5.0pt
$\ftt{fstx m n } = \ftt{ m}$
\vskip 5.0pt
$\ftt{inf } = \ftt{ add inf 1}$
\vskip 5.0pt
$\ftt{one } = \ftt{ fst (Pair 1 inf)}$
\medskip\smallskip
}
Let $(D_S,f_N)$ be as in one of the three basic cases.
Note that
$c_{\fstt{inf}} = \bot_{\fstt{int}}$ for all $c \in \fss{Sol}(\alpha)$.
From this it follows  that for all $c \in \fss{Sol}(\alpha)$
\mdisp{c_{\fstt{one}}
     \ =\ \cases{1 & if 
       $\,f_{\fstt{Pair}}(1,\bot_{\fstt{int}})\ne \bot_{\fstt{pair}} \,$,\cr
       \bot_{\fstt{int}} & otherwise . \cr
       }}
Therefore 
if $(D_B,f_K)$ is the flat extension of $(X_B,p_K)$ 
then $c_{\fstt{one}} = \bot_{\fstt{int}}$ for all
$c \in \fss{Sol}(\alpha)$.
On the other hand, if $(D_B,f_K)$ is a fully regular
extension of $(X_B,p_K)$ then $c_{\fstt{one}} = 1$ for all
$c \in \fss{Sol}(\alpha)$. 
Note that these 
equations do have a solution (i.e., $\fss{Sol}(\alpha) \ne \varnothing$)
and this fact does not depend on $(D_B,f_K)$.
\bigskip\medskip
}
\vfill\eject

\frame{20pt}{\bigskip 
{\it Example~7.1.3\enspace} Let $(X_B,p_K)$ be as in Example~2.2.1
and let
\mdisp{I\ =\ \{\,\ftt{hd},\,\ftt{hdx},\,\ftt{ones},\,\ftt{one}\,\}}
be the \tsym{S}-typed set with 
\medskip\smallskip
{\leftskip=40pt
$\,\ftt{hd}\,$ of type $\ftt{list} \to \ftt{int}$,
$\,\ftt{hdx}\,$ of type $\ftt{int}\ \ftt{list} \to \ftt{int}$,
\smallskip
$\,\ftt{ones}\,$  of type $\ftt{list}$ and $\,\ftt{one}\,$ of type $\ftt{int}$.
\bigskip
} 
Consider the system of equations $\alpha \in \fss{Eq}(I)$ given by
\medskip\smallskip
{\leftskip=35pt
$\ftt{hd xs } = \ftt{ case xs of \lcb Nil -> hd xs, Cons -> hdx\rcb}$
\vskip 5.0pt
$\ftt{hdx x xs } = \ftt{ x}$
\vskip 5.0pt
$\ftt{ones } = \ftt{ Cons 1 ones}$
\vskip 5.0pt
$\ftt{one } = \ftt{ hd ones}$
\bigskip
}
Let $(D_S,f_N)$ be as in one of the three basic cases. 
Then 
\mdisp{c_{\fstt{ones}}\ =\ f_{\fstt{Cons}}(1,c_{\fstt{ones}})} 
for all $c \in \fss{Sol}(\alpha)$ and from this it follows that
\mdisp{c_{\fstt{one}}
     \ =\ \cases{1 & if 
       $\,c_{\fstt{ones}}\ne \bot_{\fstt{list}} \,$,\cr
       \bot_{\fstt{int}} & otherwise . \cr
       }}
Thus if $(D_B,f_K)$ is the flat extension of $(X_B,p_K)$ 
then $c_{\fstt{one}} = \bot_{\fstt{int}}$ for all
$c \in \fss{Sol}(\alpha)$ (since here $c_{\fstt{ones}} = \bot_{\fstt{list}}$
for each $c \in \fss{Sol}(\alpha)$). Moreover,
if $(D_B,f_K)$ is a fully regular
extension of $(X_B,p_K)$ then $c_{\fstt{one}} = 1$ for all
$c \in \fss{Sol}(\alpha)$
(since here $c_{\fstt{ones}} \ne \bot_{\fstt{list}}$
for each $c \in \fss{Sol}(\alpha)$).
\medskip
Nevertheless, the situation here is quite different to that in Example~7.1.2.
For simplicity suppose that $(D_S,f_N)$ is as in Case~1.
Then the present system of equations has a solution if $(D_B,f_K)$ is the flat 
extension of $(X_B,p_K)$, but whether or not there are solutions for a fully 
regular extension depends on the extension. 
\medskip
In particular, if $(D_B,f_K)$ is an initial extension of $(X_B,p_K)$ then 
it is easy to see that are no solutions to the equations, since the 
equation for $\ftt{ones}$ cannot be satisfied by any finite list.
However, if the fully regular extension is obtained as an initial completion
(as in Section~4.3) then there will be a solution,
which can be considered as an infinite list in which each component is equal to $1$.
\bigskip\medskip
}
\vfill\eject

The equations in Example~7.1.1 are not first order in that they contain
the name for a function $\ftt{revs}$ which has a higher-order type.
Example~7.1.4 provides a further and very typical system of 
equations which is not
first order.

\bigskip\bigskip
\frame{20pt}{\bigskip 
{\it Example~7.1.4\enspace} Let $(X_B,p_K)$ be as in Example~2.2.1 and 
$\,I \, =\, \{\,\ftt{map},\,\ftt{mapx}\,\}\,$
be the \tsym{S}-typed set with
\medskip
{\leftskip=35pt
$\ftt{map}\,$ of type
$\,(\ftt{int} \to \ftt{int}) \ \ftt{list}\to \ftt{list}\,$, 
\smallskip
$\ftt{mapx}\,$ of type
$\,(\ftt{int} \to \ftt{int}) \ \ftt{int} \ \ftt{list}\to \ftt{list}\,$.
\medskip\smallskip
}
Let $\alpha \in \fss{Eq}(I)$ be the system of equations written `legibly' as
\medskip\smallskip
{\leftskip=35pt
$\ftt{map p ns } = \ftt{ case ns of \lcb Nil -> Nil, Cons -> mapx f\rcb}$
\vskip 5.0pt
$\ftt{mapx p m ms } = \ftt{ Cons (p m) (map p ms)}$
\medskip\smallskip
}
As usual let $(D_S,f_N)$ be as in one of the three basic cases, and
let $c \in \ass I D $; then a simple calculation shows that 
$c \in \fss{Sol}(\alpha)$ if and only if 
\medskip\smallskip
\ldisp{\qquad\quad c_{\fstt{map}}(p,z)\  =\  \cases{
   \bot_{\fstt{list}} & if $\,z = \bot_{\fstt{list}}\,$, \cr
\noalign{\smallskip}
  c_{\fstt{mapx}}(p,m',z')
      & if $\,z \ne \bot_{\fstt{list}}\,$
     and $\,z = f_{\fstt{Cons}}(m',z')\,$, \cr
\noalign{\smallskip}
         f_{\fstt{Nil}} & if $\,z = f_{\fstt{Nil}}\,$,\cr
         }}
\ldisp{\qquad\quad c_{\fstt{mapx}}(p,m,z)
\  =\  f_{\fstt{Cons}}(p(m),c_{\fstt{map}}(p,z))\ ,}
\medskip
for all $p \in D_{\fstt{int}\to \fstt{int}}$, $m \in D_{\fstt{int}}$ 
and $z \in D_{\fstt{list}}$ 
In particular, if $c \in \fss{Sol}(\alpha)$ then it follows that
\medskip\smallskip
\ldisp{\qquad\quad c_{\fstt{map}}(p,z)\  =\  \cases{
         \bot_{\fstt{list}} & if $\,z = \bot_{\fstt{list}}\,$,\cr
\noalign{\smallskip}
f_{\fstt{Cons}}(p(m),c_{\fstt{map}}(p,z'))
      & if $\,z \ne \bot_{\fstt{list}}\,$\cr
        &\qquad\quad
     and $\,z = f_{\fstt{Cons}}(m,z')\,$, \cr
\noalign{\smallskip}
   f_{\fstt{Nil}} & if $\,z = f_{\fstt{Nil}}\,$, \cr
         }}
\medskip
for all $p \in D_{\fstt{int}\to \fstt{int}}$, $z \in D_{\fstt{list}}$. 
\bigskip\smallskip
}

\bigskip\bigskip

\vfill\eject

\hfline{179}{7.2 \ COMPUTING VALUES}

\sectionhead {7.2} {Computing values}
\bigskip\medskip
Very roughly speaking, `computing values' means something like the following:
Given a system of equations $\alpha \in \fss{Eq}(I)$ and an element 
$s \in F^I_\sigma$, the value $\sem{s}^c_\sigma \in D_\sigma$ should be
determined for some, or for all, of the solutions $c \in \fss{Sol}(\alpha)$.
Now we take the standpoint that the only way of directly representing an 
element from one of the sets in the family $D_S$ is with the help of the 
ground algebra $(F_B,\fo_K)$.
This means that the only hope is to compute values which lie in the family
$X_B$. (The restriction here to ground types can hardly be avoided, since
there is no way of directly representing an element in the set $D_\sigma$ when 
$\sigma \in S \setminus B$ is a functional type. The restriction to the
family $X_B$, i.e., the exclusion of values in the family 
$\lcurl D_\beta \setminus X_\beta \,:\, \beta \in B \rcurl$, could be 
avoided by making use of a suitable extension of the ground algebra 
$(F_B,\fo_K)$, but to keep 
things as simple as possible this possibility will not be considered here.)
\medskip
A further restriction will be imposed. This is based on the plausible assumption 
that a simple explicit algorithm will be unable to distinguish between 
different solutions in $\fss{Sol}(\alpha)$, and hence only 
values $\sem{s}^c_\beta$ which are independent of 
$c \in \fss{Sol}(\alpha)$ can be computed. (Even if this assumption is not acceptable in 
general, the argument in Chapter~1 shows that it will be certainly 
satisfied by an algorithm based on interpreting equations as replacement 
rules.)
\medskip
In what follows let $I$ be a non-empty finite global \tsym{S}-typed set 
and let $\alpha \in \fss{Eq}(I)$ be a system of equations with 
$\fss{Sol}(\alpha) \ne \varnothing$. The above restrictions suggest 
looking at the family $F^\alpha_B$ with $F^\alpha_B \subset F^I_B$ 
defined for each $\beta \in B$ by

\mdisp{F^\alpha_\beta\ =\ \lcurl s \in F^I_\beta
    \,: \frm{there exists} x \in X_\beta \frm{such that}
   \sem{s}^c_\beta  = x \frm{for all} 
                          c \in \fss{Sol}(\alpha) \rcurl\ .}

In particular, $F_B \subset F^\alpha_B$, since if $s \in F_\beta$ then by Lemma~6.2.1 
$\sem{s}^c_\beta  = \sem{s}_\beta$ for all $c \in \ass I D $ 
(and so in particular for all $c \in \fss{Sol}(\alpha)$). 
Example~7.2.1 on the next page illustrates this definition using the  
equations first considered in Chapter~1.

\medskip
The problem of `computing values' can now be formulated as the following 
question: For which elements $s \in F^\alpha_\beta$ is it possible to compute
the value of $s$, i.e., the unique element $x \in X_\beta$ such that 
$\sem{s}^c_\beta = x$ for all $c \in \fss{Sol}(\alpha)$?
\medskip
It is important to note that, although 
$F^\alpha_B \subset F^I_B$ and $(D_S,f_N)$ is not involved in 
the definition of $F^I_B$, the family $F^\alpha_B$ will in general
depend on this \tsym{\Sigma}-algebra. Example~7.2.1 shows in 
particular how $F^\alpha_B$ can depend on the
\tsym{\Lambda}-algebra $(D_B,f_K)$.
\midinsert
\bigskip\medskip
\frame{20pt}{\bigskip 
{\it Example~7.2.1\enspace} 
Let $(X_B,p_K)$ be as in Example~2.2.1 and let
\mdisp{I\ =\ \{\,\ftt{fst},\,\ftt{fstx},\,\ftt{und}\,\}}
be the \tsym{S}-typed set with 
\medskip\smallskip
{\leftskip=40pt
$\,\ftt{fst}\,$ of type $\ftt{pair} \to \ftt{int}$,
$\,\ftt{fstx}\,$ of type $\ftt{int}\ \ftt{int} \to \ftt{int}$,
\smallskip
$\,\ftt{und}\,$ of type $\ftt{int}$.
\bigskip
}
Consider the system of equations $\alpha \in \fss{Eq}(I)$ given by
\medskip\smallskip
{\leftskip=35pt
$\ftt{fst p } = \ftt{ case p of \lcb Pair -> fstx\rcb}$
\vskip 5.0pt
$\ftt{fstx m n } = \ftt{ m}$
\vskip 5.0pt
$\ftt{und } = \ftt{ und}$
\medskip\smallskip
}
Let $e \in E^I_{\fstt{int}}$ be the term 
$\,\ftt{fst (Pair 1 und)}\,$. Then
\mdisp{\sem{e}^c_{\fstt{int}}
     \ =\ \cases{1 & if 
       $\,f_{\fstt{Pair}}(1,c_{\fstt{und}})\ne \bot_{\fstt{pair}} \,$,\cr
       \bot_{\fstt{int}} & otherwise , \cr
       }}
for all $c \in \fss{Sol}(\alpha)$. 
Thus if $(D_B,f_K)$ is a fully regular
extension of $(X_B,p_K)$ then $\sem{e}^c_{\fstt{int}} = 1$ for all
$c \in \fss{Sol}(\alpha)$, and hence in this case 
$e \in E^\alpha_{\fstt{int}}$. 
\medskip
On the other hand, if $(D_S,f_N)$ is as in Case~1 with
$(D_B,f_K)$ the flat extension of $(X_B,p_K)$ then there exists
a solution $c \in \fss{Sol}(\alpha)$ with 
$c_{\fstt{und}} = \bot_{\fstt{int}}$
and then $\sem{e}^c_{\fstt{int}} = \bot_{\fstt{int}}$. This implies that
here $e \notin E^\alpha_{\fstt{int}}$. 
\medskip
Of course, this example is in essence the same as Example~7.1.2.
\bigskip\medskip
}
\bigskip
\endinsert
\medskip
In what follows the \tsym{\Lambda}-algebra $(F^I_B,\fo^I_K)$ will play a
more important role than the whole of the \tsym{\Sigma}-algebra $(F^I_S,\fo^I_N)$.
This means that if $c \in \ass I D $ is an assignment then it is 
usually only the \tsym{\Lambda}-homomorphism $\sem{\cdot}^c_B$ from 
$(F^I_B,\fo^I_K)$ to $(D_B,f_K)$ which will be of interest
(and not the whole \tsym{\Sigma}-homomorphism $\sem{\cdot}^c_S$).

\medskip
It is convenient to state the problem of `computing values' in a
slightly different form, and for this the following simple fact is needed:

\proclaim{Lemma~7.2.1} The family $F^\alpha_B$ is invariant
in the \tsym{\Lambda}-algebra $(F^I_B,\fo^I_K)$. \endpro

\proof If $\kappa \in K$ is of type $\varnothing \to \beta$ then
$\fo^I_\kappa(\onept) \in F^\alpha_\beta$, since
$\sem{\,\fo^I_\kappa(\onept)\,}^c_\beta = f_\kappa(\onept)$ 
for all $c \in \ass I D $ and $f_\kappa(\onept) = p_\kappa(\onept) \in X_\beta$.
Suppose then that $\kappa \in K$ is of type $L \to \beta$ with 
$L \ne \varnothing$ and let $b \in \assb L {F^\alpha} $;
then for each $\eta \in L$ there exists $x_\eta \in X_\eta$ such that
$\sem{b(\eta)}^c_\eta = x_\eta$ for all $c \in \fss{Sol}(\alpha)$. Let
$b' \in \ass L X $ be the assignment given by $b'(\eta) = x_\eta$ for each $\eta \in L$;
then for all $c \in \fss{Sol}(\alpha)$

\mdisp{ \sem{\,\fo^I_\kappa(b)\,}^c_\beta
 \ =\ f_\kappa\bigl(\,\assb L {\sem{b}^c} \,\bigr)\ =\ f_\kappa(b')\ ,}

and $f_\kappa(b') = p_\kappa(b') \in X_\beta$;
thus $\fo^I_\kappa(b) \in F^\alpha_\beta$.
This means that the family $F^\alpha_B$ is invariant in $(F^I_B,\fo^I_K)$. \eop
\bigbreak
For each $\kappa \in K$ of type $L \to \beta$ let
$\fo^\alpha_\kappa$ denote the restriction of $\fo^I_\kappa$ to
$\assb L {F^\alpha} $. By Lemma~7.2.1 $(F^\alpha_B,\fo^\alpha_K)$
is then a \tsym{\Lambda}-algebra 
which is an extension of the ground term algebra $(F_B,\fo_K)$.
\medskip
Now for each $\beta \in B$ define a mapping 
$\sem{\cdot}^\alpha_\beta : F^\alpha_\beta \to X_\beta$
by putting $\sem{s}^\alpha_\beta = x$, where $x$ is the (unique)
element of $X_\beta$ such that $\sem{s}^c_\beta = x$ for all 
$c \in \fss{Sol}(\alpha)$. In other words, $\sem{s}^\alpha_\beta$ is
defined so that $\sem{s}^\alpha_\beta = \sem{s}^c_\beta$ for all
$c \in \fss{Sol}(\alpha)$.

\proclaim{Lemma~7.2.2} The family
$\sem{\cdot}^\alpha_B$ is a \tsym{\Lambda}-homomorphism from 
$(F^\alpha_B,\fo^\alpha_K)$ to $(X_B,p_K)$ which extends the
\tsym{\Lambda}-isomorphism 
$\sem{\cdot}_B : (F_B,\fo_K) \to (X_B,p_K)$
(i.e., $\sem{s}^\alpha_\beta = \sem{s}_\beta$ for all 
$s \in F_\beta$).
\endpro

\proof This is what the proof of Lemma~7.2.1 shows. \eop
\bigbreak

The problem of `computing values' can now be stated in terms of the following
question: For which elements $s \in F^\alpha_\beta$ is it possible to compute 
the unique ground element $s' \in F_\beta$ such that 
$\sem{s'}_\beta = \sem{s}^\alpha_\beta$? 
\medskip
To answer this question some kind of algorithm is needed which will
manipulate the elements from the sets in the family $F^I_B$: In response to the `input' 
$s$ it should either produce the `output' $s'$ (whenever possible) or produce 
no output at all. 
Such an algorithm is presented in Section~7.3, and is given in terms of 
a \tsym{\Lambda}-homomorphism from the \tsym{\Lambda}-algebra 
$(F^I_B,\fo^I_K)$ to itself. 
\medskip
In what follows let $\Phi_B : (F^I_B,\fo^I_K) \to (F^I_B,\fo^I_K)$ be a
\tsym{\Lambda}-homomorphism. 
Some of the elementary properties of such a homomorphism will now be looked at
with a view to seeing how $\Phi_B$ could form
the basis of an algorithm for `computing values'.

\proclaim{Lemma~7.2.3} $\Phi_\beta(s) = s$ for each ground term 
$s \in F_\beta$. \endpro

\proof For each $\beta \in B$ let $\Phi'_\beta$ be the restriction of
$\Phi_\beta$ to the ground terms $F_\beta$; then 
$\Phi'_B : (F_B,\fo_K) \to (F^I_B,\fo^I_K)$ is a 
\tsym{\Lambda}-homomorphism. But 
a \tsym{\Lambda}-homomorphism 
$\fss{id}'_B : (F_B,\fo_K) \to (F^I_B,\fo^I_K)$ can also 
be defined by letting $\fss{id}'_\beta(s) = s$ for all
$s \in F_\beta$. Hence by Proposition~2.2.3~(1) 
$\Phi'_B = \fss{id}'_B$, since 
$(F_B,\fo_K)$ is minimal, i.e., $\Phi_\beta(s) = s$ for each 
$s \in F_\beta$. \eop
\bigbreak
The iterates $\{\Phi^n_B\}_{n \ge 0}$ of $\Phi_B$ will also be needed: For each
$\beta \in B$ the mappings $\Phi^n_\beta : F^I_\beta \to F^I_\beta$, 
$n \ge 0$, are defined by letting $\Phi^0_\beta(s) = s$ for each
$s \in F^I_\beta$, $\Phi^1_\beta = \Phi_\beta$ and (for $n > 0$)
$\Phi^n_\beta = \Phi_\beta \, \Phi^{n-1}_\beta$. By Proposition~2.2.1
$\Phi^n_B$ is a \tsym{\Lambda}-homomorphism for each $n \ge 0$.
Now for each $\beta \in B$ let
\smallskip
\mdisp{F^\Phi_\beta \ =\ \lcurl s \in F^I_\beta \,:\,
        \Phi^n_\beta(s) \in F_\beta \frm{for some} n \ge 0 \rcurl\ .}

\proclaim{Lemma~7.2.4} The family $F^\Phi_B$ is invariant
in the \tsym{\Lambda}-algebra $(F^I_B,\fo^I_K)$. \endpro

\proof If $\kappa \in K$ is of type $\varnothing \to \beta$ then
$\fo^I_\kappa(\onept) \in F_\beta \subset F^\Phi_\beta$.
Suppose then that $\kappa \in K$ is of type $L \to \beta$ with 
$L \ne \varnothing$ and let $b \in \assb L {F^\Phi} $;
then for each $\eta \in L$ there exists $n_\eta \ge 0$ such that
$\Phi^{n_\eta}_\eta(b(\eta)) \in F_\eta$. Put $n = \max \lcurl n_\eta : \eta \in L\rcurl$;
then $\Phi^n_\eta(b(\eta)) \in F_\eta$ for each $\eta \in L$, thus
$\assb L {\Phi^b} (b)  \in \ass L F $ and hence

\mdisp{ \Phi_\beta^n(\fo^I_\kappa(b))
 \ =\ \fo^I_\kappa\bigl(\,\assb L {\Phi^b} (b) \,\bigr)
 \ =\ \fo_\kappa\bigl(\,\assb L {\Phi^b} (b) \,\bigr)
\ \in\ F_\beta\ ;}

i.e., $\fo^I_\kappa(b) \in F^\Phi_\beta$.
This means that the family $F^\Phi_B$ is invariant in $(F^I_B,\fo^I_K)$. \eop
\bigbreak

For each $\kappa \in K$ of type $L \to \beta$ let $\fo^\Phi_\kappa$ denote the 
restriction of $\fo^I_\kappa$ to $\assb L {F^\Phi} $. By
Lemma~7.2.4 $(F^\Phi_B,\fo^\Phi_K)$ is then a \tsym{\Lambda}-algebra 
which is an extension of the ground term algebra $(F_B,\fo_K)$.
\medskip
By Lemma~7.2.3 $\Phi_\beta(s) = s$ for each $s \in F_\beta$ and
hence a mapping $\Phi^\infty_\beta : F^\Phi_\beta \to F_\beta$ can 
be defined by putting $\Phi^\infty_\beta(s) = \Phi^n_\beta(s)$, where 
$n \ge 0$ is chosen so that $\Phi^n_\beta(s) \in F_\beta$.

\proclaim{Lemma~7.2.5} 
$\Phi^\infty_B : (F^\Phi_B,\fo^\Phi_K) \to (F_B,\fo_K)$
is a \tsym{\Lambda}-homomorphism. \endpro

\proof This is similar to the proof of Lemma~7.2.4. \eop
\bigbreak
The homomorphism  $\Phi_B$ is now said to be 
\indexddef{\tsym{\alpha}-consistent}{consistent homomorphism}{}{homomorphism}{consistent}
if 

\mdisp{\sem{\,\Phi_\beta(s)\,}^c_\beta \ =\ \sem{s}^c_\beta}

for all $c \in \fss{Sol}(\alpha)$ for each $s \in F^I_\beta$, $\beta \in B$.
It is important to note that in general the \tsym{\alpha}-consistency of $\Phi_B$ depends 
not only on the equations $\alpha$ but also on the \tsym{\Sigma}-algebra $(D_S,f_N)$.

\proclaim{Proposition~7.2.1} Suppose $\Phi_B$ is \tsym{\alpha}-consistent; 
then $F^\Phi_B \subset F^\alpha_B$.
Moreover, $\Phi^\infty_\beta(s)$ is the (unique) ground term which denotes
$\sem{s}^\alpha_\beta$ for all $s \in F^\Phi_\beta$, $\beta \in B$, i.e.,
\mdisp{\sem{\,\Phi^\infty_\beta(s)\,}_\beta
\ =\ \sem{s}^\alpha_\beta
}
for all $s \in F^\Phi_\beta$, $\beta \in B$. \endpro

\proof Let $s \in F^\Phi_\beta$; then for each $n > 0$ 
it follows that

\mdisp{\sem{\,\Phi^n_\beta(s)\,}^c_\beta 
\ =\ \sem{\,\Phi_\beta(\Phi^{n-1}_\beta(s))\,}^c_\beta 
\ =\ \sem{\,\Phi^{n-1}_\beta(s)\,}^c_\beta}

and thus 
$\sem{\,\Phi^\infty_\beta(s)\,}^c_\beta 
\,=\,\sem{\,\Phi^0_\beta(s)\,}^c_\beta 
\,=\,\sem{s}^c_\beta$
for each $c \in \fss{Sol}(\alpha)$. But 
$\Phi^\infty_\beta(s) \in F_\beta$ and so by Lemma~6.2.1
$\sem{\,\Phi^\infty_\beta(s)\,}^{c'}_\beta 
\,=\,\sem{\,\Phi^\infty_\beta(s)\,}_\beta$
for all $c' \in \ass I D $. Hence

\mdisp{
\sem{\,\Phi^\infty_\beta(s)\,}_\beta
\ =\ \sem{s}^c_\beta}

for all $c \in \fss{Sol}(\alpha)$. This shows that
$s \in F^\alpha_\beta$ and also that
$\sem{\,\Phi^\infty_\beta(s)\,}_\beta = \sem{s}^\alpha_\beta$. \eop
\bigbreak
If $\Phi_B$ is \tsym{\alpha}-consistent then by Proposition~7.2.1 the iterates 
$\{\Phi^n_\beta\}_{n \ge 0}$ of $\Phi_\beta$ can be used to compute the 
element denoting $\sem{s}^\alpha_\beta$ for each $s \in F^\Phi_\beta$. 
Of course, the iterates give no information when starting with an element 
$s \in F^\alpha_\beta \setminus F^\Phi_\beta$. 
\medskip
Consider for a moment the identity \tsym{\Lambda}-homomorphism 
$\fss{id}_B : (F^I_B,\fo^I_K) \to (F^I_B,\fo^I_K)$
(given by $\fss{id}_\beta(s) = s$ for all $s \in F^I_\beta$). In this case
$(F^{\fsss{id}}_B,\fo^{\fsss{id}}_K) = (F_B,\fo_K)$, and 
$\fss{id}^\infty_B$ is just the identity \tsym{\Lambda}-homomorphism from 
$(F^{\fsss{id}}_B,\fo^{\fsss{id}}_K) = (F_B,\fo_K)$ to 
itself.
Moreover, $\fss{id}_B$ is always trivially \tsym{\alpha}-consistent.
But, of course, applying Proposition~7.2.1 to
$\fss{id}_B$ does not really provide any information at all.
\medskip
The best to be hoped for is
an \tsym{\alpha}-consistent \tsym{\Lambda}-homomorphism $\Phi_B$ 
with  $F^\Phi_B = F^\alpha_B$ and such a homomorphism $\Phi_B$ will be called 
\indexddef{\tsym{\alpha}-complete}{complete homomorphism}{}{homomorphism}{complete}.
If $\Phi_B$ is \tsym{\alpha}-complete then Proposition~7.2.1 implies that 
$\Phi^\infty_\beta(s)$ is the ground element denoting
$\sem{s}^\alpha_\beta$ for all $s \in F^\alpha_\beta$. 
(Note, however, that the iterates $\{\Phi^n_\beta\}_{n \ge 0}$ cannot be used 
to effectively decide that a given term $s \in F^I_\beta$ is an 
element of $F^I_\beta \setminus F^\alpha_\beta$.)
\medskip
In Section~7.3 an \tsym{\alpha}-consistent \tsym{\Lambda}-homomorphism $\Phi_B$
is constructed for each system of equations 
$\alpha \in \fss{Eq}(I)$. In Chapter~8 it is then shown that in Case~3  
$\Phi_B$ is \tsym{\alpha}-complete.
\medskip
Finally, note that the definition of $\Phi_B$ being
\tsym{\alpha}-consistent still makes sense when 
$\fss{Sol}(\alpha) = \varnothing$ (although in this case every
\tsym{\Lambda}-homomorphism is trivially \tsym{\alpha}-consistent).
Thus in the next section, where an
\tsym{\alpha}-consistent \tsym{\Lambda}-homomorphism
$\Phi_B$ is constructed, it is not necessary to assume that  
$\fss{Sol}(\alpha) \ne \varnothing$.

\vfill\eject

\hfline{184}{7.3 \ AN ALGORITHM FOR COMPUTING VALUES}

\sectionhead {7.3} {An algorithm for computing values}
\bigskip\medskip
Again let $I$ be a non-empty finite global \tsym{S}-typed set and let  
$\alpha = s_I \in \fss{Eq}(I)$ be a system of equations. In this section  
an \tsym{\alpha}-consistent \tsym{\Lambda}-homomorphism
\mdisp{\Phi_B \,:\, (F^I_B,\fo^I_K)\, \to \,(F^I_B,\fo^I_K)}
will be defined which, as explained in Section~7.2, will form the basis of an 
algorithm for computing values. 
\medskip
The \tsym{\Lambda}-homomorphism $\Phi_B$
will depend, of course, on the system of equations $\alpha$. 
Moreover, $\Phi_B$ will also depend on $(D_S,f_N)$,
because its definition involves
choosing a support system for $(D_S,f_N)$, i.e., 
a family $N^I_B$ with $F_B \subset N^I_B \subset \kterm{F}^I_B$
such that $\sem{s}^c_\beta \ne \bot_\beta$ for all 
$s \in N^I_\beta$, $\beta \in B$ and for all $c \in \ass I D $.
\medskip
Recall that by Proposition~6.3.1 the family $\bterm{F}^I_B$ defined for each $\beta \in B$
by
\mdisp{ \bterm{F}^I_\beta\ =\ \lcurl s \in F^I_\beta \,:\,
           \Omega^I_\beta(s) \in R_\beta \rcurl}

is a support system for $(D_S,f_N)$,
where here
$\Omega^I_S : (F^I_S,\fo^I_N) \to (F^\flat_S,\fo^\flat_N)$ 
is the unique \tsym{\Sigma}-homomorphism 
such that $\,\Omega^I_\xi(\xi) \,=\, \flat_\xi\,$ for each $\xi \in I\cup P$,
and $R_B$ is the trace of $(D_B,f_K)$.
In fact, this is really the only support system of interest
(since it turns out to be the `correct' one). 
Nevertheless, in the present section the results are first formulated
in terms of an arbitrary support system.
\medskip
Before coming to the definition of the \tsym{\Lambda}-homomorphism $\Phi_B$ we must
first introduce some generalised substitution or replacement operators as well as
evaluation operators for the integer operations.
\medskip
Let $L$ be an \tsym{S}-typed set disjoint from $I$ and $P$ and 
$a \in \assb L {F^I} $ be an assignment.
Then, since $(F^{I\cup L}_S,\fo^{I\cup L}_N)$ is functionally \tsym{I\cup L\cup P}-free and
$(F^I_S,\fo^I_N)$ is a functional \tsym{\Sigma}-algebra,
there exists a unique \tsym{\Sigma}-homomorphism 
$\delta^a_S : (F^{I\cup L}_S,\fo^{I\cup L}_N) \to (F^I_S,\fo^I_N)$ 
such that $\delta^a_\xi(\xi) = \xi$ for each $\xi \in I\cup P$ and
$\delta^a_\eta(\eta) = a(\eta)$ for each $\eta \in L$. 
In particular, if $L = \varnothing$
then $\delta^\varnothing_S$ is just the identity homomorphism
$\fss{id}_S : (F^I_S,\fo^I_N) \to (F^I_S,\fo^I_N)$. 
\medskip
For each $s \in F^{I\cup L}_\sigma$ the (perhaps) more suggestive notation
\mdisp{s[a/L]}
will also be used for the element $\delta^a_\sigma(s)$ of $F^I_\sigma$.
Moreover, when working with examples it is convenient
to allow the more explicit notation

\mdisp{s[\,a(\eta_1)/\eta_1,\,\ldots,a(\eta_m)/\eta_m\,]}

where $\lvector {\eta} m $ is some enumeration of the elements of $L$. The mapping
$\delta^a_\sigma$ should be thought of as a generalised substitution or replacement 
operator. In the special case of
the `genuine' term algebras introduced in Section~6.2 it really is a replacement
operator:

\proclaim{Lemma~7.3.1} Let $a \in \assb L {E^I} $ and let
$\delta^a_S : (E^{I\cup L}_S,\blob^{I\cup L}_N) \to (E^I_S,\blob^I_N)$ 
be the special case of the homomorphism defined above.
Then for each $s \in E^{I\cup L}_\sigma$ the term $\delta^a_\sigma(s)$
is obtained by replacing for each $\eta \in L$ each occurrence of $\eta$ in $s$ by
$a(\eta)$. \endpro
\proof For each $\sigma \in S$ let ${\breve E}^I_\sigma$ be the set of terms in
$E^I_\sigma$ for which statement to be proved actually holds.
Then it is easily checked that the family ${\breve E}^I_S$ satisfies the 
hypotheses of Proposition~6.1.5 and hence ${\breve E}^I_S = E^I_S$. \eop
\bigbreak
The following fact will be required later:

\proclaim{Lemma~7.3.2} Let
$a \in \assb L {F^I} $, $c \in \ass I D $ and 
put $\,d = \assb L {\sem{a}^c} \,$, i.e., 
$d \in \ass L D $ is the assignment given by 
$\,d(\eta) = \sem{a(\eta)}^c_\eta\,$ for each $\eta \in L$. Then
\mdisp{\sem{\,s[a/L]\,}^c_\sigma \ =\ \sem{s}_\sigma^{\,c\oplus d}}
for all $s \in F^{I\cup L}_\sigma$, $\sigma \in S$. \endpro

\proof Put $\pi_S = \sem{\cdot}^c_S\,\delta^a_S$. Then by Proposition~2.2.1
$\pi_S$ is a 
\tsym{\Sigma}-homomorphism from $(F^{I\cup L}_S,\fo^{I\cup L}_N)$ to $(D_S,f_N)$ and
$\,\pi_\xi(\xi) \,=\, \sem{\,\delta_\xi^a(\xi)\,}^c_\xi 
                            \,=\, \sem{\xi}^c_\xi \,=\, c(\xi)\,$
for each $\xi \in I$, and in the same way it follows that
$\,\pi_\zeta(\zeta) \,=\, p(\zeta)$ for each $\zeta \in P$.
Moreover,
$\,\pi_\eta(\eta) \,=\, \sem{\,\delta_\eta^a(\eta)\,}^c_\eta\,=\,
\sem{a(\eta)}^c_\eta\,=\,d(\eta)\,$
for each $\eta \in L$.
But $\sem{\cdot}^{\,c \oplus d}_S$ is the unique such
\tsym{\Sigma}-homomorphism, and hence $\pi_S = \sem{\cdot}^{\,c\oplus d}_S$, i.e.,

\mdisp{\sem{\,s[a/L]\,}^c_\sigma 
        \ =\ \sem{\,\delta^a_\sigma(s)\,}^c_\sigma 
                                                 \ =\ \sem{s}^{\,c \oplus d}_\sigma}

for all $s \in F^{I\cup L}_\sigma$, $\sigma \in S$. \eop
\bigbreak
We next consider evaluation operators.
Let $\zeta \in P$ be the name of an integer operator (and so $\zeta$ is of type 
$\ftt{int}_2 \to \beta$ with $\beta$ either $\ftt{int}$ or $\ftt{bool}$, recalling that
$\ftt{int}_2$ is being used to denote the list $\,\ftt{int}\ \ftt{int}\,$).

\proclaim{Lemma~7.3.3} There exists a unique mapping
$\fss{Eval}_\zeta : F_{\fstt{int}} \times F_{\fstt{int}} \to F_\beta$
such that
\mdisp{\sem{\,\fss{Eval}_\zeta(s_1,s_2)\,}_\beta
  \ =\ \fss{o}_\zeta\,(\sem{s_1}_{\fstt{int}},\sem{s_2}_{\fstt{int}})}
for all $s_1,\, s_2 \in F_{\fstt{int}}$, where 
$\fss{o}_\zeta : X_{\fstt{int}}\times X_{\fstt{int}} \to X_{\beta}$ is the 
original integer operation corresponding to $\zeta$ (i.e., 
$\fss{o}_{\fstt{add}}(m,n) = m+n$,
$\fss{o}_{\fstt{sub}}(m,n) = m-n$ and so on).
\endpro

\proof This follows from the fact that both the mappings 
$\sem{\cdot}_{\fstt{int}}: F_{\fstt{int}} \to X_{\fstt{int}}$
and $\sem{\cdot}_{\fstt{bool}}: F_{\fstt{bool}} \to X_{\fstt{bool}}$
are bijections. \eop
\bigbreak
The mapping 
$\fss{Eval}_\zeta : E_{\fstt{int}} \times E_{\fstt{int}} \to E_\beta$
in the special case of the `genuine' term algebras 
is just the `syntactic' version of the operator
whose name is $\zeta$. This means, for example, that 
$\fss{Eval}_{\fstt{add}}(\ftt{10},\ftt{4}) \,=\, \ftt{14}$ and
$\fss{Eval}_{\fstt{eq}}(\ftt{10},\ftt{4}) \,=\, \ftt{False}$.

\proclaim{Lemma~7.3.4} For all $s_1,\, s_2 \in F_{\fstt{int}}$
\mdisp{\sem{\,\fss{Eval}_\zeta(s_1,s_2)\,}_\beta
  \ =\ f_{\fstt{int}_2\to\beta,\fstt{int}_2}
 \bigl(\,p(\zeta)\triangleleft (\sem{s_1}_{\fstt{int}},\sem{s_2}_{\fstt{int}})\,\bigr)
\ .} \endpro

\proof Let ${\hat p}(\zeta) : D_{\fstt{int}} \times D_{\fstt{int}} \to D_\beta$
be the strict extension of 
$\fss{o}_\zeta : X_{\fstt{int}} \times X_{\fstt{int}} \to X_\beta$. Then by definition 
${\hat p}(\ftt{add}) = \fss{Add}$, ${\hat p}(\ftt{sub}) = \fss{Sub}$ and so on,
and therefore by (D)
\mdisp{\fss{o}_\zeta(n_1,n_2)
\ =\ {\hat p}(\zeta)\,(n_1,n_2)
\ =\ f_{\fstt{int}_2 \to \beta,\fstt{int}_2}
\bigl(\,p(\zeta) \triangleleft (n_1,n_2)\,\bigr)
}
for all $n_1,\,n_2 \in X_{\fstt{int}}$. Taking
$n_1 = \sem{s_1}_{\fstt{int}}$ and $n_2 = \sem{s_2}_{\fstt{int}}$ gives
the result. \eop

\bigbreak

The following result is stated in terms of a concept weaker than that of a support system:
The family $N^I_B$ is said to be a 
\indexdef{pre-support system}{pre-support system}{}
if just $F_B \subset N^I_B \subset \kterm{F}^I_B$. 
(This notion has nothing to do with $(D_S,f_N)$.)
\medskip
Note that if $N^I_B$ is a support system and ${\breve N}^I_B$ a pre-support system with 
${\breve N}^I_B \subset N^I_B$ then ${\breve N}^I_B$ is also a support system.

\proclaim{Proposition~7.3.1} Let $N^I_B$ be a pre-support system.
Then there exists a unique 
\tsym{\Lambda}-homomorphism $\Phi_B : (F^I_B,\fo^I_K) \to (F^I_B,\fo^I_K)$
such that the following hold:
\bigskip
{\parindent=25pt
\item{(1)} Let $\xi \in I$ be of type $L \to \beta$; then
for all $a \in \assb L {F^I} $
\smallskip
\mdisp{\Phi_\beta(\,\fo^I_{\xi}(a) \,)\ =\ s_\xi[a/L]\ .}
\smallskip
\item{} 
In particular,  
$\,\Phi_\beta(\xi) \,=\, s_\xi\,$
if $L =\varnothing$ (in which case $\xi$ is of type $\beta$).
\medskip
\item{(2)} Let $\zeta = \ftt{case}^\theta_\beta$ for some
$\beta \in B$, $\theta \in \Bf$; then 
\smallskip
\mdisp{\Phi_\beta( \fo^I_\zeta(s \diamond a ))
  \ =\  \cases{
 \fo^I_\zeta(\Phi_\theta(s) \diamond a )
         & if $\,s \notin N^I_\theta\,$, \cr
\noalign{\smallskip}
    \fo^I_{J\to \beta,J}(a(\kappa) \triangleleft a')
 &    if $\,s = \fo^I_\kappa(a') \in N^I_\theta \,$\cr
      &\quad
     with $\,\kappa \in K\,$ of type $\,J \to \theta\,$,\cr 
}}
\medskip
\item{} 
for all $s \in F^I_\theta$, $a \in \assb {K_{\theta,\beta}} {F^I} $. (Note that, 
since $N^I_\theta \subset \kterm{F}^I_\theta$, $s$ does have a unique representation
of the form $\fo^I_\kappa(a')$ for some $\kappa \in K_\theta$ whenever $s \in N^I_\theta$.)
\medskip
\item{(3)} Let $\zeta \in P$ be the name of an integer operator
(so $\zeta$ is of type $\ftt{int}_2 \to \beta$ with
$\beta$ either $\ftt{int}$ or $\ftt{bool}$) and let 
$s_1,\,s_2 \in F^I_{\fstt{int}}$. Then
\smallskip
\mdisp{\Phi_\beta(\,\fo^I_\zeta(s_1,s_2)\,)
  \ =\ \cases{\fss{Eval}_\zeta(s_1,s_2) &
           if $\,s_1,\, s_2 \in F_{\fstt{int}}\,$, \cr
 \noalign{\vskip 2pt}
         \fo^I_\zeta(\Phi_{\fstt{int}}(s_1),
                                         \Phi_{\fstt{int}}(s_2))
            & otherwise.  }}
\bigskip
}  \endpro

\proof Let $\ell_B$ be a family of mappings satisfying the conditions in the statement
of Lemma~6.2.3.
$\Phi_\beta(s)$ will be defined by induction on 
$\ell_\beta(s)$, making use of the unique representations given in
Proposition~6.2.1~(1). Suppose first that $s \in F^I_\beta$ with 
$\ell_\beta(s) = 1$. Then either $s = \fo^I_\kappa(\onept)$ for some 
$\kappa \in K$ of type $\varnothing \to \beta$, in which case  
$\Phi_\beta(s)$ is defined to be  $s$, or $s = \xi$ for some $\xi \in I$ of 
type $\beta$ and here $\Phi_\beta(s)$ is defined to be  $s_\xi$.
\medskip
Now let $k > 1$ and suppose for each $\theta \in B$ that
$\Phi_{\theta}(t)$ has already been defined for all 
$t \in F^I_{\theta}$ with $\ell_{\theta}(t) < k$.
Consider $s \in F^I_\beta$ with $\ell_\beta(s) = k$; there are four cases:
\medskip
$1.$ The element $s$ has the form $\fo^I_\kappa(a)$, where
$\kappa \in K$ is of type $L \to \beta$ with $L \ne \varnothing$ and
$a \in \assb L {F^I} $.
Then, since $\ell_\eta (a(\eta)) < k$ for each $\eta \in L$, $\Phi_\beta(s)$ 
can be defined to be $\fo^I_\kappa(\ass L {\Phi} (a) )$.
\medskip
$2.$ The element $s$ has the form $\fo^I_\xi(a)$ with 
$\xi \in I$ of type $L \to \beta$ with $L \ne \varnothing$ and
$a \in \assb L {F^I} $. Here 
$\Phi_\beta(\,\fo^I_\xi(a)\,)$ can be defined directly using condition (1).
\medskip
$3.$ The element $s$ has the form $\fo^I_\zeta(s' \diamond a)$, where 
$\zeta = \ftt{case}^\theta_\beta$ with $\beta \in B$ and $\theta \in \Bf$ and 
where $s' \in F^I_\theta$ and $a \in \assb {K_{\theta,\beta}} {F^I} $. 
Then, since $\ell_\theta(s') < k$, condition (2) can be used to 
actually define $\Phi_\beta(\,\fo^I_\zeta(s' \diamond a)\,)$. 
\medskip
$4.$ The element $s$ has the form $\fo^I_\zeta(s_1,s_2)$ with 
$\zeta \in P$ the name of an integer operator (so $\zeta$ is of type 
$\ftt{int}_2 \to \beta$ with
$\beta$ either $\ftt{int}$ or $\ftt{bool}$) and
$s_1,\,s_2 \in F^I_{\fstt{int}}$. Then, since $\ell_{\fstt{int}}(s_j) < k$ for 
each $j = 1,\,2 $, condition (3) can be used to 
define $\Phi_\beta(\,\fo^I_\zeta(s_1,s_2)\,)$. 
\medskip\smallskip
The uniqueness of $\Phi_B$ also follows using induction on 
$\ell_\beta(s)$. \eop
\bigbreak
The \tsym{\Lambda}-homomorphism 
$\Phi_B : (F^I_B,\fo^I_K) \to (F^I_B,\fo^I_K)$ given by Proposition~7.3.1 
will be called the 
\indexdef{\tsym{\Lambda}-homomorphism defined by $\alpha$
and the pre-support system $N^I_B$}{homomorphism}{defined by equations}.

\proclaim{Proposition~7.3.2} Suppose $N^I_B$ is a support system. Then the 
\tsym{\Lambda}-homomorphism $\Phi_B$ 
defined by $\alpha$ and $N^I_B$ is \tsym{\alpha}-consistent, i.e., 
$\sem{\Phi_\beta(s)}^c_\beta = \sem{s}^c_\beta$ for all $c \in \fss{Sol}(\alpha)$ 
for each $s \in F^I_\beta$, $\beta \in B$.
\endpro

\proof The family of mappings $\ell_B$ employed  in the proof of Proposition~7.3.1 
will be needed again, as well as the unique representations given in
Proposition~6.2.1~(1). Put

\mdisp{G_\beta\ =\ \lcurl s \in F^I_\beta \,:\,
  \sem{\Phi_\beta(s)}^c_\beta = \sem{s}^c_\beta 
                          \frm{for all} c \in \fss{Sol}(\alpha) \rcurl} 

for each $\beta \in B$; the task is thus to show that $G_B = F^I_B$. First consider 
$s \in F^I_\beta$ 
with $\ell_\beta(s) = 1$. If $s = \fo^I_\kappa(\onept)$ for some 
$\kappa \in K$ of type $\varnothing \to \beta$ then $s \in G_\beta$ holds
trivially because $\Phi_\beta(s) = s$. On the other hand, if 
$s = \xi$ for some $\xi \in I$ of type $\beta$ then
$\Phi_\beta(s) = s_\xi$, and thus for each $c \in \fss{Sol}(\alpha)$ 
\mdisp{\sem{\Phi_\beta(s)}^c_\beta
 \ = \ \sem{s_\xi}^c_\beta \ =\ c(\xi)
 \ =\ \sem{\xi}^c_\beta \ =\ \sem{s}^c_\beta\ ;}
hence again $s \in G_\beta$. 
\medskip
Now let $k > 1$ and suppose for each $\theta \in B$ it is known that
$t \in G_{\theta}$ for all $t \in F^I_{\theta}$ with $\ell_{\theta}(t) < k$.
Consider $s \in F^I_\beta$ with $\ell_\beta(s) = k$; there are four cases:
\medskip
$1.$ The element $s$ has the form $\fo^I_\kappa(a)$, where
$\kappa \in K$ is of type $L \to \beta$ with $L \ne \varnothing$ and
$a \in \assb L {F^I} $. Let $c \in \fss{Sol}(\alpha)$,
then $\,\sem{\Phi_\eta(a(\eta))}^c_\eta\,=\,\sem{a(\eta)}^c_\eta$
for each $\eta \in L$, since $\ell_\eta(a(\eta)) < k$, 
and hence 
$\,\assb L {\sem{\,\ass L {\Phi} (a)\,}^c} \,=\, \assb L {\sem{a}^c} \,$. Thus
\smallskip
\ldisp{\quad\sem{\Phi_\beta(s)}^c_\beta 
  \ =\ \sem{\,\fo^I_\kappa(\ass L {\Phi} (a))\,}^c_\beta
  \ =\ f_\kappa\bigl(\,\assb L {\sem{\,\ass L {\Phi} (a)\,}^c} \, \bigr)}
\vskip-\medskipamount
\rdisp{
  \ =\ f_\kappa\bigl(\,\assb L {\sem{a}^c} \,\bigr)
  \ =\ \sem{\,\fo^I_\kappa(a)\,}^c_\beta
                               \ =\ \sem{s}^c_\beta \quad}
\smallskip
for all $c \in \fss{Sol}(\alpha)$, i.e., $s \in G_\beta$.
\medskip
$2.$ The element $s$ has the form $\fo^I_\xi(a)$ with 
$\xi \in I$ of type $L \to \beta$ for some $L \ne \varnothing$ and
with $a \in \assb L {F^I} $. 
Let $c \in \fss{Sol}(\alpha)$ and put $\,d \,=\, \assb L {\sem{a}^c} \,$;
thus by Lemma~7.3.2 
$\,\sem{\,s_\xi[a/L]\,}^c_\beta \, =\, \sem{s_\xi}_\beta^{\,c\oplus d}\,$.
Hence by Lemma~7.1.1
\smallskip
\ldisp{\quad\sem{\Phi_\beta(s)}^c_\beta 
\ =\ \sem{\,s_\xi[a/L]\,}_\beta^{\,c\oplus b}
         \  =\  \sem{s_\xi}_\beta^{\,c \oplus d} 
    \ =\ f_{\sigma,L}(c(\xi) \triangleleft d)
}
\vskip-\medskipamount
\rdisp{
\ =\ f_{\sigma,L}\bigl(c(\xi)\,
       \triangleleft\,\assb L {\sem{a}^c} \,\bigr)
\ =\ \sem{\,\fo^I_\xi(a)\,}^c_\beta\ =\ \sem{s}^c_\beta\ ,\quad }
\smallskip
and this shows that $s \in G_\beta$.
\medskip
$3.$ The element $s$ has the form $\fo^I_\zeta(s' \diamond a)$, where 
$\zeta = \ftt{case}^\theta_\beta$ with $\beta \in B$ and $\theta \in \Bf$ and 
where $s' \in F^I_\theta$ and $a \in \assb {K_{\theta,\beta}} {F^I} $. 
\medskip
Assume first that $s' \notin N^I_\theta$. Then
$\sem{\Phi_\theta(s')}^c_\theta = \sem{s'}^c_\theta$
for each $c \in \fss{Sol}(\alpha)$, since 
$\ell_\theta(s') < k$, and therefore 
\smallskip
\ldisp{\quad \sem{\Phi_\beta(s)}^c_\beta
    \  =\  \sem{\,\fo^I_\zeta(\Phi_\theta(s') \diamond a ) \,}^c_\beta}
\vskip-\medskipamount
\ldisp{\qquad\qquad
\ =\ f_{\sigma,L}\bigl(\, p(\zeta) \triangleleft (\sem{\Phi_\theta(s')}^c_\theta
           \diamond \assb {K_{\theta,\beta}} {\sem{a}^c} )\,\bigr)}
\vskip-\medskipamount
\rdisp{
\ =\ f_{\sigma,L}\bigl(\, p(\zeta\triangleleft ( \sem{s'}^c_\theta
           \diamond \assb {K_{\theta,\beta}} {\sem{a}^c} )\,\bigr)
   \ =\  \sem{\,\fo^I_\zeta(s' \diamond a )\,}^c_\beta  
                 \ =\ \sem{s}^c_\beta \quad}
\smallskip
for all $c \in \fss{Sol}(\alpha)$, which again implies that
$s \in G_\beta$.
\medskip
Assume next that $s' \in N^I_\theta$ and so
$\sem{s'}^c_\theta \ne \bot_\theta$ for each 
$c \in \ass I D $. Moreover, since
$N^I_\theta \subset \kterm{F}^I_\theta$,
$s'$ has the form $\fo^I_\kappa(a')$ with $\kappa \in K$ of type $J \to \theta$
for some $J \in {\cal F}_B$ and with $a' \in \assb {J} {F^I} $.
Then
\mdisp{f_\kappa\bigl(\,\assb {J} {\sem{a'}^{\,c\oplus b}} \,\bigr)
\ =\ \sem{\,\fo^{I\cup U}_\kappa(a')\,}^{\,c\oplus b}_\theta
\ =\ \sem{s'}^{\,c\oplus b}_\theta \,\ne \,\bot_\theta}
and therefore by Proposition~6.2.2~(2) and Proposition~5.4.1~(2) it follows that

\smallskip
\ldisp{\quad \sem{s}^c_\beta
\ =\ \sem{\,\fo^I_{\zeta}(s' \diamond a )\,}^c_\beta
\ =\ \fss{Case}^\theta_\beta\,
    \bigl(\sem{s'}^c_\theta,
         \assb {K_{\theta,\beta}} {\sem{a}^c} \bigr)
}
\vskip-\medskipamount
\ldisp{\qquad\qquad
\ =\ \fss{Case}^\theta_\beta\,
    \bigl(f_\kappa(\,\assb {J} {\sem{a'}^c} \,),
         \assb {K_{\theta,\beta}} {\sem{a}^c} \bigr)
}

\vskip-\medskipamount
\ldisp{\qquad\qquad\qquad
 \ =\  f_{J\to\beta,J}
  \bigl(\, \assb {K_{\theta,\beta}} {\sem{a}^c} \,(\kappa)\,
      \triangleleft \,\assb {J} {\sem{a'}^c} \,\bigr) 
}

\vskip-\medskipamount
\ldisp{\qquad\qquad\qquad\qquad
 \ =\    f_{J\to\beta,J}
  \bigl(\, \sem{a(\kappa)}^c_{J\to \beta}\,\triangleleft
      \,\assb {J} {\sem{a'}^c} \,\bigr) 
}
\vskip-\medskipamount
\rdisp{ 
 \ =\    f_{J\to\beta,J}
  \bigl(\, 
     \assb {\sigma\cdot J} {\sem{a(\kappa) \triangleleft a'}^c} \,\bigr) 
\ =\ \sem{\,\fo^I_{J\to \beta,J}
               (a(\kappa) \triangleleft a')\,}^c_\beta
                   \ =\ \sem{\Phi_\beta(s)}^c_\beta \quad}
\smallskip
for all $c \in \ass I D $, and thus in particular for all
$c \in \fss{Sol}(\alpha)$, i.e., $s \in G_\beta$.
\medskip
$4.$ The element $s$ has the form $\fo^I_\zeta(s_1,s_2)$ with 
$\zeta \in P$ the name of an integer operator (so $\zeta$ is of type 
$\ftt{int}_2 \to \beta$ with
$\beta$ either $\ftt{int}$ or $\ftt{bool}$) and
$s_1,\,s_2 \in F^I_{\fstt{int}}$. 
\medskip
Assume first that both $s_1$ and $s_2$ are ground terms; then by Lemma~7.3.4
\smallskip
\ldisp{\quad\sem{\Phi_\beta(s)}^c_\beta
\ =\ \sem{\,\fss{Eval}_\zeta(s_1,s_2)\,}^c_\beta
\ =\ \sem{\,\fss{Eval}_\zeta(s_1,s_2)\,}_\beta
}
\vskip-\medskipamount
\rdisp{
  \ =\ f_{\fstt{int}_2\to\beta,\fstt{int}_2}
\bigl(\,p(\zeta)\triangleleft (\sem{s_1}_{\fstt{int}},\sem{s_2}_{\fstt{int}})\,\bigr)
\ =\ \sem{\,\fo^I_\zeta(s_1,s_2)\,}^c_\beta \ =\ \sem{s}^c_\beta \quad}
\smallskip
for all $c \in \ass I D $, and thus in particular for all 
$c \in \fss{Sol}(\alpha)$, i.e., $s \in G_\beta$.
\medskip
Finally, assume that not both of $s_1$ and $s_2$ are ground terms.
Then, since $\ell_{\fstt{int}}(s_j) < k$ for each $j = 1,\,2 $, it follows that
$\sem{\Phi_{\fstt{int}}(s_j)}^c_{\fstt{int}} 
                                       = \sem{s_j}^c_{\fstt{int}}$
for all $c \in \fss{Sol}(\alpha)$, and thus 
\smallskip
\ldisp{\quad \sem{\Phi_\beta(s)}^c_\beta
   \ =\  \sem{\,\fo^I_\zeta(\Phi_{\fstt{int}}(s_1),
            \Phi_{\fstt{int}}(s_2))\,}^c_\beta}
\vskip-\medskipamount
\ldisp{\qquad\qquad
\ =\ f_{\fstt{int}_2\to\beta,\fstt{int}_2}
(\sem{\Phi_{\fstt{int}}(s_1)}^c_{\fstt{int}},\sem{\Phi_{\fstt{int}}(s_2)}^c_{\fstt{int}})
}
\vskip-\medskipamount
\rdisp{
\ =\ f_{\fstt{int}_2\to\beta,\fstt{int}_2}
(\sem{s_1}^c_{\fstt{int}},\sem{s_2}^c_{\fstt{int}})
\  =\  \sem{\,\fo^I_\zeta(s_1,s_2) \,}^c_\beta \ =\ \sem{s}^c_\beta \quad}
\smallskip
for all $c \in \fss{Sol}(\alpha)$, and so once again
$s \in G_\beta$.
\medskip\smallskip
Therefore by induction on $k$ it follows that $G_B = F^I_B$, and this shows that
$\Phi_B$ is \tsym{\alpha}-consistent. \eop
\bigbreak

If $\Phi_B : (F^I_B,\fo^I_K) \to (F^I_B,\fo^I_K)$ is a 
\tsym{\Lambda}-homomorphism then, as in Section~7.2, put

\mdisp{F^\Phi_\beta \ =\ \lcurl s \in F^I_\beta \,:\,
        \Phi^n_\beta(s) \in F_\beta \frm{for some} n \ge 0 \rcurl\ }
for each $\beta \in B$, and define
a mapping $\Phi^\infty_\beta : F^\Phi_\beta \to F_\beta$  
by letting $\Phi^\infty_\beta(s) = \Phi^n_\beta(s)$, where 
$n \ge 0$ is chosen so that $\Phi^n_\beta(s) \in F_\beta$.
If $s \in F^\Phi_\beta(s)$ then 
$\Phi^\infty_\beta(s)$
(or perhaps better $\sem{\,\Phi^\infty_\beta(s)\,}_\beta$) can be considered
as the `value' of $s$ computed by $\Phi_B$. The next result says, 
very roughly, that the `values' computed by
the \tsym{\Lambda}-homomorphism defined by $\alpha$
and $N^I_B$ do not depend on which pre-support system $N^I_B$ is used.

\proclaim{Proposition~7.3.3}
Let $N^I_B$ and ${\breve N}^I_B$ be pre-support systems and let
$\Phi_B$ (resp.\ ${\breve \Phi}_B$) be the \tsym{\Lambda}-homomorphism 
defined by $\alpha$ and $N^I_B$ (resp.\ defined by $\alpha$ and ${\breve N}^I_B$).
Then $\Phi^\infty_\beta(s) = {\breve \Phi}^\infty_\beta(s)$ 
for each $s \in F^\Phi_\beta \cap F^{\breve \Phi}_\beta$.
\endpro

\proof Let $(D_S,f_N)$ be as in Case~3 with, in addition, $(D_B,f_K)$ an initial
completion of an initial bottomed extension of $(X_B,p_K)$.
Then $\kterm{F}^I_B$ and hence also both of $N^I_B$ and ${\breve N}^I_B$ are 
support systems.
Therefore by Proposition~7.3.2 $\Phi_B$ and ${\breve \Phi}_B$ are both
\tsym{\alpha}-consistent. Moreover, Proposition~8.1.3 will show that the set
$\fss{Sol}(\alpha)$ is non-empty. Thus by Proposition~7.2.1
it follows that
\mdisp{\sem{\,\Phi^\infty_\beta(s)\,}_\beta 
\ =\ \sem{s}^\alpha_\beta\ =\ \sem{\,{\breve \Phi}^\infty_\beta(s)\,}_\beta}
for each $s \in F^\Phi_\beta \cap F^{\breve \Phi}_\beta$,
and hence that $\Phi^\infty_\beta(s) = {\breve \Phi}^\infty_\beta(s)$. \eop
\bigbreak
Of course, Proposition~7.3.3 does not say anything about the relationship
between the families $F^\Phi_B$ and $F^{\breve \Phi}_B$. 
\medskip
Let $\Phi_B$ be the \tsym{\Lambda}-homomorphism defined by $\alpha$ and 
some support system $N^I_B$.
If $\fss{Sol}(\alpha) \ne \varnothing$ then  Proposition~7.3.2 and 
Proposition~7.2.1 imply that $F^\Phi_B \subset F^\alpha_B$ and
$\,\sem{s}^\alpha_\beta\,=\,\sem{\,\Phi^\infty_\beta(s)\,}_\beta\,$
for all $s \in F^\Phi_\beta$, $\beta \in B$,
i.e., $\Phi^\infty_\beta(s)$ is the (unique) ground term which denotes
$\sem{s}^\alpha_\beta$. Note however the requirement here that
$\fss{Sol}(\alpha) \ne \varnothing$. 
If $\fss{Sol}(\alpha) = \varnothing$ then no conclusions can be made about an element
$s \in F^\Phi_\beta$.

\medskip
The \tsym{\Lambda}-homomorphism defined by $\alpha$ and 
the support system $\bterm{F}^I_B$ will now be called simply 
\indexdef{the \tsym{\Lambda}-homomorphism defined by $\alpha$}
{homomorphism}{defined by equations}.
As already mentioned, this is really the only case of interest. 
Starting on the next page some examples are given illustrating how this
\tsym{\Lambda}-homomorphism works in practice.
\vfill\eject
\bigskip\medskip
\frame{20pt}{\bigskip 
{\it Example~7.3.1\enspace} Again let $\alpha \in \fss{Eq}(I)$ be as in Example~7.1.1 
the following system of equations:
\medskip\smallskip
{\leftskip=35pt
$\ftt{revs p n }\ =\ \ftt{ case (eq n 0) of \lcb True -> Nil,}$ 
\vskip 2.5pt
$\qquad\qquad\qquad\qquad\qquad
 \ftt{False -> Cons (p n) (revs p (sub n 1))\rcb}$
\vskip 5.0pt
$\ftt{sq n }\ =\ \ftt{ mul n n}$
\vskip 5.0pt
$\ftt{shunt a b }\ = \ \ftt{ case a of \lcb Nil -> b, Cons -> shunx b\rcb}$
\vskip 5.0pt
$\ftt{shunx a n b }\ =\ \ftt{ shunt b (Cons n a)}$
\vskip 5.0pt
$\ftt{sqs n }\ =\ \ftt{ shunt (revs sq n) Nil}$
\bigskip
}
Let $\Phi_B$ be the \tsym{\Lambda}-homomorphism defined by $\alpha$ and
$e \in E^I_{\fstt{list}}$ be the term 
\mdisp{\ftt{sqs 2}}
If $(D_B,f_K)$ is a fully regular extension of $(X_B,p_K)$ (and so
$\bterm{E}^I_B = \kterm{E}^I_B$) then
the iterates $\Phi^n_{\fstt{list}}(e)$, $0 \le n \le 24$,
are those listed below:
\bigskip
{\leftskip=30pt
$\ftt{sqs 2}$
\vskip 5.0pt
$\ftt{shunt (revs sq 2) Nil}$
\vskip 5.0pt
$\ftt{case (revs sq 2) of \lcb Nil -> Nil, Cons -> shunx Nil\rcb}$
\vskip 5.0pt
$\ftt{case (case (eq 2 0) of}$
\vskip 2.5pt
$\qquad
 \ftt{\lcb True -> Nil, False -> Cons (sq 2) (revs sq (sub 2 1))\rcb)}$
\vskip 2.5pt
$\qquad\qquad\qquad\qquad 
\ftt{of \lcb Nil -> Nil, Cons -> shunx Nil\rcb}$
\vskip 5.0pt
$\ftt{case (case False of}$
\vskip 2.5pt
$\qquad
 \ftt{\lcb True -> Nil, False -> Cons (sq 2) (revs sq (sub 2 1))\rcb)}$
\vskip 2.5pt
$\qquad\qquad\qquad\qquad 
\ftt{of \lcb Nil -> Nil, Cons -> shunx Nil\rcb}$
\vskip 5.0pt
$\ftt{case (Cons (sq 2) (revs sq (sub 2 1)))}$
\vskip 2.5pt
$\qquad\qquad\qquad\qquad 
\ftt{of \lcb Nil -> Nil, Cons -> shunx Nil\rcb}$
\vskip 5.0pt
$\ftt{shunx Nil (sq 2) (revs sq (sub 2 1))}$
\vskip 5.0pt
$\ftt{shunt (revs sq (sub 2 1)) (Cons (sq 2) Nil)}$
\vskip 5.0pt
$\ftt{case (revs sq (sub 2 1)) of}$
\vskip 2.5pt
$\qquad\qquad\qquad\qquad \ftt{\lcb Nil -> Cons (sq 2) Nil,}$
\vskip 2.5pt
$\qquad\qquad\qquad\qquad \ftt{Cons -> shunx (Cons (sq 2) Nil)\rcb}$
\medskip}
\bigskip
{\it (Example~7.3.1 is continued on the next page.)}
\bigskip\medskip
}

\vfill\eject

\frame{20pt}{\bigskip 
{\it Example~7.3.1\enspace(continued)}
\medskip\smallskip
{\leftskip=30pt
\vskip 5.0pt
$\ftt{case (case (eq (sub 2 1) 0) of}$
\vskip 2.5pt
$\quad
 \ftt{\lcb True -> Nil,}$
\vskip 2.5pt
$\quad
\ftt{False -> Cons (sq (sub 2 1)) (revs sq (sub (sub 2 1) 1))\rcb)}$
\vskip 2.5pt
$\qquad\qquad\qquad\qquad \ftt{of \lcb Nil -> Cons (sq 2) Nil,}$
\vskip 2.5pt
$\qquad\qquad\qquad\qquad \ftt{Cons -> shunx (Cons (sq 2) Nil)\rcb}$
\vskip 5.0pt
$\ftt{case (case (eq 1 0) of}$
\vskip 2.5pt
$\quad
 \ftt{\lcb True -> Nil,}$
\vskip 2.5pt
$\quad
\ftt{False -> Cons (sq (sub 2 1)) (revs sq (sub (sub 2 1) 1))\rcb)}$
\vskip 2.5pt
$\qquad\qquad\qquad\qquad \ftt{of \lcb Nil -> Cons (sq 2) Nil,}$
\vskip 2.5pt
$\qquad\qquad\qquad\qquad \ftt{Cons -> shunx (Cons (sq 2) Nil)\rcb}$
\vskip 5.0pt
$\ftt{case (case False of}$
\vskip 2.5pt
$\quad
 \ftt{\lcb True -> Nil,}$
\vskip 2.5pt
$\quad
\ftt{False -> Cons (sq (sub 2 1)) (revs sq (sub (sub 2 1) 1))\rcb)}$
\vskip 2.5pt
$\qquad\qquad\qquad\qquad \ftt{of \lcb Nil -> Cons (sq 2) Nil,}$
\vskip 2.5pt
$\qquad\qquad\qquad\qquad \ftt{Cons -> shunx (Cons (sq 2) Nil)\rcb}$
\vskip 5.0pt
$\ftt{case (Cons (sq (sub 2 1)) (revs sq (sub (sub 2 1) 1))) of}$
\vskip 2.5pt
$\qquad\qquad\qquad\qquad \ftt{\lcb Nil -> Cons (sq 2) Nil,}$
\vskip 2.5pt
$\qquad\qquad\qquad\qquad \ftt{Cons -> shunx (Cons (sq 2) Nil)\rcb}$
\vskip 5.0pt
$\ftt{shunx (Cons (sq 2) Nil) (sq (sub 2 1))}$
\vskip 2.5pt
$\qquad\qquad\qquad\qquad
\ftt{(revs sq (sub (sub 2 1) 1))}$
\vskip 5.0pt
$\ftt{shunt (revs sq (sub (sub 2 1) 1))}$
\vskip 2.5pt
$\qquad\qquad\qquad\qquad
\ftt{(Cons (sq (sub 2 1)) (Cons (sq 2) Nil))}$
\vskip 5.0pt
$\ftt{case (revs sq (sub (sub 2 1) 1)) of}$
\vskip 2.5pt
$\qquad\qquad
\ftt{\lcb Nil -> (Cons (sq (sub 2 1)) (Cons (sq 2) Nil))}$
\vskip 2.5pt
$\qquad\qquad
\ftt{Cons -> shunx (Cons (sq (sub 2 1))}$
\vskip 2.5pt
$\qquad\qquad\qquad\qquad\qquad\qquad\qquad\qquad\qquad\qquad
\ftt{(Cons (sq 2) Nil))\rcb}$
\vskip 5.0pt
$\ftt{case (revs sq (sub 1 1)) of}$
\vskip 2.5pt
$\qquad\qquad
\ftt{\lcb Nil -> (Cons (sq (sub 2 1)) (Cons (sq 2) Nil))}$
\vskip 2.5pt
$\qquad\qquad
\ftt{Cons -> shunx (Cons (sq (sub 2 1))}$
\vskip 2.5pt
$\qquad\qquad\qquad\qquad\qquad\qquad\qquad\qquad\qquad\qquad
\ftt{(Cons (sq 2) Nil))\rcb}$
\bigskip
}
{\it (Example~7.3.1 is continued on the next page.)}
\bigskip\medskip
}

\vfill\eject

\frame{20pt}{\bigskip 
{\it Example~7.3.1\enspace(continued)}
\medskip\smallskip
{\leftskip=30pt
\vskip 5.0pt
$\ftt{case (revs sq 0) of}$
\vskip 2.5pt
$\qquad\qquad
\ftt{\lcb Nil -> (Cons (sq (sub 2 1)) (Cons (sq 2) Nil))}$
\vskip 2.5pt
$\qquad\qquad
\ftt{Cons -> shunx (Cons (sq (sub 2 1))}$
\vskip 2.5pt
$\qquad\qquad\qquad\qquad\qquad\qquad\qquad\qquad\qquad\qquad
\ftt{(Cons (sq 2) Nil))\rcb}$
\vskip 5.0pt
$\ftt{case (case (eq 0 0) of \lcb True -> Nil,}$
\vskip 2.5pt
$\qquad\qquad\qquad\qquad\qquad\qquad
\ftt{False -> Cons 0 (revs sq (sub 0 1))\rcb)}$
\vskip 2.5pt
$\qquad\qquad
\ftt{of \lcb Nil -> (Cons (sq (sub 2 1))}$
\vskip 2.5pt
$\qquad\qquad\qquad\qquad\qquad\qquad\qquad\qquad\qquad\qquad
\ftt{(Cons (sq 2) Nil))}$
\vskip 2.5pt
$\qquad\qquad
\ftt{Cons -> shunx (Cons (sq (sub 2 1))}$
\vskip 2.5pt
$\qquad\qquad\qquad\qquad\qquad\qquad\qquad\qquad\qquad\qquad
\ftt{(Cons (sq 2) Nil))\rcb}$
\vskip 5.0pt
$\ftt{case (case True of \lcb True -> Nil,}$
\vskip 2.5pt
$\qquad\qquad\qquad\qquad\qquad\qquad
\ftt{False -> Cons 0 (revs sq (sub 0 1))\rcb)}$
\vskip 2.5pt
$\qquad\qquad
\ftt{of \lcb Nil -> (Cons (sq (sub 2 1))}$
\vskip 2.5pt
$\qquad\qquad\qquad\qquad\qquad\qquad\qquad\qquad\qquad\qquad
\ftt{(Cons (sq 2) Nil))}$
\vskip 2.5pt
$\qquad\qquad
\ftt{Cons -> shunx (Cons (sq (sub 2 1))}$
\vskip 2.5pt
$\qquad\qquad\qquad\qquad\qquad\qquad\qquad\qquad\qquad\qquad
\ftt{(Cons (sq 2) Nil))\rcb}$
\vskip 5.0pt
$\ftt{case Nil of}$
\vskip 2.5pt
$\qquad\qquad
\ftt{\lcb Nil -> (Cons (sq (sub 2 1)) (Cons (sq 2) Nil))}$
\vskip 2.5pt
$\qquad\qquad
\ftt{Cons -> shunx (Cons (sq (sub 2 1))}$
\vskip 2.5pt
$\qquad\qquad\qquad\qquad\qquad\qquad\qquad\qquad\qquad\qquad
\ftt{(Cons (sq 2) Nil))\rcb}$
\vskip 5.0pt
$\ftt{Cons (sq (sub 2 1)) (Cons (sq 2) Nil)}$
\vskip 5.0pt
$\ftt{Cons (mul (sub 2 1) (sub 2 1)) (Cons (mul 2 2) Nil)}$
\vskip 5.0pt
$\ftt{Cons (mul 1 1) (Cons 4 Nil)}$
\vskip 5.0pt
$\ftt{Cons 1 (Cons 4 Nil)}$
\medskip
}
\medskip
Thus
$\,\Phi^n_{\fstt{list}}(e) \,=\, \ftt{Cons 1 (Cons 4 Nil)}\,$ for all $n \ge 24$. 
\bigskip\medskip
}

\vfill\eject
\bigskip\medskip
\frame{20pt}{\bigskip 
{\it Example~7.3.2\enspace} Consider the same situation as in Example~7.3.1, but now with
$(D_B,f_K)$ the flat extension of $(X_B,p_K)$ (and so
$\bterm{E}^I_B = E_B$). Then the iterates $\Phi^n_{\fstt{list}}(e)$, $0 \le n \le 21$,
are those given below.
\medskip
{\leftskip=30pt
$\ftt{sqs 2}$
\vskip 4.2pt
$\ftt{shunt (revs sq 2) Nil}$
\vskip 4.2pt
$\ftt{case (revs sq 2) of \lcb Nil -> Nil, Cons -> shunx Nil\rcb}$
\vskip 4.2pt
$\ftt{case (case (eq 2 0) of}$
\vskip 1.8pt
$\qquad
 \ftt{\lcb True -> Nil, False -> Cons (sq 2) (revs sq (sub 2 1))\rcb)}$
\vskip 1.8pt
$\qquad\qquad\qquad\qquad 
\ftt{of \lcb Nil -> Nil, Cons -> shunx Nil\rcb}$
\vskip 4.2pt
$\ftt{case (case False of}$
\vskip 1.8pt
$\qquad
 \ftt{\lcb True -> Nil, False -> Cons (sq 2) (revs sq (sub 2 1))\rcb)}$
\vskip 1.8pt
$\qquad\qquad\qquad\qquad 
\ftt{of \lcb Nil -> Nil, Cons -> shunx Nil\rcb}$
\vskip 4.2pt
$\ftt{case (Cons (sq 2) (revs sq (sub 2 1)))}$
\vskip 1.8pt
$\qquad\qquad\qquad\qquad 
\ftt{of \lcb Nil -> Nil, Cons -> shunx Nil\rcb}$
\vskip 4.2pt
$\ftt{case (Cons (mul 2 2) 
(case (eq (sub 2 1) 0) of}$
\vskip 1.8pt
$\quad
\ftt{\lcb True -> Nil,}$
\vskip 1.8pt
$\quad  
\ftt{False -> Cons (sq (sub 2 1)) (revs sq (sub (sub 2 1) 1))\rcb)}$
\vskip 1.8pt
$\qquad\qquad\qquad\qquad 
\ftt{of \lcb Nil -> Nil, Cons -> shunx Nil\rcb}$
\vskip 4.2pt
$\ftt{case (Cons 4 (case (eq 1 0) of}$
\vskip 1.8pt
$\quad
\ftt{\lcb True -> Nil,}$
\vskip 1.8pt
$\quad  
\ftt{False -> Cons (sq (sub 2 1)) (revs sq (sub (sub 2 1) 1))\rcb)}$
\vskip 1.8pt
$\qquad\qquad\qquad\qquad 
\ftt{of \lcb Nil -> Nil, Cons -> shunx Nil\rcb}$
\vskip 4.2pt
$\ftt{case (Cons 4 (case False of}$
\vskip 1.8pt
$\quad
\ftt{\lcb True -> Nil,}$
\vskip 1.8pt
$\quad  
\ftt{False -> Cons (sq (sub 2 1)) (revs sq (sub (sub 2 1) 1))\rcb)}$
\vskip 1.8pt
$\qquad\qquad\qquad\qquad 
\ftt{of \lcb Nil -> Nil, Cons -> shunx Nil\rcb}$
\vskip 4.2pt
$\ftt{case (Cons 4 (Cons (sq (sub 2 1))}$ 
\vskip 1.8pt
$\qquad\qquad
\ftt{(revs sq (sub (sub 2 1) 1)))}$
\vskip 1.8pt
$\qquad\qquad\qquad\qquad 
\ftt{of \lcb Nil -> Nil, Cons -> shunx Nil\rcb}$
\vskip 4.2pt
$\ftt{case (Cons 4 (Cons (mul (sub 2 1) (sub 2 1))}$
\vskip 1.8pt
$\qquad
\ftt{(case (eq (sub (sub 2 1) 1) 0) of}$
\vskip 1.8pt
$\qquad\qquad\qquad
\ftt{\lcb True -> Nil,}$
\vskip 1.8pt
$\qquad\qquad\qquad  
\ftt{False -> Cons (sq (sub (sub 2 1)))}$
\vskip 1.8pt
$\qquad\qquad\qquad\qquad\qquad  
\ftt{(revs sq (sub (sub (sub 2 1) 1) 1))\rcb)}$
\vskip 1.8pt
$\qquad\qquad\qquad\qquad 
\ftt{of \lcb Nil -> Nil, Cons -> shunx Nil\rcb}$
\medskip}
\smallskip
{\it (Example~7.3.2 is continued on the next page.)}
\bigskip
}
\vfill\eject

\frame{20pt}{\bigskip 
{\it Example~7.3.2\enspace(continued)}
\bigskip
{\leftskip=30pt
$\ftt{case (Cons 4 (Cons (mul 1 1)}$
\vskip 2.5pt
$\qquad
\ftt{(case (eq (sub 1 1) 0) of}$
\vskip 2.5pt
$\qquad\qquad\qquad
\ftt{\lcb True -> Nil,}$
\vskip 2.5pt
$\qquad\qquad\qquad  
\ftt{False -> Cons (sq (sub (sub 2 1)))}$
\vskip 2.5pt
$\qquad\qquad\qquad\qquad\qquad  
\ftt{(revs sq (sub (sub (sub 2 1) 1) 1))\rcb)}$
\vskip 2.5pt
$\qquad\qquad\qquad\qquad 
\ftt{of \lcb Nil -> Nil, Cons -> shunx Nil\rcb}$
\vskip 5.0pt
$\ftt{case (Cons 4 (Cons 1}$
\vskip 2.5pt
$\qquad
\ftt{(case (eq 0 0) of}$
\vskip 2.5pt
$\qquad\qquad\qquad
\ftt{\lcb True -> Nil,}$
\vskip 2.5pt
$\qquad\qquad\qquad  
\ftt{False -> Cons (sq (sub (sub 2 1)))}$
\vskip 2.5pt
$\qquad\qquad\qquad\qquad\qquad  
\ftt{(revs sq (sub (sub (sub 2 1) 1) 1))\rcb)}$
\vskip 2.5pt
$\qquad\qquad\qquad\qquad 
\ftt{of \lcb Nil -> Nil, Cons -> shunx Nil\rcb}$
\vskip 5.0pt
$\ftt{case (Cons 4 (Cons 1}$
\vskip 2.5pt
$\qquad
\ftt{(case True of}$
\vskip 2.5pt
$\qquad\qquad\qquad
\ftt{\lcb True -> Nil,}$
\vskip 2.5pt
$\qquad\qquad\qquad  
\ftt{False -> Cons (sq (sub (sub 2 1)))}$
\vskip 2.5pt
$\qquad\qquad\qquad\qquad\qquad  
\ftt{(revs sq (sub (sub (sub 2 1) 1) 1))\rcb)}$
\vskip 2.5pt
$\qquad\qquad\qquad\qquad 
\ftt{of \lcb Nil -> Nil, Cons -> shunx Nil\rcb}$
\vskip 5.0pt
$\ftt{case (Cons 4 (Cons 1 Nil)) of}$
\vskip 2.5pt
$\qquad\qquad\qquad\qquad 
\ftt{\lcb Nil -> Nil, Cons -> shunx Nil\rcb}$
\vskip 5.0pt
$\ftt{shunx Nil 4 (Cons 1 Nil)}$
\vskip 5.0pt
$\ftt{shunt (Cons 1 Nil) (Cons 4 Nil)}$
\vskip 5.0pt
$\ftt{case (Cons 1 Nil) of \lcb Nil -> Cons 4 Nil,}$
\vskip 2.5pt
$\qquad\qquad\qquad\qquad\qquad\quad 
\ftt{Cons -> shunx (Cons 4 Nil) \rcb}$
\vskip 5.0pt
$\ftt{shunx (Cons 4 Nil) 1 Nil}$
\vskip 5.0pt
$\ftt{shunt Nil (Cons 1 (Cons 4 Nil))}$
\vskip 5.0pt
$\ftt{case Nil of \lcb Nil -> (Cons 1 (Cons 4 Nil)),}$
\vskip 2.5pt
$\qquad\qquad 
\ftt{Cons -> shunx (Cons 1 (Cons 4 Nil))\rcb}$
\vskip 5.0pt
$\ftt{Cons 1 (Cons 4 Nil)}$
\bigskip}
Thus
$\,\Phi^n_{\fstt{list}}(e) \,=\, \ftt{Cons 1 (Cons 4 Nil)}\,$ for all $n \ge 21$. 
Proposition~7.3.3 implies, of course,  that the `answer' here is the same as
that for a fully regular extension. 
\bigskip\smallskip
}

\vfill\eject
\frame{20pt}{\bigskip 
{\it Example~7.3.3\enspace} Let $(X_B,p_K)$ be as in Example~2.2.1 and let
\mdisp{I \ =\ \{\,\ftt{map},\,\ftt{mapx},\,\ftt{plus},\,\ftt{hd},\,
                                      \ftt{hdx},\,\ftt{uint}\,\}}
be the \tsym{S}-typed set with 
\medskip
{\leftskip=35pt
$\ftt{map}\,$ of type
$\,(\ftt{int} \to \ftt{int}) \ \ftt{list}\to \ftt{list}\,$, 
\smallskip
$\ftt{mapx}\,$ of type
$\,(\ftt{int} \to \ftt{int}) \ \ftt{int} \ \ftt{list}\to \ftt{list}\,$,
\smallskip
$\ftt{plus}\,$ of type $\,\ftt{int}\ \ftt{int} \to \ftt{int}\,$,
\smallskip
$\ftt{hd}\,$ of type $\,\ftt{list} \to \ftt{int}\,$,
$\ \ftt{hdx}\,$ of type $\,\ftt{int}\ \ftt{list} \to \ftt{int}\,$,
\smallskip
$\ftt{uint}\,$ of type $\,\ftt{int}\,$.
\bigskip
}
Let $\alpha \in \fss{Eq}(I)$ be the following system of equations:
\medskip\smallskip
{\leftskip=35pt
$\ftt{map f ns } = \ftt{ case ns of \lcb Nil -> Nil, Cons -> mapx f\rcb}$
\vskip 5.0pt
$\ftt{mapx f m ms } = \ftt{ Cons (f m) (map f ms)}$
\vskip 5.0pt
$\ftt{plus m n } = \ftt{ add m n}$
\vskip 5.0pt
$\ftt{hd ms } = \ftt{ case ms of \lcb Nil -> uint, Cons -> hdx\rcb}$
\vskip 5.0pt
$\ftt{hdx n ns } = \ftt{ n}$
\vskip 5.0pt
$\ftt{uint } = \ftt{ uint}$
\medskip
}
\medskip
Let $\Phi_B$ be the \tsym{\Lambda}-homomorphism defined by $\alpha$
and $e \in E^I_{\fstt{list}}$ be the term 
\mdisp{\ftt{map (plus 1) (Cons 1 (Cons 2 Nil))}}
Then for any monotone regular extension $(D_B,f_K)$ of $(X_B,p_K)$
the iterates $\Phi^n_{\fstt{list}}(e)$, for $0 \le n \le 8$,
are those listed below:
\bigskip
{\leftskip=30pt
$\ftt{map (plus 1) (Cons 1 (Cons 2 Nil))}$
\vskip 5.0 pt
$\ftt{case (Cons 1 (Cons 2 Nil)) of }$
\vskip 2.5pt
$\qquad\qquad\qquad\qquad\qquad
   \ftt{\lcb Nil -> Nil, Cons -> mapx (plus 1)\rcb}$
\vskip 5.0pt
$\ftt{mapx (plus 1) 1 (Cons 2 Nil)}$
\vskip 5.0pt
$\ftt{Cons (plus 1 1) (map (plus 1) (Cons 2 Nil))}$
\vskip 5.0pt
$\ftt{Cons (add 1 1) (case (Cons 2 Nil) of}$ 
\vskip 2.5pt
$\qquad\qquad\qquad\qquad\qquad
\ftt{\lcb Nil -> Nil, Cons -> mapx (plus 1)\rcb)}$
\vskip 5.0pt
$\ftt{Cons 2 (mapx (plus 1) 2 Nil)}$
\vskip 5.0pt
$\ftt{Cons 2 (Cons (plus 1 2) (mapa (plus 1) Nil))}$
\bigskip\medskip
}
{\it (Example~7.3.3 is continued on the next page.)}
\bigskip\medskip
}

\vfill\eject
\frame{20pt}{\bigskip 
{\it Example~7.3.3 (continued) \enspace}
\bigskip\smallskip
{\leftskip=30pt
$\ftt{Cons 2 (Cons (add 1 2) (case Nil of}$ 
\vskip 2.5pt
$\qquad\qquad\qquad\qquad\qquad
\ftt{\lcb Nil -> Nil, Cons -> mapx (plus 1)\rcb))}$
\vskip 5.0pt
$\ftt{Cons 2 (Cons 3 Nil)}$
\bigskip
}
Therefore for all $n \ge 9$
\mdisp{\Phi^n_{\fstt{list}}(e) \ =\ \ftt{Cons 2 (Cons 3 Nil)}}
\medskip
Now consider the term $e' \in E^I_{\fstt{int}}$ given by 
\mdisp{\ftt{hd (map (plus 1) (Cons 1 (Cons 2 Nil)))}}
\medskip
If $(D_B,f_K)$ is here a fully regular extension of $(X_B,p_K)$
(so $\bterm{E}^I_B = \kterm{E}^I_B$) then
the iterates $\Phi^n_{\fstt{list}}(e')$, for $0 \le n \le 8$,
are those listed below:
\bigskip
{\leftskip=30pt
$\ftt{hd (map (plus 1) (Cons 1 (Cons 2 Nil)))}$
\vskip 5.0pt
$\ftt{case (map (plus 1) (Cons 1 (Cons 2 Nil))) of}$
\vskip 2.5pt
$\qquad\qquad\qquad\qquad\qquad\qquad\qquad\qquad 
\ftt{\lcb Nil -> uint, Cons -> hdx\rcb}$
\vskip 5.0pt
$\ftt{case (case (Cons 1 (Cons 2 Nil)) of}$
\vskip 2.5pt
$\qquad\qquad\qquad\qquad
\ftt{\lcb Nil -> Nil, Cons -> mapx (plus 1)\rcb) of}$
\vskip 2.5pt
$\qquad\qquad\qquad\qquad\qquad\qquad\qquad\qquad  
     \ftt{\lcb Nil -> uint, Cons -> hdx\rcb}$
\vskip 5.0pt
$\ftt{case (mapx (plus 1) 1 (Cons 2 Nil)) of}$
\vskip 2.5pt
$\qquad\qquad\qquad\qquad\qquad\qquad\qquad\qquad 
     \ftt{\lcb Nil -> uint, Cons -> hdx\rcb}$
\vskip 5.0pt
$\ftt{case (Cons (plus 1 1) (map (plus 1) (Cons 2 Nil))) of}$
\vskip 2.5pt
$\qquad\qquad\qquad\qquad\qquad\qquad\qquad\qquad 
     \ftt{\lcb Nil -> uint, Cons -> hdx\rcb}$
\vskip 5.0pt
$\ftt{hda (plus 1 1) (map (plus 1) (Cons 2 Nil))}$
\vskip 5.0pt
$\ftt{plus 1 1}$
\vskip 5.0pt
$\ftt{add 1 1}$
\vskip 5.0pt
$\ftt{2}$
\bigskip
}
Thus
$\,\Phi^n_{\fstt{int}}(e') \,=\, \ftt{2}\,$ for all $n \ge 9$. 
\bigskip\medskip
{\it (Example~7.3.3 is continued on the next page.)}
\bigskip\medskip
}

\vfill\eject
\frame{20pt}{\bigskip 
{\it Example~7.3.3 (continued) \enspace}
Suppose now $(D_B,f_K)$ is the flat extension of $(X_B,p_K)$
(so $\bterm{E}^I_B = E_B$).
In this case the iterates $\Phi^n_{\fstt{int}}(e')$, $0 \le n \le 10$, are 
the following:
\bigskip
{\leftskip=30pt
$\ftt{hd (map (plus 1) (Cons 1 (Cons 2 Nil)))}$
\vskip 5.0pt
$\ftt{case (case (Cons 1 (Cons 2 Nil)) of}$
\vskip 2.5pt
$\qquad\qquad\qquad
\ftt{\lcb Nil -> Nil, Cons -> mapx (plus 1)\rcb) of}$
\vskip 2.5pt
$\qquad\qquad\qquad\qquad\qquad\qquad\qquad\qquad 
     \ftt{\lcb Nil -> uint, Cons -> hdx\rcb}$
\vskip 5.0pt
$\ftt{case (mapx (plus 1) 1 (Cons 2 Nil)) of}$
\vskip 2.5pt
$\qquad\qquad\qquad\qquad\qquad\qquad\qquad\qquad 
     \ftt{\lcb Nil -> uint, Cons -> hdx\rcb}$
\vskip 5.0pt
$\ftt{case (Cons (plus 1 1) (map (plus 1) (Cons 2 Nil))) of}$
\vskip 2.5pt
$\qquad\qquad\qquad\qquad\qquad\qquad\qquad\qquad 
     \ftt{\lcb Nil -> uint, Cons -> hdx\rcb}$
\vskip 5.0pt
$\ftt{case (Cons (add 1 1) (case (Cons 2 Nil) of}$
\vskip 2.5pt
$\qquad\qquad\qquad
\ftt{\lcb Nil -> Nil, Cons -> mapx (plus 1)\rcb)) of}$
\vskip 2.5pt
$\qquad\qquad\qquad\qquad\qquad\qquad\qquad\qquad 
     \ftt{\lcb Nil -> uint, Cons -> hdx\rcb}$
\vskip 5.0pt
$\ftt{case (Cons 2 (mapx (plus 1) 2 Nil)) of}$
\vskip 2.5pt
$\qquad\qquad\qquad\qquad\qquad\qquad\qquad\qquad 
     \ftt{\lcb Nil -> uint, Cons -> hdx\rcb}$
\vskip 5.0pt
$\ftt{case (Cons 2 (Cons (plus 1 2) (mapx (plus 1) Nil))) of}$
\vskip 2.5pt
$\qquad\qquad\qquad\qquad\qquad\qquad\qquad\qquad 
     \ftt{\lcb Nil -> uint, Cons -> hdx\rcb}$
\vskip 5.0pt
$\ftt{case (Cons 2 (Cons (add 1 2) (case Nil of }$ 
\vskip 1.5pt
$\qquad\qquad\qquad
\ftt{\lcb Nil -> Nil, Cons -> mapx (plus 1)\rcb))) of}$
\vskip 2.5pt
$\qquad\qquad\qquad\qquad\qquad\qquad\qquad\qquad 
     \ftt{\lcb Nil -> uint, Cons -> hdx\rcb}$
\vskip 5.0pt
$\ftt{case (Cons 2 (Cons 3 Nil)) of \lcb Nil -> uint, Cons -> hdx\rcb}$
\vskip 5.0pt
$\ftt{hda 2 (Cons 3 Nil)}$
\vskip 5.0pt
$\ftt{2}$
\bigskip
}
Thus again
$\,\Phi^n_{\fstt{int}}(e') \,=\, \ftt{2}\,$ for all $n \ge 11$ (although
this time it takes longer to get to the `answer').
\bigskip\medskip
}
\vfill\eject

\frame{20pt}{\bigskip 
{\it Example~7.3.4\enspace} The following is very similar to the situation
looked at in Example~7.1.2. Let $(X_B,p_K)$ be as in Example~2.2.1 and let
\mdisp{I\ =\ \{\,\ftt{fst},\,\ftt{fstx},\,\ftt{uint}\,\}}
be the \tsym{S}-typed set with 
\medskip\smallskip
{\leftskip=40pt
$\,\ftt{fst}\,$ of type $\ftt{pair} \to \ftt{int}$,
$\,\ftt{fstx}\,$ of type $\ftt{int}\ \ftt{int} \to \ftt{int}$,
\smallskip
$\,\ftt{uint}\,$ of type $\ftt{int}$.
\bigskip
}
Consider the following system of equations $\alpha \in \fss{Eq}(I)$:
\medskip\smallskip
{\leftskip=35pt
$\ftt{fst p } = \ftt{ case p of \lcb Pair -> fstx\rcb}$
\vskip 5.0pt
$\ftt{fstx m n } = \ftt{ m}$
\vskip 5.0pt
$\ftt{uint } = \ftt{ uint}$
\medskip\smallskip
}
Let $\Phi_B$ be the \tsym{\Lambda}-homomorphism defined by $\alpha$ and 
$e \in E^I_{\fstt{int}}$ be the term 
\mdisp{\ftt{fst (Pair 1 uint)}}
If $(D_B,f_K)$ is the flat extension of $(X_B,p_K)$ (and
hence $\bterm{E}^I_B = E_B$) then 
$\,\Phi^n_{\fstt{int}}(e)\,=\, \ftt{case (Pair 1 uint) of \lcb Pair -> fstx\rcb}\,$ for all
$n \ge 1$. Thus in this case $e \notin E^\Phi_{\fstt{int}}$.
\medskip
On the other hand, if $(D_B,f_K)$ is a fully regular extension (in which case
$\bterm{E}^I_B = \kterm{E}^I_B$)
then  $\,\Phi^1_{\fstt{int}}(e) \,=\, \ftt{case (Pair 1 uint) of \lcb Pair -> fstx\rcb}\,$,
$\,\Phi^2_{\fstt{int}}(e) \,=\, \ftt{fstx 1 uint}\,$ and thus
$\,\Phi^n_{\fstt{int}}(e) \,=\, \ftt{1}\,$ for all $n \ge 3$.
This means that here $e \in E^\Phi_{\fstt{int}}$ and
$\Phi^\infty_{\fstt{int}}(e) \,=\, \ftt{1}$.
\medskip
Next let 
$\,I \,=\, \{\,\ftt{hd},\,\ftt{hdx},\,\ftt{ones}\,\}\,$ 
be the \tsym{S}-typed set with
\medskip\smallskip
{\leftskip=40pt
$\,\ftt{hd}\,$ of type $\ftt{list} \to \ftt{int}$,
$\,\ftt{hdx}\,$ of type $\ftt{int}\ \ftt{list} \to \ftt{int}$,
\smallskip
$\,\ftt{ones}\,$ of type $\ftt{list}$.
\bigskip
}
Consider the following system of equations $\alpha \in \fss{Eq}(I)$:
\medskip\smallskip
{\leftskip=35pt
$\ftt{hd xs } = \ftt{ case xs of \lcb Nil -> hd xs, Cons -> hdx\rcb}$
\vskip 5.0pt
$\ftt{hdx x xs } = \ftt{ x}$
\vskip 5.0pt
$\ftt{ones } = \ftt{ Cons 1 ones}$
\medskip\smallskip
}
Let $\Phi_B$ be the \tsym{\Lambda}-homomorphism defined by $\alpha$ and 
$e \in E^I_{\fstt{int}}$ be the term 
\mdisp{\ftt{hd ones}}
\medskip\smallskip
{\it (Example~7.3.4 is continued on the next page.)}
\bigskip\medskip
}

\vfill\eject
\frame{20pt}{\bigskip 
{\it Example~7.3.4 (continued)\enspace} 
Now if $(D_B,f_K)$ is a fully regular extension of $(X_B,p_K)$ (and so 
$\bterm{E}^I_B = \kterm{E}^I_B$) 
then the iterates $\Phi^n_{\fstt{int}}(e)$, $0 \le n \le 4$, are those 
listed below:
\medskip\smallskip
{\leftskip=30pt
$\ftt{hd ones}$
\vskip 5.0pt
$\ftt{case ones of \lcb Nil -> hd ones, Cons -> hdx\rcb}$
\vskip 5.0pt
$\ftt{case (Cons 1 ones) of \lcb Nil -> hd ones, Cons -> hdx\rcb}$
\vskip 5.0pt
$\ftt{hdx 1 ones}$
\vskip 5.0pt
$\ftt{1}$
\medskip\smallskip
}
Therefore
$\,\Phi^n_{\fstt{int}}(e) \,=\, \ftt{1}\,$ for all $n \ge 4$.
This means that here $e \in E^\Phi_{\fstt{int}}$ and
$\Phi^\infty_{\fstt{int}}(e) \,=\, \ftt{1}$. (However, although 
$e \in E^\Phi_{\fstt{int}}$, it is still possible for
$\fss{Sol}(\alpha)$ to be empty.)
\medskip
However, if $(D_B,f_K)$ is the flat extension of 
$(X_B,p_K)$ (and so $\bterm{E}^I_B = E_B$) then 
the iterates $\Phi^n_{\fstt{int}}(e)$, $n \ge 0$, are the following: 
\bigskip
{\leftskip=30pt
$\ftt{hd ones}$
\vskip 5.0pt
$\ftt{case ones of \lcb Nil -> hd ones, Cons -> hdx\rcb}$
\vskip 5.0pt
$\ftt{case (Cons 1 ones) of \lcb Nil -> hd ones, Cons -> hdx\rcb}$
\vskip 5.0pt
$\ftt{case (Cons 1 (Cons 1 ones)) of \lcb Nil -> hd ones, Cons -> hdx\rcb}$
\vskip 5.0pt
$\ftt{case (Cons 1 (Cons 1 (Cons 1 ones))) of}$
\vskip 2.5pt
$\qquad\qquad\qquad\qquad\qquad\qquad\qquad\qquad
\ftt{\lcb Nil -> hd ones, Cons -> hdx\rcb}$
\vskip 5.0pt
$\ftt{case (Cons 1 (Cons 1 (Cons 1 (Cons 1 ones)))) of}$
\vskip 2.5pt
$\qquad\qquad\qquad\qquad\qquad\qquad\qquad\qquad
\ftt{\lcb Nil -> hd ones, Cons -> hdx\rcb}$
\vskip 5.0pt
$\qquad\qquad \vdots$
\bigskip\medskip
}
Thus in this case $e \notin E^\Phi_{\fstt{int}}$.
\bigskip\medskip
}

\vfill\eject
\hfline{200}{7.4 \ REPLACEMENT RULES}

\bigskip
\sectionhead {7.4} {Replacement rules}
\bigskip\medskip

In this section we are going to introduce rules which allow valid assertions to be made 
about the solutions of equations. As mentioned at the beginning of the chapter,
these are rules which should be thought of as tools which a human being
(rather than a machine) can apply. The aim is to give the reader an idea of how 
equations can be used to make deductions about their solutions. The presentation is 
not very systematic and the material will not be required in Chapter~8.

\medskip
Consider, for instance, the system of equations $\alpha$ given in Example~7.1.1 and 
for each $n \in \Nat$ let $A_n$ be the assertion that

\mdisp{c_{\fstt{sqs}}(n)
  \ = \ 1\ 4\ 9\ \cdots\ n^2\ \frm{for all} c \in \fss{Sol}(\alpha)\ .}

The rules to be introduced below can be used to formally verify that $A_n$ is true for 
all $n \in \Nat$.
In fact, rules of this kind already occurred implicitly in the algorithm
for computing values, and were used to show that if $s \in F^{I}_\beta$ for some 
$\beta \in B$ then $\sem{\Phi_\beta(s)}^c_\beta = \sem{s}^c_\beta$ for all
$c \in \fss{Sol}(\alpha)$.
However, the power of those rules is limited: They are able to verify
that $A_n$ is true for any specific value of $n$ (for instance, that $A_{42}$ is
true) but they cannot verify the assertion that $A_n$ is true for all $n \in \Nat$.
\medskip
For the whole of the section let $I$ be a non-empty finite global \tsym{S}-typed set and 
let $\alpha = s_I \in \fss{Eq}(I)$ be a system of equations for which
$\fss{Sol}(\alpha) \ne \varnothing$.
The following situation will be considered: Let $U$ be a fixed \tsym{S}-typed set disjoint 
from $I$ and $P$. Then, given a term $s \in F^{I\cup U}_\sigma$ and an assignment
$b \in \ass U D $, we want to find rules which allow us to replace $s$ with an 
new (and hopefully simpler) term $s' \in F^{I\cup U}_\sigma$ such that
\mdisp{\sem{s'}_\sigma^{\,c \oplus b} \ =\ \sem{s}_\sigma^{\,c \oplus b}}
for all $c \in \fss{Sol}(\alpha)$. 
The elements of the set $U$ should be thought of as the names of variables
which are universally quantified by letting $b$ vary over the elements in
$\ass U D $.
\medskip
If $s,\,s' \in F^{I\cup U}_\sigma$ and $b \in \ass U D $ then we write
$s \equiv_b s'$ to mean that
$\,\sem{s}_\sigma^{\,c \oplus b} = \sem{s'}_\sigma^{\,c \oplus b}\,$
for all $c \in \fss{Sol}(\alpha)$. However, we almost always just write
$\equiv$ instead of $\equiv_b$. It will be clear from the context whether
the $b$ in $\equiv_b$ (and thus implicitly in $\equiv$) is bound or free. For example, 
three of the rules have the conditional form
\medskip
\centerline{\it if $C$ then $s \equiv s'$} 
\medskip
where $C$ is some assertion involving $b$. Here $b$ is bound and the rule is:
\medskip
\centerline{\it if $b \in \ass U D $ is such that $C$ holds then $s \equiv_b s'$}
\medskip
The second rule has the unconditional form $\,s \equiv s'\,$, by which is meant
that $\,s \equiv_b s'\,$ for all $b \in \ass U D $ (and so  $b$ is here free).
\medskip
Let $s \in F^{I\cup U}_\beta$ with $\beta \in B$; the assertion {\it\ $s$ is grounded\ }
is defined to be equivalent to the assertion 
{\it\ there exists $s_0 \in F_\beta$ such that $s \equiv s_0$\ } and in this case
$s_0$ will be referred to as the 
\indexdef{value}{value of a term}{} 
of $s$.
The assertion {\it\ $s$ is not undefined\ } is defined to be equivalent to the assertion
that {\it\ $\sem{s}_\beta^{\,c\oplus b} \ne \bot_\beta$ for all 
$c \in \fss{Sol}(\alpha)$\ }.
\medskip
Note that if $\eta \in L$ is of type $\beta \in B$ then the assertion
{\it\ $\eta$ is grounded\ } just means that $b(\eta) \in X_\beta$, and if
$s_0$ is the value of $\eta$ then $b(\eta) = \sem{s_0}_\beta$.
\medskip

Four replacement rules will be introduced. Each of these rules appeared implicitly
in Section~7.3 in the special case with $U = \varnothing$.
Only the second rule actually involves the equations $\alpha$ being considered, 
and only the third and the fourth rule depend on any property of the
\tsym{\Sigma}-algebra $(D_S,f_N)$ being employed.
\bigskip\medskip

{\bf Replacement rule 1:\ } Let $\nu \in N$ be of type $L \to \sigma$; 
the first rule is:
\bigskip
{\parindent=30pt
\item{(R1)} Let $a,\,a' \in \assb L {F^{I\cup U}} $. 
If $\,a(\eta) \equiv a'(\eta)\,$ for all $\eta \in L$ then
\mdisp{\fo^{I\cup U}_\nu(a)\ \equiv\ \fo^{I\cup U}_\nu(a')\ .}
\bigskip\medskip
}
\proclaim{Lemma~7.4.1} The replacement rule (R1) is valid. \endpro

\proof Let $c \in \fss{Sol}(\alpha)$; then by assumption
$\sem{a(\eta)}_\eta^{\,c\oplus b} = \sem{a'(\eta)}_\eta^{\,c\oplus b}$
for each $\eta \in L$, thus
$\assb L {\sem{a}^{\,c \oplus b}} = \assb L {\sem{a'}^{\,c \oplus b}} $
and therefore
\smallskip
\mdisp{\sem{\,\fo^{I\cup U}_\nu(a)\,}^{\,c \oplus b}_\sigma
   \ =\  f_\nu\bigl(\,\assb L {\sem{a}^{\,c \oplus b}} \,\bigr)
   \ =\  f_\nu\bigl(\,\assb L {\sem{a'}^{\,c \oplus b}} \,\bigr)
            \ =\  \sem{\,\fo^{I\cup U}_\nu(a')\,}^{\,c\oplus b}_\sigma\ .}
\smallskip
Hence $\,\fo^{I\cup U}_\nu(a)\, \equiv\, \fo^{I\cup U}_\nu(a')\,$. \eop

\bigbreak
The replacement rule (R1), at least when it is applied recursively,
is a form of what is called 
\indexdef{referential transparency}{referential transparency}{}.
\bigskip
{\bf Replacement rule 2:\ } The second replacement rule involves substitution.
Roughly speaking it says that a term which matches the
left-hand side of an equation can be replaced with the corresponding
right-hand side. 
It is a generalisation of the situation considered at the beginning of Section~7.3.
\medskip
Let $L$ be an \tsym{S}-typed set disjoint from $I$ and $P$ and  
$a \in \assb L {F^{I\cup U}} $ be an assignment.
Then, since 
$(F^{I\cup U}_S,\fo^{I\cup U}_N)$ is a functional \tsym{\Sigma}-algebra and
$(F^{I\cup L}_S,\fo^{I\cup L}_N)$ is functionally \tsym{I\cup L\cup P}-free,
there exists a unique \tsym{\Sigma}-homomorphism 
$\delta^a_S$ from $(F^{I\cup L}_S,\fo^{I\cup L}_N)$ to $(F^{I\cup U}_S,\fo^{I\cup U}_N)$ 
such that $\delta^a_\xi(\xi) = \xi$ for each $\xi \in I\cup P$ and
$\delta^a_\eta(\eta) = a(\eta)$ for each $\eta \in L$. 
\medskip
Let $s \in F^{I\cup L}_\sigma$; as in Section~7.3 the (perhaps) more suggestive notation
\mdisp{s[a/L]}
will also be used for the element $\delta^a_\sigma(s) \in F^{I\cup U}_\sigma$.
Moreover, when working with examples it is convenient
to allow the more explicit notation

\mdisp{s[\,a(\eta_1)/\eta_1,\,\ldots,a(\eta_m)/\eta_m\,]}

where $\lvector {\eta} m $ is some enumeration of the elements of $L$.  
The mapping
$\delta^a_\sigma$ should again be thought of as a generalised substitution or replacement 
operator:

\proclaim{Lemma~7.4.2} Let $a \in \assb L {E^{I\cup U}} $ and let
$\delta^a_S : (E^{I\cup L}_S,\blob^{I\cup L}_N) \to (E^{I\cup U}_S,\blob^{I\cup U}_N)$ 
be the homomorphism defined above using the `genuine' term algebras.
Then for each $s \in E^{I\cup L}_\sigma$ the term $\delta^a_\sigma(s)$
is obtained by replacing for each $\eta \in L$ each occurrence of $\eta$ in $s$ by
$a(\eta)$. \endpro
\proof This is the same as the proof of Lemma~7.3.1. \eop
\bigbreak
The second rule is:
\bigskip
{\parindent=30pt
\item{(R2)} Let $\xi \in I$ be of type $L \to \beta$ and  
$a \in \assb L {F^{I\cup U}} $; then 
\mdisp{ \fo^{I\cup U}_\xi(a)\ \equiv\ s_\xi[a/L]\ ,}
\item{}  where $s_\xi$ is the right-hand side of the equation for $\xi$. 
In particular, if $L = \varnothing$, in which case $\xi$ is of type $\beta$
and $a = \onept$, then  
$\,\xi \equiv s_\xi[\onept/\varnothing] = s_\xi\,$.
\bigbreak
}

The following fact (which corresponds to Lemma~7.3.2) 
is required in order to show that rule (R2) is valid:

\proclaim{Lemma~7.4.3} Let
$a \in \assb L {F^{I\cup U}} $, $c \in \ass I D $, 
and $b \in \ass U D $, and 
put $\,d = \assb L {\sem{a}^{\,c\oplus b}} \,$, i.e., 
$d \in \ass L D $ is the assignment given by 
$\,d(\eta) = \sem{a(\eta)}^{\,c\oplus b}_\eta\,$ for each $\eta \in L$. Then
\mdisp{\sem{\,s[a/L]\,}^{\,c\oplus b}_\sigma \ =\ \sem{s}_\sigma^{\,c\oplus d}}
for all $s \in F^{I\cup L}_\sigma$, $\sigma \in S$. \endpro

\proof Put $\pi_S = \sem{\cdot}^{\,c\oplus b}_S\,\delta^a_S$. Then by Proposition~2.2.1
$\pi_S$ is a 
\tsym{\Sigma}-homomorphism from $(F^{I\cup L}_S,\fo^{I\cup L}_N)$ to $(D_S,f_N)$ and
$\,\pi_\xi(\xi) \,=\, \sem{\,\delta_\xi^a(\xi)\,}^{\,c\oplus b}_\xi 
                            \,=\, \sem{\xi}^{\,c\oplus b}_\xi \,=\, c(\xi)\,$
for each $\xi \in I$, and in the same way it follows that
$\,\pi_\zeta(\zeta) \,=\, p(\zeta)$ for each $\zeta \in P$.
Moreover,
$\,\pi_\eta(\eta) \,=\, \sem{\,\delta_\eta^a(\eta)\,}^{\,c\oplus b}_\eta\,=\,
\sem{a(\eta)}^{\,c\oplus b}_\eta\,=\,d(\eta)\,$
for each $\eta \in L$.
But $\sem{\cdot}^{\,c \oplus d}_S$ is the unique such
\tsym{\Sigma}-homomorphism, and hence $\pi_S = \sem{\cdot}^{\,c\oplus d}_S$, i.e.,

\mdisp{\sem{\,s[a/L]\,}^{\,c\oplus b}_\sigma 
        \ =\ \sem{\,\delta^a_\sigma(s)\,}^{\,c\oplus b}_\sigma 
                                                 \ =\ \sem{s}^{\,c \oplus d}_\sigma}

for all $s \in F^{I\cup L}_\sigma$, $\sigma \in S$. \eop

\proclaim{Lemma~7.4.4} The replacement rule (R2) is valid. \endpro

\proof Let
$a \in \assb L {F^{I\cup U}} $, $c \in \fss{Sol}(\alpha)$ and
$b \in \ass U D $ and put $\,d \,=\, \assb L {\sem{a}^{\,c\oplus b}} \,$;
thus by Lemma~7.4.3 
$\,\sem{\,s_\xi[a/L]\,}^{\,c\oplus b}_\beta \, =\, \sem{s_\xi}_\beta^{\,c\oplus d}\,$.
If $L \ne \varnothing$ then by Lemma~7.1.1
\smallskip
\ldisp{\quad\sem{\,s_\xi[a/L]\,}_\beta^{\,c\oplus b}
         \  =\  \sem{s_\xi}_\beta^{\,c \oplus d} 
    \ =\ f_{\sigma,L}(c(\xi) \triangleleft d)
}
\vskip-\medskipamount
\rdisp{
\ =\ f_{\sigma,L}\bigl(c(\xi)\,
       \triangleleft\,\assb L {\sem{a}^{\,c\oplus b}} \,\bigr)
\ =\ \sem{\,\fo^{I\cup U}_\xi
       (a)\,}^{\,c\oplus b}_\beta\ .\quad }
\smallskip
On the other hand, if
$L = \varnothing$ then, again by Lemma~7.1.1,
\mdisp{\sem{\,s_\xi[\onept/\varnothing]\,}_\beta^{\,c\oplus b}
         \  =\  \sem{s_\xi}_\beta^{\,c \oplus d} 
    \ =\ c(\xi) \ =\ \sem{\,\xi\,}^{\,c\oplus b}_\beta\ . \quad \eop}

\eject
\bigskip\medskip
\frame{20pt}{\bigskip 
{\it Example~7.4.1\enspace} For all the examples in this section
let $\alpha = e_I$ be as in Example~7.1.1 the equations
written (without aliasing) as
\bigskip
{\leftskip=35pt
$\ftt{revs `1' `2'}$
\vskip 2.5pt
$\qquad\ =\ \ftt{ case (eq `2' 0) of \lcb True -> Nil,}$ 
\vskip 2.5pt
$\qquad\qquad\qquad
 \ftt{False -> Cons (`1' `2') (revs `1' (sub `2' 1))\rcb}$
\vskip 5.0pt
$\ftt{sq `1' }\ =\ \ftt{ mul `1' `1'}$
\vskip 5.0pt
$\ftt{shunt `1' `2' }\ = \ \ftt{ case `1' of \lcb Nil -> `2',}$ 
\vskip 2.5pt
$\qquad\qquad\qquad\qquad\qquad\qquad\qquad\qquad\qquad\qquad
\ftt{Cons -> shunx `2'\rcb}$
\vskip 5.0pt
$\ftt{shunx `1' `2' `3' }\ =\ \ftt{ shunt `3' (Cons `2') `1')}$
\vskip 5.0pt
$\ftt{sqs `1' }\ =\ \ftt{ shunt (revs sq `1') Nil}$
\bigskip\smallskip
}
and let $U = \{\ftt{n},\ftt{a},\ftt{b},\ftt{p}\}$ with $\ftt{n}$ of type $\ftt{int}$,
$\ftt{a}$ and $\ftt{b}$ of type $\ftt{list}$ and $\ftt{p}$ of type
$\ftt{int}\to\ftt{int}$. Then
\medskip\smallskip
\ldisp{\qquad 
e_{\fstt{revs}}[\,\ftt{sq} \,/\, \ftt{`1'}\,,\,\ftt{n} \,/\, \ftt{`2'}\,]}
\vskip-\medskipamount
\vskip-\smallskipamount
\ldisp{\qquad\qquad\qquad
  \ =\ \ftt{case (eq sq 0) of \lcb True -> Nil,}} 
\vskip-\medskipamount
\vskip-\smallskipamount
\rdisp{ \ftt{False -> Cons (sq n) (revs sq (sub n 1))\rcb}\qquad}
\smallskip
\ldisp{\qquad 
e_{\fstt{sq}}[\,\ftt{n} \,/\, \ftt{`1'}\,]\ =\ \ftt{mul n n}}
\smallskip
\ldisp{\qquad 
e_{\fstt{shunt}}[\,\ftt{Nil} \,/\, \ftt{`1'}\,,\,\ftt{b} \,/\, \ftt{`2'}\,]}
\vskip-\medskipamount
\vskip-\smallskipamount
\rdisp{
  \ =\ \ftt{case Nil of \lcb Nil -> Nil, Cons -> shunx b\rcb}\qquad}
\smallskip
\ldisp{\qquad 
e_{\fstt{shunt}}[\,\ftt{Cons n a} \,/\, \ftt{`1'}\,,\,\ftt{b} \,/\, \ftt{`2'}\,]}
\vskip-\medskipamount
\vskip-\smallskipamount
\rdisp{
  \ =\ \ftt{case (Cons n a) of \lcb Nil -> Nil, Cons -> shunx b\rcb}\qquad}
\smallskip
\ldisp{\qquad 
e_{\fstt{shunx}}[\,\ftt{b} \,/\, \ftt{`1'}\,,\,\ftt{n} \,/\, \ftt{`2'}
  \,,\,\ftt{a} \,/\, \ftt{`3'}\,]
\ =\ \ftt{shunt a (Cons n b)}}
\smallskip
\ldisp{\qquad 
e_{\fstt{sqs}}[\,\ftt{n} \,/\, \ftt{`1'}\,]
\ =\ \ftt{shunt (revs sq n) Nil}}
\bigskip
{\it (Example~7.4.1 is continued on the next page.)}
\bigskip\medskip
}
\bigskip\medskip

\vfill\eject

\bigskip\medskip
\frame{20pt}{\bigskip 
{\it Example~7.4.1 (continued) \enspace} 
Moreover, by rule (R2)
\medskip
\ldisp{\qquad 
\ftt{revs sq n}
  \ \equiv\ \ftt{case (eq sq 0) of \lcb True -> Nil,}} 
\vskip-\medskipamount
\vskip-\smallskipamount
\rdisp{ \ftt{False -> Cons (sq n) (revs sq (sub n 1))\rcb}\qquad}
\ldisp{\qquad 
\ftt{sq n}\ \equiv\ \ftt{mul n n}}
\ldisp{\qquad \ftt{shunt Nil b}
  \ \equiv\ \ftt{case Nil of \lcb Nil -> Nil, Cons -> shunx b\rcb}}
\ldisp{\qquad 
\ftt{shunt (Cons n a) b}}
\vskip-\medskipamount
\vskip-\smallskipamount
\rdisp{
  \ \equiv\ \ftt{case (Cons n a) of \lcb Nil -> Nil, Cons -> shunx b\rcb}\qquad}
\ldisp{\qquad 
\ftt{shunx b n a}\ \equiv\ \ftt{shunt a (Cons n b)}}
\ldisp{\qquad 
\ftt{sqs n}\ \equiv\ \ftt{shunt (revs sq n) Nil}}
\bigskip\medskip
}
\bigskip\bigskip

{\bf Replacement rule 3:\ } The third replacement rule involves the integer 
operators. 
\medskip
Let $\zeta \in P$ be the name of an integer operator (and so $\zeta$ is of type 
$\ftt{int}_2 \to \beta$ with $\beta$ either $\ftt{int}$ or $\ftt{bool}$, and where
$\ftt{int}_2$ is being used to denote the list $\,\ftt{int}\ \ftt{int}\,$).
\medskip

Recall by Lemma~7.3.3 there is a unique mapping
$\fss{Eval}_\zeta : F_{\fstt{int}} \times F_{\fstt{int}} \to F_\beta$
such that
\mdisp{\sem{\,\fss{Eval}_\zeta(s_1,s_2)\,}_\beta
  \ =\ \fss{o}_\zeta\,(\sem{s_1}_{\fstt{int}},\sem{s_2}_{\fstt{int}})}
for all $s_1,\, s_2 \in F_{\fstt{int}}$, with 
$\fss{o}_\zeta : X_{\fstt{int}}\times X_{\fstt{int}} \to X_{\beta}$ the 
integer operation corresponding to $\zeta$ (i.e., 
$\fss{o}_{\fstt{add}}(m,n) = m+n$ and so on).
The third replacement rule is:

\bigskip
{\parindent=30pt
\item{(R3)} If $s_1,\, s_2 \in F^{I\cup U}_{\fstt{int}}$ are both grounded
then
\mdisp{\fo^{I\cup U}_\zeta(s_1,s_2)
\ \equiv\ \fss{Eval}_\zeta(s'_1,s'_2)\ ,}
\item{}where $s'_j$ is the value of $s_j$ for each $j = 1,\,2$.
\bigskip
}
\proclaim{Lemma~7.4.5} The replacement rule (R3) is valid. \endpro

\proof If  $c \in \fss{Sol}(\alpha)$ then by Proposition~6.2.2~(3) and Lemma~6.2.1 
\smallskip
\ldisp{\quad \sem{\,\fo^{I\cup U}_\zeta(s_1,s_2)\,}_\beta^{\,c\oplus b}
\ =\ {\hat p}(\zeta)\bigl(\sem{s_1}_{\fstt{int}}^{\,c \oplus b}, 
\sem{s_2}_{\fstt{int}}^{\,c \oplus b}\bigr)
}
\vskip-\medskipamount
\ldisp{\qquad\quad
\ =\ {\hat p}(\zeta) \bigl(\sem{s'_1}_{\fstt{int}}^{\,c \oplus b}, 
\sem{s'_2}_{\fstt{int}}^{\,c \oplus b}\bigr)
\ =\ {\hat p}(\zeta) \bigl(\sem{s'_1}_{\fstt{int}}, 
\sem{s'_2}_{\fstt{int}}\bigr)
}
\vskip-\medskipamount
\rdisp{
\ =\ \fss{o}_\zeta(\sem{s'_1}_{\fstt{int}},\sem{s'_2}_{\fstt{int}})
\ =\ \sem{\,\fss{Eval}_\zeta(s'_1,s'_2)\,}_\beta
\ =\ \sem{\,\fss{Eval}_\zeta(s'_1,s'_2)\,}_\beta^{\,c \oplus b}
\quad}
\smallskip
(with ${\hat p}(\zeta)$ the strict extension of $\fss{o}_\zeta$)
and hence
$\,\fo^{I\cup U}_\zeta(s_1,s_2)\, \equiv\, \fss{Eval}_\zeta(s'_1,s'_2)\,$. \eop

\bigskip\bigskip
{\bf Replacement rule 4:\ }
The final replacement rule involves the $\fss{case}$ operators.
\medskip
Let $\zeta = \ftt{case}^\theta_\beta$ for some
$\beta \in B$, $\theta \in \Bf$. Let
$a \in \assb {K_{\theta,\beta}} {F^{I\cup U}} $ and
$s \in F^{I\cup U}_\theta$ with $s = \fo^{I\cup U}_\kappa(a')$,
where $\kappa \in K$ is of type $J \to \theta$ for some $J \in {\cal F}_B$ and
where $a' \in \assb {J} {F^{I\cup U}} $. Then the fourth rule is:
\bigskip
{\parindent=30pt
\item{(R4)} If $s$ is not undefined then
$\ \fo^{I\cup U}_{\zeta}(s \diamond a )
\, \equiv\, \fo^{I\cup U}_{J\to \beta,J} (a(\kappa) \triangleleft a')\ $.
\bigskip
}
Note that if $J = \varnothing$ (i.e., if $\kappa$ is of type $\varnothing\to \theta$)
then $s = \fo^{I\cup U}_\kappa(\onept)$ and in this case $s$ is automatically
not undefined, since here
$\sem{s}_\theta^{\,c \oplus b} = f_\kappa(\onept)$ for all $c \in \ass I D $, 
$b \in \ass U D $.
\bigskip\medskip

\proclaim{Lemma~7.4.6} The replacement rule (R4) is valid. \endpro 

\proof Let $c \in \fss{Sol}(\alpha)$; then by assumption
\mdisp{f_\kappa\bigl(\,\assb {J} {\sem{a'}^{\,c\oplus b}} \,\bigr)
\ =\ \sem{\,\fo^{I\cup U}_\kappa(a')\,}^{\,c\oplus b}_\theta
\ =\ \sem{s}^{\,c\oplus b}_\theta \ \ne \ \bot_\theta}
and therefore by Proposition~6.2.2~(2) and Proposition~5.4.1~(2) it follows 
that

\smallskip
\ldisp{\quad 
\sem{\,\fo^{I\cup U}_{\zeta}(s \diamond a )\,}^{\,c\oplus b}_\beta
\ =\ \fss{Case}^\theta_\beta 
 \bigl(\sem{s}^{\,c\oplus b}_\theta,
         \assb {K_{\theta,\beta}} {\sem{a}^{\,c\oplus b}} \bigr)    
}
\vskip-\medskipamount
\ldisp{\qquad\qquad
\ =\ \fss{Case}^\theta_\beta
    \bigl(
   f_\kappa(\,\assb {J} {\sem{a'}^{\,c\oplus b}} \,),
         \assb {K_{\theta,\beta}} {\sem{a}^{\,c\oplus b}} \bigr)   
}

\vskip-\medskipamount
\ldisp{\qquad\qquad\qquad\quad
 \ =\  f_{J\to\beta,J}
  \bigl(\, \assb {K_{\theta,\beta}} {\sem{a}^{\,c\oplus b}} \,(\kappa)\,
      \triangleleft \,\assb {J} {\sem{a'}^{\,c\oplus b}} \,\bigr) }

\vskip-\medskipamount
\ldisp{\qquad\qquad\qquad\qquad\qquad
 \ =\    f_{J\to\beta,J}
  \bigl(\, \sem{a(\kappa)}^{\,c\oplus b}_{J\to \beta}\,\triangleleft
      \,\assb {J} {\sem{a'}^{\,c\oplus b}} \,\bigr) 
}
\vskip-\medskipamount
\rdisp{ 
 \ =\    f_{J\to\beta,J}
  \bigl(\, 
     \assb {\sigma\cdot J} {\sem{a(\kappa) \triangleleft a'}^{\,c\oplus b}} \,\bigr) 
\ =\ \sem{\,\fo^{I\cup U}_{J\to \beta,J}
               (a(\kappa) \triangleleft a')\,}^{\,c\oplus b}_\beta
\ .\quad}
\medskip
Hence $\ \fo^{I\cup U}_{\zeta}(s \diamond a )
\, \equiv\, \fo^{I\cup U}_{J\to \beta,J} (a(\kappa) \triangleleft a')\ $. \eop

\bigbreak
If $(D_B,f_K)$ is a fully regular bottomed extension of $(X_B,p_K)$ then the condition 
in rule (R4) that $s$ is not undefined is always satisfied, since here
\mdisp{\sem{s}^d_\theta 
\ =\ \sem{\,\fo^{I\cup U}_\kappa(a')\,}^d_\theta
\ =\ f_\kappa\bigl(\,\assb {L} {\sem{a'}^d} \,\bigr)\ \ne \ \bot_\theta}
for all $d \in \ass {I\cup U} D $.
Therefore if $(D_B,f_K)$ is fully regular and $s$ has 
the form $\fo^{I\cup U}_\kappa(a')$ for some $\kappa \in K_\theta$
then this rule can always be applied
to the element $\fo^{I\cup U}_{\zeta}(s \triangleleft a )$. 
\bigbreak
Of course, in order to be able apply rule (R3) it is necessary to have suitable criteria 
for a term to be grounded, and in the same way  for rule (R4) criteria are needed
for a term to be not undefined. 
\medskip
Let us start with  some rules for establishing that a term is grounded.
First, however, a general remark.
Let $\eta \in L$ be of type $\beta \in B$ and $s \in F_\beta$; then the
statement
\medskip
\centerline{\it let $\eta$ be grounded with value $s$\ }
\medskip
is defined to be equivalent to the statement:
{\it\ let $b \in \ass U D $ be such that $b(\eta) = \sem{s}_\beta$}.
Clearly this requirement can be satisfied, i.e., there
exists an assignment $b \in \ass U D $ such that $b(\eta) = \sem{s}_\beta$.
In fact, if $\lvector {\eta} n $ are $n$ different elements of $L$,
with $\eta_j$ of type $\beta_j \in B$,  and $s_j \in F_{\beta_j}$
for each $\oneto j n $ then there
exists an assignment $b \in \ass U D $ such that $b(\eta_j) = \sem{s_j}_{\beta_j}$
for each $j$. This means that the requirement
\medskip
\centerline{\it let $\eta_j$ be grounded with value $s_j$ for each $\oneto j n $}
\medskip
can be satisfied. 
\medskip
The following rules can be useful for establishing that a term is grounded:
\medskip\smallskip
{\parindent=30pt
\item{(G1)} Let $\beta \in B$; then each $s \in F_\beta \subset F^{I\cup U}_\beta$
is grounded with value $s$.
\medskip
\item{(G2)} Let $\kappa \in K$ be of type $L \to \beta$
and let $a \in \assb L {F^{I\cup U}} $ be such that $a(\eta)$ is grounded for each
$\eta \in L$. Then the term $\fo^{I\cup U}_\kappa(a)$ is grounded with value
$\fo_\kappa(a')$, where $a' \in \ass L F $ is the assignment with
$a'(\eta) \in F_\eta$ the value of $a(\eta)$ for each $\eta \in L$.
In particular, if $\kappa \in K$ is of type $\varnothing \to \beta$ then
the term $\fo^{I\cup U}_\kappa(\onept)$ is grounded with value
$\fo_\kappa(\onept)$. 
\medskip
\item{(G3)} Let $\zeta \in P$ be the name of an integer operator 
and let $s_1,\, s_2 \in F^{I\cup U}_{\fstt{int}}$ be grounded.
Then the term
$\fo^{I\cup U}_\zeta(s_1,s_2)$ is grounded with value
$\fss{Eval}_\zeta(s'_1,s'_2)$, with $s'_j$ the value of $s_j$ for each $j = 1,\,2$. 
\bigskip
}

\proclaim{Lemma~7.4.7} The rules (G1), (G2) and (G3) are valid. \endpro

\proof It is only necessary to consider (G2), since (G1) follows immediately from
Lemma~6.2.1 and (G3) is just a disguised form of (R3).
Let $c \in \fss{Sol}(\alpha)$; then by assumption and by Lemma~6.2.1
$\sem{a(\eta)}_\eta^{\,c\oplus b} = \sem{a'(\eta)}_\eta^{\,c\oplus b}
= \sem{a'(\eta)}_\eta$
for each $\eta \in L$, thus
$\assb L {\sem{a}^{\,c \oplus b}} = \ass L {\sem{a'}} $
and therefore, again making use of Lemma~6.2.1,
\smallskip
\ldisp{\quad\sem{\,\fo^{I\cup U}_\kappa(a)\,}^{\,c \oplus b}_\beta
   \ =\  f_\kappa\bigl(\,\assb L {\sem{a}^{\,c \oplus b}} \,\bigr)
   \ =\  f_\kappa\bigl(\,\ass L {\sem{a'}} \,\bigr)}
\vskip-\medskipamount
\rdisp{ \ =\  \sem{\,\fo_\kappa(a')\,}_\beta
            \ =\  \sem{\,\fo_\kappa(a')\,}^{\,c\oplus b}_\beta\ .\quad}
\smallskip
Hence the term $\fo^{I\cup U}_\kappa(a)$ is grounded with value
$\fo_\kappa(a')$. \eop
\bigbreak

There is only one simple general rule for establishing 
that a term is not undefined, and that is given by the following trivial fact:
\medskip\smallskip
{\parindent=30pt
\item{(N1)} Every grounded term is not undefined.
\medskip\smallskip
}
In particular, (N1) and (G2) imply that if $\kappa \in K$ is of type $L \to \beta$
and $a \in \assb L {F^{I\cup U}} $ is such that $a(\eta)$ is grounded for each
$\eta \in L$ then the term $\fo^{I\cup U}_\kappa(a)$ is not undefined.
As a special case of this, if $\kappa \in K$ is of type $\varnothing \to \beta$ then the
term $\fo^{I\cup U}_\kappa(\onept)$ is not undefined.
\medskip
Less trivial rules for a term to be not undefined will involve the trace $R_B$ of 
$(D_B,f_K)$, and to be more explicit let us assume in what follows that $(D_B,f_K)$ is
\tsym{(H_B,\diamond_K)}-cored for some monotone core type $(H_B,\diamond_K)$. 
(In fact, the trace only enters implicitly below, via the
characterisation given in Proposition~3.4.3.)
\eject
The following rule can be used to establish that a term is not undefined:
\bigbreak
{\parindent=30pt
\item{(N2)} Let $\kappa \in K$ be of type $L \to \beta$ and $a \in \assb L {F^{I\cup U}} $;
let $d \in \ass L H $ be given by
\mdisp{d(\eta)
  \ =\ \cases{\natural_\eta & if $\,a(\eta)\,$ is not undefined,\cr
\noalign{\vskip 1pt}
                \flat_\eta & otherwise.
}}
\item{} If $\diamond_\kappa(d) = \natural_\beta$
then the term $\fo^{I\cup U}_\kappa(a)$ is not undefined. 
\bigskip\medskip
}

\proclaim{Lemma~7.4.8} The rule (N2) is valid. \endpro

\proof For each $\beta \in B$ let
$\varepsilon_\beta : D_\beta \to H_\beta$ be the mapping given by 

\mdisp{\varepsilon_\beta(u) 
\ =\ \cases{\natural_\beta & if $\,u \ne \bot_\beta\,$,\cr
             \flat_\beta & if $\,u = \bot_\beta\,$.}}
\medskip
Then by the definition of $(D_B,f_K)$ being
\tsym{(H_B,\diamond_K)}-cored $\varepsilon_B$ is a homomorphism from
$(D_B,f_K)$ to $(H_B,\diamond_K)$. Let $c \in \fss{Sol}(\alpha)$
and put ${\breve a} = \assb L {\sem{a}^{\,c\oplus b}} $;
if $d(\eta) = \natural_\eta$ then
\mdisp{{\breve a}(\eta)\ =\ \assb L {\sem{a}^{\,c\oplus b}} \,(\eta) 
\ =\ \sem{a(\eta)}_\eta^{\,c\oplus b}\ \ne\ \bot_\eta}  
and hence $\varepsilon_\eta({\breve a}(\eta)) = \natural_\eta$.
But $\diamond_\kappa(d) = \natural_\beta$ and $(H_B,\diamond_K)$ is monotone,
and from this it follows  that
$\diamond_\kappa(\ass L {\varepsilon} ({\breve a}) ) = \natural_\beta$.
Therefore
\mdisp{
\varepsilon_\beta\bigl(\,f_\kappa(\assb L {\sem{a}^{\,c\oplus b}} )\,\bigr)
\ =\ \varepsilon_\beta(f_\kappa({\breve a}))
\ =\ \diamond_\kappa(\ass L {\varepsilon} ({\breve a})\ =\ \natural_\beta\ , 
}
which implies that
$\,\sem{\,\fo^{I\cup U}_\kappa(a)\,}_\beta^{\,c\oplus b}
= f_\kappa(\assb L {\sem{a}^{\,c\oplus b}} ) \ne \bot_\beta\,$.
This shows that the term
$\fo^{I\cup U}_\kappa(a)$ is not undefined. \eop
\bigbreak
If $\diamond_K = \diamond^\top_K$ (which just means that $(D_B,f_K)$ is fully regular)
then the condition in (N2) is automatically satisfied. Thus in this case a term of the
form $\fo^{I\cup U}_\kappa(a)$ is always not undefined, a fact that was
already noted after the statement of rule (R4).
\bigskip
As well as the rules introduced in this section there is are two further ingredients
which are needed to make non-trivial assertions about the solutions of equations.
The first is a form of what is called
\indexddef{structural induction}{structural induction}{}{induction}{structural},
which for our purposes can be stated as follows:
For each $\beta \in B$ let $P_\beta : F_\beta \to \Bool$ be a mapping
and suppose the family $P_B$ satisfies the two conditions:
\medskip\smallskip
{\parindent=25pt
\item{(1)} If $\kappa \in K$ is of type $\varnothing \to \beta$ then
$P_\beta(\fo_\kappa(\onept)) = T$. 
\medskip
\item{(2)} If $\kappa \in K$ is of type $L \to \beta$
with $L \ne \varnothing$ and $a \in \ass L F $ is an assignment
with $P_\eta(a(\eta)) = T$ for each $\eta \in L$ then
$P_\beta(\fo_\kappa(a)) = T$.
\medskip\smallskip
}
Then $P_\beta(s) = T$ for all $s \in F_\beta$, $\beta \in B$.
This property holds, of course, because the ground term algebra $(F_B,\fo_K)$ is 
minimal (and it is, in fact, equivalent to  $(F_B,\fo_K)$ being minimal). 
\medskip
The second ingredient is needed when making assertions about terms of type
$\ftt{int}$ , and this is just the usual principle of mathematical induction (i.e., 
the particular property of the subset $\Nat$ of $\Int = X_{\fstt{int}}$).

\eject
\bigskip\bigskip
\frame{20pt}{\bigskip 
{\it Example~7.4.2\enspace} Note that the set $E_{\fstt{int}}$ is essentially equal 
to $\SynInt$, 
since each element of $E_{\fstt{int}}$ is a list whose single component is an element of
$\SynInt$. As preparation for the further analysis of the
equations in Example~7.4.1 it is useful to represent the elements in
the set $E_{\fstt{list}}$ by `real' lists from the set $\SynInt$. More precisely,
let $\varphi : \SynInt^* \to E_{\fstt{list}}$ be the mapping defined by putting
$\,\varphi(\varepsilon) \,=\, \ftt{Nil}\,$ and
\mdisp{ \varphi( \list a n )
\ =\ \ftt{Cons } a_1 \ftt{ (Cons } a_2 \ftt{ (Cons } \cdots \ftt{ (Cons } a_n 
  \ftt{ Nil)}\cdots \ftt{))}} 
whenever $n \ge 1$. Then $\varphi$ is a bijection, and it is convenient to
identify the element $e \in E_{\fstt{list}}$ with the list $\varphi^{-1}(e)$.
Moreover, the list $\list a n $ will here be written as
$\ftt{[} a_1 \ftt{,} a_2 \ftt{,}\,\ldots\, \ftt{,} a_n \ftt{]}$ (with $\varepsilon$
written as $\ftt{[]}$).
\medskip
By rule (R4)
\smallskip
\mdisp{
\ftt{case Nil of \lcb Nil -> b, Cons -> shunx b\rcb}\ \equiv\ \ftt{b}}
\smallskip
which, together with one of the deductions in Example~7.4.1, implies that
\smallskip
\mdisp{  \ftt{shunt Nil b}
\ \equiv\ \ftt{b}}
\smallskip
Suppose now that $\ftt{n}$ and
$\ftt{a}$ are both grounded. Then by (G2) $\ftt{Cons n a}$ is also grounded, and so in
particular by (N1) it is not undefined. Thus by (R4) 
\smallskip
\ldisp{\qquad\quad 
\ftt{case (Cons n a) of \lcb Nil -> b, Cons -> shunx b\rcb}}
\vskip-\medskipamount
\rdisp{ \ \equiv\ \ftt{shunx b n a}\qquad\quad}
\smallskip
Combining this with the deductions in Example~7.4.1, it follows that 
\smallskip
\mdisp{  \ftt{shunt (Cons n a) b}\ \equiv\ \ftt{shunt a (Cons n b)}}
\smallskip
Now let $\ftt{[} a_1 \ftt{,} a_2 \ftt{,}\,\ldots\, \ftt{,} a_n \ftt{]}$ and
$\ftt{[} b_1 \ftt{,} b_2 \ftt{,}\,\ldots\, \ftt{,} b_m \ftt{]}$ be elements
of $E_{\fstt{list}}$ (with $m,\, n \ge 0$) and $a_0 \in E_{\fstt{int}}$.
Let $\ftt{a}$, $\ftt{b}$ and $\ftt{n}$ 
be grounded to have respectively these three values.
Then by (G2) it follows that $\ftt{Cons n a}$ is grounded with the value
$\ftt{[} a_0 \ftt{,} a_1 \ftt{,} a_2 \ftt{,}\,\ldots\, \ftt{,} a_n \ftt{]}$ and
$\ftt{Cons n b}$ is grounded with the value
$\ftt{[} a_0 \ftt{,} b_1 \ftt{,} b_2 \ftt{,}\,\ldots\, \ftt{,} b_m \ftt{]}$.
Thus with several applications of (R1) 
\smallskip
\ldisp{\qquad  \ftt{shunt }\ftt{[} a_0 \ftt{,} a_1 \ftt{,} a_2 \ftt{,}\,\ldots\, 
   \ftt{,} a_n \ftt{] [} b_1 \ftt{,} b_2 \ftt{,}\,\ldots\, \ftt{,} b_m \ftt{]}}
\vskip-\medskipamount
\rdisp{\equiv\  \ftt{shunt }\ftt{[} a_1 \ftt{,} a_2 \ftt{,}\,\ldots\, 
   \ftt{,} a_n \ftt{] [} a_0 \ftt{,} b_1 \ftt{,} b_2 \ftt{,}\,\ldots\, \ftt{,} b_m \ftt{]}
\qquad}
\smallskip
and also with two applications of (R1) it follows that
\mdisp{\ftt{shunt }\ftt{[] [} b_1 \ftt{,} b_2 \ftt{,}\,\ldots\, \ftt{,} b_m \ftt{]}
\ \equiv\ \ftt{[} b_1 \ftt{,} b_2 \ftt{,}\,\ldots\, \ftt{,} b_m \ftt{]}}
\medskip
{\it (Example~7.4.2 is continued on the next page.)}
\bigskip\smallskip
}

\vfill\eject

\frame{20pt}{\bigskip 
{\it Example~7.4.2 (continued)\enspace} 
Define a mapping $P_{\fstt{list}} : E_{\fstt{list}} \to \Bool$ by
letting 
$P_{\fstt{list}}(\ftt{[} a_1 \ftt{,} a_2 \ftt{,}\,\ldots\, \ftt{,} a_n \ftt{]}) = T$ if and only if
\ldisp{\qquad  \ftt{shunt }\ftt{[} a_1 \ftt{,} a_2 \ftt{,}\,\ldots\, 
   \ftt{,} a_n \ftt{] [} b_1 \ftt{,} b_2 \ftt{,}\,\ldots\, \ftt{,} b_m \ftt{]}}
\vskip-\medskipamount
\rdisp{\equiv\  \ftt{[} a_n \ftt{,} a_{n-1} \ftt{,}\,\ldots\, 
   \ftt{,} a_1 \ftt{,} b_1 \ftt{,} b_2 \ftt{,}\,\ldots\, \ftt{,} b_m \ftt{]}
\qquad}
\smallskip
for all 
$\ftt{[} b_1 \ftt{,} b_2 \ftt{,}\,\ldots\, \ftt{,} b_m \ftt{]} \in E_{\fstt{list}}$.
Moreover, for each $\beta \in B \setminus \{\ftt{list}\}$ let 
$P_\beta : E_\beta \to \Bool$ be the constant mapping with the value $T$.
\smallskip
By the last assertion on the previous page $P_{\fstt{list}}(\ftt{[]}) = T$.
Now let $a$ be an element of $E_{\fstt{int}}$ and 
$s = \ftt{[} a_1 \ftt{,} a_2 \ftt{,}\,\ldots\, \ftt{,} a_n \ftt{]} \in E_{\fstt{list}}$
with $P_{\fstt{list}}(s) = T$. Then by the penultimate assertion on the previous
page
\smallskip
\ldisp{\qquad  \ftt{shunt }\ftt{[} a \ftt{,} a_1 \ftt{,} a_2 \ftt{,}\,\ldots\, 
   \ftt{,} a_n \ftt{] [} b_1 \ftt{,} b_2 \ftt{,}\,\ldots\, \ftt{,} b_m \ftt{]}}
\vskip-\medskipamount
\mdisp{\equiv\  \ftt{shunt }\ftt{[} a_1 \ftt{,} a_2 \ftt{,}\,\ldots\, 
   \ftt{,} a_n \ftt{] [} a \ftt{,} b_1 \ftt{,} b_2 \ftt{,}\,\ldots\, \ftt{,} b_m \ftt{]}
}
\vskip-\medskipamount
\rdisp{\equiv\  \ftt{[} a_n \ftt{,} a_{n-1} \ftt{,}\,\ldots\, 
   \ftt{,} a_1 \ftt{,} a \ftt{,} b_1 \ftt{,} b_2 \ftt{,}\,\ldots\, \ftt{,} b_m \ftt{]}
\qquad}
\smallskip
for all 
$\ftt{[} b_1 \ftt{,} b_2 \ftt{,}\,\ldots\, \ftt{,} b_m \ftt{]} \in E_{\fstt{list}}$.
This implies that $P_{\fstt{list}}( \blob_{\fstt{Cons}}\ a\ s) = T$.
Thus by structural induction $P_\beta(s) = T$ for all $s \in E_\beta$, $\beta \in B$,
and in particular $P_{\fstt{list}}(s) = T$ for all $s \in E_{\fstt{list}}$, i.e.,
\smallskip
\ldisp{\qquad  \ftt{shunt }\ftt{[} a_1 \ftt{,} a_2 \ftt{,}\,\ldots\, 
   \ftt{,} a_n \ftt{] [} b_1 \ftt{,} b_2 \ftt{,}\,\ldots\, \ftt{,} b_m \ftt{]}}
\vskip-\medskipamount
\rdisp{\equiv\  \ftt{[} a_n \ftt{,} a_{n-1} \ftt{,}\,\ldots\, 
   \ftt{,} a_1 \ftt{,} b_1 \ftt{,} b_2 \ftt{,}\,\ldots\, \ftt{,} b_m \ftt{]}
\qquad}
\smallskip
for all $\lvector a n ,\,\lvector b m \in E_{\fstt{int}}$, $m,\, n \ge 0$.
\medskip
The equation for $\ftt{revs}$ will now be looked at.
Let $m \in \Int$ and let $\ftt{n}$ be grounded with value 
${\underline m} \in \SynInt = E_{\fstt{int}}$.
If $m \ne 0$ then by (G1) and (R3)
\mdisp{\ftt{eq n 0}\ \equiv\ 
\fss{Eval}_{\fstt{eq}}({\underline m},\ftt{0})\ =\ \ftt{False}} 
and so by (R1)
\ldisp{\qquad 
 \ftt{case (eq n 0) of \lcb True -> Nil,}} 
\vskip-\medskipamount
\vskip-\smallskipamount
\rdisp{ \ftt{False -> Cons (sq n) (revs sq (sub n 1))\rcb}\qquad\qquad}
\vskip-\medskipamount
\ldisp{\qquad\qquad 
  \ \equiv\ \ftt{case False of \lcb True -> Nil,}} 
\vskip-\medskipamount
\vskip-\smallskipamount
\rdisp{ \ftt{False -> Cons (sq n) (revs sq (sub n 1))\rcb}\qquad}
\smallskip
Moreover, by (R4)
\ldisp{\qquad
\ftt{case False of \lcb True -> Nil,}} 
\vskip-\medskipamount
\vskip-\smallskipamount
\rdisp{ \ftt{False -> Cons (sq n) (revs sq (sub n 1))\rcb}\qquad\qquad\qquad}
\vskip-\medskipamount
\rdisp{\equiv\  \ftt{Cons (sq n) (revs sq (sub n 1))\rcb}\qquad}
\smallskip
{\it (Example~7.4.2 is continued on the next page.)}
\bigskip\medskip
}

\vfill\eject

\bigskip
\frame{20pt}{\bigskip 
{\it Example~7.4.2 (continued)\enspace} 
But this, together with one of the deductions in Example~7.4.1, implies that
\mdisp{\ftt{revs sq n}\ \equiv\  \ftt{Cons (sq n) (revs sq (sub n 1))}}
\smallskip
Now by (R3) and a deduction from Example~7.4.1 
it follows that
\mdisp{ \ftt{sq n}\ \equiv\ \ftt{mul n n}
\ \equiv\ \fss{Eval}_{\fstt{mul}}({\underline m},{\underline m})
\ =\ {\underline {m\times m}}}
and by (R3)
$\,\ftt{sub n 1}\,\equiv\,\fss{Eval}_{\fstt{sub}}({\underline m},1)
\,=\,{\underline {m-1}}\,$. 
Therefore applying (R1) three times gives us that if $m \ne 0$ then
\mdisp{\ftt{revs sq }{\underline m}
\ \equiv\  \ftt{Cons }{\underline {m\times m}} \ftt{ (revs sq } {\underline {m-1}}
\ftt{)}}
Moreover, if $m = 0$ then a much easier argument implies that
\mdisp{\ftt{revs sq }{\underline m}\ \equiv\ \ftt{Nil}}
From these two assertions it can be shown by induction on $m$ that
\mdisp{\ftt{revs sq }{\underline m}
\ \equiv\  \ftt{[}\,{\underline {m\times m}}\,\ftt{,}
\ \cdots\ {\underline {2\times 2}}\,\ftt{,}\,
 {\underline {1 \times 1}}\, \ftt{]}}
for all $m \ge 0$.
\medskip
Finally,
the equation for $\ftt{revs}$ will be looked at.
Let $m \ge 0$ and let
$\ftt{n}$ be grounded with value ${\underline m} \in \SynInt = E_{\fstt{int}}$.
Thus by the last deduction in Examples~7.4.1 and two applications of (R1)
it follows that
\smallskip
\mdisp{
\ftt{sqs }{\underline m}\ \equiv\ \ftt{shunt (revs sq }{\underline m}\ftt{ ) []}}
\smallskip
Therefore by (R1) and the deduction made on the previous page
\smallskip
\mdisp{
\ftt{sqs }{\underline m}\ \equiv\ \ftt{shunt [}\,{\underline {m\times m}}\,\ftt{,}
\ \cdots\ {\underline {2\times 2}}\,\ftt{,}\,
 {\underline {1 \times 1}}\, \ftt{] []}}
\smallskip
Hence by the deductions made about $\ftt{shunt}$ it follows that
\smallskip
\mdisp{
\ftt{sqs }{\underline m}\ \equiv\ \ftt{[}\,{\underline {1 \times 1}}
\,\ftt{,}\,{\underline {2\times 2}}\,\ftt{,}
\ \cdots\ {\underline {m\times m}}\, \ftt{]}}
\smallskip
for all $m \ge 0$, which imples that
\smallskip
\ldisp{\qquad c_{\fstt{sqs}}(m)\ =\ \sem{\,\ftt{sqs }{\underline m}\,}^c_{\fstt{list}}
}
\vskip-\medskipamount
\rdisp{
\ =\ \sem{\,\ftt{[}\,{\underline {1 \times 1}}
\,\ftt{,}\,{\underline {2\times 2}}\,\ftt{,}
\ \cdots\ {\underline {m\times m}}\, \ftt{]}\,}^c_{\fstt{list}}
\ =\ 1\ 4\ \cdots\ m^2\qquad}
\smallskip
for all $m \ge 0$ for each $c \in \fss{Sol}(\alpha)$.
\bigskip\medskip
}
\bigskip

\vfill\eject

\hfline{211}{7.5 \ NOTES}

\sectionhead {7.5} {Notes}
\bigskip\medskip
\medskip
As already mentioned in the Preface, the approach to equations presented here
is an instance of the `initial algebra semantics' philosophy propagated by the 
ADJ group, for example in the papers Goguen, Thatcher, Wagner and 
Wright (1977) and Goguen, Thatcher and Wagner (1978).
Note that in this framework the bottomed extension only really influences 
the solutions of equations via the `case' operators. 
\medskip

The algorithm presented in Section~7.3 can be seen as giving an
operational semantics to the rudimentary programming language: The meaning of a term is
the meaning of the ground term computed by the algorithm. 
In many accounts, for example Chapter~9 of Winskel (1993), the operational
semantics consists of a list of rules for evaluating terms.
The algorithm in Section~7.3 is based on these rules; in 
addition it also specifies which rule to apply when more than one  
could be applied to a given term. 
\medskip
The algorithm could of course be used to actually implement the language.
The reader interested in how functional programming languages are  
implemented in practice is recommended to consult Peyton Jones (1987) and
Peyton Jones and Lester (1991).

\vfill\eject
\hfline{212}{8.1 \ SOLUTIONS AS FIXED-POINTS}

\bigskip
{\fourteenbf Chapter 8\quad Completeness of the algorithm}
\bigskip\bigskip
For the whole of the chapter 
let $\alpha = s_I \in \fss{Eq}(I)$ be a system of equations 
(so $I$ is a non-empty finite global \tsym{S}-typed set)
and let $\Phi_B$ be the \tsym{\Lambda}-homomorphism defined by $\alpha$
(i.e., defined by $\alpha$ and the support system 
$\bterm{F}^I_B$). 
Recall from Section~7.2 that the family $F^\Phi_B \subset F^I_B$ is defined by 

\mdisp{F^\Phi_\beta \ =\ \lcurl s \in F^I_\beta \,:\,
        \Phi^n_\beta(s) \in F_\beta \frm{for some} n \ge 0 \rcurl}
for each $\beta \in B$, and if $\fss{Sol}(\alpha)$ is non-empty then
the family $F^\alpha_B \subset F^I_B$ is defined by 

\mdisp{F^\alpha_\beta\ =\ \lcurl s \in F^I_\beta
    \,: \frm{there exists} x \in X_\beta \frm{such that}
   \sem{s}^c_\beta  = x \frm{for all} 
                          c \in \fss{Sol}(\alpha) \rcurl}

for each $\beta \in B$. Two of the results of this chapter (Propositions 8.1.3 and
8.2.3) imply that in Case~3 the system of equations $\alpha$ has a solution, and that  
$\Phi_B$ is \tsym{\alpha}-complete, i.e.,
$F^\Phi_B = F^\alpha_B$.
\medskip
The main result of the chapter is Proposition~8.2.1.
This applies to both Case~3 and Case~2,
and in Case~3 it immediately implies the \tsym{\alpha}-completeness of $\Phi_B$.
The proof of this result
rely heavily on a technique which goes under the name of the 
\indexdef{method of logical relations}{method of logical relations}{}.
\medskip

\bigskip\bigskip\bigskip

\sectionhead {8.1} {Solutions as fixed-points} 
\bigskip\medskip
For the moment we continue with the set-up employed in Chapters 6 and 7.
Thus $\fss{C}$ is a bottomed concrete category with finite 
products and $({\bf D}_S,f_N)$ is a \tsym{\fss{C}}-based \tsym{\Sigma}-algebra 
such that $(D_B,f_K)$ is a monotone regular bottomed extension
of $(X_B,p_K)$. Moreover, it is assumed that the \tsym{\Sigma}-algebra $(D_S,f_N)$ has
the properties (A), (B), (C) and (D) listed in Section~5.4.
\medskip
We need to make an additional assumption about the category $\fss{C}$, namely that
constant mappings are morphisms. More precisely, if $X,\, Y \in {\cal C}$
and $y \in |Y|$ then we will assume that the mapping $x \mapsto y$ from $|X|$ to
$|Y|$ is an element of $\Hom(X,Y)$. This requirement is clearly met in each of the
three basic cases. 
\medskip
The starting point for both the proof of Proposition~8.2.3 and the analysis
of Case~2 is to represent
the set of solutions $\fss{Sol}(\alpha)$ as the set of fixed-points
of a certain  mapping $U_\alpha : \ass I D \to \ass I D $. 
In order to define $U_\alpha$ the following fact is needed:

\proclaim{Lemma~8.1.1} Let $J$ be a non-empty finite \tsym{S}-typed set disjoint from
$P$. Then for each $\sigma \in S$ and each $s \in F^{J}_\sigma$ the mapping
$\ c\ \mapsto\ \sem{s}^c_\sigma\ $  
from $\ass J D $ to $D_\sigma$ is a morphism, i.e., an element of
$\Hom(\ass J {\bf D} ,{\bf D}_\sigma)$. 
\endpro

\proof For each $s \in F^J_\sigma$ define a mapping $h_s : \ass J D \to D_\sigma$
by letting $\,h_s(c)\, =\, \sem{s}^c_\sigma\,$ for each $c \in \ass J D $, and for each 
$\sigma \in S$ let ${\grave F}^J_\sigma$ be the set of elements $s \in F^J_\sigma$ for 
which $h_s$ is a morphism, i.e., an element of $\Hom(\ass J {\bf D} ,{\bf D}_\sigma)$. 
It will be shown that the family ${\grave F}^J_S$ satisfies the hypotheses of
Proposition~6.1.5, and thus that ${\grave F}^J_S = F^J_S$.
\medskip
First consider $\xi \in J\cup P$. If $\xi \in P$ then
$\,h_\xi(c) \, =\, \sem{\xi}_\xi^c\, =\, p(\xi)\,$
for each $c \in \ass J D $; this is a constant mapping, which by assumption is a
morphism. On the other hand, if $\xi \in J$ then
$\,h_\xi(c) \, =\, \sem{\xi}_\xi^c\, =\, c(\xi)\,$
for each $c \in \ass J D $; hence $h_\xi$ is the mapping $p_\xi$ occurring in (P2)
in the definition of $\ass J {\bf D} $, and so in particular $h_s$ is a morphism.
Thus in both cases
$h_\xi \in {\grave F}^J_\xi$. This shows that $\xi \in {\grave F}^J_\xi$ for each 
$\xi \in J\cup P$, and hence that ${\grave F}^J_S$ satisfies 
condition (1) in Proposition~6.1.5.
\medskip
Next consider $\kappa \in K$ having type 
$L \to \beta$, let $a \in \assb L {{\grave F}^J} $ (which means that
$h_{a(\xi)}$ is a morphism for each $\xi \in L$)
and put 
$s = \fo^J_\kappa(a)$. Then for each $c \in \ass J D $

\mdisp{h_s(c)\ =\ \sem{\,\fo^J_\kappa(a)\,}_\beta^c
 \ =\ f_\kappa\bigl(\assb {L} {\sem{a}^c} \bigr)
\ =\ f_\kappa(\gamma(c))\ , }

where $\gamma : \ass J D \to \ass L D $ is the mapping given 
for each $c \in \ass J D $, $\xi \in L$ by

\mdisp{\gamma(c)\,(\xi) \ =\ \sem{\,a(\xi)\,}_\xi^c 
 \ =\ h_{a(\xi)}(c)\ .}

But this means that $p_\xi \circ \gamma = h_{a(\xi)}$ for each $\xi \in L$,
where $p_\xi$ is the morphism occurring in (P2) in the definition of
$\ass J {\bf D} $, and thus by Lemma~5.1.2 $\gamma$ is a morphism.
Hence $\,h_s = f_\kappa\circ\gamma\,$, as the composition of two morphisms, 
is itself a morphism, i.e., $h_s \in {\grave F}^J_\beta$.
(The reader should check that the argument above still makes sense when $L = \varnothing$.)
This shows that the family ${\grave F}^J_B$ is invariant in the \tsym{\Lambda}-algebra
$(F^J_B,\fo^J_K)$, i.e., condition (2) in Proposition~6.1.5 is satisfied. 
\medskip
Now let $\xi \in J\cup P$ be of type $\sigma = L \to \beta$,
let $L'$ be a non-empty subset of $L$ and let 
$a \in \assb {L'} {{\grave F}^J} $. Put
$s = \fo^J_{\sigma,L'}(\xi \triangleleft a) $. Then
for all $c \in \ass J D $

\mdisp{h_s(c) 
    \ =\ \sem{\,\fo^J_{\sigma,L'}(\xi \triangleleft a)\,}_{\sigma_{L\setminus L'}}^c
  \ =\ f_{\sigma,L'} \bigl(\,\sem{\xi}^c_\sigma 
                      \triangleleft \assb {L'} {\sem{a}^c} \,\bigr)
   \ =\ f_{\sigma,L'}(h_\xi(c) \triangleleft \gamma(c))\ ,}

where here
$\gamma : \ass J D \to \ass {L'} D $ is the mapping given 
for each $c \in \ass J D $, $\eta \in L'$ by

\mdisp{\gamma(c)\,(\eta) \ =\ \sem{\,a(\eta)\,}^c_\eta \ =\  h_{a(\eta)}(c)\ .} 

Thus
$\,h_s \,=\, f_{\sigma,L'}\circ \fss{t}^{L'}_\sigma \circ
(h_\xi \times \gamma) \circ \delta\,$,
where $\delta : \ass J D \to \ass J D \times \ass J D $ is the mapping given by
$\delta(c) = (c,c)$ for all $c \in \ass J D $. 
But by Lemma~5.1.2 $\delta$ is a morphism, it was shown above that $h_\xi$ is a morphism, 
and the same proof as above implies that $\gamma$ is a morphism.
Hence by Lemmas 5.1.3 and 5.2.2 $h_s$ is a morphism, i.e., 
$h_s \in {\grave F}^J_{\sigma_{L\setminus L'}}$.
Condition (3) in Proposition~6.1.5 is therefore satisfied.
\medskip
This shows that
the family ${\grave F}^J_S$ satisfies the hypotheses of Proposition~6.1.5
and hence that ${\grave F}^J_S = F^J_S$. \eop
\bigbreak
For the remainder of the section we specialise to the set-up considered at the beginning of
Section~5.4. We thus assume now that $\fss{C}$ is a bottomed concrete cartesian closed 
category and that $({\bf D}_S,f_N)$ is the \tsym{\fss{C}}-based functional
\tsym{\Sigma}-algebra derived from $({\bf D}_B,f_K)$ in 
$\fss{C}$. 
\medskip
Let us continue to assume that constant mappings are morphisms, which means that
Lemma~8.1.1 is still valid.
\medskip
Consider $\xi \in I$ of type $\sigma = L \to \beta$ with $L \ne \varnothing$;
then by Lemmas 8.1.1 and 5.1.6 the mapping
$\,(c,b) \mapsto \sem{s_\xi}_\beta^{\,c \oplus b}\,$ from
$\ass I D \times \ass L D $ to $D_\beta$ is a morphism. Thus by (E3),
and since $\varepsilon(\ass L {\bf D} ,{\bf D}_\beta) = {\bf D}_\sigma$, there
exists a unique morphism $U_\xi \in \Hom(\ass I {\bf D} ,{\bf D}_\sigma)$ such that

\mdisp{U_\xi(c)\,(b)\ =\ \sem{s_\xi}_\beta^{\,c \oplus b}}  

for all $c \in \ass I D $, $b \in \ass L D $.
Moreover, if
$\xi \in I$ is of type $\beta \in B$ then by Lemma~8.1.1 the mapping
$U_\xi : \ass I D \to D_\beta$ defined by letting
$\,U_\xi(c) \,=\, \sem{s_\xi}^{c}_\beta\,$ for each $c \in \ass I D $
is a morphism, i.e., an element of $\Hom(\ass I {\bf D} ,{\bf D}_\beta)$.
Combining these mappings then results in the
mapping $U_\alpha : \ass I D \to \ass I D $ defined by

\mdisp{U_\alpha(c)\,(\xi) \ =\ U_\xi(c) }

for all $c \in \ass I D $, $\xi \in I$.

\proclaim{Proposition~8.1.1} The mapping $U_\alpha$ is a morphism, i.e., an element of
$\Hom(\ass I {\bf D} ,\ass I {\bf D} )$. \endpro

\proof This follows immediately from Lemma~5.1.2, since
$p_\xi \circ U_\alpha = U_\xi$ for each $\xi \in I$, with $p_\xi$ the morphism
occurring in (P2) in the definition of $\ass I {\bf D} $. \eop
\bigbreak

\proclaim{Proposition~8.1.2} The set of solutions $\fss{Sol}(\alpha)$
of the system of equations $\alpha$ is exactly the set of fixed-points
of the mapping $U_\alpha : \ass I D  \to \ass I D $. In other words, 
\mdisp{\fss{Sol}(\alpha)
 \ =\ \lcurl c \in \ass I D \,:\, U_\alpha(c) = c \rcurl\ .}
\endpro

\proof By definition $U_\alpha(c) = c$ if and only if $U_\xi(c) = c(\xi)$ for each 
$\xi \in I$, and clearly $U_\xi(c) = c(\xi)$ if and only if 
$c \in \fss{Sol}(s_\xi)$. Thus $U_\alpha(c) = c$ if and only if 
$c \in \fss{Sol}(\alpha)$.  \eop
\bigbreak

\proclaim{Proposition~8.1.3} In Case~3 
the set $\fss{Sol}(\alpha)$ is non-empty. \endpro

\proof In this case $\ass I D $ is a bottomed complete poset and by Proposition~8.1.1
the mapping $U_\alpha : \ass I D \to \ass I D $ is continuous. Thus by
Propositions 4.2.9 and 8.1.2 the set of solutions $\fss{Sol}(\alpha)$ is non-empty.
\eop
\bigbreak

\vfill\eject
\hfline{215}{8.2 \ THE KLEENE SEQUENCE}

\sectionhead {8.2} {The Kleene sequence} 
\bigskip\medskip
In the present section we will work with a set-up based on that employed in Chapters 
6 and 7, but with $\fss{C}$ the bottomed concrete category with finite products occurring
in Case~2. 
\medskip
Thus in what follows $\fss{C}$ is the category having as objects the class ${\cal C}$ of 
bottomed posets and in which the underlying set $|X|$ of the object $X$ is the set 
obtained by forgetting the bottom element and the partial order, $\Hom(X,Y)$ is the set 
of all monotone mappings from $|X|$ to $|Y|$ and  
$\otimes : {\cal F}_{\cal C} \to {\cal C}$ is defined as in Section~5.1.
Of course, in this category constant mappings are morphisms.
\medskip
We assume that $({\bf D}_S,f_N)$ is a \tsym{\fss{C}}-based \tsym{\Sigma}-algebra 
such that $(D_B,f_K)$ is a monotone regular bottomed extension
of $(X_B,p_K)$, and that the \tsym{\Sigma}-algebra $(D_S,f_N)$ has
the properties (A), (B), (C) and (D). 
\medskip
In addition we assume that the family $\sqle_B$ is an ordering associated with $(D_B,f_K)$,
(where $\sqle_\sigma$ denotes the partial order order on $D_\sigma$ for each
$\sigma \in S$).
\medskip
Of course, Case~2 fits into this set-up, but it can also be used to make
statements about Case~3: Let $\fss{C}'$ be the bottomed concrete cartesian closed category 
in Case~3 and let $({\bf D}'_S,f'_N)$ be the corresponding \tsym{\fss{C}'}-based functional 
\tsym{\Sigma}-algebra. 
Then $({\bf D}'_S,f'_N)$ is also a \tsym{\fss{C}}-based \tsym{\Sigma}-algebra 
and the \tsym{\Sigma}-algebra $(D'_S,f'_N)$ satisfies all the requirements made above.
This situation will be referred to as Case~\tsym{3'}.
\medskip
The final assumption we make in this section is that a mapping $U_\alpha$ is given
which is essentially the mapping occurring in Section~8.1. More precisely, it is assumed
that a monotone mapping $U_\alpha : \ass I D \to \ass I D $ is given such that:
\bigskip
{\parindent=22pt
\item{(1)} If $\xi \in I$
is of type $\sigma = L \to \beta$ with $L \ne \varnothing$ then  
\mdisp{f_{\sigma,L}\bigl(\, U_\alpha(c)\,(\xi) \,\triangleleft\, b \,\bigr)
                          \ =\ \sem{s_\xi}_\beta^{\,c \oplus b}}  
\item{} for all $c \in \ass I D $, $b \in \ass L D $.
\medskip
\item{(2)} If $\xi \in I$ is of type $\beta \in B$ then
$\,U_\alpha(c)\,(\xi) \,=\, \sem{s_\xi}^{c}_\beta\,$ for each $c \in \ass I D $. 
\bigskip
}
In particular, in Cases 2 and \tsym{3'} these two conditions actually define the mapping 
$U_\alpha$,
which is then the mapping occurring in Section~8.1. Note that each fixed-point of 
$U_\alpha$ will be a solution of the equations; moreover, in Cases 2 and \tsym{3'} the converse 
also holds.

\medskip
Now define a sequence $\{k_n\}_{n \ge 0}$ of elements of $\ass I D $ 
by starting with $k_0 = \bot^{\!I}$ and letting $k_{n+1} = U_\alpha(k_n)$ for each 
$n \ge 0$. This sequence $\{k_n\}_{n\ge 0}$
is called the 
\indexdef{Kleene sequence}{kleene sequence}{} 
of $\alpha$.

\proclaim{Lemma~8.2.1} The sequence $\{k_n\}_{n\ge 0}$ is monotone, i.e.,
$k_n\ \ass I {\sqle} \ k_{n+1}$ for all $n \ge 0$. \endpro

\proof Clearly $k_0 = \bot^{\!I} \ \ass I {\sqle} \  k_1$, and if
$k_n\ \ass I {\sqle} \ k_{n+1}$ for some $n \ge 0$ then, since $U_\alpha$ is 
monotone, 
$k_{n+1} = U_\alpha(k_n)\ \ass I {\sqle} \ U_\alpha(k_{n+1}) = k_{n+2}$. 
Therefore by induction $k_n\ \ass I {\sqle} \ k_{n+1}$ for all $n \ge 0$. \eop
\bigbreak

In the special case when  $D_S$ is a family of complete bottomed posets and the mapping 
$U_\alpha$ is continuous then the proof of Proposition~4.2.9 shows that
the least upper bound
$k_\infty$ of the directed set $\lcurl k_n\,:\, n \ge 0 \rcurl$
is the least fixed-point of the mapping $U_\alpha$.
In particular, $k_\infty$ is a then a solution of the equations (and in Case~\tsym{3'} it is the 
least solution).
\medskip
As in Section~6.3 let $\varrho^I_S$ denote the unique \tsym{\Sigma}-homomorphism from
$(F^I_S,\fo^I_N)$ to $(D_S,f_N)$ such that $\varrho^I_\xi(\xi) = \bot_\xi$ for each 
$\xi \in I\cup P$. The next result is really a generalisation of Lemma~6.3.6:

\proclaim{Lemma~8.2.2} 
$\,\varrho^I_\beta(s) \sqle_\beta \pi_\beta(s)\,$
for all $s \in F^I_\beta$, $\beta \in B$ for any \tsym{\Lambda}-homomorphism 
$\pi_B : (F^I_B,\fo^I_K) \to (D_B,f_K)$. \endpro

\proof For each $\beta \in B$ put
$\,{\grave F}^I_\beta \,=\, \lcurl s \in F^I_\beta \,:\, 
           \varrho^I_\beta(s) \sqle_\beta \pi_\beta(s) \rcurl\,$.
Now if $\kappa \in K$ is of type $L \to \beta$ and $a \in \assb L {{\grave F}^I} $ then

\mdisp{\varrho^I_\beta(\,\fo^I_\kappa(a)\,)\ =\ f_\kappa(\assb L {\varrho^I} (a))
\ \sqle_\beta\ f_\kappa(\ass L {\pi} (a))\ =\ \pi_\beta(\,\fo^I_\kappa(a)\,)}

(since the mapping $f_\kappa$ is monotone), and thus 
$\fo^I_\kappa(a) \in {\grave F}^I_\beta$. This implies that the family ${\grave F}^I_B$ 
is invariant in $(F^I_B,\fo^I_K)$. But in the proof of Lemma~6.3.6 it was shown that 
$\varrho^I_\beta(s) = \bot_\beta$ for each
$s \in \aterm{F^I}_\beta \cup I_\beta$, and hence
$\aterm{F^I}_\beta \cup I_\beta \subset {\grave F}^I_\beta$ for each $\beta \in B$.
Therefore by Lemma~6.2.2 ${\grave F}^I_B = F^I_B$. \eop

\bigbreak

\proclaim{Lemma~8.2.3}
$\,\varrho^I_\beta(s) \,\sqle_\beta\, \varrho^I_\beta(\Phi_\beta(s))\,$
for all $s \in F^I_\beta$, $\beta \in B$. \endpro

\proof This follows immediately from Lemma~8.2.2 and Proposition~2.2.1. \eop
\bigbreak

\proclaim{Lemma~8.2.4} The sequences $\{\sem{s}^{k_n}_\beta\}_{n\ge 0}$ 
and $\{\varrho^I_\beta(\Phi^m_\beta(s))\}_{m \ge 0}$ are both monotone
for each $s \in F^I_\beta$, $\beta \in B$.  \endpro

\proof The first sequence is monotone by Lemmas 8.1.1 and 8.2.1 and the second by
Lemma~8.2.3. \eop
\bigbreak
The following is the main result of this chapter (and, in fact, of the whole study).
Note that it does not require there to be any solutions to the equations.

\proclaim{Proposition~8.2.1} (1)
For each $s \in F^I_\beta$, $\beta \in B$, and each
$n \ge 0$ there exists $m \ge 0$ such that
$\,\sem{s}^{k_n}_\beta \,\sqle_\beta\,\varrho^I_\beta(\Phi^m_\beta(s))\,$. 
\medskip
(2)\enskip
$\,\varrho^I_\beta(\Phi^n_\beta(s))\,\sqle_\beta\,\sem{s}^{k_n}_\beta\,$ 
for all $s \in F^I_\beta$, $\beta \in B$, $n \ge 0$. 
\endpro

\proof (1)\enskip This is given in Section~8.4. 
\medskip
(2)\enskip This is given at the end of the section. \eop
\bigbreak
Let $s \in F^I_\beta$ with $\beta \in B$; then Lemma~8.2.4 implies that
$\lcurl \sem{s}_\beta^{k_n}\,:\, n \ge 0 \rcurl$ and
$\lcurl \varrho^I_\beta(\Phi^n_\beta(s))\,:\, n \ge 0 \rcurl$ 
are directed subsets of $D_\beta$, and Proposition~8.2.1 implies that
they are mutually cofinal.
\medbreak
One reason why the
sequence $\{\varrho^I_\beta(\Phi^m_\beta(s))\}_{m \ge 0}$ 
is important is the following:

\proclaim{Lemma~8.2.5} For each $\beta \in B$ 
\mdisp{F^\Phi_\beta
\ =\ \lcurl s \in F^I_\beta 
\,:\, \varrho^I_\beta(\Phi^m_\beta(s)) \in X_\beta 
\frm{for some} m \ge 0 \rcurl\ .}\endpro

\proof This follows since Proposition~3.3.3, Lemma~6.3.3~(3) and Lemma~6.3.5 imply 
that $\,\lcurl t \in F^I_\beta\,:\, \varrho^I_\beta(t) \in X_\beta \rcurl
\,=\, F_\beta\,$ for each $\beta \in B$. \eop
\bigbreak

\proclaim{Proposition~8.2.2} For each $\beta \in B$
\mdisp{F^\Phi_\beta
\ =\ \lcurl s \in F^I_\beta \,:\, \sem{s}^{k_m}_\beta \in X_\beta
 \frm{for some} m \ge 0 \rcurl\ .}
\endpro

\proof By Proposition~4.1.3~(1) each element of $X_\beta$ is a maximal element of the
poset $D_\beta$, and so the result follows from
Proposition~8.2.1 and Lemma~8.2.5. \eop
\bigbreak
Until further notice let us make the additional assumptions that
$D_S$ is a family of complete bottomed posets, that the mapping $U_\alpha$
is continuous and that the mapping $c \mapsto \sem{s}^c_\beta$ is continuous
for each $s \in F^I_\beta$, $\beta \in B$.
Of course, by Lemma~8.1.1 and Proposition~8.1.1 (applied to Case~3 
with its `correct' category) this includes Case~\tsym{3'}.

\proclaim{Lemma~8.2.6} The least upper bound $k_\infty$ of the directed set 
$\lcurl k_n\,:\, n \ge 0 \rcurl$ is a solution of the equations $\alpha$. 
\endpro

\proof By Proposition~4.2.9 $k_\infty$ is the least fixed-point of the mapping 
$U_\alpha$, and hence $k_\infty$ is also a solution of the equations. \eop
\bigbreak

\proclaim{Proposition~8.2.3} The \tsym{\Lambda}-homomorphism
$\Phi_B$ is \tsym{\alpha}-complete, i.e., $F^\Phi_B = F^\alpha_B$. \endpro

\proof By Propositions 7.2.1 and 7.3.2 $F^\Phi_B \subset F^\alpha_B$, and hence
it only remains to show that $F^\alpha_B \subset F^\Phi_B$.  
Let $s \in F^\alpha_\beta$, thus in particular by Lemma~8.2.6 
$\sem{s}^{k_\infty}_\beta = x \in X_\beta$.  But by Proposition~4.1.3~(2) the set 
$\lcurl u \in D_\beta \,:\, u \sqle_\beta x \rcurl$ is finite, and hence
$\sem{s}^{k_n}_\beta = x$ for some $n \ge 0$ (since the mapping
$c \mapsto \sem{s}^c_\beta$ is continuous). 
Proposition~8.2.2 therefore implies that
$s \in F^\Phi_\beta$.  \eop
\bigbreak

\proclaim{Proposition~8.2.4} For all $s \in F^I_\beta$, $\beta \in B$ 
\mdisp{\lub \,\lcurl \varrho^I_\beta(\Phi^m_\beta(s)) \,:\, m \ge 0 \rcurl
                 \ =\ \sem{s}^{k_\infty}_\beta\ .} \endpro

\proof By Proposition~8.2.1 and Lemma~8.2.4 it immediately follows that
\mdisp{\lub \,\lcurl \varrho^I_\beta(\Phi^m_\beta(s)) \,:\, m \ge 0 \rcurl
\ =\ \lub \,\lcurl \sem{s}^{k_n}_\beta \,:\, n \ge 0 \rcurl\ ,} 
and
$\,\lub \,\lcurl \sem{s}^{k_n}_\beta \,:\, n \ge 0 \rcurl \,=\, \sem{s}^{k_\infty}_\beta\,$,
since by assumption the mapping $c \mapsto \sem{s}^c_\beta$ is continuous. \eop
\bigbreak
For each $m \ge 0$ the element $\varrho^I_\beta(\Phi^m_\beta(s))$ of $D_\beta$ can 
be considered as the partial information obtained about the `value' of $s$ after $m$ 
applications of the mapping $\Phi_\beta$. Proposition~8.2.1 thus says that this 
partial information converges to the value of $s$ with respect to the
solution $k_\infty$ of the equations (which in Case~\tsym{3'} is the least solution).
\bigbreak

It remains to prove Proposition~8.2.1~(2), and hence
the additional assumptions that $D_S$ is a family of complete bottomed posets and 
the mapping $U_\alpha$ is continuous will now be dropped.
The main part of the proof is contained in the following lemma:

\proclaim{Lemma~8.2.7} Let $c \in \ass I D $, put $c' = U_\alpha(c)$
and suppose that $c\  \ass I {\sqle} \ c'$. Then
\mdisp{\sem{\Phi_\beta(s)}^c_\beta \ \sqle_\beta \sem{s}^{c'}_\beta}
for all $s \in F^I_\beta$, $\beta \in B$.  \endpro

\proof This is more-or-less the same as the proof of Proposition~7.3.2. Let
$\ell_B$ be the family of mappings employed  in the proof of Proposition~7.3.1 and
put 

\mdisp{G_\beta\ =\ \lcurl s \in F^I_\beta \,:\,
  \sem{\Phi_\beta(s)}^c_\beta \sqle_\beta \sem{s}^{c'}_\beta  \rcurl} 

for each $\beta \in B$; the task is thus to show that $G_B = F^I_B$. First consider 
$s \in F^I_\beta$ 
with $\ell_\beta(s) = 1$. If $s = \fo^I_\kappa(\onept)$ for some 
$\kappa \in K$ of type $\varnothing \to \beta$ then $s \in G_\beta$ holds
trivially because $\Phi_\beta(s) = s$. On the other hand, if 
$s = \xi$ for some $\xi \in I$ of type $\beta$ then
$\Phi_\beta(s) = s_\xi$, and thus again $s \in G_\beta$, since
\mdisp{\sem{\Phi_\beta(s)}^c_\beta
 \ = \ \sem{s_\xi}^c_\beta \ =\ c'(\xi)
 \ =\ \sem{\xi}^{c'}_\beta \ =\ \sem{s}^{c'}_\beta\ .}
\medskip
Now let $k > 1$ and suppose for each $\theta \in B$ it is known that
$t \in G_{\theta}$ for all $t \in F^I_{\theta}$ with $\ell_{\theta}(t) < k$.
Consider $s \in F^I_\beta$ with $\ell_\beta(s) = k$; there are four cases:
\medskip
$1.$ The element $s$ has the form $\fo^I_\kappa(a)$, where
$\kappa \in K$ is of type $L \to \beta$ with $L \ne \varnothing$ and
$a \in \assb L {F^I} $. 
Then $\,\sem{\Phi_\eta(a(\eta))}^c_\eta\,\sqle_\eta\,\sem{a(\eta)}^{c'}_\eta$
for each $\eta \in L$, since $\ell_\eta(a(\eta)) < k$, 
and hence 
$\,\assb L {\sem{\,\ass L {\Phi} (a)\,}^c} \ \ass L {\sqle} \  \assb L {\sem{a}^{c'}} \,$. 
Thus $s \in G_\beta$, since
\smallskip
\ldisp{\quad\sem{\Phi_\beta(s)}^c_\beta 
  \ =\ \sem{\,\fo^I_\kappa(\ass L {\Phi} (a))\,}^c_\beta
  \ =\ f_\kappa\bigl(\,\assb L {\sem{\,\ass L {\Phi} (a)\,}^c} \, \bigr)}
\vskip-\medskipamount
\rdisp{
  \ \sqle_\beta\ f_\kappa\bigl(\,\assb L {\sem{a}^{c'}} \,\bigr)
  \ =\ \sem{\,\fo^I_\kappa(a)\,}^{c'}_\beta
                               \ =\ \sem{s}^{c'}_\beta\ . \quad}
\medskip
$2.$ The element $s$ has the form $\fo^I_\xi(a)$ with 
$\xi \in I$ of type $\sigma = L \to \beta$ for some $L \ne \varnothing$ and
with $a \in \assb L {F^I} $. Put $\,d = \assb L {\sem{a}^c} \,$; then
by Lemma~7.3.2
\mdisp{\sem{\Phi_\beta(s)}^c_\beta 
\ =\ \sem{\,s_\xi[a/L]\,}^c_\beta\ =\ \sem{s_\xi}^{\,c\oplus d}_\beta\ .}
On the other hand, 
putting $\,b = \assb L {\sem{a}^{c'}} \,$, it follows from the definition of $U_\alpha$
that
\mdisp{
\sem{s}^{c'}_\beta
\ =\ \sem{\,\fo^I_{\xi}(a)\,}^{c'}_\beta
\ =\  f_{\sigma,L}(c'(\xi) \triangleleft b )
\ =\  f_{\sigma,L}(U_\alpha(c)\,(\xi) \triangleleft b )
\ =\ \sem{s_\xi}^{\,c\oplus b}_\beta\ . 
}
But $c \oplus d\ \ass {I\cup L} {\sqle} \ c \oplus b$, since by Lemma~8.1.1
$d(\eta) = \sem{a(\eta)}^c_\eta \sqle_\eta \sem{a(\eta)}^{c'}_\eta = b(\eta)$ 
for each $\eta \in L$. Thus, again making use of Lemma~8.1.1,
$\sem{s_\xi}^{\,c\oplus d}_\beta \sqle_\beta \sem{s_\xi}^{\,c\oplus b}_\beta$,
and hence $\sem{\Phi_\beta(s)}^c_\beta \sqle_\beta \sem{s}^{c'}_\beta$,
i.e., $s \in G_\beta$.
\medskip
$3.$ The element $s$ has the form $\fo^I_\zeta(s' \diamond a)$, where 
$\zeta = \ftt{case}^\theta_\beta$ with $\beta \in B$ and $\theta \in \Bf$ and 
where $s' \in F^I_\theta$ and $a \in \assb {K_{\theta,\beta}} {F^I} $. 
Put $L = \theta\cdot K_{\theta,\beta}$ and $\sigma = L \to \beta$.
\medskip
Assume first that $s' \notin \bterm{F}^I_\theta$. Then
$\sem{\Phi_\theta(s')}^c_\theta \sqle_\theta \sem{s'}^{c'}_\theta$, because 
$\ell_\theta(s') < k$, and thus by Lemma~8.1.1 and Proposition~5.4.1
(and since $f_{\sigma,L}$ is monotone)
\smallskip
\ldisp{\quad \sem{\Phi_\beta(s)}^c_\beta
    \  =\  \sem{\,\fo^I_\zeta(\Phi_\theta(s') \diamond a ) \,}^c_\beta}
\vskip-\medskipamount
\ldisp{\qquad\quad
\ =\ f_{\sigma,L}\bigl(\, p(\zeta) \triangleleft (\sem{\Phi_\theta(s')}^c_\theta
           \diamond \assb {K_{\theta,\beta}} {\sem{a}^c} )\,\bigr)
}
\vskip-\medskipamount
\rdisp{
\ \sqle_\beta\ f_{\sigma,L}\bigl(\, p(\zeta)\triangleleft ( \sem{s'}^{c'}_\theta
           \diamond \assb {K_{\theta,\beta}} {\sem{a}^{c'}} )\,\bigr)
   \ =\  \sem{\,\fo^I_\zeta(s' \diamond a )\,}^{c'}_\beta  
                 \ =\ \sem{s}^{c'}_\beta \quad}
\smallskip
which implies that $s \in G_\beta$.
\medskip
Assume next that $s' \in \bterm{F}^I_\theta$ and so
$\sem{s'}^c_\theta \ne \bot_\theta$. Moreover, since
$\bterm{F}^I_\theta \subset \kterm{F}^I_\theta$,
$s'$ has the form $\fo^I_\kappa(a')$ with $\kappa \in K$ of type $J \to \theta$
for some $J \in {\cal F}_B$ and with $a' \in \assb {J} {F^I} $.
Therefore by (C), Proposition~5.4.1 and Lemma~8.1.1 
(and again since $f_{\sigma,L}$ is monotone) it follows that
$s \in G_\beta$, since
\smallskip
\ldisp{
\ \sem{\Phi_\beta(s)}^c_\beta
\ =\  \sem{\,\fo^I_{J\to \beta,J}(a(\kappa) \triangleleft a')\,}^c_\beta
 \ =\    f_{J\to\beta,J}
  \bigl(\, \sem{a(\kappa)}^c_{J\to \beta}\,\triangleleft
      \,\assb {J} {\sem{a'}^c} \,\bigr) 
}
\vskip-\medskipamount
\ldisp{\quad\qquad
 \ =\  f_{J\to\beta,J}
  \bigl(\, \assb {K_{\theta,\beta}} {\sem{a}^c} \,(\kappa)\,
      \triangleleft \,\assb {J} {\sem{a'}^c} \,\bigr) 
}
\vskip-\medskipamount
\ldisp{\quad\qquad\qquad
\ =\ \fss{Case}^\theta_\beta
    \bigl(
   f_\kappa(\,\assb {J} {\sem{a'}^c} \,),
         \assb {K_{\theta,\beta}} {\sem{a}^c} \bigr)    
\ =\ \fss{Case}^\theta_\beta
    \bigl(\sem{s'}^c_\theta,
         \assb {K_{\theta,\beta}} {\sem{a}^c} \bigr)   
}
\vskip-\medskipamount

\ldisp{\qquad\quad\qquad\qquad
\ =\ f_{\sigma,L}
   \bigl(\,p(\zeta)\,\triangleleft\,
(\sem{s'}^c_\theta \diamond  \assb {K_{\theta,\beta}} {\sem{a}^c} ) \,\bigr)    
}
\vskip-\medskipamount
\rdisp{
\ \sqle_\beta \ f_{\sigma,L}
   \bigl(\,p(\zeta)\,\triangleleft\,
(\sem{s'}^{c'}_\theta \diamond  \assb {K_{\theta,\beta}} {\sem{a}^{c'}} ) \,\bigr)    
   \  =\ \sem{\,\fo^I_\zeta(s' \diamond a)\,}^{c'}_\beta  
\ =\ \sem{s}^{c'}_\beta\ .\ } 

\medskip
$4.$ The element $s$ has the form $\fo^I_\zeta(s_1,s_2)$ with 
$\zeta \in P$ the name of an integer operator (so $\zeta$ is of type 
$\ftt{int}_2 \to \beta$ with $\beta$ either $\ftt{int}$ or $\ftt{bool}$) and
$s_1,\,s_2 \in F^I_{\fstt{int}}$. 
\medskip
Assume first that both $s_1$ and $s_2$ are ground terms; then $s \in G_\beta$, since
\smallskip
\ldisp{\quad\sem{\Phi_\beta(s)}^c_\beta
\ =\ \sem{\,\fss{Eval}_\zeta(s_1,s_2)\,}^c_\beta
\ =\ \sem{\,\fss{Eval}_\zeta(s_1,s_2)\,}_\beta
}
\vskip-\medskipamount
\rdisp{
  \ =\ f_{\fstt{int}_2\to\beta,\fstt{int}_2}
\bigl(\,p(\zeta)\triangleleft (\sem{s_1}_{\fstt{int}},\sem{s_2}_{\fstt{int}})\,\bigr)
\ =\ \sem{\,\fo^I_\zeta(s_1,s_2)\,}^{c'}_\beta \ =\ \sem{s}^{c'}_\beta\ .\quad}
\medskip
Finally, assume that not both of $s_1$ and $s_2$ are ground terms.
Then, since $\ell_{\fstt{int}}(s_j) < k$ for each $j = 1,\,2 $, it follows that
$\sem{\Phi_{\fstt{int}}(s_j)}^c_{\fstt{int}} 
                                      \sqle_\beta \sem{s_j}^{c'}_{\fstt{int}}$, 
and thus 
\smallskip
\ldisp{\quad \sem{\Phi_\beta(s)}^c_\beta
   \ =\  \sem{\,\fo^I_\zeta(\Phi_{\fstt{int}}(s_1),
            \Phi_{\fstt{int}}(s_2))\,}^c_\beta}
\vskip-\medskipamount
\ldisp{\qquad\qquad
\ =\ f_{\fstt{int}_2\to\beta,\fstt{int}_2}
(\sem{\Phi_{\fstt{int}}(s_1)}^c_{\fstt{int}},\sem{\Phi_{\fstt{int}}(s_2)}^c_{\fstt{int}})
}
\vskip-\medskipamount
\rdisp{
\ \sqle_\beta\ f_{\fstt{int}_2\to\beta,\fstt{int}_2}
(\sem{s_1}^{c'}_{\fstt{int}},\sem{s_2}^{c'}_{\fstt{int}})
\  =\  \sem{\,\fo^I_\zeta(s_1,s_2) \,}^{c'}_\beta \ =\ \sem{s}^{c'}_\beta\ , \quad}
\smallskip
since $f_{\fstt{int}_2\to\beta,\fstt{int}_2}$ is monotone, and so once again
$s \in G_\beta$.
\medskip\smallskip
Therefore by induction on $k$ it follows that $G_B = F^I_B$. \eop

\bigbreak
{\it Proof of Proposition~8.2.1~(2)\enskip} 
This holds when $n = 0$ by Lemma~8.2.2. Thus let $n \ge 0$ and suppose that
$\,\varrho^I_\beta(\Phi^n_\beta(s))\,\sqle_\beta\,\sem{s}^{k_n}_\beta\,$ 
for all $s \in F^I_\beta$, $\beta \in B$. Then by Lemma~8.2.7 
\mdisp{
\varrho^I_\beta(\Phi^{n+1}_\beta(s))\ =\ \varrho^I_\beta(\Phi^n_\beta(\Phi_\beta(s)))
\ \sqle_\beta\ \sem{\Phi_\beta(s)}^{k_n}_\beta
\ \sqle_\beta\ \sem{s}^{k_{n+1}}_\beta}
for all $s \in F^I_\beta$, $\beta \in B$. The result therefore follows by induction.
\eop

\vfill\eject
\hfline{221}{8.3 \ LOGICAL RELATIONS AND FUNCTIONALLY FREE ALGEBRAS}

\bigskip
\sectionhead {8.3} {Logical relations and functionally free algebras}
\bigskip\medskip
The preparations for the proof of the first part of Proposition~8.2.1 involve
looking at how a concept called the method of logical relations fits in 
with the notion of a functionally free algebra. The results in this direction
considered here (in particular Proposition~8.3.3) will then be applied in the next 
section to prove Proposition~8.2.1~(1).

\medskip
For the whole of the section let $(D_S,f_N)$ and $(Y_S,q_N)$ just be fixed, but arbitrary,
functional \tsym{\Sigma}-algebras. 
\medskip 
If $D$ and $Y$ are sets then a relation on $D \times Y$
is just a subset of  $D \times Y$. However, infix notation will always be employed, and so 
$u \preceq y$ will be used to mean that the pair $(u,y)$ belongs to the relation $\preceq$.
\medskip

Let $\preceq_B$ be a family of relations, with $\preceq_\beta$ a relation on 
the set $D_\beta \times Y_\beta$ for each $\beta \in B$.
This family is considered to be fixed for the whole of the section.
For each \tsym{B}-typed set  $L$ there is then a relation
$\ass L {\preceq} $ on the set $\ass L D \times \ass L Y $ given as usual by
$c\ \ass L {\preceq} \ a$ if and only if $c(\eta) \preceq_\eta a(\eta)$ for each
$\eta \in L$. The single assumption that will be made about the family $\preceq_B$ is 
the following:
\medskip\smallskip
{\parindent=22pt
\item{(*)} If $\kappa \in K$ is of type $L \to \beta$ and 
$c \in \ass L D $, $a \in \ass L Y $ are such that
$c\ \ass L {\preceq} \ a$ then 
\mdisp{f_\kappa(c)\ \preceq_\beta\ q_\kappa(a)\ .}
In particular, this means that $\,f_\kappa(\onept) \preceq_\beta q_\kappa(\onept)\,$
whenever $\kappa$ is of type $\varnothing \to \beta$.
\medskip\smallskip
}
The method of logical relations involves
extending the family $\preceq_B$ in an appropriate way to a family 
$\preceq_S$, with $\preceq_\sigma$ a relation on the set $D_\sigma \times Y_\sigma$
for each $\sigma \in S$. Note that if $\preceq_S$ is any such family then for each
\tsym{S}-typed set  $L$ there is the corresponding relation
$\ass L {\preceq} $ defined on the set $\ass L D \times \ass L Y $
(in the same way as above).

\proclaim{Proposition~8.3.1} The family of relations $\preceq_B$ can be
extended uniquely to a family of relations $\preceq_S$, with $\preceq_\sigma$
a relation on $D_\sigma \times Y_\sigma$ for each $\sigma \in S$,
in such a way that if $\sigma = L \to \beta$ is a 
functional type and $u \in D_\sigma$, $s \in Y_\sigma$, then
$u \preceq_\sigma s$ if and only if
\mdisp{f_{\sigma,L}(u \triangleleft c)
\ \preceq_\beta\ q_{\sigma,L}(s \triangleleft a)} 
for all $c \in \ass L D $ and all $a \in \ass L Y $ such that
$c\ \ass L {\preceq} \ a$. \endpro

\proof Let $|\cdot| : S \to \Nat$ be the mapping given in Lemma~5.1.1
with $|\beta| = 0$ for each $\beta \in B$ and such that 
$\,|L \to \beta| \, =\, 1 + \sum_{\eta \in L} |\langle \eta \rangle|\,$
for each functional type $L \to \beta$.
The family $\preceq_S$ will be defined by induction on $|\sigma|$:
If $|\sigma| = 0$ then $\sigma \in B$ and so 
$\preceq_\sigma$ is a member of the original family $\preceq_B$.  
Thus suppose that $k > 0$ and that $\preceq_\tau$ has already been defined 
for all $\tau \in S$ with $|\tau| \le k$. Consider $\sigma \in S$ with $|\sigma| = k$.
Then $\sigma = L \to \beta$ is a functional type and
$|\langle \eta \rangle| < k$ for each $\eta \in L$. This means that
$\preceq_\sigma$ can be defined
by stipulating that $u \preceq_\sigma s$ if and only if

\mdisp{f_{\sigma,L}(u \triangleleft c)
\ \preceq_\beta\ q_{\sigma,L}(s \triangleleft a)} 
for all $c \in \ass L D $ and all $a \in \ass L Y $ such that
$c(\eta) \preceq_\eta a(\eta)$ for each $\eta \in L$.  
The family $\preceq_S$ obtained in this way then has the required
property by construction. The uniqueness also follows by induction on $|\sigma|$. \eop
\bigbreak

Now for each \tsym{S}-typed set $I$ fix 
a functionally \tsym{I}-free \tsym{\Sigma}-algebra $(Z^I_S,r^I_N)$. 
If $c \in \ass I D $ then $\pi^c_S$ will denote the unique 
\tsym{\Sigma}-homomorphism from  $(Z^I_S,r^I_N)$ to 
$(D_S,f_N)$ such that $\pi^c_\xi(\xi) = c(\xi)$ for each $\xi \in I$.
Moreover, if
$a \in \ass I Y $ then $\delta^a_S$ will denote the unique 
\tsym{\Sigma}-homomorphism from  $(Z^I_S,r^I_N)$ to 
$(Y_S,q_N)$ such that $\delta^a_\xi(\xi) = a(\xi)$ for each $\xi \in I$.
\medskip
If $I$ is an \tsym{S}-typed set then 
$c \in \ass I D $ will be called 
\indexddef{compatible with $a \in \ass I Y $} 
{compatible assignment}{}{assignment}{compatible}
if 
\mdisp{\pi^c_\sigma(s)\, \preceq_\sigma\, \delta^a_\sigma(s) } 
for all $s \in Z^I_\sigma$, $\sigma \in S$.

\proclaim{Proposition~8.3.2} Let $I$ be an \tsym{S}-typed set and let
$c \in \ass I D $, $a \in \ass I Y $. 
Then $c$ is compatible with $a$ if and only if $\,c\ \ass I {\preceq} \ a\,$. \endpro

\proof Since $c(\xi) = \pi^c_\xi(\xi)$ and $a(\xi) = \delta^c_\xi(\xi)$ for each 
$\xi \in I$, it follows that
$c\ \ass I {\preceq} \ a$ whenever $c$ is compatible with $a$.
Conversely, suppose that $c\ \ass I {\preceq} \ a$.
Then, since the assignments $c$ and $a$ are fixed throughout the proof, it is convenient to 
just write $\pi_S$ instead of $\pi^c_S$ and $\delta_S$ instead of $\delta^a_S$.
Define a family $G_S \subset Z^{I}_S$ by putting
\mdisp{G_\sigma 
\ =\ \lcurl s \in Z^{I}_\sigma\,:\,
     \pi_\sigma(s) \preceq_\sigma \delta_\sigma(s) \rcurl }
for each $\sigma \in S$. 
It will be shown that the family $G_S$ satisfies the hypotheses of Proposition~6.1.5.
In particular, condition (2) in Proposition~6.1.5 is satisfied, since
by assumption $\xi \in G_\xi$ for each $\xi \in I$.
\medskip
Note that if $L$ is an \tsym{S}-typed set and $b \in \ass L G $ then 
$b(\eta) \in G_\eta$ for each $\eta \in L$
and hence $\,\pi_\eta(b(\eta))_\eta  \preceq_\eta\, \delta_\eta(b(\eta))\,$, and this means
that $\,\ass L {\pi} (b)\ \ass L {\preceq} \ \ass L {\delta} (b)\,$.
\medskip
It will next be shown that
the family $G_B$ is invariant in the \tsym{\Lambda}-algebra $(Z_B,r_K)$, and thus that
condition (1) in Proposition~6.1.5 is satisfied.
Consider $\kappa \in K$ of type $L \to \beta$ and let
$b \in \ass L G $. Then $\,\ass L {\pi} (b)\ \ass L {\preceq} \ \ass L {\delta} (b)\,$
and therefore by assumption (*) 

\mdisp{ \pi_\beta(r^{I}_\kappa(b)) 
\ =\ f_\kappa(\ass L {\pi} (b))
\ \preceq_\beta\ q_\kappa(\ass L {\delta} (b))
\ =\ \delta_\beta(r^{I}_\kappa(b))\ ,}

which shows that $\,r^{I}_\kappa(b) \in G_\beta\,$. In particular,
if $\kappa$ is of type $\varnothing \to \beta$ then
$r^{I}_\kappa(\onept) \in G_\beta$.
\medskip
Now consider  $\xi \in I$ of type $\sigma = L \to \beta$, let
 $J$ be a non-empty subset of $L$ and $b \in \ass J G $. 
To show that condition (3) in Proposition~6.1.5 is satisfied it must be verified that
$\,r^{I}_{\sigma,J}(\xi \triangleleft b) \in G_{\sigma_{L\setminus J}}\,$, and
the case with $J$ a proper subset of $L$ will be dealt with first.
Put $\tau = \sigma_{L\setminus J} = L{\setminus} J\to \beta$;
by the definition of $\preceq_\tau$ it must be shown that

\mdisp{
f_{\tau,L\setminus J}
    \bigl(\,\pi_\tau(r^{I}_{\sigma,J}(\xi \triangleleft b))\triangleleft c'\,\bigr)
\ \preceq_\beta\ q_{\tau,L\setminus J}\bigl(\,\delta_\tau(
              r^{I}_{\sigma,J}(\xi \triangleleft b)) \,\triangleleft\, a'\,\bigr)}

whenever $c' \in \ass {L\setminus J} D $ and $a' \in \ass {L\setminus J} Y $ are
such that $c'\ \ass {L\setminus J} {\preceq} \ a'$. But (as noted above)
$\,\ass I {\pi} (b)\ \ass J {\preceq} \ \ass J {\delta} (b)\,$, and hence also
$\,\ass J {\pi} (b) \oplus c'\ \ass L {\preceq} \ \ass J {\delta} (b) \oplus a'\,$.
Thus, since $\,\pi_\xi(\xi) \preceq_\xi \delta_\xi(\xi)\,$
and since $(D_S,f_N)$ and $(Y_S,q_N)$ are both functional \tsym{\Sigma}-algebras,  
it follows that
\smallskip
\ldisp{\quad
f_{\tau,L\setminus J}
   \bigl(\,\pi_\tau(r^{I}_{\sigma,J}
          (\xi \triangleleft b))\triangleleft c'\,\bigr)
\ =\ f_{\tau,L\setminus J}
   \bigl(\, f_{\sigma,J}(\pi_\sigma(\xi) \triangleleft
          \ass J {\pi} (b)   )  \triangleleft c' \,\bigr)
}
\vskip-\smallskipamount
\ldisp{\ \qquad
\ =\ f_{\sigma,L} \bigl(\,\pi_\sigma(\xi) \triangleleft
         ( \ass J {\pi} (b)  \oplus c' )\,\bigr)
\ \preceq_\beta\ q_{\sigma,L}\bigl(\,
              \delta_\sigma(\xi) \,\triangleleft\, (\ass J {\delta} (b) \oplus a')\,\bigr)
} 
\vskip-\smallskipamount
\rdisp{
\ =\ q_{\tau,L\setminus J}\bigl(\,
                q_{\sigma,J}
                    (\delta_\sigma(\xi) \triangleleft \ass J {\delta} (b) )
\,\triangleleft\, a'\,\bigr)
\ =\ q_{\tau,L\setminus J}\bigl(\,\delta_\tau(
                r^{I}_{\sigma,J}
                    (\xi \triangleleft b)) \,\triangleleft\, a'\,\bigr)
\quad} 
\smallskip
and this shows that
$\,r^{I}_{\sigma,J}(\xi \triangleleft b) \in G_\tau\,$. Finally there is the case $J = L$: 
Here it must be shown that
$\,\pi_\beta(r^{I}_{\sigma,L}(\xi \triangleleft b))
\, \preceq_\beta\, \delta_\beta(r^{I}_{\sigma,L}(\xi \triangleleft b))\,$
and this follows since
\smallskip
\ldisp{\quad
   \pi_\beta(r^{I}_{\sigma,L}(\xi \triangleleft b))
\ =\   f_{\sigma,L}(\pi_\sigma(\xi) \triangleleft
          \ass L {\pi} (b)   )  
}
\vskip-\smallskipamount
\rdisp{\ \preceq_\beta\ q_{\sigma,L}
                    (\delta_\sigma(\xi) \triangleleft \ass L {\delta} (b) )
\ =\ \delta_\beta(r^{I}_{\sigma,L}(\xi \triangleleft b))\ . \quad} 
\smallskip
Thus by Proposition~6.1.5 $G_S = Z^{I}_S$, i.e.,
$c$ is compatible with $a$. \eop
\bigbreak
Now fix an \tsym{S}-typed set $I$ and an assignment $i \in \ass I Y $.
(This is denoted by $i$ because in the application in the next section
$(Y_S,q_N)$ will contain $I$ and $i$ will be given by $i(\xi) = \xi$ for each
$\xi \in I$.) The proof of Proposition~8.1.1~(1) given there involves  showing that
certain assignments $k \in \ass I D $ are compatible with this particular assignment
$i$. Of course, Proposition~8.3.2 says
that $k$ is compatible with $i$ if and only if
$\,k\ \ass I {\preceq} \ i\,$.
However, the main technique needed for establishing compatibility turns out to be 
Proposition~8.3.3 below.
\medskip
Let $U$ be an \tsym{S}-typed set disjoint from $I$. 
An assignment $k \in \ass I D $ is then said to be
\indexddef{\tsym{U}-compatible with $i$} 
{compatible assignment}{}{assignment}{compatible}
if whenever $a \in \ass U Y $, $c \in \ass U D $ are such that $c\ \ass U {\preceq} \ a$ 
then
\mdisp{\pi_\sigma^{k \oplus c}(s)\, \preceq_\sigma\, \delta^{i\oplus a}_\sigma(s) } 

for all $s \in Z^{I\cup U}_\sigma$, $\sigma \in S$. In particular, this means that
$k$ is compatible if and only if it is \tsym{\varnothing}-compatible with $i$.
(Some care must be taken here: If the set $\ass U Y $ is empty then 
any assignment $k \in \ass I D $ is trivially \tsym{U}-compatible with $i$. Moreover,
$\ass U Y $ will be empty if and only if $Y_\eta = \varnothing$ for some
$\eta \in U$, and this is quite possible when, for example, $(Y_S,q_N)$ is some kind
of term algebra.)

\proclaim{Proposition~8.3.3} An assignment $k \in \ass I D $ is compatible with $i$
if and only if it is \tsym{U}-compatible with $i$ for all
\tsym{S}-typed sets $U$ disjoint from $I$.
\endpro

\proof If $k$ is \tsym{U}-compatible with $i$ for all
\tsym{S}-typed sets $U$ such that $U \cap I = \varnothing$ then in particular it is
\tsym{\varnothing}-compatible and thus compatible with $i$. 
\medskip
Suppose conversely that $k$ is compatible with $i$, and
let $U$ be an \tsym{S}-typed set disjoint from $I$.
Consider
$a \in \ass U Y $, $c \in \ass U D $ with $c\ \ass U {\preceq} \ a$. Then
\mdisp{(k\oplus c)\,(\eta)\ =\ c(\eta)\ \preceq_\eta\ a(\eta)\ =\ (i\oplus a)\,(\eta)}
for each $\eta \in U$ and, since $k$ is compatible with $i$, 
\mdisp{(k\oplus c)\,(\xi)\ =\ k(\xi)\ \preceq_\xi\ i(\xi)\ =\ (i\oplus a)\,(\xi)}
for each $\xi \in I$, i.e.,
$\,(k\oplus c)\,(\xi)\,\preceq_\xi\, (i\oplus a)\,(\xi)\,$ for all $\xi \in I\cup U$.
Proposition~8.3.2 (applied to the \tsym{S}-typed set $I\cup U$ and the assignments
$k \oplus c$ and $i \oplus a$) thus implies that
$\,\pi_\sigma^{k \oplus c}(s)\, \preceq_\sigma\, \delta^{i\oplus a}_\sigma(s)\,$ for all
$s \in Z^{I\cup U}_\sigma$, $\sigma \in S$, and therefore 
that $k$ is \tsym{U}-compatible with $i$. \eop 
\bigbreak
\vfill\eject

\hfline{224}{8.4 \ AN APPLICATION OF THE METHOD OF LOGICAL RELATIONS}

\bigskip
\sectionhead {8.4} {An application of the method of logical relations}
\bigskip\medskip
The set-up of Section~8.2 will be used here.
Thus $\fss{C}$ is the bottomed concrete category with finite products occurring
in Case~2, $({\bf D}_S,f_N)$ is a \tsym{\fss{C}}-based \tsym{\Sigma}-algebra 
such that $(D_B,f_K)$ is a monotone regular bottomed extension
of $(X_B,p_K)$, and such that the \tsym{\Sigma}-algebra $(D_S,f_N)$ has
the properties (A), (B), (C) and (D). 
In addition it is assumed that the family $\sqle_B$ is an ordering associated 
with $(D_B,f_K)$,
(where $\sqle_\sigma$ denotes the partial order order on $D_\sigma$ for each
$\sigma \in S$). Moreover, we are given
a monotone mapping $U_\alpha : \ass I D \to \ass I D $ satisfying the two conditions:
\medskip\smallskip
{\parindent=22pt
\item{(1)} If $\xi \in I$
is of type $\sigma = L \to \beta$ with $L \ne \varnothing$ then  
\mdisp{f_{\sigma,L}\bigl(\, U_\alpha(c)\,(\xi) \,\triangleleft\, b \,\bigr)
                          \ =\ \sem{s_\xi}_\beta^{\,c \oplus b}}  
\item{} for all $c \in \ass I D $, $b \in \ass L D $.
\medskip
\item{(2)} If $\xi \in I$ is of type $\beta \in B$ then
$\,U_\alpha(c)\,(\xi) \,=\, \sem{s_\xi}^{c}_\beta\,$ for each $c \in \ass I D $. 
\bigskip
}
This section is taken  up with the proof of Proposition~8.2.1~(1), 
which states that for each $s \in F^I_\beta$, $\beta \in B$, and each $n \in \Nat$
there exists $m \in \Nat$ such that

\mdisp{\sem{s}^{k_n}_\beta \ \sqle_\beta\ \varrho^I_\beta(\Phi^m_\beta(s))\ . }

The proof relies heavily on Proposition~8.3.3
and thus uses the method of
logical relations introduced in Section~8.3.

\medskip
The family of relations $\preceq_B$ which provide the key to the proof 
of Proposition~8.2.1~(1) is specified in the next result, which lists the important 
properties of this family:

\proclaim{Proposition~8.4.1} For each $\beta \in B$ let
$\preceq_\beta$ be the relation on $D_\beta \times F^I_\beta$ defined
by stipulating that $u \preceq_\beta s$ if and only if
$\,u \sqle_\beta \varrho^I_\beta(\Phi^m_\beta(s))\,$ for some $m \ge 0$. 
Then the family $\preceq_B$ has the following properties:
\medskip\smallskip
{\parindent=30pt
\item{(P1)} $\bot_\beta \preceq_\beta s\,$ for all $s \in F^I_\beta$, $\beta \in B$.
\medskip
\item{(P2)} Let $\kappa \in K$ be of type $L \to \beta$, let 
$c \in \ass L D $  with $f_\kappa(c) \ne \bot_\beta$
and $a \in \assb L {F^I} $. Then
$\,f_\kappa(c)\, \preceq_\beta\, \fo^I_\kappa(a)\,$
if and only if $c\ \ass L {\preceq} \ a$, i.e., if and only if
$c(\eta) \preceq_\eta a(\eta)$ for each $\eta \in L$.
In particular, if $\kappa$ is of type $\varnothing \to \beta$ then
$f_\kappa(\onept) \preceq_\beta \fo^I_\kappa(\onept)$.
\medskip
\item{(P3)} Let $\zeta \in P$ 
be the name of an integer operator (so $\zeta$ is of type 
$\ftt{int}_2 \to \beta$ with $\beta$ either $\ftt{int}$ or $\ftt{bool}$)
and for $j = 1,\,2$ let $n_j \in X_{\fstt{int}}$, 
$t_j \in F_{\fstt{int}}$ with $n_j \preceq_{\fstt{int}} t_j$. Then
$\,f_{\fstt{int}_2\to\beta,\fstt{int}_2}
\bigl(\,p(\zeta)\triangleleft (n_1,n_2)\,\bigr)
        \, \preceq_\beta\, \fss{Eval}_\zeta(t_1,t_2)\,$.
\medskip
\item{(P4)} If $u \in D_\beta$ and $s \in F^I_\beta$ then 
$\,u \preceq_\beta \Phi_\beta(s)\,$ if and only if $u \preceq_\beta s$.
\medskip
\item{(P5)} If $u \in D_\beta$ and $s \in F^I_\beta$ with
$\bot_\beta \ne u \preceq_\beta s$ then there exists
$m \ge 0$ such that $\Phi^m_\beta(s) \in \bterm{F}^I_\beta$. 
\medskip\smallskip
} \endpro

\proof (P1):\enskip This holds trivially.
\medskip
(P2):\enskip For each $m \ge 0$ put $\psi^{(m)}_B = \varrho^I_B \,\Phi^m_B$;
thus $\psi^{(m)}_B$ is a \tsym{\Lambda}-homomorphism from
$(F^I_B,\fo^I_K)$ to $(D_B,f_K)$. 
Let $\kappa \in K$ be of type $L \to \beta$, let 
$c \in \ass L D $  with $f_\kappa(c) \ne \bot_\beta$
and $a \in \assb L {F^I} $. Then 
$\,\varrho^I_\beta(\Phi^m_\beta(\fo^I_\kappa(a)))
\,=\, \psi^{(m)}_\beta(\fo^I_\kappa(a))
\, =\, f_\kappa(\assb L {\psi^{(m)}} (a) )\,$ and thus
$\,f_\kappa(c)\, \preceq_\beta\, \fo^I_\kappa(a)\,$ 
if and only if
$\,f_\kappa(c)\, \sqle_\beta\, f_\kappa(\assb L {\psi^{(m)}} (a) )\,$
for some $m \ge 0$.
But by the definition of an associated ordering this holds if and only if
$\,c\ \ass L {\sqle} \ \assb L {\psi^{(m)}} (a)\,$ for some $m \ge 0$, i.e., 
if and only if there exists $m \ge 0$ such that
$c(\eta) \preceq_\eta \varrho^I_\eta(\Phi^m_\eta(a(\eta)))$
for all $\eta \in L$.
In particular, this shows that $c\ \ass L {\preceq} \ a$ whenever
$\,f_\kappa(c)\, \preceq_\beta\, \fo^I_\kappa(a)\,$. Conversely, suppose that
$c\ \ass L {\preceq} \ a$; then for each $\eta \in L$ there exists $m_\eta \ge 0$ such that
$c(\eta) \preceq_\eta \varrho^I_\eta(\Phi^{m_\eta}_\eta(a(\eta)))$.
Put $m = \max \,\lcurl m_\eta\,: \eta \in L \rcurl$. 
Then by Lemma~8.2.3 it follows that 
$c(\eta) \preceq_\eta \varrho^I_\eta(\Phi^m_\eta(a(\eta)))$ for all $\eta \in L$, and
hence $\,f_\kappa(c)\, \preceq_\beta\, \fo^I_\kappa(a)\,$. 
\medskip
(P3):\enskip If $t \in F_{\fstt{int}}$ then $\Phi^m_{\fstt{int}}(t) = t$ for all $m \ge 0$ 
and $\varrho^I_{\fstt{int}}(t) = \sem{t}_{\fstt{int}}$. Thus if
$n \in X_{\fstt{int}}$ and $t \in F_{\fstt{int}}$ then $n \preceq_{\fstt{int}} t$
just means that $\sem{t}_{\fstt{int}} = n$.
Therefore (P3) holds more-or-less by the definition of $\fss{Eval}_\zeta$.
\medskip
(P4):\enskip This follows immediately from Lemma~8.2.3.
\medskip
(P5):\enskip If $u \in D_\beta$ and $s \in F^I_\beta$ are such that
$\bot_\beta \ne u \preceq_\beta s$ then 
$\varrho^I_\beta(\Phi^m_\beta(s)) \ne \bot_\beta$ for some $m \ge 0$. Hence
$\Phi^m_\beta(s) \in \bterm{F}^I_\beta$ for some $m \ge 0$,
since Lemma~6.3.5 implies that 
$\,\bterm{F}^I_\beta\, =\, \lcurl s \in F^I_\beta \,:\,
           \varrho^I_\beta(s) \ne \bot_\beta \rcurl\,$
for each $\beta \in B$. \eop
\bigbreak
It should be clear from the formulation of Proposition~8.4.1 that 
the results of Section~8.3 will be applied taking the functional \tsym{\Sigma}-algebra 
$(Y_S,q_N)$ there to be $(F^I_S,\fo^I_N)$.
Note then that condition (P2) is just the assumption (*) made in Section~8.3.
\medskip
Proposition~8.2.1~(1) is now an immediate corollary of the 
following result:

\proclaim{Proposition~8.4.2} For each $\beta \in B$ let $\preceq_\beta$
be a relation on the set $D_\beta \times F^I_\beta$, and suppose the family
$\preceq_B$ has the properties (P1), (P2), (P3), (P4) and (P5).
Then $\sem{s}^{k_n}_\beta \preceq_\beta s$ for all $n \ge 0$ and all $s \in F^I_\beta$,
$\beta \in B$. \endpro

\proof First note a couple of facts which follow from the properties
enjoyed by the family $\preceq_B$.

\proclaim{Lemma~8.4.2} Let $\kappa \in K$ be of type $L \to \beta$. 
\medskip
(1)\enskip If $c \in \ass L D $ and $a \in \assb L {F^I} $ are such that
$c\ \ass L {\preceq} \ a$ then $\,f_\kappa(c) \preceq_\beta \fo^I_\kappa(a)\,$.
\medskip
(2)\enskip Let $c \in \ass L D $ with $f_\kappa(c) \ne \bot_\beta$, let 
$s \in F^I_\beta$ with $f_\kappa(c) \preceq_\beta s$ and let
$m \ge 0$ be such that $\Phi^m_\beta(s) \in \bterm{F}^I_\beta$ (the existence of such an
$m$ being guaranteed by (P5)). Then
there exists a unique $a \in \assb L {F^I} $ such that
$\Phi^m_\beta(s) = \fo^I_\kappa(a)$, and then $c\ \ass L {\preceq} \ a$. \endpro

\proof (1)\enskip If $f_\kappa(c) = \bot_\beta$ then this follows
immediately from (P1), and if 
$f_\kappa(c) \neq \bot_\beta$ then it is just a part of (P2).
\medskip
(2)\enskip By Proposition~6.2.1 there exists a unique $a \in \assb L {F^I} $ such that
$\Phi^m_\beta(s) = \fo^I_\kappa(a)$,
since $\Phi^m_\beta(s) \in \bterm{F}^I_\beta \subset \kterm{F}^I_\beta$.
Then 
$\,f_\kappa(c)\, \preceq_\beta\, \Phi^m_\beta(s)\, =\, \fo^I_\kappa(a)\,$
(by (P4)), and hence (P2) implies that $c\ \ass L {\preceq} \ a$. \eop
\bigbreak

Now by Proposition~8.3.1 the family of relations $\preceq_B$ can be
extended uniquely to a family of relations $\preceq_S$, with $\preceq_\sigma$
a relation on $D_\sigma \times F^I_\sigma$ for each $\sigma \in S$,
in such a way that if $\sigma = L \to \beta \in S \setminus B$
and $u \in D_\sigma$, $s \in F^I_\sigma$, then
$u \preceq_\sigma s$ if and only if
\mdisp{f_{\sigma,L}(u \triangleleft c)
\ \preceq_\beta\ \fo^I_{\sigma,L}(s \triangleleft a)} 
for all $c \in \ass L D $ and all $a \in \assb L {F^I} $ such that
$c\ \ass L {\preceq} \ a$. 

\proclaim{Lemma~8.4.3} $\bot_\sigma \preceq_\sigma s\,$ for all
$s \in F^I_\sigma$, $\sigma \in S$. \endpro

\proof If $\sigma \in B$ then this is just (P1), so let
$\sigma = L \to \beta$ be a functional type and consider
$c \in \ass L D $, $a \in \assb L {F^I} $ with $c\ \ass L {\preceq} \ a$. Then by (P1)
and (A)
\mdisp{f_{\sigma,L}(\bot_\sigma \triangleleft c) 
\ =\ \bot_\beta
\ \preceq_\beta \ \fo^I_{\sigma,L}(s \triangleleft a)} 

from which $\bot_\sigma \preceq_\sigma s$ follows by the definition of $\preceq_\sigma$.
\eop

\bigbreak

Proposition~8.3.3 will be applied with the \tsym{S}-typed set $I$ 
occurring there being replaced by $I \cup P$.
If $J$ is  an \tsym{S}-typed set disjoint from $P$ then
$(F^J_S,\fo^J_N)$ will play the role of the
functionally free algebra $(Z^{J\cup P}_S,r^{J\cup P}_N)$
occurring in Section~8.3.
Moreover, if $b \in \assb {J\cup P} {F^J} $
then $\pi^b_S$ will again be used to denote the unique 
\tsym{\Sigma}-homomorphism from  $(F^J_S,\fo^J_N)$ to 
$(D_S,f_N)$ such that $\pi^b_\xi(\xi) = b(\xi)$ for each $\xi \in J\cup P$.
(Thus if $c \in \ass J D $ then in fact
$\sem{\cdot}^c_S = \pi^{c\oplus p}_S$.)
In the same way, if $a \in \assb {J\cup P} {F^I} $ then $\delta^a_S$ 
denotes the unique 
\tsym{\Sigma}-homomorphism from $(F^J_S,\fo^J_N)$ to 
$(F^I_S,\fo^I_N)$ such that $\delta^a_\xi(\xi) = a(\xi)$ for each $\xi \in J\cup P$.
\medskip
Consider the assignment $i \in \assb {I\cup P} {F^I} $ given by
$i(\xi) = \xi$ for each $\xi \in I\cup P$. Then clearly
$\delta^i_S : (F^I_S,\fo^I_N) \to (F^I_S,\fo^I_N)$ is just the identity
homomorphism, i.e., $\delta^i_\sigma(s) = s$ for all $s \in F^I_\sigma$, $\sigma \in S$.
The following lemma indicates how the notions introduced in Section~8.3 
could (and will) be used to prove Proposition~8.4.2:

\proclaim{Lemma~8.4.4} Let $k \in \ass I D $ and suppose $k \oplus p$ is compatible with 
the assignment $i$. Then $\,\sem{s}^{k}_\beta \preceq_\beta s\,$ for all 
$s \in F^I_\beta$, $\beta \in B$.
\endpro

\proof By definition, $k \oplus p$ being compatible with $i$ 
means that $\pi^{k\oplus p}_\sigma(s) \preceq_\sigma \delta^i_\sigma(s)$,
and thus $\sem{s}^{k}_\sigma \preceq_\sigma s$,
for all $s \in F^I_\sigma$, $\sigma \in S$.
In particular, $\sem{s}^{k}_\beta \preceq_\beta s$
for all $s \in F^I_\beta$, $\beta \in B$.
\eop
\bigbreak

\proclaim{Lemma~8.4.5} $\,p(\zeta) \preceq_\zeta \zeta\,$ for each
$\zeta \in P$. \endpro

\proof Suppose first that $\zeta \in P$ is the name of an integer operator
(so $\zeta$ is of type $\ftt{int}_2 \to \beta$ with $\beta$ either
$\ftt{int}$ or $\ftt{bool}$).
Then for any $n_j \in D_{\fstt{int}}$, $s_j \in F^I_{\fstt{int}}$
with $n_j \preceq_{\fstt{int}} s_j$ for $j = 1,\,2$ it must be shown that

\mdisp{f_{\fstt{int}_2\to\beta,\fstt{int}_2}
                                  \bigl(\,p(\zeta)\triangleleft (n_1,n_2)\,\bigr)
\ \preceq_\beta\ \fo^I_\zeta(s_1,s_2)\ .}

By (P1) this holds immediately if $n_1 = \bot_{\fstt{int}}$ or
$n_2 = \bot_{\fstt{int}}$, since then (D) implies 
$\,f_{\fstt{int}_2\to\beta,\fstt{int}_2}
                    \bigl(\,p(\zeta)\triangleleft (n_1,n_2)\,\bigr)\, =\, \bot_\beta\,$.
It can thus be assumed in what follows that 
$n_1,\, n_2 \in D_{\fstt{int}} \setminus \{\bot_{\fstt{int}}\} = X_{\fstt{int}}$.
Now $\bot_{\fstt{int}} \ne n_j \preceq_{\fstt{int}} s_j$, thus by (P5)
there exists $m_j \ge 0$ with 
$\Phi_{\fstt{int}}^{m_j}(s_j) \in \bterm{F}^I_{\fstt{int}} = F_{\fstt{int}}$.
Let 
\mdisp{m\ =\ \min \, \lcurl k \ge 0 \,:\,
 \Phi_{\fstt{int}}^k(s_j) \in F_{\fstt{int}} \,\frm{for}\, j = 1,\,2 \rcurl\ ;}
then 
$\,\Phi^{m+1}_\beta(\fo^I_\zeta(s_1,s_2))
\,=\, \fss{Eval}_\zeta(t_1,t_2)\,$
with $t_j = \Phi^m_{\fstt{int}}(s_j)$ for $j = 1,\,2$.
But by (P4) $n_j \preceq_{\fstt{int}} t_j$ for $j = 1,\,2$ and therefore by (P3)
it follows that
\mdisp{f_{\fstt{int}_2\to\beta,\fstt{int}_2}
                 \bigl(\,p(\zeta)\triangleleft (n_1,n_2)\,\bigr)
\ \preceq_\beta\ \fss{Eval}_\zeta(t_1,t_2)
\ =\ \Phi^{m+1}_\beta(\fo^I_\zeta(s_1,s_2))\ .}
Hence, again using (P4),
$\,f_{\fstt{int}_2\to\beta,\fstt{int}}
                                  \bigl(\,p(\zeta)\triangleleft (n_1,n_2)\,\bigr)
\, \preceq_\beta\, \fo^I_\zeta(s_1,s_2)\,$.
\medskip
Suppose next that $\zeta = \ftt{case}^\theta_\beta$ for some
$\beta \in B$, $\theta \in \Bf$. 
Consider assignments
$c \in \ass {K_{\theta,\beta}} D $, $a \in \assb {K_{\theta,\beta}} {F^I} $ and
elements $u \in D_\theta$, $s \in F^I_\theta$ with
$\,u \diamond c \ \ass {\theta\cdot K_{\theta,\beta}} {\preceq} \ s \diamond a\,$,
i.e., with $u \preceq_\theta s$ and
$\,c\  \ass {K_{\theta,\beta}} {\preceq} \ a$. 
Put $L = \theta\cdot K_{\theta,\beta}$.
Then, since $p(\zeta) = \fss{case}^\theta_\beta$,
it must be shown that
$\,f_{L\to\beta,L}\bigl(
   \fss{case}^\theta_\beta \triangleleft (u \diamond c)\bigr)
       \, \preceq_\beta\, \fo^I_\zeta(s \diamond a)\,$.
If $u = \bot_\theta$ then by (P1) and (C)

\mdisp{f_{L\to\beta,L}\bigl(
   \fss{case}^\theta_\beta \triangleleft (u \diamond c)\bigr)\ =\ \bot_\beta
       \ \preceq_\beta\ \fo^I_\zeta(s \diamond a)}

so in what follows it can be assumed that $u \ne \bot_\theta$.
Then $u$ has a unique representation of the form $f_\kappa(b)$ with
$\kappa \in K$ of type $J \to \theta$ for some $J \in {\cal F}_B$ and
with $b \in \ass {J} D $. Now $\bot_\theta \ne u \preceq_\theta s$, thus by (P5)
there exists $m \ge 0$ with
$\Phi_\theta^m(s) \in \bterm{F}^I_\theta$, and by taking the least such $m$ 
it can be assumed that $\Phi_\theta^i(s) \notin \bterm{F}^I_\theta$ whenever
$0 \le i < m$. Moreover, by Lemma~8.4.2~(2)
$\Phi_\theta^m(s)$ is of the form
$\fo^I_\kappa(t)$ with $t \in \assb {J} {F^I} $ and 
$b\ \ass {J} {\preceq} \ t$, and it  therefore follows that
\smallskip
\ldisp{\quad \Phi_\beta^{m+1}\bigl(\,\fo^I_\zeta(s \diamond a )\,\bigr)
\ =\ \Phi_\beta\bigl(\,\fo^I_\zeta(\Phi^m_\theta(s) \diamond a )\,\bigr)}
\vskip-\medskipamount
\rdisp{
\ =\ \Phi_\beta\bigl(\,\fo^I_\zeta(\fo^I_\kappa(t) \diamond a )\,\bigr)
  \ =\ \fo^I_{J\to\beta,J}( a(\kappa) \triangleleft t)\ .\quad}
\smallskip
But $b\ \ass {J} {\preceq} \ t$ and by assumption
$\,c(\kappa) \preceq_{J\to\beta} a(\kappa)\,$, and hence by (C)
\smallskip
\ldisp{\quad f_{L\to\beta,L}
  \bigl(\,\fss{case}^\theta_\beta \triangleleft (u \diamond c) \bigr)
\ =\  \fss{Case}^\theta_\beta (u \diamond c)
}
\vskip-\medskipamount
\rdisp{
\ =\ f_{J\to\beta,J}( c(\kappa) \triangleleft b)
  \ \preceq_\beta \ \fo^I_{J\to\beta,J}( a(\kappa) \triangleleft t)
\ =\ \Phi_\beta^{m+1}\bigl(\,\fo^I_\zeta(s \diamond a )\,\bigr)\ .\quad}
\smallskip
Thus by (P4) 
$\,f_{L\to\beta,L}\bigl(
   \fss{case}^\theta_\beta \triangleleft (u \diamond c)\bigr)
       \, \preceq_\beta\, \fo^I_\zeta(s \diamond a)\,$. \eop
\bigbreak

\proclaim{Lemma~8.4.6} Let $k \in \ass I D $ and suppose 
$k \oplus p$ is compatible with the assignment $i$. 
Then $\,U_\alpha(k)\,(\xi)\,\preceq_\xi \,\xi\,$
for each $\xi \in I$. \endpro

\proof First note that if $L$ is an \tsym{S}-typed set and $a \in \assb L {F^I} $
then , with the notation from Section~7.3,
$\,s[a/L] \,=\, \delta^{i\oplus a}_\sigma(s)\,$ for each $s \in F^{I\cup L}_\sigma$,
$\sigma \in S$. 
\medskip 
Put $k' = U_\alpha(k)$. Let $\xi \in I$ be of type $\sigma = L \to \beta$ and 
assume first that $L \ne \varnothing$. Consider $c \in \ass L D $, $a \in \assb L {F^I} $ 
with $c\ \ass L {\preceq} \ a$. Now Proposition~8.3.3 implies that 
$k \oplus p$ is \tsym{L}-compatible with the assignment $i$, and therefore
\smallskip
\ldisp{
\quad 
f_{\sigma,L}(k'(\xi)\triangleleft c)
\ =\ f_{\sigma,L}(U_\xi(k)\triangleleft c)
}

\vskip-\medskipamount
\mdisp{
\ =\ \sem{s_\xi}^{\,k\oplus c}_\beta
\ =\ \pi^{k\oplus p\oplus c}_\beta(s_\xi)
\ \preceq_\beta\ \delta^{i\oplus a}_\beta(s_\xi)
\ =\ s_\xi[a/L] 
}
\vskip-\medskipamount
\rdisp{
 \ =\ \Phi_\beta(\fo^I_\xi(a))
\ =\ \Phi_\beta\bigl(\,\fo^I_{\sigma,L}(\xi \triangleleft a)\,\bigr)
\ .\quad }
\smallskip
Thus by (P4) 
$\,f_{\sigma,L}(k'(\xi)\triangleleft c)
\,\preceq_\beta \,\fo^I_{\sigma,L}(\xi \triangleleft a)\,$
also holds, which by the definition
of $\preceq_\sigma$ implies that $k'(\xi) \preceq_\xi \xi$.
The case with $L = \varnothing$ is somewhat easier, since here 
(without making use of Proposition~8.3.3) it follows that
\mdisp{
k'(\xi)\ =\ U_\xi(k)\ =\ \sem{s_\xi}^k_\beta
\ =\ \pi^{k\oplus p}_\beta(s_\xi)
\ \preceq_\beta\ \delta^i_\beta(s_\xi)
\ =\ s_\xi  \ =\ \Phi_\beta(\xi)\ ,}
and so again by (P4) $k'(\xi) \preceq_\xi \xi$. \eop
\bigbreak

\proclaim{Lemma~8.4.7} Let $k \in \ass I D $ and suppose 
$k \oplus p$ is compatible with the assignment $i$. 
Then $U_\alpha(k) \oplus p$ is also compatible with $i$. \endpro

\proof Again put $k' = U_\alpha(k)$. By Lemma~8.4.5 
$\,p(\zeta) \,\preceq_\zeta \,\zeta\, =\, i(\zeta)\,$ for each
$\zeta \in P$ and by Lemma~8.4.5 
$\,k'(\xi) \,\preceq_\xi\, \xi\, =\, i(\xi)\,$ for each $\xi \in I$.
But this just means that
$\,k'\oplus p \ \ass {I\cup P} {\preceq} \ i\,$, and therefore by Proposition~8.3.2
$k' \oplus p$ is compatible with $i$. 
\eop

\proclaim{Lemma~8.4.8} Let $\bot^{\!I} \in \ass I D $ be 
the bottom assignment  given by $\bot^{\!I}(\xi) = \bot_\xi$ 
for each $\xi \in I$. Then
$\bot^{\!I} \oplus p$ is compatible with $i$. \endpro

\proof By Lemma~8.4.3 
$\,\bot^{\!I}(\xi) \, =\,  \bot_\xi \, \preceq_\xi\, \xi\,$ 
for each $\xi \in I$, and thus by
Lemma~8.4.5 and Proposition~8.3.2 it follows, exactly as in the proof of Lemma~8.4.6,
that $\bot^{\!I} \oplus p$ is compatible with $i$. \eop
\bigbreak
Now by definition $k_0 = \bot^{\!I}$
and $k_{n+1} = U_\alpha(k_n)$ for each $n \in \Nat$.
Therefore by Lemmas 8.4.8 and 8.4.7 and induction it follows that
$k_n \oplus p$ is compatible with the assignment $i$ for each $n \in \Nat$.
This, together with Lemma~8.4.4, completes the proof of Proposition~8.4.2. \eop
\bigskip\bigskip\bigskip\bigskip

\sectionhead {8.5} {Notes}
\bigskip\medskip
The proof of Proposition~8.2.1 is adapted from the proofs of Lemmas 11.14
and 11.23 in Winskel (1993). For more information regarding the method of 
logical relations, 
on which these proofs are based, see Mitchell's survey article
Mitchell (1990).

\vfill\eject

\hfline{229}{BIBLIOGRAPHY}

\def\ref#1#2#3{\filbreak \hangindent=15pt \hangafter=1 #1, #2, #3 \filbreak}

{\fourteenbf Bibliography}
\bigskip\bigskip
\ref {Banaschewski, B.\ and E.\ Nelson (1982)}
     {Completions of partially ordered sets}
     {{\it SIAM J.\ Computing\/}, 11, 521--528.}
\bigskip
\ref {Barr, M.\ and C.\ Wells (1990)}
     {Category theory for computing science}
     {Prentice Hall, New York.} 
\bigskip
\ref {Bauer, F.L., R.\ Berghammer {\it et al\/} (1985)}
     {The Munich Project CIP, Vol.\ 1: The Wide Spektrum Language CIP-L} 
     {{\it LNCS}, Vol.\ 183, Springer-Verlag.}
\bigskip
\ref {Bird, R.\ and Wadler, P. (1988)} 
     {Introduction to functional programming}
     {Prentice-Hall, Hemel Hempstead.} 
\bigskip
\ref {Birkhoff, G. (1935)}
     {On the structure of abstract algebras}
     {{\it Proc.\ Camb.\ Phil.\ Soc.}, 31, 433--454.} 
\bigskip
\ref {Birkhoff, G. (1967)}
     {Lattice Theory}
     {\ Third (New) Edition,\ {\it American Mathematical Society Colloquium
     Publications}, Vol.\ XXV.}
\bigskip
\ref {Birkhoff, G.\ and J.D.\ Lipson (1970)}
     {Heterogeneous algebras}
     {{\it J.\ Combinatorial Th.}, 8, 115--133.} 
\bigskip
\ref {Bourbaki, N. (1970)} 
     {Alg\`ebre} 
     {Chapitre 1, Hermann, Paris.}
\bigskip
\ref {Burstall, R.M.\ and P.J.\ Landin (1969)}
     {Programs and their proofs}
     {in: B.\ Meltzer and D.\ Michie, {\it Machine Intelligence, Vol.\ 4}, 
     17--43, Edinburgh Univ.\ Press.} 
\bigskip
\ref {Cohn, P.M. (1965)}
     {Universal Algebra}
     {Harper and Row, New York.} 
\bigskip
\ref {Courcelle, B.\ and M.\ Nivat (1976)}
     {Algebraic families of interpretations}
     {in:\ {\it Proc.\ Annual Symposium on Foundations of Computer Science},
     137--146, Houston.}
\bigskip
\ref {Field, A.J.\ and P.G.\ Harrison (1988)}
     {Functional programming}
     {Addison-Wesley, Wokingham, Berks.}
\bigskip
\ref {Glaser, H., Henkin, C.\ and D.\ Till (1984)}
     {\ Principles of functional programming}
     {\ Prentice-Hall, Englewood Cliffs, N.J.} 
\bigskip
\ref {Goguen, J.A., J.W.\ Thatcher, E.G.\ Wagner and J.B.\ Wright (1977)}
     {Initial algebra semantics and continuous algebras}
     {{\it Journal ACM}, Vol.\ 24, 68--95.} 

\bigskip
\ref {Goguen, J.A., J.W.\ Thatcher and  E.G.\ Wagner (1978)}
     {An initial algebra approach to the specification correctness,
      and implementation of abstract data types}
     {in: R.T.~Yeh, {\it Current Trends in Programming Methodology,
      Vol.\ 4, Data Structuring}, 80--149, Prentice-Hall,
      Englewood Cliffs, N.J.} 
\bigskip

\ref {Gr\"atzer, G. (1968)}
     {Universal Algebra}
     {van Nostrand, Princeton, N.J.} 
\bigskip
\ref {Henderson, P. (1980)}
     {Functional programming}
     {Prentice-Hall, Englewood Cliffs, N.J.}
\bigskip
\ref {Higgins, P.J. (1963)}
     {Algebras with a scheme of operators}
     {{\it Mathematische Nachrichten}, 27, 115--132.} 
\bigskip
\ref {Holyer, I. (1993)}
     {Functional programming with Miranda}
     {Pitman, London.}
\bigskip
\ref {Huet, G.\ and D.C.\ Oppen (1980)}
     {Equations and rewrite rules: a survey} 
     {in: R.V.\ Book {\it Formal Language Theory: Perspectives and Open 
     Problems}, Academic Press.}
\bigskip
\ref {Kleene, S.C. (1952)}
     {Introduction to Metamathematics}
     {North-Holland, Amsterdam.}
\bigskip
\ref {{\L}ukasiewicz, J. (1925)}
     {?} {?}
\bigskip
\ref {Mac Lane, S.\ and G.\ Birkhoff (1967)}
     {Algebra}
     {The MacMillan Company, London.} 
\bigskip
\ref {MacLennan, B.J. (1990)}
     {Functional programming}
     {Addison-Wesley, Reading, Mass.}
\bigskip
\ref {Maibaum, T.S.E. (1972)}
     {The characterization of the derivation trees of context-free sets 
     of terms as regular sets}
     {{\it Proc.\ 13th Ann.\ IEEE Symp.\ on Switching and Automata Theory}, 
     224--230.} 
\bigskip
\ref {Markowsky, G. (1977)}
     {Categories of chain-complete posets}
     {{\it Theoret.\ Computer Science\/}, 4, 125--135.}
\bigskip
\ref {Markowsky, G.\ and B.\ Rosen  (1976)}
     {Bases for chain-complete posets}
     {{\it IBM J.\ Research Develop.\/}, 20, 138--147.}
\bigskip

\ref {Mitchell, J.C. (1990)}
     {Type systems for programming languages} 
     {Chapter 8 of 
     {\it Handbook of Theoretical Computer Science}, Elsevier, Amsterdam.}
 \bigskip
\ref {Morris, F.L. (1973)}
     {Advice on structuring compilers and proving them correct}
     {{\it Proc.\ Symp.\ on Principles of Programming Languages}, Boston,
      144--152.} 
\bigskip
\ref {Morris, J.H. (1968)}
     {Lambda-calculus models of programming languages}
     {Proj.\ MAC, Rep.\ MAC-TR-57, MIT, Cambridge, Mass.}
\bigskip

\ref {Paulson, L.C. (1991)} 
     {ML for the working mathematician}
     {Cambridge University Press.} 
\bigskip

\ref {Peyton Jones, S.L. (1987)} 
     {The Implementation of Functional Programming Languages}
     {Prentice-Hall, Hemel Hempstead.} 
\bigskip
\ref {Peyton Jones, S.L.\ and D.R.\ Lester (1991)} 
     {Implementing functional languages}
     {Prentice Hall, Hemel Hempstead.} 
\bigskip

\ref {Reade, C. (1989)} 
     {Elements of functional programming}
     {Addison-Wesley, Reading, Mass.}
\bigskip
\ref {Rosen, B.K. (1973)} 
     {Tree-manipulating systems and Church-Rosser theorems}
     {{\it Journal of the ACM\/}, 20, 160--187.}
\bigskip
\ref {Scott, D.S. (1976)}
     {Data types as lattices}
     {{\it SIAM J.\ Computing\/}, 5, 522--587.}
 \bigskip

\ref {Scott, D.S.\ and C.\ Gunter (1990)}
     {\ Semantic domains}
     {Chapter 12 of 
     {\it Handbook of Theoretical Computer Science}, Elsevier, Amsterdam.}
 \bigskip

\ref {Turner, D.A. (1985)}
     {Miranda: A non-strict functional language with polymorphic types}
     {in: {\it Proceedings IFIP International Conference on Functional
     Programming Languages and Computer Architecture},
     LNCS, Vol.\ 201, Springer-Verlag.} 
\bigskip
\ref {Whitehead, A.N. (1898)}
     {A treatise on universal algebra}
     {Cambridge University Press. Reprinted in 1960 by Hafner, New York.} 
\bigskip
\ref {Wikstr\"om, \AA. (1987)} 
     {Functional programming using standard ML}
     {Prentice-Hall, Hemel Hempstead.} 
\bigskip
\ref {Winskel, G. (1993)} 
     {The Formal Semantics of Programming Languages}
     {The MIT Press, Cambridge, Mass.}
\bigskip

\ref {Wirsing, M. (1990)}
     {Algebraic specification} 
     {Chapter 13 of 
     {\it Handbook of Theoretical Computer Science}, Elsevier, Amsterdam.}
\bigskip
\ref {Wright, J.B., E.G.\ Wagner and J.W.\ Thatcher (1978)}
     {A uniform approach to inductive posets and inductive closure}
     {{\it Theoret.\ Computer Science\/}, 7, 57--77.}
\bigskip

\vfill\eject

\def\indexitemt#1#2#3#4{\smallskip\hbox to 7.5truecm{#1\hss #3}}
\def\indexitem#1#2#3#4{\par\hbox to 7.5truecm{\quad#2\hss #3}}

\def\indexpage#1#2{\hbox to 16truecm{\vtop{#1}\hss\vtop{#2}}}

\hfline{232}{INDEX}

{\fourteenbf Index}
\bigskip\bigskip

\indexpage
{
\indexitemt{abstractor} {} {171} {7.1}
\indexitemt{algebra} {} {12, 25} {1.2}
\indexitem{algebra} {adapted to typed set} {150} {6.1}
\indexitem{algebra} {associated} {37, 43, 150, 152} {6.1}
\indexitem{algebra} {bottomed} {58} {3.2}
\indexitem{algebra} {category-based} {126} {5.2}
\indexitem{algebra} {containing a typed set} {37} {2.3}
\indexitem{algebra} {disjoint from a typed set} {46} {2.5}
\indexitem{algebra} {extension of} {46} {2.5}
\indexitem{algebra} {free} {37} {2.3}
\indexitem{algebra} {functionally free} {145, 147} {6.0}
\indexitem{algebra} {functional} {145, 147} {6.0}
\indexitem{algebra} {ground term} {56} {3.1}
\indexitem{algebra} {heterogenous} {54} {2.7}
\indexitem{algebra} {initial} {12, 33} {1.2}
\indexitem{algebra} {initial completion of} {104} {4.4}
\indexitem{algebra} {minimal} {30, 38} {2.2}
\indexitem{algebra} {multi-sorted} {54} {2.7}
\indexitem{algebra} {reachable} {54} {2.7}
\indexitem{algebra} {regular} {33, 38} {2.3}
\indexitem{algebra} {term} {40, 44, 45} {2.4}
\indexitem{algebra} {term generated} {54} {2.7}
\indexitem{algebra} {tree} {51} {2.6}
\indexitem{algebra} {unambiguous} {33, 38} {2.3}
\indexitem{algebras} {isomorphic} {28} {2.2}
\indexitemt{application operator} {} {90, 93, 117} {5.1}
\indexitem{application operator} {total} {128} {5.2}
\indexitemt{assignment} {} {21} {2.1}
\indexitem{assignment} {compatible} {222, 223} {8.3}
\indexitemt{associated algebra} {} {37, 43, 150, 152} {6.1}
\indexitemt{associated ordering} {} {83} {4.1}
\indexitem{associated ordering} {monotone} {87} {4.1}
\indexitemt{associated subalgebra} {} {29} {2.2}
\bigskip
\indexitemt{base element} {} {76} {3.5}
\indexitemt{basic term algebra} {} {156} {6.2}
\indexitemt{bottom element} {} {57} {3.2}
\indexitem{bottom element} {of a poset} {83} {4.1}
}
{
\indexitemt{bottomed algebra} {} {58} {3.2}
\indexitem{bottomed algebra} {cored} {70} {3.4}
\indexitem{bottomed algebra} {fully regular} {65} {3.3}
\indexitem{bottomed algebra} {initial} {60} {3.2}
\indexitem{bottomed algebra} {minimal} {60} {3.2}
\indexitem{bottomed algebra} {monotone} {75} {3.5}
\indexitem{bottomed algebra} {regular} {64} {3.3}
\indexitem{bottomed algebra} {strongly monotone} {75} {3.5}
\indexitem{bottomed algebra} {structurally monotone} {72} {3.4}
\indexitem{bottomed algebra} {unambiguous} {61} {3.2}
\indexitem{bottomed algebras} {isomorphic} {59} {3.2}
\indexitemt{bottomed category} {} {122} {5.1}
\indexitemt{bottomed extensions} {conjugate} {} {3.2}
\indexitem{bottomed extensions} {conjugate} {59} {3.2}
\indexitemt{bottomed extension} {} {57, 59} {3.2}
\indexitem{bottomed extension} {cored} {70} {3.4}
\indexitem{bottomed extension} {flat} {57} {3.2}
\indexitem{bottomed extension} {initial} {62} {3.2}
\indexitem{bottomed extension} {minimal} {62} {3.2}
\indexitem{bottomed extension} {monotone} {75} {3.5}
\indexitem{bottomed extension} {regular} {14, 64} {1.2}
\indexitemt{bottomed family} {} {59} {3.2}
\indexitemt{bottomed ground term algebra} {} {66} {3.3}
\indexitemt{bottomed homomorphism} {} {59} {3.2}
\indexitemt{bottomed isomorphism} {} {59} {3.2}
\indexitemt{bottomed poset} {} {83} {4.1}
\indexitemt{bottomed set} {} {57} {3.2}
\indexitemt{bottomed subalgebra} {minimal} {} {3.2}
\indexitem{bottomed subalgebra} {minimal} {60} {3.2}
\indexitemt{bottom} {} {57} {3.2}
\indexitemt{built-in function} {} {132} {5.3}
\bigskip
\indexitemt{canonical finite products} {} {115} {5.1}
\indexitemt{cartesian closed category} {} {116} {5.1}
\indexitemt{cartesian product} {} {21} {2.1}
\indexitemt{case operator} {} {132} {5.3}
\indexitemt{category-based algebra} {} {126} {5.2}
\indexitem{category-based algebra} {functional} {130} {5.2}
}
\vfill\eject
\indexpage
{
\indexitemt{category} {} {} {5.1}
\indexitem{category} {cartesian closed} {116} {5.1}
\indexitem{category} {concrete} {113} {5.1}
\indexitemt{chain in a poset} {} {101} {4.3}
\indexitem{chain in a poset} {finite} {101} {4.3}
\indexitemt{codomain} {} {20} {2.1}
\indexitemt{cofinal directed set} {} {98} {4.3}
\indexitemt{compatible assignment} {} {222, 223} {8.3}
\indexitemt{compatible family of relations} {} {76} {3.5}
\indexitemt{complete homomorphism} {} {170, 182} {7.0}
\indexitemt{complete poset} {} {90} {4.2}
\indexitemt{completion of a poset} {} {96} {4.3}
\indexitem{completion of a poset} {ideal} {97, 102} {4.3}
\indexitem{completion of a poset} {initial} {96} {4.3}
\indexitemt{concrete category} {} {113} {5.1}
\indexitemt{conjugate bottomed extensions} {} {59} {3.2}
\indexitemt{conjugate extension of a poset} {} {96} {4.3}
\indexitemt{consistent homomorphism} {} {170, 182} {7.0}
\indexitemt{constructor name} {} {55} {3.1}
\indexitemt{continuous extension property} {} {97} {4.3}
\indexitemt{continuous mapping} {} {90} {4.2}
\indexitemt{core of bottomed algebra} {} {65} {3.3}
\indexitem{core type} {monotone} {72} {3.4}
\indexitemt{core type} {} {70} {3.4}
\indexitemt{cored bottomed algebra} {} {70} {3.4}
\indexitemt{cored bottomed extension} {} {70} {3.4}
\indexitemt{currying operator} {} {90, 94} {4.2}
\bigskip
\indexitemt{denotational semantics} {} {4} {1.1}
\indexitemt{derived functional algebra} {} {130} {5.2}
\indexitemt{directed set} {} {90} {4.2}
\indexitem{directed set} {cofinal} {98} {4.3}
\indexitemt{directed sets} {} {} {4.3}
\indexitem{directed sets} {mutually cofinal} {98} {4.3}
\indexitemt{domain} {} {20} {2.1}
\indexitemt{domain theory} {} {108} {4.4}
}
{
\indexitemt{embedding} {} {89} {4.2}
\indexitemt{enumerated signature} {} {25} {2.2}
\indexitemt{enumeration} {} {} {2.5}
\indexitem{enumeration} {extension of} {48} {2.5}
\indexitemt{equation} {} {172} {7.1}
\indexitemt{equations} {system of} {172} {7.1}
\indexitemt{exponential object} {} {117} {5.1}
\indexitemt{extension} {} {} {2.5}
\indexitem{extension} {of a signature} {46} {2.5}
\indexitem{extension} {of a term algebra specifier} {47} {2.5}
\indexitem{extension} {of an algebra} {46} {2.5}
\indexitem{extension} {of an enumeration} {48} {2.5}
\indexitem{extension} {strict} {132} {5.3}
\indexitemt{extension homomorphism} {} {59} {3.2}
\indexitemt{extension of a poset} {} {96} {4.3}
\indexitemt{extensionality property} {} {102} {4.3}
\bigskip
\indexitemt{family of relations} {} {} {3.5}
\indexitem{family of relations} {compatible} {76} {3.5}
\indexitem{family of relations} {full compatible} {76} {3.5}
\indexitemt{family} {} {20} {2.1}
\indexitem{family} {bottomed} {59} {3.2}
\indexitem{family} {contained in a family} {20} {2.1}
\indexitem{family} {invariant} {28} {2.2}
\indexitemt{fat bottomed homomorphism} {} {75} {3.5}
\indexitemt{first order functional type} {} {125} {5.2}
\indexitemt{first refinement of a weak ordering} {} {86} {4.1}
\indexitemt{flat extension} {} {57} {3.2}
\indexitemt{free algebra} {} {37} {2.3}
\indexitemt{full compatible family of relations} {} {76} {3.5}
\indexitemt{fully regular bottomed algebra} {} {65} {3.3}
\indexitemt{function hierarchy} {full continuous} {} {5.4}
\indexitem{function hierarchy} {full continuous} {144} {5.4}
\indexitem{function hierarchy} {full set-theoretic} {144} {5.4}
\indexitemt{functional algebra} {} {145, 147} {6.0}
\indexitemt{functional application operator} {} {117} {5.1}
\indexitemt{functional category-based algebra} {} {130} {5.2}
}
\vfill\eject
\indexpage
{
\indexitemt{functional signature} {} {130} {5.2}
\indexitemt{functional type} {} {125} {5.2}
\indexitem{functional type} {first order} {125} {5.2}
\indexitem{functional type} {higher order} {125} {5.2}
\indexitemt{functionally free algebra} {} {145, 147} {6.0}
\indexitemt{function} {} {20} {2.1}
\indexitem{function} {built-in} {132} {5.3}
\indexitem{function} {primitive} {132} {5.3}
\bigskip
\indexitemt{global typed set} {} {171} {7.1}
\indexitemt{ground signature} {} {55} {3.1}
\indexitemt{ground term algebra} {} {56} {3.1}
\indexitem{ground term algebra} {bottomed} {66} {3.3}
\indexitemt{ground type} {} {55} {3.1}
\bigskip
\indexitemt{heterogenous algebra} {} {54} {2.7}
\indexitemt{higher order functional type} {} {125} {5.2}
\indexitemt{homomorphism} {} {25, 123} {5.2}
\indexitem{homomorphism} {bottomed} {59} {3.2}
\indexitem{homomorphism} {complete} {170, 182} {7.0}
\indexitem{homomorphism} {consistent} {170, 182} {7.0}
\indexitem{homomorphism} {defined by equations} {187, 189} {7.3}
\indexitem{homomorphism} {extension} {59} {3.2}
\indexitem{homomorphism} {fat bottomed} {75} {3.5}
\bigskip
\indexitemt{ideal completion of a poset} {} {97, 102} {4.3}
\indexitemt{ideal of a poset} {102} {} {4.3}
\indexitem{ideal of a poset} {directed} {102} {4.3}
\indexitem{ideal of a poset} {principal} {102} {4.3}
\indexitemt{image} {} {20} {2.1}
\indexitemt{induction} {} {} {7.4}
\indexitem{induction} {structural} {207} {7.4}
\indexitemt{initial algebra} {} {12, 33} {1.2}
\indexitemt{initial bottomed algebra} {} {60} {3.2}
\indexitemt{initial bottomed extension} {} {62} {3.2}
\indexitemt{initial completion of a poset} {} {96, 104} {4.3}
\indexitemt{initial natural number triple} {} {18} {2.0}
\indexitemt{initial set of functional types} {} {123} {5.2}
\indexitemt{integer operator} {} {132} {5.3}
\indexitemt{invariant family} {} {28} {2.2}
}
{
\indexitemt{inverse of a morphism} {} {114} {5.1}
\indexitemt{isomorphic algebras} {} {28} {2.2}
\indexitemt{isomorphic bottomed algebras} {} {59} {3.2}
\indexitemt{isomorphic posets} {} {89} {4.2}
\indexitemt{isomorphism} {} {28} {2.2}
\indexitem{isomorphism} {bottomed} {59} {3.2}
\indexitem{isomorphism} {in a category} {114} {5.1}
\bigskip
\indexitemt{kernel of a homomorphism} {} {75} {3.5}
\indexitemt{Kleene sequence} {} {215} {8.2}
\bigskip
\indexitemt{labelled tree} {} {51} {2.6}
\indexitemt{least upper bound} {} {90} {4.2}
\bigskip
\indexitemt{magma} {} {54} {2.7}
\indexitemt{mapping} {} {20} {2.1}
\indexitem{mapping} {continuous} {90} {4.2}
\indexitem{mapping} {monotone} {87, 89} {4.1}
\indexitem{mapping} {strict} {57} {3.2}
\indexitem{mapping} {typed} {21} {2.1}
\indexitemt{method of logical relations} {} {212} {8.1}
\indexitemt{minimal algebra} {} {30, 38} {2.2}
\indexitemt{minimal bottomed algebra} {} {60} {3.2}
\indexitemt{minimal bottomed extension} {} {62} {3.2}
\indexitemt{minimal bottomed subalgebra} {} {60} {3.2}
\indexitemt{minimal subalgebra} {} {30} {2.2}
\indexitemt{monotone associated ordering} {} {87} {4.1}
\indexitemt{monotone bottomed algebra} {} {75} {3.5}
\indexitemt{monotone bottomed extension} {} {75} {3.5}
\indexitemt{monotone core type} {} {72} {3.4}
\indexitemt{monotone mapping} {} {87, 89} {4.1}
\indexitemt{morphisms in a category} {} {113} {5.1}
\indexitemt{multi-sorted algebra} {} {54} {2.7}
\indexitemt{mutually cofinal directed sets} {} {98} {4.3}
\indexitemt{natural number triple} {} {18} {2.0}
\indexitem{natural number triple} {initial} {18} {2.0}
\bigskip
\indexitemt{objects of a category} {} {113} {5.1}
\indexitemt{operational semantics} {} {4} {1.1}
}
\vfill\eject
\indexpage
{
\indexitemt{operator} {} {} {5.1}
\indexitem{operator} {application} {90, 93, 117} {5.1}
\indexitem{operator} {case} {132} {5.3}
\indexitem{operator} {currying} {90, 94} {4.2}
\indexitem{operator} {integer} {132} {5.3}
\indexitem{operator} {uncurrying} {90, 94} {4.2}
\indexitemt{operator name} {} {25} {2.2}
\indexitemt{order isomorphism} {} {89} {4.2}
\indexitemt{ordering} {} {} {4.1}
\indexitem{ordering} {associated with extension} {83} {4.1}
\indexitem{ordering} {weak} {85} {4.1}
\bigskip
\indexitemt{partial order} {} {81} {4.0}
\indexitem{partial order} {product} {23} {2.1}
\indexitemt{partially ordered set} {} {81} {4.0}
\indexitemt{Peano triple} {} {18} {2.0}
\indexitemt{pervasive signature} {} {32} {2.2}
\indexitemt{poset} {} {81} {4.0}
\indexitem{poset} {bottomed} {83} {4.1}
\indexitem{poset} {complete} {90} {4.2}
\indexitem{poset} {completion of} {96} {4.3}
\indexitem{poset} {extension of} {96} {4.3}
\indexitemt{pre-support system} {} {186} {7.3}
\indexitemt{preinvariant set} {} {50} {2.6}
\indexitemt{primitive function} {} {132} {5.3}
\indexitemt{primitive type} {} {37} {2.3}
\indexitemt{product partial order} {} {23} {2.1}
\indexitemt{product type} {} {73} {3.4}
\indexitemt{product} {} {} {2.1}
\indexitem{product} {cartesian} {21} {2.1}
\bigskip
\indexitemt{reachable algebra} {} {54} {2.7}
\indexitemt{referential transparency} {} {201} {7.4}
\indexitemt{regular algebra} {} {33, 38} {2.3}
\indexitemt{regular bottomed algebra} {} {64} {3.3}
\indexitemt{regular bottomed extension} {} {14,, 64} {1.2}
\bigskip
\indexitemt{sensible signature} {} {54} {2.7}
\indexitemt{set of functional types} {} {} {5.2}
\indexitem{set of functional types} {initial} {123} {5.2}
}
{
\indexitemt{set} {} {} {3.2}
\indexitem{set} {bottomed} {57} {3.2}
\indexitem{set} {directed} {90} {4.2}
\indexitem{set} {of functional types} {123} {5.2}
\indexitem{set} {preinvariant} {50} {2.6}
\indexitem{set} {typed} {20} {2.1}
\indexitemt{signature} {} {12, 25} {1.2}
\indexitem{signature} {enumerated} {25} {2.2}
\indexitem{signature} {extension of} {46} {2.5}
\indexitem{signature} {functional} {130} {5.2}
\indexitem{signature} {ground} {55} {3.1}
\indexitem{signature} {pervasive} {32} {2.2}
\indexitem{signature} {sensible} {54} {2.7}
\indexitem{signature} {single-sorted} {32} {2.2}
\indexitemt{single-sorted signature} {} {32} {2.2}
\indexitemt{solution} {} {172} {7.1}
\indexitemt{standard term algebra} {} {40, 44} {2.4}
\indexitemt{strict extension} {} {132} {5.3}
\indexitemt{strict mapping} {} {57} {3.2}
\indexitemt{strongly monotone bottomed algebra} {} {75} {3.5}
\indexitemt{structural induction} {} {207} {7.4}
\indexitemt{structurally monotone bottomed algebra} {} {72} {3.4}
\indexitemt{subalgebra} {} {28} {2.2}
\indexitem{subalgebra} {associated with invariant family} {29} {2.2}
\indexitem{subalgebra} {minimal} {30} {2.2}
\indexitemt{subposet} {} {96} {4.3}
\indexitemt{support system} {} {146, 164} {6.0}
\indexitemt{supports case operators} {} {134} {5.3}
\indexitemt{supports integer operators} {} {132} {5.3}
\indexitemt{system of equations} {} {172} {7.1}
\bigskip
\indexitemt{term algebra specifier} {} {40, 44} {2.4}
\indexitem{term algebra specifier} {extension of} {47} {2.5}
\indexitemt{term algebra} {} {40, 44, 45} {2.4}
\indexitem{term algebra} {basic} {156} {6.2}
\indexitem{term algebra} {standard} {40, 44} {2.4}
\indexitemt{term generated algebra} {} {54} {2.7}
\indexitemt{terminal object of a category} {} {115} {5.1}
\indexitemt{total application operator} {} {128} {5.2}
\indexitemt{trace of bottomed algebra} {} {67} {3.3}
}
\vfill\eject
\indexpage
{
\indexitemt{tree algebra} {} {51} {2.6}
\indexitemt{tree} {} {} {2.6}
\indexitem{tree} {labelled} {51} {2.6}
\indexitemt{type} {} {25} {2.2}
\indexitem{type} {functional} {125} {5.2}
\indexitem{type} {ground} {55} {3.1}
\indexitem{type} {of element in a typed set} {20} {2.1}
\indexitem{type} {of operator name} {25} {2.2}
\indexitem{type} {of tree} {51} {2.6}
\indexitem{type} {primitive} {37} {2.3}
\indexitemt{type frame} {} {144} {5.4}
\indexitemt{typed mapping} {} {21} {2.1}
\indexitemt{typed set} {} {20} {2.1}
\indexitem{typed set} {global} {171} {7.1}
\indexitemt{typing} {} {20} {2.1}
}
{
\indexitemt{unambiguous algebra} {} {33, 38} {2.3}
\indexitemt{unambiguous bottomed algebra} {} {61} {3.2}
\indexitemt{uncurrying operator} {} {90, 94} {4.2}
\indexitemt{underlying set} {} {113} {5.1}
\indexitemt{upper bound} {} {90} {4.2}
\indexitem{upper bound} {least} {90} {4.2}
\bigskip
\indexitemt{value of a term} {} {200} {7.4}
\bigskip
\indexitemt{weak ordering} {} {85} {4.1}
}
\bye